**The General Traveling Salesman Problem, Version 5**

**Howard Kleiman**

# PREFACE

This work consists of a number of algorithms to solve the general traveling salesman problem. Although the algorithms - other than the heuristics that can be obtained from them - have comparatively slow running times, they have one great advantage. An algorithm whose nodes satisfy a Euclidean relationship are not always useful in practical situations. Thus, when considering routes between nodes, the costs may be more important than the distances. For instance, we may use different means of transport going between different transportation. In the simplest case, we may move an item using trucks, airplanes, trains, ships along even the *same* tour. In this case, costs are more important than distance. The same is true in examples in biotech and other sciences.

TABLE OF CONTENTS

Chapter 1

The Symmetric GTSP, n even

I. INTRODUCTION.

Let M be an $n \times n$ symmetric cost matrix where $n$ is even. We present an algorithm that extends the concept of admissible permutation and the modified Floyd-Warshall algorithm given in math. CO/0212305 that was used to obtain near-optimal or optimal solutions to the asymmetric traveling salesman problem. Using the modified F-W, we obtain a derangement, $D_{minimal}$, that we patch into a tour, $T_{UPPERBOUND}$. Once we have obtained $T_{UPPERBOUND}$, we consider it to be a circuit consisting of edges. From it, we extract two sets of alternate edges. Each set forms a perfect matching. We denote by $\sigma_{T_{UPPERBOUND}}$ the perfect matching whose edges have a smaller value. $\sigma_{T_{UPPERBOUND}}$ can also be expressed as a product of $\frac{n}{2}$ pair-wise disjoint 2-cycles. We then construc $\sigma_{T_{UPPERBOUND}}^{-1} M^-$. Applying the modified F-W algorithm to it, using sieve criteria, we can obtain paths (called either acceptable or 2-circuit paths) that can form circuits. We prove that every tour whose value is less than $|T_{UPPERBOUND}|$ and is constructed from circuits obtained by using F-W on $\sigma_{T_{UPPERBOUND}}^{-1} M^-$ can be obtained by patching acceptable and 2-circuit cycles. Let $S$ be a set of cycles that can be patched to form a tour. Formulae are derived that give the number of points contained in such a set of cycles as well as a bound for $|S|$. To be more precise, we use PHASES 1 and 2 given in arXiv.org/math. CO/0212305 with the following modification: Since a 2-cycle in a derangement is an edge, we must check to see if an arc extending a path in the algorithm leads to an arc symmetric to an arc in the current derangement. We disallow such paths. In PHASE 2, we use the modified Floyd-Warshall algorithm to obtain an approximation to a minimally-valued derangement. During this process, we *delete* all arcs that are symmetric to the current derangement, $D_i$, when we apply the modified F-W algorithm to M.

II. DEFINITIONS.

An *edge* is an undirected line segment connecting two points, say $a$ and $b$. A *circuit* is a set of edges and points such that precisely two edges are incident to each point. A *tour* in an $n \times n$ symmetric cost matrix M is a circuit containing precisely $n$ points and $n$ edges where each edge has a value or cost. The *value of a tour* is the sum of the values of its edges. An *arc* is an edge that has been given an orientation or direction. Arc $(b\ a)$ is *symmetric* to $(a\ b)$. A *cycle* is a directed circuit. The *value of a cycle* is the sum of the values of its arcs. A *permutation* is a rearrangement of some or all of a fixed set of points, say $V = \{1, 2, \ldots, n\}$. A *derangement* is a permutation that moves all of the points in $V$. A *perfect matching or PM* is a set of $\frac{n}{2}$ edges that are pair-wise disjoint. *Henceforth, it is equivalent to a set of pair-wise disjoint 2-cycles where each edge is equivalent to a pair of symmetric entries in M, i.e. to a derangement consisting of 2-cycles*. $\sigma_{ABSOLUTE}$ is a minimally valued derangement obtained from a minimally valued perfect matching of the symmetric cost matrix M. $\sigma^{-1}M$ is a permutation of the columns of M by the derangement $\sigma^{-1}$. $\sigma^{-1}M^-$ is the matrix obtained from $\sigma^{-1}M$ by subtracting the value in each diagonal entry $(a\ a)$ from all entries in row $a$. A *path* in $\sigma^{-1}M^-$ is a sequence of arcs of the form $[a_1\ a_2\ \ldots\ a_r]$ where $a_i \neq a_j$, $i = 1, 2, \ldots, r$; $j = 1, 2, \ldots, r$. A *cycle* in $\sigma^{-1}M^-$ is a path that becomes a cycle. If $C$ is a cycle in $\sigma^{-1}M^-$, then $\sigma C$ always yields a derangement. In the symmetric case, this isn't satisfactory since a 2-cycle of $\sigma C$ is an edge. Given a symmetric cost matrix M, our purpose is to obtain a *circuit* consisting of edges. We thus define acceptable and 2-circuit paths. An *acceptable path* is a path satisfying the following condition: It yields an *acceptable path of edges* of the form $[a_1\ \sigma(a_2)\ a_2\ \sigma(a_3)\ a_3\ \ldots\ \sigma(a_r)a_r]$ where no two points belong to the same 2-cycle of $\sigma$. Due to symmetry in M, every non-diagonal entry in M defines an edge having double the value of the entry. Thus, by construction, since both $[a_i\sigma(a_{i+1})]$ and $[\sigma(a_{i+1})a_{i+1}]$ lie in M, they may be thought of as edges if we wish to do so. An *unlinked 2-circuit path* is a path in which we allow precisely *one* pair of points to belong to the same 2-cycle of $\sigma$. In a *linked 2-circuit path*, we allow precisely *two* pairs of points to belong to the same 2-cycles provided that *the points of the 2-cycles interlace*. i.e., as we traverse the path, we don't have a point from a 2-cycle followed by the other point of the 2-cycle. An *acceptable cycle* is an acceptable path that forms a cycle. An *unlinked 2-circuit* *cycle* is an unlinked 2-circuit path that forms a cycle. A *linked 2-circuit cycle* is a linked 2-circuit path that forms a cycle. A *tour* is a circuit in M containing $n$ points and edges. Since we consider a perfect matching as a set of 2-cycles, we here consider a tour as a *directed* set of edges. At the same time, when applying our algorithm, we use the fact that it *is* constructed of edges. $T_{FWTSPOPT}$ denotes a minimally-valued tour that can be obtained by applying the modified F-W algorithm to M. An *optimal tour* in M which may require the use of paths not constructed by using minimal paths obtained from F-W is denoted by $T_{TSPOPT}$. Let $C$ be a cycle obtained from $\sigma^{-1}M^-$ where $\sigma$ is a product of $\frac{n}{2}$ pair-wise disjoint 2-cycles. Then $|C|$ is the sum of the values of the arcs of $C$. Let $a_i$ be a point of a cycle, $C$, constructed in $\sigma^{-1}M^-$ such that as we traverse $C$ in a clock-wise direction, the partial sum of the arcs of the arcs obtained so far is no greater than $|C|$. Then $a_i$ is a *determining point* of $C$.

A set of pair-wise disjoint acceptable cycles such that no two cycles have a point in the same 2-cycle of $\sigma_T^{-1}M^-$ is called an *acceptable permutation*. $pt(C)$ is the number of points in the permutation cycle $C$.

III. USEFUL THEOREMS.

Theorem 1.1 *Let* $C = (a_1 a_2 \ldots a_n)$ *be a cycle of length n. Assume that the weight* $w(a_i, C(a_i))$ $(i=1,2,\ldots,n)$ *corresponds to the arc* $(a_i, C(a_i))$ *of C. Then if*

$$W = \sum_{i=1}^{i=n} w(a_i, C(a_i)) \leq 0$$

*there exists at least one vertex* $a_{i*}$ *with* $1 \leq a_{i*} \leq n$ *such that*

$$S_m = \sum_{j=0}^{j=m} w(a_{i*+j}, C(a_{i*+j})) \leq 0 \qquad (A)$$

*where* $m = 0,1,2,\ldots,n-1$ *and* $i*+j$ *is modulo* $n$.

Proof. We prove the theorem by induction. Let k = 2. We thus have a 2-cycle. If both arcs have non-positive value, then the theorem is proved. If the non-positive arc, $(a_1\ a_2)$, has a smaller weight than a positive one, then the sum of the weights of the two arcs is positive. This can't be the case. Thus, the sum of the two weights is non-positive. Now let the theorem always be true when our cycle has k arcs. Suppose the cycle has k+1 arcs. In what follows, assume that if a value is 0, its sign is negative. Then one of the following is true: (a) there exists a pair of consecutive arcs both of whose values have the same sign or (b) the signs of values of the arcs consecutively alternate in sign. First, we consider (a). Without loss of generality, let the two arcs be $(a_1\ a_2)$ and $(a_2\ a_3)$. Assume that each one has a non-positive value. We now define the arc $(a_1\ a_3)$ where $w(a_1\ a_3) = w(a_1\ a_2) + w(a_2\ a_3)$. Now replace arcs $(a_1\ a_2)$ and $(a_2\ a_3)$ by $(a_1\ a_3)$. The result is a cycle $C'$ containing $k$ arcs. By induction, the theorem holds for $C'$. Now replace $(a_1\ a_3)$ by $(a_1\ a_2)$ and $(a_2\ a_3)$. Let $a_{i*}$ be a determining vertex of $C'$. Then the path, $P_{i*}$, from $a_{i*}$ to $a_1$ is non-positive. If both $w(a_1\ a_2)$ and $w(a_2\ a_3)$ are non-negative, then both $P_{i*} \cup w(a_1\ a_2)$ and $P_{i*} \cup w(a_1\ a_2) \cup w(a_2\ a_3)$ are non-positive. Thus, the theorem is valid in this case. Now assume that both $w(a_1\ a_2)$ and $w(a_2\ a_3)$ are positive. But, since the theorem is valid for $C'$, $P_{i*} \cup w(a_1\ a_3)$ is non-positive. This assures us that each of $P_{i*} \cup w(a_1\ a_2)$ and $P_{i*} \cup w(a_1\ a_2) \cup w(a_2\ a_3)$ is non-positive. Therefore, the theorem holds when the values of $(a_1\ a_2)$ and $(a_2\ a_3)$ both have the same sign. We now consider case (b). Here, the signs alternate between positive and negative. Since the arcs lie on a cycle, we may assume that the first sign is negative. If k+1 is odd, it is always true that at least one pair of consecutive arcs has the same sign. Thus, assume that k+1 is even. Then, starting with a non-positively-valued arc, we can arrange the arcs of the cycle in pairs where the first arc has a non-positive value, while the second one has a positive value. Suppose the sum of each pair of arcs is positive. Then the sum of the values of the cycle is positive. Therefore, there exists at least one pair of arcs the sum of whose values is non-positive. Remembering that the first arc of each pair is non-positive, we follow the same procedure as in (a): $C'$ is a non-positive cycle containing k arcs. Therefore, $C$ is also positive.

**Corollary 1.1a** *Suppose that C is a cycle such that*

$$W = \sum_{i=1}^{i=n} w(a_i, C(a_i)) \leq N \qquad (B)$$

*Then there exists a determining vertex $a_{i*}$ such that each partial sum, $S_m$, has the property that*

$$S_m = \sum_{j=0}^{j=m} w(a_{i*+j}, C(a_{i*+j})) \leq N \qquad (C)$$

*always holds. Here $m = 0,1,2, \ldots, n-1$ while $i* + j$ is modulo n.*

Proof. Subtract $N$ from both sides of (B). Now let the weight of arc $(a_n, C(a_n))$ become $w(a_n, C(a_n)) - N$. From theorem 1.3,

$$W^* = W - N = \sum_{i=1}^{i=n} w(a_i, C(a_i)) \leq 0 \quad (D)$$

Therefore, we can obtain a determining vertex $a_{i*}$ having the property that every partial sum having $a_{i*}$ as it initial vertex is non-positive. It follows that if we restore $w(a_n, C(a_n))$ to its original value, every partial sum with initial vertex $a_{i*}$ is less than or equal to $N$.

**Corollary 1.1b** *Let C be a positively-valued cycle of length n obtained from a cost matrix W whose entries may be positive, negative, or zero. Then there exists at least one determining vertex, say $a_{i*}$, of C such that each subpath $S_m$ having initial vertex $a_{i*}$ has the property that*

$$S_m = \sum_{j=0}^{j=m} w(a_{i*+j}, C(a_{i*+j})) \geq 0 \qquad (E)$$

*where $m = 0,1,2, \ldots, n-1$ and $i* + j$ is modulo n.*

Proof. Let $N = 0$ in $(B)$. Then multiply each term of $(B)$ by $-1$. It follows that we obtain $.-W = \sum_{i=1}^{i=n} -w(a_i, C(a_i)) \leq 0$. The rest of the proof is similar to that of Corollary 1.1a.

Example 1. We now give an example of how to obtain $i = i'$.

Let $C = (a_1 \ a_2 \ \ldots \ a_n)$

$a_1 = -7, \ a_2 = -10, \ a_3 = +1, \ a_4 = +2, \ a_5 = -7, \ a_6 = +4, \ a_7 = -9,$

$a_8 = +11, \ a_9 = -2, \ a_{10} = -1, \ a_{11} = -4, \ a_{12} = -4, \ a_{13} = -8, \ a_{14} = +9,$

$a_{15} = +9, \ a_{16} = +21, \ a_{17} = +1, \ a_{18} = -2, \ a_{19} = -1, \ a_{20} = -3,$

$a_{21} = -3, a_{22} = -12, a_{23} = +6, a_{24} = +2, a_{25} = +3$

-7 −10 +1 +2 −7 +4 −9 +11 −2 −1 −4 −4 −8 +9 +9 +21 +1

-2 -1 -3 -3 -12 +6 +2 +3

We now add terms with like signs going from left to right. We place the ordinal number of the first number in each sum above it.

   1   3  5  6  7    8   9   14   18   23

-17  +3  -7  +4  -9   +11  -19  +40  -21  +11

We next add the positive number to the right of each negative number to the negative number.

  1   5        18

-14  -3  +2  +21  -10

We now add terms with like signs going from left to right.

   1       18

-17   +23   -10

Finally, assuming that all points lie on a circle, we add like terms going from left to right. We thus obtain

 18       18

-27  +23  =  -4

This tells us that $i'$ is the eighteenth ordinal number; its value is –2.

Thus, the partial sums are: -2, -3, -6, -9, -21, -15, -13, -10, -17, -27, -26, -24, -31, -27, -36, -25, -27, -28, -32 –36, -44, -35, -26, -5, -4

Theorem 1.2 The Floyd-Warshall Algorithm

*If we perform a triangle operation for successive values j = 1,2,...,n, each entry $d_{ik}$ of an n X n cost matrix M becomes equal to the value of the shortest path from i to k provided that M contains no negative cycles.*

The version given here is modeled on theorem 6.4 in [4].

Proof. We shall show by induction that that after the triangle operation for $j = j_0$ is executed, $d_{ik}$ is the value of the shortest path with intermediate vertices $v \leq j_0$, for all $i$ and $k$. The theorem holds for $j_0 = 1$ since $v = 0$. Assume that the inductive hypothesis is true for $j = j_0 - 1$ and consider the triangle operation for $j = j_0$:

$$d_{ik} = min\{d_{ik}, d_{ij_0} + d_{j_0k}\}.$$

If the shortest path from $i$ through $k$ with $v \leq j_0$ doesn't pass through $j_0$, $d_{ik}$ will be unchanged by this operation, the first argument in the min-operation will be selected, and $d_{ik}$ will still satisfy the inductive hypothesis. On the other hand if the shortest path from $i$ to $k$ with intermediate vertices $v \leq j_0$ does pass through $j_0$, $d_{ik}$ will be replaced by $d_{ij_0} + d_{j_0k}$. By the inductive hypothesis, $d_{ij_0}$ and $d_{j_0k}$ are both optimal values with intermediate vertices $v \leq j_0 - 1$. Therefore, $d_{ij_0} + d_{j_0k}$ is optimal with intermediate vertices $v \leq j_0$.

We now give an example of how F-W works.

Example 2. Let $d(1, 3) = 5$, $d(3, 7) = -2$, $d(1, 7) = 25$. Then $d(1, 3) + d(3, 7) < d(1, 7)$. Note however, that the intermediate vertex 3 comes from the fact that we have reached column $j = 3$ in the algorithm. We now substitute $d(1, 3) + d(3, 7) = 3$ for the entry in (1, 7). Suppose now that $d(1, 10) = 7$ while $d(7, 10) = -5$.

$$d(1, 3) + d(3, 7) + d(7,10) = -2 < d(1, 10) = 7.$$

Theorem 1.3 *Let M be a value matrix containing both positive and negative values. Suppose that M contains one or more negative cycles. Then if a negative path P becomes a non-simple path containing a negative cycle, C, as a subpath, C is obtainable as an independent cycle in the modified F-W algorithm using fewer columns than the number used by P to construct C.*

Proof. Using the modified F-W algorithm, no matter which vertex, $a_i$, of $C$ is first obtained in $P$, it requires the same number of columns to return to $a_i$ and make $P$ a non-simple path. Since $b_1 \neq a_1$, It requires at least one column to go from $b_i$ to $a_1$, concluding the proof.

In the following example, we construct $P$ and $C$.

Example 3.

$P = [1^{-20} 3^5 7^{-5} 13^{12} 15^1 19^3 20^{-18} 18^1 14^1 6^3 7^{-5}]$,

$C = (20\ 18\ 14\ 6\ 7\ 13\ 15\ 19)$.

In what follows, Roman numerals represent numbers of iterations.

| $P$ | $C$ |
|---|---|
| I.  [1 3] + 3 | I.  [20 18] + 18 |
|     [1 7] + 4 | ------------------------ |
|     [1 13] + 6 | II.  [20 14] + 16 |
|     [1 15] + 2 | ------------------------ |
|     [1 19] + 4 | III.  [20 6] + 12 |
|     [1 20] + 1 |     [20 7] + 1 |
| ------------------------ |     [20 13] + 6 |
| II.  [1 18] + 18 |     [20 15] + 2 |
| ------------------------ |     [20 19] + 4 |
| III.  [1 14] + 16 |     [20 20] + 1 |
| ------------------------ | ------------------------ |
| IV.  [1 6] + 12 |     + 60 |
|     [1 7] + 1 | |
| ------------------------ | |
|     + 67 | |

`Theorem 1.4` *Let c be any real number, $S_a = \{a_i \mid i = 1,2,\ldots,n\}$, a set of real numbers in increasing order of value. For $i = 1, 2, \ldots, n$, let $b_i = a_i + c$. Then $S_b = \{b_i \mid i = 1,2,\ldots,n\}$ preserves the ordering of $S_a$.*

Proof. The theorem merely states that adding a fixed number to a set ordered according to the value of its elements retains the same ordering as that of the original set.

*Comment*. This is very useful when we are dealing with entries of the value matrix which have not been changed during the algorithm. However, as the algorithm goes on, if we have entry(i, j) and d(i, j), we must go through all values of j = j' where the (i, j')-th entry's value in the current $\sigma_\alpha^{-1} M^-(k)$ is less than the value of the (i, j')-th entry in $D^{-1}M^-$ (The terminology will become clearer in the examples given.)

`Theorem 1.5`. *Let $D = \prod_{i=1}^{i=r} C_i$ be a derangement in $S_n$. Assume that $a_i$ $(i = 1,2,\ldots,r)$ is a point on $C_i$. Then $s = (a_1\ a_2\ \ldots\ a_r)$ has the property that $Ds$ is an $n$-cycle, $H$.*

Proof. Consider the arcs of $H$ obtained by the action of $D$ on $s$:
$(a_1\ a_2) \to (a_1\ D(a_2))$, $(a_2\ a_3) \to (a_2\ D(a_3))$, $\ldots$, $(a_r\ a_1) \to (a_r\ D(a_1))$. Thus,

$$H = (a_1 D(a_2) \ldots a_2 D(a_3) \ldots a_3 D(a_4) \ldots a_{r-1} D(a_r) \ldots a_r D(a_1) \ldots a_1)$$

`Theorem 1.6` *Let $T$ be a tour containing an even number of edges. Let $\sigma_T$ be the set of smaller-valued alternating edges of $T$ that may be considered a set of $\frac{n}{2}$ pair-wise disjoint 2-cycles. Then there always exists an acceptable cycle, $s$, in $\sigma_T^{-1} M^-$ containing $\frac{n}{2}$ points such that $\sigma_T s = T^{-1} = T$.*

Proof. Let $T = (a_1\ a_2\ a_3\ \ldots\ a_{2n})$. $\sigma_T s = T \Rightarrow s = \sigma_T T$. Since $\sigma_T$ consisting of alternating edges of $T$, assume it is $S = \{[a_i\ a_{i+1}] / i = 1,3,5,\ldots,2n-1\}$. Consider the cycle $s = (a_1\ a_{2n-1}\ a_{2n-3}\ \ldots\ a_3)$ in $\sigma_T^{-1} M^-$. Applying $\sigma_T$ to it, where the arc $(a_i\ a_j)$ is mapped into $[a_i\ \sigma_T(a_j)][\sigma_T(a_j)\ a_j]$ we obtain $[a_1\ \underline{a_{2n}}\ a_{2n-1}\ \underline{a_{2n-2}}\ a_{2n-3}\ \ldots\ a_3\ \underline{a_2}]$. Thus, $s$ is an acceptable cycle containing $\frac{n}{2}$ points that yields the tour $T^{-1}$ where – due to symmetry - $|T^{-1}| = |T|$. Furthermore, since $T^{-1}$ is a circuit consisting of edges, $T^{-1} = T$.

Theorem 1.7 *Let n = 2m. A perfect matching always exists if a finite-valued hamilton cycle exists.*

Theorem 1.8 *Let n = 2m. Assume that $\sigma_{ABSOLUTE}$ is the smallest-valued PM that we've obtained. Then*

*a necessary and sufficient condition for an acceptable cycle $C = (a_1\ a_2\ ...\ a_m)$ in $\sigma_{ABSOLUTE}^{-1} M^{-1}$ to yield a perfect matching is that the cycle*

$C' = (\sigma_{ABSOLUTE}(a_2)\ \sigma_{ABSOLUTE}(a_1)\ \sigma_{ABSOLUTE}(a_m)\ ...\ \sigma_{ABSOLUTE}(a_3))$ *is disjoint from $C$.*

*Furthermore, the arcs of the permutation obtained from C and $C = (a_1\ a_2\ ...\ a_m)$ yield a perfect matching, while*

$$T = (a_1\ \sigma_{ABSOLUTE}(a_2)\ a_2\ \sigma_{ABSOLUTE}(a_3)\ a_3\ ...\ \sigma_{ABSOLUTE}(a_m)\ a_m\ \sigma_{ABSOLUTE}(a_1))$$

*yields a tour.*

Corollary 1.8a *If the length of $C$ is $m' < m$, then $T$ is a circuit of edges of length 2m'.*

Proof. From theorem 1.6, we know that if $n = 2m$, then we obtain a tour. It follows from theorems 1.6 and 1.8, if $m' < m$, we obtain a cycle, $C$, consisting of edges. The number of edges cannot be $n$ since $2m' < 2m = n$. Thus, $C$ cannot be a tour.

*Note 1.* Although it might be time consuming, if we can obtain a set of disjoint cycles, $C_i$, covering all $n$ vertices, we could obtain a minimally-valued derangement. This would be the best possible lower bound for $\sigma_{TSPOPT}$, i.e., a minimal derangement obtained using the modified F-W algorithm, is the smallest possible derangement obtainable: It is a lower bound for both $\sigma_{TSPOPT}$ as well as for $\sigma_{FWTSPOPT}$.

Theorem 1.9. *Every acceptable cycle in $\sigma^{-1} M^-$ yields a unique perfect matching.*

Proof. We first note that $\sigma(\sigma(a)) = a$. Let $C = (a_1\ a_2\ ...\ a_m)$. Define $C' = (\sigma(a_2)\ \sigma(a_1)\ \sigma(a_m)\ ...\ \sigma(a_3))$

$C'$ is an acceptable cycle since it has the same number of points as $C$ and its points are in the same 2-cycles as those of $C$. Consider the directed edges obtained from $C$:

$\{(a_1\ \sigma(a_2)), (a_2\ \sigma(a_3)),\ ...\ ,(a_m\ \sigma(a_1))\}$. They are pair-wise disjoint and contain precisely the points in the 2-cycles containing points of $C$. On the other hand, the directed edges from $C'$ are

$\{(\sigma(a_2)\ a_1), (\sigma(a_1)\ a_m),\ ...\ ,((\sigma_3)\ a_2)\}$. The second set contains precisely all directed edges that are symmetric to those in the first set. Thus, we have obtained a new set of 2-cycles that replace those in which $C$ had points. By the definition of acceptable permutation, we have changed the edges only in the 2-cycles of $\sigma$ containing a point of $C$. Thus, $\sigma CC' = \sigma^*$, a unique perfect matching.

`Corollary 1.9a` *Theorem 1.9 is also true for 2-circuit cycles.*

Proof. The proof is the same as for theorem 1.9 because all pairs of non-linking points of the cycle occur in the "companion cycle". Furthermore, as with acceptable cycles, a 2-circuit cycle yields two circuits of edges where *each* circuit has the same value whether its edges are directed in a clockwise or counter-clockwise direction.

`Theorem 1.10` *Let $\sigma$ be a perfect matching. Then a necessary and sufficient condition that it is a minimal-valued perfect matching is that $\sigma^{-1}M^-$ contains no negatively-valued acceptable cycle.*

Proof. Suppose $C$ is a negatively-valued acceptable cycle. Then $\sigma C$ yields a perfect matching of smaller value than $\sigma$. On the other hand, if no acceptable cycle exists, then $\sigma$ is a minimally-valued perfect matching.

In what follows, $\sigma_{ABSOLUTE}$ is a minimally-valued perect matching of M expressed as a product of 2-cycles.

`Theorem 1.11` *Let $\sigma$ be a PM. Then the following hold:*

*(1) An acceptable circuit containing an even number of vertices is obtainable from an acceptable cycle in $\sigma^{-1}M^-$.*

*(2) A 2- circuit cycle in $\sigma^{-1}M^-$ contains in its alternating path precisely two circuits each of which has an odd number of edges.*

*(3) Tthe first vertex of each of the two circuits along the alternating path belongs to the same 2-cycle of $\sigma$.*

*Furthermore, if $s$ is a permutation such that $\sigma s$ is a tour, then each of the cycles of $s$ is either an acceptable or a 2-circuit cycle.*

Proof. An acceptable cycle always yields an acceptable circuit containing twice the number of edges as the cycle contains. To insure that a path is a 2 circuit path, it must have distinct vertices in $P_{20i}$, while its alternating path contains precisely one circuit of the following kind:
$P = (a\ q) = [a\ \underline{a}'\ b\ \underline{b}'... \underline{a}\ ...\ d\ \underline{d}'\ e\ \underline{e}'\ f\ \underline{f}'\ ...\ p\ \underline{p}'\ q]$. Our circuit consists of the edges
$[a\ \underline{a}'\ b\ \underline{b}'\ ...\ \underline{a}]$. The arc $(a\ b)$ yields the edge $[a\ \underline{a}']$. But $(b\ a')$ is an arc of $\sigma$ when we consider it as a directed derangement. Thus, $(a'\ b)$ is an arc symmetric to the arc $(b\ a')$ of $\sigma$. It follows that any underlined vertex in $P$ is followed by its companion in a 2-cycle of $\sigma$. Now consider the cycle
$C = (a\ b\ ...\ ?\ ...\ d\ e\ f\ ...\ p\ q\ ...\ r)$ that defines the alternating path of edges
$P' = [a\ \underline{a}'\ b\ \underline{b}'... \underline{a}\ ?\ ...\ d\ \underline{d}'\ e\ \underline{e}'\ f\ \underline{f}'\ ...\ p\ \underline{p}'\ q\ ...\ r\ \underline{??}a]$. What are the vertices represented by "?" and "??" ? Let *s* be the companion of $a$ in a unique 2-cycle of $\sigma$. Then "??" represents *s*. But what represents "?" ? It could be *s* since that would allow $C$ to consist of distinct vertices, while allowing us to obtain two circuits of edges from $P'$. Could it be another vertex? But if we had three or more circuits, the last vertex of the final circuit would have to be *s* which is impossible. Thus, there can be only two circuits – each containing an odd number of vertices – in $P$. One of the edges of each circuit isn't actually obtained from a cycle. However, since it always is an edge of $\sigma$ and thus has a value of 0 in $\sigma^{-1}M^-$, we can assume that it is part of the circuit. Actually, the missing edge tells us precisely which edge of another circuit can be deleted. On the other hand, by including it in theory, the formula for the number of edges in all circuits - ($n + 2r - 2$) - can be used without modification.

*Note.* In theorem 1.11, we have shown that if a cycle of $s$ is not acceptable, then it must be a 2 circuit cycle: No cycle can yield three or more circuits. We obtain a tour by constructing a tree of circuits in which circuits are linked by a common edge. (In the case of a 2 circuit cycle, we delete the "phantom edge" of a circuit.).

In what follows, if $C$ is an acceptable cycle, $C_{circuit}$ is the circuit of edges obtainable from $C$.

What follows is another – more detailed - version of theorem 1.11.

Let $\sigma$ be a PM such that $|\sigma_{ABSOLUTE}| \leq |\sigma| < |T_{UPPERBOUND}|$. Furthermore, suppose that there exists a tour $T$ such that $|T| < |T_{UPPERBOUND}|$. Then

(a) there exists an acceptable cycle $t$ such that $\sigma\, t = T$

(b) there exists a permutation $t'$ such that $\sigma_{ABSOLUTE}\, t' = T$ where $t'$ consists of a set of disjoint cycles each of which is either acceptable and yields a circuit containing an even number of vertices, or else is a 2 circuit cycle yielding a set of two or more disjoint circuits each of which contains an odd number of points. In the latter case, the circuits can be obtained directly from 2-circuit paths.   .

(c) Each circuit, $C_{icircuit}$, consists of edges with at most one identical to the edge of a fixed cycle $C_{jcircuit}$ where $C_{icircuit} \neq C_{jcircuit}, j \neq i$, while every circuit has an edge identical to at least one circuit.

(d) Defining a common edge as a "link" between two circuits, the circuits form a tree.

(e) The total number of edges (and vertices) in all of the circuits obtained is $n + 2r - 2$ where $r$ is the number of circuits.

Proof. (a) From theorem 1.6, there exists a acceptable cycle $t$ whose value is non-positive such that $T_{UPPERBOUND}\, t^{-1} = \sigma$.

Furthermore, each point of $t^{-1}$ is contained in precisely one 2-cycle of $\sigma$.

(b) If a cycle, $C'$, of $t'$ is acceptable, then, from theorem 1.10, it yields a circuit containing an even number of points. We know that $t'$ exists and contains disjoint cycles. Any cycle $C''$ that is not acceptable must still yield circuits that contain all of its vertices of the cycle. Otherwise, we couldn't obtain a tour. This can only be the case if it yields two or more circuits each of which (from theorem 1.10) contains an odd number of points.

(c) In order to obtain a tour from a set of circuits, $\{C_{icircuit}\}$, containing all $n$ vertices, each circuit, $C_{icircuit}$, must contain a path $[a \ ... \ b]$ such that the edge $[a \ b]$ is identical to precisely one edge $[a \ b]$ of some circuit $C_{jcircuit}$. It can't contain two or edges identical to those in $C_{jcircuit}$, since then we obtain at least two circuits. On the other hand, if a circuit doesn't have at least one edge in common with another circuit, then we again must obtain at least two circuits.

(d) If the circuits form a tree whose "links" are identical edges, then we can obtain a tour by deleting common edges. Otherwise, if they don't form a tree, we would obtain more than one circuit..

(e) Let $r$ be the number of circuits, $C_{icircuit}$. We thus delete $2r-2$ common edges (and vertices) from the $r$ circuits forming the tree. After deleting the $2r-2$ common edges, we obtain a tour containing $n$ edges. Thus, the total number of edges (and vertices) in all circuits is $n + 2r - 2$.

Theorem 1.13 *Let $T$ be a tour where $\sigma_T$ is its set of alternating edges the sum of whose values is no greater than $|\frac{T}{2}|$. Suppose that $S$ is a set of acceptable and 2-circuit cycles obtained from $\sigma_T^{-1}M^-$ .Let $a$ be the number of acceptable cycles and $t$ the number of 2-circuit cycles in $S$. Assume that*

$$C_{PATH} = [a \ \underline{f_1} \ f_1 \ ... \ \underline{l_1} \ l_1 \ \underline{a} \ b \ \underline{f_2} \ f_2 \ ... \ \underline{l_2} \ l_2 \ \underline{b} \ a]$$

*is a 2-circuit path obtainable from a 2-circuit cycle. For the purpose of discussion, from the 2-circuit permutation cycle $C$, define each of $(a \ f_1 ... l_1)$ and $(b \ f_2 ... l_2)$ as a cycle rather than as a circuit. Assume*

*that two cycles of $S$ are linked if precisely one point in the first cycle belongs to the same 2-cycle of $\sigma_T$ to which a point in the second cycle belongs. We now add the condition that the points $a$ and $b$ of $C$ cannot be linked. Then necessary and sufficient conditions for the cycles in S to yield a tour $T'$ with $|T'|<|T|$ are the following:*

*(1) The number of points moved by the cycles in $S$ is $\dfrac{n}{2} + 3t + a - 1$.*

*(2) Each cycle in $S$ is linked to at least one cycle of $S$.*

*(3) Each pair of cycles in $S$ has at most one point in common*

*(4) The cycles of $S$ form a tree by linking.*

*(5) The sum of the values of the cycles in $S$ is smaller than $|T|$.*

Proof. When two cycles are linked, their points of linkage yield a common edge of circuits that can be constructed from the cycles. Depending upon whether or not the two cycles can be linked by going in a clockwise or counterclockwise direction of the circuit formed from the second cycle, two cycles may or may not have a point in common. The important thing is that only one *common* edge can be deleted from two linked cycles. Since the linked cycles form a tree, after deleting common edges from the circuits obtained by the cycles, we obtain a circuit containing all of the points of $S$. Let $p$ be the number of points moved by a 2-circuit cycle. Then the number of points in the two circuits obtained from the cycle is $2p-2$. The reason is as follows:

The circuits we obtain are ($a$ $\underline{f_1}$ $f_1$ ... $\underline{l_1}$ $l_1$), ( $b$ $\underline{f_2}$ $f_2$ ... $\underline{l_2}$ $l_2$). We thus omit the following arcs:

$\underline{a}\,b$, $\underline{b}\,a$. Let $A_i$, $i=1,2, \ldots, a$, be the subset of acceptable permutation cycles in $S$, while $T_j$, $j=1,2,\ldots,s$, is the subset of permutation 2-circuit cycles. We are assuming that we can obtain a tour by deleting arcs. Thus, the number of edges contained in all circuits constructed from the cycles in $S$ is $\sum_{i=1}^{a} 2|A_i| + \sum_{j=1}^{s}(2|T_j| - 2)$. In order to construct a tour from these cycles, we must construct $a + 2t - 1$ linkages of $a + 2t$ circuits. Each linkage deletes two edges. Thus, we must delete $2(a + 2t - 1)$ edges. Since a tour contains $n$ edges, we obtain the equation

$$\sum_{i=1}^{a} 2|A_I| + \sum_{j=1}^{s}(2|T_I| - 2) - 2(a + 2t - 1)$$

$$= 2(\sum_{i=1}^{a}|A_i| + \sum_{j=1}^{s}(|T_j|) - 2t - 2(a + 2t - 1) = n$$

$$\Rightarrow 2(\sum_{i=1}^{a}|A_i| + \sum_{j=1}^{s}(|T_j|)) = n + 2t + 2(a + 2t - 1) = n + 6t + 2a - 2$$

Therefore, the number of points in the corresponding cycles in $S$ is $\dfrac{n}{2} + 3t + a - 1$.

Next, each cycle yields one or more circuits composed of edges. The edges are of two kinds: (1) Those that we have obtained from arcs of the cycles. (2) Those we obtained from the edges (2-cycles) of $\sigma_T$. By definitions, each of the edges obtained from $\sigma_T$ has a value of zero. Furthermore, all of the edges that are deleted belong to $\sigma_T$. It follows that the value of one (or two) *directed* circuits corresponds to the value of a permutation cycle lying in $S$ implying that the sum of the values of the permutation cycles in $S$ corresponds to the sum of the values of the *directed* circuits obtained from corresponding permutation cycles.

`Corollary 1.13a` *Let $S$ be a set of cycles that yields a tour. Denote by $S'$ the set obtained by deleting $X$ acceptable and $Y$ 2-circuit cycles from $S$. Define $t' = t - Y$, $a' = a - X$. Then if $p'$ is the total number of points moved by the cycles in $S'$, $p' < \dfrac{n}{2} + 3t' + a' - 1$.*

Proof. Each 2-circuit cycle moves at least four points, while each acceptable cycle moves at least two points. Thus, $(3t - 3Y) < (3t - 4Y)$, $(a - X) < (a - 2X)$.

**Lemma 1.14** *Let $a_1, a_2, \ldots, a_r$ be an ordered set, $S$, of integers whose sum is N. Furthermore, suppose $a_r + a_1 = 0$. Then there exists an integer $a_{i'}$ such that each partial sum*

$$\sum_{j=i'}^{j=i} a_j, i = i', i'+1, \ldots, r, 1, 2, \ldots, i'-1 \text{ has a value no greater than N.}$$

Proof. If $S$ defines a set of sequential values on a cycle $C = (a_1\ a_2\ \ldots\ a_r)$ where we define $|(a_r\ a_1)| = 0$, then as shown in theorem 3, we can always obtain a determining point $a_{i'}$ such that each partial sum is no greater than N.

Using this lemma, we obtain theorem 14.

Theorem 1.14 *Suppose we construct a tree by consecutively linking acceptable and 2-circuit cycles obtained from $\sigma_T^{-1} M^-$ whose values are, respectively, $|C_1|, |C_2|, \ldots |C_r|$ where $|C_i| < |T| - |\sigma_T|, i = 1, 2, \ldots, r$. Define a "phantom cycle" $C_{r1}$ whose value is zero. Then if there exists a tour of value less than $|T|$, there exists a cycle, $C_{i'}$ such that each partial sum*

$$\sum_{j=i'}^{j=i} C_j, i = i', i'+1, \ldots, r-1, r, 1, 2, \ldots, i'-1 \text{ has a value less than } |T|.$$

Proof. The proof follows directly from lemma 1.

Define a phantom acceptable cycle as a 2-cycle of value zero that may be used to link cycles forming a tour. Such a cycle doesn't exist but it may have useful properties in constructing a tour. It always must be eliminated in the course of the algorithm to obtain a tour. In what follows - if necessary – we may assume that the cycles in our hypothesis include at most one phantom cycle. It is discarded once the tour is obtained. Also, $L_i$ denotes a set of linked cycles.

Theorem 1.15 *Let $S_{FW}$ be the set of all acceptable permutations of value no greater than $|T_{UPPERBOUND}|$ whose cycles are obtained from $\sigma_T^{-1} M^-$ using the modified F-W algorithm, while $S_{ALL}$ is the set of all acceptable permutations of value no greater than $|T_{UPPERBOUND}|$. Then the following hold:*

*If we can obtain all elements of $S_{FW}$ ($S_{ALL}$) in polynomial time, we can always obtain $T_{FWTSPOPT}$ ($T_{TSPOPT}$) in polynomial time.*

Proof. Corresponding to each acceptable cycle $C$ is an acceptable cycle $C'$ each of whose arcs yields an arc symmetric to an arc obtained from an arc of $C$. Thus, $\sigma_T\ C\ C' = \sigma'$, a perfect matching. There are $\frac{n}{2}$ 2-cycles in $\sigma_T$. It follows that we require no more than an acceptable permutation containing at most $\frac{n}{2}$ points to obtain a set of arcs no two of which belong to the same 2-cycle. These arcs yield a perfect matching by changing each arc into an edge. Given each acceptable permutation, $P$, there exists a unique perfect matching $\sigma_P$ that contains $\frac{n}{2}$ edges. These edges are a set of alternating edges of a minimally-valued tour $T_{\sigma_P}$. Since it is a minimally-valued tour, it can be obtained as the product $\sigma_P s$ where $s$ is an acceptable permutation whose cycles are obtainable from $\sigma_P^{-1} M^-$. On the other hand, $T^* = \sigma_{FWTSPOPT}$, has a value no greater than $|T_{UPPERBOUND}|$. Therefore, it contains a set of alternating edges of value no greater than $|T_{UPPERBOUND}|$ that is a perfect matching, $\sigma_{T^*}$. This perfect matching must be an element of $S_{FW}$. It follows that we can obtain $s$ in polynomial time by applying the modified F-W algorithm to $\sigma_{T^*}^{-1} M^-$.

Using rooted trees whose roots are the numerically smallest of the determining vertices of the acceptable cycles obtained using the modified F-W algorithm, an analogous proof holds for (2).

Theorem 1.16 *Let $S$ be a set of acceptable or 2-circuit cycles obtained from $\sigma_T^{-1} M^-$ such that the following hold for the cycles in $S$:*

*(1) The respective non-linking points of the cycles belong to the same 2-cycles.*

*(2) The respective linking points of the cycles belong to the same 2-cycles.*

*(3) The cycles all have the same value.*

*Then if one of them belongs to a set of cycles yielding a tour, they all belong to respective sets of cycles that yield a tour having the same value.*

Proof. Consider an arbitrary cycle, $C$, that can be linked to one of the cycles in $S$. By the conditions given in the hypotheses, they all can be linked to $C$. Since all of the cycles in $S$ have the same value, the tours containing them all have the same value.

Theorem 1.17 *Let $\sigma$ be a product of $\dfrac{n}{2}$ disjoint 2-cycles that is obtained from a perfect matching. Suppose that $T$ is a tour. Let $s$ be a permutation such that $\sigma s = T$. Then $s$ contains at least one point in each 2-cycle of $T$ such that each cycle of $s$ is either an acceptable or 2-circuit cycle.*

Proof. We consider alternate possibilities. First, $s$ must have at least one point in each 2-cycle in order to obtain a tour. Each arc, $(a\ b)$, in a cycle can represent a sequence of two edges $[a\ \sigma(b)][\sigma(b)\ b]$. In this way, each acceptable cycle of $j$ arcs represents a circuit containing $2j$ edges. Suppose a cycle is *not* acceptable. Using this procedure, there is only one way in which it can contribute cycles: Suppose the cycle is of one of the following form: (1) $(a_1\ a_2\ ...\ a_r\ b_1\ a_{r+1}\ a_{r+2}\ ...\ a_s)$ where all of the $a_i$'s belong to distinct 2-cycle, while $b_1$ is the other point in the 2-cycle $(a_1\ b_1)$ of $\sigma$. Then we obtain

$$[a_1\ \underline{\sigma(a_2)}\ a_2\ \underline{\sigma(a_3)}\ ...\ a_r\ \underline{\sigma(b_1)}\ b_1\ \underline{\sigma(a_{r+1})}\ a_{r+1}\ \underline{\sigma(a_{r+2})}\ a_{r+2}\ \underline{\sigma(a_{r+3})}\ ...\ a_s\ \underline{\sigma(a_1)}]$$
$$= [a_1\ \underline{\sigma(a_2)}\ ...\ a_r\ \underline{a_1}\ b_1\ \underline{\sigma(a_{r+1})}\ a_{r+1}\ \underline{\sigma(a_{r+2})}\ a_{r+2}\ \underline{\sigma(a_{r+3})}\ ...\ a_s\ \underline{b_1}]$$
$$= [a_1\ \underline{\sigma(a_2)}\ ...\ a_r][b_1\ \underline{\sigma(a_{r+1})}\ a_{r+1}\ \underline{\sigma(a_{r+2})}\ a_{r+2}\ \underline{\sigma(a_{r+3})}\ ...\ a_s]$$

Thus, we obtain two circuits. Furthermore, we cannot obtain *more than* two circuits. If the two circuits have no edge in common they are said to be unlinked. If they have precisely one edge in common, they are linked. If they are linked, they can become one circuit by deleting the common edge. Suppose we have two cases

$(a_i\ b_i)$, $i = 1, 2$ where the points *are not interlaced* (say $a_1, b_2, a_2, b_1$) as they are with linked circuits. Then we have

$$[a_1\ ...\ \underline{a_1}\ b_1\ \sigma(a_{r+1})\ a_{r+1}\ ...\ \underline{a_{r+1}}\ ...\ b_1\ a_1]$$
$$= [a_1\ ...\ \underline{a_1}][b_1\ ...\ a_{r+1}\ ...\ \underline{a_{r+1}}\ ...\ b_1]$$

But $[b_1\ ...\ a_{r+1}\ ...\ \underline{a_{r+1}}\ ...\ b_1]$ is *not* a circuit because we cannot express it as a sequence of pair-wise disjoint edges such that the first repetition of a point occurs only when we reach the initial point $b_1$.

*Note*. Using the determining vertices of cycles obtained using the modified F-W algorithm, we can employ rooted trees together with the end-points of the respective paths used to obtain cycles, to obtain *all* acceptable cycles. They are then added to the list of acceptable cycles obtained using the F-W algorithm. We would thus construct any further acceptable permutations to obtain $S_{ALL}$. Another approach is to obtain *all* paths using the modified F-W algorithm. This consists of the following: Every time we wish to replace a path by a shorter one by changing an entry in some $P_i$, we copy the original path from $P_i$. Also, if a path can no longer be extended because its value is greater than a path previously obtained, we save it and try to extend it to a cycle. After we have obtained all acceptable cycles using the modified F-W algorithm, we delete those paths that don't have an initial vertex that is the determining vertex of an acceptable cycle. We then try to obtain an acceptable cycle from each of the remaining paths by using a rooted tree for each one. The following is a sequence that might shorten our search:

1. Using the determining vertices of the cycles obtained using modified F-W, we construct determining vertex rooted trees, $T_\alpha$, whose branches correspond to paths.

2. No path can contain more than $\frac{n}{2} + 2$ arcs.

3. Given a set of determining pairs $S = \{(a_i, b_i)\}$, a path stops once we have reached all points $b_i$ in row $a_i$.

4. If all points, $b_i$, have already occurred in such a path, we delete it and go on to the next one.

5. To reduce the amount of backtracking, we place the value of each non-negative cycle in the rows of *VALUES* corresponding to the points of such a cycle obtained using the modified F-W algorithm. .

6. Using $T_\alpha$, we place an entry in a row of the new $P_{in}$ for each occurrence of this entry in a branch of the tree.

7. Given a proposed arc $(c_i\ a)$ in row $d$ if an entry already exists in $(d\ a)$, we check to see if the value

$|(d\ ...\ c_i\ a)| - |(d\ a)|$ is one of numbers in row $d$ of *VALUES*, we must backtrack to see if $a$ has already occurred in the given branch. If it *doesn't* occur, and the entry of the arc symmetric to $(c_i\ a)$ doesn't lie in row $d$ containing arcs of all branches of our rooted tree, we don't have to backtrack. Otherwise, we must do so. Although not perfect, this should significantly reduce the amount of backtracking necessary.

This saves us the trouble of backtracking in at least some of the paths that eventually lead to acceptable cycles. A couple of final thoughts. It may make a difference if the tour is obtainable by patching together only acceptable cycles or a linked cycle is included. In the latter case, it may be more likely that our acceptable permutation in theorem 1.13 contains more than one cycle. Secondly, if a tour of smaller value exists, it may be more likely that the acceptable permutation contains only one cycle. Lastly, we might use our algorithm that requires obtaining not only acceptable but also unlinked and linked cycles. Using a larger cycle and a few smaller ones, we try to obtain a reasonably small-valued tour. Once we have such a tour, say $T_{large\ cycle}$, we obtain $\sigma_{T_{large\ cycle}}$. Using the modified F-W algorithm, we then obtain all acceptable permutations, $p_i$, of value less than $|\sigma_{T_{large\ cycle}}|$. Using each $p_i^{-1}M^-$, we try to obtain a still smaller tour by constructing an acceptable cycle containing precisely $\frac{n}{2}$ points.

As a final thought, we can use variations of the modified F-W to obtain near-optimal cycles. Thus, we only extend paths that have more vertices than the path that replaced it. For each determinig vertex, each path chosen is a line on a separate $P_{in}$ generally containing rows with an identical number: one row for each path obtained. Corresponding to each such row is a row in a corresponding $\sigma_T^{-1}M^-$ in which each new value is placed. The main rule hear is that we *can't* change any number in the path in its $P_{in}$ row. For instance, each point in a path can be printed in boldface. The same can be done to the entries corresponding to each entry in $P_{in}$ that describes a subpath of the path.

Let $\sigma$ be a product of $\frac{n}{2}$ pair-wise disjoint 2-cycle in an $n \times n$ symmetric cost matrix M that is the unique PM obtained from a tour $T$. Define $VALUES$ to be a blank list where we will place the sorted values of negative cycles obtained using the modified Floyd-Warshall algorithm. Let $P_{in}$ be the $n \times n$ matrix containing the paths obtained during the $i-th$ iteration of F-W. The following theorem gives necessary and sufficient conditions for each path to yield the same number of edges as the number of arcs comprising it.

Theorem 1.18. *Let $T$ be a tour containing 2m points and $\sigma_T$ the minimally-valued perfect matching associated with it. Let $S = \{C_1, C_2, \ldots, C_r\}$ be the set of acceptable and 2-circuit cycles obtainable from $\sigma_T^{-1} M^-$. Suppose that a subset, $S'$, of $S$ can be linked to form a tour, $T'$ where $|T'| < |T|$. Then if we define a "phantom cycle" $C, C \not\subset S, |C| = 0$, we can always obtain an order of linking the cycles (including at most one "phantom cycle") such that each subset of the ordering has a value less than $|T| - |\sigma_T|$*

Proof. Given any ordered set of numbers having a given sum where we can assume that the numbers lie on a circle, there always exists at least one "determining" number such that each partial sum of the numbers is no greater than the sum.

Theorem 1.19 *Let $\sigma_{ABSOLUTE}$ be a minimally-valued perfect matching in M whi;e $T$ is a tour that is an upperbound for $T_{OPT}$. Then the number of acceptable permutations whose value is less than $|T|$ is the same in $\sigma_{ABSOLUTE}^{-1} M^-$ as in $\sigma_T^{-1} M^-$.*

Proof. The number of perfect matchings less in value than $T$ is fixed. But each perfect matching corresponds to a unique acceptable permutation less than $|T|$ in value. We can obtain all such acceptable permutations using cycles in $\sigma_{ABSOLUTE}^{-1} M^-$ and $\sigma_T^{-1} M^-$, respectively.

*Note.* The value of theorem 1.19 is that an acceptable permutation obtained from cycles in $\sigma_{ABSOLUTE}^{-1} M^-$ is always non-negative. Thus, we don't require Condition (A), i. e., if we can obtain all acceptable permutations each of whose value is less than $T$ using cycles obtained from $\sigma_{ABSOLUTE}^{-1} M^-$ in polynomial time, then we can obtain am optimal tour, $T_{OPT}$, in polynomial time.

Theorem 1.20 *Assume that Condition A is valid. Using the modified F-W algorithm to obtain $T_{FWOPT}$, if we are unable to construct an acceptable cycle or a 2-circuit path in $\sigma_T^{-1} M^-$ containing at least one point of a 2-cycle of $\sigma_T$, then*

*(1) no tour of smaller value than $T$ exists consisting only of acceptable cycles,*

*or,*

*(2) if we enlarge our algorithm to include all 2-circuit paths, either one of them can be extended to construct a 2-circuit cycle containing $a$ or $b$ or else $T_{OPT} = T_{FWOPT} = T$.*

Proof. In order to construct $T_{FWOPT}$, we require a set of acceptable and 2-circuit cycles that can be be linked together to form a tour. Since we don't have a point in a 2-cycle, say $(a\ b)$ in any of the cycles obtained, we can't obtain a tour whose value is less than $T$. But, using the modified F-W algorithm, we can't obtain a tour consisting only of acceptable cycles unless at least one of the cycles containis $a$ or $b$. Thus, $T_{FWOPT} = T$. Furthermore, since no cycle contains $a$ or $b$, there exists no cycle containtaining $a$ or $b$ satisfying Condition A. It follows that in this case, $T_{OPT} = T_{FWOPT} = T$. If there exists a 2-circuit cycle containing $a$ or $b$, at least one of these points must lie in a 2-circuit path, say $P$. However, $P$, may not satisfy Condition A. To remedy this, if we obtain *all* of the 2-circuit paths satsifying Condition A, at least one of them must link up to $P$ to form a 2-circuit cycle unless no tour exists of smaller value than $T$. Thus, if we extend each of the 2-circuit paths to obtain 2-circuit cycles that satsify Condition A, the theorem is valid.

*Note.* Let $\sigma_{ABSOLUTE}$ be a perfect matching whose value is no greater than any other perfect matching obtainable from M. $T_{FWOPT}$ be an upper bound for $T_{OPT}$ Then if we can obtain all acceptable permutations $P$ having the property that

$|P| \leq |T_{FWOPT}|$ in polynomial time, then we can obtain an optimal tour in polynomial time. The reason for this is that given $T_{OPT}$, one set of alternate edges of $T_{OPT}$, say $|\sigma_{T_{OPT}}| \leq |T_{OPT}|$. But $\sigma_{T_{OPT}}$ can be represented as a product of pair-wise disjoint 2-cycles. Such a product is obtainable from a product of pair-wise disjoint acceptable polynomials. (Here pair-wise disjoint means that no pair of cycles has a point or points lying in the same 2-cycle.) If it requires at most polynomial time to obtain all of the cycles in $\sigma_{ABSOLUTE}^{-1} M^{-1}$ whose value is no greater than $T_{FWOPT}$ as well as the products of disjoint cycles obtainable from pair-wise disjoint products of cycles, one of such products must yield the 2-cycle equvalence of $\sigma_{T_{OPT}}$. But $\sigma_{T_{OPT}}^{-1} M^-$ contains a minmal-valued cycle containing $\frac{n}{2}$ points (acceptable cycle), $\frac{n}{2} + 2$ points (2-circuit cycle). One of them yields $T_{OPT}$.

Theorem 1.21 *Let $C = (d_1 \, d_2 \, ... \, d_n)$ be a cycle with $d_1$ a determining vertex of $C$. Then if each partial sum of arcs $s_i = |(d_1 \, d_2)| + |(d_2 \, d_3)| + \, ... \, + |d_{i-1} \, d_i|$ has a non-positive value, $d_i$ is also a determining vertex.*

Proof.
$|d_1 \, d_2| + |d_2 \, d_3| + \, ... \, |d_{i-1} \, d_i| + |d_i \, d_{i+1}| = s_i + |d_i \, d_{i+1}| < |T| - |\sigma_T| \Rightarrow |d_i \, d_{i+1}| < |T| - |\sigma_T| - s_i$

.In general,
$|(d_1 \, d_2)| + |(d_2 \, d_3)| + \, ... \, + |(d_i \, d_{i+1})| \, ... \, + |(d_{i+j} \, d_{i+j+1})| = s_i + |(d_i \, d_{i+1})| \, ... \, + |(d_{i+j} \, d_{i+j+1})| < |T| - |\sigma_T| =$
$|(d_i \, d_{i+1})| + \, ... \, |(d_{i+j} \, d_{i+j+1})| < |T| - |\sigma_T| - s_i$

It follows that since each $s_i$ is non-positive, $d_i$ is a determining vertex.

Theorem 1.22 *Suppose that $u$ and $\sigma(u)$ define a 2-cycle in $\sigma_T^{-1} M^-$ where $u$ is not a determining vertex of an acceptable or 2-circuit cycle $C = (d_1 d_2 \, ... \, d_j u \, d_{j+2} \, ... \, d_m)$. Let $C_{comp}$ be the companion matrix of $C$. Then there either exists a determining point $d_{min}$, of $C$ or a determining point* $d_{comp}$ of $C_{comp}$ such that either every partial sum from $d_{min}$ to $u$ or from $d_{comp}$ to $\sigma_T(d_{j+2})$ has a non-positive value. Furthermore, at least one of the above paths in $C$ or $C_{comp}$ is acceptable.

Proof. Let $C = (d_1 \ldots d_j u d_{j+2} \ldots d_{j+r} d_{j+r+1} \ldots d_m)$ be an acceptable or 2-circuit cycle in $\sigma_T^{-1} M^-$ such that $u$ is not a determining point while $d_1$ is one. Assume $|C| < |T| - |\sigma_T|$. Let $P_{1i} = [d_1 \ldots d_i]$ be a maximum-valued subpath of all subpaths of $P_{1u} = [d_1 \ldots u]$ emanating from $d_1$. Denote it by $P_{max}$ The arc $(d_i\ d_{i+1})$ cannot have a positive value because that would imply that $P_{1(i+1)}$ would have a greater value than $P_{1i}$. Call $d_{i+1}, d_{min}$, if $|P_{max}| > 0$; otherwise $d_{min} = d_1$ since every partial sum $|P_{1i}|$, $1 < i < u$ is non-positive. Thus, assume that $|P_{max}| > 0$. Define $P_{(i+1)u} = (d_{i+1} \ldots u) = P_{min}$. We now prove that $d_{min}$ is a determining point of $C$. Suppose that some subpath $(d_{i+1}\ d_{i'})$ of $(d_{i+1}\ u)$ has a positive value. Then $P_{max} + (d_{i+1}\ d_{i'})$ has a greater value than $P_{max}$. Thus, all of the subpaths from $d_{i+1}$ to $u$ are non-positive. Since $u$ is not a determining vertex, there exists a subpath $(u \ldots d_{j+r})$ of $(u \ldots d_m)$ with $|(u \ldots d_{j+r})| \geq |T| - |\sigma_T|$. But each subpath $(d_{min}\ d_{j+r'})$, of $(d_{min} \ldots d_m)$ has the property that its value is less than $|T| - |\sigma_T| - |P_{max}|$. This follows from the fact the each subpath $(d_1 \ldots d_{j+r'}) = P_{max} + (d_{min} \ldots d_{j+r'})$. Now consider a subpath of the form $(d_{min} \ldots d_m\ d_1 \ldots d_{i'})$, $i' \leq i+1$. By definition, $|P_{max}| \geq |(d_1 \ldots d_{i'})|$. Therefore, $(d_{min} \ldots d_m d_1 \ldots d_{i'}) < |T| - |\sigma_T|$ implying that $d_{min}$ is a determining point of $C$. We now prove that either the path $(d_{min} \ldots u)$ in $C$ or a corresponding path $(d_{min comp} \ldots \sigma_T(d_{j+2}))$ in $C_{comp}$ is acceptable.

$$C_{comp} = (\sigma_T(d_2)\ \sigma_T(d_1)\ \sigma_T(d_m) \ldots \sigma_T(d_{j+r+1})\ \sigma_T(d_{j+r}) \ldots \sigma_T(d_{j+2})\ \sigma_T(u)\ \sigma_T(d_j) \ldots \sigma_T(d_3))$$

Before going further, we note that since $(u\ \sigma_T(d_{j+2})) = (\sigma_T^2(d_{j+2})\ \sigma_T(u)) = (d_{j+2}\ \sigma_T(u))$, these arcs are symmetric. Therefore, they have the same value. The only reason why their values in $\sigma_T^{-1} M^-$ are different is that the diagonal values in their respective rows are different. Continuing, if $\sigma_T(d_{j+2})$ is a determining point, we need go no further since we obtain the same circuit of edges or set of two circuits of edges from $C_{comp}$ as we do from $C$. If $\sigma_T(d_{j+2})$ is not a determining point of $C_{comp}$, as we did earlier, we can prove that we can obtain a set of non-positive partial sums from $\sigma_T(d_{comp})$ to $\sigma_T(d_{j+2})$. Furthermore, $\sigma_T(d_{comp})$ is a determining point of $C_{comp}$. We now prove that at least one of the paths each of whose partial sums is non-positive is acceptable. Suppose that $C$ is a 2-circuit cycle containing both points of a 2-cycle, $(a\ b)$, of $\sigma_T$ where both points lie in the interval $(d_{min}\ u)$. If we correspond a point $v$ of $C$ to $\sigma_T(v)$ of $C_{comp}$, the points of $C_{comp}$ run in a counter-clockwise direction from those of $C$. In general, since $(a\ b)$ is a 2-cycle of $\sigma_T$, $\sigma_T(a) = b$, $\sigma_T(b) = a$. Thus, *both* $a$ and $b$ occur in $C_{comp}$. Consider the following :

(1) From $C$ : $(d_{min}\ d_{i+1}\ ...\ a\ ...\ b\ ...\ u\ d_{j+2}\ ...\ d_m\ d_1\ ...\ )$.

(2) From $C_{comp}$ :

$(\sigma_T(d_{i+1})\ \sigma_T(d_{min})\ ...\ \sigma_T(d_1)\ \sigma_T(d_m)\ ...\ \sigma_T(d_{j+2})\ \sigma_T(u)\ ...\ a\ ...\ b\ ...\ \sigma_T(d_{i+2}))$.

Now consider the case where $a$ lies in $(d_{min}\ ...\ u)$ while $b$ doesn't:

(3) $C = (d_{min}\ d_{i+1}\ ...\ a\ ...\ u\ ...\ b\ ...\ d_{i-1})$

Then

(4) $C_{comp} = (\sigma_T(d_{i+1})\ \sigma_T(d_{min})\ ...\ a\ ...\ \sigma_T(d_{j+2})\ \sigma_T(u)\ ...\ b\ ...\ \sigma_T(d_{i+2}))$

Thus, if one point of $(a\ b)$ occurs in $C$ while the other doesn't, the same is true for $C_{comp}$. It follows that even if $C$ is a linked 2-circuit cycle, at least one of the paths $(d_{min}\ ...\ u)$ or $(d_{comp}\ ...\ \sigma_T(d_{j+2}))$ is acceptable. In the latter case, $\sigma_T(u)$ is the successor point of $\sigma_T(d_{j+2})$.

Theorem 1.23 *Let $(a\ b)$ be a 2-cycle of $\sigma_T$. Then if the arc $(a\ u)$ is not a determining point of a cycle, $C$, containing $a$, $(\sigma_T(u)\ \sigma_T(a))$ is not a determining point of its companion cycle $C_{comp}$.*

Proof. Since M is symmetric, the sequence of edges $/\sigma_T(a)\ a\ \sigma_T(u)\ u| = |a\ \sigma_T(a)\ u\ \sigma_T(u)|$

Theorem 1.24 *Let $T$ be an upper bound for an optimal tour in M. Then, if - using the modified F-W algorithm on $\sigma_T^{-1} M^-$ - we are unable to find an acceptable or 2-circuit cycle that contains a point in a fixed 2-cycle, $II$, of $\sigma_T$, $T = T_{OPT}$.*

Proof. If $T_{OPT}$ has a smaller value than $T$, then, using the modified F-W algorithm on $\sigma_T^{-1} M^-$ we can find a set, $S$, of acceptable and 2-circuit cycles at least one of which contains a point in $II$.

Theorem 1.25 *Let $D$ be a derangement each of whose cycles contains points that are in 2-cycles disjoint from the 2-cycles of each other cycle. Assume that the circuits of edges obtained from the cycles of $D$ cover $n$ points. Suppose that the cycles of $D$ are defined in the following way:*

*(1) the number of points of the acceptable cycles are $\sum_{i=1}^{a} p_{a_i}$; (2) the number of points of its unlinked 2-circuit cycles are $\sum_{i=1}^{i=u} p_{u_i}$; the number of points of its linked two circuits are $\sum_{i=1}^{i=l} p_{l_i}$. Then the number of points in all of its cycles is $\frac{n}{2} + u + 2l$.*

Proof. An acceptable cycle containing $r$ points yields a circuit of edges containing $2r$ points. A 2-circuit cycle containing $s$ points yields two disjoint circuits containing $2s - 2$ points. Since the number of points in a derangement on $n$ points is $n$, we obtain the following formula:

$$n = 2(\sum_{i=1}^{i=a} p_{a_i}) + 2(\sum_{i=1}^{i=u} p_{u_i} - 1) + 2(\sum_{i=1}^{i=l} p_{l_i} - 2)$$

implying that the number of points $p$ in the derangement is

$$\frac{n}{2} = \sum_{i=1}^{i=a} p_{a_i} + \sum_{i=1}^{i=t}(p_{t_i} - 1) = \sum_{i=1}^{i=a} p_{a_i} + \sum_{i=1}^{i=u} p_{u_i} + \sum_{i=1}^{l} p_{l_i} - u - 2l \Rightarrow p = \frac{n}{2} + u + 2l$$

Corollary 1.25a *The number of points in a set of cycles that yield a non-tour derangement is always fewer than a set that yields a tour.*

Proof. We prove that unless our derangement is a tour, then

$$\frac{n}{2} + 3t + a - 1 = \frac{n}{2} + 3(u + l) + a - 1 > \frac{n}{2} + u + 2l.$$

$$\frac{n}{2} + 3(u + l) + a - 1 > \frac{n}{2} + u + 2l \Rightarrow 2u + l + a > 1.$$

The only cases in which the inequality on the right is not valid is if either $l = 1$ or $a = 1$. In either case, we obtain a tour.

Theorem 1.26. *Let $S = \{C_1, C_2, \ldots, C_r\}$ be a set of cycles obtained by applying modified F-W to $\sigma_T^{-1} M^-$ whose elements satisfy the following conditions:*

*(1) $|C_1| = |C_2| = \ldots = |C_r|$,*

*(2) the respective linking points of each one lies in exactly the same set of 2-cycles as each of the others,*

*(3) their respective sets of non-linking points are identical.*

*Then each one of them can be patched to exactly the same set of cycles to yield one or more tours having the same value.*

Proof. Since their respective values are the same, their respective linking points lie in the same set of 2-cycles and their respective sets of linking points are the same, the theorem is valid.

Corollary 1.26a *In order to obtain the smallest-value tour, we may eliminate condition (1) of theorem 1.26, and select a cycle of smallest value that satsifies (2) and (3) of $S$.*

Proof. Each of the cycles in $S$ yields a tour of the same value. By choosing a cycle of smaller value that satisfies (2) and (3), we can obtain a tour of smaller value than obtained using a cycle in $S$.

Corollary 1.26b *Suppose that resepctive sets of acceptable cycles, $S_i$, $i = 1, 2, \ldots$ each satisfies conditions (1) and (2). Furthermore, aussme that we randomly choose $S_{i_\alpha}$ and $S_{i_\beta}$ and that cycles $C_{i_\alpha}$ and $C_{i_\beta}$ have points in pair-wise disjoint 2-cycles, $C_{i_\alpha} \in S_{i_\alpha}$, $C_{i_\beta} \in S_{i_\beta}$. Then any acceptable* *permutation each of whose cycles come from an element of a fixed subset of { $S_i$ | i = 1, 2, ...} yields the same acceptable permutation as any other choice of cycles from the same set of elements.*

Proof. The proof follows directly from the theorem.

*Note.* This theorem will allow us to eliminate all but one product of a set of cycles from $\prod_{i=1}^{i=r} S_i$ satsifying these conditions, i.e., it can cut down the running time for obtaining all perfect matchings substantially each of whose values is less than $T_{UPPERBOUND}$.

Let $\sigma$ be a product of $\frac{n}{2}$ pair-wise disjoint 2-cycle in an $n \, X \, n$ symmetric cost matrix M that is the unique PM obtained from a tour $T$. Define $VALUES$ to be a blank list where we will place the sorted values of negative cycles (including there suffixes) obtained using the modified Floyd-Warshall algorithm. Let $P_{in}$ be the $n \, X \, n$ matrix containing the paths obtained during the $i-th$ iteration of F-W. The following theorem gives necessary and sufficient conditions for each path to yield the same number of edges as the number of arcs comprising it.

*Note.* Using this corollary, we can obtain

`Theorem 1.27` *Let each path, $Q$, in $P_{in}$ satisfy $|Q| < |T| - |\sigma_T|$. Assume that the following are valid:*

*If $Q = (d \, ... \, c')$ is an acceptable path, and we wish to consider adjoining the arc $(c' \, a)$ to $Q$, then suppose*

*(1) either (i) $\sigma(a)$ is not the $\sigma(c')-th$ entry in row d of $P_{in}$, or (ii) if it is, then $(\sigma(a) \, \sigma(c'))$ is not an arc of $Q$;*

*(2) any of the following hold: (i) the entry $(d \, a)$ of $P_{in}$ is blank; (ii) if the entry $(d \, a)$ exists in $P_{in}$, and $|(d \, a)|$ is the value of $(d \, a)$ in $\sigma^{-1}M^-(in)$, then $|(d \, ... \, c' \, a)| - |(d \, a)|$ is not a number in the $a-th$ row of $VALUES$; (iii)*

*VALUES contains no cycle with a determining point $a$ whose value together with its suffixes corresponds to $a$, i.e., the points obtained from backtracking from $a$ show that the path obtained belongs to no suffix.*

*Then $Q' = (d \ ... \ c' \ a)$ is an acceptable path that yields the same number of edges as it has arcs.*

Proof. Assuming the our conclusion is true for $Q$, we must prove the following for $Q'$: (a) it contains no pair of symmetric arcs and (b) it is a simple path. (a) Given an arc $(c' \ a)$, its symmetric arc is $(\sigma(a) \ \sigma^{-1}(c'))$. But $\sigma$ is a product of $\frac{n}{2}$ pair-wise disjoint 2-cycles. Thus, $\sigma^2(c') = c' \Rightarrow \sigma(c') = \sigma^{-1}(c')$. We now prove that $Q'$ contains no cycle. From the nature of F-W, this only occurs if the path $Q'$ contains a negative cycle, say $C_{NEG} = (a \ b' \ ... \ c')$. From theorem 1.1, the negative cycle must occur in row $a$ before it can occur in row $d$. If that is the case, we obtain it before it can occur in $Q'$. But once it is obtained, we place its value in *VALUES* along with its unique suffix. .

If 2(i) is the case, then $a$ can't be a point of $Q$. 2(ii): if a negative circuit containing $a$ exists, then we must have already obtained it. Furthermore, $|(d \ ... \ c' \ a)| - |(d \ a)|$ must be its value. If this number along with its suffix isn't in *VALUES*, then such a negative circuit doesn't exist. 2(iii): Suppose the value of $|(d \ ... \ c' \ a)| - |(d \ a)|$ is in *VALUES*. Since every negative cycle has a unique suffix corresponding to its value and determining (and initial) point, if when backtracking from $a$ we obtain a point not in its suffix, then we may add $(c' \ a)$ to $(d \ ... \ c')$. On the other hand, if backtracking yields the suffix of $a$, then we can't add the arc $(c' \ a)$.

*Note.* Theorem 1.27 allows us to test all possible extensions of paths without doing extensive backtracking. The only case where such backtracking might be necessary is when the entry $(\sigma_T(a) \ \sigma_T(c'))$ occurs.

`Corollary 1.27a` *When a path satisfying (1) and (2) becomes a permutation cycle, $C$, the number of points in $C$ equals the number of arcs in $C$.*

Proof. Follows directly from theorem 1.2.

`Corollary 1.27b` *Suppose we choose a set $S$ of pair-wise disjoint cycles that such*

*(a) each cycle in $S$ satisfies theorem 1.2,*

*(b) each 2-cycle of $\sigma$ contains at least one point of a cycle in $S$.*

*Then when the product of the cycles of $S$ are multiplied by $\sigma$, we obtain a derangement.*

Proof. (a) The arcs of the cycles in $S$ yield directed edges no two of which are symmetric. Thus, they are distinct edges. (b) By hypothesis, we have obtained a set of distinct edges such that each 2-cycle has at least one point of an edge incident to it. Let the cycles in $S$ form the permutation $s$. Consider $\sigma s$. It must be a permutation. Furthermore, it can't have any identity elements. This could only occur if both points of a 2-cycle of $\sigma$ haven't been acted upon by $s$ or if a new set of symmetric arcs has been formed. From (b), the former can't occur, while (a) rules out the latter. It follows that we have obtained a derangement of edges.

`Corollary 1.27c` *Suppose the respective cycles in $S$ have sets of points in pair-wise disjoint sets of 2-cycles such that*

*(a) the arcs in each cycle in $S$ satisfy the hypotheses of theorem 1.2,*

*(b) the set of 2-cycles containing one or more points of a cycle in $S$ is disjoint from the sets of 2-cycles containing points of the other cycles in $S$,*

*(c) each 2-cycle of $\sigma$ contains at least one point of a cycle in $S$.*

*Then when the product of the cycles of $S$ are multiplied by $\sigma$, we obtain a derangement.*

*Note*. Alternately, we can directly construct the derangement by transforming each cycle into a circuit containing twice as many edges.

IV. THE ALGORITHM

PHASES 1 and 2 are explained in the examples in V.

PHASE 3

Suppose we have obtained $m$ acceptable or 2-circuit cycles by applying the modified F-W to $\sigma_T^{-1} M^-$. Then we construct an $m \times \frac{n}{2}$ matrix the headings of whose columns are the 2-cycles of $\sigma_T$. We then place the respective linking points and pairs of non-linking points in the columns of the 2-cycles containing them. In an unlinked 2-circuit cycle, we print $L_1$ after a linking point in the first circuit and $L_2$ after a linking point in the second circuit. Using theorem 26, we note that that if two cycles with the same value have their respective linking points and non-linking points in exactly the same columns, then we may delete one of them. The reason is that each of them can be linked to the same cycle. The tree obtained will yield a circuit or circuits of directed edges whose total value is the same in both cases. The same is true of a pair of companion cycles or a cycle being represented by more than one determining point. Thus, we need keep only one of such a set of cycles. We next assign a row number to each cycle such that cycles containing points in the largest number of 2-cycle columns have their points placed first. The cycles are placed in increasing order of magnitude. Their values as well as the number of columns in which it has points are placed next to their row numbers. Thus, cycles containing $\frac{n}{2} - 1$ points are entered in the matrix first, then those containing $\frac{n}{2}$ points, etc.. We now check the columns of our matrix to see which column has the fewest entries. Assume that the column having the fewest entries is headed by $(a\ \sigma_T(a))$. By cutting and pasting, we place each cycle having an entry in that column in the top rows of the matrix. These cycles are placed in order of increasing value. We now use theorem 1.14. Thus, any set of linked cycles has the property that the sum of the values of its cycles is less than $|T| - |\sigma_T|$. We assume that – if necessary – a phantom cycle may be included in such a set. Now let $C_1^*$ be the cycle of smallest value that contains a point in column $(a\ \sigma_T(a))$. Starting with the smallest-valued cycle, $C_1$, in the set of cycles having the most points, we first check to see if it can be linked to a point in $(a\ \sigma_T(a))$. If it can be linked, we link it. The value of the linked cycles must be less than $|T| - |\sigma_T|$. Now suppose that it can't be linked, We assume that a tour must include it. Therefore, from this point on, each set of linked cycles must have a value no greater than $|T| - |\sigma_T| - |C_1^*|$. We next want to obtain all cycles that can be linked to $C_1$. The simplest way to do this is check the columns *not* containing a point of $C_1$. It is worth noting that if $/C_1/ = s$, then if $D_1$ can be linked to $C_1$, it can have no have points in no more than $\frac{n}{2} - s + 1$ columns. Thus, we need search the columns only of rows with cycles no greater than entries in $\frac{n}{2} - s + 1$ columns. $D_i$ may be linked to $C_1$ with a phantom cycle. We also note that $/C_1/ + /D_1/ < /T/ - /\sigma_T/ - |C_1^*|$. On the other hand, if a cycle has $r$ points in columns *not containing* points in $C_1$, we check to see its column number (next to its value and row number). If it is $r + 1$ and the column contains a linking point in both its row and the row of $C_1$, then it can be linked to $C_1$ provided its value is no greater than $/T/ - /\sigma_T/ - |C_1^*|$ or $|T| - |\sigma_T|$ as the case may be. Once a phantom cycle has been used, we place a $P$ in any cycle that may be linked to it as well as its descendants so that we can tell if a branch contains one. As mentioned earlier, only one phantom cycle can occur in a branch. However, no branch that ends with a $"P"$ cycle can yield a tour. If a cycle occurs that links to $C_1$, we no longer use $"P"$ with it and any descendants it may have. Also, if $C_1^*$ is linked to a structure, then our new upper bound for linking cycles to that structure becomes $/T/ - /\sigma_T/$ once again. If a tour is obtained, its value becomes a new upper bound for the construction of all branches. Any branch that has a value that isn't smaller than the new upper bound is being form less than $/T/ - /\sigma_T/$ in value. Also, we use our formula for points and our inequality for the number of cycles in the tree to insure that our tree has the possibility of yielding a tour. At the bottom of our list of cycles, we add rows in which of which we place a tree being formed. Suppose $C_1$ and $D_j$ are linked. The new tree formed is called $TREE_{i,j}$. We denote the linking of another cycle to $TREE_{i,j}$ as $TREE_{i,j,k}$, etc.. In the column, say $(a\ b)$, where 2-cycles are deleted to form an addition to a tree, we place an L after the point (or points) of respective cycles that have been placed in the column. If we are linking an unlinked 2-circuit cycle, we place LL after a point where a 2-cycle has been deleted to extend a branch of the tree. We continue this procedure with $C_i$, $i = 1,2, ...$ as well as with $C_i^*, i = 1,2, ...$. Each row represents a separate tree being formed to construct a possible tour. Before doing a linking, we check to see if the new cycle being linked is $C_i^{'}$. If it is, we adjust the value of the tree to which it is being attached. Each of the new rows corresponds to a different tree. If we have not obtained a tour using each $C_2'$, we cannot obtain a tour using the modified F-W.

We now discuss obtaining an optimal tour. We use the entries in $\sigma_{FWOPT}^{-1} M^-$ to construct cycles. Our first task is to construct *all* cycles that have the same determining point as each $C_i'$ (they all have the same determining point). Let $d'$ be the determining point of $C_i'$. We construct a tree whose root is $d'$. We add arcs to $d'$ to form branches each of whose value is less than $|T_{FWOPT}| - |\sigma_{T_{FWOPT}}|$. For simplicity, we henceforth denote $T_{FWOPT}$ by $T$ and $\sigma_{T_{FWOPT}}$ by $\sigma_T$. Using $TREE_{d'}$, we can obtain any cycles whose determining point is $d'$. However, there may be points in a cycle containing $d'$ that are not determining points. From theorem 1.22, we may obtain such cycles by obtaining all acceptable non-positive paths that terminate in $d'$. From that point onward, our branch may have any value less than $|T| - |\sigma_T|$. This greatly reduces the number of branches we have to check. However, a problem exists in that if the cycle containing $d'$ is a 2-circuit, we can't use the root of the tree as one of the two vertices belonging to the same 2-cycle. We thus adopt the following convention: the first point encountered we print in red inside a red square. The second point is printed in black inside a black square. If it is a linked 2-circuit, the first point of the second 2-cycle that is inside the two squares is written in black inside a red circle, while the second one is written in red inside a black circle. To ne more precise: let a point reached in back-tracking be $p$. Assume that we have already passed a point, $u$, a black point in side a black square. Then if we have not reached $\sigma_T(u)$, we accept $\sigma_T(p)$ as point of a linked 2-circuit cycle. In this case, we print $p$ as a black numeral in a red circle, $\sigma_T(p)$ as a red numeral in a black circle. If $p$ has passed $\sigma_T(u)$, we delete $\sigma_T(p)$ and the arc of which it is the terminal point. In the first case, the links connecting the black point - black circle points that follow $\sigma_T(p)$ are dashed black line segments. From that point on, the branch can only have acceptable points. Suppose the root of the current tree we are constructing is $i$. The next question is how we extend each branch $(i \ldots p)$. Our method is to obtain all 2-arcs paths from $p$. We then backtrack, arc by arc, from $p$ to see if, using a extension of $p$, we have obtained an interior cycle or the companion point to a point in one of the 2-arc paths. Suppose we have backtracked to point $q$. If $q$ exists in a 2-arc path, we delete it as well as any arcs extending from it. If the point $\sigma_T(q)$ exists in one of the 2-arc paths, we check to see if $q$ is a red point in a red square. If so, $\sigma_T(q)$ is printed as a black point in a black square. If $q$ is a black point in a red circle, we check to see if $\sigma_T(q)$ is in italics. If it isn't, we print $\sigma_T(q)$ as a red point in a black circle. If it is, we delete $\sigma_T(q)$ as well as all arcs emanating from it. We continue this way until we have obtained all possible branches. Suppose the root of the tree is $i$. If we reach a point , $i$, we have obtained a cycle. If the cycle doesn't contain $d'$, we no longer extend the branch ending in $i$. If the cycle contains $d'$, we backtrack to the root, $i$, of the tree printing the last point of the cycle, then the next to last point, etc,. until we reach $i$. When we have constructed all non-positive trees using all allowable points as roots, we have obtained all cycles containing $d'$ where $d'$ isn't a determining point. Using these cycles together with those having $d'$ as a determining point, we call the set of such cycles $S_{d'}$ Any tour of smaller value than $T_{FWOPT}$ must either be derivable from a cycle in $S_{d'}$ or else it can be derived from an element, $C$, of $S_{d'}$ together with one or more cycles that can be linked to it or extended from a branch of $TREE_{S_{d'}}$. We now construct acceptable or 2-circuit paths using the set of 2-cycles not containing a point of $C$. In our constructions, we allow one linking point to link to a linking point in $C$. If $C$ is an unlinked 2-circuit cycle, we allow two points lying in different "circuits" of $C$. Once we have obtained all cycles, we attempt to construct a tour by linking them to each other and $C$. We do this with each cycle in $S_{d'}$. Either we obtain a tour smaller in value than $T_{FWOPT}$ or $T_{OPT} = T_{FWOPT}$.

*Note.* It is worth noting that we may have variations of PHASES 1 and 2. For instance, we can go directly to PHASE 2 to obtain a derangement whose cycles can be patched together to yield an upper bound for $T_{FWOPT}$ and $T_{OPT}$. Alternately, we can use Helsgaun's algorithm [1] to obtain one.

V. EXAMPLES

To illustrate PHASE 1, we give the following example. The symmetric cost matrix, M, has been chosen randomly from positive integers no greater than 99.

```
Example 4
```

M

|    | 1  | 2  | 3  | 4  | 5  | 6  | 7  | 8  | 9  | 10 | 11 | 12 | 13 | 14 | 15 | 16 | 17 | 18 | 19 | 20 |    |
|----|----|----|----|----|----|----|----|----|----|----|----|----|----|----|----|----|----|----|----|----|----|
| 1  | ∞  | 26 | 4  | 30 | 74 | 5  | 4  | 38 | 28 | 78 | 81 | 7  | 97 | 10 | 94 | 40 | 98 | 49 | 40 | 70 | 1  |
| 2  | 26 | ∞  | 69 | 30 | 41 | 80 | 50 | 74 | 1  | 60 | 9  | 9  | 31 | 87 | 89 | 91 | 6  | 82 | 23 | 85 | 2  |
| 3  | 4  | 69 | ∞  | 23 | 7  | 61 | 60 | 98 | 99 | 90 | 84 | 57 | 4  | 56 | 66 | 30 | 51 | 3  | 25 | 47 | 3  |
| 4  | 30 | 30 | 23 | ∞  | 33 | 59 | 5  | 92 | 26 | 48 | 84 | 18 | 57 | 28 | 47 | 25 | 81 | 48 | 70 | 17 | 4  |
| 5  | 74 | 41 | 7  | 33 | ∞  | 82 | 29 | 80 | 5  | 87 | 87 | 97 | 55 | 45 | 72 | 94 | 20 | 9  | 90 | 20 | 5  |
| 6  | 5  | 80 | 61 | 59 | 82 | ∞  | 1  | 6  | 43 | 9  | 39 | 41 | 56 | 45 | 62 | 38 | 50 | 52 | 41 | 50 | 6  |
| 7  | 4  | 30 | 60 | 5  | 29 | 1  | ∞  | 34 | 78 | 49 | 73 | 10 | 56 | 36 | 87 | 31 | 45 | 59 | 88 | 42 | 7  |
| 8  | 38 | 74 | 98 | 92 | 80 | 6  | 34 | ∞  | 14 | 55 | 43 | 91 | 85 | 93 | 75 | 17 | 64 | 78 | 60 | 41 | 8  |
| 9  | 28 | 1  | 99 | 26 | 5  | 43 | 78 | 14 | ∞  | 50 | 28 | 81 | 98 | 95 | 25 | 31 | 73 | 63 | 87 | 56 | 9  |
| 10 | 78 | 60 | 90 | 48 | 87 | 9  | 49 | 55 | 50 | ∞  | 37 | 24 | 95 | 59 | 30 | 25 | 8  | 90 | 64 | 36 | 10 |
| 11 | 81 | 9  | 84 | 84 | 87 | 39 | 73 | 43 | 28 | 37 | ∞  | 34 | 61 | 14 | 11 | 3  | 3  | 74 | 22 | 26 | 11 |
| 12 | 7  | 9  | 57 | 18 | 97 | 41 | 10 | 91 | 81 | 24 | 34 | ∞  | 84 | 62 | 56 | 34 | 95 | 17 | 71 | 30 | 12 |

| 13 | 97 | 31 | 4 | 57 | 55 | 56 | 56 | 85 | 98 | 93 | 61 | 84 | ∞ | 66 | 66 | 30 | 49 | 82 | 23 | 86 | 13 |
| 14 | 10 | 87 | 56 | 28 | 45 | 45 | 36 | 93 | 95 | 59 | 14 | 62 | 66 | ∞ | 25 | 17 | 47 | 47 | 55 | 5 | 14 |
| 15 | 94 | 89 | 66 | 47 | 72 | 62 | 87 | 75 | 25 | 30 | 11 | 56 | 66 | 25 | ∞ | 89 | 85 | 87 | 8 | 2 | 15 |
| 16 | 40 | 91 | 30 | 25 | 94 | 38 | 31 | 17 | 31 | 25 | 3 | 34 | 30 | 17 | 89 | ∞ | 54 | 18 | 92 | 90 | 16 |
| 17 | 98 | 6 | 51 | 81 | 20 | 50 | 45 | 64 | 73 | 8 | 3 | 95 | 49 | 47 | 85 | 54 | ∞ | 66 | 48 | 90 | 17 |
| 18 | 49 | 82 | 3 | 48 | 9 | 52 | 59 | 78 | 63 | 90 | 74 | 17 | 82 | 47 | 87 | 18 | 66 | ∞ | 87 | 16 | 18 |
| 19 | 40 | 23 | 25 | 70 | 90 | 41 | 88 | 60 | 87 | 64 | 22 | 71 | 23 | 55 | 8 | 92 | 48 | 87 | ∞ | 22 | 19 |
| 20 | 70 | 85 | 47 | 17 | 20 | 50 | 42 | 41 | 56 | 36 | 26 | 30 | 86 | 5 | 2 | 90 | 90 | 16 | 22 | ∞ | 20 |
|  | 1 | 2 | 3 | 4 | 5 | 6 | 7 | 8 | 9 | 10 | 11 | 12 | 13 | 14 | 15 | 16 | 17 | 18 | 19 | 20 |  |

MIN(M)

|  | 1 | 2 | 3 | 4 | 5 | 6 | 7 | 8 | 9 | 10 | 11 | 12 | 13 | 14 | 15 | 16 | 17 | 18 | 19 | 20 |  |
|---|---|---|---|---|---|---|---|---|---|---|---|---|---|---|---|---|---|---|---|---|---|
| 1 | 3 | 7 | 6 | 12 | 14 | 2 | 9 | 4 | 8 | 16 | 19 | 18 | 20 | 5 | 10 | 11 | 15 | 13 | 17 | 1 | 1 |
| 2 | 9 | 17 | 11 | 12 | 19 | 1 | 4 | 7 | 13 | 5 | 10 | 3 | 8 | 6 | 20 | 18 | 14 | 15 | 16 | 2 | 2 |
| 3 | 18 | 13 | 1 | 5 | 4 | 19 | 16 | 20 | 17 | 14 | 12 | 7 | 6 | 15 | 2 | 11 | 10 | 8 | 9 | 3 | 3 |
| 4 | 7 | 20 | 12 | 3 | 16 | 9 | 14 | 1 | 2 | 5 | 15 | 10 | 18 | 13 | 6 | 19 | 17 | 11 | 8 | 4 | 4 |
| 5 | 9 | 3 | 18 | 17 | 20 | 7 | 4 | 2 | 14 | 13 | 15 | 1 | 8 | 6 | 10 | 11 | 19 | 16 | 12 | 5 | 5 |
| 6 | 7 | 1 | 8 | 10 | 16 | 11 | 12 | 19 | 9 | 14 | 17 | 20 | 18 | 13 | 4 | 3 | 15 | 2 | 5 | 6 | 6 |
| 7 | 6 | 1 | 4 | 12 | 5 | 2 | 16 | 8 | 14 | 20 | 17 | 10 | 13 | 18 | 3 | 11 | 9 | 15 | 19 | 17 | 7 |
| 8 | 6 | 9 | 16 | 7 | 1 | 20 | 11 | 10 | 19 | 17 | 2 | 15 | 18 | 5 | 13 | 12 | 4 | 14 | 3 | 8 | 8 |
| 9 | 2 | 5 | 8 | 15 | 4 | 1 | 11 | 16 | 6 | 10 | 20 | 18 | 17 | 7 | 12 | 19 | 14 | 13 | 3 | 9 | 9 |
| 10 | 17 | 6 | 12 | 16 | 15 | 20 | 11 | 4 | 7 | 9 | 8 | 14 | 2 | 19 | 1 | 5 | 3 | 18 | 13 | 10 | 10 |
| 11 | 16 | 17 | 2 | 15 | 14 | 19 | 20 | 9 | 12 | 10 | 6 | 8 | 13 | 7 | 18 | 1 | 3 | 4 | 5 | 11 | 11 |
| 12 | 1 | 2 | 7 | 18 | 4 | 10 | 20 | 11 | 16 | 6 | 15 | 3 | 14 | 19 | 9 | 13 | 8 | 17 | 5 | 12 | 12 |
| 13 | 3 | 19 | 16 | 2 | 17 | 5 | 6 | 7 | 4 | 11 | 14 | 15 | 18 | 12 | 8 | 20 | 10 | 1 | 9 | 13 | 13 |
| 14 | 20 | 1 | 11 | 16 | 15 | 4 | 7 | 5 | 6 | 17 | 18 | 19 | 3 | 10 | 12 | 13 | 2 | 8 | 9 | 14 | 14 |
| 15 | 20 | 19 | 11 | 9 | 14 | 10 | 4 | 12 | 6 | 3 | 13 | 5 | 8 | 17 | 7 | 18 | 2 | 16 | 1 | 15 | 15 |
| 16 | 11 | 8 | 14 | 18 | 4 | 10 | 13 | 3 | 7 | 9 | 12 | 6 | 1 | 17 | 15 | 20 | 2 | 19 | 5 | 16 | 16 |
| 17 | 11 | 2 | 10 | 5 | 7 | 14 | 19 | 13 | 6 | 3 | 16 | 8 | 18 | 9 | 4 | 15 | 20 | 12 | 1 | 17 | 17 |
| 18 | 3 | 5 | 20 | 12 | 16 | 14 | 4 | 1 | 6 | 7 | 9 | 17 | 11 | 8 | 2 | 13 | 15 | 19 | 10 | 18 | 18 |
| 19 | 15 | 11 | 20 | 2 | 13 | 3 | 1 | 6 | 17 | 14 | 8 | 10 | 4 | 12 | 9 | 18 | 7 | 5 | 16 | 19 | 19 |
| 20 | 15 | 14 | 18 | 4 | 5 | 19 | 11 | 12 | 10 | 8 | 7 | 3 | 6 | 9 | 1 | 2 | 13 | 16 | 17 | 20 | 20 |
|  | 1 | 2 | 3 | 4 | 5 | 6 | 7 | 8 | 9 | 10 | 11 | 12 | 13 | 14 | 15 | 16 | 17 | 18 | 19 | 20 |  |

$D_0 = (1^{26}\ 2^{69}\ 3^{23}\ 4^{63}\ 5^{82}\ 6^1\ 7^{34}\ 8^{14}\ 9^{50}\ 10^{37}\ 11^{34}\ 12^{84}\ 13^{66}\ 14^{25}\ 15^{89}\ 16^{54}\ 17^{66}\ 18^{87}\ 19^{22}\ 20^{20}\ )$

We now change $D_0$ into a form that can be used in our algorithm. In place of $a^{d(a\ D_0(a))}$, we put $a^{d(a\ MIN(M)(a,1)) - d(a\ D_0(a))}$ yielding

$D_0 = (1^{-22}\ 2^{-68}\ 3^{-20}\ 4^{-26}\ 5^{-77}\ 6^0\ 7^{-33}\ 8^{-8}\ 9^{-49}\ 10^{-29}\ 11^{-31}\ 12^{-77}\ 13^{-62}\ 14^{-20}\ 15^{-87}\ 16^{-51}\ 17^{-63}\ 18^{-84}\ 19^{-14}\ 20^{-68}\ )$

We place $D_0$ in *row form*, i.e., we place the numbera from 1 to n consecutively in the top row, while their functional values are placed in the bottom row.

$D_0 =$

|   | -22 | -68 | -20 | -28 | -77 | 0 | -33 | -8 | -49 | -29 | -31 | -77 | -62 | -11 | -87 | -51 | -63 | -84 | -14 | -68 |
|---|---|---|---|---|---|---|---|---|---|---|---|---|---|---|---|---|---|---|---|---|
|   | 1 | 2 | 3 | 4 | 5 | 6 | 7 | 8 | 9 | 10 | 11 | 12 | 13 | 14 | 15 | 16 | 17 | 18 | 19 | 20 |
|   | 2 | 3 | 4 | 5 | 6 | 7 | 8 | 8 | 10 | 11 | 12 | 13 | 14 | 15 | 16 | 17 | 18 | 19 | 20 | 1 |

$D_0^{-1}$

|   | 1 | 2 | 3 | 4 | 5 | 6 | 7 | 8 | 9 | 10 | 11 | 12 | 13 | 14 | 15 | 16 | 17 | 18 | 19 | 20 |
|---|---|---|---|---|---|---|---|---|---|---|---|---|---|---|---|---|---|---|---|---|
|   | 20 | 19 | 18 | 17 | 16 | 15 | 14 | 13 | 12 | 11 | 10 | 9 | 8 | 7 | 6 | 5 | 4 | 3 | 2 | 1 |

Let the maximum number of blocks in a row used in a trial be $T = [log\ n] + 1$. Our first step is to sort all vertices in an ascending manner with respect to $d(a,\ D_i(a)) - d(a,\ MIN(M)(a,1))$. Let $v$ be the point with the smallest difference value. Our initial point in the first trial is $v$. We choose $(v\ MIN(M)(v,1))$ as our first arc. When we reach a point previously chosen, *if the point is the initial point of the cycle*, we conclude the trial using the cycle obtained. Otherwise, if each element of the block following it has a negative value, we can choose vertices in it until we reach a vertex that hasn't been chosen. If all of the vertices in the block have been previously chosen, we can try the next block. We may continue this procedure until we have gone through at most $T$ blocks. Each point chosen must have a negative value associated with it. If we reach an arc $A = (a\ D_i(a)),\ i = 0,1,2,...$, we may choose any other point that lies in the same block. If no other vertex lies in that block, we may choose any vertex in the next block. If we still cannot find a vertex that has not been repeated, we check each possible cycle to see which one has the smallest value, say $s_i$. This completes one trial. The first arc chosen in the second trial is $(v\ MIN(M)(v,2))$. The maximum number of trials is $T$ Place the edges obtained from $s_i$ into the row form for $D_i$ in place of the edges with corresponding initial vertices. The new derangement obtained is $D_{i+1}$ Now sort the vertices of $D_{i+1}$ The first vertex obtained in the sorting is a point used in the second trial. These comments roughly sketch the algorithm in the asymmetric case.

We now go over changes we must make to the algorithm used in the asymmetric case.

PHASE 1.

(1) Before choosing an arc $A = (a\ b)$ in a permutation to be applied to $D_i$, we check the row form of $D_i$ to see if the arc $(D_i(b)\ a)$ is in the $D_i(b)$ column. If it is, we can't use arc $A$ in the permutation.

(2) During each trial, as we place an arc $(a\ b)$ in a permutation, we place both the arc $(a\ D_i(b))$ and the arc $(D_i(b)\ a)$ in a list sorted in increasing order of magnitude of the initial integer of the arc. Before a new arc $(a\ b)$ is added to a permutation, we check to see if $(D_i(b)\ a)$ is in the list. If it is, we can't place $(a\ b)$ in the permutation.

TRIAL 1. MIN(M)(15,1) = 20.

(15 20) → (15 19) -87

| | | |
|---|---|---|
| (19 15) → (19 14) -14 | 2. (2 9) | 12. |
| (14 20) repeat | 3. (3 16) | 13. (13 3) |
| (14 1) → (14 20) -5 | | 14. (14 1) |
| (20 15) not allowed | 5. (5 18) | 15. (15 20) |

(20  14) → (20  13) -65

(13  3) → (13  2) -62

(2  9) → (2  8) -68

(8  6) → (8  5) -8

(5  9) repeat

(5  3) repeat

(5  18) → (5  17) -73

(17  11) → (17  10) -63

(10  17) → (10  16) -29

(16  10) repeat

(16  8) → (16  7) -14

(7  6) repeat

(7  1) repeat

(7  4) → (7  3) -29

(3  13) not allowed

(3  6) repeat

(3  5) repeat

16. (16  8)

7. (7  4)    17. (17  11)

8. (8  6)

19. (19  15)

10. (10  17)    20. (20  14)

We can't proceed further since $[log\ 20] + 1 = 3$. We thus must form a cycle using (3  16).

(3  16) → (3  15) +7

Since all of the differences are negative except the last one, and 7 is small, the permutation with the smallest difference value is (15  19  14  20  13  2  8  5  17  10  16  7  3): -516

TRIAL 2. MIN(15, 2) = 19

(15  19) → (15  18) -79

(18 3) → (18 2) -84                    2. (2 9)

(2 8)(2 9) → (2 8) -68

(8 6) → (8 5) -8                                               14. (14 16)

(5 18) → (5 17) -62                    5. (5 18)               15. (15 19)

(10 17) → (10 16) -29                                          16. (16 8)

(16 8) → (16 7) -37                    7. (7 1)                17. (17 11)

(7 1) → (7 20) -30                                             18. (18 3)

(20 15) → (20 14) -68

(14 16) → (14 15) -8                   10. (10 17)             20. (20 15)

Our cycle is (15 18 2 8 5 17 10 16 7 20 14): -536

TRIAL 3. MIN(M)(15, 3) = 11

(15 11) → (15 10) -78

(10 17) → (10 16) -29

(16 11) repeat

(16 8) → (16 7) -14                    5. (5 9)                15. (15 11)

(7 6) → (7 5) -33                                              16. (16 8)

(5 9) → (5 8) -77                      7. (7 6)

(8 6) repeat                           8. (8 16)

(8 9) arc of $D_0$

(8 16) → (18 15) +3                    10. (10 17)

Our cycle is (15 10 16 7 5 8): -228

$s$ is the cycle obtained in Trial 2: (15 18 2 8 5 17 10 16 7 20 14). Our new edges are (15 19), (18 3), (2 9), (8 6), (5 18), (17 11), (10 17), (7 1), (20 15). $D_1 = D_0 s$. We thus obtain:

```
     -22  0 -20 -28 -4   0  -3  0 -49   0  -31 -77 -62 -12 -6 -14   0   0 -14   0

      1   2   3   4   5   6   7   8   9  10   11  12  13  14 15  16  17  18  19  20
```

$D_1 =$

```
      2   9   4   5  18   7   1   6  10  17   12  13  14  16 19   8  11   3  20  15

      1   2   3   4   5   6   7   8   9  10   11  12  13  14 15  16  17  18  19  20
```

$D_1^{-1} =$

```
      7   1  18   3   4   8   6  18   2   9   17  11  12  13 20  14  10   5  15  19
```

Choose 12.

TRIAL 1.  MIN(M)(12,1) = 1                           .

(12  1)  →  (12  7)  -77                                              12. (12  1)

(7  6)  not allowed                                  3. (3  13)

(7  1)  arc repeat

(7  4)  →  (7  3)  +1

(3  18)  not allowed

(3  13)  →  (3  12)  -19                             7. (7  4)

                                          .

Our cycle is (12  7  3): -95.

Trial 2.  MIN(M)(12,2) = 2                                    1. (1  3)

(12  2) → (12  1) -78                                         12. (12  2)

(1  3) → (1  18) -22

(18  3)  arc

(18  5)  not allowed

(18  20) → (18  19) +13

(19  15)  not allowed

(19  11) → (19  17) 0                   8.                    18. (18  20)

(17  11)  arc repeat                    9.                    19. (19  13)

(17  2)  repeat

We can't go further, d(17  13) - d(17  11) = +46. d(19  13) - d(19  20) = 23 - 22 = 1.

d(18  13) - d(18  3) = 82 - 3 = 79. d(1  13) - d(1  2) = 97 - 26 = 51. Thus, our smallest valued cycle is (12  1  18  19): -86:

Trial 3.  MIN(M)(12,3) = 7

                                                              12. (12  7)

(12  7) → (12  6) -74                   3. (3  13)

(6  7)  arc repeat

(6  1) → (6  7) +4

(7  6)  not allowed                     6. (6  1)

(7  1)  arc                             7. (7  4)

(7 4) → (7 3) +1

(3 18) not allowed

(3 13) → (3 12) -19

Our smallest-valued cycle is obtained from Trial 1: (12 7 3): -95. Our new edges are (12 1), (7 4), (3 13).:

| -22 | 0 | -1 | -28 | -4 | 0 | -4 | 0 | -49 | 0 | -31 | -3 | -62 | -12 | -6 | -14 | 0 | 0 | -14 | 0 |
|---|---|---|---|---|---|---|---|---|---|---|---|---|---|---|---|---|---|---|---|
| 1 | 2 | 3 | 4 | 5 | 6 | 7 | 8 | 9 | 10 | 11 | 12 | 13 | 14 | 15 | 16 | 17 | 18 | 19 | 20 |

$D_2 =$

| 2 | 9 | 13 | 5 | 18 | 7 | 4 | 6 | 10 | 17 | 12 | 1 | 14 | 16 | 19 | 8 | 11 | 3 | 20 | 15 |
|---|---|---|---|---|---|---|---|---|---|---|---|---|---|---|---|---|---|---|---|

| 1 | 2 | 3 | 4 | 5 | 6 | 7 | 8 | 9 | 10 | 11 | 12 | 13 | 14 | 15 | 16 | 17 | 18 | 19 | 20 |
|---|---|---|---|---|---|---|---|---|---|---|---|---|---|---|---|---|---|---|---|

$D_2^{-1}$

| 12 | 1 | 18 | 7 | 4 | 8 | 6 | 16 | 2 | 9 | 17 | 11 | 3 | 13 | 20 | 14 | 10 | 5 | 15 | 19 |
|---|---|---|---|---|---|---|---|---|---|---|---|---|---|---|---|---|---|---|---|

Choose 13.

Trial 1. MIN(M)(13,1) = 3                           1. (1 3)

(13 3) not allowed

Trial 2. MIN(M)(13,2) = 19                          13. (13 19)

(13 19) → (13 15) -43

(15 20) not allowed                                 15. (15 11)

 (15 19) arc

(15 11) → (15 17) +3                                17. (17 2)

(17 11) repeat                                      18. (18 20)

(17 2) → (17 1) +3

(1 3) → (1 18) -22

(18 3) arc

(18 5) not allowed

(18 20) → (18 19) +13

(19 15) not allowed

(19 11) repeat

(19 20) arc

(19 2) repeat positive

We can go no further  d(19 14) - d(19 11) = +33. d(18 14) - d(18 3) = +44. . d(1 14) - d(1 2) = -16.. d(17 14) - d(17 11) = +44. d(15 14) - d(15 19) = +17.  Our smallest-valued cycle is (13 15 17 1): -62

Trial 3.  MIN(M)(13,3) = 16                              1. (1 3)                  .

(13 16) → (13 14) -36                                                             .

(14 20) → (14 19) -12                                                      13. (13 15)

(19 15) not allowed                                                         14. (14 20)

(19 11) → (19 17) 0

(17 11) arc

(17 2) → (17 1) +3                                                          17. (17 2)

(1 3) → (1 18) -22

(18 3) arc                                                                        19. (19 11)

(18 5) not allowed

(18 20) repeat positive

We can go no further. d(18 14) - d(18 3) = 44. d(1 14) - d(1 2) = -16. d(17 14) - d(17 11) = 44.

d(19 14) - d(19 20) = 55 - 22 = 33. Our best cycle is (13 14 19 17 1): -61

Our smallest-valued cycle occurs in Trial 3: (13 15 17 1): -62. Our new edges are: (13 19), (15 11), (17 2), (1 14).

$D_3 =$

| -6 | 0 | -1 | -28 | -4 | 0 | -4 | 0 | -49 | 0 | -31 | 0 | -19 | -12 | -9 | -14 | -3 | 0 | -14 | 0 |
|---|---|---|---|---|---|---|---|---|---|---|---|---|---|---|---|---|---|---|---|
| 1 | 2 | 3 | 4 | 5 | 6 | 7 | 8 | 9 | 10 | 11 | 12 | 13 | 14 | 15 | 16 | 17 | 18 | 19 | 20 |
| 14 | 9 | 13 | 5 | 18 | 7 | 4 | 6 | 10 | 17 | 12 | 1 | 19 | 16 | 11 | 8 | 2 | 3 | 20 | 15 |

$D_3^{-1}$

| 1 | 2 | 3 | 4 | 5 | 6 | 7 | 8 | 9 | 10 | 11 | 12 | 13 | 14 | 15 | 16 | 17 | 18 | 19 | 20 |
|---|---|---|---|---|---|---|---|---|---|---|---|---|---|---|---|---|---|---|---|
| 12 | 17 | 18 | 7 | 4 | 8 | 6 | 16 | 2 | 9 | 15 | 11 | 3 | 1 | 20 | 14 | 10 | 5 | 13 | 19 |

Choose 9.

TRIAL 1.  MIN(M)(9,1) = 2.

(9 2) not allowed

TRIAL 2.  MIN(M)(9 2) = 5                      1. (1 3)

(9 5) → (9 4) -45

(4 7) not allowed

(4 20) → (4 19) -16                      4. (4 20)

(19 15) → (19 20) -14

(20 15) arc repeat

(20 14) → (20 1) +3

(1  3)  → (1  18) -6

(18  3)  arc

(18  5)  not allowed

(18  20) repeat positive

We can go no further.

d(20 10) - d(20  15) = +34

Our best cycle is (9  4  19  20): -41

TRIAL 3. MIN(M)(9,3) = 8.

(9  8)  → (9  16) -35

(16  11) → (16  15) -14

(15  20)  not allowed

(15  19) → (15  13) -3

(13  3)  not allowed

(13  19)  arc repeat

(13  16) → (13  14) +7

(14  20)  repeat

(14  1)  not allowed

(14  11) repeat positive

9. (9  5)            19. (19  15)

20. (20  14)

13. (13  16)

15. (15  19)

16. (16  11)

9. (9  8)

We can go no further. d(15  10) - d(15  11) = +19.  Our smallest-valued cycle is (9  16  15): -30.

It follows that $s$ is obtained from TRIAL 2: (9  4  19  20): -41. Our new edges are (9  5), (4  20), (19  15),

(20  10) .

| -6 | 0 | -1 | -12 | -4 | 0 | -4 | 0 | -4 | 0 | -31 | 0 | -19 | -12 | -9 | -14 | -3 | 0 | 0 | -34 |
|---|---|---|---|---|---|---|---|---|---|---|---|---|---|---|---|---|---|---|---|
| 1 | 2 | 3 | 4 | 5 | 6 | 7 | 8 | 9 | 10 | 11 | 12 | 13 | 14 | 15 | 16 | 17 | 18 | 19 | 20 |

$D_4 =$

| 14 | 9 | 13 | 20 | 18 | 7 | 4 | 6 | 5 | 17 | 12 | 1 | 19 | 16 | 11 | 8 | 2 | 3 | 15 | 10 |
|---|---|---|---|---|---|---|---|---|---|---|---|---|---|---|---|---|---|---|---|
| 1 | 2 | 3 | 4 | 5 | 6 | 7 | 8 | 9 | 10 | 11 | 12 | 13 | 14 | 15 | 16 | 17 | 18 | 19 | 20 |

$D_4^{-1} =$

| 12 | 17 | 18 | 7 | 9 | 8 | 6 | 16 | 2 | 20 | 15 | 11 | 3 | 1 | 19 | 14 | 10 | 5 | 13 | 4 |
|---|---|---|---|---|---|---|---|---|---|---|---|---|---|---|---|---|---|---|---|

Choose 20.

TRIAL 1. MIN(M)(20,1) = 15

(20  15)  → (20  19) -34

(19  15)  not allowed

(19  11)  → (19  15) +14

(15  20)  not allowed

(15  19)  not allowed                                                                 19. (19  11)

(15  11)  arc repeat                                                                  20. (20  15)

Can't go further. Have gone through $[\log n] + 1$ columns of MIN((M). Furthermore, path is no longer negative. No negative cycles can be obtained from any subpath.

TRIAL 2. MIN(M)(20,2) = 14

(20  14)  → (20  1) -31                           1. (1  3)           11. (11  16)

(1  3)  → (1  18) -6

(18  3)  arc

(18  5)  not allowed                              4. (4  12)          14. (14  11)

(18  20)  →  (18  4) +13

(4  7)  not allowed

(4  20)  arc repeat

(4  12)  →  (4  11) +1                                           18. (18  20)

(11  16)  →  (11  14) -31

(14  20)  not allowed                                            20. (20  14)

(14  1)  not allowed

(14  11)  →  (14  15) -3

(15  20)  repeat

(15  19)  not allowed

(15  11)  arc repeat

Can't go further. d(15  10) - d(15  11) = +19

Our smallest-valued cycle in this tiral is (20  1  18  4  11  14  15): -38

TRIAL 3.  MIN(M)(20,3) = 18.

(20  18)  →  (20  5) -20                                         11. (11  16)

(5  9)  not allowed

(5  3)  →  (5  18) -2                             .

(18  3)  arc                              4. (4  12)             14. (14  11)

(18  5)  not allowed                      5. (5  3)

(18  20)  not allowed

(18  12)  →  (18  11) +14

(11  16)  →  (11  14) -31

(14  20)  repeat

(14 1) not allowed                                                              .                             20. (20 18)

(14 11) → (14 15) -3

(15 20) repeat

(15 19) → not allowed

(15 11) arc

Can't go further. The smallest-valued cycle obtainable in this trial is (20 5 18 11 14 15): -14

It follows that our smallest-valued cycle is (20 1 18 4 11 14 15): -38. Our new edges are: (20 14), (1 3), (18 20), (4 12), (11 16), (14 11), (15 10).

$D_5 =$

| 0 | 0 | -1 | -13 | -4 | 0 | -4 | 0 | -4 | 0 | 0 | 0 | -19 | -9 | -28 | -14 | -3 | -13 | 0 | -3 |
|---|---|---|---|---|---|---|---|---|---|---|---|---|---|---|---|---|---|---|---|
| 1 | 2 | 3 | 4 | 5 | 6 | 7 | 8 | 9 | 10 | 11 | 12 | 13 | 14 | 15 | 16 | 17 | 18 | 19 | 20 |
| 3 | 9 | 13 | 12 | 18 | 7 | 4 | 6 | 5 | 17 | 16 | 1 | 19 | 11 | 10 | 8 | 2 | 20 | 15 | 14 |

$D_5^{-1} =$

| 1 | 2 | 3 | 4 | 5 | 6 | 7 | 8 | 9 | 10 | 11 | 12 | 13 | 14 | 15 | 16 | 17 | 18 | 19 | 20 |
|---|---|---|---|---|---|---|---|---|---|---|---|---|---|---|---|---|---|---|---|
| 12 | 17 | 1 | 7 | 9 | 8 | 6 | 16 | 2 | 15 | 14 | 4 | 3 | 20 | 19 | 14 | 10 | 5 | 13 | 18 |

Choose 15.

TRIAL 1. MIN(M)(15,1) = 20

(15 20) → (15 18) -28                                              1. (1 7)

(18 3) → (18 1) -13

(1 3) arc

(1 7) → (1 6) 0

(6 7) arc repeat                                              15. (15 20)

(6 20) repeat

(6 8) not allowed                                             18. (18 3)

Can't go further. d(6 10) - d(6 7) = +8

Our smallest-valued negative cycle is (15 18 1 6): -33.

TRIAL 2.  MIN(M)(15,2) = 19

(15 19) → (15 13) -22

(13 3) not allowed

(13 19) arc repeat

(13 16) → (13 11) +7

(11 16) arc

(11 17) → (11 10) 0

(10 17) arc

(10 6) → (10 8) +1

(8 6) arc repeat

(8 9) → (8 2) +8

(2 9) arc

(2 17) repeat positive

Can't go further.

No negative cycle exists.

Our smallest-valued negative cycle occurs in Trial 1: (15 18 1 6): -33.

New edges are: (15 20), (18 3), (1 7), (6 10).

$D_6 =$

|   | 1 | 2 | 3 | 4 | 5 | 6 | 7 | 8 | 9 | 10 | 11 | 12 | 13 | 14 | 15 | 16 | 17 | 18 | 19 | 20 |
|---|---|---|---|---|---|---|---|---|---|----|----|----|----|----|----|----|----|----|----|----|
| top | 0 | 0 | -1 | -13 | -4 | -8 | -4 | 0 | -4 | 0 | 0 | 0 | -19 | -9 | 0 | -14 | -3 | 0 | 0 | -3 |
| bottom | 7 | 9 | 13 | 12 | 18 | 10 | 4 | 6 | 5 | 17 | 16 | 1 | 19 | 11 | 20 | 8 | 2 | 3 | 15 | 14 |

$D_6^{-1} =$

| 1 | 2 | 3 | 4 | 5 | 6 | 7 | 8 | 9 | 10 | 11 | 12 | 13 | 14 | 15 | 16 | 17 | 18 | 19 | 20 |
|---|---|---|---|---|---|---|---|---|----|----|----|----|----|----|----|----|----|----|----|
| 12 | 17 | 18 | 7 | 9 | 8 | 1 | 16 | 2 | 6 | 14 | 4 | 3 | 20 | 19 | 11 | 10 | 5 | 13 | 15 |

(1 7 4 12)(2 9 5 18 3 13 19 15 20 14 11 16 8 6 10 17): 157

Choose 13.

Trial 1. MIN(M)(13,1) = 3

(13  3)  not allowed

(13  19)  arc

Can't go further. Thus, using Phase 1, we cannot obtain a derangement containing $n$ edges that has a value less than that of $D_6$. We now go on to Phase 2.

PHASE 2

*Boldface entries* denote arcs that yield edges symmetric to edges in $D_i$. Construct negative cycles containing only non-boldface entries. When we no longer can obtain such a negative cycle, we have obtained $D_{n\,edges}$. Using patching, we construct the smallest-valued tour obtainable, say $T_{UPPERBOUND}$.

Every tour $T$ of even length has the property that $T^{-1}M^-$ contains two cycles of length $\frac{n}{2}$. The arc of each cycle corresponds to an arc symmetric to an arc of $T$. The sum of the values of one of the arcs of one of the cycles is negative, while the sum of the arcs of the other cyle is positive. Let $N$ be the negatively-valued cycle. Alternately, we can choose alternating arcs of $T$. By construction, each set of arcs is a perfect matching. We choose the set, $N$, the sum of whose values of arcs is smaller. $N$ can also be thought of as a derangement on $n$ points, i.e., a product of 2-cycles. Call this derangement $\sigma_T$. For simplicity, in what follows, we denote $T_{UPPERBOUND}$ by $T$.

$$D_6^{-1}M^-$$

|   | 7 | 9 | 13 | 12 | 18 | 10 | 4 | 6 | 5 | 17 | 16 | 1 | 19 | 11 | 20 | 8 | 2 | 3 | 15 | 14 |   |
|---|---|---|---|---|---|---|---|---|---|---|---|---|---|---|---|---|---|---|---|---|---|
|   | 1 | 2 | 3 | 4 | 5 | 6 | 7 | 8 | 9 | 10 | 11 | 1 | 13 | 14 | 15 | 16 | 17 | 18 | 19 | 20 |   |
| 1 | 0 | 24 | 93 | 3 | 45 | 74 | 26 | 1 | 70 | 94 | 36 | ∞ | 36 | 77 | 66 | 34 | 22 | 0 | 90 | 6 | 1 |
| 2 | 49 | 0 | 30 | 8 | 81 | 49 | 29 | 79 | 40 | 5 | 90 | 25 | 22 | 8 | 84 | 73 | ∞ | 68 | 88 | 86 | 2 |
| 3 | 56 | 95 | 0 | 53 | -1 | 86 | 19 | 57 | 3 | 47 | 26 | 0 | 21 | 80 | 43 | 95 | 65 | ∞ | 62 | 52 | 3 |
| 4 | -13 | 18 | 49 | 0 | 30 | 30 | ∞ | 41 | 15 | 63 | 7 | 12 | 52 | 66 | -1 | 74 | 12 | 5 | 29 | 10 | 4 |
| 5 | 20 | -4 | 46 | 88 | 0 | 78 | 24 | 73 | ∞ | 11 | 85 | 65 | 81 | 78 | 11 | 71 | 32 | -2 | 63 | 36 | 5 |
| 6 | -8 | 34 | 47 | 32 | 43 | 0 | 50 | ∞ | 73 | 41 | 29 | -4 | 32 | 30 | 41 | -3 | 71 | 52 | 43 | 36 | 6 |
| 7 | ∞ | 73 | 51 | 5 | 54 | 44 | 0 | -4 | 24 | 40 | 26 | -1 | 83 | 68 | 37 | 29 | 25 | 55 | 82 | 31 | 7 |
| 8 | 28 | 8 | 79 | 85 | 72 | 49 | 86 | 0 | 74 | 58 | 11 | 32 | 54 | 37 | 35 | ∞ | 68 | 92 | 69 | 87 | 8 |
| 9 | 73 | ∞ | 93 | 76 | 58 | 45 | 21 | 38 | 0 | 68 | 26 | 23 | 82 | 23 | 51 | 9 | -4 | 94 | 20 | 90 | 9 |
| 10 | 41 | 42 | 87 | 16 | 82 | ∞ | 40 | 1 | 79 | 0 | 17 | 70 | 56 | 29 | 28 | 47 | 52 | 82 | 22 | 51 | 10 |
| 11 | 70 | 25 | 58 | 31 | 71 | 34 | 81 | 36 | 84 | 0 | 0 | 78 | 19 | ∞ | 23 | 40 | 6 | 81 | 8 | 11 | 11 |
| 12 | 3 | 74 | 77 | ∞ | 10 | 17 | 11 | 34 | 90 | 88 | 27 | 0 | 84 | 27 | 23 | 84 | 2 | 50 | 49 | 55 | 12 |
| 13 | 33 | 75 | ∞ | 61 | 59 | 70 | 34 | 33 | 32 | 26 | 7 | 74 | 0 | 38 | 63 | 62 | 28 | -19 | 43 | 43 | 13 |

| | 1 | 2 | 3 | 4 | 5 | 6 | 7 | 8 | 9 | 10 | 11 | 12 | 13 | 14 | 15 | 16 | 17 | 18 | 19 | 20 | |
|---|---|---|---|---|---|---|---|---|---|---|---|---|---|---|---|---|---|---|---|---|---|
| 14 | 22 | 81 | 52 | 48 | 33 | 45 | 14 | 31 | 31 | 33 | 3 | -4 | 41 | 0 | -9 | 79 | 73 | 42 | 11 | ∞ | 14 |
| 15 | 85 | 23 | 64 | 54 | 85 | 28 | 45 | 60 | 70 | 83 | 87 | 92 | 6 | 9 | 0 | 73 | 87 | 64 | ∞ | 23 | 15 |
| 16 | 14 | 14 | 13 | 17 | 1 | 8 | 8 | 21 | 77 | 37 | ∞ | 23 | 75 | -14 | 73 | 0 | 74 | 13 | 72 | 0 | 16 |
| 17 | 39 | 67 | 43 | 89 | 60 | 2 | 75 | 44 | 14 | ∞ | 48 | 92 | 42 | -3 | 84 | 58 | 0 | 45 | 79 | 41 | 17 |
| 18 | 56 | 60 | 79 | 14 | ∞ | 87 | 45 | 49 | 6 | 63 | 15 | 46 | 84 | 71 | 13 | 75 | 79 | 0 | 84 | 44 | 18 |
| 19 | 80 | 79 | 15 | 63 | 79 | 56 | 62 | 33 | 82 | 40 | 84 | 32 | ∞ | 14 | 14 | 52 | 15 | 17 | 0 | 47 | 19 |
| 20 | 37 | 51 | 81 | 25 | 11 | 31 | 12 | 45 | 15 | 85 | 85 | 65 | 17 | 21 | ∞ | 16 | 80 | 42 | -3 | 0 | 20 |
| | 1 | 2 | 3 | 4 | 5 | 6 | 7 | 8 | 9 | 10 | 11 | 12 | 13 | 14 | 15 | 16 | 17 | 18 | 19 | 20 | |

j = 1

(6 1)(1 8) = (6 8): -7; (6 1)(1 18): -8

j = 12

(6 12)(12 1) = (6 1): -1; (6 12)(12 17) = (6 17): -2; (14 12)(12 1) = (14 1): -1;

(14 12)(12 17) = (14 17): -2; (14 12 17): -5

j = 17

(14 17)(17 14) = (14 14): -5

CYCLE (14 12 17): -5

$D_6(14\ 12\ 17) = D_7$

Our new edges are (14 1), (12 2), (17 11). Thus,

      1  2  3  4  5  6  7  8  9  10  11  12  13  14  15  16  17  18  19  20

$D_7 =$

$$D_7^{-1} = \begin{pmatrix} 7 & 9 & 13 & 12 & 18 & 10 & 4 & 6 & 5 & 17 & 16 & 2 & 19 & 1 & 20 & 8 & 11 & 3 & 15 & 14 \\ 1 & 2 & 3 & 4 & 5 & 6 & 7 & 8 & 9 & 10 & 11 & 12 & 13 & 14 & 15 & 16 & 17 & 18 & 19 & 20 \\ 14 & 12 & 18 & 7 & 9 & 8 & 1 & 16 & 2 & 6 & 17 & 4 & 3 & 20 & 19 & 11 & 10 & 5 & 13 & 15 \end{pmatrix}$$

$$D_7^{-1} M^-$$

| | 7 | 9 | 13 | 12 | 18 | 10 | 4 | 6 | 5 | 17 | 16 | 2 | 19 | 1 | 20 | 8 | 11 | 3 | 15 | 14 | |
|---|---|---|---|---|---|---|---|---|---|---|---|---|---|---|---|---|---|---|---|---|---|
| | 1 | 2 | 3 | 4 | 5 | 6 | 7 | 8 | 9 | 10 | 11 | 12 | 13 | 14 | 15 | 16 | 17 | 18 | 19 | 20 | |
| 1 | 0 | 24 | 93 | 3 | 45 | 74 | 26 | 1 | 70 | 94 | 36 | 22 | 36 | ∞ | 66 | 34 | 77 | 0 | 90 | 6 | 1 |
| 2 | 49 | 0 | 30 | 8 | 81 | 49 | 29 | 79 | 40 | 5 | 90 | ∞ | 22 | 25 | 84 | 73 | 8 | 68 | 88 | 86 | 2 |
| 3 | 56 | 95 | 0 | 53 | -1 | 86 | 19 | 57 | 3 | 47 | 26 | 65 | 21 | 0 | 43 | 95 | 80 | ∞ | 62 | 52 | 3 |
| 4 | -13 | 18 | 49 | 0 | 30 | 30 | ∞ | 41 | 15 | 63 | 7 | 12 | 52 | 12 | -1 | 74 | 66 | 5 | 29 | 10 | 4 |
| 5 | 20 | -4 | 46 | 88 | 0 | 78 | 24 | 73 | ∞ | 11 | 85 | 32 | 81 | 65 | 11 | 71 | 78 | -2 | 63 | 36 | 5 |
| 6 | -8 | 34 | 47 | 32 | 43 | 0 | 50 | ∞ | 73 | 41 | 29 | 71 | 32 | -4 | 41 | -3 | 30 | 52 | 43 | 36 | 6 |
| 7 | ∞ | 73 | 51 | 5 | 54 | 44 | 0 | -4 | 24 | 40 | 26 | 25 | 83 | -1 | 37 | 29 | 68 | 55 | 82 | 31 | 7 |
| 8 | 28 | 8 | 79 | 85 | 72 | 49 | 86 | 0 | 74 | 58 | 11 | 68 | 54 | 32 | 35 | ∞ | 37 | 92 | 69 | 87 | 8 |
| 9 | 73 | ∞ | 93 | 76 | 58 | 45 | 21 | 38 | 0 | 68 | 26 | -4 | 82 | 23 | 51 | 9 | 23 | 94 | 20 | 90 | 9 |
| 10 | 41 | 42 | 87 | 16 | 82 | ∞ | 40 | 1 | 79 | 0 | 17 | 52 | 56 | 70 | 28 | 47 | 29 | 82 | 22 | 51 | 10 |
| 11 | 70 | 25 | 58 | 31 | 71 | 34 | 81 | 36 | 84 | 0 | 0 | 6 | 19 | 78 | 23 | 40 | ∞ | 81 | 8 | 11 | 11 |
| 12 | 1 | 72 | 75 | ∞ | 8 | 15 | 9 | 32 | 88 | 86 | 25 | 0 | 82 | -2 | 21 | 82 | 25 | 48 | 47 | 53 | 12 |
| 13 | 33 | 75 | ∞ | 61 | 59 | 70 | 34 | 33 | 32 | 26 | 7 | 28 | 0 | 74 | 63 | 62 | 38 | -19 | 43 | 43 | 13 |

| 14 | 26 | 85 | 56 | 52 | 37 | 49 | 18 | 35 | 35 | 37 | 7 | 77 | 45 | 0 | -5 | 83 | 4 | 46 | 15 | ∞ | 14 |
| --- | --- | --- | --- | --- | --- | --- | --- | --- | --- | --- | --- | --- | --- | --- | --- | --- | --- | --- | --- | --- | --- |
| 15 | 85 | 23 | 64 | 54 | 85 | 28 | 45 | 60 | 70 | 83 | 87 | 87 | 6 | 92 | 0 | 73 | 9 | 64 | ∞ | 23 | 15 |
| 16 | 14 | 14 | 13 | 17 | 1 | 8 | 8 | 21 | 77 | 37 | ∞ | 74 | 75 | 23 | 73 | 0 | -14 | 13 | 72 | 0 | 16 |
| 17 | 45 | 73 | 49 | 95 | 66 | 8 | 81 | 50 | 20 | ∞ | 54 | 6 | 48 | 98 | 90 | 64 | 0 | 51 | 85 | 47 | 17 |
| 18 | 56 | 60 | 79 | 14 | ∞ | 87 | 45 | 49 | 6 | 63 | 15 | 79 | 84 | 46 | 13 | 75 | 71 | 0 | 84 | 44 | 18 |
| 19 | 80 | 79 | 15 | 63 | 79 | 56 | 62 | 33 | 82 | 40 | 84 | 15 | ∞ | 32 | 14 | 52 | 14 | 17 | 0 | 47 | 19 |
| 20 | 37 | 51 | 81 | 25 | 11 | 31 | 12 | 45 | 15 | 85 | 85 | 80 | 17 | 65 | ∞ | 16 | 21 | 42 | -3 | 0 | 20 |
|  | 1 | 2 | 3 | 4 | 5 | 6 | 7 | 8 | 9 | 10 | 11 | 12 | 13 | 14 | 15 | 16 | 17 | 18 | 19 | 20 |  |

Any arc represented by an entry in $D_7^{-1}M^-$ that is symmetric to an arc in $D_7$ is printed in bf. In general, if $(a\ D_7(b))$ is an arc of $D_7$, then the arc $(D_7(b)\ D_7^{-1}(a))$ in $D_7^{-1}M^-$ yields an edge $(D_7(b)\ a)$ that is symmetric to $(a\ D_7(b))$.

We use the row forms of $D_7$ and $D_7^{-1}$ to obtain the values of row entries in $D_7^{-1}M^-$ in increasing order of magnitude. As usual, entries that are in bf indicate arcs that are symmetric to an arc of $D_7$. We now obtain paths in $D_7^{-1}M^-$ using all non-boldface entries

j = 1

(6 1)(1 8) = (6 8): -7; (6 1)(1 18) = (6 18): -8; (12 1)(1 8) = (12 8): 2; (12 1)(1 18) = (12 18): 1

j = 14

(12 14)(14 15) = (12 15): -7

j = 15

(12 15)(15 13) = (12 13): <u>-1</u>

j = 17

(16 17)(17 12) = (16 12): -11

j = 20

(4 20)(20 19) = (4 19): -10

$P_{20}$

|    | 1 | 2 | 3 | 4 | 5 | 6 | 7 | 8 | 9 | 10 | 11 | 12 | 13 | 14 | 15 | 16 | 17 | 18 | 10 | 20 |    |
|----|---|---|---|---|---|---|---|---|---|----|----|----|----|----|----|----|----|----|----|----|----|
| 1  |   |   |   |   |   |   |   |   |   |    |    |    |    |    |    |    |    |    |    |    | 1  |
| 2  |   |   |   |   |   |   |   |   |   |    |    |    |    |    |    |    |    |    |    |    | 2  |
| 3  |   |   |   |   |   |   |   |   |   |    |    |    |    |    |    |    |    |    |    |    | 3  |
| 4  |   |   |   |   |   |   |   |   |   |    |    |    |    |    |    |    |    |    |    |    | 4  |
| 5  |   |   |   |   |   |   |   |   |   |    |    |    |    |    |    |    |    |    |    |    | 5  |
| 6  | 6 |   |   |   |   |   |   | 1 |   |    |    |    |    |    |    |    |    | 1  |    |    | 6  |
| 7  |   |   |   |   |   |   |   |   |   |    |    |    |    |    |    |    |    |    |    |    | 7  |
| 8  |   |   |   |   |   |   |   |   |   |    |    |    |    |    |    |    |    |    |    |    | 8  |
| 9  |   |   |   |   |   |   |   |   |   |    |    |    |    |    |    |    |    |    |    |    | 9  |
| 10 |   |   |   |   |   |   |   |   |   |    |    |    |    |    |    |    |    |    |    |    | 10 |
| 11 |   |   |   |   |   |   |   |   |   |    |    |    |    |    |    |    |    |    |    |    | 11 |

| | 1 | 2 | 3 | 4 | 5 | 6 | 7 | 8 | 9 | 10 | 11 | 12 | 13 | 14 | 15 | 16 | 17 | 18 | 19 | 20 | |
|---|---|---|---|---|---|---|---|---|---|---|---|---|---|---|---|---|---|---|---|---|---|
| 12 | 12 | | | | | | 1 | | | | | 15 | 12 | 14 | | | 1 | | | 12 |
| 13 | | | | | | | | | | | | | | | | | | | | | 13 |
| 14 | | | | | | | | | | | | | | | | | | | | | 14 |
| 15 | | | | | | | | | | | | | | | | | | | | | 15 |
| 16 | | | | | | | | | | | | 17 | | | | 16 | | | | | 16 |
| 17 | | | | | | | | | | | | | | | | | | | | | 17 |
| 18 | | | | | | | | | | | | | | | | | | | | | 18 |
| 19 | | | | | | | | | | | | | | | | | | | | | 19 |
| 20 | | | | | | | | | | | | | | | | | | | | | 20 |
| | 1 | 2 | 3 | 4 | 5 | 6 | 7 | 8 | 9 | 10 | 11 | 12 | 13 | 14 | 15 | 16 | 17 | 18 | 19 | 20 | |

$$D_7^{-1}M^-(20)$$

| | 7 | 9 | 13 | 12 | 18 | 10 | 4 | 6 | 5 | 17 | 16 | 2 | 19 | 1 | 20 | 8 | 11 | 3 | 15 | 14 | |
|---|---|---|---|---|---|---|---|---|---|---|---|---|---|---|---|---|---|---|---|---|---|
| | 1 | 2 | 3 | 4 | 5 | 6 | 7 | 8 | 9 | 10 | 11 | 12 | 13 | 14 | 15 | 16 | 17 | 18 | 19 | 20 | |
| 1 | 0 | | | | | | | | | | | | | ∞ | | | | | 6 | | 1 |
| 2 | | 0 | | 8 | | | | | | | | ∞ | | | | | | | | | 2 |
| 3 | | | 0 | -1 | | | | | | | | | | | | | | ∞ | | | 3 |
| 4 | -13 | | | 0 | | ∞ | | | | | | | | | | | | | | | 4 |
| 5 | | -4 | | | 0 | | | | ∞ | | | | | | | | | | | | 5 |
| 6 | -8 | | | | | 0 | | -7 | | | | | | | -3 | | | -8 | | | 6 |
| 7 | ∞ | | | | | | 0 | | | | | | | -1 | | | | | | | 7 |
| 8 | | | | | | | | 0 | | | 11 | | | | | ∞ | | | | | 8 |
| 9 | | ∞ | | | | | | | 0 | | -4 | | | | | | | | | | 9 |
| 10 | | | | | | ∞ | | 1 | | 0 | | | | | | | | | | | 10 |
| 11 | | | | | | | | | | 0 | 0 | | | | | | | ∞ | | | 11 |
| 12 | | | | ∞ | | | 9 | *2* | | | | 0 | <u>-1</u> | | -7 | | | *1* | | | 12 |

| | 1 | 2 | 3 | 4 | 5 | 6 | 7 | 8 | 9 | 10 | 11 | 12 | 13 | 14 | 15 | 16 | 17 | 18 | 19 | 20 | |
|---|---|---|---|---|---|---|---|---|---|---|---|---|---|---|---|---|---|---|---|---|---|
| 13 | | | ∞ | | | | | | | | | 0 | | | | | -19 | | | 1 | 13 |
| 14 | | | | | | | | | | | 0 | -5 | | | | | | | ∞ | | 14 |
| 15 | | | | | | | | | | | 6 | | 0 | | | | | ∞ | | | 15 |
| 16 | | | | | | | | | ∞ | | | | | 0 | -14 | | | | | | 16 |
| 17 | | | | | 5 | | | ∞ | | | | | | | 0 | | | | | | 17 |
| 18 | | | ∞ | | | 6 | | | | | | | | | | | 0 | | | | 18 |
| 19 | | 15 | | | | | | | | | ∞ | | | | | | | | 0 | | 19 |
| 20 | | | | | | | | | | | | | | | ∞ | | | -3 | 0 | | 20 |
| | 1 | 2 | 3 | 4 | 5 | 6 | 7 | 8 | 9 | 10 | 11 | 12 | 13 | 14 | 15 | 16 | 17 | 18 | 19 | 20 | |

The path (12 ... 13) can't be extended. Thus, $D_7$ is the smallest-valued derangement we can obtain using the modified F-W algorithm.

We now construct an upper bound for $|T_{OPT}|$. Given a fixed vertex from $S = \{1, 2, ..., n\}$, we use an abbreviated method to construct the smallest-valued tour $T_{UPPERBOUND}$ obtainable by patching the two cycles of $D_7$.

When doing patching of cycles, we must include cases where we traverse a cycle counter-clockwise. In general,

(1) Two adjacent points of each cycle must be included – one for entering, the other for leaving, the cycle.

(2) When adding an arc, $(a\ b)$, if $a_1$ and $a_2$ are the points on the cycle adjacent to $a$, we must obtain $d(a\ b) - d(a\ a_1)$ and $d(a\ b) - d(a\ a_2)$ in order to determine the gain or loss in value of a path that will eventually become a tour. In order to find the value of arcs, we use MIN(M) together with the row forms for $D_7$ and $D_7^{-1}$ For instance, if we wish to find the minimum-valued entry in row 5 of $D_7^{-1}M^-$, we first obtain MIN(M)(5,1) = $m$. We then obtain $D_7^{-1}(m)$. All of the other arcs in the cycles retain their original values. By determining which case leads to the smallest possible gain in value yields a tour of smallest value obtainable using this particular method.

We must start our path using each of the $n$ possible vertices in constructing tours. Instead of giving through all twenty constructions, we give a case that yields the smallest-valued tour. Also, from a theorem given in Feller [ ], for a large value of $n$, a randomly chosen permutation contains approximately $\log n$ cycles.

*Note.* Before constructing a tour by patching together cycles of $D_{\text{3-cycle 2}}$, we use the following rule: if the number of cycles in our derangement is greater than $\dfrac{\log n}{\log 2}$, after we have obtained $n$ branches, we sort the values of the next set of $2n$ branches. We continue our construction using the $n$ smallest-valued branches of the sorted set. This continues until the end of the algorithm. We now prove the following theorem that may reduce the running time of the patching algorithm.

Theorem 1.28 *Let two branches obtained from our algorithm go through the same cycles and have the following properties: (1) the terminal vertex of both branches is $a$. (2) If the arcs deleted are of the respective forms*

$$\begin{array}{cc} \Gamma: & \Delta: \\ \mathbf{A} \downarrow \mathbf{B} & \mathbf{C} \downarrow \mathbf{D} \\ b \to a \leftarrow c & b \to a \leftarrow c \end{array}$$

*with* $\max\{\mathbf{A}, \mathbf{B}\} \leq \min\{\mathbf{C}, \mathbf{D}\}$, *then we can continue our algorithm using only* $\Gamma$.

Proof. Our paths from $b, a$ of $\Gamma$ will always be no greater than if we had used $b, a$ of $\Delta$.

*Note.* Alternately, if a derangement has more than $\log n$ cycles, we might use Helsgaun's algorithm to obtain $T_{UPPERBOUND}$. However, even with our algorithm, by choosing the $n$ smallest-valued branches out of the $2n$ possibilties as we add one arc arc to each node, it would change the running time of this phase of the algorithm to

$O((n \log n)^2)$ instead of $O(n^2)$. Instead of going through all $n$ possibilities, we start with a point that gives us the smallest-valued tour. Let our initial vertex be 2.

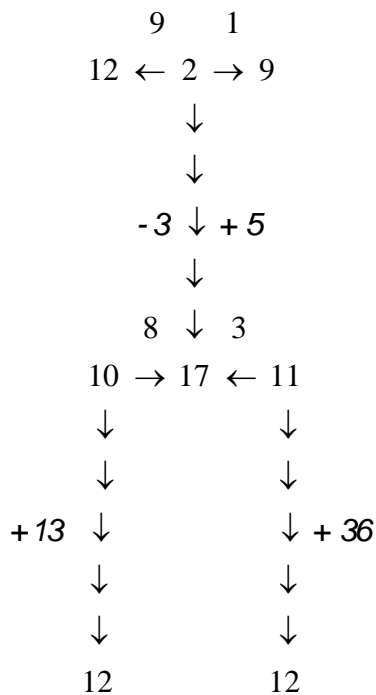

The numbers in italics are the sums of the gains or losses in values of a branch as an arc of a cycle is removed and replaced by an arc going to another cycle. Thus, the branch 12 – c-c- 2 – 17 has a value 3 less than the value of the path 12 – cc- 2 – 9. The horizontal arrows indicate the arcs of cycles that will be replaced by vertical-arrow arcs if we go in a clockwise or counter-clockwise direction. When constructing a tour, the direction of the arrow tells us to proceed in the direction the arrow is pointing (clockwise or counter-clockwise) as we traverse a cycle. An arrow pointing to the left tells us to proceed in a counter-clockwise direction from the terminal point of the arrow (the arrow-head) to the initial point of the arrow, i. e., from 12 to 2 going in a counter-clockwise manner. Using the above diagram we obtain the following tour:

$$T_{165} = (12\ 4\ 7\ 1\ 14\ 20\ 15\ 19\ 13\ 3\ 18\ 5\ 9\ 2\ 17\ 11\ 16\ 8\ 6\ 10)$$

The above diagram tells us that in constructing $T_{165}$ from $D_7$, we have added 13 to obtain the tour.

We now go on to PHASE 3(a).

$$\sigma_T = (1\ 7)(2\ 9)(3\ 13)(4\ 12)(5\ 18)(6\ 10)(8\ 16)(11\ 17)(14\ 20)(15\ 19).$$

$$\sigma_T^{-1}M^-$$

|    | 7 | 9 | 13 | 12 | 18 | 10 | 1 | 16 | 2 | 6 | 17 | 4 | 3 | 20 | 19 | 8 | 11 | 5 | 15 | 14 |    |
|----|---|---|----|----|----|----|---|----|---|---|----|---|---|----|----|---|----|---|----|----|----|
|    | 1 | 2 | 3  | 4  | 5  | 6  | 7 | 8  | 9 | 10| 11 | 12| 13| 14 | 15 | 16| 17 | 18| 19 | 20 |    |
| 1  | 0 | 24| 93 | 3  | 45 | 74 | ∞ | 36 | 22| 1 | 94 | 26| 0 | 66 | 36 | 34| 77 | 70| 90 | 6  | 1  |
| 2  | 49| 0 | 30 | 8  | 81 | 59 | 25| 90 | ∞ | 79| 5  | 29| 68| 84 | 22 | 73| 8  | 40| 88 | 86 | 2  |
| 3  | 56| 95| 0  | 53 | -1 | 86 | 0 | 26 | 65| 57| 47 | 19| ∞ | 43 | 21 | 94| 80 | 3 | 62 | 52 | 3  |
| 4  |-13| 8 | 39 | 0  | 30 | 30 | 12| 7  | 12| 41| 63 | ∞ | 5 | -1 | 52 | 74| 66 | 15| 29 | 10 | 4  |
| 5  | 20| -4| 46 | 88 | 0  | 78 | 65| 85 | 32| 73| 11 | 24| -2| 11 | 81 | 71| 78 | ∞ | 63 | 36 | 5  |
| 6  | -8| 34| 47 | 32 | 43 | 0  | -4| 29 | 71| ∞ | 41 | 50| 52| 41 | 32 | -3| 30 | 73| 53 | 36 | 6  |
| 7  | ∞ | 74| 52 | 6  | 55 | 45 | 0 | 27 | 26| -3| 41 | 1 | 56| 38 | 84 | 30| 69 | 25| 83 | 32 | 7  |
| 8  | 17| -3| 68 | 74 | 61 | 38 | 21| 0  | 57| -11| 47| 75| 81| 24 | 43 | ∞ | 26 | 63| 58 | 76 | 8  |
| 9  | 77| ∞ | 97 | 80 | 62 | 49 | 27| 30 | 0 | 42| 72 | 25| 98| 55 | 86 | 13| 27 | 4 | 24 | 94 | 9  |
| 10 | 40| 41| 86 | 15 | 81 | ∞  | 69| 16 | 51| 0 | -1 | 39| 81| 27 | 55 | 46| 28 | 78| 21 | 50 | 10 |
| 11 | 70| 25| 58 | 31 | 71 | 34 | 78| 0  | 6 | 36| 0  | 81| 81| 23 | 19 | 40| ∞  | 84| 8  | 11 | 11 |
| 12 | -8| 63| 66 | ∞  | -1 | 6  |-11| 16 | -9| 23| 77 | 0 | 39| 12 | 53 | 73| 16 | 79| 38 | 44 | 12 |
| 13 | 52| 94| ∞  | 80 | 78 | 89 | 93| 26 | 27| 52| 45 | 53| 0 | 82 | 19 | 81| 57 | 51| 62 | 62 | 13 |
| 14 | 31| 90| 61 | 57 | 42 | 54 | 5 | 12 | 82| 40| 42 | 23| 51| 0  | 50 | 88| 9  | 40| 20 | ∞  | 14 |
| 15 | 79| 17| 58 | 48 | 79 | 22 | 86| 81 | 81| 54| 77 | 39| 58| -6 | 0  | 67| 3  | 64| ∞  | 17 | 15 |
| 16 | 14| 14| 13 | 17 | 1  | 8  | 23| ∞  | 74| 21| 37 | 8 | 13| 73 | 75 | 0 | -14| 77| 72 | 0  | 16 |

| 17 | 42 | 70 | 46 | 92 | 63 | 5 | 95 | 51 | 3 | 47 | ∞ | 78 | 48 | 87 | 45 | 61 | 0 | 17 | 82 | 44 | 17 |
|---|---|---|---|---|---|---|---|---|---|---|---|---|---|---|---|---|---|---|---|---|---|
| 18 | 50 | 54 | 73 | 8 | ∞ | 81 | 40 | 9 | 73 | 43 | 57 | 39 | -6 | 7 | 78 | 69 | 65 | 0 | 78 | 38 | 18 |
| 19 | 80 | 79 | 15 | 63 | 79 | 56 | 32 | 84 | 15 | 33 | 40 | 62 | 17 | 14 | ∞ | 52 | 14 | 82 | 0 | 47 | 19 |
| 20 | 37 | 51 | 81 | 25 | 11 | 31 | 65 | 85 | 80 | 45 | 85 | 12 | 42 | ∞ | 17 | 36 | 21 | 15 | -3 | 0 | 20 |
|  | 1 | 2 | 3 | 4 | 5 | 6 | 7 | 8 | 9 | 10 | 11 | 12 | 13 | 14 | 15 | 16 | 17 | 18 | 19 | 20 |  |

As we consider adding an arc $(c'\ a)$ to an acceptable path, we should use the criteria given in theorem 1.27 to determine whether or not we can add it. As described earlier we use MIN(M) and the row form of $\sigma_T$ to obtain the values of arcs. We now construct paths that we place in $P_{20}$. Corresponding values are placed in $\sigma_T^{-1} M^-(20)$. The paths written in italics represent 2-circuit unlinked paths that are directly obtainable while we are constructing acceptable paths.

j = 1

(4 1)(1 4) = (4 4): -10

CYCLE (1 4): -10

(4 1)(1 10) = (4 10): -12; (4 1)(1 20) = (4 20): -7; (6 1)(1 4) = (6 4): -5; (6 1)(1 10) = (6 10):

-7;

(6 1)(1 13) = (6 13): -8; (6 1)(1 20) = (6 20): -2; (12 1)(1 10) = (12 10): -7;

(12 1)(1 13) = -8; (12 1)(1 20) = (12 20): -2

*(12 1 4): -5;*

j = 2

(5 2)(2 4) = (5 4): 4; (5 2)(2 11) = (5 11): 1; (5 2)(2 17) = (5 17): 4; (8 2)(2 4) = (8 4): 5;

(8 2)(2 11) = (8 11): 2; (8 2)(2 17) = (8 17): 5

j = 4

(2 4)(4 1) = (2 1): -5; (5 4)(4 1) = (5 1): -9; (5 4)(4 14) = (5 14): 3;

(6 4)(4 2) = (6 2): 3; (6 4)(4 8) = (6 8): 2; (6 4)(4 9) = (6 9): 7; (6 4)(4 14) = (6 14): -6;

(7 4)(4 1) = (7 1): -7; (7 4)(4 14) = (7 14): 5; (18 4)(4 1) = (18 1): -5; (18 4)(4 14) = (18 14): 7

j = 5

(3 5)(5 2) = (3 2): -5; (12 5)(5 2) = (12 2): -5; (16 5)(5 2) = (16 2): -3; (16 5)(5 13) = (16 13): -1

*(3 5 13): -3*

j = 6

(12 6)(6 16) = (12 16): 3; (16 6)(6 1) = (16 1): 0; (17 6)(6 1) = (17 1): -3,

(17 6)(6 7) = (17 7): 1

j = 7

(3 7)(7 4) = (3 4): 6; (3 7)(7 10) = (3 10): -3; (3 7 12): 1; (6 7)(7 12) = (6 12): -3;

(12 7)(7 10) = (12 10): -14;

(12 7)(7 12) = (12 12): -10

CYCLE (12 7): -10

(14 7)(7 10) = (14 10): 2; (14 7)(7 12) = (14 12): 6; (17 7)(7 12) = (17 12): 2

*(12 7 4): -5; (6 7 10): -7*

j = 8

(6 8)(8 2) = (6 2): -1; (11 8)(8 2) = (11 2): -3;

(11 8)(8 10) = (11 10): -11

j = 9

(12 9)(9 18) = (12 18): -5

j = 10

(1 10)(10 11) = (1 11): 0;

(4 10)(10 4): 3

CYCLE (4 1 10): 3

(4 10)(10 11) = (4 11): -13; (7 10)(10 11) = (7 11): -4; (8 10)(10 4) = (8 4): 4;

(8 10)(10 8) = (8 8): 5

CYCLE (8 10): 5

(11 10)(10 11) = (11 11): -13

CYCLE (11 8 10): -13

(12 10)(10 11) = (12 11): -15; (12 10)(10 19) = (12 19): 7; (14 10)(10 11) = (14 11): 1

*(12 7 10 4): 1*

j = 11

(2 11)(11 8) = (2 8): 5; (4 11)(11 8) = (4 8): -13; (4 11)(11 9) = (4 9): -7;

(4 11)(11 15) = (4 15): 6; (4 11)(11 19) = (4 19): -5; (5 11)(11 8) = (5 8): 1;

(5 11)(11 9) = (5 9): <u>7</u>;

(8 11)(11 8) = (8 8): -12

CYCLE (8 10 11): -12

(8 11)(11 15) = (8 15): 7; (8 11)(11 19) = (8 19): -4; (8 11)(11 20) = (8 20): -1;

(12 11)(11 8) = (12 8): <u>-15</u>; (12 11)(11 15) = (12 15): 4; (12 11)(11 19): -7;

(14 11)(11 8) = (14 8): <u>1</u>; (14 11)(11 9) = (14 9): <u>7</u>

j = 12

(3 12)(12 6) = (3 6): <u>7</u>; (3 12)(12 9) = (3 9): <u>-8</u>; (6 12)(12 1) = (6 1): <u>-11</u>;

(6 12)(12 5) = (6 5): <u>-4</u>; (6 12)(12 9) = (6 9): <u>-12</u>; (7 12)(12 5) = (7 5): <u>0</u>;

(7 12)(12 6) = (7 6): <u>7</u>;

(7 12)(12 7) = (7 7): -10

CYCLE (7 12): -10

(7 12)(12 9) = (7 9): <u>-8</u>; (14 12)(12 1) = (14 1): <u>-2</u>; (14 12)(12 5) = (14 5): <u>7</u>;

(14 12)(12 9) = (14 9): <u>-3</u>; (16 12)(12 9) = (16 9): <u>-1</u>; (17 12)(12 5) = (17 5): 1

*(7 12 1): <u>-7</u>*

j = 14

(4 14)(14 7) = (4 7): <u>4</u>; (4 14)(14 17) = (4 17): 8; (5 14)(14 7) = (5 7): <u>8</u>;

(6 14)(14 17) = (6 17): 3; (15 14)(14 7) = (15 7): <u>-7</u>;

j = 16

(6 16)(16 6) = (6 6): 5

CYCLE (6 16): 5

(6 16)(16 17) = (6 17): -17; (6 16)(16 20) = (6 20): -3; (12 16)(16 5) = (12 5): -3;

(12 16)(16 17) = (12 17): -11

j = 17

(6 17)(17 6) = (6 6): -12

CYCLE (6 16 17): -12

(6 17)(17 9) = (6 9): -14; (15 17)(17 6) = (15 6): 8;

(15 17)(17 9) = (15 9): 8; (16 17)(17 6) = (16 6): -9; (16 17)(17 9) = (16 9): -9;

(16 17)(17 18) = (16 18): 3

j = 18

(9 18)(18 13) = (9 13): -2; (12 18)(18 13) = (12 13): -11; (16 18)(18 13) = (16 13): -3

j = 20

(1 20)(20 19) = (1 19): 3; (4 20)(20 5) = (4 5): 4; (4 20)(20 12) = (4 12): 5;

(4 20)(20 18) = (4 18): 8; (4 20)(20 19) = (4 19): -10;

(12 20)(20 19) = (12 19): -5

CYCLES

(4 1): -10; (4 1 10): 3; (6 16): 2; (6 16 17): -12; (7 12): -10; (8 10): 5; (8 10 11): -12;

(12 6 16): 4; (12 7): -10

## 2-CIRCUIT PATHS

(3  5  13): -3; (7  12  1): -7; (12  7  4): -5; (12  7  10  4): 1; (12  1  4): -5;

## *VALUES OF CYCLES*

| 1 | -10: 4; |
|---|---|
| 2 | |
| 3 | -3:  5 13 |
| 4 | -10: 1; |
| 5 | -3:  13 3 |
| 6 | -12:  16 17 |
| 7 | -10: 12 |
| 8 | -12:  10 11 |
| 9 | |
| 10 | -12:  11 8 |
| 11 | -12:  8 10 |
| 12 | -10: 7 |
| 13 | -3:  3 5 |
| 14 | |
| 15 | |
| 16 | -12:  17 6 |
| 17 | -12:  6 16 |
| 18 | |
| 19 | |
| 20 | |

## VALUES OF 2–CIRCUIT PATHS

| 1  |                    |
|----|--------------------|
| 2  |                    |
| 3  | -3: 5 13           |
| 4  |                    |
| 5  |                    |
| 6  |                    |
| 7  | -7: 12 1           |
| 8  |                    |
| 9  |                    |
| 10 |                    |
| 11 |                    |
| 12 | -5: 1 4; -5: 7 4   |
| 13 |                    |
| 14 |                    |
| 15 |                    |

| 16 | |
|---|---|
| 17 | |
| 18 | |
| 19 | |
| 20 | |

$$P_{20}$$

|  | 7 | 9 | 13 | 12 | 18 | 10 | 1 | 16 | 2 | 6 | 17 | 4 | 3 | 20 | 19 | 8 | 11 | 5 | 15 | 14 |  |
|---|---|---|---|---|---|---|---|---|---|---|---|---|---|---|---|---|---|---|---|---|---|
|  | 1 | 2 | 3 | 4 | 5 | 6 | 7 | 8 | 9 | 10 | 11 | 12 | 13 | 14 | 15 | 16 | 17 | 18 | 19 | 20 |  |
| 1 |  |  |  |  |  |  |  |  |  | 1 | 10 |  |  |  |  |  |  |  | 20 | 1 | 1 |
| 2 | 4 |  |  | 2 |  |  |  | 11 |  |  | 2 |  |  |  |  |  |  |  |  |  | 2 |
| 3 |  | 5 |  | 7 | 3 |  | 3 |  |  | 7 |  | 7 |  |  |  |  |  |  |  |  | 3 |
| 4 | 4 |  |  |  | 20 |  | 14 | 11 | 11 | 1 | 10 | 20 |  | 4 | 11 |  | 14 | 20 | 20 | 1 | 4 |
| 5 | 4 | 5 |  | 2 |  |  | 14 | 11 | 11 |  | 2 |  |  | 4 |  |  | 2 |  |  |  | 5 |
| 6 | 6 | 8 |  | 1 |  |  | 6 | 4 | 17 |  |  | 7 | 1 | 4 |  | 6 | 16 |  |  | 16 | 6 |
| 7 | 4 |  |  | 7 | 12 | 12 |  |  | 12 | 7 |  | 7 |  | 4 |  |  |  |  |  |  | 7 |
| 8 |  | 8 |  | 10 |  |  |  |  |  | 8 | 2 |  |  |  | 11 |  | 2 |  | 20 | 11 | 8 |
| 9 |  |  |  |  |  |  |  |  |  |  |  |  | 18 |  |  |  |  | 9 |  |  | 9 |
| 10 |  |  |  |  |  |  |  |  |  |  |  |  |  |  |  |  |  |  |  |  | 10 |
| 11 |  | 8 |  |  |  |  | 11 |  | 8 |  |  |  |  |  |  |  |  |  |  |  | 11 |
| 12 | 12 | 5 |  |  | 12 | 12 | 12 | 11 | 12 | 7 | 10 |  | 18 | 12 | 11 | 6 | 16 | 9 | 11 | 1 | 12 |
| 13 |  |  |  |  |  |  |  |  |  |  |  |  |  |  |  |  |  |  |  |  | 13 |
| 14 | 12 |  |  |  | 12 |  | 14 | 11 | 12 | 7 | 10 | 7 |  |  |  |  |  |  |  |  | 14 |
| 15 |  |  |  |  |  | 17 | 14 |  | 17 |  |  |  |  | 15 |  |  | 15 | 17 |  |  | 15 |
| 16 | 6 | 5 |  |  | 16 | 17 |  |  | 17 |  |  |  | 18 |  |  |  | 16 | 17 |  |  | 16 |
| 17 | 6 |  |  |  | 12 | 17 | 6 |  |  |  |  | 7 |  |  |  |  |  |  |  |  | 17 |

| 18 | 4 |  |  | 18 |  |  |  |  |  |  |  |  |  | 4 |  |  |  |  |  | 18 |
|---|---|---|---|---|---|---|---|---|---|---|---|---|---|---|---|---|---|---|---|---|
| 19 |  |  |  |  |  |  |  |  |  |  |  |  |  |  |  |  |  |  |  | 19 |
| 20 |  |  |  |  |  |  |  |  |  |  |  |  |  |  |  |  |  |  |  | 20 |
|  | 1 | 2 | 3 | 4 | 5 | 6 | 7 | 8 | 9 | 10 | 11 | 12 | 13 | 14 | 15 | 16 | 17 | 18 | 19 | 20 |  |

$$\sigma_T^{-1} M^-(20)$$

|  | 7 | 9 | 13 | 12 | 18 | 10 | 1 | 16 | 2 | 6 | 17 | 4 | 3 | 20 | 19 | 8 | 11 | 5 | 15 | 14 |  |
|---|---|---|---|---|---|---|---|---|---|---|---|---|---|---|---|---|---|---|---|---|---|
|  | 1 | 2 | 3 | 4 | 5 | 6 | 7 | 8 | 9 | 10 | 11 | 12 | 13 | 14 | 15 | 16 | 17 | 18 | 19 | 20 |  |
| 1 | 0 |  |  |  |  |  |  |  | *1* | *0* |  |  |  |  |  |  |  |  | <u>3</u> |  | 1 |
| 2 | <u>-5</u> | 0 |  | *8* |  |  |  | <u>5</u> |  | *5* |  |  | *7* |  |  |  | *8* |  |  |  | 2 |
| 3 |  | <u>-5</u> | 0 | <u>6</u> |  |  |  |  | *-3* |  | *1* |  |  |  |  |  |  |  |  |  | 3 |
| 4 |  |  |  | 0 | <u>4</u> |  |  | <u>-13</u> | <u>-7</u> | *-12* | *-13* | <u>5</u> |  | <u>4</u> | 6 |  | *8* | <u>8</u> | <u>-10</u> | *-7* | 4 |
| 5 | <u>-9</u> |  |  | *4* | 0 |  | *8* | <u>1</u> | <u>7</u> |  | *1* |  |  | *3* |  |  | *4* |  |  |  | 5 |
| 6 |  | <u>-1</u> |  | *-5* | *-4* | 0 |  | *2* | <u>-14</u> |  |  | *-3* | *-8* | *-6* |  |  | *-17* |  |  | *-3* | 6 |
| 7 |  |  |  | <u>0</u> | <u>7</u> | 0 |  | <u>-8</u> | *-3* | *-4* | *-2* |  | *5* |  |  |  |  |  |  |  | 7 |
| 8 |  |  |  | <u>4</u> |  |  |  | 0 |  |  | *2* |  |  |  | *7* |  | *5* |  | *-4* | *-1* | 8 |
| 9 |  |  |  |  |  |  |  |  | 0 |  |  |  | <u>-2</u> |  |  |  |  |  |  |  | 9 |
| 10 |  |  |  |  |  |  |  |  |  | 0 |  |  |  |  |  |  |  |  |  |  | 10 |
| 11 |  | <u>-3</u> |  |  |  |  |  |  |  | *-11* | 0 |  |  |  |  |  |  |  |  |  | 11 |
| 12 |  | <u>-5</u> |  | <u>-5</u> |  |  |  | *-15* |  | <u>-11</u> |  | 0 | <u>-11</u> |  |  | *3* | <u>-11</u> | *-5* | <u>-3</u> | *-2* | 12 |
| 13 |  |  |  |  |  |  |  |  |  |  |  |  | 0 |  |  |  |  |  |  |  | 13 |
| 14 | <u>-2</u> |  |  |  | <u>7</u> |  |  | <u>1</u> | <u>-3</u> | *2* | *1* | *6* |  | 0 |  |  |  |  |  |  | 14 |
| 15 |  |  |  |  |  | <u>-1</u> |  |  |  |  |  |  |  |  | 0 |  |  |  |  |  | 15 |
| 16 | <u>0</u> | <u>-3</u> |  |  | <u>-9</u> |  |  | <u>-11</u> |  |  |  |  | <u>-3</u> |  |  | 0 |  |  |  |  | 16 |
| 17 | <u>-3</u> |  |  | <u>1</u> |  | *1* |  |  |  |  |  | *2* |  |  |  |  | 0 |  |  |  | 17 |
| 18 | <u>-5</u> |  |  | ∞ |  |  |  |  |  |  |  |  |  | *7* |  |  |  | 0 |  |  | 18 |
| 19 |  |  |  |  |  |  |  |  |  |  |  |  |  |  |  |  |  |  | 0 |  | 19 |

| 20 |   |   |   |   |   |   |   |   |   |   |   |   |   |   |   |   |   |   |   | 0 | 20 |
|----|---|---|---|---|---|---|---|---|---|---|---|---|---|---|---|---|---|---|---|---|----|
|    | 1 | 2 | 3 | 4 | 5 | 6 | 7 | 8 | 9 | 10| 11| 12| 13| 14| 15| 16| 17| 18| 19| 20|    |

j = 1

(2 1)(1 10) = (2 10): -4; (2 1)(1 13) = (2 13): -5; (16 1)(1 4) = (16 4): -14;

(16 1)(1 13) = (16 13): 0; (16 1)(1 20) = (16 20): 6; (17 1)(1 4) = (17 4): 0; (17 1)(1 13) = (17 13): -3; (17 1)(1 20) = (17 20): 3;

(18 1)(1 13) = (18 13): -5; (18 1)(1 20) = (18 20): 1

j = 2

(3 2)(2 4) = 3; (3 2) (2 11) = (3 11): 0; (3 2)(2 17) = (3 17): 3;  (6 2)(2 11) = (6 11): 4;

(11 2)(2 4) = 5;

(11 2)(2 11) = (11 11): 2

CYCLE (11 8 2): 2

*(11 2)(2 17) = (11 17): 5*

j = 4

(3 4)(4 1) = (3 1): -10; (3 4)(4 14) = (3 14): 2; (8 4)(4 1) = (8 1): -9; (8 4)(4 14) = 3;

j = 5

(4 5)(5 2) = (4 2): 0; (4 5)(5 13) = (4 13): 2; (6 5)(5 2) = (6 2): -8; (7 5)(5 2) = (7 2): -4;

(7 5)(5 13) = (7 13): -2; (14 5)(5 2) = (14 2): 3; (14 5)(5 13) = (14 13): 5;

(17 5)(5 2) = (17 2): -3;

j = 6

(7 6)(6 7) = (7 7): 3

CYCLE (7 12 6): 3

(7 6)(6 16) = (7 16): 4; (15 6)(6 1) = (15 1): 0; (15 6)(6 16) = (15 16): 5; (16 6)(6 1) = (16 1): -17; (16 6)(6 7) = (16 7): -13;

(16 6)(6 16) = (16 16): -12

CYCLE (16 17 6): -12

*(7 12 6 1): -1*

j = 7

(5 7)(7 10) = (5 10): 5; (12 7)(7 10) = (12 10): -14; (15 7)(7 10) = (15 10): -4;

(15 7)(7 12) = (15 12): 0

j = 8

(2 8)(8 2) = (2 2): 3

CYCLE (2 11 8): 3

(2 8)(8 10): -6; (4 8)(8 2) = (4 2): -16; (5 8)(8 10) = (5 10): -10; (12 8)(8 2) = (12 2): -18

j = 9

(4 9)(9 16) = (4 16): 6; (4 9)(9 18) = (4 18): -3; (6 9)(9 18) = (6 18): -10;

(7 9)(9 18) = (7 18): -4; (14 9)(9 18) = (14 18): 1;

(16 9)(9 16) = (16 16): 2

CYCLE (16 17 9): 2

(16 9)(9 18) = (16 18): -7

j = 10

*(12 7 10 4): 1*

j = 11

(3 11)(11 8) = (3 8): 0; (12 11)(11 14) = (12 14): 8; (12 11)(11 19) = (12 19): -7

j = 13

(12 13)(13 15): 8

*(16 5 2 11 8): 2*

j = 14

(8 14)(14 7) = (8 7): 8; (16 14)(14 19) = (16 19): 5

j = 15

(12 15)(15 14) = 2

j = 16

(7 16)(16 17) = (7 17): -10; (7 16)(16 20) = (7 20): 4

j = 18

(4 18)(18 13) = (4 13): -9;

(4 18)(18 4) = (4 4): 5

CYCLE (4 1 10 11 9 18): 5

(6 18)(18 13) = (6 13): -16; (7 18)(18 4) = (7 4): 4; (7 18)(18 8) = (7 8): 5;

(7 18)(18 13) = (7 13): -10; (7 18)(18 14) = (7 14): 3; (14 18)(18 13) = (14 13): -5;

(14 18)(18 14) = (14 14): 8

CYCLE (14 7 12 9 18): 8

(16 18)(18 13) = (16 13): -13

j = 19

(4 19)(19 3) = (4 3): 5; (4 19)(19 17) = (4 17): 4

j = 20

(5 20)(20 5) = (5 5): 8

CYCLE (5 2 4 1 20): 8

(5 20)(20 19) = (5 19): -6; (6 20)(20 19) = (6 19): -6; (7 20)(20 19) = (7 19): 1;

(16 20)(20 19) = (16 19): -3; (17 20)(20 19) = (17 19): 0

CYCLES

(2 11 8): 3; (5 2 4 1 20): 8; (7 12 6): 3; (11 8 2): 2; (14 7 12 9 18): 8; (16 17 6): -12; (16 17 9): 2

## 2-CIRCUIT PATHS

(7 12 6 1): -1; (11 2 17): 5; (12 7 10 4): 1; (16 5 2 11 8): 2;

## *VALUES OF CYCLES*

| 1  | -10: 4;      |
|----|--------------|
| 2  |              |
| 3  | -3: 5 13     |
| 4  | -10: 1;      |
| 5  | -3: 13 3;    |
| 6  | -12: 16 17,  |
| 7  | -10, -1      |
| 8  | -12, -8      |
| 9  |              |
| 10 | -12, -8      |
| 11 | -12, -8      |
| 12 | -10, -1      |
| 13 | -3: 3 5      |
| 14 |              |
| 15 |              |
| 16 | -12: 17 6    |

| 17 | -12: 6 16 |
|---|---|
| 18 | |
| 19 | |
| 20 | |

## VALUES OF 2 – CIRCUIT PATHS

| 1 | |
|---|---|
| 2 | |
| 3 | -3: 5 13 |
| 4 | |
| 5 | |
| 6 | |
| 7 | -1: 12 6 1 |
| 8 | |
| 9 | |
| 10 | |
| 11 | 5: 2 17 |
| 12 | -5: 1 4; -5: 7 4; 1: 7 10 4 |
| 13 | |
| 14 | |
| 15 | |

| 16 | 2:  ...2 11 8 |
|----|---------------|
| 17 | |
| 18 | |
| 19 | |
| 20 | |

$P_{40}$

|  | 7 | 9 | 13 | 12 | 18 | 10 | 1 | 16 | 2 | 6 | 17 | 4 | 3 | 20 | 19 | 8 | 11 | 5 | 15 | 14 |  |
|---|---|---|----|----|----|----|---|----|---|----|----|----|----|----|----|---|----|---|----|----|---|
|  | 1 | 2 | 3 | 4 | 5 | 6 | 7 | 8 | 9 | 10 | 11 | 12 | 13 | 14 | 15 | 16 | 17 | 18 | 19 | 20 |  |
| 1 |   |   |   |   |   |   |   |   |   | 1  | 10 |    |    |    |    |   |    |   | 20 | 1  | 1 |
| 2 | 4 |   |   | 2 |   |   |   | 11 | 11 | 8 | 2  |    | 1  |    |    |   |    |   |    |    | 2 |
| 3 | 4 | 5 |   | 2 | 3 |   | 3 | 11 |   | 7 | 2  | 7  |    |    |    |   | 2  |   |    |    | 3 |
| 4 | 4 | 8 | 19 |  | 20 |   | 14 | 11 | 11 | 1 | 10 | 20 | 18 | 4  | 11 | 9 | 19 | 9 | 20 | 1  | 4 |
| 5 | 4 | 5 |   | 2 |   |   | 14 | 11 | 11 | 8 | 2  |    | 4  |    |    |   | 2  |   | 20 | 1  | 5 |
| 6 | 6 | 5 |   | 1 | 12 |   | 6 | 4  | 17 |   | 10 | 7  | 18 | 4  |    | 6 | 16 | 9 | 20 | 16 | 6 |
| 7 | 4 | 5 |   | 18 | 12 |   |   | 18 | 12 | 7 |    | 7  | 18 | 18 |    |   | 16 | 9 | 20 | 16 | 7 |
| 8 | 4 | 8 |   | 10 |   |   | 14 |    |    | 8 | 2  |    |    | 4  | 11 |   | 2  |   | 20 | 11 | 8 |
| 9 |   |   |   |   |   |   |   |   |   |   |    |    | 18 |    |    |   |    | 9 |    |    | 9 |
| 10 |  |   |   |   |   |   |   |   |   |   |    |    |    |    |    |   |    |   |    |    | 10 |
| 11 |  | 8 |   | 2 |   |   |   | 11 |   | 8 |    |    |    |    |    |   |    |   |    |    | 11 |
| 12 | 12 | 8 |   |   | 12 | 12 | 12 | 11 | 12 | 7 | 10 |    | 18 | 11 | 11 | 6 | 16 | 9 | 11 | 1  | 12 |
| 13 |  |   |   |   |   |   |   |   |   |   |    |    |    |    |    |   |    |   |    |    | 13 |
| 14 | 12 | 5 |   |   | 12 |   | 14 | 11 | 12 | 7 | 10 | 7  | 18 |    |    |   |    | 9 |    |    | 14 |

| 15 | 6 |   |   |   | 17 | 14 |   | 17 | 7 |   | 7 |   | 15 |   | 6 | 15 | 17 |   |   | 15 |
|---|---|---|---|---|---|---|---|---|---|---|---|---|---|---|---|---|---|---|---|---|
| 16 | 6 | 5 |   | 1 | 16 | 17 | 6 |   | 17 |   |   |   | 18 | 18 |   |   | 16 | 9 | 20 | 1 | 16 |
| 17 | 6 | 5 |   | 1 | 12 | 17 | 6 |   |   |   | 7 | 1 |   |   |   |   |   |   | 20 | 1 | 17 |
| 18 | 4 |   |   | 18 |   |   |   |   |   |   |   | 1 | 4 |   |   |   |   |   |   | 1 | 18 |
| 19 |   |   |   |   |   |   |   |   |   |   |   |   |   |   |   |   |   |   |   |   | 19 |
| 20 |   |   |   |   |   |   |   |   |   |   |   |   |   |   |   |   |   |   |   |   | 20 |
|   | 1 | 2 | 3 | 4 | 5 | 6 | 7 | 8 | 9 | 10 | 11 | 12 | 13 | 14 | 15 | 16 | 17 | 18 | 19 | 20 |   |

$$\sigma_T^{-1} M^- (40)$$

|   | 7 | 9 | 13 | 12 | 18 | 10 | 1 | 16 | 2 | 6 | 17 | 4 | 3 | 20 | 19 | 8 | 11 | 5 | 15 | 14 |   |
|---|---|---|---|---|---|---|---|---|---|---|---|---|---|---|---|---|---|---|---|---|---|
|   | 1 | 2 | 3 | 4 | 5 | 6 | 7 | 8 | 9 | 10 | 11 | 12 | 13 | 14 | 15 | 16 | 17 | 18 | 19 | 20 |   |
| 1 | 0 |   |   |   |   |   |   |   | -4 | 0 |   | -5 |   |   |   |   |   | *3* |   |   | 1 |
| 2 | -5 | 0 |   |   |   | 5 |   | -6 | 5 |   |   | *7* |   |   | 8 |   |   |   |   |   | 2 |
| 3 | <u>-10</u> | -5 | 0 | *3* |   | <u>0</u> |   | -3 | 0 | *1* |   | *2* |   |   | *3* |   |   |   |   |   | 3 |
| 4 |   | <u>-16</u> | <u>5</u> | 0 | <u>4</u> |   | 4 | <u>-13</u> | <u>-7</u> | -12 | -13 | 5 | <u>-9</u> | 4 | 6 | 6 | <u>4</u> | -3 | <u>-10</u> | -7 | 4 |
| 5 | -9 |   |   | 4 | 0 |   | 8 | *1* | *7* | -10 | *1* |   |   | *3* |   |   | 4 |   | <u>-6</u> |   | 5 |
| 6 |   | <u>-8</u> |   | -5 | -4 | 0 |   | 2 | -14 | -12 | 4 | -3 | <u>-16</u> | -6 |   |   | -17 | -10 |   | -3 | 6 |
| 7 |   | <u>-4</u> |   | 4 | 0 | 7 | 0 | <u>5</u> | -8 | -3 | -4 | -2 | <u>-10</u> | *3* |   | 4 | -10 | -4 | <u>1</u> | 4 | 7 |
| 8 | <u>-9</u> |   |   | <u>4</u> |   |   |   | 0 |   | 2 |   |   | *3* | 7 |   | *5* |   | -4 | -1 |   | 8 |
| 9 |   |   |   |   |   |   |   |   | 0 |   |   | -2 |   |   |   |   |   |   |   |   | 9 |
| 10 |   |   |   |   |   |   |   |   | 0 |   |   |   |   |   |   |   |   |   |   |   | 10 |
| 11 |   | -3 |   | 5 |   |   |   |   | -11 | 0 |   |   |   |   |   |   |   |   |   |   | 11 |
| 12 |   | <u>-18</u> |   | -5 |   |   |   | -15 | -11 |   |   | 0 | -11 | *2* | 8 | 3 | -11 | -5 | -7 | -2 | 12 |
| 13 |   |   |   |   |   |   |   |   |   |   |   |   | 0 |   |   |   |   |   |   |   | 13 |
| 14 | -2 | <u>1</u> |   |   | 7 |   |   | *1* | -3 | 2 | *1* | 6 | 6 | 0 |   |   | *1* | *1* |   |   | 14 |
| 15 | <u>0</u> |   |   |   | 8 | -1 |   |   |   |   |   | 0 |   |   | 0 | 6 | *3* |   |   |   | 15 |
| 16 | <u>-17</u> | -3 |   | -14 | 1 | -9 | -13 |   | -11 |   |   |   | <u>-13</u> | <u>0</u> |   | 0 |   | -7 | -3 |   | 16 |
| 17 | -3 | <u>-3</u> |   |   | *1* | 5 | *1* |   | *3* | -2 |   | 2 | -3 |   |   |   | 0 | -5 | <u>-3</u> | 6 | 17 |

| | 1 | 2 | 3 | 4 | 5 | 6 | 7 | 8 | 9 | 10 | 11 | 12 | 13 | 14 | 15 | 16 | 17 | 18 | 19 | 20 | |
|---|---|---|---|---|---|---|---|---|---|---|---|---|---|---|---|---|---|---|---|---|---|
| 18 | -5 | | | | ∞ | | | | | | | | *1* | *7* | | | | 0 | | *1* | 18 |
| 19 | | | | | | | | | | | | | | | | | | | 0 | | 19 |
| 20 | | | | | | | | | | | | | | | | | | | | 0 | 20 |
| | 1 | 2 | 3 | 4 | 5 | 6 | 7 | 8 | 9 | 10 | 11 | 12 | 13 | 14 | 15 | 16 | 17 | 18 | 19 | 20 | |

j = 1

(3 1)(1 10) = (3 10): -9; (3 1)(1 20) = (3 20): -4; (8 1)(1 13) = (8 13): -9; (8 1)(1 20) = (8 20): -3

(15 1)(1 4) = (15 4): 3; (15 1)(1 13) = (15 13):1 -3; (15 1)(1 20) = (15 20): 3;

(16 1)(1 13) = (16 13): -20; (16 1)(1 20) = (16 20): -11

*(3 5 2 4 1 13): -10*

j = 2

(4 2)(2 4) = (4 4): -9

CYCLE (4 1 10 11 8 2): -9

(6 2)(2 11) = (6 11): -3

(17 2)(2 17) = (17 17): 5

CYCLE (17 6 7 12 5 2): 5

*(12 7 10 11 8 2 4): -10; (17 6 7 12 5 2 11): 2*

j = 4

*(7 12 9 18 4 1): -9; (16 17 6 1 4 8): -10*

j = 6

(15 6)(6 16) = (15 16): 5; (16 6)(6 1) = (16 1): -17

j = 7

*4 14 7 12): 5*

j = 8

(3 8)(8 10) = (3 10): -11; (7 8)(8 10) = (7 10): -6

(14 8)(8 2) = (14 2): -5

j = 9

(16 9)(9 16) = (16 16)

CYCLE (16 17 6 7 12 9): -8

(16 9)(9 18) = (16 18): -7

(18 9)(9 18) = (18 18): 5

CYCLE (18 4 1 10 11 9): 5

j = 10

(7 10)(10 11) = (7 11): -7

j = 11

(6 11)(11 8) = (6 8): -3; (6 11)(11 19) = (6 19): 5

j = 13

(6  13)(13  15) = (6  15): 3;  (16  13)(13  15) = (16  15): 6

j = 14

(7  14)(14  7) = (7  7): 8

CYCLE  (7  12  9  18  14): 8

(12  14)(14  7) = (12  7): 7

j = 16

(15  16)(16  5) = (15  5): 6

j = 17

(6  17)(17  6) = (6  6): 5

CYCLE  (6  7  12  5  2  17): 5

j = 20

(3  20)(20  19) = (3  19): -7;  (8  20)(20  5) = (8  5): 8;  (8  20)(20  19) = (8  19): -6

(15  20)(20  19) = (15  19): 3;  (16  20)(20  19) = (16  19): -14

CYCLES

(4  1  10  11  8  2): -8;  (6  7  12  5  2  17): 5;  (7  12  9  18  14): 8;  (16  17  6  7  12  9): -8;

(12  7  4  1  20): 0;  (17  6  7  12  5  2): 5;  (18 4  1  10  11  9): 1.

2-CIRCUIT PATHS

(3  5  2  4  1  13): -10;  (4  14  7  12): 5;  (7  12  9  8  4  1): -9;  (12  7  10  11  8  2  4): -10;

(16 17 6 1 4 8): -10; (17 6 7 12 5 2 11): 2

*VALUES OF CYCLES*

| 1 | -10: 4; -8: ...8 2 4; |
|---|---|
| 2 | -8: ...10 11 8 |
| 3 | -3: 5 13; 8: ...1 20 19 |
| 4 | -10: 1; -8 ...3 5 2 |
| 5 | -10: 13 3 |
| 6 | -12: 16 17; -8: ...9 16 17; |
| 7 | -10: 12; -8: ...16 17 6; |
| 8 | -12: 10 11; |
| 9 | -8: ...6 7 12; |
| 10 | -12: 11 8; |
| 11 | -12: 8 10; |
| 12 | -10: 7; -8: ...17 6 7; 8: ...15 14 7 |
| 13 | -3: 3 5 |
| 14 | |

| 15 | |
|---|---|
| 16 | -12: 17 6;  -8: ...7 12 9; |
| 17 | -12,  6 16;  -8: ...12 9 16 |
| 18 | |
| 19 | |
| 20 | |

## VALUES OF 2 – CIRCUIT PATHS

| 1 | |
|---|---|
| 2 | |
| 3 | -10: ...4 1 13;  -3:  5 13 |
| 4 | 5: ...14 7 12 |
| 5 | |
| 6 | |
| 7 | -9: ...8 4 1;  -1:  12 6 1 |
| 8 | |
| 9 | |
| 10 | |
| 11 | 5:  2 17 |
| 12 | -10: ...8 2 4;  -5:  1 4;  -5:  7 4;  1:  7 10 4 |
| 13 | |
| 14 | |
| 15 | |
| 16 | -10: ...1 4 8;  2: ...2 11 8 |
| 17 | 2: ...5 2 11 |
| 18 | |
| 19 | |
| 20 | |

$P_{60}$

| | 7 | 9 | 13 | 12 | 18 | 10 | 1 | 16 | 2 | 6 | 17 | 4 | 3 | 20 | 19 | 8 | 11 | 5 | 15 | 14 | |
|---|---|---|---|---|---|---|---|---|---|---|---|---|---|---|---|---|---|---|---|---|---|
| | 1 | 2 | 3 | 4 | 5 | 6 | 7 | 8 | 9 | 10 | 11 | 12 | 13 | 14 | 15 | 16 | 17 | 18 | 19 | 20 | |
| 1 | | | | | | | | | | 1 | 10 | | | | | | | | 20 | 1 | 1 |
| 2 | 4 | | | 2 | | | | 11 | 11 | 8 | 2 | | 1 | | | | | | | | 2 |
| 3 | 4 | 5 | | 2 | 3 | | 3 | 11 | | 8 | 2 | 7 | | | | | 2 | | 20 | 1 | 3 |
| 4 | 4 | 8 | 19 | | 20 | | 14 | 11 | 11 | 1 | 10 | 20 | 18 | 4 | 11 | 9 | 19 | 9 | 20 | 1 | 4 |
| 5 | 4 | 5 | | 2 | | | 14 | 11 | 11 | 8 | 2 | | | 4 | | | 2 | | 20 | 1 | 5 |
| 6 | 6 | 5 | | 1 | 12 | | 6 | 11 | 17 | | 2 | 7 | 18 | 4 | 13 | 6 | 2 | 9 | 11 | 16 | 6 |
| 7 | 4 | 5 | | 18 | 12 | 7 | | 18 | 12 | 8 | 10 | 7 | 18 | 18 | | 6 | 16 | 9 | 20 | 16 | 7 |
| 8 | 4 | 8 | | 10 | 20 | | 14 | | | 8 | 2 | | 1 | 4 | 11 | | 2 | | 20 | 1 | 8 |
| 9 | | | | | | | | | | | | | 18 | | | | | 9 | | | 9 |
| 10 | | | | | | | | | | | | | | | | | | | | | 10 |
| 11 | | 8 | | 10 | | | | 11 | | 8 | | | | | | | 2 | | | | 11 |
| 12 | 12 | 8 | | | 12 | 12 | 12 | 11 | 12 | 7 | 10 | | 18 | 11 | 11 | 6 | 16 | 9 | 11 | 1 | 12 |
| 13 | | | | | | | | | | | | | | | | | | | | | 13 |
| 14 | 12 | 8 | | | 12 | | 14 | 11 | 12 | 7 | 10 | 7 | 18 | | | | | 9 | | | 14 |

| | 1 | 2 | 3 | 4 | 5 | 6 | 7 | 8 | 9 | 10 | 11 | 12 | 13 | 14 | 15 | 16 | 17 | 18 | 19 | 20 | |
|---|---|---|---|---|---|---|---|---|---|---|---|---|---|---|---|---|---|---|---|---|---|
| 15 | 6 | | | 1 | 16 | 17 | 14 | | 17 | 7 | | 7 | 1 | 15 | | 6 | 15 | 17 | 20 | 1 | 15 |
| 16 | 6 | 5 | | 1 | 16 | 17 | 6 | | 17 | | | | 1 | 15 | 13 | | 16 | 9 | 20 | 1 | 16 |
| 17 | 6 | 5 | | 1 | 12 | 17 | 6 | | | | | 7 | 1 | | | | | | 20 | 1 | 17 |
| 18 | 4 | | | 18 | | | | | | | | | | 4 | | | | | | 1 | 18 |
| 19 | | | | | | | | | | | | | | | | | | | | | 19 |
| 20 | | | | | | | | | | | | | | | | | | | | | 20 |
| | 1 | 2 | 3 | 4 | 5 | 6 | 7 | 8 | 9 | 10 | 11 | 12 | 13 | 14 | 15 | 16 | 17 | 18 | 19 | 20 | |

$$\sigma_T^{-1} M^-(60)$$

| | 7 | 9 | 13 | 12 | 18 | 10 | 1 | 16 | 2 | 6 | 17 | 4 | 3 | 20 | 19 | 8 | 11 | 5 | 15 | 14 | |
|---|---|---|---|---|---|---|---|---|---|---|---|---|---|---|---|---|---|---|---|---|---|
| | 1 | 2 | 3 | 4 | 5 | 6 | 7 | 8 | 9 | 10 | 11 | 12 | 13 | 14 | 15 | 16 | 17 | 18 | 19 | 20 | |
| 1 | 0 | | | | | | | | -4 | 0 | | | -5 | | | | | | 3 | | 1 |
| 2 | -5 | 0 | | | | | | 5 | | -6 | 5 | | | 7 | | | 8 | | | | 2 |
| 3 | - | -5 | 0 | 3 | | | | 0 | | - | 0 | 1 | | 2 | | | 3 | | <u>-7</u> | -4 | 3 |
| 4 | | - | 5 | 0 | 4 | | 4 | - | -7 | - | - | 5 | <u>-9</u> | 4 | 6 | 6 | 4 | -3 | - | -7 | 4 |
| 5 | -9 | | | 4 | 0 | | 8 | 1 | 7 | - | 1 | | | 3 | | | 4 | -6 | | | 5 |
| 6 | | -8 | | -5 | -4 | 0 | | <u>-3</u> | - | - | -3 | -3 | - | -6 | 3 | | - | - | 5 | -3 | 6 |
| 7 | | -4 | | 4 | 0 | 7 | 0 | 5 | -8 | -6 | -7 | -2 | - | 3 | | 4 | - | -4 | 1 | 4 | 7 |
| 8 | -9 | | | 4 | <u>8</u> | | | 0 | | | 2 | | -9 | 3 | 7 | | 5 | | <u>-6</u> | -3 | 8 |
| 9 | | | | | | | | | 0 | | | | -2 | | | | | | | | 9 |
| 10 | | | | | | | | | 0 | | | | | | | | | | | | 10 |
| 11 | | -3 | | 5 | | | | | - | 0 | | | | | | | | | | | 11 |
| 12 | | - | | -5 | | | <u>7</u> | - | | - | | 0 | <u>-</u> | 8 | | 3 | - | -5 | -7 | -2 | 12 |
| 13 | | | | | | | | | | | | | 0 | | | | | | | | 13 |
| 14 | -2 | <u>-5</u> | | 7 | | | 1 | -3 | 2 | 1 | 6 | 6 | 0 | | | 1 | 1 | | | | 14 |
| 15 | 0 | | | 1 | <u>6</u> | 8 | -1 | | | | | 0 | 1 | | 0 | 5 | 3 | | <u>3</u> | 6 | 15 |
| 16 | - | -3 | | - | 1 | | - | | -9 | | | 1 | - | 0 | 6 | 0 | | -5 | <u>-</u> | - | 16 |
| 17 | -3 | -3 | | | 1 | 5 | 1 | | 3 | -2 | | 2 | -3 | | | | 0 | -5 | -3 | 6 | 17 |
| 18 | -5 | | | | | | | | | | | | 1 | 7 | | | | 0 | | 1 | 18 |
| 19 | | | | | | | | | | | | | | | | | | | 0 | | 19 |
| 20 | | | | | | | | | | | | | | | | | | | | 0 | 20 |

|   | 1 | 2 | 3 | 4 | 5 | 6 | 7 | 8 | 9 | 10 | 11 | 12 | 13 | 14 | 15 | 16 | 17 | 18 | 19 | 20 |   |
|---|---|---|---|---|---|---|---|---|---|----|----|----|----|----|----|----|----|----|----|----|---|

j = 2

(14 2)(2 4) = (14 4): 3

j = 4

(14 4)(4 14) = (14 14): 2

CYCLE (14 7 10 1 8 2 4): 2

j = 5

(12 7)(7 12) = (12 12): 8

CYCLE (12 9 18 13 15 14 7): 8

(15 5)(5 2) = (15 2): 2

j = 8

*[6 7 12 5 2 11 8 10]: -14*

j = 13

(12 13)(13 15) = (12 15): 8

j = 15

(12 15)(15 14) = (12 14): 2

j = 19

(3 19)(19 3) = (3 3): 8

CYCLE (3 5 2 4 1 20 19): 8

(3 19)(19 17) = (3 17): <u>7</u>; (8 19)(19 17) = (8 17): <u>7</u>; (16 19)(19 3) = (16 3): <u>1</u>

CYCLES

(3 5 2 4 1 20 19): 8; (12 9 18 13 15 14 7); (14 7 10 11 8 2 4): 6.

2-CIRCUIT PATHS

(6 7 12 5 2 11 8 10): -14

(15 17 6 16 5 2) can be extended to (15 17 6 16 5 2 11). (16 17 6 1 20 19 3) can be extended to (16 17 6 1 20 19 3 5). Neither of these two paths nor any of the other paths can be extended further.

Our next step is to obtain 2-circuit cycles using the smallest-valued 2-circuit path emanating from each of our twenty row points.

*VALUES OF 2 – CIRCUIT PATHS*

| | |
|---|---|
| 1 | |
| 2 | |
| 3 | -10: …4 1 13; -3: 5 13 |
| 4 | 5: …14 7 12 |
| 5 | |
| 6 | -14: …11 8 10 |
| 7 | -9: …8 4 1; -1: 12 6 1 |
| 8 | |
| 9 | |
| 10 | |
| 11 | 5: 2 17 |
| 12 | -10: …8 2 4; -5: 1 4; -5: 7 4; 1: 7 10 4 |
| 13 | |
| 14 | |
| 15 | |
| 16 | -10: …1 4 8; 2: …2 11 8 |
| 17 | 2: …5 2 11 |

| 18 | |
|---|---|
| 19 | |
| 20 | |

*VALUES OF CYCLES*

| 1 | -10: 4; -8: ...8 2 4; |
|---|---|
| 2 | -8: ...10 11 8 |
| 3 | -3: 5 13 |
| 4 | -10: 1; -8 ...3 5 2 |
| 5 | -10: 13 3 |
| 6 | -12: 16 17; -8: ...9 16 17 |
| 7 | -10: 12; -8: ...16 17 6; |
| 8 | -12: 10 11; |
| 9 | -8: ...6 7 12; |
| 10 | -12: 11 8; |
| 11 | -12: 8 10; |
| 12 | -10: 7; -8: ...17 6 7 |
| 13 | -3: 3 5 |

| 14 | |
|---|---|
| 15 | |
| 16 | -12: 17 6;  -8: ...7 12 9; |
| 17 | -12,  6 16;  -8: ...12 9 16 |
| 18 | |
| 19 | |
| 20 | |

## VALUES OF 2 – CIRCUIT PATHS

| 1 | |
|---|---|
| 2 | |
| 3 | -10: ...4 1 13;  -3:  5 13 |
| 4 | 5: ...14 7 12 |
| 5 | |
| 6 | 6: ...11 8 10 |
| 7 | -9: ...8 4 1;  -7:  12 1;  -1:  12 6 1 |
| 8 | |
| 9 | |
| 10 | |
| 11 | 5:  2 17 |
| 12 | -10: ...8 2 4;  -5:  1 4;  -5:  7 4;  1:  7 10 4 |
| 13 | |
| 14 | |
| 15 | |
| 16 | -10: …1 4 8;  2: …2 11 8 |
| 17 | 2: …5 2 11 |
| 18 | |
| 19 | |
| 20 | |

We now use theorem 1.22 to obtain negative paths that lead to points that cannot be determining vertices of 2-circuit cycles. In particular, there exist no determining points from cycles in rows 13 and 19. These points generally also occur least frequently in our set of cycles.

j = 1

(4 1)(1 10) = (4 10): -12; (4 1)(1 13) = (4 13): -13*; (4 1)(1 20) = (4 20): -7; (6 1)(1 4) = (6 4): -5;

*(6 1)(1 10) = (6 10): -7*; (6 1)(1 13) = (6 13): -8*; (6 1)(1 20) = (6 20): -2; *(12 1)(1 4) = (12 4): -5;*

(12 1)(1 10) = (12 10): -7; (12 1)(1 13) = (12 13): -8*; (12 1)(1 20) = -2

j = 7

(12 7)(7 10) = (12 10): -14

j = 9

(12 9)(9 18) = (12 18): -5

j = 10

(7 10)(10 11) = (7 11): -4; (8 10)(10 11) = (8 11): -12

j = 14

(15 14)(14 7) = (15 7): -1

j = 16

(6 16)(16 5) = (6 5): -2; (6 16)(16 17) = (6 17): -17; (6 16)(16 20) = (6 20): -3

j = 17

(16 17)(17 6) = (16 6): -9; (16 17)(17 9) = (16 9): -11

$P_{20}$

|    | 1  | 2 | 3 | 4 | 5  | 6  | 7  | 8 | 9  | 10 | 11 | 12 | 13 | 14 | 15 | 16 | 17 | 18 | 19 | 20 |    |
|----|----|---|---|---|----|----|----|---|----|----|----|----|----|----|----|----|----|----|----|----|----|
| 1  |    |   |   |   |    |    |    |   |    |    |    |    |    |    |    |    |    |    |    |    | 1  |
| 2  |    |   |   |   |    |    |    |   |    |    |    |    |    |    |    |    |    |    |    |    | 2  |
| 3  |    |   |   |   |    |    |    |   |    |    |    |    |    |    |    |    |    |    |    |    | 3  |
| 4  | 4  |   |   |   |    |    |    |   |    | 1  |    |    | 1* |    |    |    |    |    |    |    | 4  |
| 5  |    |   |   |   |    |    |    |   |    |    |    |    |    |    |    |    |    |    |    |    | 5  |
| 6  | 6  |   |   | 1 | 16 |    | 6  |   |    | 1  |    | 7  | 1* |    |    | 6  | 16 |    |    | 16 | 6  |
| 7  |    |   |   |   |    |    |    |   |    | 7  | 10 |    |    |    |    |    |    |    |    |    | 7  |
| 8  |    |   |   |   |    |    |    |   |    | 8  | 10 |    |    |    |    |    |    |    |    |    | 8  |
| 9  |    |   |   |   |    |    |    |   |    |    |    |    |    |    |    |    |    |    |    |    | 9  |
| 10 |    |   |   |   |    |    |    |   |    |    |    |    |    |    |    |    |    |    |    |    | 10 |
| 11 |    |   |   |   |    |    |    |   |    |    |    |    |    |    |    |    |    |    |    |    | 11 |
| 12 | 12 |   |   | 1 |    | 12 |    |   | 12 | 1  |    |    | 1* |    |    |    | 9  |    | 1  |    | 12 |
| 13 |    |   |   |   |    |    |    |   |    |    |    |    |    |    |    |    |    |    |    |    | 13 |
| 14 |    |   |   |   |    |    |    |   |    |    |    |    |    |    |    |    |    |    |    |    | 14 |
| 15 |    |   |   |   |    |    | 14 |   |    |    |    |    |    | 15 |    |    |    |    |    |    | 15 |
| 16 |    |   |   |   |    | 17 |    |   | 17 |    |    |    |    |    |    |    | 16 |    |    |    | 16 |
| 17 |    |   |   |   |    |    |    |   |    |    |    |    |    |    |    |    |    |    |    |    | 17 |
| 18 |    |   |   |   |    |    |    |   |    |    |    |    |    |    |    |    |    |    |    |    | 18 |
| 19 |    |   |   |   |    |    |    |   |    |    |    |    |    |    |    |    |    |    |    |    | 19 |

| 20 | | | | | | | | | | | | | | | | | | | | | 20 |
|---|---|---|---|---|---|---|---|---|---|---|---|---|---|---|---|---|---|---|---|---|---|
| | 1 | 2 | 3 | 4 | 5 | 6 | 7 | 8 | 9 | 10 | 11 | 12 | 13 | 14 | 15 | 16 | 17 | 18 | 19 | 20 | |

$$\sigma_T^{-1}M^-\,(20)$$

| | 7 | 9 | 13 | 12 | 18 | 10 | 1 | 16 | 2 | 6 | 17 | 4 | 3 | 20 | 19 | 8 | 11 | 5 | 15 | 14 | |
|---|---|---|---|---|---|---|---|---|---|---|---|---|---|---|---|---|---|---|---|---|---|
| | 1 | 2 | 3 | 4 | 5 | 6 | 7 | 8 | 9 | 10 | 11 | 12 | 13 | 14 | 15 | 16 | 17 | 18 | 19 | 20 | |
| 1 | | | | | | | | | | | | | | | | | | | | | 1 |
| 2 | | | | | | | | | | | | | | | | | | | | | 2 |
| 3 | | | | | | | | | | | | | | | | | | | | | 3 |
| 4 | | | | | | | | | | -12 | | | -13 | | | | | | | | 4 |
| 5 | | | | | | | | | | | | | | | | | | | | | 5 |
| 6 | | | | -5 | -2 | | | | | | | -3 | -8* | | | | -17 | | | -3 | 6 |
| 7 | | | | | | | | | | | -4 | | | | | | | | | | 7 |
| 8 | | | | | | | | | | -12 | | | | | | | | | | | 8 |
| 9 | | | | | | | | | | | | | | | | | | | | | 9 |
| 10 | | | | | | | | | | | | | | | | | | | | | 10 |
| 11 | | | | | | | | | | | | | | | | | | | | | 11 |

| | 1 | 2 | 3 | 4 | 5 | 6 | 7 | 8 | 9 | 10 | 11 | 12 | 13 | 14 | 15 | 16 | 17 | 18 | 19 | 20 | |
|---|---|---|---|---|---|---|---|---|---|---|---|---|---|---|---|---|---|---|---|---|---|
| 12 | | | | | | | | | -14 | | | | -8* | | | | | -5 | | -2 | 12 |
| 13 | | | | | | | | | | | | | | | | | | | | | 13 |
| 14 | | | | | | | | | | | | | | | | | | | | | 14 |
| 15 | | | | | | | -1 | | | | | | | | | | | | | | 15 |
| 16 | | | | | | -9 | | | -11 | | | | | | | | | | | | 16 |
| 17 | | | | | | | | | | | | | | | | | | | | | 17 |
| 18 | | | | | | | | | | | | | | | | | | | | | 18 |
| 19 | | | | | | | | | | | | | | | | | | | | | 19 |
| 20 | | | | | | | | | | | | | | | | | | | | | 20 |
| | 1 | 2 | 3 | 4 | 5 | 6 | 7 | 8 | 9 | 10 | 11 | 12 | 13 | 14 | 15 | 16 | 17 | 18 | 19 | 20 | |

*VALUES OF SPECIAL 2 – CIRCUIT CYCLES*

| | |
|---|---|
| 1 | |
| 2 | |
| 3 | |
| 4 | |
| 5 | |
| 6 | |
| 7 | |
| 8 | |
| 9 | |
| 10 | |
| 11 | |
| 12 | |
| 13 | |
| 14 | |
| 15 | |
| 16 | |
| 17 | |
| 18 | |
| 19 | |
| 20 | |

j = 5

(6 5)(5 2) = (6 2): -6

j = 6

(16 6)(6 1) = (16 1): -17; (16 6)(6 7) = (16 7): -13

j = 7

(15 10)(10 11) = (15 11): -5

j = 11

(15 11)(11 8) = (15 8): -5

j = 12

(15 12)(12 5) = (15 5): -1; (15 12)(12 9) = (15 9): -9; (16 12)(12 5) = (16 5): -13;

(16 12)(12 9) = (16 9): -21

$P_{40}$

|    | 1  | 2 | 3 | 4 | 5  | 6  | 7  | 8  | 9  | 10 | 11 | 12 | 13 | 14 | 15 | 16 | 17 | 18 | 19 | 20 |    |
|----|----|---|---|---|----|----|----|----|----|----|----|----|----|----|----|----|----|----|----|----|----|
| 1  |    |   |   |   |    |    |    |    |    |    |    |    |    |    |    |    |    |    |    |    | 1  |
| 2  |    |   |   |   |    |    |    |    |    |    |    |    |    |    |    |    |    |    |    |    | 2  |
| 3  |    |   |   |   |    |    |    |    |    |    |    |    |    |    |    |    |    |    |    |    | 3  |
| 4  | 4  |   |   |   |    |    |    |    |    | 1  |    |    | 1* |    |    |    |    |    |    |    | 4  |
| 5  |    |   |   |   |    |    |    |    |    |    |    |    |    |    |    |    |    |    |    |    | 5  |
| 6  | 6  | 5 |   | 1 | 16 |    | 6  |    |    |    |    | 7  | 1* |    |    | 6  | 16 |    |    | 16 | 6  |
| 7  |    |   |   |   |    |    |    |    |    | 7  | 10 |    |    |    |    |    |    |    |    |    | 7  |
| 8  |    |   |   |   |    |    |    |    |    | 8  | 10 |    |    |    |    |    |    |    |    |    | 8  |
| 9  |    |   |   |   |    |    |    |    |    |    |    |    |    |    |    |    |    |    |    |    | 9  |
| 10 |    |   |   |   |    |    |    |    |    |    |    |    |    |    |    |    |    |    |    |    | 1  |
| 11 |    |   |   |   |    |    |    |    |    |    |    |    |    |    |    |    |    |    |    |    | 1  |
| 12 | 12 |   |   | 1 |    |    | 12 |    | 12 | 1  |    |    | 1* | 4  |    |    |    | 9  |    | 1  | 1  |
| 13 |    |   |   |   |    |    |    |    |    |    |    |    |    |    |    |    |    |    |    |    | 1  |
| 14 |    |   |   |   |    |    |    |    |    |    |    |    |    |    |    |    |    |    |    |    | 1  |
| 15 |    |   |   |   | 12 |    | 14 | 11 | 12 | 7  | 10 | 7  |    | 15 |    |    |    |    |    |    | 1  |
| 16 | 6  |   |   | 7 | 12 | 17 | 6  |    | 12 |    |    | 7  |    |    |    |    | 16 |    |    |    | 1  |
| 17 |    |   |   |   |    |    |    |    |    |    |    |    |    |    |    |    |    |    |    |    | 1  |
| 18 |    |   |   |   |    |    |    |    |    |    |    |    |    |    |    |    |    |    |    |    | 1  |

| 19 |   |   |   |   |   |   |   |   |   |    |    |    |    |    |    |    |    |    |    |    | 19 |
|----|---|---|---|---|---|---|---|---|---|----|----|----|----|----|----|----|----|----|----|----|----|
| 20 |   |   |   |   |   |   |   |   |   |    |    |    |    |    |    |    |    |    |    |    | 20 |
|    | 1 | 2 | 3 | 4 | 5 | 6 | 7 | 8 | 9 | 10 | 11 | 12 | 13 | 14 | 15 | 16 | 17 | 18 | 19 | 20 |    |

$$\sigma_T^{-1} M^- \ (40)$$

|  | 7 | 9 | 13 | 12 | 18 | 10 | 1 | 16 | 2 | 6 | 17 | 4 | 3 | 20 | 19 | 8 | 11 | 5 | 15 | 14 |  |
|--|---|---|----|----|----|----|---|----|---|---|----|---|---|----|----|---|----|---|----|----|--|
|  | 1 | 2 | 3 | 4 | 5 | 6 | 7 | 8 | 9 | 10 | 11 | 12 | 13 | 14 | 15 | 16 | 17 | 18 | 19 | 20 |  |
| 1 |   |   |   |   |   |   |   |   |   |    |    |    |    |    |    |    |    |    |    |    | 1 |
| 2 |   |   |   |   |   |   |   |   |   |    |    |    |    |    |    |    |    |    |    |    | 2 |
| 3 |   |   |   |   |   |   |   |   |   |    |    |    |    |    |    |    |    |    |    |    | 3 |
| 4 |   |   |   |   |   |   |   |   |   | -12 |   |   | -13 |   |   |   |   |    |    |    | 4 |
| 5 |   |   |   |   |   |   |   |   |   |    |    |    |    |    |    |    |    |    |    |    | 5 |
| 6 |   | <u>-6</u> |   | -5 | <u>-2</u> |   |   |   |   |    |    | -3 | -8* |   |   |   | -17 |    |    | -3 | 6 |
| 7 |   |   |   |   |   |   |   |   |   |    | -4 |   |    |    |    |   |    |    |    |    | 7 |
| 8 |   |   |   |   |   |   |   |   |   |    | -12 |  |    |    |    |   |    |    |    |    | 8 |
| 9 |   |   |   |   |   |   |   |   |   |    |    |   |    |    |    |   |    |    |    |    | 9 |
| 10 |  |   |   |   |   |   |   |   |   |    |    |   |    |    |    |   |    |    |    |    | 10 |

| | 1 | 2 | 3 | 4 | 5 | 6 | 7 | 8 | 9 | 10 | 11 | 12 | 13 | 14 | 15 | 16 | 17 | 18 | 19 | 20 | |
|---|---|---|---|---|---|---|---|---|---|---|---|---|---|---|---|---|---|---|---|---|---|
| 11 | | | | | | | | | | | | | | | | | | | | | 11 |
| 12 | | | | | | | | | -14 | | | -8* | -6 | | | | -5 | | -2 | | 12 |
| 13 | | | | | | | | | | | | | | | | | | | | | 13 |
| 14 | | | | | | | | | | | | | | | | | | | | | 14 |
| 15 | | | | | -1 | | -1 | -5 | -9 | -4 | -5 | 0 | | | | | | | | | 15 |
| 16 | -17 | | | | -13 | -9 | -13 | | -21 | | | -12 | | | | | | | | | 16 |
| 17 | | | | | | | | | | | | | | | | | | | | | 17 |
| 18 | | | | | | | | | | | | | | | | | | | | | 18 |
| 19 | | | | | | | | | | | | | | | | | | | | | 19 |
| 20 | | | | | | | | | | | | | | | | | | | | | 20 |
| | 1 | 2 | 3 | 4 | 5 | 6 | 7 | 8 | 9 | 10 | 11 | 12 | 13 | 14 | 15 | 16 | 17 | 18 | 19 | 20 | |

*VALUES OF SPECIAL 2−CIRCUIT CYCLES*

| | |
|---|---|
| 1 | |
| 2 | |
| 3 | |
| 4 | |
| 5 | |
| 6 | |
| 7 | |
| 8 | |
| 9 | |
| 10 | |
| 11 | |
| 12 | |
| 13 | |
| 14 | |
| 15 | |
| 16 | |
| 17 | |
| 18 | |
| 19 | |

| 20 | |
|----|--|

j = 1

(16 1)(1 4) = (16 4): -14; (16 1)(1 13) = (16 13): -17*; (16 1)(1 20) = (16 20): -11

j = 2

(6 2)(2 11) = (6 11): -1

j = 4

(16 4)(4 2) = (16 2): <u>-6</u>

j = 5

(6 5)(5 2) = (6 2): <u>-6</u>; (15 5)(5 2) = (15 2): <u>-5</u>; (16 5)(5 2) = (16 2): <u>-17</u>

j = 8

(15 8)(8 2) = (15 2): <u>-8</u>

j = 9

(15 9)(9 18) = (15 18): -5

(16 9)(9 16) = (16 16): -8

CYCLE (16 17 6 7 12 9): -8

j = 13

(16 13)(13 15) = (16 15): 2*

j = 15

(16 15)(15 14) = (16 14): <u>-4*</u>

j = 18

(15 18)(18 13) = (15 13): <u>-11\*</u>

j = 20

(16 20)(20 5) = (16 5): <u>0</u>; (16 20)(20 19) = (16 19): <u>-14\*</u>

$P_{60}$

|    | 1  | 2 | 3 | 4 | 5  | 6  | 7  | 8  | 9  | 10 | 11 | 12 | 13 | 14 | 15 | 16 | 17 | 18 | 19 | 20 |    |
|----|----|---|---|---|----|----|----|----|----|----|----|----|----|----|----|----|----|----|----|----|----|
| 1  |    |   |   |   |    |    |    |    |    |    |    |    |    |    |    |    |    |    |    |    | 1  |
| 2  |    |   |   |   |    |    |    |    |    |    |    |    |    |    |    |    |    |    |    |    | 2  |
| 3  |    |   |   |   |    |    |    |    |    |    |    |    |    |    |    |    |    |    |    |    | 3  |
| 4  | 4  |   |   |   |    |    |    |    |    | 1  |    |    | 1* |    |    |    |    |    |    |    | 4  |
| 5  |    |   |   |   |    |    |    |    |    |    |    |    |    |    |    |    |    |    |    |    | 5  |
| 6  | 6  | 5 |   | 1 | 16 |    | 6  |    |    |    | 2  | 7  | 1* |    |    | 6  | 16 |    |    | 16 | 6  |
| 7  |    |   |   |   |    |    |    |    |    | 7  | 10 |    |    |    |    |    |    |    |    |    | 7  |
| 8  |    |   |   |   |    |    |    |    |    | 8  | 10 |    |    |    |    |    |    |    |    |    | 8  |
| 9  |    |   |   |   |    |    |    |    |    |    |    |    |    |    |    |    |    |    |    |    | 9  |
| 10 |    |   |   |   |    |    |    |    |    |    |    |    |    |    |    |    |    |    |    |    | 1  |
| 11 |    |   |   |   |    |    |    |    |    |    |    |    |    |    |    |    |    |    |    |    | 1  |
| 12 | 12 |   |   | 1 |    |    | 12 |    | 12 | 1  |    |    | 1* | 4  |    |    |    | 9  |    | 1  | 1  |
| 13 |    |   |   |   |    |    |    |    |    |    |    |    |    |    |    |    |    |    |    |    | 1  |
| 14 |    |   |   |   |    |    |    |    |    |    |    |    |    |    |    |    |    |    |    |    | 1  |
| 15 |    | 8 |   |   | 12 |    | 14 | 11 | 12 | 7  | 10 | 7  | 18 | 15 |    |    |    | 9  |    |    | 1  |
| 16 | 6  | 5 |   | 1 | 12 | 17 | 6  |    | 12 |    |    | 7  | 1* | 15 | 13 |    | 16 |    | 20 |    | 1  |
| 17 |    |   |   |   |    |    |    |    |    |    |    |    |    |    |    |    |    |    |    |    | 1  |

| | 1 | 2 | 3 | 4 | 5 | 6 | 7 | 8 | 9 | 10 | 11 | 12 | 13 | 14 | 15 | 16 | 17 | 18 | 19 | 20 | |
|---|---|---|---|---|---|---|---|---|---|---|---|---|---|---|---|---|---|---|---|---|---|
| 18 | | | | | | | | | | | | | | | | | | | | | 1 |
| 19 | | | | | | | | | | | | | | | | | | | | | 1 |
| 20 | | | | | | | | | | | | | | | | | | | | | 2 |
| | 1 | 2 | 3 | 4 | 5 | 6 | 7 | 8 | 9 | 10 | 11 | 12 | 13 | 14 | 15 | 16 | 17 | 18 | 19 | 20 | |

$$\sigma_{T_{165}}^{-1} M^{-} (60)$$

| | 7 | 9 | 13 | 12 | 18 | 10 | 1 | 16 | 2 | 6 | 17 | 4 | 3 | 20 | 19 | 8 | 11 | 5 | 15 | 14 | |
|---|---|---|---|---|---|---|---|---|---|---|---|---|---|---|---|---|---|---|---|---|---|
| | 1 | 2 | 3 | 4 | 5 | 6 | 7 | 8 | 9 | 10 | 11 | 12 | 13 | 14 | 15 | 16 | 17 | 18 | 19 | 20 | |
| 1 | | | | | | | | | | | | | | | | | | | | | 1 |
| 2 | | | | | | | | | | | | | | | | | | | | | 2 |
| 3 | | | | | | | | | | | | | | | | | | | | | 3 |
| 4 | | | | | | | | | | -12 | | | -13 | | | | | | | | 4 |
| 5 | | | | | | | | | | | | | | | | | | | | | 5 |
| 6 | | -6 | | -5 | -2 | | | | | | | -3 | -8* | | | | -17 | | | -3 | 6 |
| 7 | | | | | | | | | | | -4 | | | | | | | | | | 7 |
| 8 | | | | | | | | | | | -12 | | | | | | | | | | 8 |
| 9 | | | | | | | | | | | | | | | | | | | | | 9 |

| | 1 | 2 | 3 | 4 | 5 | 6 | 7 | 8 | 9 | 10 | 11 | 12 | 13 | 14 | 15 | 16 | 17 | 18 | 19 | |
|---|---|---|---|---|---|---|---|---|---|---|---|---|---|---|---|---|---|---|---|---|
| 10 | | | | | | | | | | | | | | | | | | | | 10 |
| 11 | | | | | | | | | | | | | | | | | | | | 11 |
| 12 | | | | | | | | -14 | | | -8* | -6 | | | -5 | | | -2 | | 12 |
| 13 | | | | | | | | | | | | | | | | | | | | 13 |
| 14 | | | | | | | | | | | | | | | | | | | | 14 |
| 15 | | -8 | | -1 | | -1 | -5 | -9 | -4 | -5 | 0 | -11 | | | -5 | | | | | 15 |
| 16 | -17 | -17 | | -14 | -13 | -9 | -13 | | -21 | | -12 | -4* | 2* | | | | | -14 | -11 | 16 |
| 17 | | | | | | | | | | | | | | | | | | | | 17 |
| 18 | | | | | | | | | | | | | | | | | | | | 18 |
| 19 | | | | | | | | | | | | | | | | | | | | 19 |
| 20 | | | | | | | | | | | | | | | | | | | | 20 |
| | 1 | 2 | 3 | 4 | 5 | 6 | 7 | 8 | 9 | 1 | 1 | 1 | 1 | 1 | 1 | 1 | 1 | 1 | 2 | |

*VALUES OF SPECIAL 2 – CIRCUIT CYCLES*

| | |
|---|---|
| 1 | |
| 2 | |
| 3 | |
| 4 | |
| 5 | |
| 6 | |
| 7 | |
| 8 | |
| 9 | |
| 10 | |
| 11 | |
| 12 | |
| 13 | |
| 14 | |
| 15 | |
| 16 | -8: ...7 12 9; |
| 17 | |

| 18 | |
|---|---|
| 19 | |
| 20 | |

j = 1

(15 2)(2 4) = (15 4): 0

j = 13

(15 13)(13 15) = (15 15): 8

CYCLE (15 14 7 12 9 18 13): 8

j = 19

(16 19)(19 3) = (16 3): <u>1</u>

We have only one path that might be extended: (16 17 6 1 20 19 3): 1. Since the path contains 19, we can extend it to become a 2-circuit path. Its possible extensions follow:

*16 17 6* **1** *20 19 3* **7** *12 9 16*
*16 17 6* **1** *20 19* **3 7** *12 9 18* **13** *16 …* **13**

can't be extended to a cycle whose value is less than 9. We thus obtain the 2-circuit cycle

*( 16 17 6* **1** *20 19 3* **7** *12 9) : 6* . We place it in *VALUES SPECIAL 2-CIRCUIT CYCLES.*

|    | 7 | 9 | 13 | 12 | 18 | 10 | 1 | 16 | 2 | 6 | 17 | 4 | 3 | 20 | 19 | 8 | 11 | 5 | 15 | 14 |    |
|    | 1 | 2 | 3  | 4  | 5  | 6  | 7 | 8  | 9 | 10| 11 | 12| 13| 14 | 15 | 16| 17 | 18| 19 | 20 |    |
|----|---|---|----|----|----|----|---|----|---|---|----|---|---|----|----|---|----|---|----|----|----|
| 1  | 0 |   |    |    |    |    |   |    |   | -4| 0  |   | -5|    |    |   |    |   | 3  |    | 1  |
| 2  | -5| 0 |    |    |    |    |   | 5  |   | -6| 5  |   | 7 |    |    |   | 8  |   |    |    | 2  |
| 3  | - | -5| 0  | 3  |    |    |   | 0  |   | - | 0  | 1 |   | 2  |    |   | 3  |   | -7 | -4 | 3  |
| 4  | - | - | 5  | 0  | 4  |    | 4 | -  | -7| - | -  | 5 | -9| 4  | 6  | 6 | 4  | -3| -  | -7 | 4  |
| 5  | -9|   |    | 4  | 0  |    | 8 | 1  | 7 | - | 1  |   |   | 3  |    |   | 4  |   | -6 |    | 5  |
| 6  |   | -8|    | -5 | -4 | 0  |   | -3 | - | - | -3 | -3| - | -6 | 3  |   | -  | - | 5  | -3 | 6  |
| 7  |   | -4|    | 4  | 0  | 7  | 0 | 5  | -8| -6| -7 | -2| - | 3  |    | 4 | -  | -4| 1  | 4  | 7  |
| 8  | -9|   |    | 4  | 8  |    |   | 0  |   |   | 2  |   | -9| 3  | 7  |   | 5  |   | -6 | -3 | 8  |
| 9  |   |   |    |    |    |    |   |    | 0 |   |    |   | -2|    |    |   |    |   |    |    | 9  |
| 10 |   |   |    |    |    |    |   |    |   | 0 |    |   |   |    |    |   |    |   |    |    | 10 |
| 11 |   | -3|    | 5  |    |    |   |    |   | - | 0  |   |   |    |    |   |    |   |    |    | 11 |
| 12 |   | - |    | -5 |    |    | - |    |   | - |    | 0 | - | 8  |    | 3 | -  | -5| -7 | -2 | 12 |
| 13 |   |   |    |    |    |    |   |    |   |   |    |   | 0 |    |    |   |    |   |    |    | 13 |
| 14 | -7| -5|    | 6  | 7  |    | 1 | -3 | 2 | 1 | 6  | 6 | 0 |    |    | 1 | 1  |   |    |    | 14 |
| 15 | 0 | 2 |    | 1  | 6  | 8  | -1|    |   |   |    | 0 | 1 |    | 0  | 5 | 3  |   | 3  | 6  | 15 |
| 16 | - | -3| 1  | -  | 1  |    | - |    | -9|   |    | 1 | - | 0  | 6  | 0 |    | -5| -  | -  | 16 |

| | 1 | 2 | 3 | 4 | 5 | 6 | 7 | 8 | 9 | 10 | 11 | 12 | 13 | 14 | 15 | 16 | 17 | 18 | 19 | 20 | |
|---|---|---|---|---|---|---|---|---|---|---|---|---|---|---|---|---|---|---|---|---|---|
| 17 | *-3* | *-3* | | | *1* | *5* | *1* | | *3* | *-2* | | *2* | *-3* | | | | 0 | -5 | *-3* | *6* | 17 |
| 18 | *-5* | | | | | | | | | | | *1* | *7* | | | | | 0 | | *1* | 18 |
| 19 | | | | | | | | | | | | | | | | | | | 0 | | 19 |
| 20 | | | | | | | | | | | | | | | | | | | | 0 | 20 |
| | 1 | 2 | 3 | 4 | 5 | 6 | 7 | 8 | 9 | 10 | 11 | 12 | 13 | 14 | 15 | 16 | 17 | 18 | 19 | 20 | |

Since we cannot extend any path further, we now will use the smallest-valued path given in each row to obtain 2-circuit cycles. Before doing so, we number the cycles in order of increasing value. Since two companion cycles yield the same circuit or circuits, we delete one of them to make our search easier. Also, we may obtain the same circuit starting with different determining vertices. In this case we also eliminate one. However, as we mentioned in describing our algorithm, the simplest procedure is to place all cycles obtained – acceptable or 2-circuit - in an $m \; X \; \dfrac{n}{2}$ matrix. If any set of cycles all have their linking points in the same set of 2-cycle columns while their respective non-linking points are the same, we can delete all but one of these cycles. Using such a procedure, we obtain the following set of acceptable cycles.

1. (16 17 6): -12   2. (4 1): -10   3. (4 1 10 11 8 2): -8   4. (4 1 20 12 7): 0   5. (11 8 2): 2

6. (6 7 12): 3   7. (6 16): 5   8. (6 7 12 5 2 17): 5   9. (4 1 20 5 2): 8   10.

(12 9 18 13 15 14 7): 8

We now search for 2-circuit cycles.

## UNLINKED 2 - CIRCUIT PATHS

|    | 1 | 2 | 3 | 4 | 5 | 6 | 7  | 8  | 9 | 10 | 11 | 12 | 13 | 14 | 15 | 16 | 17 | 18 | 19 | 20 |    |
|----|---|---|---|---|---|---|----|----|---|----|----|----|----|----|----|----|----|----|----|----|----|
| 1  |   |   |   |   |   |   |    |    |   |    |    |    |    |    |    |    |    |    |    |    | 1  |
| 2  |   |   |   |   |   |   |    |    |   |    |    |    |    |    |    |    |    |    |    |    | 2  |
| 3  | 4 | 5 |   | 2 | 3 |   |    |    |   |    |    | 1  |    |    |    |    |    |    |    |    | 3  |
| 4  |   |   |   |   |   |   | 14 |    |   |    |    | 7  |    | 4  |    |    |    |    |    |    | 4  |
| 5  |   |   |   |   |   |   |    |    |   |    |    |    |    |    |    |    |    |    |    |    | 5  |
| 6  |   | 5 |   |   | 12|   | 6  | 11 |   | 8  | 2  | 7  |    |    |    |    |    |    |    |    | 6  |
| 7  |   |   |   |   |   |   |    |    |   |    |    |    |    |    |    |    |    |    |    |    | 7  |
| 8  |   |   |   |   |   |   |    |    |   |    |    |    |    |    |    |    |    |    |    |    | 8  |
| 9  |   |   |   |   |   |   |    |    |   |    |    |    |    |    |    |    |    |    |    |    | 9  |
| 10 |   |   |   |   |   |   |    |    |   |    |    |    |    |    |    |    |    |    |    |    | 10 |
| 11 |   | 8 |   |   |   |   |    | 11 |   |    |    |    |    |    |    |    | 2  |    |    |    | 11 |
| 12 |   | 8 |   | 2 |   |   | 12 | 11 |   | 7  | 10 |    |    |    |    |    |    |    |    |    | 12 |
| 13 |   |   |   |   |   |   |    |    |   |    |    |    |    |    |    |    |    |    |    |    | 13 |
| 14 |   |   |   |   |   |   |    |    |   |    |    |    |    |    |    |    |    |    |    |    | 14 |
| 15 |   |   |   |   |   |   |    |    |   |    |    |    |    |    |    |    |    |    |    |    | 15 |
| 16 | 6 |   |   | 1 |   | 17|    | 4  |   |    |    |    |    |    |    |    | 16 |    |    |    | 16 |

| | 1 | 2 | 3 | 4 | 5 | 6 | 7 | 8 | 9 | 10 | 11 | 12 | 13 | 14 | 15 | 16 | 17 | 18 | 19 | 20 | |
|---|---|---|---|---|---|---|---|---|---|---|---|---|---|---|---|---|---|---|---|---|---|
| 17 | | 5 | | | 12 | 17 | 6 | | | 2 | 7 | | | | | | | | | | 17 |
| 18 | | | | | | | | | | | | | | | | | | | | | 18 |
| 19 | | | | | | | | | | | | | | | | | | | | | 19 |
| 20 | | | | | | | | | | | | | | | | | | | | | 20 |
| | 1 | 2 | 3 | 4 | 5 | 6 | 7 | 8 | 9 | 10 | 11 | 12 | 13 | 14 | 15 | 16 | 17 | 18 | 19 | 20 | |

UNLINKED $\sigma_T^{-1} M^-$

| | 7 | 9 | 13 | 12 | 18 | 10 | 1 | 16 | 2 | 6 | 17 | 4 | 3 | 20 | 19 | 8 | 11 | 5 | 15 | 14 | |
|---|---|---|---|---|---|---|---|---|---|---|---|---|---|---|---|---|---|---|---|---|---|
| | 1 | 2 | 3 | 4 | 5 | 6 | 7 | 8 | 9 | 10 | 11 | 12 | 13 | 14 | 15 | 16 | 17 | 18 | 19 | 20 | |
| 1 | | | | | | | | | | | | | | | | | | | | | 1 |
| 2 | | | | | | | | | | | | | | | | | | | | | 2 |
| 3 | -10 | -5 | | 3 | -1 | | | | | | | | -10 | | | | | | | | 3 |
| 4 | | | | | | | 4 | | | | | 5 | | -1 | | | | | | | 4 |
| 5 | | | | | | | | | | | | | | | | | | | | | 5 |
| 6 | | -8 | | | -4 | | -4 | -3 | | -14 | -3 | -3 | | | | | | | | | 6 |
| 7 | | | | | | | | | | | | | | | | | | | | | 7 |
| 8 | | | | | | | | | | | | | | | | | | | | | 8 |
| 9 | | | | | | | | | | | | | | | | | | | | | 9 |
| 10 | | | | | | | | | | | | | | | | | | | | | 10 |

| | 1 | 2 | 3 | 4 | 5 | 6 | 7 | 8 | 9 | 10 | 11 | 12 | 13 | 14 | 15 | 16 | 17 | 18 | 19 | 20 | |
|---|---|---|---|---|---|---|---|---|---|---|---|---|---|---|---|---|---|---|---|---|---|
| 11 | | -3 | | | | | 0 | | | | | | | | | | 5 | | | | 11 |
| 12 | | -18 | | -10 | | | -11 | -15 | | -14 | -15 | | | | | | | | | | 12 |
| 13 | | | | | | | | | | | | | | | | | | | | | 13 |
| 14 | | | | | | | | | | | | | | | | | | | | | 14 |
| 15 | | | | | | | | | | | | | | | | | | | | | 15 |
| 16 | -17 | | | -14 | | -9 | | -7 | | | | | | | | | -14 | | | | 16 |
| 17 | | -3 | | | 1 | 5 | 1 | | | | 2 | 2 | | | | | | | | | 17 |
| 18 | | | | | | | | | | | | | | | | | | | | | 18 |
| 19 | | | | | | | | | | | | | | | | | | | | | 19 |
| 20 | | | | | | | | | | | | | | | | | | | | | 20 |
| | 1 | 2 | 3 | 4 | 5 | 6 | 7 | 8 | 9 | 10 | 11 | 12 | 13 | 14 | 15 | 16 | 17 | 18 | 19 | 20 | |

j = 4

(12 4)(4 13) = (12 13): -5; (12 4)(4 14) = (12 14): -11;

(12 4)(4 18) = (12 18): 5; (12 4)(4 20) = (12 20): 0

(12 *7* 10 11 8 2 4 *1*): -23;  (12 7 10 11 8 *2* 4 *9*): 2

j = 8

(16 8)(8 2) = (16 2): <u>-10</u>

j = 10

(16 17 *6* 1 4 8 *10*): -18

j = 11

(17 11)(11 8) = (17 8): <u>2</u>

(17 6 7 12 5 *2* 11 *9*): <u>8</u>

j = 12

(4 12)(12 5) = (4 5): <u>4</u>; (14 12)(12 9) = <u>-4</u>

(4 14 *7* 12 *1*): <u>-3</u>

j = 17

(11 8 *2* 17 *9*): <u>8</u>

*UNLINKED 2 - CIRCUITS $P_{20}$*

|    | 1 | 2 | 3 | 4 | 5  | 6  | 7  | 8  | 9  | 10 | 11 | 12 | 13 | 14 | 15 | 16 | 17 | 18 | 19 | 20 |    |
|----|---|---|---|---|----|----|----|----|----|----|----|----|----|----|----|----|----|----|----|----|----|
| 1  |   |   |   |   |    |    |    |    |    |    |    |    |    |    |    |    |    |    |    |    | 1  |
| 2  |   |   |   |   |    |    |    |    |    |    |    |    |    |    |    |    |    |    |    |    | 2  |
| 3  | 4 | 5 |   | 2 | 3  |    |    |    |    |    |    | 1  |    |    |    |    |    |    |    |    | 3  |
| 4  |   |   |   |   | 12 |    | 14 |    | 12 |    |    | 7  |    | 4  |    |    |    |    |    |    | 4  |
| 5  |   |   |   |   |    |    |    |    |    |    |    |    |    |    |    |    |    |    |    |    | 5  |
| 6  |   | 5 |   |   | 12 |    | 6  | 11 |    | 8  | 2  | 7  |    |    |    |    |    |    |    |    | 6  |
| 7  |   |   |   |   |    |    |    |    |    |    |    |    |    |    |    |    |    |    |    |    | 7  |
| 8  |   |   |   |   |    |    |    |    |    |    |    |    |    |    |    |    |    |    |    |    | 8  |
| 9  |   |   |   |   |    |    |    |    |    |    |    |    |    |    |    |    |    |    |    |    | 9  |
| 10 |   |   |   |   |    |    |    |    |    |    |    |    |    |    |    |    |    |    |    |    | 10 |
| 11 |   | 8 |   |   |    |    |    | 11 |    |    |    |    |    |    |    |    | 2  |    |    |    | 11 |
| 12 | 4 | 8 |   | 2 |    |    | 12 | 11 |    | 7  | 10 |    | 4  | 4  |    |    | 4  |    | 4  |    | 12 |
| 13 |   |   |   |   |    |    |    |    |    |    |    |    |    |    |    |    |    |    |    |    | 13 |
| 14 |   |   |   |   |    |    |    |    |    |    |    |    |    |    |    |    |    |    |    |    | 14 |
| 15 |   |   |   |   |    |    |    |    |    |    |    |    |    |    |    |    |    |    |    |    | 15 |
| 16 | 6 | 8 |   | 1 |    | 17 |    | 4  |    |    |    |    |    |    |    |    | 16 |    |    |    | 16 |
| 17 |   | 5 |   |   | 12 | 17 | 6  | 11 |    |    | 2  | 7  |    |    |    |    |    |    |    |    | 17 |
| 18 |   |   |   |   |    |    |    |    |    |    |    |    |    |    |    |    |    |    |    |    | 18 |

| | 1 | 2 | 3 | 4 | 5 | 6 | 7 | 8 | 9 | 10 | 11 | 12 | 13 | 14 | 15 | 16 | 17 | 18 | 19 | 20 | |
|---|---|---|---|---|---|---|---|---|---|---|---|---|---|---|---|---|---|---|---|---|---|
| 19 | | | | | | | | | | | | | | | | | | | | | 19 |
| 20 | | | | | | | | | | | | | | | | | | | | | 20 |
| | 1 | 2 | 3 | 4 | 5 | 6 | 7 | 8 | 9 | 10 | 11 | 12 | 13 | 14 | 15 | 16 | 17 | 18 | 19 | 20 | |

$$\text{UNLINKED } \sigma_{T_{165}}^{-1} M^{-}(20)$$

| | 7 | 9 | 13 | 12 | 18 | 10 | 1 | 16 | 2 | 6 | 17 | 4 | 3 | 20 | 19 | 8 | 11 | 5 | 15 | 14 | |
|---|---|---|---|---|---|---|---|---|---|---|---|---|---|---|---|---|---|---|---|---|---|
| | 1 | 2 | 3 | 4 | 5 | 6 | 7 | 8 | 9 | 10 | 11 | 12 | 13 | 14 | 15 | 16 | 17 | 18 | 19 | 20 | |
| 1 | | | | | | | | | | | | | | | | | | | | | 1 |
| 2 | | | | | | | | | | | | | | | | | | | | | 2 |
| 3 | -10 | -5 | | 3 | -1 | | | | | | | | -10 | | | | | | | | 3 |
| 4 | | | | | <u>4</u> | | 4 | | <u>-4</u> | | | 5 | | -1 | | | | | | | 4 |
| 5 | | | | | | | | | | | | | | | | | | | | | 5 |
| 6 | | -8 | | | -4 | | -4 | -3 | | -14 | -3 | -3 | | | | | | | | | 6 |
| 7 | | | | | | | | | | | | | | | | | | | | | 7 |
| 8 | | | | | | | | | | | | | | | | | | | | | 8 |
| 9 | | | | | | | | | | | | | | | | | | | | | 9 |
| 10 | | | | | | | | | | | | | | | | | | | | | 10 |

| | 1 | 2 | 3 | 4 | 5 | 6 | 7 | 8 | 9 | 10 | 11 | 12 | 13 | 14 | 15 | 16 | 17 | 18 | 19 | 20 | |
|---|---|---|---|---|---|---|---|---|---|---|---|---|---|---|---|---|---|---|---|---|---|
| 11 | | -3 | | | | | 0 | | | | | | | | | 5 | | | | | 11 |
| 12 | | -18 | | -10 | | | -11 | -15 | | -14 | -15 | | -5 | -11 | | | | *5* | | *0* | 12 |
| 13 | | | | | | | | | | | | | | | | | | | | | 13 |
| 14 | | | | | | | | | | | | | | | | | | | | | 14 |
| 15 | | | | | | | | | | | | | | | | | | | | | 15 |
| 16 | -17 | | | -14 | | -9 | | -7 | | | | | | | | | -14 | | | | 16 |
| 17 | | -3 | | | 1 | 5 | 1 | <u>2</u> | | | 2 | 2 | | | | | | | | | 17 |
| 18 | | | | | | | | | | | | | | | | | | | | | 18 |
| 19 | | | | | | | | | | | | | | | | | | | | | 19 |
| 20 | | | | | | | | | | | | | | | | | | | | | 20 |
| | 1 | 2 | 3 | 4 | 5 | 6 | 7 | 8 | 9 | 10 | 11 | 12 | 13 | 14 | 15 | 16 | 17 | 18 | 19 | 20 | |

j = 5

(4 5)(5 2) = (4 2): <u>0</u>; (4 5)(5 13) = (4 13): 2

j = 8

(17 *6* 7 12 5 2 11 8 *10*): -9

j = 9

(4 9)(9 18) = (4 18): 0

j = 18

(4 18)(18 13) = (4 13): <u>-6</u>

We have only two paths that might be extended: [4 14 7 12 5 2] and [4 14 7 12 5 13 18].

(**4** 14 7 **12** 5 2) : 8; [**4** 14 7 **12** 5 2 11 8 10] : -6; [**4** 14 7 **12** 5 2 17] : 8; [**4** 14 7 **12** 9 18 13] : -6

We thus have only one unlinked 2-circuit cycle: (4 14 7 12 5 2): 8.

We now choose smallest value of each linked 2-circuit path in each row to search for linked 2-circuit cycles.

j = 1

(4 1)(1 4) = (4 4): 0

CYCLE (4 14 *7* 12 *1*): 0

(4 1)(1 10) = (4 10): -2; (4 1)(1 13) = (4 13): -3

j = 9

(12 9)(9 18) = (12 18): 6

j = 10

(4 10)(10 11) = (4 11): -1

j = 11

(4 11)(11 8) = (4 8): <u>-1</u>

j = 18

(12 18)(18 13) = (12 13): <u>0</u>

*LINKED 2 - CIRCUITS* $P_{20}$

|   | 1  | 2 | 3 | 4 | 5 | 6 | 7  | 8  | 9 | 10 | 11 | 12 | 13 | 14 | 15 | 16 | 17 | 18 | 19 | 20 |   |
|---|----|---|---|---|---|---|----|----|---|----|----|----|----|----|----|----|----|----|----|----|---|
| 1 |    |   |   |   |   |   |    |    |   |    |    |    |    |    |    |    |    |    |    |    | 1 |
| 2 |    |   |   |   |   |   |    |    |   |    |    |    |    |    |    |    |    |    |    |    | 2 |
| 3 |    |   |   |   |   |   |    |    |   |    |    |    |    |    |    |    |    |    |    |    | 3 |
| 4 | 12 |   |   |   |   |   | 14 | 11 |   | 1  | 10 | 7  | 1  | 4  |    |    |    |    | 1  |    | 4 |
| 5 |    |   |   |   |   |   |    |    |   |    |    |    |    |    |    |    |    |    |    |    | 5 |
| 6 |    |   |   |   |   |   |    |    |   |    |    |    |    |    |    |    |    |    |    |    | 6 |

| | 1 | 2 | 3 | 4 | 5 | 6 | 7 | 8 | 9 | 10 | 11 | 12 | 13 | 14 | 15 | 16 | 17 | 18 | 19 | 20 | |
|---|---|---|---|---|---|---|---|---|---|---|---|---|---|---|---|---|---|---|---|---|---|
| 7 | | | | | | | | | | | | | | | | | | | | | 7 |
| 8 | | | | | | | | | | | | | | | | | | | | | 8 |
| 9 | | | | | | | | | | | | | | | | | | | | | 9 |
| 10 | | | | | | | | | | | | | | | | | | | | | 10 |
| 11 | | 8 | | | | | 11 | 17 | | | | | | | | 2 | | | | | 11 |
| 12 | | 8 | | 2 | | | 12 | 11 | 4 | 7 | 10 | | 18 | | | | 9 | | | | 12 |
| 13 | | | | | | | | | | | | | | | | | | | | | 13 |
| 14 | | | | | | | | | | | | | | | | | | | | | 14 |
| 15 | | | | | | | | | | | | | | | | | | | | | 15 |
| 16 | 6 | | | 1 | | 17 | | 4 | | 8 | | | | | | 16 | | | | | 16 |
| 17 | | 5 | | | 12 | 17 | 6 | 11 | | 8 | 2 | 7 | | | | | | | | | 17 |
| 18 | | | | | | | | | | | | | | | | | | | | | 18 |
| 19 | | | | | | | | | | | | | | | | | | | | | 19 |
| 20 | | | | | | | | | | | | | | | | | | | | | 20 |
| | 1 | 2 | 3 | 4 | 5 | 6 | 7 | 8 | 9 | 10 | 11 | 12 | 13 | 14 | 15 | 16 | 17 | 18 | 19 | 20 | |

## LINKED 2 - CIRCUITS $\sigma_{T_{165}}^{-1} M^{-}$ (20)

| | 7 | 9 | 13 | 12 | 18 | 10 | 1 | 16 | 2 | 6 | 17 | 4 | 3 | 20 | 19 | 8 | 11 | 5 | 15 | 14 | |
|---|---|---|---|---|---|---|---|---|---|---|---|---|---|---|---|---|---|---|---|---|---|
| | 1 | 2 | 3 | 4 | 5 | 6 | 7 | 8 | 9 | 10 | 11 | 12 | 13 | 14 | 15 | 16 | 17 | 18 | 19 | 20 | |
| 1 | | | | | | | | | | | | | | | | | | | | | 1 |
| 2 | | | | | | | | | | | | | | | | | | | | | 2 |
| 3 | | | | | | | | | | | | | | | | | | | | | 3 |
| 4 | -3 | | | | | | 4 | -1 | | 0 | -1 | 5 | -3 | -1 | | | | | | 3 | 4 |
| 5 | | | | | | | | | | | | | | | | | | | | | 5 |
| 6 | | | | | | | | | | | | | | | | | | | | | 6 |
| 7 | | | | | | | | | | | | | | | | | | | | | 7 |
| 8 | | | | | | | | | | | | | | | | | | | | | 8 |
| 9 | | | | | | | | | | | | | | | | | | | | | 9 |

| | 1 | 2 | 3 | 4 | 5 | 6 | 7 | 8 | 9 | 10 | 11 | 12 | 13 | 14 | 15 | 16 | 17 | 18 | 19 | 20 | |
|---|---|---|---|---|---|---|---|---|---|---|---|---|---|---|---|---|---|---|---|---|---|
| 10 | | | | | | | | | | | | | | | | | | | | | 10 |
| 11 | | -3 | | | | | 0 | <u>8</u> | | | | | | | | 5 | | | | | 11 |
| 12 | | -18 | | -10 | | | -11 | -15 | <u>2</u> | -14 | -15 | | <u>0</u> | | | | 6 | | | | 12 |
| 13 | | | | | | | | | | | | | | | | | | | | | 13 |
| 14 | | | | | | | | | | | | | | | | | | | | | 14 |
| 15 | | | | | | | | | | | | | | | | | | | | | 15 |
| 16 | -17 | | | -14 | | -9 | | -7 | | <u>-18</u> | | | | | | -14 | | | | | 16 |
| 17 | | -3 | | | 1 | 5 | 1 | 2 | | <u>-9</u> | 2 | 2 | | | | | | | | | 17 |
| 18 | | | | | | | | | | | | | | | | | | | | | 18 |
| 19 | | | | | | | | | | | | | | | | | | | | | 19 |
| 20 | | | | | | | | | | | | | | | | | | | | | 20 |
| | 1 | 2 | 3 | 4 | 5 | 6 | 7 | 8 | 9 | 10 | 11 | 12 | 13 | 14 | 15 | 16 | 17 | 18 | 19 | 20 | |

j = 8

(4 8)(8 2) = (4 2): <u>-4</u>

We cannot extend (12 13): <u>0.</u>

We thus can obtain (4 2)(2 4) = (4 4): 4. There are no additional extensions of (4 2): <u>-4</u>.

We thus have obtained only one linked 2-circuit cycle: (4  14  *7*  12  1  *10*  11  8  2): 4.

We now construct a matrix, $M_{2-CYCLES}$ , whose columns consist of the 2-cycles of $\sigma_{T_{165}}^{-1} M^{-}$ . In order to do so, we search for a point that occurs least often among all of the acceptable and 2-circuit cycles. If more than one such point occurs, we pick the one whose minimum-valued entry is greater than that of the others, say $p$ . From theorems 1.3 and 1.26, given the 2-cycle $(p\ q)$, every cycle that contains $p$ corresponds to a cycle containing $q$ in both value and length. (Thus, we are essentially talking about a property of 2-cycles). Since $p$ may be contained in more than one cycle, we place such cycle (or cycles) at the top of the list of cycles whose points will be placed in $M_{2-CYCLES}$. The remaining cycles are listed in terms of the number of 2-cycles containing its points. Each set of such cycles is listed in order of increasing magnitude.

Let our 2-cycles be named as follows:

1. (1  7)  2. (2  9)  3. (3  13)  4. (4  12)  5. (5  18)  6. (6  10)  7. (8  16)  8. (11  17)  9. (14  20)  10. (15  19)

After removing companion and duplicate cycles, only one cycle has a point from 3 or 10. 3 contains the point 13 whose row has as its minimum value, 19, at the entry (13  15). On the other hand, row 19 of 10 has a minimum value of 14. Thus, we choose $p = 13$.

*(12  9  18  13  15  14  7): 8. Thus, from our formula, $p = \dfrac{n}{2} + 3t + a - 1$, for the number of points in a tour, the maximum number of points that an acceptable cycle can have is four, while a 2-circuit cycle has at most six. . The following acceptable cycles have no more than four points:

(1)  (4  1): -10   (2)  (16  17  6): -12   (3)  (2  11  8): 2   (4)  (7  12  6): 3

Two 2-circuit cycles contain no more than six points: (5)  (4  14  7  12  1): 0   (6)  (4  14  7  12  5  2): 8

(16  17  6)(3  5  2  4  1  20  19): -4  yields the derangement.

$$D = (16\ 11\ 17\ 10\ 6\ 8)(3\ 18\ 5\ 9\ 2\ 12\ 4\ 7\ 1\ 14\ 20\ 15\ 19\ 13):\ 152$$

Note that because only one cycle is negative while the other is positive, we can't guarantee that this is a minimally-valued derangement for *all* tours – including $T_{OPT}$. The only way to prove that that is the case is to obtain *all* acceptable cycles using trees. As we shall see, our method for obtaining $T_{OPT}$ doesn't require obtaining a minimum weight derangement. Since a computer program requires the matrix, $M_{2-CYCLES}$, containing cycles in $\sigma_T^{-1} M^-$, we now use it to obtain $T_{FWOPT}$.

|     | CYCLES |     |     |     |     |     |     |     |     |     |     |
| --- | --- | --- | --- | --- | --- | --- | --- | --- | --- | --- | --- |
|     | I | II | III | IV | V | VI | VII | VIII | IX | X |     |
|     | 7 | 9 | 13 | 12 | 18 | 10 | 16 | 17 | 20 | 19 |     |
|     | 1 | 2 | 3 | 4 | 5 | 6 | 8 | 11 | 14 | 15 |     |
| *1* | 7 | 9 | 13 | 12 | 18 |   |   |   | 14 | 15 | *1* |
| *2* |   |   |   |   |   | 6 | 16 | 17 |   |   | *2* |
| *3* | 1 |   |   | 4 |   |   |   |   |   |   | *3* |
| *4* |   | 2 |   |   |   |   | 8 | 11 |   |   | *4* |
| *5* |   |   |   | 12 |   | 6 |   | 17 |   |   | *5* |
| *6* | 7,1 |   |   | 4,12 |   |   |   |   | 14 |   | *6* |
| *7* | 7 | 2 |   | 4,12 | 5 |   |   |   | 14 |   | *7* |
| *8* |   |   |   |   |   |   |   |   |   |   | *8* |
|     | 1 | 2 | 3 | 4 | 5 | 6 | 8 | 11 | 14 | 15 |     |

As can be seen from the table, precisely two cycles can be linked to 1: 4 and 5. Both are acceptable cycles. However, the total number of points linked is 10. If a tour had been obtained the number would be 11. Thus, $T_{FWOPT} = T_{165}$

PHASE 3(b)

We now give the procedures necessary to obtain an optimal tour, $T_{OPT}$. For simplicity, instead of using Roman numerals to represent 2-cycles, we use arabic numerals printed in boldface. Also, we may denote $T_{FWOPT}$ as $T$ and $\sigma_{FWOPT}$ as $\sigma_T$.

We now describe the general algorithm for obtaining $T_{OPT}$.

First, we must determine if there exists any 2-cycle, $(a\ b)$, such that the following is true:

(a) a minimally-valued entry of row $a$ or row $b$ has a value at least as great as $|T| - |\sigma_T|$.

If (a) is the case, and row $a$ is the row of $\sigma_T^{-1} M^-$, then we can directly construct paths that are acceptable and non-positive until they reach $a$ as described in theorem 22. In example 4, row 13 is such a row.

However, case (b) may occur:

(b) Given every 2-cycle $(a\ b)$ the minimally-valued entry of both row $a$ and row $b$ have values less than $|T| - |\sigma_T|$.

We consider case (b) first. From theorem 9 and its corollary, we can obtain from a cycle containing point $a$ of a 2-cycle, a companion cycle that has the same number of points, value and type (acceptable or 2-circuit) as well as point $b$. Given the above statement, let $(a\ b)$ be a 2-cycle such that point $a$ occurs in the smallest number of cycles obtained using the modified F-W to obtain $T_{FWOPT}$. We first obtain all cycles containing $a$ where $a$ is a determining point of the cycle. Thus, we let $a$ be the root of a tree whose branches are either acceptable or 2-circuit. We use the following procedure to ensure that this is the case. Let $B = (a\ ...\ t)$ be a branch of our $a$-root tree. In general, we have the following cases: (i) If $B$ is acceptable, all of its points are red numerals inside red circles. Its points are connected by red arcs. (ii) (a) If it is an unlinked 2-circuit path, $a$ is a red numeral enclosed in a red square, while $\sigma_T(a)$ is a black numeral enclosed in a black square. All of the numerals of $B$ except $a$ are red and enclosed in red circles until we reach $\sigma_T(a)$. Furthermore they are connected by red arcs. All of the numerals after $\sigma_T(a)$ are black numerals enclosed in black circles. They are connected by black arcs. (ii)(b) If It is a linked 2-circuit path, we modify (ii)(a) in the following ways: There exists a point $p$ in the interval $(a\ \sigma_T(a))$ where $\sigma_T(p)$ occurs in the interval $(\sigma_T(a)\ t)$. In this case we print the numeral $p$ in black. It remains enclosed in a red circle. Similarly, we print $\sigma_T(p)$ as a red numeral inside a black circle. Up until this point, other than that, the conditions are the same as in (ii)(a). However, the points in the subtree having $\sigma_T(p)$ as its root are all connected with dashed black arcs and *printed in italics*. This prevents us from having more than two linked circuits. Using this procedure, we obtain all cycles having $a$ as a determining point. We note again that all branches obtained have a value less than $|T| - |\sigma_T|$.

We have discussed the structure of $B = (a\ ...\ t)$. We now give rules that may shorten the running time of this protion of our algorithm.

(A) Construct a subtree, $SUB_t$, whose root is $t$. Each of the paths of $SUB_t$ extends $B$ by at most two points. A path of $SUB_t$ cannot be extended further if no arc exists such that $|B| + |SUB_t| \geq |T| - |\sigma_T|$. Furthermore, if the terminal point of a path of length 2 is $\sigma_T(t)$, and its color is black, while $t$ is also black, we delete it and the arc terminating in it. If $t$ is red and the arc preceding $\sigma_T(t)$ is red, we delete as in the previous case. If $\sigma(t) = a$, we have obtained a cycle. In all other cases, we proceed to the next step.

(B) Sort the points of each node level in increasing order of magnitude. (a subpath of node level $i, i = 1,2$ has $i$ arcs.)

(C) Backtrack from $t$ along $B$ one point at a time.

   (i) If a given point, $p$, is identical to a point, $p'$, in $SUB_t$, we delete the point as well as any arc incident to it, i.e., any arc of which it an initial or terminal point.

   (ii) If $p \neq a$ is a red point and $\sigma_T(p)$ is a red point in $SUB_t$, delete $\sigma_T(p)$ as well as any arc of $SUB_t$ incident to it.

   (iii) If $\sigma_T(a)$ is a red point of $SUB_t$, print it as a black point placed inside a black square. The subtree emanating from $\sigma_T(a)$ contains only black points and arcs.

   (iv) If $p' = \sigma_T(p)$ is a non-italic black point, we reprint it as a red point in a black circle, while $p$ is printed as a black point in a red circle. All links emanating from $p'$ are printed as dashed black lines. (This prevents us from obtaining more than two pair of interlacing points indicating a linked 2-circuit path.)

   (v) If $p' = \sigma_T(p)$ has a dashed black link terminating in it, we delete it as well as any arc incident to it.

Using these rules, we place the points of each cycle, $C_i$, obtained in the top rows of a matrix similar to $M_{2-CYCLES}$, say

$M_{2-CYCLES,i}$

The rules that were given in (b) apply to (a) with the following exceptions:

(I) We must construct all trees whose roots are initial points of at least one arc having a non-positive value.

(II) The initial point of a 2-circuit cycle won't be $a$. In fact, since $a$ is not a determining vertex, if it is contained in a 2-circuit cycle, we may obtain the *terminal point* of a 2-cycle first rather than its initial point.

(III) Generally, the 2-cycle we obtain doesn't contain the root of the given tree.

We now deal with case (a) where $a$ is not a determining point of a cycle. If $T_{OPT} \neq T_{FWOPT}$, it must contain at least one point of 2-cycle $(a\ b)$ (in this case 3). We use theorem 1.22 to see if any further cycle exists containing $a$ can be obtained from arcs in $\sigma_{FWOPT}^{-1} M^-$. (We already have found one, namely cycle $1$, in the above table). To save space, we only show trees that contain such cycles. We use the same system of constructing and coloring points and arcs with some changes that must be made in the case of 2-circuit cycles. Our main difference with the normal case, i.e., when we obtained all acceptable and 2-cricuit cycles where $a$ is a determining point, is that when we use theorem 1.22, $a$ is not a determining case. Thus, when we search for 2-circuit cycles, we don't have $a$ as a unique point that is always the initial point of an unlinked cycle, or the first cycle of a linked 2-circuit cycle. In fact it isn't an initial point of any cycle. Also, there are generally a number of 2-circuit paths each having a different initial point of a cycle. We adapt to this situation in the following way: After we have obtained an acceptable path that has non-positive subpaths, say $P_{i,a} = (i\ ...\ a)$, after reaching $a$ we revert to our usual procedure of permitting subpaths to have values less than $|T| - |\sigma_T|$. We continue using our previous method of first constructing paths containing two arcs before backtracking. As previously, let $p$ be a point we've arrived at by backtracking, while $p'$ is a point in one of our 2-arc paths. If $p' = \sigma_T(p)$, we print $\sigma_T(p)$ as a black point in a black square, while $p$ is printed as a red point in a red square. All points after $p'$ are printed as black points connect by black arcs. Suppose we have extended $P_{i,a}$ to a point $q'$ and that we backtrack to a point $q$ lying in $(p' \ldots p)$ where $q' = \sigma_T(q)$. In this case, we print $q'$ as a red point lying in a red circle, while $q$ is printed as a black point lying in a red circle. As in the previous case, if $P' = (i \ldots p \ldots q \ldots p' \ldots a \ldots q' \ldots)$, all arcs following $q'$ are printed as black dashed lines, while all point are printed in italics. At this point, we note that once we have obtained $p' = \sigma_T(p)$, although $p$ is not necessarily a determining point of a cycle, when printed as a red point in a red square, it plays the same roll as far as our algorithm is concerned.

Our next step is to link acceptable and 2-circuit cyles to try to obtain a tour whose value is less than $|T| - |\sigma_T|$. We check the value and the number of points of each cycle containing $a$. Let $C_{a,i}$ be one of these cycles. Assume it is either acceptable or a linked 2-circuit cycle. We will use theorem 14 to treat the values of cycles as a set of numbers. In particular, if we allow any set of numbers to be thought of as values of arcs on a circle by adding a "phantom arc" to complete the circle we can always obtain a determining point such that every subset of points is no greater than the sum of the arcs. In the case of linking cycles, we substitute values of cycles as they are added to tree of linked cycles. We thus try to construct acceptable or 2-circuit cycles *either* containing only points *not* in the 2-cycles of the points of $C_{a,i}$ or containing at most one linking point to it. If it is an unlinked 2-circuit cycle, we allow two linking points to each of its "circuits". In the process of obtaining the cycles $C_{a,i}$ $i = 1, 2, \ldots$, we obtained all negative cycles. We check to see if any of them satisfy the criteria mentioned above. Let the set $N = \{ C_j \, / \, |C_j| < 0 \}$ consist of all such negative cycles. Given each cycle, $C_{a,i}$, we construct a unique matrix, $M_{i, 2-cycle}$, containing $\frac{n}{2}$ columns where each column corresponds to a 2-cycle. We place the points of $C_{a,i}$ in the first row of $M_{i, 2-cycle}$. We then place the cycles of $N$ in the second, third, ..., $(|N|+1)$ row in increasing order of magnitude. Using negative cycles of value less than $|T| - |\sigma_T| - |C_{a,i}|$, we try to link cycles such that each new subtree of cycles has the same upper bound. If at some point, we can link $C_{a,i}$ to a subtree, then our upper bound becomes $|T| - |\sigma_T|$. We allow a "phantom cycle" to be added so that our subtree can be expanded. (We note that the "phantom tree" cannot permanently remain in the subtree because in reality if a phantom tree exists at the end of

our construction, we will have created *two* subtrees.) Call the subtrees obtained in a sequence $SUB_{i,1}$, $SUB_{i,2}$, .... If no negative cycle can be obtained, then $V_1 = |C_{a,i}|$ Then each expanded subtree cannot have a value greater than $min\ \{|T|-|\sigma_T|, |T|-|\sigma_T|-|C_{a,i}|\}$. We now note that the sum of its points and those of $C_{a,i}$, say $p$, must satsify $p \leq \frac{n}{2} + 3t + a - 1$. We also check the number of cycles linked using $a + l + 2u \leq N_1$ where $N_1$ is the number of linking points on $C_{a,i}$. As earlier, $a$ is the number of acceptable cycles, $l$ is the number of linked and $u$ is the number of unlinked 2-circuit cycles. Once we have obtained all new positive cycles satisfying the above criteria, we do the following:

We place each cycle $C_{a,i}\ i = 1, 2,\ ...$ in the top rows of a new $M_{2-CYCLES}$.

Next, we place the points of the non-positive cycles in $N$ as mentioned earlier.

Lastly, we place the new cycles we have obtained that could link to it in the remaining rows of $M_{i,2-CYCLES}$ according to the number of 2-cycles that their points are in. As previously, the cycles are arranged in two ways:

(1) the number of 2-cycles their points are in.  (2) Given a set of cycles whose points are in the same number of cycles, we order these cycles in terms of increasing order of magnitude.

Choosing one $C_{a,i}$ at a time, and using at most one "phantom" cycle whose value is zero, if a tour exists, from theorem 1.14, we can always choose a subtree constructed such that the phantom cycle disappears, while each "partial subtree" has a value less than $|T|-|\sigma_T|$ or $|T|-|\sigma_T|-|C_{a,i}|$ depending upon whether or not $C_{a,i}$ belongs to the partial subtree. If we can't find a tour using $C_{a,1}$, we choose $C_{a,2}$. We continue this procedure until we have chosen each $C_{a,i}$. If we cannot obtain a tour using all of the $C_{a,i}$, then $T_{OPT} = T_{FWOPT}$. If we obtain one or more tours, we choose the one having the smallest value, say $T_{OPT1}$. In that case, $T_{OPT} = T_{OPT1}$.

In the current case, we have obtained three cycles containing 13 from non-positive root trees with roots 12, 15 and 16 to obtain the following cycles: (*1*) (12 9 18 13 15 14 7): 8, (*2*) (15 14 7 12 9 18 13): 8.

*( 3 )* (16 17 6 1 13 15 14 7 12 9). In this simple case, we can see immediately that the first two cycles are identical but have different determining points. For large cost matrices M, it is best to place all cycles that have the same number of points and the same value together in a subset in $M_{a,i}$ and, using theorem 26, delete all but one of those that also have linking points in the same set of 2-cycles and exactly the same non-linking points. The third cycle can be written as the following unlinked 2-circuit cycle: (7 12 9 16 17 6 1 13 15 14): 6.

*( 1 )*: Cycles are obtained from the branches of trees confined to 2-cycles 6, 7 and 8. At least one cycle contains a linking point of *( 1 )*. No 2-circuit cycle can exist. *( 3 )*: No cycles. From *( 1 )*, we obtain

*2*: (6 16 17): -12, *3*: (8 10 11): -12.

$$M_{13,1}$$

|   | 1 | 2 | 3 | 4 | 5 | 6 | 7 | 8 | 9 | 10 |   |
|---|---|---|---|---|---|---|---|---|---|---|---|
| *1* | 7 | 9 | 13 | 12 | 18 |   |   |   | 14 | 15 | *1* |
| *2* |   |   |   |   |   | 6 | 16 | 17 |   |   | *2* |
| *3* |   |   |   |   |   | 10 | 8 | 11 |   |   | *3* |
|   | I | II | III | IV | V | VI | VII | VIII | IX | X |   |

Using theorem 1.26, we delete row *3*.

Since we can't obtain a linking of *( 1 )* and *2*, we can't obtain a tour of smaller value than $T_{FWOPT}$.

Therefore, $T_{OPT} = T_{FWOPT}$.

Example 5.

In this example, we chose a non-random matrix, 3-cycle M, containing a derangement consisting of five cycles. That derangement is (1 3 7)(2 5 11 14 19)(4 16 8)(6 18 13)(10 17 12). We assume that this derangement was obtained at the end of PHASE 1. In PHASE 2, we don't use arcs that yield directed edges that are symmetric to arcs in the derangement. We thus always obtain a derangement containing $n$ edges.. We now give PHASES 2 and 3.

3-cycle M

|    | 1  | 2  | 3  | 4  | 5  | 6  | 7  | 8  | 9  | 10 | 11 | 12 | 13 | 14 | 15 | 16 | 17 | 18 | 19 | 20 |    |
|----|----|----|----|----|----|----|----|----|----|----|----|----|----|----|----|----|----|----|----|----|----|
| 1  | ∞  | 26 | 1  | 30 | 74 | 5  | 3  | 38 | 28 | 78 | 81 | 7  | 97 | 10 | 94 | 40 | 98 | 49 | 40 | 70 | 1  |
| 2  | 26 | ∞  | 69 | 30 | 1  | 80 | 50 | 74 | 1  | 60 | 3  | 9  | 31 | 87 | 89 | 91 | 6  | 82 | 23 | 85 | 2  |
| 3  | 1  | 69 | ∞  | 23 | 7  | 61 | 2  | 98 | 99 | 90 | 84 | 57 | 4  | 56 | 66 | 30 | 51 | 3  | 25 | 47 | 3  |
| 4  | 30 | 30 | 23 | ∞  | 33 | 59 | 5  | 3  | 26 | 48 | 84 | 18 | 57 | 28 | 47 | 1  | 81 | 48 | 70 | 17 | 4  |
| 5  | 74 | 1  | 7  | 33 | ∞  | 82 | 29 | 80 | 5  | 87 | 2  | 97 | 3  | 45 | 72 | 94 | 20 | 9  | 90 | 20 | 5  |
| 6  | 5  | 80 | 61 | 59 | 82 | ∞  | 1  | 6  | 43 | 9  | 39 | 41 | 3  | 45 | 62 | 38 | 50 | 1  | 41 | 50 | 6  |
| 7  | 3  | 30 | 2  | 5  | 29 | 1  | ∞  | 34 | 78 | 49 | 73 | 10 | 56 | 36 | 87 | 31 | 45 | 59 | 88 | 42 | 7  |
| 8  | 38 | 74 | 98 | 3  | 80 | 6  | 34 | ∞  | 14 | 55 | 43 | 91 | 85 | 93 | 75 | 2  | 64 | 78 | 60 | 1  | 8  |
| 9  | 28 | 1  | 99 | 26 | 5  | 43 | 78 | 14 | ∞  | 50 | 28 | 81 | 98 | 95 | 3  | 31 | 73 | 63 | 87 | 2  | 9  |
| 10 | 78 | 60 | 90 | 48 | 87 | 9  | 49 | 55 | 50 | ∞  | 37 | 1  | 95 | 59 | 30 | 25 | 3  | 90 | 64 | 36 | 10 |
| 11 | 81 | 3  | 84 | 84 | 2  | 39 | 73 | 43 | 28 | 37 | ∞  | 3  | 61 | 14 | 11 | 3  | 3  | 74 | 22 | 26 | 11 |
| 12 | 7  | 9  | 57 | 18 | 97 | 41 | 10 | 91 | 81 | 1  | 3  | ∞  | 84 | 62 | 56 | 34 | 2  | 17 | 71 | 30 | 12 |
| 13 | 97 | 31 | 4  | 57 | 5  | 3  | 56 | 85 | 98 | 93 | 61 | 84 | ∞  | 66 | 66 | 30 | 49 | 2  | 23 | 86 | 13 |

| | 1 | 2 | 3 | 4 | 5 | 6 | 7 | 8 | 9 | 10 | 11 | 12 | 13 | 14 | 15 | 16 | 17 | 18 | 19 | 20 | |
|---|---|---|---|---|---|---|---|---|---|---|---|---|---|---|---|---|---|---|---|---|---|
| 14 | 10 | 87 | 56 | 28 | 45 | 45 | 36 | 93 | 95 | 59 | 14 | 62 | 66 | ∞ | 25 | 14 | 47 | 47 | 2 | 5 | 14 |
| 15 | 94 | 89 | 66 | 47 | 72 | 62 | 87 | 75 | 3 | 30 | 11 | 56 | 66 | 25 | ∞ | 89 | 85 | 87 | 8 | 2 | 15 |
| 16 | 40 | 91 | 30 | 1 | 94 | 38 | 31 | 2 | 31 | 25 | 3 | 34 | 30 | 14 | 89 | ∞ | 54 | 18 | 92 | 90 | 16 |
| 17 | 98 | 6 | 51 | 81 | 20 | 50 | 45 | 64 | 73 | 3 | 3 | 2 | 49 | 47 | 85 | 54 | ∞ | 66 | 48 | 90 | 17 |
| 18 | 49 | 82 | 3 | 48 | 9 | 1 | 59 | 78 | 63 | 90 | 74 | 17 | 2 | 47 | 87 | 18 | 66 | ∞ | 87 | 16 | 18 |
| 19 | 40 | 23 | 25 | 70 | 90 | 41 | 88 | 60 | 87 | 64 | 22 | 71 | 23 | 2 | 8 | 92 | 48 | 87 | ∞ | 22 | 19 |
| 20 | 70 | 85 | 47 | 17 | 20 | 50 | 42 | 1 | 2 | 36 | 26 | 30 | 86 | 5 | 2 | 90 | 90 | 16 | 22 | ∞ | 20 |
| | 1 | 2 | 3 | 4 | 5 | 6 | 7 | 8 | 9 | 10 | 11 | 12 | 13 | 14 | 15 | 16 | 17 | 18 | 19 | 20 | |

3-cycle MIN(M)

| | 1 | 2 | 3 | 4 | 5 | 6 | 7 | 8 | 9 | 10 | 11 | 12 | 13 | 14 | 15 | 16 | 17 | 18 | 19 | 20 | |
|---|---|---|---|---|---|---|---|---|---|---|---|---|---|---|---|---|---|---|---|---|---|
| 1 | 3 | 7 | 6 | 12 | 14 | 2 | 9 | 4 | 8 | 16 | 19 | 18 | 20 | 5 | 10 | 11 | 15 | 13 | 17 | 1 | 1 |
| 2 | 9 | 5 | 11 | 17 | 12 | 19 | 1 | 4 | 13 | 7 | 10 | 3 | 8 | 6 | 20 | 18 | 14 | 15 | 16 | 2 | 2 |
| 3 | 1 | 7 | 18 | 13 | 5 | 4 | 19 | 16 | 20 | 17 | 14 | 12 | 6 | 15 | 2 | 11 | 10 | 8 | 9 | 3 | |
| 4 | 16 | 8 | 7 | 20 | 12 | 3 | 9 | 14 | 1 | 2 | 5 | 15 | 10 | 18 | 13 | 6 | 19 | 17 | 11 | 4 | 4 |
| 5 | 2 | 11 | 13 | 9 | 3 | 18 | 17 | 20 | 7 | 4 | 14 | 15 | 1 | 8 | 6 | 10 | 19 | 16 | 12 | 5 | |
| 6 | 7 | 18 | 13 | 1 | 8 | 10 | 16 | 11 | 19 | 9 | 14 | 17 | 20 | 12 | 4 | 3 | 15 | 2 | 5 | 6 | 6 |
| 7 | 6 | 3 | 1 | 4 | 12 | 5 | 2 | 16 | 8 | 14 | 20 | 17 | 10 | 18 | 3 | 11 | 9 | 15 | 19 | 17 | 7 |
| 8 | 20 | 16 | 4 | 6 | 9 | 7 | 1 | 11 | 10 | 19 | 17 | 2 | 15 | 18 | 5 | 13 | 12 | 14 | 3 | 8 | 8 |
| 9 | 2 | 29 | 15 | 5 | 8 | 4 | 1 | 11 | 16 | 6 | 10 | 18 | 17 | 7 | 12 | 19 | 14 | 13 | 3 | 9 | 9 |
| 10 | 12 | 17 | 6 | 16 | 15 | 20 | 11 | 4 | 7 | 9 | 8 | 14 | 2 | 19 | 1 | 5 | 3 | 18 | 13 | 10 | 10 |
| 11 | 5 | 2 | 12 | 16 | 17 | 15 | 14 | 19 | 20 | 9 | 10 | 6 | 8 | 13 | 7 | 19 | 1 | 3 | 4 | 11 | 11 |
| 12 | 10 | 11 | 17 | 1 | 2 | 7 | 18 | 4 | 20 | 16 | 6 | 15 | 3 | 14 | 19 | 9 | 13 | 8 | 5 | 12 | 12 |

| 13 | 18 | 6  | 3  | 5  | 19 | 16 | 2  | 17 | 7  | 4  | 11 | 14 | 15 | 8  | 12 | 20 | 10 | 1  | 9  | 13 | 13 |
| 14 | 19 | 20 | 1  | 11 | 16 | 15 | 4  | 7  | 5  | 6  | 17 | 18 | 3  | 10 | 12 | 13 | 2  | 8  | 9  | 14 | 14 |
| 15 | 20 | 9  | 19 | 11 | 14 | 10 | 4  | 12 | 6  | 3  | 13 | 5  | 8  | 17 | 7  | 18 | 2  | 16 | 1  | 15 | 15 |
| 16 | 8  | 11 | 14 | 18 | 4  | 10 | 13 | 3  | 7  | 9  | 12 | 6  | 1  | 17 | 15 | 20 | 2  | 19 | 5  | 16 | 16 |
| 17 | 12 | 10 | 11 | 2  | 5  | 7  | 14 | 19 | 13 | 6  | 3  | 16 | 8  | 18 | 9  | 4  | 15 | 20 | 1  | 17 | 17 |
| 18 | 6  | 13 | 3  | 5  | 20 | 12 | 16 | 14 | 4  | 1  | 7  | 9  | 17 | 11 | 8  | 2  | 15 | 19 | 10 | 18 | 18 |
| 19 | 14 | 15 | 11 | 20 | 2  | 13 | 3  | 1  | 6  | 17 | 8  | 10 | 4  | 12 | 9  | 18 | 7  | 5  | 16 | 19 | 19 |
| 20 | 15 | 14 | 18 | 4  | 5  | 19 | 11 | 12 | 10 | 8  | 7  | 3  | 6  | 9  | 1  | 2  | 13 | 16 | 17 | 20 | 20 |
|    | 1  | 2  | 3  | 4  | 5  | 6  | 7  | 8  | 9  | 10 | 11 | 12 | 13 | 14 | 15 | 16 | 17 | 18 | 19 | 20 |    |

In what follows, we note that in the permutations of the cost matrix, we only place values in entries that correspond to the inital and terminal points of a path of length at least two. Also, the symbol ∞ is only included to ensure that the author has permuted columns correctly. It doesn't have to be included in a computer program. The entries in boldface denote arcs that are symmetric to respective arcs of the derangement. Entries in italics indicate the value of a path that has been extended. Underlined entries denote paths that may possibly be extended.

$$D^{-1}_{3\text{-}cycle\,2}M^-$$

|   | 3 | 5 | 7 | 16 | 11 | 18 | 1 | 4 | 2 | 17 | 14 | 10 | 6 | 19 | 20 | 8 | 12 | 13 | 15 | 9 |   |
|---|---|---|---|----|----|----|---|---|---|----|----|----|---|----|----|---|----|----|----|---|---|
|   | 1 | 2 | 3 | 4  | 5  | 6  | 7 | 8 | 9 | 10 | 11 | 12 | 13 | 14 | 15 | 16 | 17 | 18 | 19 | 20 |   |
| 1 | 0 | 73 | 2 | 39 | 80 | 48 | ∞ | 29 | 26 | 97 | 10 | 77 | 4 | 39 | 69 | 37 | 6 | 96 | 93 | 27 | 1 |
| 2 | 68 | 0 | 49 | 90 | 2 | 81 | 25 | 29 | ∞ | 5 | 87 | 59 | 79 | 22 | 84 | 73 | 8 | 30 | 88 | 0 | 2 |
| 3 | ∞ | 5 | 0 | 28 | 82 | 1 | -1 | 21 | 69 | 49 | 56 | 88 | 59 | 23 | 45 | 96 | 55 | 2 | 64 | 97 | 3 |
| 4 | 22 | 32 | 4 | 0 | 83 | 47 | 29 | ∞ | 30 | 80 | 28 | 47 | 58 | 69 | 16 | 2 | 17 | 56 | 46 | 25 | 4 |
| 5 | 5 | ∞ | 27 | 92 | 0 | 7 | 72 | 31 | 1 | 18 | 45 | 85 | 80 | 78 | 18 | 78 | 95 | 1 | 79 | 3 | 5 |
| 6 | 60 | 81 | 0 | 37 | 38 | 0 | 4 | 58 | 80 | 49 | 45 | 8 | ∞ | 40 | 49 | 5 | 40 | 2 | 61 | 42 | 6 |
| 7 | -1 | 26 | ∞ | 28 | 70 | 56 | 0 | 2 | 30 | 42 | 36 | 46 | -2 | 85 | 39 | 31 | 7 | 53 | 84 | 75 | 7 |
| 8 | 95 | 77 | 31 | -1 | 40 | 75 | 35 | 0 | 74 | 61 | 93 | 52 | 3 | 57 | -2 | ∞ | 88 | 82 | 72 | 11 | 8 |

| | 1 | 2 | 3 | 4 | 5 | 6 | 7 | 8 | 9 | 10 | 11 | 12 | 13 | 14 | 15 | 16 | 17 | 18 | 19 | 20 | |
|---|---|---|---|---|---|---|---|---|---|---|---|---|---|---|---|---|---|---|---|---|---|
| 9 | 95 | 1 | 74 | 27 | 24 | 59 | 24 | 22 | 0 | 69 | 94 | 46 | 39 | 83 | -2 | 10 | 77 | 94 | -1 | ∞ | 9 |
| 10 | 87 | 84 | 46 | 22 | 34 | 87 | 75 | 45 | 60 | 0 | 59 | ∞ | 6 | 61 | 33 | 52 | -2 | 92 | 27 | 47 | 10 |
| 11 | 67 | -15 | 56 | -14 | ∞ | 57 | 64 | 67 | -11 | -11 | 14 | 20 | 22 | 5 | 9 | 26 | -14 | 44 | -6 | 11 | 11 |
| 12 | 56 | 96 | 9 | 33 | 1 | 16 | 6 | 17 | 9 | 1 | 62 | 0 | 40 | 70 | 29 | 90 | ∞ | 83 | 55 | 80 | 12 |
| 13 | 1 | 2 | 53 | 27 | 58 | -1 | 94 | 54 | 31 | 46 | 66 | 90 | 0 | 20 | 83 | 82 | 81 | ∞ | 63 | 95 | 13 |
| 14 | 54 | 43 | 34 | 15 | 12 | 45 | 8 | 26 | 87 | 45 | ∞ | 57 | 43 | 0 | 3 | 91 | 60 | 64 | 23 | 93 | 14 |
| 15 | 64 | 70 | 85 | 87 | 9 | 85 | 92 | 45 | 89 | 83 | 25 | 28 | 60 | 6 | 0 | 73 | 54 | 64 | ∞ | 1 | 15 |
| 16 | 28 | 92 | 29 | ∞ | 1 | 16 | 38 | -1 | 91 | 52 | 17 | 23 | 36 | 90 | 88 | 0 | 32 | 28 | 87 | 29 | 16 |
| 17 | 49 | 18 | 43 | 52 | 1 | 64 | 96 | 79 | 6 | ∞ | 47 | 1 | 48 | 46 | 88 | 62 | 0 | 47 | 83 | 71 | 17 |
| 18 | 1 | 7 | 57 | 16 | 72 | ∞ | 47 | 46 | 82 | 64 | 47 | 88 | -1 | 85 | 14 | 76 | 15 | 0 | 85 | 61 | 18 |
| 19 | 17 | 82 | 80 | 84 | 14 | 79 | 32 | 62 | 17 | 40 | 4 | 54 | 33 | ∞ | 14 | 32 | 63 | 15 | 0 | 79 | 19 |
| 20 | 45 | 18 | 40 | 88 | 24 | 14 | 68 | 15 | 85 | 88 | 5 | 34 | 48 | 20 | ∞ | -1 | 28 | 84 | 0 | 0 | 20 |
| | 1 | 2 | 3 | 4 | 5 | 6 | 7 | 8 | 9 | 10 | 11 | 12 | 13 | 14 | 15 | 16 | 17 | 18 | 19 | 20 | |

We now try to obtain a negative cycle, $s$, such that $D_{3\text{-cycle }2}s$ is a derangement containing $n$ edges.

j = 4

(11 4)(4 3) = (11 3): <u>-10</u>

j = 9

(11 9)(9 2) = (11 2): <u>-10</u>; (11 9)(9 19) = (11 19): -12

j = 10

(11 10)(10 13) = (11 13): -5

j = 13

(11 13)(13 1) = (11 1): <u>-4</u>

$P_{20}$

|    | 1  | 2 | 3 | 4  | 5 | 6 | 7 | 8 | 9  | 10 | 11 | 12 | 13 | 14 | 15 | 16 | 17 | 18 | 19 | 20 |    |
|----|----|---|---|----|---|---|---|---|----|----|----|----|----|----|----|----|----|----|----|----|----|
| 1  |    |   |   |    |   |   |   |   |    |    |    |    |    |    |    |    |    |    |    |    | 1  |
| 2  |    |   |   |    |   |   |   |   |    |    |    |    |    |    |    |    |    |    |    |    | 2  |
| 3  |    |   |   |    |   |   |   |   |    |    |    |    |    |    |    |    |    |    |    |    | 3  |
| 4  |    |   |   |    |   |   |   |   |    |    |    |    |    |    |    |    |    |    |    |    | 4  |
| 5  |    |   |   |    |   |   |   |   |    |    |    |    |    |    |    |    |    |    |    |    | 5  |
| 6  |    |   |   |    |   |   |   |   |    |    |    |    |    |    |    |    |    |    |    |    | 6  |
| 7  |    |   |   |    |   |   |   |   |    |    |    |    |    |    |    |    |    |    |    |    | 7  |
| 8  |    |   |   |    |   |   |   |   |    |    |    |    |    |    | 8  |    |    |    | 15 |    | 8  |
| 9  |    |   |   |    |   |   |   |   |    |    |    |    |    |    |    |    |    |    |    |    | 9  |
| 10 |    |   |   |    |   |   |   |   |    |    |    |    |    |    |    |    |    |    |    |    | 10 |
| 11 | 13 | 9 | 4 | 11 |   |   |   |   | 11 | 11 |    |    | 10 |    |    |    |    |    | 9  |    | 11 |
| 12 |    |   |   |    |   |   |   |   |    |    |    |    |    |    |    |    |    |    |    |    | 12 |
| 13 |    |   |   |    |   |   |   |   |    |    |    |    |    |    |    |    |    |    |    |    | 13 |
| 14 |    |   |   |    |   |   |   |   |    |    |    |    |    |    |    |    |    |    |    |    | 14 |

| | 1 | 2 | 3 | 4 | 5 | 6 | 7 | 8 | 9 | 10 | 11 | 12 | 13 | 14 | 15 | 16 | 17 | 18 | 19 | 20 | |
|---|---|---|---|---|---|---|---|---|---|---|---|---|---|---|---|---|---|---|---|---|---|
| 15 | | | | | | | | | | | | | | | | | | | | | 15 |
| 16 | | | | | | | | | | | | | | | | | | | | | 16 |
| 17 | | | | | | | | | | | | | | | | | | | | | 17 |
| 18 | | | | | | | | | | | | | | | | | | | | | 18 |
| 19 | | | | | | | | | | | | | | | | | | | | | 19 |
| 20 | | | | | | | | | | | | | | | | | | | | | 20 |
| | 1 | 2 | 3 | 4 | 5 | 6 | 7 | 8 | 9 | 10 | 11 | 12 | 13 | 14 | 15 | 16 | 17 | 18 | 19 | 20 | |

$$D_{3\text{-cycle }2}^{-1}M^{-}(20)$$

| | 3 | 5 | 7 | 16 | 11 | 18 | 1 | 4 | 2 | 17 | 14 | 10 | 6 | 19 | 20 | 8 | 12 | 13 | 15 | 9 | |
|---|---|---|---|---|---|---|---|---|---|---|---|---|---|---|---|---|---|---|---|---|---|
| | 1 | 2 | 3 | 4 | 5 | 6 | 7 | 8 | 9 | 10 | 11 | 12 | 13 | 14 | 15 | 16 | 17 | 18 | 19 | 20 | |
| 1 | 0 | | | | | | ∞ | | | | | | | | | | | | | | 1 |
| 2 | | 0 | | | | | | | ∞ | | | | | | | | | | | 0 | 2 |
| 3 | ∞ | | 0 | | | | | | | | | | | | | | | | | | 3 |
| 4 | | | | 0 | | | | | ∞ | | | | | | | | | | | | 4 |
| 5 | | ∞ | | | 0 | | | | | | | | | | | | | | | | 5 |
| 6 | | | | | | 0 | | | | | | | ∞ | | | | | | | | 6 |
| 7 | | | ∞ | | | | 0 | | | | | | | | | | | | | | 7 |
| 8 | | | | | | | | 0 | | | | | | | | ∞ | | | | | 8 |
| 9 | | | | | | | | | 0 | | | | | | | | | | | ∞ | 9 |
| 10 | | | | | | | | | | 0 | | ∞ | | | | | | | | | 10 |

| | 1 | 2 | 3 | 4 | 5 | 6 | 7 | 8 | 9 | 10 | 11 | 12 | 13 | 14 | 15 | 16 | 17 | 18 | 19 | 20 | |
|---|---|---|---|---|---|---|---|---|---|---|---|---|---|---|---|---|---|---|---|---|---|
| 11 | | <u>-4</u> | -10 | -10 | | ∞ | | | | | 0 | -5 | | | | | | -12 | | | 11 |
| 12 | | | | | | | | | | 0 | | | | | | ∞ | | | | | 12 |
| 13 | | | | | | | | | | 0 | | | | | | | ∞ | | | | 13 |
| 14 | | | | | | | | | ∞ | | 0 | | | | | | | | | | 14 |
| 15 | | | | | | | | | | | | | 0 | | | | | ∞ | | | 15 |
| 16 | | | | ∞ | | | | | | | | | | | 0 | | | | | | 16 |
| 17 | | | | | | | | ∞ | | | | | | | | 0 | | | | | 17 |
| 18 | | | | | ∞ | | | | | | | | | | | | 0 | | | | 18 |
| 19 | | | | | | | | | | | | ∞ | | | | | | 0 | | | 19 |
| 20 | | | | | | | | | | | | | | ∞ | | | | | 0 | | 20 |
| | 1 | 2 | 3 | 4 | 5 | 6 | 7 | 8 | 9 | 10 | 11 | 12 | 13 | 14 | 15 | 16 | 17 | 18 | 19 | 20 | |

j = 2

(11 2)(2 5) = (11 5): -8; (11 2)(2 10) = (11 10): -5; (11 2)(2 17) + (11 17): -2

j = 3

(11 3)(3 6) + (11 6): -9

j = 5

(11 5)(5 18) = (11 18): -7; (11 5)(5 20) = (11 20): -5

j = 6

(11 6)(6 7) = (11 7): -5; (11 6)(6 12) = (11 12): -1; (11 6)(6 16) = (11 16): -9

j = 7

(11 7)(7 8) = (11 8): -3; (11 7)(7 13) = (11 13): -7

j = 8

(11 8)(8 15) = (11 15): -5

j = 20

(11 20)(20 16) = (11 16): <u>-6</u>

$P_{40}$

|   | 1 | 2 | 3 | 4 | 5 | 6 | 7 | 8 | 9 | 10 | 11 | 12 | 13 | 14 | 15 | 16 | 17 | 18 | 19 | 20 |   |
|---|---|---|---|---|---|---|---|---|---|----|----|----|----|----|----|----|----|----|----|----|---|
| 1 |   |   |   |   |   |   |   |   |   |    |    |    |    |    |    |    |    |    |    |    | 1 |
| 2 |   |   |   |   |   |   |   |   |   |    |    |    |    |    |    |    |    |    |    |    | 2 |
| 3 |   |   |   |   |   |   |   |   |   |    |    |    |    |    |    |    |    |    |    |    | 3 |
| 4 |   |   |   |   |   |   |   |   |   |    |    |    |    |    |    |    |    |    |    |    | 4 |
| 5 |   |   |   |   |   |   |   |   |   |    |    |    |    |    |    |    |    |    |    |    | 5 |
| 6 |   |   |   |   |   |   |   |   |   |    |    |    |    |    |    |    |    |    |    |    | 6 |
| 7 |   |   |   |   |   |   |   |   |   |    |    |    |    |    |    |    |    |    |    |    | 7 |
| 8 |   |   |   |   |   |   |   |   |   |    |    |    |    |    |    |    |    |    |    |    | 8 |
| 9 |   |   |   |   |   |   |   |   |   |    |    |    |    |    |    |    |    |    |    |    | 9 |
| 10 | | | | | | | | | | | | | | | | | | | | | 10 |
| 11 | 18 | 9 | 4 | 11 | 2 | 3 | 6 | 7 | 11 | 2 | | 6 | 7 | | 8 | 20 | 2 | 5 | | 5 | 11 |
| 12 | | | | | | | | | | | | | | | | | | | | | 12 |
| 13 | | | | | | | | | | | | | | | | | | | | | 13 |
| 14 | | | | | | | | | | | | | | | | | | | | | 14 |
| 15 | | | | | | | | | | | | | | | | | | | | | 15 |
| 16 | | | | | | | | | | | | | | | | | | | | | 16 |
| 17 | | | | | | | | | | | | | | | | | | | | | 17 |
| 18 | | | | | | | | | | | | | | | | | | | | | 18 |
| 19 | | | | | | | | | | | | | | | | | | | | | 19 |
| 20 | | | | | | | | | | | | | | | | | | | | | 20 |
| | 1 | 2 | 3 | 4 | 5 | 6 | 7 | 8 | 9 | 10 | 11 | 12 | 13 | 14 | 15 | 16 | 17 | 18 | 19 | 20 | |

$$D_{3\text{-}cycle\,2}^{-1} M^{-}(20)$$

| | 3 | 5 | 7 | 16 | 11 | 18 | 1 | 4 | 2 | 17 | 14 | 10 | 6 | 19 | 20 | 8 | 12 | 13 | 15 | 9 | |
|---|---|---|---|---|---|---|---|---|---|---|---|---|---|---|---|---|---|---|---|---|---|
| | 1 | 2 | 3 | 4 | 5 | 6 | 7 | 8 | 9 | 10 | 11 | 12 | 13 | 14 | 15 | 16 | 17 | 18 | 19 | 20 | |
| 1 | 0 | | | | | | ∞ | | | | | | | | | | | | | | 1 |
| 2 | | 0 | | | | | | | ∞ | | | | | | | | | | | 0 | 2 |
| 3 | ∞ | | 0 | | | | | | | | | | | | | | | | | | 3 |
| 4 | | | | 0 | | | | ∞ | | | | | | | | | | | | | 4 |
| 5 | | ∞ | | | 0 | | | | | | | | | | | | | | | | 5 |
| 6 | | | | | | 0 | | | | | | | ∞ | | | | | | | | 6 |
| 7 | | | ∞ | | | | 0 | | | | | | | | | | | | | | 7 |
| 8 | | | | | | | | 0 | | | | | | | | ∞ | | | | | 8 |
| 9 | | | | | | | | | 0 | | | | | | | | | | | ∞ | 9 |

| | 1 | 2 | 3 | 4 | 5 | 6 | 7 | 8 | 9 | 10 | 11 | 12 | 13 | 14 | 15 | 16 | 17 | 18 | 19 | 20 | |
|---|---|---|---|---|---|---|---|---|---|---|---|---|---|---|---|---|---|---|---|---|---|
| 10 | | | | | | | | | | 0 | | ∞ | | | | | | | | | 10 |
| 11 | -4 | -10 | -10 | | -8 | -9 | -5 | -3 | | -5 | 0 | -1 | -7 | | -5 | -6 | -2 | -7 | -12 | -5 | 11 |
| 12 | | | | | | | | | | | 0 | | | | | ∞ | | | | | 12 |
| 13 | | | | | | | | | | | | 0 | | | | | ∞ | | | | 13 |
| 14 | | | | | | | | ∞ | | | | | 0 | | | | | | | | 14 |
| 15 | | | | | | | | | | | | | | | 0 | | | | ∞ | | 15 |
| 16 | | | ∞ | | | | | | | | | | | | | 0 | | | | | 16 |
| 17 | | | | | | | | ∞ | | | | | | | | | 0 | | | | 17 |
| 18 | | | | | ∞ | | | | | | | | | | | | | 0 | | | 18 |
| 19 | | | | | | | | | | | | | ∞ | | | | | | 0 | | 19 |
| 20 | | | | | | | | | | | | | | | ∞ | | | | | 0 | 20 |
| | 1 | 2 | 3 | 4 | 5 | 6 | 7 | 8 | 9 | 10 | 11 | 12 | 13 | 14 | 15 | 16 | 17 | 18 | 19 | 20 | |

We cannot extend the negative path (11 ... 16) to a negative path smaller than itself.

A tour of smallest value occurs when our initial vertex is 11.

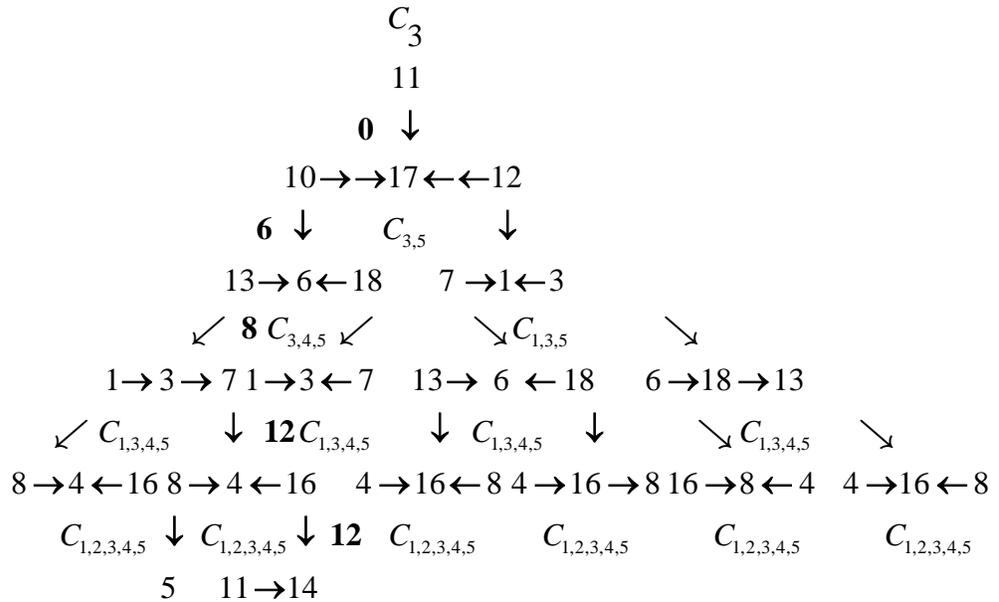

$T_{UPPERBOUND}$ = (11 17 12 10 6 18 13 3 1 7 4 8 16 14 19 15 20 9 2 5 ) = $|D_{n\,edges}|$ + 12 = 68 .

For simplicity, we denote $T_{UPPERBOUND}$ as $T$ and $\sigma_{T_{UPPERBOUND}}$ as $\sigma_T$. At the start of PHASE 3(a), when constructing acceptable paths, $P_{in}$ and $\sigma_T^{-1}M^-(20)$ are blank matrices. We described how we would use the modified F-W algorithm to obtain acceptable and 2-circuit cycles earlier.

$$T^{-1}M^-$$

| | 7 | 5 | 1 | 8 | 11 | 18 | 4 | 16 | 2 | 6 | 17 | 10 | 3 | 19 | 20 | 14 | 12 | 13 | 15 | 9 | |
|---|---|---|---|---|---|---|---|---|---|---|---|---|---|---|---|---|---|---|---|---|---|
| | 1 | 2 | 3 | 4 | 5 | 6 | 7 | 8 | 9 | 10 | 11 | 12 | 13 | 14 | 15 | 16 | 17 | 18 | 19 | 20 | |
| 1 | 0 | 71 | ∞ | 35 | 78 | 46 | 27 | 40 | 23 | 2 | 95 | 75 | -2 | 37 | 67 | 7 | 4 | 94 | 91 | 25 | 1 |
| 2 | 49 | 0 | 25 | 73 | 2 | 81 | 29 | 91 | ∞ | 79 | 5 | 59 | 68 | 22 | 84 | 86 | 8 | 30 | 88 | 0 | 2 |
| 3 | 1 | 6 | 0 | 97 | 83 | 2 | 22 | 30 | 68 | 60 | 50 | 89 | ∞ | 24 | 46 | 55 | 56 | 3 | 65 | 98 | 3 |
| 4 | 2 | 30 | 27 | 0 | 81 | 45 | ∞ | 1 | 27 | 56 | 78 | 45 | 20 | 67 | 14 | 25 | 15 | 54 | 44 | 23 | 4 |
| 5 | 27 | ∞ | 72 | 78 | 0 | 7 | 31 | 94 | -1 | 80 | 18 | 85 | 5 | 88 | 18 | 43 | 95 | 1 | 70 | 3 | 5 |
| 6 | 0 | 81 | 4 | 5 | 38 | 0 | 58 | 38 | 79 | ∞ | 49 | 8 | 60 | 40 | 49 | 44 | 40 | 2 | 61 | 42 | 6 |
| 7 | ∞ | 24 | -2 | 29 | 68 | 54 | 0 | 26 | 25 | -4 | 40 | 44 | -3 | 83 | 37 | 31 | 5 | 51 | 82 | 73 | 7 |
| 8 | 32 | 78 | 36 | ∞ | 41 | 76 | 1 | 0 | 72 | 4 | 62 | 53 | 96 | 58 | -1 | 91 | 89 | 83 | 73 | 12 | 8 |
| 9 | 77 | 4 | 27 | 13 | 27 | 62 | 25 | 30 | 0 | 42 | 72 | 49 | 98 | 86 | 1 | 94 | 80 | 97 | 2 | ∞ | 9 |
| 10 | 40 | 78 | 69 | 46 | 28 | 81 | 39 | 16 | 51 | 0 | -6 | ∞ | 81 | 55 | 27 | 50 | -8 | 86 | 21 | 41 | 10 |
| 11 | 70 | -1 | 78 | 40 | ∞ | 71 | 81 | 0 | 0 | 36 | 0 | 34 | 81 | 19 | 23 | 11 | 0 | 58 | 8 | 25 | 11 |

| | 1 | 2 | 3 | 4 | 5 | 6 | 7 | 8 | 9 | 10 | 11 | 12 | 13 | 14 | 15 | 16 | 17 | 18 | 19 | 20 | |
|---|---|---|---|---|---|---|---|---|---|---|---|---|---|---|---|---|---|---|---|---|---|
| 12 | 9 | 96 | 6 | 90 | 1 | 16 | 17 | 33 | 8 | 40 | 1 | 0 | 56 | 70 | 29 | 61 | ∞ | 83 | 55 | 80 | 12 |
| 13 | -2 | 1 | 14 | 81 | 57 | -2 | 53 | 26 | 27 | -1 | 45 | 89 | 0 | 19 | 82 | 62 | 80 | ∞ | 62 | 94 | 13 |
| 14 | 2 | 43 | 17 | 91 | 12 | 45 | 26 | 12 | 85 | 43 | 45 | 57 | 54 | 0 | 3 | ∞ | 59 | 64 | 23 | 93 | 14 |
| 15 | 2 | 70 | 18 | 73 | 9 | 85 | 45 | 87 | 87 | 60 | 83 | 28 | 64 | 6 | 0 | 23 | 54 | 64 | ∞ | 1 | 15 |
| 16 | -1 | 80 | -6 | -12 | -11 | 4 | -13 | ∞ | 77 | 24 | 40 | 11 | 16 | 78 | 76 | 0 | 20 | 16 | 75 | 17 | 16 |
| 17 | 12 | 18 | 10 | 62 | 1 | 64 | 79 | 52 | 4 | 48 | ∞ | 1 | 49 | 46 | 88 | 45 | 0 | 47 | 83 | 71 | 17 |
| 18 | 14 | 7 | 4 | 76 | 72 | ∞ | 46 | 16 | 80 | -1 | 64 | 88 | 1 | 85 | 14 | 45 | 15 | 0 | 85 | 61 | 18 |
| 19 | -5 | 82 | 6 | 54 | 14 | 79 | 62 | 84 | 15 | 33 | 40 | 56 | 17 | ∞ | 14 | -6 | 63 | 15 | 0 | 79 | 19 |
| 20 | 9 | 18 | 13 | -1 | 24 | 14 | 15 | 88 | 83 | 48 | 88 | 34 | 45 | 20 | ∞ | 3 | 28 | 84 | 0 | 2 | 20 |
| | 1 | 20 | 3 | 4 | 5 | 6 | 7 | 8 | 9 | 10 | 11 | 12 | 13 | 14 | 15 | 16 | 17 | 18 | 19 | 50 | |

In general, by using MIN(M) and giving the inverse of a derangement, we can obtain the values of any row in ascending order of magnitude.

$$T^{-1} = \begin{pmatrix} 1 & 2 & 3 & 4 & 5 & 6 & 7 & 8 & 9 & 10 & 11 & 12 & 13 & 14 & 15 & 16 & 17 & 18 & 19 & 20 \\ 3 & 9 & 13 & 7 & 2 & 10 & 1 & 4 & 20 & 12 & 5 & 17 & 18 & 16 & 19 & 8 & 11 & 6 & 14 & 15 \end{pmatrix}$$

In this case, suppose we want to find the third smallest value in row 5. MIN(M)(5, 3) = 13. . $T^{-1}(13) = 18$

This, (5 18) has the third smallest value in row 5 of $T^{-1}M^{-}$. Since $n$ is even, by choosing those arcs in $T^{-1}M^{-}$ that represent the inverse, $T^{-1}$, of $T$, we obtain two disjoint cycles - $s_{-}$ and $s_{+}$ - each of which moves $\frac{n}{2}$ points. Furthermore, the sum of their respective values have the same absolute value but different signs. This makes sense since $Ts_{-}s_{2} = T^{-1}$. On the other hand, $T \, s_{-}$ is a product of $\frac{n}{2}$ pair-wise disjoint cycles, i. e., a PM. We now construct $Ts_\_ := \sigma_T$. We give the inverse of $\sigma_T$ in order to use it and MIN(M) to obtain the smallest entry values in each row.

$$\sigma_T^{-1} = \begin{array}{cccccccccccccccccccc} 1 & 2 & 3 & 4 & 5 & 6 & 7 & 8 & 9 & 10 & 11 & 12 & 13 & 14 & 15 & 16 & 17 & 18 & 19 & 20 \\ 3 & 9 & 1 & 7 & 11 & 10 & 4 & 16 & 2 & 6 & 5 & 17 & 18 & 19 & 20 & 8 & 12 & 13 & 14 & 15 \end{array}$$

*Note.* As mentioned earlier, we can also obtain $\sigma_T$ as the smaller-valued set of alternating edges of $T$. When represented as a set of edges, $\sigma_T$ is a perfect matching.

Before proceeding further, we denote the 2-cycles of $\sigma_T$ in the following way:

1 (1 3); 2 (2 9); 3 (4 7); 4 (5 11); 5 (6 10); 6 (8 16); 7 (12 17); 8 (13 18); 9 (14 19); 10 (15 20).

A cycle may be obtained more than once using different determining vertices. Using theorem 1.26, we can eliminate all but one of such cycles (including companion cycles) when we place cycles in the matrix $M_{a,i}$. The remaining times that it occurs are included in our list of determining points. These can be useful when constructing trees in 3(b).

$$\sigma_T^{-1} M^-$$

| | 3 | 9 | 1 | 7 | 11 | 10 | 4 | 16 | 2 | 6 | 5 | 17 | 18 | 19 | 20 | 8 | 12 | 13 | 14 | 15 | |
|---|---|---|---|---|---|---|---|---|---|---|---|---|---|---|---|---|---|---|---|---|---|
| | 1 | 2 | 3 | 4 | 5 | 6 | 7 | 8 | 9 | 10 | 11 | 12 | 13 | 14 | 15 | 16 | 17 | 18 | 19 | 20 | |
| 1 | 0 | 27 | ∞ | 2 | 80 | 77 | 29 | 39 | 25 | 4 | 73 | 97 | 48 | 39 | 69 | 37 | 6 | 96 | 9 | 93 | 1 |
| 2 | 68 | 0 | 25 | 49 | 2 | 59 | 29 | 90 | ∞ | 79 | 0 | 5 | 81 | 22 | 84 | 73 | 8 | 30 | 86 | 88 | 2 |
| 3 | ∞ | 98 | 0 | 1 | 83 | 89 | 22 | 29 | 68 | 60 | 6 | 50 | 2 | 24 | 46 | 97 | 56 | 3 | 55 | 65 | 3 |
| 4 | 18 | 21 | 25 | 0 | 79 | 43 | ∞ | -4 | 25 | 53 | 28 | 76 | 43 | 65 | 12 | -2 | 13 | 52 | 23 | 42 | 4 |
| 5 | 5 | 3 | 72 | 27 | 0 | 85 | 31 | 92 | -1 | 80 | ∞ | 18 | 7 | 88 | 18 | 78 | 95 | 1 | 43 | 70 | 5 |
| 6 | 52 | 34 | -4 | -8 | 30 | 0 | 50 | 29 | 71 | ∞ | 73 | 41 | -8 | 32 | 41 | -3 | 32 | -6 | 36 | 53 | 6 |
| 7 | -3 | 73 | -2 | ∞ | 68 | 44 | 0 | 26 | 25 | -4 | 24 | 40 | 54 | 83 | 37 | 29 | 5 | 51 | 31 | 82 | 7 |
| 8 | 96 | 12 | 36 | 32 | 41 | 53 | 1 | 0 | 72 | 4 | 78 | 62 | 76 | 58 | -1 | ∞ | 89 | 83 | 91 | 73 | 8 |

| | 1 | 2 | 3 | 4 | 5 | 6 | 7 | 8 | 9 | 10 | 11 | 12 | 13 | 14 | 15 | 16 | 17 | 18 | 19 | 20 | |
|---|---|---|---|---|---|---|---|---|---|---|---|---|---|---|---|---|---|---|---|---|---|
| 9 | 98 | ∞ | 27 | 77 | 27 | 49 | 25 | 30 | 0 | 42 | 4 | 72 | 62 | 86 | 1 | 13 | 80 | 97 | 94 | 2 | 9 |
| 10 | 81 | 41 | 69 | 40 | 29 | ∞ | 39 | 16 | 51 | 0 | 78 | -6 | 81 | 55 | 45 | 46 | -8 | 86 | 50 | 21 | 10 |
| 11 | 82 | 26 | 79 | 71 | ∞ | 35 | 82 | 1 | 1 | 37 | 0 | 1 | 72 | 20 | 24 | 41 | 1 | 59 | 12 | 9 | 11 |
| 12 | 55 | 79 | 5 | 8 | 0 | -1 | 16 | 32 | 7 | 39 | 95 | 0 | 15 | 69 | 28 | 89 | ∞ | 82 | 60 | 54 | 12 |
| 13 | 2 | 96 | 95 | 54 | 59 | 91 | 55 | 28 | 29 | 1 | 3 | 47 | 0 | 21 | 84 | 83 | 82 | ∞ | 64 | 64 | 13 |
| 14 | 54 | 93 | 8 | 34 | 12 | 57 | 26 | 15 | 85 | 43 | 43 | 45 | 45 | 0 | 3 | 91 | 60 | 64 | ∞ | 23 | 14 |
| 15 | 64 | 1 | 92 | 85 | 9 | 28 | 45 | 87 | 87 | 60 | 70 | 83 | 85 | 6 | 0 | 73 | 54 | 64 | 23 | ∞ | 15 |
| 16 | 28 | 29 | 38 | 29 | 1 | 23 | -1 | ∞ | 89 | 36 | 92 | 64 | 64 | 46 | 88 | 0 | 52 | 47 | 45 | 83 | 16 |
| 17 | 49 | 71 | 96 | 43 | 1 | 1 | 79 | 52 | 4 | 48 | 18 | ∞ | 64 | 46 | 88 | 62 | 0 | 47 | 45 | 83 | 17 |
| 18 | 1 | 61 | 47 | 57 | 72 | 88 | 46 | 16 | 80 | -1 | 7 | 64 | ∞ | 85 | 14 | 76 | 15 | 0 | 45 | 85 | 18 |
| 19 | 23 | 85 | 12 | 86 | 20 | 62 | 68 | 90 | 21 | 39 | 88 | 46 | 85 | ∞ | 20 | 58 | 69 | 21 | 0 | 6 | 19 |
| 20 | 45 | 0 | 68 | 40 | 24 | 34 | 15 | 88 | 83 | 48 | 18 | 88 | 14 | 20 | ∞ | -1 | 28 | 84 | 3 | 0 | 20 |
| | 1 | 2 | 3 | 4 | 5 | 6 | 7 | 8 | 9 | 10 | 11 | 12 | 13 | 14 | 15 | 16 | 17 | 18 | 19 | 20 | |

We now construct acceptable paths to obtain acceptable and 2-circuit cycles. We are assuming that the value of a positive cycle is no greater than $|T| - |\sigma_T| - 1$ = 11

$\sigma_T$:  1 2 3 4 5 6 8 11 14 15
           7 9 13 12 18 10 16 17 20 19

$\sigma_T^{-1}$:  7 9 10 12 13 16 17 18 19 20
                1 2 6 4 3 8 11 5 15 14

j = 1

(5 1)(1 4) = (5 4): 7; (5 1)(1 10) = (5 10): 9; (5 1)(1 17) = (5 17): 11; (7 1)(1 17) = (7 17): 3;

(7 1)(1 19) = (7 19): 6; (13 1)(1 4) = (13 4): 4; (13 1)(1 17) = (13 17): 8;

(13 1)(1 19) = (13 19): 11; (18 1)(1 4) = (18 4): 3; (18 1)(1 17) = (18 17): 7;

(18 1)(1 19) = (18 19): 10

j = 2

(5 2)(2 5) = (5 5): 5

CYCLE P = [5 9 2 11 5]: (5 2): 5

(5 2)(2 12) = (5 12): 8; (15 2)(2 5) = (15 5): 3; (15 2)(2 11) = (15 11): 1;

(15 2)(2 12) = (15 12): 6; (15 2)(2 17) = (15 17): 9; (20 2)(2 5) = (20 5): 3;

(20 2)(2 11) = (20 11): 0; (20 2)(2 12) = (20 12): 5; (20 2)(2 17) = (20 17): 8

j = 3

(6 3)(3 11) = (6 11): 2; (7 3)(3 11) = (7 11): 4; (7 3)(3 13) = (7 13): 0; (7 3)(3 18) = (7 18): 1;

(12 3)(3 4) = (12 4): 6; (12 3)(3 13) = (12 13): 7; (12 3)(3 18) = (13 18): 8;

(14 3)(3 4) = (14 4): 9; (14 3)(3 13) = (14 13): 10; (14 3)(3 18) = (14 18): 11

j = 4

(1 4)(4 8) = (1 8): -2; (1 4)(4 16) = (1 16): 0; (3 4)(4 8) = (3 8): -3; (3 4)(4 16) = (3 16): 0;

(5 4)(4 8) = (5 8): 3; (5 4)(4 16) = (5 16): 5; (6 4)(4 8) = (6 8): -12; (6 4)(4 16) = (6 16): -10;

(12 4)(4 8) = (12 8): 2; (12 4)(4 16) = (12 16): 4; (13 4)(4 8) = (13 8): 0;

(13 4)(4 16) = (13 16): 2; (14 4)(4 8) = (14 8): 5; (14 4)(4 16) = (14 16): 7;

(18 4)(4 8) = (18 8): -1; (18 4)(4 16) = (18 16): 1

j = 5

(2 5)(5 1) = (2 1): 7;

(2 5)(5 2) = (2 2): 5

CYCLE P = [5 9 2 11 5]: (2 5): 5

(2 5)(5 13) = (2 13): 3; (2 5)(5 18) = (2 18): 3; (12 5)(5 1) = (12 1): <u>5</u>;

(12 5)(5 2) = (12 2): <u>3</u>; (12 5)(5 9) = (12 9): -1; (12 5)(5 13) = (12 13): 1; (12 5)(5 18) = (12 18): 1;

(15 5)(5 1) = (15 1): <u>8</u>; (15 5)(5 13) = (15 13): 4; (15 5)(5 18) = (15 18): 4;

(16 5)(5 1) = (16 1): <u>6</u>; (16 5)(5 2) = (16 2): <u>4</u>; (16 5)(5 9) = (16 9): 0; (16 5)(5 13) = (16 13): 2;

(16 5)(5 18) = (16 18): 2; (17 5)(5 1) = (17 1): <u>6</u>; (17 5)(5 2) = (17 2): <u>4</u>;

(17 5)(5 9) = (17 9): 0; (17 5)(5 13) = (17 13): 2; (17 5)(5 18) = (17 18): 2;

(20 5)(5 1) = (20 1): <u>7</u>; (20 5)(5 13) = (20 13): 3; (20 5)(5 18) = (20 18): 3

j = 6

(12 6)(6 3) = (12 3): <u>-5</u>; (12 6)(6 4) = (12 4): <u>-9</u>; (12 6)(6 13) = (12 13): -9;

(12 6)(6 16) = (12 16): -4; (12 6)(6 18) = (12 18): -7; (17 6)(6 3) = (17 3): <u>-3</u>;

(17 6)(6 4) = (17 4): <u>-7</u>; (17 6)(6 13) = (17 13): -8; (17 6)(6 16) = (17 16): -2;

(17 6)(6 18) = (17 18): -5

j = 7

(8 7)(7 1) = (8 1): <u>-2</u>; (8 7)(7 3) = (8 3): <u>-1</u>; (8 7)(7 10) = (8 10): -3; (8 7)(7 17) = (8 17): 6;

(16 7)(7 1) = (16 1) <u>-4</u>; (16 7)(7 3) = (16 3): <u>-3</u>; (16 7)(7 10) = (16 10): -5; (16 7)(7 17): 4

j = 8

(1 8)(8 10) = (1 10): 2; (1 8)(8 15) = (1 15): -3; (3 8)(8 10) = (3 10): 1; (3 8)(8 15) = (3 15): -4;

(4 8)(8 10) = (4 10): 0; (5 8)(8 7) = (5 7): <u>4</u>; (5 8)(8 10) = (5 10): 7; (5 8)(8 15) = (5 15): 2;

(6 8)(8 15) = (6 15): -13; (11 8)(8 7) = (11 7): <u>2</u>; (11 8)(8 10) = (11 10): 5;

(11 8)(8 15) = (11 15): 0; (12 8)(8 10) = (12 10): 6; (12 8)(8 15) = (12 15): 1;

(13 8)(8 15) = (13 15): -1; (14 8)(8 10) = (14 10): 9

j = 9

(5 9)(9 15) = (5 15): 0; (5 9)(9 20) = (5 20): 1;

(11 9)(9 11) = (11 11): 5

CYCLE P = [11 <u>2</u> 9 <u>5</u> 11]: (11 9): 5

(11 9)(9 20) = (11 20): 3; (12 9)(9 20) = (12 20): 1; (17 9)(9 15) = (17 15): 1;

(17 9)(9 20) = (17 20): 2

j = 10

(1 10)(10 12) = (1 12): -4; (1 10)(10 17) = (1 17): -6; (3 10)(10 12) = (3 12): -5;

(3 10)(10 17) = (3 17): -7; (4 10)(10 12) = (4 12): -6; (4 10)(10 17) = (4 17): -8;

(5 10)(10 12) = (5 12): 1; (5 10)(10 17) = (5 17): -1; (7 10)(10 12) = (7 12): -10;

(7 10)(10 17) = (7 17): -12; (8 10)(10 12) = (8 12): -9; (8 10)(10 17) = (8 17): -11;

(11 10)(10 12) = (11 12): -1; (11 10)(10 17) = (11 17): -3; (13 10)(10 12) = (13 12): -5;

(13 10)(10 17) = (13 17): -7; (14 10)(10 12) = (14 12): 3; (14 10)(10 17) = (14 17): 1;

(16 10)(10 12) = (16 12): -11; (16 10)(10 17) = (16 17): -13; (18 10)(10 12) = (18 12): -7;

(18 10)(10 17) = (18 17): -9

j = 11

(3 11)(11 9) = (3 9): <u>7</u>; (6 11)(11 9) = (6 9): <u>3</u>; (6 11)(11 12) = (6 12): 3;

(6 11)(11 17) = (6 17): 3; (6 11)(11 20) = (6 20): 11; (7 11)(11 8) = (7 8): <u>5</u>;

(7 11)(11 9) = (7 9): <u>5</u>; (9 11)(11 8) = (9 8): <u>5</u>;

(9 11)(11 9) = (9 9): 5

CYCLE P = [9 <u>5</u> 11 <u>2</u> 9]: (9 11): 5

(9 11)(11 12) = (9 12): 5; (9 11)(11 17): 5; (13 11)(11 9) = (13 9): <u>4</u>; (15 11)(11 8) = (15 8): <u>2</u>;

(15 11)(11 12) = (15 12): 2; (15 11)(11 17) = (15 17): 2; ((18 11)(11 8) = (18 8): <u>8</u>;

(18 11)(11 9) = (18 9): <u>8</u>; (20 11)(11 8) = (20 8): <u>1</u>; (20 11)(11 12) = (20 12): 1;

(20 11)(11 17) = (20 17): 1;

(20 11)(11 20) = (20 20): 9

CYCLE P = [20 <u>9</u> 2 <u>5</u> 11 <u>15</u> 20]: (20 2 11): 9

j = 12

(1 12)(12 5) = (1 5): <u>-4</u>; (1 12)(12 9) = (1 9): <u>3</u>;

(3 12)(12 3) = (3 3): 0

CYCLE P = [3 <u>7</u> 4 <u>16</u> 8 <u>6</u> 10 <u>17</u> 12 <u>1</u> 3]: (4 8 10 12): 2

(3 12)(12 5) = (3 5): <u>-5</u>; (3 12)(12 9) = (3 9): <u>5</u>; (4 12)(12 3) = (4 3): <u>-1</u>;

(4 12)(12 4) = (4 4): 2

CYCLE P = [4 <u>16</u> 8 <u>6</u> 10 <u>17</u> 12 <u>7</u> 4]: (4 8 10 12): 2

(4 12)(12 5) = (4 5): <u>-6</u>; (4 12)(12 9) = (4 9): <u>1</u>;

(5 12)(12 5) = (5 5): 1

CYCLE P = [5 3 1 7 4 16 8 6 10 17 12 11 5]: (5 1 4 8 10 12): 1

(7 12)(12 3) = (7 3): -5; (7 12)(12 5) = (7 5): -10; (7 12)(12 9) = (7 9): -3;

(8 12)(12 3) = (8 3): -4; (8 12)(12 5) = (8 5): -9; (8 12)(12 9) = (8 9): -2;

(9 12)(12 5) = (9 5): 5; (10 12)(12 3) = (10 3): -1; (10 12)(12 4) = (10 4): 2;

(10 12)(12 5) = (10 5): -6; (10 12)(12 9) = (10 9): 1; (11 12)(12 4) = (11 4): 7;

(11 12)(12 9) = (11 9): 6; (13 12)(12 5) = (13 5): -2; (13 12)(12 9) = (13 9): 2;

(14 12)(12 5) = (14 5): 6; (14 12)(12 9) = (14 9): 10; (15 12)(12 3) = (15 3): 7;

(16 12)(12 3) = (16 3): -6; (16 12)(12 5) = (16 5): -11; (16 12)(12 9) = (16 9): -4;

(18 12)(12 3) = (18 3): -2; (18 12)(12 4) = (18 4): 1; (18 12)(12 5) = (18 5): -7;

(18 12)(12 9) = (18 9): 0; (20 12)(12 3) = (20 3): 10

j = 13

(2 13)(13 1) = (2 1): 5; (2 13)(13 10) = (2 10): 4; (3 13)(13 10) = (3 10): 3;

(3 13)(13 11) = (3 11): 5; (6 13)(13 1) = (6 1): -6; (6 13)(13 11) = (6 11): -5;

(7 13)(13 11) = (7 11): 3; (12 13)(13 1) = (12 1): -7; (12 13)(13 10) = (12 10): -8;

(15 13)(13 1) = (15 1): 6; (15 13)(13 10) = (15 10): 5; (17 13)(13 1) = (17 1): -6;

(17 13)(13 11) = (17 11): -5; (20 13)(13 1) = (20 1): 5; (20 13)(13 10) = (20 10): 4

j = 14

(15 14)(14 15) = (15 15): 9

CYCLE P = [15 19 14 20 15]: (15 14): 9

j = 15

(1 15)(15 2) = (1 2): -2;  (3 15)(15 2) = (3 2): -3;

(5 15)(15 5) = (5 5): 9

CYCLE P = [5 2 9 20 15 11 5]: (5 9 15): 9

(5 15)(15 14) = (5 14): 6; (6 15)(15 14) = (6 14): -7; (8 15)(15 2) = (8 2): 0;

(8 15)(15 14) = (8 14): 5; (9 15)(15 14) = (9 14): 7; (11 15)(15 2) = (11 2): 1;

(12 15)(15 2) = (12 2): 2; (12 15)(15 14) = (12 14): 7; (13 15)(15 2) = (13 2): 0;

(13 15)(15 14) = (13 14): 5; (14 15)(15 2) = (14 2): 4;

(14 15)(15 14) = (14 14): 9

CYCLE P = [14 20 15 19 14]: (14 15): 9

j = 16

(5 16)(16 5) = (5 5): 6

CYCLE P = [5 3 1 7 4 8 16 11 5]: (5 1 4 16): 6

(6 16)(16 5) = (6 5): -11; (17 16)(16 5) = (17 5): -1; (17 16)(16 7) = (17 7): -3;

(20 16)(16 5) = (20 5): 0; (20 16)(16 7) = (20 7): -2

j = 17

(1 17)(17 5) = (1 5): -5; (1 17)(17 9) = (1 9): -2; (2 17)(17 6) = (2 6): 9;

(3 17)(17 5) = (3 5): -6; (3 17)(17 5) = (3 5): -6; (3 17)(17 9) = (3 9): -3;

(4 17)(17 5) = (4 5): -7; (4 17)(17 9) = (4 9): -4;

(5 17)(17 5) = (5 5): 0

CYCLE P = [5 3 1 7 4 16 8 6 10 12 17 11 5]: (5 1 4 8 10 17): 0

(7 17)(17 5) = (7 5): -11; (7 17)(17 9) = (7 9): -8; (8 17)(17 5) = (8 5): -10;

(8 17)(17 9) = (8 9): -7; (9 17)(17 6) = (9 6): 6;

(9 17)(17 9) = (9 9): 9

CYCLE P = [9 5 11 12 17 2 9]: (9 11 17): 9

(10 17)(17 5) = (10 5): -7; (13 17)(17 5) = (13 5): -6; (13 17)(17 9) = (13 9): -3;

(14 17)(17 5) = (14 5): 2; (14 17)(17 9) = (14 9): 5; (15 17)(17 6) = (15 6): 10;

(16 17)(17 5) = (16 5): -12; (16 17)(17 9) = (16 9): -9; (18 17)(17 5) = (18 5): -8;

(18 17)(17 9) = (18 9): -5; (20 17)(17 6) = (20 6): 9

j = 18

(2 18)(18 1) = (2 1): 4; (2 18)(18 10) = (2 10): 2; (5 18)(18 1) = (5 1): 2;

(5 18)(18 10) = (5 10): 0; (15 18)(18 1) = (15 1): 5; (20 18)(18 1) = (20 1): 4;

(20 18)(18 10) = (20 10): 2

j = 19

(20 19)(19 20) = (20 20): 9

CYCLE P = [20 14 19 15 20]: (20 19): 9

j = 20

(5 20)(20 16) = (5 16): <u>0</u>; (5 20)(20 14) = (5 14): <u>4</u>; (6 20)(20 2) = (6 2): <u>1</u>;

(6 20)(20 19) = (6 19): <u>4</u>; (9 20)(20 16) = (9 16): <u>1</u>; (9 20)(20 19) = (9 19): <u>5</u>;

(11 20)(20 19) = (11 19): <u>6</u>; (12 20)(20 19) = (12 19): <u>4</u>;

(19 20)(20 19) = (19 19): 9

CYCLE P = [19 15 20 14 19]: (19 20): 9

$P_{20}$

|    | 1  | 2  | 3  | 4  | 5  | 6  | 7 | 8  | 9  | 10 | 11 | 12 | 13 | 14 | 15 | 16 | 17 | 18 | 19 | 20 |    |
|----|----|----|----|----|----|----|---|----|----|----|----|----|----|----|----|----|----|----|----|----|----|
| 1  |    | 15 | 12 | 1  | 17 |    |   | 4  | 17 | 8  |    | 10 |    |    | 8  | 4  | 10 |    |    |    | 1  |
| 2  | 18 |    |    |    | 2  | 17 |   |    |    | 18 |    |    | 5  |    |    |    | 1  | 5  | 1  |    | 2  |
| 3  |    | 15 |    | 3  | 17 |    |   | 4  | 17 | 13 | 13 | 10 | 3  |    | 8  | 4  | 10 |    |    |    | 3  |
| 4  |    |    | 12 |    | 17 |    | 8 | 4  | 17 | 8  |    | 10 |    |    |    |    | 10 |    |    |    | 4  |
| 5  | 18 | 5  |    |    |    |    | 8 | 4  | 5  | 18 |    | 10 |    | 20 | 9  | 20 | 10 | 5  |    | 9  | 5  |
| 6  | 13 | 20 | 6  | 6  | 16 |    | 8 | 4  | 11 | 8  | 13 | 11 | 6  | 15 | 8  | 4  |    |    | 20 | 11 | 6  |
| 7  |    |    | 12 |    | 17 |    |   | 11 | 17 | 7  | 13 | 10 | 3  |    |    |    | 10 | 3  | 1  |    | 7  |
| 8  | 7  | 15 | 12 |    | 17 |    | 8 |    | 17 | 7  |    | 10 |    | 15 | 8  |    | 10 |    |    |    | 8  |
| 9  |    | 15 |    |    | 12 | 17 |   | 11 |    |    | 9  | 11 |    | 15 | 9  | 20 | 11 |    | 20 | 9  | 9  |
| 10 |    |    | 12 | 12 | 17 |    |   |    | 12 |    |    | 10 |    |    |    |    | 10 |    |    |    | 10 |
| 11 |    | 15 |    | 12 |    |    | 8 | 11 | 12 | 8  |    | 10 |    |    | 8  |    | 10 | 5  | 20 | 9  | 11 |
| 12 | 13 | 15 | 6  | 6  | 12 | 12 |   | 4  | 5  | 13 |    |    | 6  | 15 | 8  | 6  |    | 6  | 20 | 9  | 12 |

| 13 | 13 | 15 |   | 1 | 17 |   |   | 4 | 17 | 13 | 13 |   |   | 15 | 8 | 4 | 10 |   | 1 |   | 13 |
|---|---|---|---|---|---|---|---|---|---|---|---|---|---|---|---|---|---|---|---|---|---|
| 14 |   | 15 | 14 | 12 | 17 |   |   | 4 |   | 8 |   | 10 | 3 |   | 14 | 4 | 10 | 3 |   |   | 14 |
| 15 | 18 | 15 | 12 |   | 2 | 17 |   | 11 | 17 | 13 | 2 | 11 | 5 |   |   |   | 11 | 5 |   | 11 | 15 |
| 16 | 7 | 5 | 12 |   | 17 |   | 16 |   | 17 | 7 |   | 10 | 5 |   |   |   | 10 | 5 |   | 11 | 16 |
| 17 | 13 | 5 | 6 | 6 | 16 | 17 | 16 |   | 5 |   | 13 |   | 6 |   | 9 | 6 |   | 6 |   | 9 | 17 |
| 18 | 18 |   | 12 | 12 | 17 |   |   | 4 | 17 | 18 | 18 | 10 |   |   |   | 4 | 10 |   | 1 |   | 18 |
| 19 |   |   |   |   |   |   |   |   |   |   |   |   |   |   |   |   |   |   |   |   | 19 |
| 20 | 18 | 20 | 12 |   | 16 | 17 | 16 | 11 |   | 18 | 2 | 11 | 5 |   |   | 20 | 11 | 5 |   |   | 20 |
|   | 1 | 2 | 3 | 4 | 5 | 6 | 7 | 8 | 9 | 10 | 11 | 12 | 13 | 14 | 15 | 16 | 17 | 18 | 19 | 20 |   |

$$\sigma_T^{-1} M^-(20)$$

|   | 3 | 9 | 1 | 7 | 11 | 10 | 4 | 16 | 2 | 6 | 5 | 17 | 18 | 19 | 20 | 8 | 12 | 13 | 14 | 15 |   |
|---|---|---|---|---|---|---|---|---|---|---|---|---|---|---|---|---|---|---|---|---|---|
|   | 1 | 2 | 3 | 4 | 5 | 6 | 7 | 8 | 9 | 10 | 11 | 12 | 13 | 14 | 15 | 16 | 17 | 18 | 19 | 20 |   |
| 1 | 0 | -2 | ∞ | 2 | -5 |   |   |   | -2 | 2 |   | -4 |   |   | -3 |   | -6 |   |   | 9 | 1 |
| 2 | 4 | 0 |   |   | 2 | 9 |   |   | ∞ | 2 | 0 | 5 | 3 |   |   |   | 8 | 3 |   |   | 2 |
| 3 | ∞ | -3 | 0 | 1 | -6 |   |   |   | -3 | 1 | 5 | -5 | 2 |   | -4 | 0 | -7 | 3 |   |   | 3 |
| 4 |   |   |   | 0 | -7 |   | ∞ | -4 | -4 | 0 |   | -6 |   |   |   | -2 | -8 |   |   |   | 4 |
| 5 | 2 | 3 |   | 6 | 0 |   | 4 | 3 | -1 | 0 | ∞ | 1 | 7 | 4 | 0 | 0 | -1 | 1 |   | 1 | 5 |
| 6 | -6 | 1 | -4 | -8 | -11 | 0 |   | -12 | 3 | ∞ | -5 | -14 | -8 | -7 | -13 | -10 | -16 | -6 | 4 | 11 | 6 |
| 7 | -3 |   | -5 | ∞ | -11 |   | 0 | 5 | -8 | -4 | 3 | -10 | 0 |   |   | -12 | 1 | 6 |   |   | 7 |
| 8 | -2 | 0 | -4 |   | -10 |   | 1 | 0 | -7 | -3 |   | -9 |   | 5 | -1 | ∞ | -11 |   |   |   | 8 |
| 9 |   | ∞ |   |   | 5 | 6 |   | 5 | 0 |   | 4 | 5 |   | 7 | 1 | 1 | 5 |   | 5 | 2 | 9 |
| 10 |   |   | -1 | 2 | -7 | ∞ |   |   | 1 | 0 |   | -6 |   |   |   |   | -8 |   |   |   | 10 |

| | 1 | 2 | 3 | 4 | 5 | 6 | 7 | 8 | 9 | 10 | 11 | 12 | 13 | 14 | 15 | 16 | 17 | 18 | 19 | 20 | |
|---|---|---|---|---|---|---|---|---|---|---|---|---|---|---|---|---|---|---|---|---|---|
| 11 | | | | | ∞ | | 1 | 1 | **5** | 0 | -1 | | | 0 | | -3 | | **6** | 3 | 1 | 11 |
| 12 | **-7** | 2 | **-5** | **-9** | 0 | -1 | | 2 | -1 | **-8** | | 0 | -9 | **7** | 1 | -4 | ∞ | -7 | **4** | 1 | 12 |
| 13 | 2 | **0** | | 4 | **-6** | | | **-3** | 1 | 3 | -5 | 0 | **5** | -1 | 2 | -7 | ∞ | 11 | | | 13 |
| 14 | | **4** | 8 | 9 | **2** | | | 5 | **5** | 9 | | 3 | 10 | 0 | 3 | 7 | 1 | 11 | ∞ | | 14 |
| 15 | **5** | 1 | **7** | | 3 | 3 | | 2 | | **-3** | 1 | 6 | 4 | 6 | 0 | | 2 | 4 | | ∞ | 15 |
| 16 | **-4** | 4 | **-6** | | -12 | -1 | ∞ | **-9** | -5 | | -11 | | | | 0 | | -13 | 2 | | | 16 |
| 17 | **-6** | 4 | **-3** | **-7** | **-1** | 1 | **-3** | | 0 | | **-5** | ∞ | -8 | | | -2 | 0 | -5 | | | 17 |
| 18 | 1 | | **-2** | **1** | **-8** | | | 8 | **-5** | -1 | 7 | -7 | ∞ | | | | -9 | 0 | 10 | | 18 |
| 19 | | | | | | | | | | | | | | ∞ | | | | | 0 | 6 | 19 |
| 20 | 4 | 0 | 10 | | 0 | 9 | **-2** | | | 2 | 0 | 5 | 3 | | ∞ | -1 | 8 | 3 | 3 | 0 | 20 |
| | 1 | 2 | 3 | 4 | 5 | 6 | 7 | 8 | 9 | 10 | 11 | 12 | 13 | 14 | 15 | 16 | 17 | 18 | 19 | 20 | |

C followed by a set 2-cycles (boldface) represents 2-cycles each of which contains a point in a given cycle. DC followed by a set of 2-cycles represents 2-cycles none of whose points occur in the given cycle.

The set of 2-cycles was given earlier. Using theorem 26 and its corollary, we delete cycles that have points in the same set of 2-cycles and values that are greater than or equal to that of other cycles in the set.

## ACCEPTABLE CYCLES

1. C 2, 4; 5; (11 9): 5. 2. C 9, 10; 9; (14 15). 3. C 2, 4, 10; 9; (5 9 15). 4. C 2, 4, 7; 9; (9 11 17).

5. C 3, 5, 6, 7; 2; (4 8 10 12). 6. C 1, 3, 5, 6, 7; 0; (3 4 8 10 12).

7. DC 2, 8, 9, 10; 1; (5 1 4 8 10 12).

## 2-CIRCUIT UNLINKED PATHS

1. C 5, 8; -7; [6 18 *10*]. 2. C 5, 7; -7; [*10* 17 6]. 3. C 3, 6; -3; [*4* 8 7]. 4. C 1, 3; -1; [7 1 *4*].

5. C 2, 4; 1; [*2* 5 *9*]. 6. C 2, 10; 2; [*9* 15 2]. 7. C 4, 7; 2; [*11* 17 *5*]. 8. C 1, 8; 3; [*3* 13 *1*].

9. C 2, 4; 3; [*5* 2 *11*]; 10. C 2, 4; 3; [*5* 9 *11*]. 11. C 3, 5, 6; -8; [*6* 4 8 *10*].

12. C 3, 5, 7; -2; [*7* 10 12 *4*]. 13. C 2, 4, 10; 10; [*15* 2 11 *20*]. 14. C 1, 3, 5, 6, 7; 1; [*1* 4 8 10 12 *3*].

We now obtain paths used to construct $\sigma_1^{-1} M^-(40)$.

j = 1

(2 1)(1 4) = (2 4): 6; (5 1)(1 4) = (5 4): 4; (5 1)(1 19) = (5 19): 11; (6 1)(1 17) = (6 17): 0;

(8 1)(1 19) = (8 19): 7; (12 1)(1 19) = 2; (15 1)(1 4) = (15 4): 7; (15 1)(1 17) = (15 17): 10

(15 1)(1 19) = (15 19): 10; (16 1)(1 19) = (16 19): 5;

(17 1)(1 17) = (17 17): 0

CYCLE P = [17 <u>10</u> 6 <u>18</u> 13 <u>3</u> 1 <u>12</u> 17]: (17 6 13 1): 0

(17 1)(1 19) = (17 19): 3; (20 1)(1 4) = (20 4): 6

j = 2

(1 2)(2 11) = (1 11): -2; (3 2)(2 11) = (3 11): -3; (8 2)(2 11) = (8 11): 0;

(12 2)(2 11) = (12 11): 2;

(12 2)(2 12) = (12 12): 7

CYCLE P = [12 7 4 16 8 20 15 9 2 17 12]: (12 4 8 15 2): 7

(13 2)(2 11) = (13 11): 0; (17 2)(2 12) = (17 12): 7

(17 2)(2 17) = (17 17): 10

CYCLE P = [17 <u>10</u> 6 <u>8</u> 16 <u>11</u> 5 <u>9</u> 2 <u>12</u> 17]: (17 6 16 5 2): 10

j = 3

(7 3)(3 11) = (7 11): 1;  (7 3)(3 13) = (7 13): -3;  (7 3)(3 18) = (7 18): -2;

(8 3)(3 13) = (8 13): -2;  (8 3)(3 18) = (8 18): -3;  (10 3)(3 4) = (10 4): 0;

(10 3)(3 11) = (10 11): 5;  (10 3)(3 13) = (10 13): 0;  (10 3)(3 18) = (10 18): 2;

(12 3)(3 11) = (12 11): 1;  (16 3)(3 11) = (16 11): 0;  (16 3)(3 13) = (16 13): -4;

(16 3)(3 18) = (16 18): -3;  (18 3)(3 4) = (18 4): -1;  (18 3)(3 11) = (18 11): 4;

(18 3)(3 18) = (18 18): 1

CYCLE  P  =  [18 6 10 17 12 1 3 13 18]: (18 10 12 3): 1

j = 4

(2 4)(4 8) = (2 8): 2;  (2 4)(4 16) = (2 16): 4;  (5 4)(4 8) = (5 8): 0;  (10 4)(4 8) = (10 8): -4;

(10 4)(4 16) = (10 16): -2;  (12 4)(4 8) = (12 8): -13;  (12 4)(4 15) = (12 15): 3;

(12 4)(4 16) = (12 16): -11;  (15 4)(4 16) = (15 16): 5;  (17 4)(4 8) = (17 8): -11;

(17 4)(4 15) = (17 15): 5;  (17 4)(4 16) = (17 16): -9;

(17 4)(4 17) = (17 17): 6

CYCLE  P  =  [17 10 6 7 4 12 17]: (17 6 4): 6

(15 4)(4 8) = (15 8): -3;  (15 4)(4 16) = (15 16): -1;  (18 4)(4 8) = (18 8): -3;

(18 4)(4 16) = (18 16): -1

j = 5

(1 5)(5 1) = (1 1): 0

CYCLE P = [1 7 4 16 8 6 10 12 17 11 5 3 1]: (1 4 8 10 17 5): 0

(3 5)(5 9) = (3 9): -7; (4 5)(5 1) = (4 1): -2; (4 5)(5 2) = (4 2): -4; (4 5)(5 9) = (4 9): -8;

(4 5)(5 13) = (4 13): 0; (4 5)(5 2) = (4 2): -4; (4 5)(5 15) = (4 15): 11; (4 5)(5 18) = (4 18): -6;

(6 5)(5 2) = (6 2): -8; (6 5)(5 9) = (6 9): -12; (6 5)(5 18) = (6 18): -8; (7 5)(5 1) = (7 1): -6;

(7 5)(5 2) = (7 2): -8; (7 5)(5 9) = (7 9): -12; (7 5)(5 13) = (7 13): -4; (7 5)(5 15) = (7 15): 7;

(7 5)(5 18) = -10; (8 5)(5 1) = (8 1): -5; (8 5)(5 2) = (8 2): -7; (8 5)(5 9) = (8 9): -11;

(8 5)(5 13) = (8 13): -3; (8 5)(5 18) = (8 18): -9; (9 5)(5 1) = (9 1): 10; (10 5)(5 1) = (10 1): -2;

(10 5)(5 2) = (10 2): -4; (10 5)(5 9) = (10 9): -8; (10 5)(5 15) = (10 15): 11;

(10 5)(5 18) = (10 18): -6; (13 5)(5 2) = (13 2): -3; (13 5)(5 9) = (13 9): -7;

(13 5)(5 13) = (13 13): 1

CYCLE P = [13 3 1 7 4 16 8 6 10 12 17 11 5 18 13]: (13 1 4 8 10 17 5): 1

(14 5)(5 2) = (14 2): 4; (14 5)(5 9) = (14 9): 1; (14 5)(5 13) = (14 13): 9;

(14 5)(5 18) = (14 18): 3; (16 5)(5 1) = (16 1): -7; (16 5)(5 2) = (16 2): -9;

(16 5)(5 9) = (16 9): -13; (16 5)(5 13) = (16 13): -5; (16 5)(5 15) = (16 15): 6;

(16 5)(5 18) = (16 18): -11; (17 5)(5 2) = (17 2): 2; (17 5)(5 9) = (17 9): -2;

(18 5)(5 1) = (18 1): -3; (18 5)(5 2) = (18 2): -5; (18 5)(5 9) = (18 9): -9;

(18 5)(5 15) = (18 15): 10;

(18 5)(5 18) = (18 18): 10

CYCLE P = [18 6 10 12 17 11 5 13 18]: (18 10 17 5): -7

(20 5)(5 9) = (20 9): -1; (20 5)(5 18) = (20 18): 1

j = 6

(9 6)(6 7) = (9 7): 7; (9 6)(6 13) = (9 13): 9; (9 6)(6 18) = (9 18): 7; (15 6)(6 3) = (15 3): <u>-1</u>;

(15 6)(6 4) = (15 4): <u>-5</u>; (15 6)(6 13) = (15 13): -5; (15 6)(6 16) = (15 16): 0;

(15 6)(6 18) = (15 18): -3

j = 7

(9 7)(7 1) = (9 1): <u>4</u>; (9 7)(7 3) = (9 3): <u>-2</u>; (17 7)(7 1) = (17 1): <u>-6</u>; (17 7)(7 3) = (17 3): <u>-5</u>;

(17 7)(7 17) = (17 17): 2

CYCLE P = [17 <u>10</u> 6 <u>8</u> 16 <u>4</u> 7 <u>12</u> 17]: (17 6 16 7): 2

(20 7)(7 1) = (20 1): <u>-5</u>; (20 7)(7 3) = (20 3): <u>-4</u>; (20 7)(7 10) = (20 10): -6

j = 8

(2 8)(8 15) = (2 15): 1; (10 8)(8 4) = (10 4): <u>0</u>; (10 8)(8 15) = (10 15): -5;

(10 8)(8 10) = (10 10): 0

CYCLE P = [10 <u>17</u> 12 <u>7</u> 4 <u>16</u> 8 <u>6</u> 10]: (10 12 4 8): 0

(15 8)(8 7) = (15 7): <u>3</u>;

(15 8)(8 15) = (15 15): -4

CYCLE P = [15 <u>11</u> 5 <u>13</u> 18 <u>3</u> 1 <u>7</u> 4 <u>16</u> 8 <u>20</u> 15]: (15 5 18 1 4 8): -4

(17 8)(8 15) = (17 15): -12; (18 8)(8 2) = (18 2): <u>9</u>; (18 8)(8 15) = (18 15): -4

j = 9

(1 9)(9 16) = (1 16): 11; (1 9)(9 20) = (1 20): 0; (4 9)(9 11) = (4 11): -4;

(4 9)(9 20) = (4 20): -6; (7 9)(9 11) = (7 11): -8; (7 9)(9 20) = (7 20): -10;

(7 9)(9 15) = (7 15): -11; (7 9)(9 20) = (7 20): -10; (8 9)(9 11) = (8 11): -3;

(8 9)(9 15) = (8 15): -6; (8 9)(9 20) = (8 20): -9; (10 9)(9 11) = (10 11): -4;

(10 9)(9 15) = (10 15): -7; (10 9)(9 20) = (10 20): -6; (13 9)(9 15) = (13 15): -6;

(13 9)(9 20) = (13 20): -5; (14 9)(9 11) = (14 11): 5; (14 9)(9 15) = (14 15): 2;

(14 9)(9 20) = (14 20): 3; (16 9)(9 15) = (16 15): -12; (16 9)(9 20) = (16 20): -11;

(17 9)(9 15) = (17 15): -1; (17 9)(9 20) = (17 20): 0; (18 9)(9 15) = (18 15): -8;

(18 9)(9 20) = (18 20): -7

j = 10

(1 10)(10 12) = (1 12): -4; (1 10)(10 17) = (1 17): -6; (2 10)(10 12) = (2 12): -4;

(2 10)(10 17) = (2 17): -6; (3 10)(10 17) = (3 17): -5; (5 10)(10 12) = (5 12): -6;

(5 10)(10 17) = (5 17): -8;

(12 10)(10 12) = (12 12): 2

CYCLE P = [12 11 5 <u>18</u> 13 <u>6</u> 10 <u>17</u> 12]: (12 5 13 10): 2

j = 11

(3 11)(11 19) = (3 19): 9; (7 11)(11 8) = (7 8): <u>-7</u>; (7 11)(11 19) = (7 19): 4;

(8 11)(11 8) = (8 8): -4

CYCLE P = [8 <u>4</u> 7 <u>6</u> 10 <u>12</u> 17 <u>2</u> 9 <u>5</u> 11 <u>16</u> 8]: (8 7 10 17 9 11): -4

(8 11)(11 20) = (8 20): 4;

(12 11)(11 12) = (12 12): 2

CYCLE P = [12 10 6 1 3 7 4 16 8 20 15 9 2 5 11 17 12]: (12 6 3 4 8 15 2 11): 2

(17 11)(11 9) = (17 9): -4;

(17 11)(11 17) = (17 17): -4

CYCLE P = [17 10 6 18 13 5 11 12 17]: (17 6 13 11): -4

j = 12

(2 12)(12 3) = (2 3): 1; (2 12)(12 4) = (2 4): 4; (5 12)(12 3) = (5 3): -1;

(5 12)(12 5) = (5 5): -4

CYCLE P = [5 13 18 6 10 17 12 11 5]: (5 18 10 12): -4

(6 12)(12 6) = (6 6): -8

CYCLE P = [6 18 13 5 11 17 12 10 6]: (6 13 11 12): -8

(15 12)(12 3) = (15 3): 7

j = 13

(10 13)(13 10) = (10 10): 1

CYCLE P = [10 17 12 1 3 18 13 6 10]: (10 12 3 13): 1

(15 13)(13 1) = (15 1): -3; (15 13)(13 10) = (15 10): -4; (20 13)(13 1) = (20 1): -8;

(20 13)(13 14) = (20 14): 11

j = 15

(2 15)(15 14) = (2 14): <u>7</u>; (3 15)(15 14) = (3 14): <u>0</u>;

(2 15)(15 2) = (2 2): 2

CYCLE P = [2 <u>11</u> 5 <u>13</u> 18 <u>3</u> 1 <u>7</u> 4 <u>16</u> 8 <u>20</u> 15 <u>9</u> 2]: (2 5 18 1 4 8 15): 2

(6 15)(15 14) = (6 14): <u>-5</u>; (7 15)(15 14) = (7 14): <u>-5</u>; (8 15)(15 5) = (8 5): <u>1</u>;

(8 15)(15 14) = (8 14): <u>-2</u>; (10 15)(15 14) = (10 14): <u>-1</u>;

(14 15)(15 14) = (14 14): 8

CYCLE P = [14 <u>1</u> 3 <u>7</u> 4 <u>16</u> 8 <u>6</u> 10 <u>12</u> 17 <u>11</u> 5 <u>2</u> 9 <u>20</u> 15 <u>19</u> 14]: (14 3 4 8 10 17 5 9 15): 8

(16 15)(15 14) = (16 14): <u>-6</u>; (17 15)(15 2) = (17 2): <u>-11</u>; (17 15)(15 14) = (17 14): <u>-6</u>;

(18 15)(15 14) = (18 14): <u>-2</u>

j = 16

(5 16)(16 5) = (5 5): 1

CYCLE P = [5 <u>2</u> 9 <u>15</u> 20 <u>8</u> 16 <u>11</u> 5]: (5 9 20 16): 1

(5 16)(16 7) = (5 7): <u>-1</u>; (9 16)(16 5) = (9 5): <u>2</u>; (9 16)(16 7) = (9 7): <u>0</u>;

(10 16)(16 7) = (10 7): <u>-8</u>; (17 16)(16 5) = (17 5): <u>-8</u>

j = 17

(3 17)(17 5) = (3 5): <u>-4</u>;

(5 17)(17 5) = (5 5): -7

CYCLE P = [5 <u>13</u> 18 <u>6</u> 10 <u>12</u> 17 <u>11</u> 5]: (5 18 10 17): -7

(5 17)(17 9) = (5 9): <u>-7</u>

j = 18

(4 18)(18 1) = (4 1): -5; (4 18)(18 15) = (4 15): 8; (6 18)(18 1) = (6 1): -7;

(7 18)(18 1) = (7 1): -9; (8 18)(18 1) = (8 1): -8; (10 18)(18 1) = (10 1): -5;

(10 18)(18 10) = (10 10): -7

CYCLE P = [10 12 17 11 5 13 18 6 10]: (10 17 5 18): -7

(15 18)(18 1) = (15 1): 3; (15 18)(18 10) = (15 10): 1; (16 18)(18 1) = (16 1): -10

j = 19

(6 19)(19 20) = (6 20): 9

j = 20

(1 20)(20 16) = (1 16): -1; (1 20)(20 19) = (1 19): 3; (3 20)(20 7) = (3 7): 10;

(3 20)(20 16) = (3 16): -6; (3 20)(20 19) = (3 19): 0;

(7 20)(20 7) = (7 7): 5

CYCLE P = [7 6 10 12 17 11 5 2 9 15 20 4 7]: (7 10 17 5 9 20): 5

(7 20)(20 16) = (7 16): -11; (7 20)(20 19) = (7 19): -7; (8 20)(20 19) = (8 19): -6;

(10 20)(20 16) = (10 16): -7; (10 20)(20 19) = (10 19): -3;

(16 20)(20 16) = (16 16): -12

CYCLE P = [16 4 7 6 10 12 17 11 5 2 9 15 20 8 16]: (16 7 10 17 5 9 20): -12

(16 20)(20 19) = (16 19): -8; (17 20)(20 19) = (17 19): 0; (18 20)(20 16) = (18 16): -8;

(18 20)(20 19) = (18 19): 0

ACCEPTABLE CYCLES

1. DC 8; 8; (14 3 4 8 10 17 5 9 15). 2. DC 8, 9; 2; (12 6 3 4 8 15 2 11).

3. DC 1, 8, 9; -12; (16 7 10 17 5 9 20). 4. DC 2, 9, 10; 1; (13 1 4 8 10 17 5).

5. DC 5, 7, 9; 2; (2 5 18 1 4 8 15). 6. DC 1, 8, 9, 10; -4; (8 7 10 17 9 11).

7. DC 2, 5, 7, 9; -4; (15 5 18 1 4 8). 8. DC 2, 8, 9, 10; 0; (1 4 8 10 17 5).

9. DC 1, 5, 6, 8, 9; 5; (7 10 17 5 9 20). 10. C 2, 4, 5, 6, 7; 10; (17 6 16 5 2).

11. C 2, 3, 6, 7, 10; 7; (12 4 8 15 2); *12. C 4, 5, 7, 8; -7; (18 10 17 5). 13. C 1, 5, 7, 8; 0; (17 16 13 1).

14. C 2, 4, 6, 10; 1; (5 9 20 16). *15. C 4, 5, 7, 8; 2; (12 5 13 10). 16. C 3, 5, 7, 6; (17 6 16 7).

17. C 3, 5, 7; 6; (17 6 4).

2-CIRCUIT UNLINKED PATHS

1. DC 1, 4, 8, 9; -5; [*12* 6 4 8 15 2 *17*]. 2. C 2, 4, 5, 7, 8; 6; [*2* 5 13 10 17 *9*].

3. C 2, 3, 5, 6, 7; 6; [*8* 7 10 17 9 *16*]. 4. C 2, 4, 5, 6, 7; 7; [*17* 6 16 5 2 *12*].

5. C 1, 5, 7, 8; -1; [*12* 6 13 1 *17*]. 6. C 1, 3, 5, 7; -4; [*7* 10 12 3 *4*]. 7. C 4, 5, 7 ,8; -3;

 [*17* 6 13 11 *12*];

8. C 4, 5, 7, 8; -1; [*18* 10 17 5 *13*]. 9. C 1, 5, 7, 8; 0; [*18* 10 12 3 *13*]. 10. C 2, 4, 6, 10; 0;

[*20* 16 5 9 *15*].

11. C 4, 5, 7, 8; 10; [*5* 18 10 17 *11*]. 12. C 1, 6, 8; -2; [*6* 13 1 *10*].

$P_{40}$

|    | 1  | 2  | 3  | 4  | 5  | 6  | 7  | 8  | 9  | 10 | 11 | 12 | 13 | 14 | 15 | 16 | 17 | 18 | 19 | 20 |    |
|----|----|----|----|----|----|----|----|----|----|----|----|----|----|----|----|----|----|----|----|----|----|
| 1  |    | 15 | 12 | 1  | 17 |    |    | 4  | 17 | 8  | 2  | 10 |    |    | 8  | 20 | 10 |    | 20 | 9  | 1  |
| 2  | 18 |    | 12 | 12 | 2  | 17 |    | 4  |    | 18 |    | 10 | 5  | 15 | 8  | 4  | 10 | 5  | 1  |    | 2  |
| 3  |    | 15 |    | 3  | 17 |    |    | 4  | 5  | 13 | 2  | 10 | 3  | 15 | 8  | 20 | 10 |    | 20 | 9  | 3  |
| 4  | 18 | 5  | 12 |    | 17 |    | 8  | 4  | 5  | 8  | 9  | 10 | 5  |    | 18 |    | 10 | 5  |    | 9  | 4  |
| 5  | 18 | 5  | 12 | 1  |    |    | 16 | 4  | 17 | 18 |    | 10 |    | 20 | 9  | 20 | 10 | 5  | 1  | 9  | 5  |
| 6  | 18 | 5  | 6  | 6  | 16 |    | 8  | 4  | 5  | 8  | 13 | 11 | 6  | 15 | 8  | 4  | 1  | 5  | 20 | 11 | 6  |
| 7  | 18 | 5  | 12 |    | 17 |    |    | 11 | 5  | 7  | 9  | 10 | 5  | 15 | 9  | 20 | 10 | 5  | 20 | 9  | 7  |
| 8  | 18 | 5  | 12 |    | 17 |    | 8  |    | 5  | 7  | 9  | 10 | 5  | 15 | 9  |    | 10 | 5  | 20 | 9  | 8  |
| 9  | 18 | 15 | 7  |    | 16 | 17 | 16 | 11 |    |    | 9  | 11 | 6  | 15 | 9  | 20 | 11 | 6  | 20 | 9  | 9  |
| 10 | 18 | 5  | 12 | 8  | 17 |    | 16 | 4  | 5  |    | 3  | 10 | 3  | 15 | 9  | 20 | 10 | 5  | 20 | 9  | 10 |
| 11 |    | 15 |    | 12 |    |    | 8  | 11 | 12 | 8  |    | 10 |    |    | 8  |    | 10 | 5  | 20 | 9  | 11 |
| 12 | 13 | 15 | 6  | 6  | 12 | 12 |    | 4  | 5  | 13 | 3  |    | 6  | 15 | 8  | 4  |    | 6  | 1  | 9  | 12 |

|    | 1  | 2  | 3  | 4  | 5  | 6  | 7  | 8  | 9  | 10 | 11 | 12 | 13 | 14 | 15 | 16 | 17 | 18 | 19 | 20 |    |
|----|----|----|----|----|----|----|----|----|----|----|----|----|----|----|----|----|----|----|----|----|----|
| 13 | 13 | 5  |    | 1  | 17 |    |    | 4  | 5  | 13 | 2  |    |    | 15 | 9  | 4  | 10 |    | 1  | 9  | 13 |
| 14 |    | 5  | 14 | 12 | 17 |    |    | 4  | 5  | 8  | 9  | 10 | 5  |    | 9  | 4  | 10 | 5  |    | 9  | 14 |
| 15 | 13 | 15 | 6  | 6  | 2  | 17 | 16 | 11 | 17 | 13 | 2  | 11 | 6  |    |    | 6  | 11 | 6  |    | 11 | 15 |
| 16 | 18 | 5  | 12 |    | 17 |    | 16 |    | 5  | 7  | 3  | 10 | 5  | 15 | 8  |    | 10 | 5  | 1  | 9  | 16 |
| 17 | 7  | 15 | 7  | 6  | 16 | 17 |    | 4  | 11 |    | 13 | 2  | 6  | 15 | 9  | 4  |    | 6  | 1  | 9  | 17 |
| 18 | 5  | 8  | 12 | 3  | 17 |    |    | 4  | 5  | 18 | 3  | 10 |    | 15 | 9  | 20 | 10 |    | 1  | 9  | 18 |
| 19 |    |    |    |    |    |    |    |    |    |    |    |    |    |    |    |    |    |    |    |    | 19 |
| 20 | 13 | 20 | 7  | 1  | 16 | 17 | 16 | 11 | 5  | 7  | 2  | 11 | 5  | 13 |    | 20 | 11 | 5  |    |    | 20 |
|    | 1  | 2  | 3  | 4  | 5  | 6  | 7  | 8  | 9  | 10 | 11 | 12 | 13 | 14 | 15 | 16 | 17 | 18 | 19 | 20 |    |

$$\sigma_T^{-1} M^-(40)$$

|    | 3  | 9  | 1  | 7  | 11 | 10 | 4  | 16 | 2  | 6  | 5  | 17 | 18 | 19 | 20 | 8  | 12 | 13 | 14 | 15 |    |
|----|----|----|----|----|----|----|----|----|----|----|----|----|----|----|----|----|----|----|----|----|----|
|    | 1  | 2  | 3  | 4  | 5  | 6  | 7  | 8  | 9  | 10 | 11 | 12 | 13 | 14 | 15 | 16 | 17 | 18 | 19 | 20 |    |
| 1  | 0  | -2 | ∞  | 2  | -5 |    |    | -2 | 2  | -2 | -4 |    | -3 | -1 | -6 |    | 3  | 0  |    |    | 1  |
| 2  | 4  | 0  | 1  | 4  | 2  | 9  |    | 2  | ∞  | 2  | 0  | -4 | 3  | 7  | 1  | 4  | -6 | 3  |    |    | 2  |
| 3  | ∞  | -3 | 0  | 1  | -4 |    |    | -3 | 1  | -3 | -5 | 2  | 2  | -6 | -6 | -5 | 3  | 0  | -5 |    | 3  |
| 4  | -5 |    |    | 0  | -7 |    | ∞  | -4 | -8 | 0  | -4 | -6 | 0  |    | 8  | -2 | -8 | -6 | -3 | -6 | 4  |
| 5  | 2  | 3  | -1 | 4  | 0  |    | -1 | 3  | -1 | 0  | ∞  | -4 | 7  | 4  | 0  | 0  | -8 | 1  | 11 | 1  | 5  |
| 6  | -7 | -12| -4 | -8 | -11| 0  |    | -12| -12| ∞  | -5 | -7 | -8 | -7 | -13| -10| -1 | -10| 3  | 9  | 6  |
| 7  | -9 | -8 | -5 | ∞  | -11|    | 0  | -7 | -12| -4 | -8 | -10| -4 |    | -11| -11| -12| -10| -7 | -10| 7  |
| 8  | -8 | -7 | -4 |    | -10|    | 1  | 0  | -9 | -3 | -5 | -9 | -3 | 5  | -8 | ∞  | -11| -9 | -6 | -9 | 8  |
| 9  | 4  | ∞  | 3  |    | 2  | 6  | 0  | 5  | 0  |    | 4  | 5  | 6  | 7  | 1  | 1  | 5  | 7  | 5  | 2  | 9  |
| 10 | -5 | -4 | -1 | 0  | -7 | ∞  | -8 | -4 | -8 | 0  | -4 | -6 | 0  | -1 | -7 | -7 | -8 | -6 | -8 | -6 | 10 |
| 11 |    |    |    | ∞  |    |    | 1  | 1  | 5  | 0  | -1 |    |    | 0  |    | -3 |    | 6  | 3  |    | 11 |
| 12 | -7 | 2  | -5 | -9 | 0  | -1 |    | 2  | -1 | -8 | 1  | 0  | -9 | 7  | 1  | -4 | ∞  | -7 | 2  | 1  | 12 |
| 13 | 2  | -3 |    | 4  | -6 |    |    | -7 | 1  | 0  | -5 | 0  | 5  | -6 | 2  | -7 | ∞  | -2 | -5 |    | 13 |

| | 1 | 2 | 3 | 4 | 5 | 6 | 7 | 8 | 9 | 10 | 11 | 12 | 13 | 14 | 15 | 16 | 17 | 18 | 19 | 20 | |
|---|---|---|---|---|---|---|---|---|---|---|---|---|---|---|---|---|---|---|---|---|---|
| 14 | | *4* | *8* | *9* | *2* | | | *5* | *1* | *9* | *4* | *3* | *9* | *0* | *2* | *7* | *1* | *3* | ∞ | *3* | 14 |
| 15 | *-3* | *1* | *-1* | *-5* | *3* | *3* | *3* | *2* | | *-3* | *3* | *-3* | *-5* | *6* | *0* | *0* | *2* | *-3* | | ∞ | 15 |
| 16 | -10 | -9 | -6 | | -12 | | -1 | ∞ | -13 | -5 | 0 | -11 | -5 | -6 | -12 | 0 | -13 | -11 | -8 | -11 | 16 |
| 17 | -6 | -11 | -3 | -7 | -8 | 1 | | -11 | 0 | | -5 | ∞ | -8 | -6 | -12 | -9 | 0 | -5 | 0 | -3 | 17 |
| 18 | *1* | *9* | *-2* | *1* | *-8* | | | *-3* | *-9* | *-1* | *4* | *-7* | ∞ | *-2* | *-8* | *-8* | *-9* | *0* | *0* | *-3* | 18 |
| 19 | | | | | | | | | | | | | ∞ | | | | | | 0 | 6 | 19 |
| 20 | *-5* | *0* | *-4* | *1* | *0* | *9* | *-2* | | *-1* | *-6* | *0* | *5* | *1* | *11* | ∞ | *-1* | *3* | *1* | *3* | *0* | 20 |
| | 1 | 2 | 3 | 4 | 5 | 6 | 7 | 8 | 9 | 10 | 11 | 12 | 13 | 14 | 15 | 16 | 17 | 18 | 19 | 20 | |

j = 1

(4 1)(1 4) = (4 4): -3

CYCLE P = [4 16 8 6 10 12 17 11 5 13 18 3 1 7 4]: (4 8 10 17 5 13 1): -3

(6 1)(1 17) = (6 17): -3; (6 1)(1 19) = (6 19): 0; (9 1)(1 4) = (9 4): 6; (10 1)(1 4) = (10 4): -3;

(10 1)(1 10) = (10 10): -1

CYCLE P = [10 12 17 11 5 13 18 3 1 6 10]: (10 17 5 18 1): -1

(20 1)(1 17) = (20 17): 1

j = 2

(17 2)(2 5) = (17 5): -9; (17 2)(2 11) = (17 11): -11;

(17 2)(2 17) = (17 17): -3

CYCLE P = [17 10 6 7 4 16 8 20 15 9 2 12 17]: (17 6 4 8 15 2): -3

j = 3

(2 3)(3 4) = (2 4): 2; (5 3)(3 4) = (5 4): 0; (9 3)(3 13) = (9 13): 0;

(9 3)(3 18) = (9 18): 1; (20 3)(3 13) = (20 13): -2; (20 3)(3 18) = (20 18): -1

j = 4

(2 4)(4 8) = (2 8): 0; (2 4)(4 16) = (2 16): 2; (5 4)(4 8) = (5 8): -4; (10 4)(4 8) = (10 8): -7;

(15 4)(4 8) = (15 8): 1; (15 4)(4 16) = (15 16): 3

j = 5

(3 5)(5 9) = (3 9): -5;

(9 5)(5 9) = (9 9): 1

CYCLE P = [9 15 20 8 16 11 5 2 9]: (9 20 16 5): 1

(17 5)(5 18) = (17 18): -8

j = 7

(9 7)(7 1) = (9 1): -3; (9 7)(7 3) = (9 3): -2; (9 7)(7 10) = (9 10): -4; (10 7)(7 1) = (10 1): -11;

(10 7)(7 3) = (10 3): -10;

(10 7)(7 10) = (10 10): -12

CYCLE P = [10 12 17 11 5 2 9 15 20 8 16 4 7 6 10]: (10 17 5 9 20 16 7): -12

(17 7)(7 1) = (17 1): -13; (17 7)(7 3) = (17 3): -12

j = 8

(2 8)(8 15) = (2 15); -1; (5 8)(8 15) = (5 15): -5;

(10 8)(8 10) = (10 10): -3

CYCLE P = [10 12 17 11 5 13 18 3 1 7 4 16 8 6 10]: (10 17 5 18 1 4 8): -3

(10 8)(8 15) = (10 15): -8;

(15 8)(8 15) = (15 15): 2

CYCLE P = [15 9 2 11 5 13 18 3 1 7 4 16 8 20 15]: (15 2 18 1 4 8): 2

j = 10

(9 10)(10 12) = (9 12): -10; (9 10)(10 17) = (9 17): -12; (20 10)(10 12) = (20 12): -12;

(20 10)(10 17) = (20 17): -14

j = 11

(17 11)(11 17) = (17 17): -10

CYCLE P = [17 10 6 7 4 16 8 20 15 9 2 5 11 12 17]: (17 6 4 8 15 2 11): -10

j = 12

(9 12)(12 3) = (9 3): -5; (9 12)(12 5) = (9 5): -10;

(9 12)(12 9) = (9 9): -3

CYCLE P = [9 15 20 8 16 4 7 6 10 17 12 2 9]: (9 20 16 7 10 12): -3

(20 12)(12 3) = (20 3): -7; (20 12)(12 5) = (20 5): -12; (20 12)(12 9) = (20 9): -5

j = 13

(20 13)(13 11) = (20 11): -2

j = 15

(2 15)(15 2) = (2 2): 0

CYCLE P = [2 11 5 13 18 6 10 17 12 1 3 7 4 16 8 20 15 9 2]: (2 5 18 10 12 3 4 8 15): 0 *

(5 15)(15 2) = (5 2): <u>-4</u>;

(5 15)(15 5) = (5 5): -4

CYCLE P = [5 <u>13</u> 18 <u>6</u> 10 <u>17</u> 12 <u>1</u> 3 <u>7</u> 4 <u>16</u> 8 <u>20</u> 15 <u>11</u> 5]: (5 8 10 12 3 4 8 15): -4

(10 15)(15 2) = (10 2): <u>-7</u>; (10 15)(15 14) = (10 14): <u>-2</u>;

j = 16

(3 16)(16 7) = (3 7): <u>-6</u>;

(5 16)(16 5) = (5 5): -1

CYCLE P = [5 <u>13</u> 18 <u>6</u> 10 <u>17</u> 12 <u>1</u> 3 <u>7</u> 4 <u>8</u> 16 <u>11</u> 5]: (5 18 10 12 3 4 16): -1

(7 16)(16 7) = (7 7): -12

CYCLE P = [7 <u>6</u> 10 <u>12</u> 17 <u>11</u> 5 <u>2</u> 9 <u>15</u> 20 <u>8</u> 16 <u>4</u> 7]: (7 10 17 5 9 20 16): -12

(18 16)(16 7) = (18 7): <u>-9</u>

j = 17

(6 17)(17 6) = (6 6): 0

CYCLE P = [6 <u>7</u> 4 <u>8</u> 16 <u>11</u> 5 <u>13</u> 18 <u>3</u> 1 <u>12</u> 17 <u>10</u> 6]: (6 4 16 5 18 1 17): 0

(9 17)(17 5) = (9 5): <u>-11</u>;

(9 17)(17 9) = (9 9): -8

CYCLE P = [9 <u>15</u> 20 <u>8</u> 16 <u>4</u> 7 <u>6</u> 10 <u>12</u> 17 <u>2</u> 9]: (9 20 16 7 10 17): -8

(20 17)(17 5) = (20 5): <u>-13</u>; (20 17)(17 9) = (20 9): <u>-10</u>

j = 18

(17 18)(18 17) = (17 17): 7

CYCLE P = [17 <u>10</u> 6 <u>7</u> 4 <u>16</u> 8 <u>20</u> 15 <u>9</u> 2 <u>11</u> 5 <u>13</u> 18 <u>12</u> 17]: (17 6 4 8 15 2 5 18): 7

j = 19

(4 19)(19 3) = (4 3): <u>9</u>; (6 19)(19 20) = (6 20): 8; (13 19)(19 3) = (13 3): <u>10</u>;

j = 20

(8 20)(20 19) = (8 19): <u>-6</u>

<div align="center">ACCEPTABLE CYCLES</div>

1. DC 9; 0; (2 5 18 10 12 3 4 8 15). 2. DC 1,9; 7; (17 6 4 8 15 2 5 18).

3. DC 1, 8, 9; -12; (10 17 5 9 20 16 7). 4. DC 2, 8, 9; -3; (4 8 10 17 5 15 1). 5.

5. DC 2, 9, 10; -3; (10 17 5 18 1 4 8). 8. DC 5, 7, 9; 2; (15 2 5 18 1 4 8).

9. DC 1, 4, 8, 9; -8; (9 20 16 7 10 17). 10. C 1, 4, 5, 7, 8; 4; (10 17 5 18 1).

11. C 2, 4, 6, 10; 1; (9 20 16 5).

<div align="center">2-CIRCUIT UNLINKED PATHS</div>

1. DC 1, 8, 9; [*17* 6 4 8 15 2 11 *12*]. 2. DC 2, 6, 9, 10; -7; [*7* 10 17 5 18 1 *4*].

3. DC 2, 6, 7, 9, 10; -5; [*6* 4 5 18 1 *10*]. 4. DC 2, 3, 6, 9, 10; 5; [*5* 18 10 12 3 *11*].

5. C 1, 4, 5, 7, 8; 5; [*9* 20 16 5 *2*].

$P_{60}$

|    | 1  | 2  | 3  | 4  | 5  | 6  | 7  | 8  | 9  | 10 | 11 | 12 | 13 | 14 | 15 | 16 | 17 | 18 | 19 | 20 |    |
|----|----|----|----|----|----|----|----|----|----|----|----|----|----|----|----|----|----|----|----|----|----|
| 1  |    | 15 | 12 | 1  | 17 |    |    | 4  | 17 | 8  | 2  | 10 |    |    | 8  | 20 | 10 |    | 20 | 9  | 1  |
| 2  | 18 |    | 12 | 12 | 2  | 17 |    | 4  |    | 18 |    | 10 | 5  | 15 | 8  | 4  | 10 | 5  | 1  |    | 2  |
| 3  |    | 15 |    | 3  | 17 |    | 16 | 4  | 5  | 13 | 2  | 10 | 3  | 15 | 8  | 20 | 10 |    | 20 | 9  | 3  |
| 4  | 18 | 15 | 19 |    | 17 |    | 8  | 4  | 5  | 8  | 9  | 10 | 5  |    | 18 |    | 10 | 5  | 20 | 9  | 4  |
| 5  | 18 | 15 | 12 | 3  |    |    | 16 | 4  | 17 | 18 |    | 10 |    | 15 | 8  | 20 | 10 | 5  | 1  | 9  | 5  |
| 6  | 18 | 5  | 6  | 6  | 16 |    | 8  | 4  | 5  | 8  | 13 | 11 | 6  | 15 | 8  | 4  | 1  | 5  | 1  | 19 | 6  |
| 7  | 18 | 5  | 12 |    | 17 |    |    | 11 | 5  | 7  | 9  | 10 | 5  | 15 | 9  | 20 | 10 | 5  | 20 | 9  | 7  |
| 8  | 18 | 5  | 12 |    | 17 |    | 8  |    | 5  | 7  | 9  | 10 | 5  | 15 | 9  |    | 10 | 5  | 20 | 9  | 8  |
| 9  | 7  | 15 | 12 |    | 17 | 17 | 16 | 11 |    | 7  | 9  | 10 | 3  | 15 | 9  | 20 | 10 | 3  | 20 | 9  | 9  |
| 10 | 18 | 15 | 7  | 1  | 17 |    | 16 | 4  | 5  |    | 3  | 10 | 3  | 15 | 8  | 20 | 10 | 5  | 20 | 9  | 10 |
| 11 |    | 15 |    | 12 |    |    | 8  | 11 | 12 | 8  |    | 10 |    |    | 8  |    | 10 | 5  | 20 | 9  | 11 |
| 12 | 13 | 15 | 6  | 6  | 12 | 12 |    | 4  | 5  | 13 | 3  |    | 6  | 15 | 8  | 4  |    | 6  | 1  | 9  | 12 |
| 13 | 13 | 5  | 19 | 1  | 17 |    |    | 4  | 5  | 13 | 2  |    |    | 15 | 9  | 4  | 10 |    | 1  | 9  | 13 |

| 14 |   | 15 | 14 | 3 | 17 |   |   | 4 | 5 | 8 | 2 | 10 | 5 |   | 14 | 4 | 10 | 5 |   | 9 | 14 |
| 15 | 13 | 15 | 6 | 6 | 2 | 17 | 8 | 4 | 17 | 8 | 2 | 11 | 6 |   |   | 4 | 11 | 6 |   | 11 | 15 |
| 16 | 18 | 5 | 12 |   | 17 |   | 16 |   | 5 | 7 | 3 | 10 | 5 | 15 | 8 |   | 10 | 5 | 20 | 9 | 16 |
| 17 | 7 | 15 | 7 | 6 | 2 | 17 | 16 | 4 | 11 |   | 2 | 2 | 6 | 15 | 8 | 4 |   | 6 | 1 | 9 | 17 |
| 18 | 5 | 8 | 12 | 3 | 17 |   | 16 | 4 | 5 | 18 | 3 | 10 |   | 15 | 9 | 20 | 10 |   | 1 | 9 | 18 |
| 19 |   |   |   |   |   |   |   |   |   |   |   |   |   |   |   |   |   |   |   |   | 19 |
| 20 | 7 | 20 | 12 | 1 | 17 | 17 | 16 | 11 | 17 | 7 | 13 | 10 | 3 | 13 |   | 20 | 10 | 3 |   |   | 20 |
|   | 1 | 2 | 3 | 4 | 5 | 6 | 7 | 8 | 9 | 10 | 11 | 12 | 13 | 14 | 15 | 16 | 17 | 18 | 19 | 20 |   |

$$\sigma_T^{-1} M^-(60)$$

|   | 3 | 9 | 1 | 7 | 11 | 10 | 4 | 16 | 2 | 6 | 5 | 17 | 18 | 19 | 20 | 8 | 12 | 13 | 14 | 15 |   |
|---|---|---|---|---|---|---|---|---|---|---|---|---|---|---|---|---|---|---|---|---|---|
|   | 1 | 2 | 3 | 4 | 5 | 6 | 7 | 8 | 9 | 10 | 11 | 12 | 13 | 14 | 15 | 16 | 17 | 18 | 19 | 20 |   |
| 1 | 0 | -2 | ∞ | 2 | -5 |   |   | -2 | 2 | -2 | -4 |   |   | -3 | -1 | -6 |   | 3 | 0 |   | 1 |
| 2 | 4 | 0 | 1 | 4 | 2 | 9 |   | 0 | ∞ | 2 | 0 | -4 | 3 | 5 | -1 | 2 | -6 | 3 |   |   | 2 |
| 3 | ∞ | -3 | 0 | 1 | -4 |   | -6 |   | -5 | 1 | -3 | -5 | 2 | 2 | -6 | -6 | -5 | 3 | 0 | -5 | 3 |
| 4 | -5 | 9 | 9 | 0 | -7 |   | ∞ | -4 | -8 | 0 | -4 | -6 | 0 |   | 8 | -2 | -8 | -6 | -3 | -6 | 4 |
| 5 | 2 | -4 | -1 | 0 | 0 |   | -1 | -4 | -1 | 0 | ∞ | -4 | 7 | 1 | -5 | 0 | -8 | 1 | 11 | 1 | 5 |
| 6 | -7 | -12 | -4 | -8 | -11 | 0 |   | -12 | -12 | ∞ | -5 | -7 | -8 | -7 | -13 | -10 | -7 | -10 | 2 | 8 | 6 |
| 7 | -9 | -8 | -5 | ∞ | -11 |   | 0 | -7 | -12 | -4 | -8 | -10 | -4 |   | -11 | -11 | -12 | -10 | -7 | -10 | 7 |
| 8 | -8 | -7 | -4 |   | -10 |   | 1 | 0 | -9 | -3 | -5 | -9 | -3 | 5 | -8 | ∞ | -11 | -9 | -6 | -9 | 8 |
| 9 | -3 | ∞ | -2 |   | -11 | 6 | 0 | 5 | 0 | -4 | 4 | -10 | -3 | 7 | 1 | 1 | -12 | -2 | 5 | 2 | 9 |
| 10 | -11 | -7 | -10 | -3 | -7 | ∞ | -8 | -7 | -8 | 0 | -4 | -6 | 0 | -2 | -8 | -7 | -8 | -6 | -8 | -6 | 10 |
| 11 |   |   |   |   | ∞ |   |   | 1 | 1 | 5 | 0 | -1 |   |   | 0 |   | -3 |   | 6 | 3 | 11 |
| 12 | -7 | 2 | -5 | -9 | 0 | -1 |   | 2 | -1 | -8 | 1 | 0 | -9 | 7 | 1 | -4 | ∞ | -7 | 2 | 1 | 12 |
| 13 | 2 | -3 | 10 | 4 | -6 |   |   | -7 | 1 | 0 | -5 | 0 | 5 | -6 | 2 | -7 | ∞ | -2 | -5 |   | 13 |

| | 1 | 2 | 3 | 4 | 5 | 6 | 7 | 8 | 9 | 10 | 11 | 12 | 13 | 14 | 15 | 16 | 17 | 18 | 19 | 20 | |
|---|---|---|---|---|---|---|---|---|---|---|---|---|---|---|---|---|---|---|---|---|---|
| 14 | | *4* | *8* | *9* | *2* | | | *5* | *1* | *9* | *4* | *3* | *9* | *0* | *2* | *7* | *1* | *3* | ∞ | *3* | 14 |
| 15 | -3 | *1* | -1 | -5 | *3* | *3* | *3* | -9 | | -5 | *3* | -3 | -5 | 6 | 0 | -7 | *2* | -3 | 6 | ∞ | 15 |
| 16 | -_10_ | -9 | -6 | | -_12_ | | -1 | ∞ | -_13_ | -5 | *0* | -_11_ | -5 | -6 | -_12_ | *0* | -_13_ | -_11_ | -8 | -_11_ | 16 |
| 17 | -6 | -_11_ | -_12_ | -7 | -9 | *1* | | -_11_ | -9 | | -_11_ | ∞ | -8 | -6 | -_12_ | -9 | 0 | -8 | *0* | -3 | 17 |
| 18 | *1* | 9 | -2 | *1* | -8 | | -_9_ | -3 | -5 | -1 | *4* | -7 | ∞ | -2 | -8 | -8 | -9 | 0 | *0* | -3 | 18 |
| 19 | | | | | | | | | | | | | ∞ | | | | | | *0* | 6 | 19 |
| 20 | -5 | *0* | -_7_ | *1* | -_13_ | 9 | -2 | | -_10_ | -6 | -_2_ | -_12_ | -5 | 11 | ∞ | -1 | -_14_ | -3 | *3* | 0 | 20 |
| | 1 | 2 | 3 | 4 | 5 | 6 | 7 | 8 | 9 | 10 | 11 | 12 | 13 | 14 | 15 | 16 | 17 | 18 | 19 | 20 | |

j = 1

(10 1)(1 10) = (10 10): -7

CYCLE P = [10 _12_ 17 _1_ 5 _2_ 9 _15_ 20 _8_ 16 _4_ 7 _3_ 1 _6_ 10]: (10 17 5 9 20 16 7 1): -7

j = 2

(5 2)(2 5) = (5 5): -2

CYCLE P = [5 _13_ 18 _6_ 10 _17_ 12 _1_ 3 _7_ 4 _16_ 8 _20_ 15 _9_ 2 _11_ 5]: (5 18 10 12 3 4 8 15 2): -2

j = 3

(4 3)(3 4) = (4 4): 10

CYCLE P = [4 _16_ 8 _6_ 10 _12_ 17 _11_ 5 _2_ 9 _15_ 20 _14_ 19 _1_ 3 _7_ 4]: (4 8 10 17 5 9 20 19 3): 10

(10 3)(3 13) = (10 13): -8; (10 3)(3 18) = (10 18): -7; (17 3)(3 13) = (17 13): -10;

(17 3)(3 18) = (17 18): -9; (20 3)(3 18) = (20 18): -4

j = 5

(9 5)(5 1) = (9 1): -_6_;

(9 5)(5 9) = (9 9): -12

CYCLE P = [9 15 20 8 16 4 7 6 10 12 17 11 5 2 9]: (9 20 16 7 10 17 5): -12

(9 5)(5 13) = (9 13): -4; (9 5)(5 18) = (9 18): -10; (20 5)(5 1) = (20 1): -8; (20 5)(5 2) = (20 2): -10;

(20 5)(5 9) = (20 9): -14; (20 5)(5 13) = (20 13): -6; (20 5)(5 18) = (20 18): -12

j = 7

(3 7)(7 3) = (3 3): -8

CYCLE P = [3 18 13 6 10 12 17 11 5 2 9 15 20 8 16 4 7 1 3]: (3 13 10 17 5 9 20 16 7): -8

(18 7)(7 1) = (18 1): -12; (18 7)(7 3) = (18 3): -11;

j = 9

(20 9)(9 20) = (20 20): -12

CYCLE P = [20 8 16 4 7 6 10 12 17 11 5 2 9 15 20]: (20 16 7 10 17 5 9): -12

j = 13

(10 13)(13 10) = (10 10): -7

CYCLE P = [10 12 17 11 5 2 9 15 20 8 16 4 7 1 3 18 13 6 10]: (10 17 5 9 20 16 7 3 13): -7

j = 18

(9 18)(18 1) = (9 1): -9;

(10 18)(18 10) = (10 10): -8

CYCLE P = [10 12 17 11 5 2 9 15 20 8 16 4 7 1 3 13 18 6 10]: (10 17 5 9 20 16 7 3 18): -8

(17 18)(18 17) = (17 17): 6

CYCLE P = [17 <u>10</u> 6 <u>7</u> 4 <u>8</u> 16 <u>4</u> 7 <u>1</u> 3 <u>13</u> 18 <u>12</u> 17]: (17 6 4 16 7 3 18): 6

(20 18)(18 1) = (20 1): <u>-11</u>

## ACCEPTABLE CYCLES

1. DC 9; -8; (10 17 5 9 20 16 7 3 18). 2. DC 8; 10; (4 8 10 17 5 9 20 18 3).

3. DC 8, 9; -7; (10 17 5 9 20 16 7 1). 4. DC 1, 5, 8; -12; (9 20 16 7 10 17 5).

## 2-CIRCUIT UNLINKED PATHS

1. DC 9; -4; [*5* 18 10 12 3 4 8 15 2 *11*]. 2. DC 1, 8, 9; 2; [*20* 16 7 10 17 5 9 *15*].

3. DC 1, 2, 9; 2; [*20* 16 7 10 17 5 18 *15*]. 4. DC 1, 2, 8, 9; 5; [*20* 16 7 10 17 5 18 *15*].

5. DC 2, 3 5, 6, 9; 1; [*3* 13 10 17 5 *1*].

$P_{80}$

|    | 1  | 2  | 3  | 4  | 5  | 6  | 7  | 8  | 9  | 10 | 11 | 12 | 13 | 14 | 15 | 16 | 17 | 18 | 19 | 20 |    |
|----|----|----|----|----|----|----|----|----|----|----|----|----|----|----|----|----|----|----|----|----|----|
| 1  |    | 15 | 12 | 1  | 17 |    |    | 4  | 17 | 8  | 2  | 10 |    |    | 8  | 20 | 10 |    | 20 | 9  | 1  |
| 2  | 18 |    | 12 | 12 | 2  | 17 |    | 4  |    | 18 |    | 10 | 5  | 15 | 8  | 4  | 10 | 5  | 1  |    | 2  |
| 3  |    | 15 |    | 3  | 17 |    | 16 | 4  | 5  | 13 | 2  | 10 | 3  | 15 | 8  | 20 | 10 |    | 20 | 9  | 3  |
| 4  | 18 | 15 | 19 |    | 17 |    | 8  | 4  | 5  | 8  | 9  | 10 | 5  |    | 18 |    | 10 | 5  | 20 | 9  | 4  |
| 5  | 18 | 15 | 12 | 3  |    |    | 16 | 4  | 17 | 18 |    | 10 |    | 15 | 8  | 4  | 10 | 5  | 1  | 9  | 5  |
| 6  | 18 | 5  | 6  | 6  | 16 |    | 8  | 4  | 5  | 8  | 13 | 11 | 6  | 15 | 8  | 4  | 1  | 5  | 1  | 19 | 6  |
| 7  | 18 | 5  | 12 |    | 17 |    |    | 11 | 5  | 7  | 9  | 10 | 5  | 15 | 9  | 20 | 10 | 5  | 20 | 9  | 7  |
| 8  | 18 | 5  | 12 |    | 17 |    | 8  |    | 5  | 7  | 9  | 10 | 5  | 15 | 9  |    | 10 | 5  | 20 | 9  | 8  |
| 9  | 18 | 15 | 12 |    | 17 | 17 | 16 | 11 |    | 7  | 18 | 10 | 5  | 15 | 9  | 20 | 10 | 3  | 20 | 9  | 9  |
| 10 | 18 | 15 | 7  | 1  | 17 |    | 16 | 4  | 5  |    | 3  | 10 | 3  | 15 | 8  | 20 | 10 | 3  | 20 | 9  | 10 |

| | 1 | 2 | 3 | 4 | 5 | 6 | 7 | 8 | 9 | 10 | 11 | 12 | 13 | 14 | 15 | 16 | 17 | 18 | 19 | 20 | |
|---|---|---|---|---|---|---|---|---|---|---|---|---|---|---|---|---|---|---|---|---|---|
| 11 | | 15 | | 12 | | | 8 | 11 | 12 | 8 | | 10 | | | 8 | | 10 | 5 | 20 | 9 | 11 |
| 12 | 13 | 15 | 6 | 6 | 12 | 12 | | 4 | 5 | 13 | 3 | | 6 | 15 | 8 | 4 | | 6 | 1 | 9 | 12 |
| 13 | 13 | 5 | 19 | 1 | 17 | | | 4 | 5 | 13 | 2 | | | 15 | 9 | 4 | 10 | | 1 | 9 | 13 |
| 14 | | 15 | 14 | 3 | 17 | | | 4 | 5 | 8 | 2 | 10 | 5 | | 14 | 4 | 10 | 5 | | 9 | 14 |
| 15 | 13 | 15 | 7 | 6 | 2 | 17 | 8 | 4 | 17 | 8 | 2 | 11 | 6 | | | 4 | 11 | 6 | | 11 | 15 |
| 16 | 18 | 5 | 12 | | 17 | | 16 | | 5 | 7 | 3 | 10 | 5 | 15 | 8 | | 10 | 5 | 20 | 9 | 16 |
| 17 | 7 | 15 | 7 | 6 | 2 | 17 | 16 | 4 | 11 | | 2 | 2 | 3 | 15 | 8 | 4 | | 3 | 1 | 9 | 17 |
| 18 | 7 | 8 | 7 | 3 | 17 | | 16 | 4 | 5 | 18 | 3 | 10 | | 15 | 9 | 20 | 10 | | 1 | 9 | 18 |
| 19 | | | | | | | | | | | | | | | | | | | | | 19 |
| 20 | 18 | 5 | 12 | 1 | 17 | 17 | 16 | 11 | 5 | 7 | 13 | 10 | 5 | 13 | | 20 | 10 | 5 | | | 20 |
| | 1 | 2 | 3 | 4 | 5 | 6 | 7 | 8 | 9 | 10 | 11 | 12 | 13 | 14 | 15 | 16 | 17 | 18 | 19 | 20 | |

$$\sigma_T^{-1} M^- (80)$$

| | 3 | 9 | 1 | 7 | 11 | 10 | 4 | 16 | 2 | 6 | 5 | 17 | 18 | 19 | 20 | 8 | 12 | 13 | 14 | 15 | |
|---|---|---|---|---|---|---|---|---|---|---|---|---|---|---|---|---|---|---|---|---|---|
| | 1 | 2 | 3 | 4 | 5 | 6 | 7 | 8 | 9 | 10 | 11 | 12 | 13 | 14 | 15 | 16 | 17 | 18 | 19 | 20 | |
| 1 | 0 | -2 | ∞ | 2 | -5 | | | -2 | 2 | -2 | -4 | | | -3 | -1 | -6 | | 3 | 0 | | 1 |
| 2 | 4 | 0 | 1 | 4 | 2 | 9 | | 0 | ∞ | 2 | 0 | -4 | 3 | 5 | -1 | 2 | -6 | 3 | | | 2 |
| 3 | ∞ | -3 | 0 | 1 | -4 | | -6 | | -5 | 1 | -3 | -5 | 2 | 2 | -6 | -6 | -5 | 3 | 0 | -5 | 3 |
| 4 | -5 | 9 | 9 | 0 | -7 | | ∞ | -4 | -8 | 0 | -4 | -6 | 0 | | 8 | -2 | -8 | -6 | -3 | -6 | 4 |
| 5 | 2 | -4 | -1 | 0 | 0 | | -1 | -4 | -1 | 0 | ∞ | -4 | 7 | 1 | -5 | -2 | -8 | 1 | 11 | 1 | 5 |
| 6 | -7 | -12 | -4 | -8 | -11 | 0 | | -12 | -12 | ∞ | -5 | -7 | -8 | -7 | -13 | -10 | -7 | -10 | 2 | 8 | 6 |
| 7 | -9 | -8 | -5 | ∞ | -11 | | 0 | -7 | -12 | -4 | -8 | -10 | -4 | | -11 | -11 | -12 | -10 | -7 | -10 | 7 |
| 8 | -8 | -7 | -4 | | -10 | | 1 | 0 | -9 | -3 | -5 | -9 | -3 | 5 | -8 | ∞ | -11 | -9 | -6 | -9 | 8 |
| 9 | <u>-9</u> | ∞ | -2 | | -11 | 6 | 0 | 5 | 0 | -4 | <u>-3</u> | -10 | -4 | 7 | 1 | 1 | -12 | -10 | 5 | 2 | 9 |

| | 1 | 2 | 3 | 4 | 5 | 6 | 7 | 8 | 9 | 10 | 11 | 12 | 13 | 14 | 15 | 16 | 17 | 18 | 19 | 20 | |
|---|---|---|---|---|---|---|---|---|---|---|---|---|---|---|---|---|---|---|---|---|---|
| 10 | -11 | -7 | -10 | -3 | -7 | ∞ | -8 | -7 | -8 | 0 | -4 | -6 | -8 | -2 | -8 | -7 | -8 | -7 | -8 | -6 | 10 |
| 11 | | | | | ∞ | | | 1 | 1 | 5 | 0 | -1 | | | 0 | | -3 | | 6 | 3 | 11 |
| 12 | -7 | 2 | -5 | -9 | 0 | -1 | | 2 | -1 | -8 | 1 | 0 | -9 | 7 | 1 | -4 | ∞ | -7 | 2 | 1 | 12 |
| 13 | 2 | -3 | 10 | 4 | -6 | | | -7 | 1 | 0 | -5 | 0 | 5 | -6 | 2 | -7 | ∞ | -2 | -5 | | 13 |
| 14 | | 4 | 8 | 9 | 2 | | 5 | 1 | 9 | 4 | 3 | 9 | 0 | 2 | 7 | 1 | 3 | ∞ | 3 | | 14 |
| 15 | -3 | 1 | -1 | -5 | 3 | 3 | -8 | -9 | | -5 | 3 | -3 | -5 | 6 | 0 | -7 | 2 | -3 | 6 | ∞ | 15 |
| 16 | -10 | -9 | -6 | | -12 | -1 | ∞ | -13 | -5 | 0 | -11 | -5 | -6 | -12 | 0 | -13 | -11 | -8 | | -11 | 16 |
| 17 | -6 | -11 | -12 | -7 | -9 | 1 | | -11 | -9 | | -11 | ∞ | -10 | -6 | -12 | -9 | 0 | -9 | 0 | -3 | 17 |
| 18 | -12 | 9 | -11 | 1 | -8 | | -9 | -3 | -5 | -1 | 4 | -7 | ∞ | -2 | -8 | -8 | -9 | 0 | 0 | -3 | 18 |
| 19 | | | | | | | | | | | | | | ∞ | | | | | 0 | 6 | 19 |
| 20 | -11 | -10 | -7 | -3 | -13 | 9 | -2 | | -14 | -6 | -2 | -12 | -6 | 11 | ∞ | -1 | -14 | -12 | 3 | 0 | 20 |
| | 1 | 2 | 3 | 4 | 5 | 6 | 7 | 8 | 9 | 10 | 11 | 12 | 13 | 14 | 15 | 16 | 17 | 18 | 19 | 20 | |

We obtain precisely two acceptable cycles by extending two paths. Using a corollary to theorem 1.26, only one cycle remains:

(18 3)(3 18) = (18 18): -8

CYCLE P = [18 6 10 12 17 11 5 2 9 15 20 8 16 4 7 1 3 13 18]: (18 10 17 5 9 20 16 7 3): -8

<p align="center">ACCEPTABLE CYCLES</p>

1. DC 9; -8; (18 10 17 5 9 20 16 7 3).

Before proceeding further, we first obtain a 2-cycle one of whose linking points, say $a$, is contained in the smallest number of cycles. Once we have done this, we place each cycle, $C_{a,i}$ containing $a$ in the top row of an $m \; X \; (\frac{n}{2} + 3)$ matrix, $M_{a,i}$. We then place all acceptable cycles obtained in the succeeding rows according to the number of 2-cycles containing its points. At most one of these points can belong to $C_{a,i}$. Those cycles contained in the same number of 2-cycles are then sorted in increasing order of value. The larger the number of 2-cycles, the smaller its row number in $M_{a,i}$. If $r$ cycles have points (inluding non-linking points) contained in the same 2-cycles, we delete all but one of these cycles. Also, if the above is true except that their values are different, we can delete those with the larger value. We then use our formulas to limit the number of searches necessary to obtain a tour of smaller value than $|T_{UPPERBOUND}|$.

It is best to place a cycle directly into the matrix $M_{a,i}$, sort it according to the number of cycles its points are in and its value, and then immediately delete those that satisfy the conditions of theorem 1.26 or its corollary. However, to show how many cycles can be directly eliminated, we first give the full set of cycles before deleting those that satisfy the conditions of theorem 1.26 or its corollary/

IX.

(1) DC 9; -8; (10 17 5 9 20 16 7 3 18). (2) DC 9; -7; (3 13 10 17 5 9 20 16 7).

(3) DC 9; -2; (5 18 10 12 3 4 8 15 2). (4) DC 9; 0; (2 5 18 10 12 3 4 8 15).

(5) DC 8; 8; (14 3 4 8 10 17 5 9 15). (6) DC 8; 10; (4 8 10 17 5 9 20 19 3)

(7) DC 9; 10; (4 8 10 17 5 9 20 19).

---------------------------------------------------------------------------------------------------------------

VIII

(1) DC 8, 9; -7; (10 17 5 9 20 16 7 1). (2) DC 4, 9; -2; (9 20 16 7 10 12 3 18).

(3) DC 8,9; 2; (12 6 3 4 8 15 2 11). (4) DC 1, 9; 7; (17 6 4 8 15 2 5 18).

---------------------------------------------------------------------------------------------------------

VII

(1) DC 1, 8, 9; -12; (9 20 16 7 10 17 5). (2) DC 2, 8, 9; -3; (4 8 10 17 5 15 1).

(3) DC 2, 9, 10; 0; (6 4 16 5 18 1 17). (4) DC 2, 9, 10; 1; (13 1 4 8 10 17 5).

(5) DC 5, 7, 9; 2; (2 5 18 1 4 8 15).

-----------------------------------------------------------------------------------------------------------------

VI

(1) DC 1, 4, 8, 9; -8; (9 20 16 7 10 17). (2) DC 1, 8, 9, 10; -4; (8 7 10 17 9 11).

(3) DC 1, 4, 8, 9; -3; (9 20 16 7 10 12). (4) DC 2, 8, 9, 10; 0; (1 4 8 10 17 5).

(5) DC 2, 8, 9, 10; 1; (5 1 4 8 10 12). (6) DC 2, 5, 7, 9; 8; (15 5 18 1 4 8).

-----------------------------------------------------------------------------------------------------------------

V

(1) C 1, ,3 , 5, 6, 7; 0; (3 4 8 10 12). (2) C 1, 4, 5, 7, 8; 4; (10 17 5 18 1).

(3) C 1, 5, 6, 8, 9; 4; (10 17 5 18 1). (4) C 2, 3, 6, 7, 10; 7; (12 4 8 15 2).

(5) C 2, 4, 5, 6, 7; 10; (17 6 16 5 2).

-------------------------------------------------------------------------------------------

IV

(1) C 4, 5, 7, 8; -7; (18 10 17 5). (2) C 1, 5, 7, 8; 0; (17 16 13 1). (3) C 2, 4, 6, 10; 1; (9 20 16 5).

(4) C 3, 5, 6, 7; 2; (4 8 10 12). (5) C 4, 5, 7, 8; 2; (12 5 13 10. (6) C 3, 5, 6, 7; 2; (17 6 16 7).

-----------------------------------------------------------------------------------------------------------------------

III

(1) C 3, 5, 7; 6; (17 6 4). (2) C 2, 4, 10; 9; (5 9 15). (3) C 2, 4, 7; 9; (9 11 17).

-----------------------------------------------------------------------------------------------------------------

II

(1) C 2, 4,; 5; (11 9). (2) C 9, 10; 9; (14 15).

Using theorem 1.26 and its corollary, the thirty-seven cycles given above reduces to twenty-six:

1. DC 9; -8; (18 10 17 5 9 20 16 7 3). 2. DC 8; 8; (14 3 4 8 10 17 5 9 15).

3. DC 8, 9; -7; (10 17 5 9 20 16 7 1). 4. DC 1, 9; 7; (17 6 4 8 15 2 5 18).

5. DC 1, 8, 9; -12; (9 20 16 7 10 17 5). 6. DC 2, 8, 9; -3; (4 8 10 17 5 15 1).

7. DC 2, 9, 10; -3; (10 17 5 18 1 4 8). 8. DC 5, 7, 9; 2; (2 5 18 1 4 8 15).

9. DC 1, 4, 8, 9; -8; (9 20 16 7 10 17). 10. DC 1, 8, 9, 10; -4; (8 7 10 17 9 11).

11. DC 2, 8, 9, 10; 0; (1 4 8 10 17 5). 12. DC 1, 5, 6, 8, 9; 5; (7 10 17 5 9 20).

13. DC 2, 4, 8 ,9 ,10; 0; (3 4 8 10 12). 14. DC 2, 3, 6, 9, 10; 4; (10 17 5 18 1).

15. DC 1, 4, 5, 8, 8; 7; (12 4 8 15 2). 16. DC 1, 3, 8, 9, 10; 10; (17 6 16 5 2).

17. C 4, 5, 7, 8; -7; (18 10 17 5). 18. C 1, 5, 7, 8; 0; (17 16 13 1). 19. C 2, 4, 6, 10; 1; (5 9 20 16).

20. C 3, 5, 6, 7; 2; (4 8 10 12). 21. C 4, 5, 7, 8; 2; (12 5 13 10). 22. C 3, 5, 6, 8; 2; (17 6 16 7).

23. C 2, 4, 7; 9; (9 11 17). 24. C 2, 4, 10; 9; (5 9 15). 4.

25. C 2, 4; 5; (11 9): 26. C 9, 10; 9; (14 15).

Only two cycles contain a point in 9: 2. DC 8; 8; (14 3 4 8 10 17 5 9 15) , 26. C 9, 10; 9; (14 15).

2. contains points in each 2-cycle except 8. If it can be linked to another acceptable cycle to form a tour, it must be a 2-cycle containing a point in 8. No 2-cycle satisfies this condition. We therefore must find an acceptable cycle or cycles that can be linked to 26.to obtain a tour.

| | | | 3: 1<br>(1 3) | 7: 3<br>(2 9) | 9: 2<br>(4 7) | 10: 5<br>(5 11) | 11: 4<br>(6 10) | 16: 6<br>(8 16) | 17: 7<br>(12 17) | 18: 8<br>(13 18) | 19: 9<br>(14 19) | 20: 10<br>(15 20) | |
|---|---|---|---|---|---|---|---|---|---|---|---|---|---|
| VALUE | NOT IN 2- | | 1 | 2 | 3 | 4 | 5 | 6 | 7 | 8 | 9 | 10 | |
| 9 | all but 9,10 | 26. | | | | | | | | | 14 | 15 | 26.. |
| -8 | 9 | 1. | 3 | 9 | 7 | 5 | 10 | 16 | 17 | 18 | | 20 | 1. |
| -7 | 8, 9 | 3. | 1 | 9 | 7 | 5 | 10 | 16 | 17 | | | 20 | 3. |
| 7 | 1, 9 | 4. | | 2 | 4 | 5 | 6 | 8 | 17 | 18 | | 15 | 4. |
| -12 | 1,8,9 | 5. | | 9 | 7 | 5 | 10 | 16 | 17 | | | 20 | 5. |
| -3 | 2,8,9 | 6. | 1 | | 4 | 5 | 10 | 8 | 17 | | | 15 | 6. |
| -3 | 2,9,10 | 7. | 1 | | 4 | 5 | 10 | 8 | 17 | 18 | | | 7. |
| 2 | 5,7,9 | 8. | 1 | 2 | 4 | 5 | | 8 | | 18 | | 15 | 8. |
| -8 | 1,4,8,9 | 9. | | 9 | 7 | | 10 | 16 | 17 | | | 20 | 9. |
| -4 | 1,8,9,10 | 10. | | 9 | 7 | 11 | 10 | 8 | 17 | | | | 10. |
| | | | 1 | 2 | 3 | 4 | 5 | 6 | 7 | 8 | 9 | 10 | |

| | | | 1 | 2 | 3 | 4 | 5 | 6 | 7 | 8 | 9 | 10 | |
|---|---|---|---|---|---|---|---|---|---|---|---|---|---|
| 0 | 2,8,9,10 | 11. | 1 | | 4 | 5 | 10 | 8 | 17 | | | | 11. |
| 0 | 2,4,8,9,10 | 12. | 3 | | 4 | | 10 | 8 | 12 | | | | 12. |
| 4 | 2,3,6,9,10 | 13. | | | | | | | | | | | 13. |
| 5 | 1,5,6,8,9 | 14. | | 9 | 7 | 5 | 10 | | 17 | | | 20 | 14. |
| 7 | 1,4,5,8,9 | 15. | | 2 | 4 | | | 8 | 12 | | | 15 | 15. |
| 10 | 1,3,8,9,10 | 16. | | 2 | | 5 | 6 | 16 | 17 | | | | 16. |
| | IN 2- | | | | | | | | | | | | |
| -7 | 4,5,7,8 | 17. | | | | 5 | 10 | | 17 | 18 | | | 17. |

| 0 | 1,5,7,8 | 18. | 1 |   |   |   |   | 16 | 17 | 13 |   |   | 18. |
|---|---------|-----|---|---|---|---|---|----|----|----|---|---|-----|
| 1 | 2,4,6,10 | 19. |   | 9 |   | 5 |   | 16 |    |    |   | 20 | 19. |
| 2 | 3,5,6,7 | 20. |   |   | 4 |   | 10 | 8 | 12 |    |   |   | 20. |
| 2 | 4,5,7,8 | 21. |   |   |   | 5 | 10 |   | 12 | 13 |   |   | 21. |
| 2 | 3,5,6,8 | 22. |   |   | 7 |   |   | 6 | 17 | 16 |   |   | 22. |
| 9 | 2,4,10 | 23. |   | 9 |   | 5 |   |   |    |    |   | 15 | 23. |
| 9 | 2,4,7 | 24. |   | 9 |   | 11 |   |   | 17 |    |   |   | 24. |
| 5 | 2,4 | 25. |   | 9 |   | 11 |   |   |    |    |   |   | 25. |
|   |     |     | 1 | 2 | 3 | 4 | 5 | 6 | 7 | 8 | 9 | 10 |    |

Using $M_{14,1}$,, $1. \oplus 26.$ yields a tour whose value is 1. In linking cycles, we use theorem 1.14. Thus, we define "paths of cycle-trees" with the help of defining at most one "phantom cycle". This allows us to link cycles as though they were arcs of a path. We now see if we can obtain a tour of value no greater than zero. Since any such tour must contain a link of value 9 (26.), Until we link up to cycle 26., our first cycle can have a value no greater than –9. We thus start with 5. whose value is –12. The number of points in a linking point cycle can be no greater than 4, while its value even if it is a phantom cycle can be now greater than 3.

No cycles containing no more than four points and a value no greater than 3 can be linked to 5. . No cycle (phantom or otherwise) other than 5. has a value less than –8. It follows that

$1. \oplus 26.$ yields *[ 18 6 10 12 17 11 5 2 9 15 19 14 20 8 16 4 7 1 3 13 ]* . The latter tour, $T_{57}$, is our new $T_{UPPERBOUND}$.

We new construct 2-circuit cycles in the hope of obtaining a smaller-valued tour.

In order to better understand our method of patching circuits, we demonstrate it for .

[*18* *6* *10* *12* *17* *11* *5* *2* *9* *15* *20* *8* *16* *4* *7* *1* *3* *13* *18*]
$$\downarrow \quad \searrow$$
$$19 \rightarrow 14$$

From this point on, we are interested only in tours of value no greater than 56. Thus, the value in $\sigma_T^{-1} M^-$ of the sum of all cycles in a tour must be no greater than 0. Once we have obtained all cycles using the modified F-W algorithm, we link cycles using all of our formulas:

$$p = \frac{n}{2} + 3t + a - 1, |SUBTREE| < |T| - |\sigma_T|, a + l + 2u \leq N.$$

Here $N$ is the number of linking points in the cycle containing the point $a*$. (We recall that $a$ is the point contained in the smallest number of cycles. However, to avoid confusion with the symbol $a$ as the number of acceptable cycles in a tour, we henceforth use $a*$ to represent a point in the 2-cycle whose points occur in the smallest number of cycles.) From theorem 1.14, we can link cycles such that no partial sum has a sum of values as great as $|T| - |\sigma_T|$.

2-CIRCUIT UNLINKED PATHS

1. C 5, 8; -7; [*6* 18 *10*]. 2. C 5, 7; -7; [*10* 17 *6*]. 3. C 3, 6; -3; [*4* 8 *7*]. 4. C 1, 3; -1; [*7* 1 *4*].

5. C 2, 4; 1; [*2* 5 *9*]. 6. C 2, 10; 2; [*9* 15 *2*]. 7. C 4, 7; 2; [*11* 17 *5*]. 8. C 1, 8; 3; [*3* 13 *1*].

9. C 2, 4; 3; [*5* 2 *11*]; 10. C 2, 4; 3; [*5* 9 *11*]. 11. C 3, 5, 6; -8; [*6* 4 8 *10*].

12. C 3, 5, 7; -2; [*7* 10 12 *4*]. 13. C 2, 4; 10; [*15* 2 11 *20*]. 14. C 1, 3, 5, 6, 7; 1; [*1* 4 8 10 12 *3*].

2-CIRCUIT UNLINKED PATHS

1. DC 1, 4, 8, 9; -5; [*12* 6 4 8 15 2 *17*]. 2. C 2, 4, 5, 7, 8; 6; [*2* 5 13 10 17 *9*].

3. C 2, 3, 5, 6, 7; 6; [*8* 7 10 17 9 *16*]. 4. C 2, 4, 5, 6, 7; 7; [*17* 6 16 5 2 *12*].

5. C 1, 5, 7, 8; -1; [*12* 6 13 1 *17*]. 6. C 1, 3, 5, 7; -4; [*7* 10 12 3 *4*]. 7. C 4, 5, 7 ,8; -3; [*17* 6 13 11 *12*];

8. C 4, 5, 7, 8; -1; [*18* 10 17 5 *13*]. 9. C 1, 5, 7, 8; 0; [*18* 10 12 3 *13*]. 10. C 2, 4, 6, 10; 0; [*20* 16 5 9 *15*].

## 2-CIRCUIT UNLINKED PATHS

1. DC 1, 8, 9; [*17* 6 4 8 15 2 11 *12*]. 2. DC 2, 6, 9, 10; -7; [*7* 10 17 5 18 1 *4*].

3. DC 2, 6, 7, 9, 10; -5; [*6* 4 5 18 1 *10*]. 4. DC 2, 3, 6, 9, 10; 5; [*5* 18 10 12 3 *11*].

5. C 1, 4, 5, 7, 8; 5; [*9* 20 16 5 *2*].

## 2-CIRCUIT UNLINKED PATHS

1. DC 9; -4; [*5* 18 10 12 3 4 8 15 2 *11*]. 2. DC 1, 8, 9; 2; [*20* 16 7 10 17 5 9 *15*].

3. DC 1, 2, 9; 2; [*20* 16 7 10 17 5 18 *15*]. 4. DC 1, 2, 8, 9; 5; [*20* 16 7 10 17 5 18 *15*].

5. DC 2, 3 5, 6, 9; 1; [*3* 13 10 17 5 *1*].

We now present the set of 2-circuit paths of smallest value for each row number heading a path.

1. C 1, 3, 5, 6, 7; 1; [*1* 4 8 10 12 *3*].

2. C 2, 4; 1; [*2* 5 *9*].

3. DC 2, 3, 5, 6, 9; 1; [*3* 13 10 17 5 *1*].

4. C 3, 6; -3; [*4* 8 *7*].

5. DC 9; -4; [*5* 18 10 12 3 4 8 15 2 *11*].

6. C 3, 5, 6; -8; [*6* 4 8 *10*].

7. DC 2, 6, 9, 10; -7; [*7* 10 17 5 18 1 *4*].

8. C 2, 3, 5, 6, 7; 6; [*8* 7 10 17 9 *16*].

9. C 2, 10; 2; [*9* 15 *2*].

10. C 5, 7; -7; [*10* 17 *6*].

11. C 4, 7; 2; [*11* 17 *5*].

12. DC 1, 4, 8, 9; -5; [*12* 6 4 8 15 2 *17*].

15. C 2, 4; 10; [*15* 2 11 *20*].

17. DC 1, 8, 9; -10; [*17* 6 4 8 15 2 11 *12*].

18. C 4, 5, 7, 8; -1; [*18* 10 17 5 *13*]

20. C 2, 4, 6, 10; 0; [*20* 16 5 9 *15*].

<p style="text-align:center">UNLINKED 2-CIRCUIT PATHS</p>

(1) If *( a b )* represents an acceptable path and we wish to use the arc *( b c )* to extend it to an unlinked 2-circuit path, the entry *( a c )* must be ∞.

We now proceed to obtain 2-circuit cycles using the 2-circuit paths we obtained earlier. For easier comprehension, given a fixed column, linked paths are written in italics as they are obtained.

j = 1

(3 1)(1 19) = (3 19): 10

j = 2

(9 2)(2 11) = (9 11): 2; (9 2)(2 12) = (9 12): 7; (9 2)(2 17) = (9 17): 10

j = 3

(1 3)(3 13) = (1 13): 3; (1 3)(3 18) = (1 18): 4

j = 4

(7 4)(4 8) = (7 8): -11; (7 4)(4 15) = (7 15): 5

j = 5

(9 5)(5 13) = (9 13): 11; (9 5)(5 18) = (9 18): 3; (11 5)(5 1) = (11 1): 7;

(11 5)(5 2) = (11 2): 5; (11 5)(5 9) = (11 9): 1; (11 5)(5 13) = (11 13): 9;

(11 5)(5 18) = (11 18): 3

j = 6

(10 6)(6 3) = (10 3): -11; (10 6)(6 4) = (10 4): -15; (10 6)(6 13) = (10 13): -15;

(10 6)(6 16) = (10 16): -10; (10 6)(6 18) = (10 18): -13

j = 7

(4 7)(7 1) = (4 1): -6; (4 7)(7 3) = (4 3): -5; (4 7)(7 10) = (4 10): -7; (4 7)(7 17) = (4 17): 2

j = 8

(7 8)(8 15) = (7 15): 4

j = 9

(2 9)(9 15) = (2 15): 2; (2 9)(9 20) = (2 20): 3; (11 9)(9 15) = (11 15): 2;

(11 9)(9 20) = (11 20): 3

j = 10

(4 10)(10 12) = (4 12): -13; (4 10)(10 17) = (4 17): -15; (6 10)(10 12) = (6 12): -14;

(6 10)(10 17) = (6 17): -16

j = 11

(9 11)(11 8) = (9 8): <u>3</u>; (9 11)(11 12) = (9 12): 3; (9 11)(11 17) = ( 17): 3

[1 4 8 10 12 3]     [5 18 10 12 3 4 8 15 2 11]

*(1 11)(11 17) = (1 17): 8;  (5 11)(11 9) = (5 9): <u>-3</u>*

j = 12

(4 12)(12 3) = (4 3): <u>-8</u>; (4 12)(12 5) = (4 5): <u>-13</u>; (4 12)(12 9) = (4 9): <u>-6</u>;

(4 12)(12 13) = (4 13): 2;  (9 12)(12 3) = (9 3): <u>8</u>; (9 12)(12 4) = (9 4) <u>11</u>;

(9 12)(12 6) = (9 6): <u>2</u>; (17 12)(12 3) = (17 3): <u>-5</u>; (17 12)(12 13) = (17 13): 5

[17 6 4 8 15 2 11 12]

*(17 12)(12 5) = (17 5): <u>-10</u>; (17 12)(12 7) = (17 7): <u>6</u>; (17 12)(12 9) = (17 9): <u>-3</u>*

j = 13

(1 13)(13 11) = (1 11): <u>6</u>; (4 13)(13 11) = (4 11): <u>5</u>; (10 13)(13 1) = (10 1): <u>-13</u>;

(10 13)(13 11) = (10 11): <u>-12</u>; (10 13)(13 14) = (10 14): 6; (18 13)(13 1) = (18 1): <u>1</u>

j = 14

(10 14)(14 15) = (10 15): 9

j = 15

(2 15)(15 14) = (2 14): <u>8</u>; (7 15)(15 2) = (7 2): <u>5</u>; (7 15)(15 14) = (7 14): <u>10</u>;

(10 15)(15 2) = (10 2): <u>10</u>; (11 15)(15 14) = (11 14): <u>8</u>; (20 15)(15 14) = (20 14): <u>6</u>

j = 16

(8 16)(16 5) = (8 5): <u>7</u>; (10 16)(16 5) = (10 5): <u>-9</u>; (10 16)(16 7) = (10 7): <u>-11</u>

j = 17

(4 17)(17 5) = (4 5): <u>-14</u>; (4 17)(17 9) = (4 9): <u>-11</u>; (6 17)(17 5) = (6 5): <u>-15</u>;

(6 17)(17 9) = (6 9): <u>-12</u>; (6 17)(17 11) = (6 11): <u>2</u>; (12 17)(17 5) = (12 5): <u>-4</u>

j = 18

(9 18)(18 1) = (9 1): <u>6</u>; (9 18)(18 10) = (9 10): <u>4</u>; (10 18)(18 8) = (10 8): <u>3</u>;

(10 18)(18 15) = (10 15): <u>1</u>; (11 18)(18 1) = (11 1): <u>4</u>; (11 18)(18 10) = (11 10): <u>2</u>

j = 20

(2 20)(20 16) = (2 16): <u>2</u>; (2 20)(20 19) = (2 19): <u>6</u>; (11 20)(20 19) = (11 19): <u>6</u>;

(15 20)(20 16) = (15 16): <u>9</u>

## 2-CIRCUIT UNLINKED CYCLES

1. DC 2, 9, 10; -10; (*7* 10 17 5 18 1 *4* 8) 2. DC 2, 4, 9, 10; 5; (*1* 4 8 10 12 *3* 13)

3. C 3, 5, 6, 7; -5; (*4* 8 *7* 10 12 5) 4. C 3, 5, 6, 7; -15; (*6* 4 8 *10* 17) 5. C 2, 4, 7, 10; 7;

(*9* 15 *2* 11 17) 6. C 2, 4, 7, 10; 10; (*9* 15 *2* 11 12) 7. C 5, 7, 8; -14; (*10* 17 *6* 13)

8. C C 2, 4, 10; 3; (*2* 5 *9* 15) 9. C 2, 4, 10; 3; (*2* 5 *9* 20) 10. C 2, 4, 10; 3; (*9* 15 *2* 5)

11. C 2, 4, 10; 3; (*9* 15 *2* 11) 12. C 2, 4, 7; 5; (*11* 17 *5* 9) 13. C 4, 7, 8; 10; (*11* 17 *5* 18)

## 2-CIRCUIT LINKED PATHS

1. DC 9; -3; [5 18 10 12 3 4 8 15 *2* 11 *9*]. 2. DC 9; -3; [5 18 10 *12* 3 4 8 15 2 11 *17*].

3. DC 1, 8, 9; -10; [17 6 4 8 15 2 *11* 12 *5*]. 4. DC 1, 8, 9; -3; [17 6 4 8 15 *2* 11 12 *9*].

5. DC 1, 8, 9; 6; [17 6 *4* 8 15 2 11 12 *7*]. 6. DC 1, 4, 8, 9; -1; [12 6 4 8 15 *2* 17 *9*].

7. DC 2, 8, 9, 10; 8; [1 4 8 10 *12* 3 11 *17*]. 8. C 2, 4, 6 10; 1; [20 16 5 *9* 15 *2*].

9. C 2, 4; 5; [2 *5* 9 *11*].

## 2-CIRCUIT CYCLES OBTAINED FROM LINKED PATHS

1. DC 9; -2; (5 18 10 *12* 3 4 8 15 2 11 *17*). 2. DC 9; 11; [17 6 *4* 8 15 2 11 12 *7*].

$P_{100}$ *2–CIRCUIT UNLINKED*

|   | 1 | 2 | 3 | 4 | 5 | 6 | 7 | 8 | 9 | 10 | 11 | 12 | 13 | 14 | 15 | 16 | 17 | 18 | 19 | 20 |   |
|---|---|---|---|---|---|---|---|---|---|----|----|----|----|----|----|----|----|----|----|----|---|
| 1 |   |   | 12| 1 |   |   |   | 4 | 11| 8  | 13 | 10 | 3  |    |    |    |    | 3  |    |    | 1 |
| 2 |   |   |   |   | 2 |   |   | 5 |   |    |    |    |    | 15 | 9  | 20 |    |    | 20 | 9  | 2 |
| 3 | 5 |   |   |   | 17|   |   |   | 13|    |    | 3  |    |    |    |    | 10 |    | 1  |    | 3 |
| 4 | 7 |   | 12|   | 17|   | 8 | 4 | 12| 7  | 13 | 10 | 12 |    |    |    | 10 |    |    |    | 4 |
| 5 |   | 15| 12| 3 |   |   |   | 4 |   | 18 | 2  | 10 |    |    | 8  |    |    | 5  |    |    | 5 |
| 6 |   |   |   | 6 | 17|   |   | 4 |   | 8  | 17 | 10 |    |    |    |    | 10 |    |    |    | 6 |
| 7 | 18| 15|   | 1 | 17|   |   | 4 |   | 7  |    |    |    | 15 | 8  |    | 10 | 5  |    |    | 7 |
| 8 |   |   |   |   |   |   | 8 |   | 17| 7  |    |    |    |    | 9  | 10 |    |    |    |    | 8 |

| | 1 | 2 | 3 | 4 | 5 | 6 | 7 | 8 | 9 | 10 | 11 | 12 | 13 | 14 | 15 | 16 | 17 | 18 | 19 | 20 | |
|---|---|---|---|---|---|---|---|---|---|---|---|---|---|---|---|---|---|---|---|---|---|
| 9 | 18 | 15 | 12 | 12 | 2 | 12 | | 11 | | 18 | 2 | 11 | 5 | | 9 | | 11 | 5 | | | 9 |
| 10 | 13 | 15 | | | | 17 | | 18 | | | 13 | | 6 | 13 | 18 | 6 | 10 | 6 | | | 10 |
| 11 | 5 | | | 17 | | | 5 | 18 | | | 5 | 15 | 9 | | 11 | 5 | 20 | 9 | | | 11 |
| 12 | | 15 | | 6 | 17 | 12 | | 4 | | | | | | | 8 | | 2 | | | | 12 |
| 13 | | | | | | | | | | | | | | | | | | | | | 13 |
| 14 | | | | | | | | | | | | | | | | | | | | | 14 |
| 15 | | 15 | | | | | | | 2 | | | | | | | | | | | 11 | 15 |
| 16 | | | | | | | | | | | | | | | | | | | | | 16 |
| 17 | | 15 | 12 | 6 | | 17 | | 4 | | | 2 | 11 | 12 | | 8 | | | | | | 17 |
| 18 | 13 | | | 17 | | | | 18 | | | 5 | | | | 10 | | | | | | 18 |
| 19 | | | | | | | | | | | | | | | | | | | | | 19 |
| 20 | | | | 16 | | | 5 | | | | | 15 | 9 | 20 | | | | | | | 20 |
| | 1 | 2 | 3 | 4 | 5 | 6 | 7 | 8 | 9 | 10 | 11 | 12 | 13 | 14 | 15 | 16 | 17 | 18 | 19 | 20 | |

$\sigma_1^{-1} M^{-}(100)$ 2 − CIRCUIT UNLINKED

| | 1 | 2 | 3 | 4 | 5 | 6 | 7 | 8 | 9 | 10 | 11 | 12 | 13 | 14 | 15 | 16 | 17 | 18 | 19 | 20 | |
|---|---|---|---|---|---|---|---|---|---|---|---|---|---|---|---|---|---|---|---|---|---|
| 1 | | | 1 | 2 | | | | -2 | <u>8</u> | 2 | <u>6</u> | -4 | *3* | | | | *4* | | | | 1 |
| 2 | | | | | 2 | | | <u>6</u> | 1 | | *5* | | | <u>8</u> | *2* | *2* | | | <u>6</u> | *3* | 2 |
| 3 | 1 | | | | -4 | | | | 3 | | | 2 | | | | | -5 | | *10* | | 3 |
| 4 | <u>-6</u> | | <u>-8</u> | | <u>-14</u> | -3 | -4 | <u>-11</u> | -7 | <u>5</u> | -13 | *2* | | | | -15 | | | | | 4 |
| 5 | | -4 | -1 | 0 | | | -4 | | 0 | -4 | -6 | | | -5 | | | 1 | | | | 5 |
| 6 | | | | -8 | <u>-15</u> | | -12 | <u>-12</u> | -8 | <u>2</u> | -14 | | | | | -16 | | | | | 6 |
| 7 | -9 | <u>5</u> | | -7 | -11 | | | -11 | | -4 | | | | <u>10</u> | *4* | | -12 | -10 | | | 7 |
| 8 | | | | | <u>7</u> | | 1 | | -7 | -3 | | | | | 6 | -11 | | | | | 8 |

| | 1 | 2 | 3 | 4 | 5 | 6 | 7 | 8 | 9 | 10 | 11 | 12 | 13 | 14 | 15 | 16 | 17 | 18 | 19 | 20 | |
|---|---|---|---|---|---|---|---|---|---|---|---|---|---|---|---|---|---|---|---|---|---|
| 9 | 6 | 2 | 8 | 11 | *4* | *2* | | *3* | | *4* | *2* | *3* | *11* | | 1 | *3* | *3* | *5* | | | 9 |
| 10 | -13 | 10 | -11 | -15 | -9 | -7 | -11 | *3* | | -12 | | -15 | 6 | 1 | -10 | -8 | -13 | | | | 10 |
| 11 | 4 | | | 2 | | | 1 | *2* | | | *9* | 8 | *2* | | 1 | *3* | 6 | *3* | | | 11 |
| 12 | | -13 | | -9 | -4 | -1 | | -13 | | | | | | -14 | -5 | | | | | | 12 |
| 13 | | | | | | | | | | | | | | | | | | | | | 13 |
| 14 | | | | | | | | | | | | | | | | | | | | | 14 |
| 15 | | 1 | | | | | | | | 1 | | | | | 9 | | | | | 10 | 15 |
| 16 | | | | | | | | | | | | | | | | | | | | | 16 |
| 17 | 7 | -11 | -5 | -7 | | 1 | | -11 | | -11 | -10 | *5* | | -12 | | | | | | | 17 |
| 18 | | | | -8 | | | | -1 | | -1 | | | | -9 | | | | | | | 18 |
| 19 | | | | | | | | | | | | | | | | | | | | | 19 |
| 20 | | | | 0 | | | -1 | | | | | 6 | 0 | -1 | | | | | | | 20 |
| | 1 | 2 | 3 | 4 | 5 | 6 | 7 | 8 | 9 | 10 | 11 | 12 | 13 | 14 | 15 | 16 | 17 | 18 | 19 | 20 | |

j = 1

(4 1)(1 19) = (4 19): 3; (9 1)(1 4) = (9 4): 8; (10 1)(1 19) = (10 19): -4; (11 1)(1 4) = (11 4): 6

j = 3

(4 3)(3 13) = (4 13): -4; (4 3)(3 18) = (4 18): -5; (9 3)(3 7) = (9 7): 10; (17 3)(3 13) = (17 13): -1;

(17 3)(3 18) = (17 18): -2

j = 4

(9 4)(4 16) = (9 16): 6; (10 4)(4 2) = (10 2): 6; (10 4)(4 8) = (10 8): -19; (10 4)(4 9) = (10 9): 10;

(10 4)(4 15) = (10 15): -3; (10 4)(4 16) = (10 16): -17

j = 5

(4 5)(5 1) = (4 1): <u>-9</u>; (4 5)(5 2) = (4 2): <u>-11</u>; (4 5)(5 9) = (4 9): -15; (4 5)(5 13) = (4 13): -7;

(6 5)(5 1) = (6 1): <u>-10</u>; (6 5)(5 2) = (6 2): <u>-12</u>; (6 5)(5 9) = (6 9): -16; (6 5)(5 13) = (6 13): 3;

(10 5)(5 2) = (10 2): <u>-6</u>; (10 5)(5 9) = (10 9): -10; (12 5)(5 1) = (12 1): <u>1</u>; (12 5)(5 13) = (12 13): 3

[17 6 7 8 15 2 11 12 5]

*(17 5)(5 1) = (17 1): <u>-5</u>; (17 5)(5 13) = (17 13): -3*

j = 6

(9 6)(6 3) = (9 3): <u>-2</u>; (9 6)(6 4) = (9 4): <u>-6</u>; (9 6)(6 13) = (9 13): -6; (9 6)(6 16) = (9 16): -1

j = 7

[17 6 7 8 15 2 11 12 7]

*(17 7)(7 3) = (17 3): <u>4</u>; (17 7)(7 10) = (17 10): <u>2</u>*

j = 8

(2 8)(8 7) = (2 7): <u>3</u>; (2 8)(8 10) = (2 10): 10; (9 8)(8 7) = (9 7): <u>4</u>; (10 8)(8 2) = (10 2): <u>-7</u>;

(10 8)(8 15) = (10 15): -20

j = 9

(1 9)(9 15) = (1 15): 9; (1 9)(9 20) = (1 20): 10; (4 9)(9 15) = (4 15): -14; (4 9)(9 20) = (4 20): -13;

(6 9)(9 15) = (6 15): -15; (6 9)(9 20) = (6 20): -14; (10 9)(9 20) = (10 20): -8

[12 6 4 8 15 2 17 9]

*(12 9)(9 11) = (12 11): 3*

j = 10

(2 10)(10 12) = (2 12): 4; (2 10)(10 17) = (2 17): 2; (9 10)(10 12) = (9 12): -2;

(9 10)(10 17) = (9 17): -4

j = 11

(1 11)(11 9) = (1 9): 7; (10 11)(11 9) = (10 9): -11

j = 12

[11 17 5 18 10 12]                    [10 17 6 13 11 12]

*(11 12)(12 3) = (11 3): 1; (11 12)(12 4) = (11 4): 4; (10 12)(12 3) = (10 3): -6;*

*(10 12)(12 4) = (10 4): -3; (10 12)(12 9) = (10 9): -4*

j = 13

(6 13)(13 11) = (6 11): -5; (9 13)913 1) = (9 1): -4

j = 15

(4 15)(15 14) = (4 14): -8; (6 15)(15 14) = (6 14): -9; (6 15)(15 19) = (6 19): 8;

(10 15)(15 2) = (10 2): -19; (10 15)(15 5) = (10 5): -11; (10 15)(15 14) = (10 14): -14

j = 16

(2 16)(16 7) = (2 7): 1; (15 16)(16 7) = (15 7): 8

j = 17

[1 4 8 10 13 3 13 11 17]

*(1 17)(17 9) = (1 9):* <u>11</u>

j = 18

(4 18)(18 11) = (4 11): <u>2</u>

j = 20

(4 20)(20 13) = (4 13): <u>1</u>; (6 20)(20 19) = (6 19): <u>-11</u>

## 2-CIRCUIT UNLINKED CYCLES

1. DC 1, 3, 4, 9; 0; (9 15 2 5 18 10 17). 2. DC 1, 3, 6, 9; 5; (9 15 2 5 18 10 12).

3. C 3, 5, 6, 7; -15; (10 17 6 4 8). 4. C 1, 5, 7, 8; -9; (10 17 6 13 1). 5. C 1, 3, 6; -4; (4 8 7 1).

## 2-CIRCUIT LINKED PATHS

1. DC 9; -3; [17 *6* 4 8 15 2 11 12 3 18 *10*]. 2. DC 9; -2; [17 *6* 4 8 15 2 11 12 3 13 *10*].

3. DC 9; 5; [7 10 17 *5* 18 1 4 8 15 2 *11*]. 4. DC 9; 9; [1 4 *8* 10 12 3 13 11 9 20 *16*].

5. DC 9; 10; [7 10 *17* 5 18 1 4 8 15 2 *12*]. 6. DC 1, 8, 9; -15; [4 *8* 10 17 5 9 20 *16*].

7. DC 1, 8, 9; -14; [4 *8* 7 10 17 5 9 20 *16*]. 8. DC 1, 8, 9; [12 6 4 8 15 *2* 17 5 *9*].

9. DC 1, 8, 9; 1; [6 *4* 8 10 17 5 9 20 *7*]. 10. DC 2, 8, 9, 10; 7; [1 4 8 10 *12* 3 11 *17*].

11. DC 1, 8, 9, 10; 10; [8 7 10 17 *9* 16 5 *2*]. 12. C 4, 5, 7, 8; -11; [10 *17* 6 13 11 *12*].

13. C 4, 5, 6, 7; 9; [10 *17* 6 16 5 *12*].

## 2-CIRCUIT CYCLES OBTAINED FROM LINKED PATHS

1. DC  9; -11; (17 *6* 4 8 15 2 11 12 3 18 *10*).   2. DC 9; -10; [17 *6* 4 8 15 2 11 12 3 13 *10*].

$P_{120}$ *2–CIRCUIT UNLINKED*

|   | 1 | 2 | 3 | 4 | 5 | 6 | 7 | 8 | 9 | 10 | 11 | 12 | 13 | 14 | 15 | 16 | 17 | 18 | 19 | 20 |   |
|---|---|---|---|---|---|---|---|---|---|----|----|----|----|----|----|----|----|----|----|----|---|
| 1 |   |   | 12 | 1 |   |   |   | 4 | 11 | 8 | 13 | 10 | 3 |   | 9 |   |   | 3 |   | 9 | 1 |
| 2 |   |   | 6 | 6 | 2 | 12 | 16 | 11 | 5 | 8 |   | 10 | 6 | 15 | 9 | 20 | 10 | 6 | 20 | 9 | 2 |
| 3 | 5 |   |   |   | 17 |   |   |   | 13 |   |   | 3 |   |   |   | 10 |   | 1 |   |   | 3 |
| 4 | 5 | 5 | 12 |   | 17 |   | 8 | 4 | 5 | 7 | 18 | 10 | 20 | 15 | 9 |   | 10 | 3 | 1 | 9 | 4 |
| 5 |   | 15 | 12 | 3 |   |   |   | 4 |   | 18 | 2 | 10 |   |   | 8 |   |   | 5 |   |   | 5 |
| 6 | 5 | 5 |   | 6 | 17 |   |   | 4 | 5 | 8 | 13 | 10 | 5 | 15 | 9 |   | 10 |   | 15 | 9 | 6 |
| 7 | 18 | 15 |   | 1 | 17 |   |   | 4 |   | 7 |   |   |   | 15 | 8 |   | 10 | 5 |   |   | 7 |
| 8 |   |   |   |   |   | 8 |   | 17 | 7 |   |   |   |   |   | 9 | 10 |   |   |   |   | 8 |
| 9 | 13 | 15 | 6 | 6 | 2 | 12 | 8 | 11 |   | 18 | 2 | 11 | 6 |   | 9 | 6 | 10 | 5 |   |   | 9 |

|    | 1  | 2  | 3  | 4 | 5  | 6  | 7 | 8 | 9  | 10 | 11 | 12 | 13 | 14 | 15 | 16 | 17 | 18 | 19 | 20 |    |
|----|----|----|----|---|----|----|---|---|----|----|----|----|----|----|----|----|----|----|----|----|----|
| 10 | 13 | 15 | 6  | 6 | 15 | 17 |   | 4 | 11 |    | 13 |    | 6  | 15 | 8  | 6  | 10 | 6  | 1  | 9  | 10 |
| 11 | 5  |    |    | 1 | 17 |    |   | 5 | 18 |    |    | 5  | 15 | 9  |    | 11 | 5  | 20 | 9  | 11 |
| 12 | 5  | 15 |    | 6 | 17 | 12 |   | 4 |    |    |    |    |    | 8  |    | 2  |    |    |    | 12 |
| 13 |    |    |    |   |    |    |   |   |    |    |    |    |    |    |    |    |    |    |    | 13 |
| 14 |    |    |    |   |    |    |   |   |    |    |    |    |    |    |    |    |    |    |    | 14 |
| 15 |    | 15 |    |   |    | 8  |   |   |    | 2  |    |    |    |    | 20 |    |    |    | 11 | 15 |
| 16 |    |    |    |   |    |    |   |   |    |    |    |    |    |    |    |    |    |    |    | 16 |
| 17 |    | 15 | 12 | 6 |    | 17 |   | 4 |    | 18 | 2  | 11 | 3  |    | 8  |    |    | 3  |    |    | 17 |
| 18 | 13 |    |    |   | 17 |    |   |   | 18 |    |    |    | 5  |    |    |    | 10 |    |    |    | 18 |
| 19 |    |    |    |   |    |    |   |   |    |    |    |    |    |    |    |    |    |    |    | 19 |
| 20 |    |    |    |   | 16 |    |   | 5 |    |    |    |    |    | 15 | 9  | 20 |    |    |    |    | 20 |
|    | 1  | 2  | 3  | 4 | 5  | 6  | 7 | 8 | 9  | 10 | 11 | 12 | 13 | 14 | 15 | 16 | 17 | 18 | 19 | 20 |    |

$\sigma_1^{-1} M^{-} (120)\ 2-CIRCUIT\ UNLINKED$

|   | 1   | 2    | 3  | 4  | 5    | 6  | 7  | 8    | 9    | 10 | 11 | 12   | 13 | 14  | 15   | 16 | 17   | 18  | 19 | 20   |   |
|---|-----|------|----|----|------|----|----|------|------|----|----|------|----|-----|------|----|------|-----|----|------|---|
| 1 |     |      | 1  | 2  |      |    |    | -2   | 7    | 2  | 6  | -4   | 3  |     | 9    |    |      | 4   |    | 10   | 1 |
| 2 |     |      |    | 2  |      | 1  |    | 1    |      |    |    |      |    | 8   | 2    | 2  |      | -1  | 6  | 3    | 2 |
| 3 | 1   |      |    | -4 |      |    |    |      | 3    |    |    |      | 2  |     |      |    | -5   |     | 10 |      | 3 |
| 4 | -9  | -13  | -8 |    | -14  |    | -3 | -4   | -15  | -7 | 2  | -13  | 1  | -8  | -14  |    | -15  | -5  | 3  | -13  | 4 |
| 5 |     | -4   | -1 | 0  |      |    |    | -4   |      | 0  | -4 | -6   |    |     | -5   |    | 1    |     |    |      | 5 |
| 6 | -10 | -12  |    | -8 | -15  |    |    | -12  | -16  | -8 | -5 | -14  | -8 | -9  | -15  |    | -16  |     | 8  | -14  | 6 |
| 7 | -9  | 5    |    | -7 | -11  |    |    | -11  |      | -4 |    |      |    | 10  | 4    |    | -12  | -10 |    |      | 7 |

| | 1 | 2 | 3 | 4 | 5 | 6 | 7 | 8 | 9 | 10 | 11 | 12 | 13 | 14 | 15 | 16 | 17 | 18 | 19 | 20 | |
|---|---|---|---|---|---|---|---|---|---|---|---|---|---|---|---|---|---|---|---|---|---|
| 8 | | | | *7* | 1 | | -7 | -3 | | | | | | | 6 | -11 | | | | | 8 |
| 9 | <u>-4</u> | 2 | <u>-2</u> | <u>-6</u> | *4* | 2 | <u>*4*</u> | *3* | | *4* | *2* | *3* | *11* | | 1 | -1 | -4 | 5 | | | 9 |
| 10 | -13 | <u>-*19*</u> | -11 | -15 | <u>-*11*</u> | -7 | -11 | -19 | -10 | | -12 | | -15 | <u>-*14*</u> | -20 | -17 | -8 | -13 | -4 | -8 | 10 |
| 11 | *4* | | | 6 | 2 | | | *1* | *2* | | | 9 | 8 | *2* | | 1 | *3* | 6 | *3* | | 11 |
| 12 | <u>1</u> | -13 | | -9 | -4 | -1 | | -13 | | | | | | -14 | | -5 | | | | | 12 |
| 13 | | | | | | | | | | | | | | | | | | | | | 13 |
| 14 | | | | | | | | | | | | | | | | | | | | | 14 |
| 15 | | 1 | | | | | | | 1 | | | | | | *9* | | | | | 10 | 15 |
| 16 | | | | | | | | | | | | | | | | | | | | | 16 |
| 17 | *7* | -11 | -5 | -7 | | 1 | | -11 | | -11 | -10 | -1 | | -12 | | | -2 | | | | 17 |
| 18 | | | | -8 | | | | -1 | | -1 | | | | -9 | | | | | | | 18 |
| 19 | | | | | | | | | | | | | | | | | | | | | 19 |
| 20 | | | | | 0 | | -1 | | | | | *6* | 0 | -1 | | | | | | | 20 |
| | 1 | 2 | 3 | 4 | 5 | 6 | 7 | 8 | 9 | 10 | 11 | 12 | 13 | 14 | 15 | 16 | 17 | 18 | 19 | 20 | |

j = 1

(4 1)(1 4) = (4 4): -7

CYCLE P = [4 <u>16</u> 8 <u>4</u> 7 <u>6</u> 10 <u>12</u> 17 <u>11</u> 5 <u>3</u> 1 <u>7</u> 4]

$P_1$ = (4 <u>16</u> 8), $P_2$ = (7 <u>6</u> 10 <u>12</u> 17 <u>11</u> 5 <u>3</u> 1 <u>7</u> 4)

(4 1)(1 19) = (4 19): 0; (6 1)(1 19) = (6 19): -1; (9 1)(1 19) = (9 19): 5;

(12 1)(1 19) = (12 19): 10

j = 2

(4 2)(2 11) = (4 11): -3; (10 2)(2 11) = (10 11): -19

j = 3

(9 3)(3 13) = (9 13): 0; (9 3)(3 18) = (9 18): 1

j = 4

(9 4)(4 8) = (9 8): -10; (9 4)(4 16) = (9 16): -8

j = 5

(10 5)(5 9) = (10 9): -12

j = 7

(2 7)(7 1) = (2 1): -2; (2 7)(7 3) = (2 3): -1; (2 7)(7 10) = (2 10): -3; (2 7)(7 17) = (2 17): 6

j = 10

(2 10)(10 12) = (2 12): -9; (2 10)(10 17) = (2 17): -11

j = 11

(10 11)(11 19) = (10 19): -7

j = 12

(2 12)(12 3) = (2 3): -4; (2 12)(12 13) = (2 13): 6

j = 14

(6 14)(14 3) = (6 3): -1

$P_{140}$ 2–CIRCUIT UNLINKED

|   | 1 | 2 | 3 | 4 | 5 | 6 | 7 | 8 | 9 | 10 | 11 | 12 | 13 | 14 | 15 | 16 | 17 | 18 | 19 | 20 |   |
|---|---|---|---|---|---|---|---|---|---|----|----|----|----|----|----|----|----|----|----|----|---|
| 1 |   |   | 12 | 1 |   |   |   | 4 | 11 | 8 | 13 | 10 | 3 |   | 9 |   |   | 3 |   | 9 | 1 |
| 2 | 7 |   | 12 | 6 | 2 | 12 | 16 | 11 | 5 | 7 |   | 10 | 12 | 15 | 9 | 20 | 10 | 6 | 20 | 9 | 2 |
| 3 | 5 |   |   |   | 17 |   |   |   |   | 13 |   |   | 3 |   |   |   | 10 |   | 1 |   | 3 |
| 4 | 5 | 5 | 12 |   | 17 |   | 8 | 4 | 5 | 7 | 2 | 10 | 20 | 15 | 9 |   | 10 | 3 | 1 | 9 | 4 |
| 5 |   | 15 | 12 | 3 |   |   |   | 4 |   | 18 | 2 | 10 |   |   | 8 |   |   | 5 |   |   | 5 |
| 6 | 5 | 5 | 14 | 6 | 17 |   |   | 4 | 5 | 8 | 13 | 10 | 5 | 15 | 9 |   | 10 |   | 1 | 9 | 6 |
| 7 | 18 | 15 |   | 1 | 17 |   |   | 4 |   | 7 |   |   | 15 | 8 |   | 10 | 5 |   |   |   | 7 |

| | | | | | | | | | | | | | | | | | | | | |
|---|---|---|---|---|---|---|---|---|---|---|---|---|---|---|---|---|---|---|---|---|
| 8 | | | | | | 8 | | 17 | 7 | | | | | 9 | 10 | | | | | 8 |
| 9 | 13 | 15 | 6 | 6 | 2 | 12 | 8 | 4 | | 18 | 2 | 11 | 3 | | 9 | 4 | 10 | 3 | 1 | 9 |
| 10 | 13 | 15 | 6 | 6 | 15 | 17 | | 4 | 5 | | 2 | | 6 | 15 | 8 | 6 | 10 | 6 | 11 | 9 | 10 |
| 11 | 5 | | | 1 | 17 | | | | 5 | 18 | | | 5 | 15 | 9 | | 11 | 5 | 20 | 9 | 11 |
| 12 | 5 | 15 | | 6 | 17 | 12 | | 4 | | | | | | | 8 | | 2 | | 1 | | 12 |
| 13 | | | | | | | | | | | | | | | | | | | | | 13 |
| 14 | | | | | | | | | | | | | | | | | | | | | 14 |
| 15 | | 15 | | | | 8 | | | | 2 | | | | | | 20 | | | | 11 | 15 |
| 16 | | | | | | | | | | | | | | | | | | | | | 16 |
| 17 | | 15 | 12 | 6 | | 17 | | 4 | | 18 | 2 | 11 | 3 | | 8 | | | 3 | | | 17 |
| 18 | 13 | | | | 17 | | | | | 18 | | | 5 | | | | 10 | | | | 18 |
| 19 | | | | | | | | | | | | | | | | | | | | | 19 |
| 20 | | | | | 16 | | | | 5 | | | | | 15 | 9 | 20 | | | | | 20 |
| | 1 | 2 | 3 | 4 | 5 | 6 | 7 | 8 | 9 | 10 | 11 | 12 | 13 | 14 | 15 | 16 | 17 | 18 | 19 | 20 | |

$$\sigma_1^{-1}M^{-}(140)\ 2-CIRCUIT\ UNLINKED$$

| | 1 | 2 | 3 | 4 | 5 | 6 | 7 | 8 | 9 | 10 | 11 | 12 | 13 | 14 | 15 | 16 | 17 | 18 | 19 | 20 | |
|---|---|---|---|---|---|---|---|---|---|---|---|---|---|---|---|---|---|---|---|---|---|
| 1 | | | 1 | 2 | | | -2 | *7* | 2 | 6 | -4 | *3* | | 9 | | | 4 | | 10 | 1 |
| 2 | -2 | | -4 | | 2 | | *1* | | 1 | -3 | | -9 | 6 | *8* | 2 | 2 | -11 | -1 | 6 | *3* | 2 |
| 3 | 1 | | | -4 | | | | | 3 | | | 2 | | | | | -5 | | 10 | | 3 |
| 4 | -9 | -13 | -8 | | -14 | -3 | -4 | -15 | -7 | -3 | -13 | *1* | -8 | -14 | | -15 | -5 | *0* | -13 | 4 |
| 5 | | -4 | -1 | 0 | | | -4 | | 0 | -4 | -6 | | | -5 | | | 1 | | | 5 |
| 6 | -10 | -12 | -1 | -8 | -15 | | | -12 | -16 | -8 | -5 | -14 | -8 | -9 | -15 | | -16 | | -1 | -14 | 6 |

| | 1 | 2 | 3 | 4 | 5 | 6 | 7 | 8 | 9 | 10 | 11 | 12 | 13 | 14 | 15 | 16 | 17 | 18 | 19 | 20 | |
|---|---|---|---|---|---|---|---|---|---|---|---|---|---|---|---|---|---|---|---|---|---|
| 7 | -9 | 5 | | -7 | -11 | | | -11 | | -4 | | | | 10 | 4 | | -12 | -10 | | | 7 |
| 8 | | | | 7 | | 1 | | -7 | -3 | | | | | | 6 | -11 | | | | | 8 |
| 9 | -4 | 2 | -2 | -6 | 4 | 2 | 4 | -10 | | 4 | 2 | 3 | 3 | | 1 | -8 | -4 | 1 | 5 | | 9 |
| 10 | -13 | -19 | -11 | -15 | -11 | -7 | -11 | -19 | -12 | | -19 | | -15 | -14 | -20 | -17 | -8 | -13 | -7 | -8 | 10 |
| 11 | 4 | | | 6 | 2 | | | 1 | 2 | | | 9 | 8 | 2 | | 1 | 3 | 6 | 3 | | 11 |
| 12 | 1 | -13 | | -9 | -4 | -1 | | -13 | | | | | | -14 | | -5 | | 10 | | | 12 |
| 13 | | | | | | | | | | | | | | | | | | | | | 13 |
| 14 | | | | | | | | | | | | | | | | | | | | | 14 |
| 15 | | 1 | | | | | | | 1 | | | | | | 9 | | | | | 10 | 15 |
| 16 | | | | | | | | | | | | | | | | | | | | | 16 |
| 17 | 7 | -11 | -5 | -7 | | 1 | | -11 | | | -11 | -10 | -1 | | -12 | | | -2 | | | 17 |
| 18 | | | | -8 | | | | | -1 | | | -1 | | | | -9 | | | | | 18 |
| 19 | | | | | | | | | | | | | | | | | | | | | 19 |
| 20 | | | | | 0 | | | -1 | | | | | | 6 | 0 | -1 | | | | | 20 |
| | 1 | 2 | 3 | 4 | 5 | 6 | 7 | 8 | 9 | 10 | 11 | 12 | 13 | 14 | 15 | 16 | 17 | 18 | 19 | 20 | |

We have only three paths that can be extended: (2 1): -2; (2 3): -4; (6 3): -1. In order to save space, we examine each of these possibilities using rooted trees, $\sigma_1^{-1}M^-$ (140) $2-CIRCUIT\ UNLINKED$ and $\sigma_1^{-1}M^-$.

(2 1)(1 10) doesn't yield a path smaller than one obtained earlier. We now examine (2 3):

$$
\begin{array}{c}
\mathbf{2} \\
\downarrow \\
5 \\
\downarrow \\
\mathbf{9} \to 20 \to 16 \to 7 \to 12 \to 3 \\
\downarrow \underline{-2} \\
\underline{\underline{13}}
\end{array}
$$

$$
\begin{array}{c}
\mathbf{6} \\
\downarrow \\
4 \\
\downarrow \\
8 \\
\downarrow \\
\mathbf{10} \to 17 \to 5 \to 9 \to 15 \to 14 \to 3 \\
\downarrow \underline{2} \\
\underline{\underline{18}}
\end{array}
$$

We thus cannot obtain any further cycles.

### 2-CIRCUIT UNLINKED CYCLES

1. DC 2, 14, 19; -7; (*4* 8 *7* 10 17 5 1).

### 2-CIRCUIT LINKED PATHS

1. DC 9; 11; [9 *15* 2 11 12 6 3 13 1 19 *20*]. 2. DC 8, 9; 5; [12 *6* 4 8 15 2 17 5 1 *10*].

3. DC 1, 8, 9; -18; [10 *17* 6 4 8 15 2 11 *12*]. 4. DC 1, 4, 8, 9; -14; [10 *17* 6 4 8 15 2 *12*].

### 2-CIRCUIT LINKED CYCLE OBTAINED FROM LINKED PATH

1. DC 8, 9; -1; (12 *6* 4 8 15 2 17 5 1 *10*).

Before going on, it must be noted that we haven't fully followed the idea of the modified F-W algorithm. Namely, by choosing the smallest 2-circuit unlinked path in each row to check for unlinked 2-circuit cycles, we arbitrarily chose the smallest-valued unlinked path in each row. This doesn't guarantee that we will obtain the smallest possible unlinked 2-circuit cycle with a particular row value as a determining vertex. In order to do so, we would have had to apply the F-W algorithm to *each* possible unlinked 2-circuit path. Let $n_i$ be the number of acceptable unlinked 2-circuit paths obtained in row $i$, while $MAX(UNLINKED) = \max\{n_i \mid i = 1, 2, ..., n\}$. In order to obtain all such possibilities, we would have had to apply the F-W algorithm to $MAX(UNLINKED)$ cost matrices whose entries consist of unlinked 2-circuit paths. Similarly, in order to obtain *all* possibilities for linked 2-circuit cycles, we would have had to apply the algorithm to $MAX(LINKED)$ cost matrices. We now add 2-circuit cycles to the set of acceptable cycles obtained earlier. Unlike acceptable cycles, 2-circuit cycles have the property that certain numbers in 2-cycles do not yield edges that correspond to 2-cycles of $\sigma_1$. Thus, the two numbers in such a 2-cycle belong to *different* circuits. Such numbers are either printed in boldface or in italics. Assume that $C$ is a 2-circuit cycle. Let $p$ be the number of points in $C$, while $N$ is the number of its linking points. Then, if $C$ is an unlinked cycle, $N = p - 2$; if $C$ is linked, then $N = p - 4$ implying if $C$ is unlinked, $p \geq 4$, linked, $p \geq 5$.

To recapitulate, a *linking point* of a cycle, $C$, is the only point of a 2-cycle of $\sigma_1$ contained in $C$. This distinguishes it from a *non-linking point* of $C$ which is one of the two points of a 2-cycle of $\sigma_1$ contained in $C$. DC represents "doesn't contain", C, "contains". Each Roman numeral represents $\frac{n}{2}$ - (the number of linking point 2-cycles in $C$). A number in italics and before C or DC represents a 2-cycle that contains no linking points. When used in patching cycle Q to cycle Q*, the non-linking points must always be 2-cycles that have no linking points in Q*. The numbers in boldface directly after C or DC represent 2-cycles each of which contains a linking point. A *linking point 2-cycle* is a 2-cycle containing a linking point. We denote the patching of cycles $C_1$ and $C_2$ by $C_1 \oplus C_2$.

Each cycle in $S$ contains at most one point belonging to a 2-cycle, say $t$, where another cycle of $S$ contains a point in $t$. Furthermore, $t$ must belong to a *linking edge* (2-cycle) of the two cycles.

Theorem 1.29 *While constructing a tour, assume that we have patched $j+1$ cycles to an initial cycle $C_i$ to obtain a tree, $T_{j+1}$, formed from a set of linked cycles $C_{j+1}$ linked to $C_i$. Let $N_j$ represent the number of 2-cycles of $\sigma_1$ that contain no points in $C_i$, while $N_{j+1}$ is the number of 2-cycles in $\sigma_1$ that contain at least one point in $C_{j+1}$ but none in $C_i$. Then the respective number of remaining acceptable, unlinked and linked cycles needed to obtain a tour satisfies the inequality $a_{j+} + u_{j+} + 2l_{j+} \leq N_{j+1}$.*

*Furthermore, the number of cycles needed to construct a tour is no greater than $a_{j+1} + u_{j+1} + l_{j+1} + N_1 - N_{j+1} - l_{j+1} + 1$. The addition of 1 signifies the inclusion of $C_i$.*

Proof. Let $C_1, C_2, \ldots, C_j$ be the sequence of patching of $j$ cycles added one at a time to a tree of cycles $T_j$ whose root is $C_i$. If $N_j$ represents the number of 2-cycles of $\sigma_1$ that contain no points in $C_j$, then $SN_j$ is the set of 2-cycles containing no points in $C_j$. We now note the following: if $C_a$ is an acceptable cycle that can be patched to $C_j$, then it contains one point in $C_j$ and *at least* one point in a 2-cycle of $SN_j$. Next, assume that $C_u$ is an unlinked 2-circuit cycle that can be patched to $C_j$. Then it has two points that are patched to $C_j$ and *at least* one 2-cycle and perhaps one or more points that belong to $SN_j$. Finally, if $C_l$ is a linked 2-circuit cycle, two pairs of its points are 2-cycles that belong to $SN_j$ and possibly more points also do so, while one point belongs to a 2-cycle in $C_j$. More generally, it follows that if $C_{j+1}$ is the patching following $C_j$ after patching one of the cycles described above, the following inequality holds: $a_{j+1} + u_{j+1} + 2l_{j+1} \leq N_{j+1}$. Here $a_{j+1}$, $u_{j+1}$ and $l_{j+1}$ are the respective number of acceptable cycles, unlinked and linked cycles used in the first $j+1$ patchings to form $T_{j+1}$. Denote by $a_{j+}, u_{j+}, l_{j+}$, the respective number of cycles remaining that can be used to construct a tour by patching. Then $a_{j+} + u_{j+} + 2l_{j+} \leq N_1 - N_{j+1}$. We have already patched $a_{j+1} + u_{j+1} + l_{j+1}$ cycles. We have only $N_1 - N_{j+1}$ linking point 2-cycles left that contain no points in $C_{j+1}$. It follows that the maximum number of cycles that can exist in a tour is no greater than

$a_{j+1} + u_{j+1} + l_{j+1} + N_1 - N_{j+1} - l_{j+1} + 1$

*Note 1*. If $C_i$ is an unlinked cycle, then it yields *two* circuits that can't be linked. If a tour exists containing it, there must also exist a patching of cycles containing two linking points – one each to a linking point of each circuit. If we can find no satisfactory cycle of this nature, then no tour exists containing $C_i$.

*Note 2*. We always use the value of the smallest-valued tour obtained thus far as an upper bound for values of any new tour obtained.. When we obtain a tour of smaller value, we substitute its value for the previous one.

*Note 3.* If we obtain a patched set of cycles that contains at least one point in each 2-cycle of $\sigma_1$ but whose set of points doesn't satisfy the formula for the number of points in a set of cycles that yields a tour, $p = \frac{n}{2} + 3t + a - 1$, then we have obtained a derangement, i.e., a set of pair-wise disjoint circuits containing all $n$ points. Here $t$ is the number of 2-circuit cycles, $a$, the number of acceptable cycles. An interesting question is: Using this portion of our F-W algorithm, are there criteria for a derangement to be a lower bound for an optimal tour? There is one obvious case where that is true: if we obtain a minimal-valued derangement consisting of a single cycle, then it must yield an optimal tour. By definition, the F-W algorithm yields the smallest-valued path between two points. Perhaps other cases can be discovered.

Here we have a new cycle containing a point in 9:  DC 1, 8;  10, (4 8 10 17 5 9 20 19).

1. (1 3); 2: (2 9); 3: (4 7); 4: (5 11); 5: (6 10); 6: (8 16); 7: (12 17); 8: (13 18) ; 9: (14 19); 10: (15 20).

No 2-circuit cycle contains a point in 9. Thus, we simply place our new 2-circuit cycles in $M_{26,1}$ and renumber its entries after using theorem 1.26 and its corollary to delete cycles.  Also, X denotes a pair of non-linking points.

Underlining of a 2-cycle means that the 2-cycle contains a point of one of the two circuits formed from an unlinked 2-circuit cycle. We only do it after the symbol $\mathbf{C}$.

1. *5, 7* DC 9; -11; (17 *6* 4 8 15 2 11 12 3 18 *10*).

2. DC 9; -8; (10 17 5 9 20 16 7 3 18). 3. *4, 7* DC 9; -2; (5 18 10 *12* 3 4 8 15 2 11 *17*).

4. DC 8; 8; (14 3 4 8 10 17 5 9 15). 5. DC 8, 9; -7; (10 17 5 9 20 16 7 1).

6. DC 4, 9; -2; (9 20 16 7 10 12 3 18). 7. *5, 7* DC 8, 9; -1; (12 *6* 4 8 15 2 17 5 1 *10*);

8. DC 1, 9; 7; (17 6 4 8 15 2 5 18). 9. DC 1, 8; 10; (4 8 10 17 5 9 20 19).

10. DC 1, 8, 9; -12; (9 20 16 7 10 17 5). 11. *3* DC 2, 9, 10; -10; (*7* 10 17 5 18 1 *4* 8).

12. DC 2, 8, 9; -3; (4 8 10 17 5 15 1). 13. DC 2, 9, 10; 0; (6 4 16 5 18 1 17).

14. DC 5, 7, 9; 2; (2 5 18 1 4 8 15). 15. *3, 7* DC 1, 8, 9; 11; (17 6 *4* 8 15 2 11 12 *7*).

16. DC 1, 4, 8, 9; -8; (9 20 16 7 10 17). 17. *3* DC 2, 8, 9, 10; -7; (*4* 8 *7* 10 17 5 1).

18. DC 1, 8, 9, 10; -4; (8 7 10 17 9 11). 19. DC 2, 8, 9, 10; 0; (1 4 8 10 17 5).

20. *2* DC 1, 3, 6, 9; 0; (*9* 15 *2* 5 18 10 17). 21. DC 1, 6, 8, 9; 5; (7 10 17 5 9 20).

22. *1* DC 2, 4, 9, 10; 5; (*1* 4 8 10 12 *3* 13). 23. DC 2, 5, 7, 9; 8; (15 5 18 1 4 8).

24. C 1, ,3 , 5, 6, 7; 0; (3 4 8 10 12). 25. C 1, 4, 5, 7, 8; 4; (10 17 5 18 1).

26. C 2, 3, 6, 7, 10; 7; (12 4 8 15 2). 27. C 2, 4, 5, 6, 7; 10; (17 6 16 5 2).

28. C 4, 5, 7, 8; -7; (18 10 17 5). 29. *3* C 4, 5, 6, 7; -5; (*4* 8 *7* 10 12 5).

30. C 1, 5, 7, 8; 0; (17 16 13 1). 31. C 2, 4, 6, 10; 1; (9 20 16 5). 32. C 3, 5, 6, 7; 2; (4 8 10 12).

33. *5* C 3, 6, 7; -15; (*6* 4 8 *10* 17). 34. *5* C 1, 7, 8; -9; (*10* 17 *6* 13 1). 35. C 3, 5, 7; 6; (17 6 4).

36. C 2, 4, 10; 9; (5 9 15). 37. C 2, 4, 7; 9; (9 11 17). 38. *5* C 7, 8; -14; (*10* 17 *6* 13).

39. *3* C 1, 6; -4; (4 8 *7* 1). 40. *2* C 4, 10; 3; (*2* 5 *9* 15). 41. *4* C 2, 7; 5; (*11* 17 *5* 9).

42. C 2, 4,; 5; (11 9). 43. C 9, 10; 9; (14 15). 44. *4* C *7*, 8; 10; (*11* 17 *5* 18).

| VALUE | NOT IN 2- | | 3: 1 (1 3) | 7: 3 (2 9) | 9: 2 (4 7) | 10: 5 (5 11) | 11: 4 (6 10) | 16: 6 (8 16) | 17: 7 (12 17) | 18: 8 (13 18) | 19: 9 (14 19) | 20: 10 (15 20) | |
|---|---|---|---|---|---|---|---|---|---|---|---|---|---|
| | | | 1 | 2 | 3 | 4 | 5 | 6 | 7 | 8 | 9 | 10 | |
| 8 | 8 | 4. | 3 | 9 | 4 | 5 | 10 | 8 | 17 | | 19 | 20 | |
| 9 | all but | 43. | | | | | | | | | 14 | 15 | 42. |
| 10 | 1,8 | 9. | | 9 | 4 | 5 | 10 | 8 | 17 | | 19 | 20 | |
| -11 | 9 | 1. | 3 | 2 | 4 | 11 | X | 8 | X | 18 | | 15 | 1. |
| -8 | 9 | 2. | 3 | 9 | 7 | 5 | 10 | 16 | 17 | 18 | | 20 | 2. |
| -2 | 9 | 3. | 3 | 2 | 4 | X | 10 | 8 | X | 18 | | 15 | 3. |
| -7 | 8,9 | 5. | 1 | 9 | 7 | 5 | 10 | 16 | 17 | | | 20 | 3. |
| -2 | 4,9 | 6. | 3 | 9 | 7 | | 10 | 16 | 12 | 18 | | 20 | |
| -1 | 8,9 | 7. | 1 | 2 | 4 | 5 | X | 8 | X | | | 15 | |
| 7 | 1,9 | 8. | | 2 | 4 | 5 | 6 | 8 | 17 | 18 | | 15 | 4. |
| -12 | 1,8,9 | .10. | | 9 | 7 | 5 | 10 | 16 | 17 | | | 20 | 6. |
| -10 | 2,9,10 | 11. | 1 | | X | 5 | 10 | 8 | 17 | 18 | | | 7. |
| -3 | 2,8,9 | 12. | 1 | | 4 | 5 | 10 | 8 | 17 | | | 15 | 8. |
| 0 | 2,9,10 | 13. | 1 | | 4 | 5 | 6 | 16 | 17 | 18 | | | 9. |
| 2 | 5,7,9 | 14. | 1 | 2 | 4 | 5 | | 8 | | 18 | | 15 | 10. |
| 11 | 1,8,9 | 15. | | 2 | X | 11 | 6 | 8 | X | | | 15 | 11. |
| -8 | 1,4,8,9 | 16. | | 9 | 7 | | 10 | 16 | 17 | | | 20 | 12. |
| | | | 1 | 2 | 3 | 4 | 5 | 6 | 7 | 8 | 9 | 10 | |

| | | | 1 | 2 | 3 | 4 | 5 | 6 | 7 | 8 | 9 | 10 | |
|---|---|---|---|---|---|---|---|---|---|---|---|---|---|
| -7 | 2,8,9,10 | 17. | 1 | -7 | X | 5 | 10 | 8 | 17 | | | | 13. |
| -4 | 1,8,9,10 | 18. | | 9 | 7 | 11 | 10 | 8 | 17 | | | | 14. |
| 0 | 1,3,6,9 | 19. | | X | | 5 | 10 | | 17 | 18 | | 15 | 15. |
| 0 | 2,8,9,10 | 20. | 1 | | 4 | 5 | 10 | 8 | 17 | | | | 16. |

|   |   |   | 1 | 2 | 3 | 4 | 5 | 6 | 7 | 8 | 9 | 10 |   |
|---|---|---|---|---|---|---|---|---|---|---|---|----|---|
|   |   |   | 1 | 2 | 3 | 4 | 5 | 6 | 7 | 8 | 9 | 10 |   |
| 5 | 1,6,8,9 | 21. |   | 9 | 7 | 5 | 10 |   | 17 |   |   | 20 |   |
| 5 | 2,4,9,10 | 22. | X |   | 4 |   | 10 | 8 | 12 | 13 |   |   |   |
| 8 | 2,5,7,9 | 23. | 1 |   | 4 | 5 |   | 8 |   | 18 |   | 15 |   |
| 0 | 2,4,8,9,10 | 24. | 3 |   | 4 |   | 10 | 8 | 12 |   |   |   | 17 |
| 4 | 2,3,6,9,10 | 25. |   |   |   | 5 | 10 |   | 17 | 18 |   |   | .18. |
| 7 | 1,4,5,8,9 | 26. |   | 2 | 4 |   |   | 8 | 12 |   |   | 15 | 20. |
| 10 | 1,3,8,9,10 | 27. |   | 2 |   | 5 | 6 | 16 | 17 |   |   |   | 21. |
|   | IN |   |   |   |   |   |   |   |   |   |   |   |   |
| -7 | 4,5,7,8 | 28. |   |   |   | 5 | 10 |   | 17 | 18 |   |   | 22. |
| -5 | 4,5,6,7 | 29. |   |   | X | 5 | 10 | 8 | 12 |   |   |   | 23. |
| 0 | 1,5,7,8 | 30. | 1 |   |   |   |   | 16 | 17 | 13 |   |   | 24. |
| 1 | 2,4,6,10 | 31. |   | 9 |   | 5 |   | 16 |   |   |   | 20 | 25. |
| 2 | 3,5,6,7 | 32. |   |   | 4 |   | 10 | 8 | 12 |   |   |   | 26. |
| -15 | 3,6,7 | 33. |   |   | 4 |   | X | 8 | 17 |   |   |   | 30. |
| -9 | 1,7,8 | 34. | 1 |   |   |   | X |   | 17 | 13 |   |   | 31. |
| 6 | 3,5,7 | 35. |   |   | 4 |   | 6 |   | 17 |   |   |   | 32. |
| 9 | 2,4,7 | 36. |   | 9 |   | 11 |   |   | 17 |   |   |   | 33. |
| 9 | 2,4,10 | 37. |   | 9 |   | 5 |   |   |   |   |   | 15 | 34. |
| -14 | 7,8 | 38. |   |   |   |   | X |   | 17 | 13 |   |   | 35. |
|   |   |   | 1 | 2 | 3 | 4 | 5 | 6 | 7 | 8 | 9 | 10 |   |
|   |   |   | 1 | 2 | 3 | 4 | 5 | 6 | 7 | 8 | 9 | 10 |   |
| -4 | 1,6 | 39. | 1 |   | X |   |   | 8 |   |   |   |   | 36. |
| 3 | 4,10 | 40. |   | X |   | 5 |   |   |   |   |   | 15 | 37. |
| 5 | 2,4 | 41. |   | 9 |   | 11 |   |   |   |   |   |   | 39. |
| 5 | 2,7 | 42. |   | 9 |   | X |   |   | 17 |   |   |   |   |

| 10 | 7,8 | 44. |   |   |   | X |   |   | 17 | 18 |   |    | 41.. |
|----|-----|-----|---|---|---|---|---|---|----|----|---|----|------|
|    |     |     | 1 | 2 | 3 | 4 | 5 | 6 | 7  | 8  | 9 | 10 |      |

We now try to improve on the upper bound, $T_{57}$, obtained earlier. We thus require a tree of linked cycles of value no greater than zero. We again use the method of obtaining a "path" of subtrees always of value no greater than 0. As before, we allow at most one "phantom cycle" to be used on the condition that it be linked to a usbtree before a tour is achieved. We first start with 4. Its value is 8. Thus, it can't be the initial cycle of a tree that yields a tour. However, only one 2-cycle doesn't contain one of its points: 8. Any cycle linking to it can't be 2-circuit, because every 2-circuit cycle containins at least four points. From our formula $p = \frac{n}{2} + 3t + a - 1 = 9 + a$ for the number of points in a set of cycles yielding a tour if $a = 2, p = 11$. Thus, we require an acceptable 2-cycle containing two points. No 2-cycle contains a point in 8. We can't use a phantom cycle to help obtain a tour, since such a cycle must contain at least one point not in point in 8. Thus, 4. can't belong to a set of cycles yielding a tour of non-positive value. Next, consider 43. In this case, we have a cycle whose value is 9. Furthermore, it only has points in cycles 9 and 10. Our first cycle must have a non-positive value, and – if it is not linked by a phantom cycle- must have a point in 9 or 10 but not both. 1. satsifies these conditions. Furthermore, $1. \oplus 43.$ is a tour having the value –2. It thus becomes a new upperbound for both $T_{FWOPT}$ and $T_{OPT}$. If a smaller-valued tour exists, it must have a value no greater than –3. We now go to 9. . Since |9.| = 10, we require a first cycle, $C_1$, whose value is no greater than –13 until we link 9. to the tree containing it. In that case, succeeding trees need have a value no greater than –3. We have two cycles whose value is no greater than –13: 33. and 38. whose respective values are –15 and –14. If we use a phantom cycle after our first cycle, we require a 2-cycle whose points lie in 1 and 8. No such 2-cycle exists. 9. is acceptable. If 9. is to be linked to a tree containing the first cycle, the tree requires a point in 1 or 8 and no more than one point in 9. All of the points of 33. lie in 2-cycles of 9. . Thus, 33. can't be linked to 9. 38. contains a pair of non-linking points in 5. Since 9. also contains a point in 5, 9. can't be linked to a tree containing 38. . It follows that

$$T_{FWOPT} = 1. \oplus 43. = (17\ 10\ 12\ 1\ 3\ 13\ 18\ 6\ 7\ 4\ 16\ 8\ 20\ 14\ 19\ 15\ 9\ 2\ 5\ 11): 54$$

3(b). Our next step is to obtain $T_{OPT}$. For simplicity, we henceforth call $T_{FWOPT}$ by $T$, $\sigma_{T_{FWOPT}}$ by $\sigma_T$. After constructing $\sigma_T^{-1} M^-$, we use entries from the latter matrix to obtain acceptable and 2-circuit paths of value no greater than $|T| - |\sigma_T| - 1 = 21$ that form branches of a tree whose root is 19. We then collect the cycles obtained. If we obtain a tour of value no greater than 21, we make it the new upper bound for $T_{OPT}$, say $T_{OPT}^{(i)}$. Since there may exist cycles where 19 is *not* a determining vertex, we use theorem 1.22 to obtain such cycles: we construct non-positive paths as branches of trees that are acceptable until we reach the point 19. Any path that can be extended beyond 19 can have a value no greater than $|T| - |\sigma_T| - 1 = 21$. Once we have all cycles containing 19, if no cycle is negative, we don't have to use phantom cycles in linking cycles. We thus concentrate on the values of cycles to be linked to a tree as well as the 2-cycles used in linking. Assuming that we sequential obtain upper bound tours $T_{OPT}^{(i)}$, $i = 1, 2, \ldots, r$, $T_{OPT} = T_{OPT}^{(r)}$,

$\sigma_T$ = (1 3)(2 5)(4 16)(6 7)(8 20)(9 15)(10 12)(11 17)(13 18)(14 19).

$$\sigma_T^{-1} M^-$$

| | 3 | 5 | 1 | 16 | 2 | 7 | 6 | 20 | 15 | 12 | 17 | 10 | 18 | 19 | 9 | 4 | 11 | 13 | 14 | 8 | |
|---|---|---|---|---|---|---|---|---|---|---|---|---|---|---|---|---|---|---|---|---|---|
| | 1 | 2 | 3 | 4 | 5 | 6 | 7 | 8 | 9 | 10 | 11 | 12 | 13 | 14 | 15 | 16 | 17 | 18 | 19 | 20 | |
| 1 | 0 | 73 | ∞ | 39 | 25 | 2 | 4 | 69 | 93 | 6 | 97 | 77 | 48 | 39 | 27 | 29 | 80 | 96 | 9 | 37 | 1 |
| 2 | 68 | 0 | 25 | 90 | ∞ | 49 | 79 | 84 | 88 | 8 | 5 | 59 | 81 | 22 | 0 | 29 | 2 | 30 | 86 | 73 | 2 |
| 3 | ∞ | 6 | 0 | 29 | 68 | 1 | 60 | 46 | 65 | 56 | 50 | 89 | 2 | 24 | 98 | 22 | 83 | 3 | 55 | 97 | 3 |
| 4 | 22 | 32 | 29 | 0 | 29 | 4 | 58 | 16 | 46 | 17 | 80 | 47 | 47 | 69 | 25 | ∞ | 83 | 56 | 27 | 2 | 4 |

| | 1 | 2 | 3 | 4 | 5 | 6 | 7 | 8 | 9 | 10 | 11 | 12 | 13 | 14 | 15 | 16 | 17 | 18 | 19 | 20 | |
|---|---|---|---|---|---|---|---|---|---|---|---|---|---|---|---|---|---|---|---|---|---|
| 5 | 7 | ∞ | 74 | 94 | 0 | 28 | 81 | 19 | 71 | 96 | 19 | 86 | 8 | 89 | 4 | 32 | 1 | 2 | 44 | 79 | 5 |
| 6 | 60 | 81 | 4 | 37 | 79 | 0 | ∞ | 49 | 61 | 40 | 49 | 8 | 0 | 40 | 42 | 58 | 38 | 2 | 44 | 5 | 6 |
| 7 | 1 | 28 | 2 | 30 | 29 | ∞ | 0 | 41 | 86 | 9 | 44 | 48 | 58 | 87 | 77 | 4 | 72 | 55 | 35 | 33 | 7 |
| 8 | 97 | 77 | 37 | 1 | 73 | 33 | 5 | 0 | 74 | 90 | 63 | 54 | 77 | 59 | 13 | 2 | 42 | 84 | 92 | ∞ | 8 |
| 9 | 96 | 2 | 25 | 28 | -2 | 75 | 40 | -1 | 0 | 78 | 70 | 47 | 60 | 84 | ∞ | 23 | 25 | 95 | 92 | 11 | 9 |
| 10 | 89 | 86 | 77 | 24 | 59 | 48 | 8 | 35 | 29 | 0 | 2 | ∞ | 89 | 63 | 49 | 47 | 36 | 94 | 58 | 54 | 10 |
| 11 | 81 | -1 | 78 | 0 | 0 | 70 | 36 | 23 | 8 | 0 | 0 | 34 | 71 | 19 | 25 | 81 | ∞ | 58 | 11 | 40 | 11 |
| 12 | 56 | 96 | 6 | 33 | 8 | 9 | 40 | 29 | 55 | ∞ | 1 | 0 | 16 | 70 | 80 | 17 | 2 | 83 | 61 | 90 | 12 |
| 13 | 2 | 3 | 95 | 28 | 29 | 54 | 1 | 84 | 64 | 82 | 47 | 91 | 0 | 21 | 96 | 55 | 59 | ∞ | 64 | 83 | 13 |
| 14 | 54 | 43 | 8 | 15 | 85 | 34 | 43 | 3 | 23 | 60 | 45 | 57 | 45 | 0 | 93 | 26 | 12 | 64 | ∞ | 91 | 14 |
| 15 | 63 | 69 | 91 | 86 | 86 | 84 | 59 | -1 | ∞ | 53 | 82 | 27 | 84 | 5 | 0 | 44 | 8 | 63 | 22 | 72 | 15 |
| 16 | 29 | 93 | 39 | ∞ | 90 | 30 | 37 | 89 | 88 | 33 | 53 | 24 | 17 | 91 | 30 | 0 | 2 | 29 | 16 | 1 | 16 |
| 17 | 48 | 17 | 95 | 51 | 3 | 42 | 47 | 87 | 82 | -1 | ∞ | 0 | 63 | 45 | 70 | 78 | 0 | 46 | 44 | 61 | 17 |
| 18 | 1 | 7 | 47 | 16 | 80 | 57 | -1 | 14 | 85 | 15 | 64 | 88 | ∞ | 85 | 61 | 46 | 72 | 0 | 45 | 76 | 18 |
| 19 | 23 | 88 | 38 | 90 | 21 | 86 | 39 | 20 | 6 | 69 | 46 | 62 | 85 | ∞ | 85 | 68 | 20 | 21 | 0 | 58 | 19 |
| 20 | 46 | 19 | 69 | 89 | 84 | 41 | 49 | ∞ | 1 | 29 | 89 | 35 | 15 | 21 | 1 | 16 | 25 | 85 | 4 | 0 | 20 |
| | 1 | 2 | 3 | 4 | 5 | 6 | 7 | 8 | 9 | 10 | 11 | 12 | 13 | 14 | 15 | 16 | 17 | 18 | 19 | 20 | |

We now exhibit trees whose branches are acceptable or 2-circuit paths. The value of each branch can never be greater than 21.

We obtain the following cycles containing 19 from trees 19(a) and 19(b):

1. (19 9 20): 21  2. (19 9 5 1): 20  3. (19 9 5 18 1): 16  4. (19 9 5 18 1 6 20): 18

5. (19 9 5 17 10 7 16 20): 21  6. (19 9 5 18 1 10 11 4 20): 21  7. (19 9 8): 21

8. (19 9 8 7 1): 20  9. (19 9 5 18 7 1): 15  10. (19 9 5 18 7 16 20): 14

11. (19 9 5 18 7 1 10 11 4 20): 20.

11. yields a tour whose value is 20. Thus, it becomes a new upper bound. Using theorem 1.22, we now construct non-positively valued acceptable paths in $\sigma_T^{-1} M^-$. In order to obtain one or more cycles whose value is no greater than 19, we note that from the non-positive acceptable tree, only one branch could have been extended to a non-positive 2-circuit path: [9 5 17 12 11]: 0. This can be extended to [9 5 17 12 11 4]: 0. The latter path cannot be extended further. This tells us that no non-positive cycle exists. Thus, any cycle that can be used – together with one of the cycles above - to construct a tour whose value is no greater than 19 must have a positive value. It follows that we need only to be concerned with cycles above that have a value no greater than 18. Also, since we are unable to obtain a non-positive path whose terminal point is 19 or 14, from theorem 1.22, we can assume that all cycles containing 19 have 19 as a determining point.

$$M_{19}$$

| VALUE | IN | | (1 3) | (2 5) | (4 16) | (6 7) | (8 20) | (9 15) | (10 12) | (11 17) | (13 18) | (14 19) |
|---|---|---|---|---|---|---|---|---|---|---|---|---|
|  |  |  | 1 | 2 | 3 | 4 | 5 | 6 | 7 | 8 | 9 | 10 |
| 16 | 1,2,6,9,10 | 3. | 1 | 5 |  |  |  | 9 |  |  | 18 | 19 |
| 18 | 1,2,4,5,6,9,10 | 4. | 1 | 5 |  | 6 | 20 | 9 |  |  | 18 | 19 |
| 15 | 1,2,4,6,9,10 | 9. | 1 | 5 |  | 7 |  | 9 |  |  | 18 | 19 |
| 14 | 2,3,4,5,9,10 | 10. |  | 5 | 16 | 7 | 20 | 9 |  |  | 18 | 19 |
|  |  |  | 1 | 2 | 3 | 4 | 5 | 6 | 7 | 8 | 9 | 10 |

Given 3., if a path is acceptable, we may choose at most one linking point from the cycle in question. If it is a 2-circuit path, we may choose at most one linking point from the red portion of the path and one from the black portion. We exhibit the respective trees associated with each of them.

1: (1 3)  2: (2 5)  3: (4 16)  4: (6 7)  5: (8 20)  6: (9 15)  7: (10 12)  8: (11 17)  9: (13 18)  10: (14 19)

3: 1  5: 2  7: 4  12: 7  15: 6  16: 3  17: 8  18: 9  19: 10  20: 5

3. ( 19 9 5 18 1)   CYCLES: *1.* (10 11) 2   *2.* (11 2 17 10): 2   *3.* (11 2 17 12): 2   *4.* (11 5 17 10): 2

*5.* (11 5 17 12): 2   *6.* (12 17): 2   *7.* (20 9 8 4): 3   *8.* (20 9 8 16): 3

$$M_{19,3.}$$

| VALUE | IN 2-CYCLES | | (1 3) 1 | (2 5) 2 | (4 16) 3 | (6 7) 4 | (8 20) 5 | (9 15) 6 | (10 12) 7 | (11 17) 8 | (13 18) 9 | (14 19) 10 | |
|---|---|---|---|---|---|---|---|---|---|---|---|---|---|
| 16 | 1,2,6,9,10 | *3.* | 1 | 5 | | | | 9 | | | 18 | 19 | *3.* |
| 2 | 7,8 | *1.* | | | | | | | 10 | 11 | | | *1.* |
| 2 | 2,7,8 | *2.* | | 2 | | | | | 10 | 11,17 | | | *2.* |
| 2 | 2,7,8 | *3.* | | 2 | | | | | 12 | 11,17 | | | *3.* |
| 2 | 2,7,8 | *4.* | | 5 | | | | | 10 | 11,17 | | | *4.* |
| 2 | 2,7,8 | *5.* | | 5 | | | | | 12 | 11,17 | | | *5.* |
| 2 | 7,8 | *6.* | | | | | | | 12 | 17 | | | *6.* |
| 3 | 3,5,6 | *7.* | | | 4 | | 20,8 | 9 | | | | | *7.* |
| 3 | 3,5,6 | *8.* | | | 16 | | 20,8 | 9 | | | | | *8.* |
| | | | 1 | 2 | 3 | 4 | 5 | 6 | 7 | 8 | 9 | 10 | |

Since there exists no cycle that has a point in a 2-cycle of 4., we can't obtain a tour. We now exhibit the trees that might – together with 4. – obtain a tour.

4. (19 9 5 18 1 6 20):  CYCLE: *1.* (12 17): 1

$$M_{19,5.}$$

| VALUE | IN 2-CYCLES | | (1 3) 1 | (2 5) 2 | (4 16) 3 | (6 7) 4 | (8 20) 5 | (9 15) 6 | (10 12) 7 | (11 17) 8 | (13 18) 9 | (14 19) 10 |
|---|---|---|---|---|---|---|---|---|---|---|---|---|

| 18 | 1,2,4,5,6,9,10 | 4. | 1 | 5 |  | 6 | 20 | 9 |  |  | 18 | 19 | 4. |
| 1 | 7,8 | 1. |  |  |  |  |  |  | 12 | 17 |  |  | 1. |
|  |  |  | 1 | 2 | 3 | 4 | 5 | 6 | 7 | 8 | 9 | 10 |  |

We thus can't obtain a tour containing 4. .We now exhibit the trees associated with 10. .

9. (19 9 5 18 7 1)  15 CYCLES: *1.* (5  17): 4   *2.* (10  11): 2   *3.* (11 2 17 10): 2   *4.* (11 2 17 12): 2

*5.* (11 5 17 10): 2   *6.* (11 5 17 12): 2   *7.* (17 12): 1

$$M_{19,10.}$$

| VALUE | IN 2-CYCLES |  | (1 3) 1 | (2 5) 2 | (4 16) 3 | (6 7) 4 | (8 20) 5 | (9 15) 6 | (10 12) 7 | (11 17) 8 | (13 18) 9 | (14 19) 10 |  |
|---|---|---|---|---|---|---|---|---|---|---|---|---|---|
| 15 | 1,2,4,6,9,10 | 9. | 1 | 5 |  | 7 |  | 9 |  |  | 18 | 19 | 10. |
| 4 | 2,8 | 1. |  | 5 |  |  |  |  |  | 17 |  |  | 1. |
| 2 | 7,8 | 2. |  |  |  |  |  |  | 10 | 11 |  |  | 2. |
| 2 | 2,7,8 | 3. |  | 2 |  |  |  |  | 10 | 11,17 |  |  | 3. |
| 2 | 2,7,8 | 4. |  | 2 |  |  |  |  | 12 | 11,17 |  |  | 4. |
| 2 | 2,7,8 | 5. |  | 5 |  |  |  |  | 10 | 11,17 |  |  | 5. |
| 2 | 2,7,8 | 6. |  | 5 |  |  |  |  | 12 | 11,17 |  |  | 6. |
| 1 | 7,8 | 7. |  |  |  |  |  |  | 12 | 17 |  |  | 7. |
|  |  |  | 1 | 2 | 3 | 4 | 5 | 6 | 7 | 8 | 9 | 10 |  |

We can't obtain a tour using 9. .  We now exhibit trees associated with 10.

. 10. (19 9 5 18 7 16 20)  CYCLES: *1.* (17 10 11 2): 2   *2.* (17 10 11 5): 2   *3.* (17 12 11 2): 3

*4.* (17 12 11 5): 3

$$M_{19,12.}$$
(1 3)    (2 5)    (4 16)    (6 7    (8 20)    (9 15)    (10 12)    (11 17)    (13 18)    (14 19)

| VALUE | IN 2-CYCLES | | 1 | 2 | 3 | 4 | 5 | 6 | 7 | 8 | 9 | 10 | |
|---|---|---|---|---|---|---|---|---|---|---|---|---|---|
| 14 | 1,2,6,9,10 | 10. | | 5 | 16 | 7 | 20 | 9 | | | 18 | 19 | 12. |
| 2 | 2,7,8 | 1. | 2 | | | | | | 10 | 11,17 | | | 1. |
| 2 | 2,7,8 | 2. | 5 | | | | | | 10 | 11,17 | | | 2. |
| 3 | 2,7,8 | 3. | 2 | | | | | | 12 | 11,17 | | | 3. |
| 3 | 2,7,8 | 4. | 2 | | | | | | 12 | 11,17 | | | 4. |
| | | | 1 | 2 | 3 | 4 | 5 | 6 | 7 | 8 | 9 | 10 | |

We are unable to obtain a point in 1. Therefore, we can't obtain a tour containing 10. . It follows that the cycle given in 11. yields an optimal tour.

$$T_{OPT} = (\ 19\ 15\ 9\ 2\ 5\ 13\ 18\ 6\ 7\ 3\ 1\ 12\ 10\ 17\ 11\ 16\ 4\ 8\ 20\ 15\ ): 52$$

**Theorem 1.30** *Let $PM_1$ and $PM_2$ represent two perfect matchings obtained from an $n \times n$ cost matrix M. Suppose that their respective sets of arcs are disjoint. Then we can always obtain a tour by alternatingly choosinfg an arc from $PM_1$ followed by one from $PM_2$.*

Proof. Let

$PM_1 = \{[\ a_1\ a_2\ ], [\ a_3\ a_4\ ], \ldots, [\ a_{2n-1}\ a_{2n}\ ]\}$, $PM_2 = \{[\ a_1\ a_{r_1}\ ], [\ a_2\ a_{r_2}\ ], \ldots, [\ a_{2n-1}\ a_{r_{2n-1}}\ ], [\ a_{2n}\ a_{r_{2n}}\ ]\}$.

The $r_j s$ are simply a derangement of $a_1, a_2, \ldots, a_{2n}$. For simplicity, we choose the $r_j s$ from the points of a cycle such that the edges in the set $PM_2$ are all distinct from those in $PM_1$. Thus, let $r_1 = 4, r_2 = 5, r_3 = 1, r_4 = 6, r_5 = 3, r_6 = 2$. We thus obtain $C = (\ a_1\ a_4\ a_3\ a_5\ a_6\ a_2\ )$.

# Chapter 2

## The Symmetric GTSP, n odd

I. Introduction. In this chapter, we deal with a symmetric $n \times n$ cost matrix, M, where $n$ is odd.

In order to understand our procedure, we first define an *almost perfect matching* (APM). Let $T$ be a tour in M. Let $(a\ b)$ be an arc in $T$ such that the path $P = (c\ a\ b)$ has the smallest value of all 2-arc paths in M. (We obtain $P$ by adding the smallest value of an arc from $c$ to $a$ to the smallest value of an arc emanating from $a$  We do this for all columns $a$.) Choose the two sets of alternating edges in $T - \{a\}$ where $T$ has been written such that $a$ is its last point. Choose the set of edges, $SMALLER$, the sum of whose values is smaller. We next change each edge in $SMALLER$ into a 2-cycle. Since $SMALLER$ contains $n-1$ points, it doesn't contain one point: $a$. $a$ becomes an identity point. Then $SMALLER \cup \{a\}$ is an APM. It is worth noting that we have tried to approximate the even case by eliminating the *smallest-valued* arc. We now modify our definition of an acceptable path to one in which the inclusion of point $a$ is equivalent to choosing a point in a 2-cycle. We show why this is appropriate. W.l.o.g., suppose that $P = [a_1\ a_3\ ...\ a\ a_i\ a_{i+2}\ ...\ a_r]$ is an acceptable path according to our definition. Then by changing our path of arcs into one of edges, we obtain
$P_{EDGES} = [a_1\ \underline{a_4}\ a_3\ \underline{a_6}\ a_5\ ...\ a\ \underline{a_{i+1}}\ a_i\ \underline{a_{i+3}}\ a_{i+2}\ ...\ \underline{a_{r+1}}\ a_r]$. Thus, instead of having $2r$ points, $P_{EDGES}$ has $2r-1$ points. This property applies to the number of points obtained in 2-circuit paths as well. Now we discus an acceptable circuit, say $C = (a_1\ a_2\ ...\ a_n)$. Thus, if we substitute APM for PM in our theorems in chapter one, we obtain analogous results. One difference in working with APM's is that we must look out for 2-cycles containing $a$. Thus, if we are considering extending our path with the arc $(a_i\ a)$, we check $P_{in}$ to see if $(a\ a_i)$ already exists in determining point row of our path. Also, since $a'$ doesn't belong to a 2-cycle, it is always a non-linking point. One change must be noted in our formula for the number of points in the cycles that can be linked together to obtain a tour. Although the fixed point used in our algorithms doesn't belong to a 2-cycle, it nevertheless appears in each tour. To take this into account, our formula becomes $p = \dfrac{n+1}{2} + 3t + a - 1$.

II. Theorems

**Theorem 2.1** *Assume M is an $n \times n$ cost matrix where n is odd. Let $T$ be a tour in M and $\sigma_T$ an APM obtained from $T$ with a fixed point $a'$. Then no acceptable or 2-circuit path in $\sigma_T^{-1} M^-$ can contain a subpath of form $[a\ a'\ \sigma_T(a)]$.*

Proof. Since $a'$ is a fixed point, $[a\ a']$ is a directed edge. If $[a\ a'\ \sigma_T(a)]$ belongs to a cycle, $C$, then when we transform $C$ into a circuit (or two circuits) of edges, we obtain $[a\ a'\ a\ \sigma_T(a)]$. But $[a\ a'\ a]$ is one edge instead of two. Thus, we can't obtained a tour using $C$.

**Theorem 2.2** *Let $(j\ i)$ be the smallest-valued arc in $\sigma_T^{-1} M^-$ that terminates in $i$, while $(i\ k)$ is the smallest arc emanating from $i$. Define $\max = \{maximum\ value\ of\ |(j\ i)| + |(i\ k)|,\ i = 1, 2, \ldots, n\}$ Denote by $\min$ the smallest-valued negative path obtainable in $\sigma_T^{-1} M^-$. Then if $\max + \min \geq |T| - |\sigma_T|$, thee exists no tour of smaller value than $T$.*

Proof In order for us to construct a tour of msaller vlaue than $T$ in $\sigma_T^{-1} M^-$, there must exist a cycle, say $C$, that contains either $i$ or $\sigma_T(i)$. If $C$ contains $i$, then its companion cycle, $C_{COMP}$, contains $\sigma_T(i)$. $||C| = |C_{COMP}|$. Thus, if we construct a tour by linking circuits, at least one cycle yielding a circuit must contain a point from the 2-cycle

$(i\ \sigma_T(i))$. Such a cycle must have a value less than $|T| - |\sigma_T|$. Given our hypothesis, this is impossible. Thus, we can't obtain a tour of value less than $T$.

III. Examples

Suppose that we must consider an entry of the form $|(a\ \ldots\ a_n\ \ldots\ x)|$ vis-a-vis $|(a\ x)|$. We use the following rule: If $|(a\ x)| + |(a_n\ a_1)|$ is greater than $|(a\ \ldots\ a_n\ \ldots\ x)|$, we choose $|(a\ \ldots\ a_n\ \ldots\ x)|$. Otherwise, $|(a\ x)|$ remains the entry.

Example 6.

M

|   | 1 | 2 | 3 | 4 | 5 | 6 | 7 | 8 | 9 |   |
|---|---|---|---|---|---|---|---|---|---|---|
| 1 | ∞ | 27 | 31 | 6 | 39 | 79 | 8 | 11 | 41 |   |
| 2 | 27 | ∞ | 70 | 42 | 31 | 1 | 10 | 32 | 90 | 2 |
| 3 | 31 | 70 | ∞ | 7 | 24 | 24 | 62 | 91 | 99 | 3 |
| 4 | 6 | 42 | 7 | ∞ | 58 | 57 | 31 | 4 | 48 | 4 |
| 5 | 39 | 31 | 24 | 58 | ∞ | 60 | 93 | 49 | 19 | 5 |
| 6 | 79 | 1 | 24 | 57 | 60 | ∞ | 29 | 26 | 49 | 6 |
| 7 | 8 | 10 | 62 | 31 | 63 | 29 | ∞ | 18 | 30 | 7 |
| 8 | 11 | 32 | 91 | 4 | 49 | 46 | 18 | ∞ | 6 | 8 |
| 9 | 41 | 90 | 99 | 48 | 19 | 99 | 30 | 6 | ∞ | 9 |
|   | 1 | 2 | 3 | 4 | 5 | 6 | 7 | 8 | 9 |   |

MIN(M)

|   | 1 | 2 | 3 | 4 | 5 | 6 | 7 | 8 | 9 |   |
|---|---|---|---|---|---|---|---|---|---|---|
| 1 | 4 | 7 | 8 | 1 | 3 | 5 | 9 | 6 |   | 1 |
| 2 | 6 | 7 | 1 | 5 | 8 | 4 | 3 | 9 |   | 2 |
| 3 | 4 | 5 | 6 | 1 | 7 | 2 | 9 | 8 |   | 3 |
| 4 | 8 | 1 | 3 | 7 | 2 | 9 | 6 | 5 |   | 4 |
| 5 | 9 | 3 | 2 | 1 | 8 | 4 | 6 | 7 |   | 5 |
| 6 | 2 | 3 | 8 | 7 | 9 | 4 | 5 | 1 |   | 6 |
| 7 | 1 | 2 | 8 | 6 | 9 | 4 | 3 | 5 |   | 7 |
| 8 | 4 | 9 | 1 | 7 | 6 | 2 | 5 | 3 |   | 8 |
| 9 | 8 | 5 | 7 | 1 | 4 | 6 | 2 | 3 |   | 9 |
|   | 1 | 2 | 3 | 4 | 5 | 6 | 7 | 8 | 9 |   |

$T_{UPPERBOUND} = T = D_4 = (1\ 7\ 2\ 6\ 3\ 5\ 9\ 8\ 4) = (6\ 3\ 5\ 9\ 8\ 4\ 1\ 7\ 2)$. We will add another column and row to M to obtain the $10 \times 10$ cost matrix $M_e$. This will allow us to construct a pseudo-perfect matching from $T$, i.e., the new column and row won't be symmetric. However, they will have one symmetric pair of points. We now use $(6\ 3\ 5\ 9\ 8\ 4\ 1\ 7\ 2)$ to create the following APM's.

|   | 24 | 19 | 4 | 8 |   |
|---|---|---|---|---|---|
| (6 3) | (5 9) | (8 4) | (1 7) | 2 |   |
|   | 24 | 6 | 6 | 10 |   |
| (3 5) | (9 8) | (4 1) | (7 2) |   |   |

We apply the permutation $p_1 = (3\ 5)(9\ 8)(4\ 1)(7\ 2)$ to the columns of M.

It follows that 6 is our identity point.

$$p_1^{-1} M^-$$

|   | 4 | 7 | 5 | 1 | 3 | 6 | 2 | 9 | 8 |   |
|---|---|---|---|---|---|---|---|---|---|---|
|   | 1 | 2 | 3 | 4 | 5 | 6 | 7 | 8 | 9 |   |
| 1 | 0 | 2 | 33 | ∞ | 25 | 73 | 21 | 35 | 5 | 1 |
| 2 | 32 | 0 | 21 | 17 | 60 | -9 | ∞ | 80 | 22 | 2 |
| 3 | -17 | 38 | 0 | 7 | ∞ | 0 | 46 | 75 | 67 | 3 |
| 4 | ∞ | 25 | 52 | 0 | 1 | 51 | 36 | 42 | -2 | 4 |
| 5 | 34 | 69 | ∞ | 15 | 0 | 36 | 7 | -5 | -5 | 5 |
| 6 | 57 | 29 | 60 | 79 | 24 | ∞ | 1 | 49 | 26 | 6 |
| 7 | 21 | ∞ | 53 | -2 | 52 | 19 | 0 | 20 | 8 | 7 |
| 8 | -2 | 12 | 43 | 5 | 85 | 40 | 26 | 0 | ∞ | 8 |
| 9 | 42 | 24 | 13 | 35 | 93 | 93 | 84 | ∞ | 0 | 9 |
|   | 1 | 2 | 3 | 4 | 5 | 6 | 7 | 8 | 9 |   |

j = 1

(3 1)(1 2) = (3 2): -15; (3 1)(1 7) = (3 7): 4; (8 1)(1 2) = (8 2): 0; (3 1)(1 9) = (3 9): -12

*(3 1)(1 5) = (3 5): 8; (8 1)(1 9) = (8 9): 3*

j = 2(

(1 2)(2 6) = (1 6): -7

(3 2)(2 3) = (3 3): 6

CYCLE (3 1 2): 6

(3 2)(2 6) = (3 6): -24; (8 2)(2 6) = -9

j = 4

(7 4)(4 5) = (7 5): -1; (7 4)(4 9) = (7 9): -4; (8 4)(4 5) = (8 5): 6

*(3 4)(4 5) = (3 5): 8*

j = 5

(4 5)(5 7) = (4 7): 8; (4 5)(5 8) = (4 8): -4; (4 5)(5 9) = (4 9): -4

(7 5)(5 7) = (7 7): 6

CYCLE (7 4 5): 6

(7 5)(5 8) = (7 8): -6; (7 5)(5 9) = (7 9): -6

(8 5)(5 8) = (8 8): 1

CYCLE (8 4 5): 1

*(8 5)(5 9) = (8 9): 1*

j = 6

*(3 6)(6 5) = (3 5): <u>0</u>*

j = 7

(5 7)(7 4) = (5 4): <u>5</u>; (6 7)(7 4) -1

j = 8

(4 8)(8 4) = (4 4): <u>1</u>

CYCLE (4 5 8): 1

(5 8)(8 1) = (5 1): -7; (5 8)(8 4) = 0

j = 9

(3 9)(9 3) = (3 3): 1

CYCLE (3 1 9): 1

(5 9)(9 3) = (5 3): 8

$$P_9$$

|   | 1 | 2 | 3 | 4 | 5 | 6 | 7 | 8 | 9 |   |
|---|---|---|---|---|---|---|---|---|---|---|
| 1 |   | 1 |   |   |   | 2 |   |   |   | 1 |
| 2 |   |   |   |   |   |   |   |   |   | 2 |
| 3 | 3 | 1 |   | 2 |   | 2 | 1 |   | 1 | 3 |
| 4 |   |   |   |   | 4 |   | 5 | 5 | 5 | 4 |
| 5 | 8 | 8 | 9 | 8 |   |   | 5 | 5 | 5 | 5 |
| 6 |   |   |   | 7 |   |   | 6 |   |   | 6 |
| 7 |   |   |   | 7 | 4 |   |   | 5 | 5 | 7 |
| 8 | 8 | 1 |   |   | 4 | 2 |   |   |   | 8 |
| 9 |   |   |   |   |   |   |   |   |   | 9 |
|   | 1 | 2 | 3 | 4 | 5 | 6 | 7 | 8 | 9 |   |

$$p_I^{-1}M^-(9)$$

|   | 4 | 7 | 5 | 1 | 3 | 6 | 2 | 9 | 8 |   |
|---|---|---|---|---|---|---|---|---|---|---|
|   | 1 | 2 | 3 | 4 | 5 | 6 | 7 | 8 | 9 |   |
| 1 | 0 |   |   | ∞ |   | -7 |   |   |   | 1 |
| 2 |   | 0 |   |   |   |   | ∞ |   |   | 2 |
| 3 |   | *-15* | 0 |   | ∞ | *-24* | *4* |   | *-12* | 3 |
| 4 | ∞ |   |   | 0 |   | *8* |   | *-4* | *-4* | 4 |
| 5 | <u>-7</u> |   | ∞ | <u>0</u> | 0 |   |   |   |   | 5 |
| 6 |   |   |   | <u>-1</u> |   | ∞ |   |   |   | 6 |
| 7 |   | ∞ |   | -4 | -1 |   |   | -6 | -6 | 7 |
| 8 |   | *0* |   |   | *6* | *-9* |   | 0 | ∞ | 8 |
| 9 |   |   |   |   |   |   |   | ∞ | 0 | 9 |
|   | 1 | 2 | 3 | 4 | 5 | 6 | 7 | 8 | 9 |   |

ACCEPTABLE CYCLES

(3 1 2): 6; (3 1 2 6): 0; (7 4 5): 6; (8 4 5): 1; (4 5 8): 1; (3 1 9): 1

ACCEPTABLE 2-CIRCUIT PATHS

[3 1 5]: 8; [8 1 9]: 3; [8 4 5 9]: 1; [5 9 3]: 8; [3 4 5]: 8

Since the are only three are only three underlined entries in $p_I^{-1}M^-$ (9), we save time by trying to extend each path individually.

[5 8 1]: -7 extends to [5 8 1 2]: -5. [5 8 1 2] can be extended to [5 8 1 2 6]: -14 which cannot be extended further.

[5 8 4]: 0 can't be extended further.

[6 7 4]: -1 can be extended to [6 7 4 5] and [6 7 4 9].

[6 7 4 5}: 0. [6 7 4 5] can be extended to [6 7 4 5 8]: -5 and [6 7 4 5 9]. Neither can be extended further.

[6 7 4 9]: -3. [6 7 4 9] can't be extended further.

When $n$ is odd, the general formula for obtaining the number of points in the cycles that can be linked to form a tour is $p = \dfrac{n+1}{2} + 3t + a - 1$. Thus, our formula for obtaining the number of points in a tour using two acceptable 2-cycles yields $\dfrac{n+1}{2} + a - 1 = 5 + 2 - 1 = 6$. But we have only one cycle that contains 6. It has four points in it. It follows that if a tour can be obtained by linking two acceptable cycles, the other cycle must contain two points. Since no cycle contains two points, we can't obtain a tour using only acceptable cycles. We now proceed to obtain two circuit cycles using 2-circuit paths.

$$P_{18}$$

|   | 4 | 7 | 5 | 1 | 3 | 6 | 2 | 9 | 8 |   |
|---|---|---|---|---|---|---|---|---|---|---|
|   | 1 | 2 | 3 | 4 | 5 | 6 | 7 | 8 | 9 |   |
| 1 |   |   |   |   |   |   |   |   |   | 1 |
| 2 |   |   |   |   |   |   |   |   |   | 2 |
| 3 | 3 |   |   |   | <u>1</u> |   |   |   |   | 3 |
| 4 |   |   |   |   |   |   |   |   |   | 4 |
| 5 |   |   | <u>9</u> |   |   |   |   | 5 |   | 5 |
| 6 |   |   |   |   |   |   |   |   |   | 6 |
| 7 |   |   |   |   |   |   |   |   |   | 7 |
| 8 |   |   |   | 8 | 4 |   |   |   | <u>5</u> | 8 |
| 9 |   |   |   |   |   |   |   |   |   | 9 |
|   | 1 | 2 | 3 | 4 | 5 | 6 | 7 | 8 | 9 |   |

$$\sigma_T^{-1}M^-(18)$$

|   | 4 | 7 | 5 | 1 | 3 | 6 | 2 | 9 | 8 |   |
|---|---|---|---|---|---|---|---|---|---|---|
|   | 1 | 2 | 3 | 4 | 5 | 6 | 7 | 8 | 9 |   |
| 1 |   |   |   |   |   |   |   |   |   | 1 |
| 2 |   |   |   |   |   |   |   |   |   | 2 |
| 3 | X |   |   |   | 8 |   |   |   |   | 3 |
| 4 |   |   |   |   |   |   |   |   |   | 4 |
| 5 |   |   | 8 |   |   |   |   | X |   | 5 |
| 6 |   |   |   |   |   |   |   |   |   | 6 |
| 7 |   |   |   |   |   |   |   |   |   | 7 |
| 8 |   |   |   | X | X |   |   |   | 1 | 8 |
| 9 |   |   |   |   |   |   |   |   |   | 9 |
|   | 1 | 2 | 3 | 4 | 5 | 6 | 7 | 8 | 9 |   |

j = 3

(5 3)(3 1) = (5 1): -9

j = 5

(3 5)(5 8) = (3 8): 3; (3 5)(5 9) = (3 9): 3

j = 8

(3 8)(84) = (3 4): 8

$P_{18}$

|   | 1 | 2 | 3 | 4 | 5 | 6 | 7 | 8 | 9 |   |
|---|---|---|---|---|---|---|---|---|---|---|
| 1 |   |   |   |   |   |   |   |   |   | 1 |
| 2 |   |   |   |   |   |   |   |   |   | 2 |
| 3 | 3 |   |   | <u>8</u> | 1 |   |   | 5 | 5 | 3 |
| 4 |   |   |   |   |   |   |   |   |   | 4 |
| 5 | <u>3</u> |   | 9 |   |   |   |   |   | 5 | 5 |
| 6 |   |   |   |   |   |   |   |   |   | 6 |
| 7 |   |   |   |   |   |   |   |   |   | 7 |
| 8 |   |   |   | 8 | 4 |   |   |   | 5 | 8 |
| 9 |   |   |   |   |   |   |   |   |   | 9 |
|   | 1 | 2 | 3 | 4 | 5 | 6 | 7 | 8 | 9 |   |

$\sigma_T^{-1}M^-(18)$

|   | 4 | 7 | 5 | 1 | 3 | 6 | 2 | 9 | 8 |   |
|---|---|---|---|---|---|---|---|---|---|---|
|   | 1 | 2 | 3 | 4 | 5 | 6 | 7 | 8 | 9 |   |
| 1 |   |   |   |   |   |   |   |   |   | 1 |
| 2 |   |   |   |   |   |   |   |   |   | 2 |
| 3 | X |   |   | <u>8</u> | 8 |   |   | 3 | 3 | 3 |
| 4 |   |   |   |   |   |   |   |   |   | 4 |
| 5 | <u>-9</u> |   | 8 |   |   |   |   |   | X | 5 |
| 6 |   |   |   |   |   |   |   |   |   | 6 |
| 7 |   |   |   |   |   |   |   |   |   | 7 |
| 8 |   |   |   | X | X |   |   |   | 1 | 8 |
| 9 |   |   |   |   |   |   |   |   |   | 9 |
|   | 1 | 2 | 3 | 4 | 5 | 6 | 7 | 8 | 9 |   |

Since we have only two underlined entries in $\sigma_T^{-1} M^-(18)$, we deal with each one individually.

[3 1 5 8 4] can't be extended. [5 9 3 1][1 2] = [5 9 3 1 2]: -7.

[5 9 3 1 2][2 6] = [5 9 3 1 2 6]: -17. (5 9 3 1 2 6): 8 is an unlinked 2-circuit cycle.

The path [5 9 3 1 2 6] can't be extended further. The cycle $C = $ (5 9 3 1 2 6) contains our fixed point as well as a point in each 2-cycle. From our formula for the number of points in a tour: $p = \dfrac{n+1}{2} + 3t + a - 1 = 5 + 3t + a - 1 = 4 + 3t + a$. $C$ yields two circuits. The first one has 9 as a linking point; the second, 1,2,6. But we have only one 2-circuit cycle. Thus, $t = 1$, $p = 4 + 3 + a$. We thus require either an acceptable cycle containing either a linking point in each "circuit" or else two 2-cycles each of which has a linking point from each of the two "circuits". We have only one 2-circuit cycle. Thus, we require either an acceptable 2-cycle containing a linking point to each of the two circuits. We have no 2-circuit cycles. We have only one 2-circuit cycle. Thus, we can't construct a tour of value less than $/D_4/$. It follows that

$T_{FWOPT} = D_4 = (1\ 7\ 2\ 6\ 3\ 5\ 9\ 8\ 4): 102$. Our next step is to obtain

(1) all cycles with a determining point 6,

(2) all cycles containing 6 where 6 is not a determining point.

We now exhibit the trees obtained. As in the previous chapter, in cases given in (2), the initial point of the first pair of points lying on a branch where both belong to the same 2-cycle is printed in red . It is inside a red square. The second point of the pair is printed in black and is inside a black square. One exception is made: If the initial point is both a point of an interlacing pair encountered on a branch as well as the initial point of the first pair encountered on two different branches, then it is printed in black inside a red square.

As we note from the trees obtained, there exists no 2-circuit cycle. Thus, $T_{OPT} = T_{FWOPT} = D_4$.

The next example involves constructing clusters in the upper triangle of the cost matrix. We define a *cluster* as a set of objects which are similar between each pair of objects and dissimilar to objects belonging to other clusters. In the case of example 7, we partition the upper triangle of the cost matrix M into four polygons and a single entry:

I, II, III, IV, V. In I, assume that we get from one entry to another from the entry (2 8) by assuming that I is a portion of a chess board such that we move from (2 8) to any other entry in one by the moves of a pawn that can move one ox at a time going east, west north south. In each move, we add 1 to the value 47. In II, each entry is at most $+[\log 60]$ different from that of (3 5). In III, each entry is at most $\pm [\log 73]$ different from the entry in (4 13). In IV, each entry is at most $\pm [\log 30]$ different from the entry in (8 13). V. is the entry (3 15). Each of the entries in any of the four polygons is closer to its fundamental entry than to any entry in the other three polygons.

Example 7.

M

| | 1 | 2 | 3 | 4 | 5 | 6 | 7 | 8 | 9 | 10 | 11 | 12 | 13 | 14 | 15 | |
|---|---|---|---|---|---|---|---|---|---|---|---|---|---|---|---|---|
| 1 | ∞ | 53 | 52 | 51 | 50 | 49 | 48 | 48 | 48 | 48 | 50 | 51 | 52 | 53 | 54 | 1 |
| 2 | 53 | ∞ | 66 | 66 | 50 | 49 | 48 | 47 | 48 | 49 | 50 | 51 | 52 | 53 | 54 | 2 |
| 3 | 52 | 66 | ∞ | 62 | 65 | 64 | 48 | 48 | 48 | 49 | 50 | 75 | 69 | 77 | 36 | 3 |
| 4 | 51 | 66 | 62 | ∞ | 63 | 62 | 65 | 49 | 49 | 49 | 50 | 77 | 73 | 76 | 71 | 4 |
| 5 | 50 | 50 | 65 | 63 | ∞ | 66 | 65 | 50 | 73 | 76 | 74 | 71 | 77 | 75 | 76 | 5 |
| 6 | 49 | 49 | 64 | 62 | 66 | ∞ | 65 | 51 | 70 | 77 | 73 | 37 | 23 | 30 | 69 | 6 |
| 7 | 48 | 48 | 48 | 65 | 65 | 65 | ∞ | 73 | 74 | 71 | 75 | 26 | 39 | 27 | 72 | 7 |
| 8 | 48 | 47 | 48 | 49 | 50 | 51 | 73 | ∞ | 27 | 31 | 73 | 31 | 30 | 35 | 73 | 8 |
| 9 | 48 | 48 | 48 | 49 | 73 | 70 | 74 | 27 | ∞ | 30 | 38 | 29 | 26 | 36 | 77 | 9 |
| 10 | 48 | 49 | 49 | 49 | 76 | 77 | 71 | 31 | 30 | ∞ | 36 | 35 | 73 | 71 | 74 | 10 |
| 11 | 50 | 50 | 50 | 50 | 74 | 73 | 76 | 73 | 38 | 36 | ∞ | 30 | 26 | 37 | 35 | 11 |
| 12 | 51 | 51 | 75 | 77 | 71 | 37 | 26 | 31 | 29 | 35 | 30 | ∞ | 29 | 36 | 30 | 12 |
| 13 | 52 | 52 | 69 | 73 | 77 | 23 | 39 | 30 | 26 | 73 | 26 | 29 | ∞ | 30 | 36 | 13 |
| 14 | 53 | 63 | 77 | 76 | 75 | 30 | 27 | 35 | 36 | 71 | 37 | 36 | 30 | ∞ | 32 | 14 |
| 15 | 54 | 54 | 36 | 71 | 76 | 69 | 72 | 73 | 77 | 74 | 35 | 30 | 36 | 32 | ∞ | 15 |
| | 1 | 2 | 3 | 4 | 5 | 6 | 7 | 8 | 9 | 10 | 11 | 12 | 13 | 14 | 15 | |

We define a *cluster* as a set of objects that are similar among themselves but dissimilar to objects belonging to another cluster. In the case of example 7, we partition the upper triangle of the cost matrix into five polygons: I, II, III, IV, V. In I, assume that we get from one entry to another passing through the entry (2 8) by assuming that I is a portion of a chess board such that we can move from (2 8) to any other entry in I by the moves of a pawn that can move north, south, east, or west. In each move, we add 1 to the "distance" of moves from (2 8). In II, each entry is at most $+[\log 62]$ different from 62. In III, each entry is at most $\pm[\log 73]$ different from 73. In IV, the difference is at most $\pm[\log 30]$ from 30.

V is the entry (3  15). In example 7, the difference between any pair of entries in a cluster is less than the difference between an entry in one cluster and an entry in any other cluster. In MIN(M), if any one of the first $[\log n] + 1$ entries in a row occurs consecutively more than once, we consider each such set of consecutive column numbers as *one* entry in the row when performing PHASE 1.

M

|    | 1  | 2  | 3  | 4  | 5  | 6  | 7  | 8  | 9  | 10 | 11 | 12 | 13 | 14 | 15 |    |
|----|----|----|----|----|----|----|----|----|----|----|----|----|----|----|----|----|
| 1  | ∞  | 53 | 52 | 51 | 50 | 49 | 48 | 48 | 48 | 48 | 50 | 51 | 52 | 53 | 54 | 1  |
| 2  | 53 | ∞  | 66 | 66 | 50 | 49 | 48 | 47 | 48 | 49 | 50 | 51 | 52 | 53 | 54 | 2  |
| 3  | 52 | 66 | ∞  | 62 | 65 | 64 | 48 | 48 | 48 | 49 | 50 | 75 | 69 | 77 | 36 | 3  |
| 4  | 51 | 66 | 62 | ∞  | 63 | 62 | 65 | 49 | 49 | 49 | 50 | 77 | 73 | 76 | 71 | 4  |
| 5  | 50 | 50 | 65 | 63 | ∞  | 66 | 65 | 50 | 73 | 76 | 74 | 71 | 77 | 75 | 76 | 5  |
| 6  | 49 | 49 | 64 | 62 | 66 | ∞  | 65 | 51 | 70 | 77 | 73 | 37 | 23 | 30 | 69 | 6  |
| 7  | 48 | 48 | 48 | 65 | 65 | 65 | ∞  | 73 | 74 | 71 | 75 | 26 | 39 | 27 | 72 | 7  |
| 8  | 48 | 47 | 48 | 49 | 50 | 51 | 73 | ∞  | 27 | 31 | 73 | 31 | 30 | 35 | 73 | 8  |
| 9  | 48 | 48 | 48 | 49 | 73 | 70 | 74 | 27 | ∞  | 30 | 38 | 29 | 26 | 36 | 77 | 9  |
| 10 | 48 | 49 | 49 | 49 | 76 | 77 | 71 | 31 | 30 | ∞  | 36 | 35 | 73 | 71 | 74 | 10 |
| 11 | 50 | 50 | 50 | 50 | 74 | 73 | 76 | 73 | 38 | 36 | ∞  | 30 | 26 | 37 | 35 | 11 |
| 12 | 51 | 51 | 75 | 77 | 71 | 37 | 26 | 31 | 29 | 35 | 30 | ∞  | 29 | 36 | 30 | 12 |
| 13 | 52 | 52 | 69 | 73 | 77 | 23 | 39 | 30 | 26 | 73 | 26 | 29 | ∞  | 30 | 36 | 13 |
| 14 | 53 | 63 | 77 | 76 | 75 | 30 | 27 | 35 | 36 | 71 | 37 | 36 | 30 | ∞  | 32 | 14 |
| 15 | 54 | 54 | 36 | 71 | 76 | 69 | 72 | 73 | 77 | 74 | 35 | 30 | 36 | 32 | ∞  | 15 |
|    | 1  | 2  | 3  | 4  | 5  | 6  | 7  | 8  | 9  | 10 | 11 | 12 | 13 | 14 | 15 |    |

MIN(M)

|    | 1   | 2    | 3    | 4   | 5  | 6  | 7  | 8  | 9  | 10 | 11 | 12 | 13 | 14 |    |
|----|-----|------|------|-----|----|----|----|----|----|----|----|----|----|----|----|
| 1  | 7*  | 8*   | 9*   | 10* | 6  | 5  | 11 | 4  | 12 | 3  | 13 | 2  | 14 | 15 | 1  |
| 2  | 8   | 7*   | 9*   | 6   | 10 | 5  | 11 | 12 | 13 | 1  | 14 | 15 | 3  | 4  | 2  |
| 3  | 7*  | 8*   | 9*   | 10  | 11 | 1  | 2  | 5  | 6  | 4  | 12 | 15 | 14 | 13 | 3  |
| 4  | 8*  | 9*   | 10*  | 11  | 1  | 6  | 7  | 2  | 5  | 13 | 3  | 14 | 15 | 12 | 4  |
| 5  | 1*  | 2*   | 8*   | 3   | 6  | 7  | 4  | 11 | 12 | 9  | 15 | 14 | 13 | 10 | 5  |
| 6  | 13  | 14   | 12   | 1   | 2  | 8  | 4  | 7  | 5  | 3  | 11 | 9  | 10 | 15 | 6  |
| 7  | 12  | 14   | 13   | 1   | 2  | 3  | 4  | 6  | 5  | 8  | 15 | 11 | 10 | 9  | 7  |
| 8  | 9   | 13   | 10   | 12  | 14 | 2  | 1  | 3  | 4  | 5  | 11 | 6  | 7  | 15 | 8  |
| 9  | 13  | 8    | 10   | 12  | 14 | 11 | 1  | 2  | 3  | 4  | 5  | 15 | 6  | 7  | 9  |
| 10 | 9   | 8    | 12   | 11  | 1  | 2  | 3  | 4  | 15 | 6  | 7  | 14 | 13 | 5  | 10 |
| 11 | 13  | 120  | 15   | 10  | 14 | 9  | 1  | 2  | 3  | 4  | 8  | 5  | 6  | 7  | 11 |
| 12 | 7   | 9*   | 13*  | 11  | 15 | 8  | 10 | 14 | 6  | 1  | 2  | 3  | 5  | 4  | 12 |
| 13 | 6   | 9*   | 11*  | 12  | 8  | 14 | 15 | 7  | 1  | 2  | 4  | 5  | 10 | 3  | 13 |
| 14 | 7   | 6*   | 13*  | 15  | 8  | 9  | 12 | 11 | 1  | 2  | 4  | 3  | 5  | 10 | 14 |
| 15 | 12  | 14   | 11   | 13  | 1  | 2  | 6  | 7  | 10 | 5  | 3  | 8  | 9  | 4  | 15 |
|    | 1   | 2    | 3    | 4   | 5  | 6  | 7  | 8  | 9  | 10 | 11 | 12 | 13 | 14 | 15 |

$$D_0 = \begin{array}{cccccccccccccccc} -5 & -19 & -14 & -14 & -16 & -42 & -47 & 0 & -4 & -6 & -4 & -3 & -4 & -2 & -24 \\ 1 & 2 & 3 & 4 & 5 & 6 & 7 & 8 & 9 & 10 & 11 & 12 & 13 & 14 & 15 \\ 2 & 3 & 4 & 5 & 6 & 7 & 8 & 9 & 10 & 11 & 12 & 13 & 14 & 15 & 1 \end{array}$$

TRIAL 1. MIN(M)(7 1) = 12

(7 12) → (7 11): -47

(11 13) → (11 12): -4

(12 9) → (12 8): 0

(8 10) → (8 9): 4

(9 8) → (9 7): -3

(7 11 12 8 9): -50

TRIAL 2. MIN(M)(7 2) = 14

(7 14) → (7 13): -46

(13 9) → (13 8): -3

(8 13) → (8 12): 4

(12 7) → (12 6): -3

(6 12) → (6 11): -28

(11 15) → (11 14): 5

(14 6) → (14 5): -2

(5 8) → (5 7): -16

(7 13 8 12 6 11 14 5): -89

TRIAL 3. MIN(M)(6 1) = 13

(6 13) → (6 12): -42

(12 7) → (12 6): -3

(6 12): -45

The smallest-valued cycle is obtained in trial 2: (7 13 8 12 6 11 14 5): -89

$D_1 = $

| -5 | -19 | -14 | -14 | 0 | -14 | -1 | -3 | -4 | -6 | -5 | 0 | -4 | -3 | -24 |
|---|---|---|---|---|---|---|---|---|---|---|---|---|---|---|
| 1 | 2 | 3 | 4 | 5 | 6 | 7 | 8 | 9 | 10 | 11 | 12 | 13 | 14 | 15 |
| 2 | 3 | 4 | 5 | 8 | 12 | 14 | 13 | 10 | 11 | 15 | 7 | 9 | 6 | 1 |

$D_1^{-1} = $

| 1 | 2 | 3 | 4 | 5 | 6 | 7 | 8 | 9 | 10 | 11 | 12 | 13 | 14 | 15 |
|---|---|---|---|---|---|---|---|---|---|---|---|---|---|---|
| 15 | 1 | 2 | 3 | 4 | 14 | 12 | 5 | 13 | 9 | 10 | 6 | 8 | 7 | 11 |

The initial arcs for our trials are: (15 12), (15 14), (15 11).

TRIAL 1. MIN(M)(15 1) = 12

(15 12) → (15 6): -24

(6 13) → (6 8): -14

(8 9) → (8 13): -3

(13 11) → (13 10): 0

(10 8) → (10 5): -5

(5 1) → (5 15): -46

TRIAL 2. MIN(15 2) = 14

(15 14) → (15 7): -22

(7 12) → (7 6): -1

(6 13) → (6 8): -14

(8 9) → (8 13): -3

(13 11) → (13 10): 0

(10 8) → (10 5): -5

(5 1) → (5 15): 0

(15 7 6 8 13 10 5): -45

TRIAL 3. MIN(M)(15 3) = 11

(15 11) → (15 10): -19

(10 8) → (10 5): -5

(5 1) → (5 15): 0

(15 10 5): -24

TRIAL 1 yields the smallest-valued cycle: (15 6 8 13 10 5): -46.

$D_2 =$

| | -5 | -19 | -14 | -14 | 0 | 0 | -1 | 0 | -4 | -1 | -9 | 0 | -3 | -3 | 0 |
|---|---|---|---|---|---|---|---|---|---|---|---|---|---|---|---|
| | 1 | 2 | 3 | 4 | 5 | 6 | 7 | 8 | 9 | 10 | 11 | 12 | 13 | 14 | 15 |
| | 2 | 3 | 4 | 5 | 1 | 13 | 14 | 9 | 10 | 8 | 15 | 7 | 11 | 6 | 12 |

$D_2^{-1} =$

| | 1 | 2 | 3 | 4 | 5 | 6 | 7 | 8 | 9 | 10 | 11 | 12 | 13 | 14 | 15 |
|---|---|---|---|---|---|---|---|---|---|---|---|---|---|---|---|
| | 5 | 1 | 2 | 3 | 4 | 14 | 12 | 10 | 8 | 9 | 13 | 15 | 6 | 7 | 11 |

Our next set of initial arcs are: (2 8), (2 7*), (2 9*), (2 6).

TRIAL 1.  MIN(M)(2 1) = 8

(2 8) → (2 10): -19

(10 12) → (10 15): 5

(15 14) → (15 7): 2

(7 13) → (7 6): 21

TRIAL 2(a)*.  MIN(M)(2 2) = 7

(2 7) → (2 12): -18

(12 9) → (12 8): 3

(8 13) → (8 6): 3

(6 14) → (6 7): 7

(7 1) → (7 5): 21

NO CYCLE.

TRIAL 2(b)*  MIN(M)(2 3) = 9

(2 9) → (2 8): -18

(8 13) → (8 6): 3

(6 14) → (6 7): 7

(7 1) → (7 5): 21

NO CYCLE.

TRIAL 3.  MIN(M)(2 4) = 6

(2 6) → (2 14): -16

(14 13) → (14 6): 3

(6 14) → (6 7): 7

(7 1) → (7 5): 21

NO CYCLE.

We thus cannot proceed beyond $D_2$. We thus must apply the modified F-W algorithm to see if we can obtain a negative cycle in $D_2^{-1}M^-$.

M

|    | 1  | 2  | 3  | 4  | 5  | 6  | 7  | 8  | 9  | 10 | 11 | 12 | 13 | 14 | 15 |    |
|----|----|----|----|----|----|----|----|----|----|----|----|----|----|----|----|----|
| 1  | ∞  | 53 | 52 | 51 | 50 | 49 | 48 | 48 | 48 | 48 | 50 | 51 | 52 | 53 | 54 | 1  |
| 2  | 53 | ∞  | 66 | 66 | 50 | 49 | 48 | 47 | 48 | 49 | 50 | 51 | 52 | 53 | 54 | 2  |
| 3  | 52 | 66 | ∞  | 62 | 65 | 64 | 48 | 48 | 48 | 49 | 50 | 75 | 69 | 77 | 76 | 3  |
| 4  | 51 | 66 | 62 | ∞  | 63 | 62 | 65 | 49 | 49 | 49 | 50 | 77 | 73 | 76 | 71 | 4  |
| 5  | 50 | 50 | 65 | 63 | ∞  | 66 | 65 | 50 | 73 | 76 | 74 | 71 | 77 | 75 | 76 | 5  |
| 6  | 49 | 49 | 64 | 62 | 66 | ∞  | 65 | 51 | 70 | 77 | 73 | 37 | 23 | 30 | 69 | 6  |
| 7  | 48 | 48 | 48 | 65 | 65 | 65 | ∞  | 73 | 74 | 71 | 75 | 26 | 39 | 27 | 72 | 7  |
| 8  | 48 | 47 | 48 | 49 | 50 | 51 | 73 | ∞  | 27 | 31 | 73 | 31 | 30 | 35 | 73 | 8  |
| 9  | 48 | 48 | 48 | 49 | 73 | 70 | 74 | 27 | ∞  | 30 | 38 | 29 | 26 | 36 | 77 | 9  |
| 10 | 48 | 49 | 49 | 49 | 76 | 77 | 71 | 31 | 30 | ∞  | 36 | 35 | 73 | 71 | 74 | 10 |
| 11 | 50 | 50 | 50 | 50 | 74 | 73 | 76 | 73 | 38 | 36 | ∞  | 30 | 26 | 37 | 35 | 11 |
| 12 | 51 | 51 | 75 | 77 | 71 | 37 | 26 | 31 | 29 | 35 | 30 | ∞  | 29 | 36 | 30 | 12 |
| 13 | 52 | 52 | 69 | 73 | 77 | 23 | 39 | 30 | 26 | 73 | 26 | 29 | ∞  | 30 | 36 | 13 |
| 14 | 53 | 63 | 77 | 76 | 75 | 30 | 27 | 35 | 36 | 71 | 37 | 36 | 30 | ∞  | 32 | 14 |
| 15 | 54 | 54 | 36 | 71 | 76 | 69 | 72 | 73 | 77 | 74 | 35 | 30 | 36 | 32 | ∞  | 15 |
|    | 1  | 2  | 3  | 4  | 5  | 6  | 7  | 8  | 9  | 10 | 11 | 12 | 13 | 14 | 15 |    |

$$D_2^{-1}M^-$$

| | 2 | 3 | 4 | 5 | 1 | 13 | 14 | 9 | 10 | 8 | 15 | 7 | 11 | 6 | 12 | |
|---|---|---|---|---|---|---|---|---|---|---|---|---|---|---|---|---|
| | 1 | 2 | 3 | 4 | 5 | 6 | 7 | 8 | 9 | 10 | 11 | 12 | 13 | 14 | 15 | |
| 1 | 0 | -1 | -2 | -3 | ∞ | -1 | 0 | -5 | -5 | -5 | 1 | -5 | -3 | -4 | -2 | 1 |
| 2 | ∞ | 0 | 0 | -16 | -13 | -14 | -13 | -18 | -17 | -19 | -12 | -18 | -16 | -17 | -15 | 2 |
| 3 | 4 | ∞ | 0 | 3 | -10 | 7 | 15 | -14 | -13 | -14 | 14 | -14 | -12 | 2 | 13 | 3 |
| 4 | -1 | -1 | ∞ | 0 | -12 | 10 | 13 | -14 | -14 | -14 | 8 | 2 | -13 | -1 | 14 | 4 |
| 5 | 0 | 15 | 13 | ∞ | 0 | 27 | 25 | 23 | 26 | 0 | 26 | 15 | 24 | 16 | 21 | 5 |
| 6 | 26 | 41 | 39 | 43 | 19 | 0 | 7 | 47 | 54 | 28 | 46 | 42 | 50 | ∞ | 14 | 6 |
| 7 | 21 | 21 | 38 | 38 | 21 | 12 | 0 | 47 | 44 | 46 | 45 | ∞ | 48 | 38 | -1 | 7 |
| 8 | 21 | 21 | 22 | 23 | 21 | 3 | 8 | 0 | 4 | ∞ | 46 | 46 | 46 | 24 | 4 | 8 |
| 9 | 18 | 18 | 19 | 43 | 18 | -4 | 6 | ∞ | 0 | -3 | 47 | 44 | 8 | 40 | -1 | 9 |
| 10 | 18 | 18 | 18 | 76 | 17 | 42 | 41 | -1 | ∞ | 0 | 43 | 40 | 5 | 46 | 4 | 10 |
| 11 | 15 | 15 | 15 | 39 | 15 | -9 | 2 | 3 | 1 | 38 | 0 | 41 | ∞ | 38 | -5 | 11 |
| 12 | 45 | 49 | 51 | 45 | 25 | 3 | 10 | 3 | 9 | 5 | 4 | 0 | 4 | 11 | ∞ | 12 |
| 13 | 26 | 43 | 47 | 51 | 26 | ∞ | 4 | 0 | 46 | 4 | 10 | 13 | 0 | -3 | 3 | 13 |
| 14 | 33 | 47 | 46 | 45 | 23 | 0 | ∞ | 6 | 41 | 5 | 2 | -3 | 7 | 0 | 6 | 14 |
| 15 | 24 | 6 | 41 | 46 | 24 | 6 | 2 | 47 | 44 | 43 | ∞ | 42 | 5 | 39 | 30 | 15 |
| | 1 | 2 | 3 | 4 | 5 | 6 | 7 | 8 | 9 | 10 | 11 | 12 | 13 | 14 | 15 | |

j = 2

(1 2)(2 4) = (1 4): -17; (1 2)(2 6) = (1 6): -15; (1 2)(2 7) = (1 7): -15; (1 2)(2 9) = (1 9): -18;

(1 2)(2 10) = (1 10): -20; (1 2)(2 11) = (1 11): -13; (1 2)(2 12) = (1 12): -19;

(1 2)(2 13) = (1 13): -17; (1 2)(2 14) = (1 14): -18; (1 2)(2 15) = (1 15): -16

j = 4

(1 4)(4 1) = (1 1): -18

CYCLE (1 2 4): -18

$$D_3^{-1}M^-$$

|    | 3 | 5 | 4 | 2 | 1 | 13 | 14 | 9 | 10 | 8 | 15 | 7 | 11 | 6 | 12 |    |
|----|---|---|---|---|---|----|----|---|----|---|----|---|----|---|----|----|
|    | 1 | 2 | 3 | 4 | 5 | 6  | 7  | 8 | 9  | 10| 11 | 12| 13 | 14| 15 |    |
| 1  | 0 | -2 | -1 | 1 | ∞ | 0 | 1 | -4 | -4 | -4 | 2 | -4 | -2 | -3 | -1 | 1 |
| 2  | 0 | 0 | 0 | ∞ | 3 | 2 | 3 | -2 | -1 | -3 | 4 | --2 | 0 | -1 | 1 | 2 |
| 3  | ∞ | 3 | 0 | 4 | -10 | 7 | 15 | -14 | -13 | -14 | 14 | -14 | -12 | 2 | 13 | 3 |
| 4  | 0 | 1 | ∞ | 0 | -11 | 11 | 14 | -13 | -13 | -13 | 9 | 3 | -12 | 0 | 15 | 4 |
| 5  | 15 | ∞ | 13 | 0 | 0 | 27 | 25 | 23 | 26 | 0 | 26 | 15 | 24 | 16 | 21 | 5 |
| 6  | 41 | 43 | 39 | 26 | 19 | 0 | 7 | 47 | 54 | 28 | 46 | 42 | 50 | ∞ | 14 | 6 |
| 7  | 21 | 38 | 38 | 21 | 21 | 12 | 0 | 47 | 44 | 46 | 45 | ∞ | 48 | 38 | -1 | 7 |
| 8  | 21 | 23 | 22 | 21 | 21 | 3 | 8 | 0 | 4 | ∞ | 46 | 46 | 46 | 24 | 4 | 8 |
| 9  | 18 | 43 | 19 | 18 | 18 | -4 | 6 | ∞ | 0 | -3 | 47 | 44 | 8 | 40 | -1 | 9 |
| 10 | 18 | 76 | 18 | 18 | 17 | 42 | 41 | -1 | ∞ | 0 | 43 | 40 | 5 | 46 | 4 | 10 |
| 11 | 15 | 39 | 15 | 15 | 15 | -9 | 2 | 3 | 1 | 38 | 0 | 41 | ∞ | 38 | -5 | 11 |
| 12 | 49 | 45 | 51 | 45 | 25 | 3 | 10 | 3 | 9 | 5 | 4 | 0 | 4 | 11 | ∞ | 12 |
| 13 | 43 | 51 | 47 | 26 | 26 | ∞ | 4 | 0 | 46 | 4 | 10 | 13 | 0 | -3 | 3 | 13 |
| 14 | 47 | 45 | 46 | 33 | 23 | 0 | ∞ | 6 | 41 | 5 | 2 | -3 | 7 | 0 | 6 | 14 |
| 15 | 6 | 46 | 41 | 24 | 24 | 6 | 2 | 47 | 44 | 43 | ∞ | 42 | 5 | 39 | 30 | 15 |
|    | 1 | 2 | 3 | 4 | 5 | 6 | 7 | 8 | 9 | 10 | 11 | 12 | 13 | 14 | 15 |    |

j = 1

(1 3)(3 8) = (1 8): -15; (1 3)(3 9) = (1 9): -14; (1 3)(3 10) = (1 10): -15;

(1 3)(3 12) = (1 12): -15; (1 3)(3 13) = (1 13): -13

j = 8

(1 8)(8 6) = (1 6): <u>-12</u>; (1 8)(8 7) = (1 7): <u>-7</u>; (1 8)(8 15) = (1 15): -11; (3 8)(8 6) = (3 6): <u>-11</u>;

(3 8)(8 7) = (3 7): <u>-6</u>; (3 8)(8 15) = (3 15): -10; (4 8)(8 6) = (4 6): <u>-10</u>; (4 8)(8 7) = (4 7): <u>-5</u>;

j = 9

(2 9)(9 6) = (2 6): <u>-5</u>; (2 9)(9 10) = (2 10): -4; (3 9)(9 6) = (3 6): <u>-17</u>; (3 9)(9 7) = (3 7): <u>-7</u>;

(3 9)(9 10) = (3 10): -16; (4 9)(9 6) = (4 6): <u>-17</u>; (4 9)(9 10) = (4 10): -16

j = 10

(3 10)(10 15) = (3 15): -12; (4 10)(10 15) = (4 15): -12

j = 12

(1 12)(12 11) = (1 11): <u>-11</u>; (3 12)(12 6) = (3 6): <u>-11</u>; (3 12)(12 11) = (3 11): <u>-10</u>;

(3 12)(12 14) = (3 14): -3

j = 13

(4 13)(13 7) = (4 7): <u>-8</u>; (4 13)(13 11) = (4 11): <u>-2</u>

j = 15

(1 15)(15 1) + (1 1): -5

CYCLE (1 3 8 15): -5

$P_{20}$

|    | 1 | 2 | 3 | 4 | 5 | 6  | 7  | 8 | 9 | 10 | 11 | 12 | 13 | 14 | 15 |    |
|----|---|---|---|---|---|----|----|---|---|----|----|----|----|----|----|----|
| 1  |   |   | 1 |   |   | 8  | 8  | 3 | 3 | 3  | 12 | 3  | 3  |    | 8  | 1  |
| 2  |   |   |   |   |   | 9  |    |   | 2 | 9  |    |    |    |    |    | 2  |
| 3  |   |   |   |   |   | 12 | 9  | 3 | 3 | 9  | 12 | 3  |    | 12 | 10 | 3  |
| 4  |   |   |   |   |   | 9  | 13 | 4 | 4 | 9  | 13 |    | 4  |    | 10 | 4  |
| 5  |   |   |   |   |   |    |    |   |   |    |    |    |    |    |    | 5  |
| 6  |   |   |   |   |   |    |    |   |   |    |    |    |    |    |    | 6  |
| 7  |   |   |   |   |   |    |    |   |   |    |    |    |    |    |    | 7  |
| 8  |   |   |   |   |   |    |    |   |   |    |    |    |    |    |    | 8  |
| 9  |   |   |   |   |   |    |    |   |   |    |    |    |    |    |    | 9  |
| 10 |   |   |   |   |   |    |    |   |   |    |    |    |    |    |    | 10 |
| 11 |   |   |   |   |   |    |    |   |   |    |    |    |    |    |    | 11 |
| 12 |   |   |   |   |   |    |    |   |   |    |    |    |    |    |    | 12 |
| 13 |   |   |   |   |   |    |    |   |   |    |    |    |    |    |    | 13 |
| 14 |   |   |   |   |   |    |    |   |   |    |    |    |    |    |    | 14 |
| 15 |   |   |   |   |   |    |    |   |   |    |    |    |    |    |    | 15 |
|    | 1 | 2 | 3 | 4 | 5 | 6  | 7  | 8 | 9 | 10 | 11 | 12 | 13 | 14 | 15 |    |

$$D_4^{-1}M^-$$

|   | 4 | 5 | 9 | 2 | 1 | 13 | 14 | 12 | 10 | 8 | 15 | 7 | 11 | 6 | 3 |   |
|---|---|---|---|---|---|---|---|---|---|---|---|---|---|---|---|---|
|   | 1 | 2 | 3 | 4 | 5 | 6 | 7 | 8 | 9 | 10 | 11 | 12 | 13 | 14 | 15 |   |
| 1 | 0 | -1 | -3 | 2 | ∞ | 1 | 2 | 0 | -3 | -3 | 3 | -3 | -1 | -2 | 1 | 1 |
| 2 | 0 | 0 | -2 | ∞ | 3 | 2 | 3 | 1 | -1 | -3 | 4 | --2 | 0 | -1 | 0 | 2 |
| 3 | 14 | 17 | 0 | 18 | 4 | 21 | 29 | 27 | 1 | 0 | 0 | 0 | 2 | 16 | ∞ | 3 |
| 4 | ∞ | 1 | -13 | 0 | -11 | 11 | 14 | 15 | -13 | -13 | 9 | 3 | -12 | 0 | 0 | 4 |
| 5 | 13 | ∞ | 23 | 0 | 0 | 27 | 25 | 21 | 26 | 0 | 26 | 15 | 24 | 16 | 15 | 5 |
| 6 | 39 | 43 | 47 | 26 | 19 | 0 | 7 | 14 | 54 | 28 | 46 | 42 | 50 | ∞ | 41 | 6 |
| 7 | 38 | 38 | 47 | 21 | 21 | 12 | 0 | -1 | 44 | 46 | 45 | ∞ | 48 | 38 | 21 | 7 |
| 8 | 18 | 19 | -4 | 17 | 17 | -1 | 4 | 0 | 0 | ∞ | 42 | 42 | 42 | 20 | 17 | 8 |
| 9 | 19 | 43 | ∞ | 18 | 18 | -4 | 6 | -1 | 0 | -3 | 47 | 44 | 8 | 40 | 18 | 9 |
| 10 | 18 | 76 | -1 | 18 | 17 | 42 | 41 | 4 | ∞ | 0 | 43 | 40 | 5 | 46 | 18 | 10 |
| 11 | 15 | 39 | 3 | 15 | 15 | -9 | 2 | -5 | 1 | 38 | 0 | 41 | ∞ | 38 | 15 | 11 |
| 12 | 51 | 45 | 3 | 45 | 25 | 3 | 10 | ∞ | 9 | 5 | 4 | 0 | 4 | 11 | 49 | 12 |
| 13 | 47 | 51 | 0 | 26 | 26 | ∞ | 4 | 3 | 46 | 4 | 10 | 13 | 0 | -3 | 43 | 13 |
| 14 | 46 | 45 | 6 | 33 | 23 | 0 | ∞ | 6 | 41 | 5 | 2 | -3 | 7 | 0 | 47 | 14 |
| 15 | 35 | 40 | 41 | 18 | 18 | 0 | -4 | 24 | 38 | 37 | ∞ | 36 | -1 | 33 | 0 | 15 |
|   | 1 | 2 | 3 | 4 | 5 | 6 | 7 | 8 | 9 | 10 | 11 | 12 | 13 | 14 | 15 |   |

j = 1

(4 3)(3 12) = (4 12): -13; (4 3)(3 13) = (4 13): -11; (8 3)(3 9) = (8 9): -3;

(8 3)(3 13) = (8 13): -2

j = 8

(9 8)(8 6) = (9 6): -5; (11 8)(8 3) = (11 3): -9; (11 8)(8 7) = (11 7): -1

j = 9

(1 9)(9 6) = (1 6): -7; (1 9)(9 8) = (1 8): -4; (1 9)(9 10) = (1 10): -6; (2 9)(9 6) = (2 6): -5;

(2 9)(9 8) = (2 8): -2; (2 9)(9 10) = (2 10): -4

j = 10

(4 10)(10 8) = (4 8): -9; (4 10)(10 13) = (4 13): -8

j = 12

(4 12)(12 6) = (4 6): -10; (4 12)(12 7) = (4 7): -3; (4 12)(12 11) = (4 11): -9

j = 13

(4 13)(13 7) = (4 7): -8

$P_{20}$

|   | 1 | 2 | 3 | 4 | 5 | 6 | 7 | 8 | 9 | 10 | 11 | 12 | 13 | 14 | 15 |   |
|---|---|---|---|---|---|---|---|---|---|----|----|----|----|----|----|---|
| 1 |   |   |   |   |   | 9 |   | 9 | 1 | 9  |    |    |    |    |    | 1 |
| 2 |   |   |   |   |   | 9 |   | 9 | 2 | 9  |    |    |    |    |    | 2 |
| 3 |   |   |   |   |   |   |   |   |   |    |    |    |    |    |    | 3 |
| 4 |   |   | 4 |   |   | 12| 13| 10|   | 4  | 12 | 3  | 10 |    |    | 4 |
| 5 |   |   |   |   |   |   |   |   |   |    |    |    |    |    |    | 5 |
| 6 |   |   |   |   |   |   |   |   |   |    |    |    |    |    |    | 6 |
| 7 |   |   |   |   |   |   |   |   |   |    |    |    |    |    |    | 7 |
| 8 |   |   | 8 |   |   |   |   |   | 3 |    |    |    | 3  |    |    | 8 |
| 9 |   |   |   |   |   | 8 |   | 9 |   |    |    |    |    |    |    | 9 |
| 10|   |   |   |   |   |   |   |   |   |    |    |    |    |    |    | 10|
| 11|   |   | 8 |   |   |   | 8 | 11|   |    |    |    |    |    |    | 11|
| 12|   |   |   |   |   |   |   |   |   |    |    |    |    |    |    | 12|
| 13|   |   |   |   |   |   |   |   |   |    |    |    |    |    |    | 13|
| 14|   |   |   |   |   |   |   |   |   |    |    |    |    |    |    | 14|
| 15|   |   |   |   |   |   |   |   |   |    |    |    |    |    |    | 15|
|   | 1 | 2 | 3 | 4 | 5 | 6 | 7 | 8 | 9 | 10 | 11 | 12 | 13 | 14 | 15 |   |

$$D_4^{-1} M^-$$

|    | 1 | 2  | 3   | 4  | 5   | 6   | 7  | 8  | 9   | 10 | 11 | 12  | 13  | 14 | 15 |    |
|----|---|----|-----|----|-----|-----|----|----|-----|----|----|-----|-----|----|----|----|
| 1  | 0 | -1 | *-3*  |    | ∞   | -7  |    | -4 | -3  | -3 |    | *-3*  | *-1*  | *-2* |    | 1  |
| 2  | 0 | 0  | -2  | ∞  |     | -5  |    | -2 | *-1*  | -4 |    | -2  |     | *-1* |    | 2  |
| 3  |   |    | 0   |    |     |     |    |    |     |    | 0  |     |     |    | ∞  | 3  |
| 4  | ∞ |    | *-13* | 0  | -11 | -10 | -8 | -9 | *-13* | -9 | -9 | *-13* | -12 |    |    | 4  |
| 5  |   | ∞  |     | 0  | 0   |     |    |    |     |    |    |     |     |    |    | 5  |
| 6  |   |    |     |    |     | 0   | 7  |    |     |    |    |     |     | ∞  |    | 6  |
| 7  |   |    |     |    |     |     | 0  | -1 |     |    |    | ∞   |     |    |    | 7  |
| 8  |   |    | *-4*  |    | *-1*  |     |    | 0  | 0   | ∞  | -2 |     |     |    |    | 8  |
| 9  |   |    | ∞   |    | -4  |     | -1 |    | 0   | -3 |    |     |     | 18 |    | 9  |
| 10 |   |    | -1  |    |     |     |    |    | ∞   | 0  |    |     | -8  |    |    | 10 |
| 11 |   |    |     |    |     | -9  |    | -5 |     |    | 0  |     | ∞   |    |    | 11 |
| 12 |   |    |     |    |     |     |    | ∞  |     | 5  |    | 0   |     |    |    | 12 |
| 13 |   |    |     |    |     | ∞   |    |    |     |    |    |     | 0   | -3 |    | 13 |
| 14 |   |    |     |    |     |     | ∞  |    |     |    |    | -2  |     | 0  |    | 14 |
| 15 |   |    |     |    |     |     |    | -4 |     |    | ∞  |     | -1  |    | 0  | 15 |
|    | 1 | 2  | 3   | 4  | 5   | 6   | 7  | 8  | 9   | 10 | 11 | 12  | 13  | 14 | 15 |    |

j = 8

(1 8)(8 3) = (1 3): -8; (2 8)(8 3) = (2 3): -6

We now try to extend each of these paths:

(1 9 8 3)(3 10) = (1 10): -8; (1 9 8 3)(3 12) = (1 12): -8; (1 9 8 3)(3 13) = (1 13): -8

(2 9 8 3)(3 10) = (2 10): -6; (2 9 8 3)(3 12) = (2 12): -6; (2 9 8 3)(3 13) = (2 9 8 3 13): -6

None of the above path can be extended. It follows that $D_4$ is a minimally-valued derangement obtained in PHASE 1.

$$D_4 = (1\ 4\ 2\ 5)(3\ 9\ 10\ 8\ 12\ 7\ 14\ 6\ 13\ 11\ 15): 560$$

We now use a tree whose initial point is $4$ to obtain a tour from $D_4$.

$$1 \to 4$$
$$\downarrow -2$$
$$3 \to 9 \leftarrow 10$$
$$\downarrow 2$$
$$1$$

Our tour is

$$T = T_{UPPERBOUND} = (4\ 9\ 10\ 8\ 12\ 7\ 14\ 6\ 13\ 11\ 15\ 3\ 1\ 5\ 2): 562.$$

We now give a conjecture, Condition (B), for odd $n$, analogous to Condition (A) for even $n$

Condition(B) .

Let $T = (a_1\ a_2\ \ldots\ a_n)$ be an upper bound for $T_{OPT}$ where $n$ is odd. Without loss of generality, let $a_n$ be a fixed point while $|\prod_{j=1}^{j=\frac{n-1}{2}} (a_{2j-1}\ a_{2j})| > |\prod_{j=1}^{j=\frac{n-1}{2}} (a_{2j}\ a_{2j+1})|$ where $\sigma_T = \prod_{j=1}^{j=\frac{n-1}{2}} (a_{2j}\ a_{2j+1})$. Let P be an acceptable or 2-circuit path in $\sigma_T^{-1} M^-$. Then the following hold:

(1) If $a_n$ isn't an interior point of $P$, then $|P| < |T| - |(a_n\ a_1)| - |\sigma_T|$.

(2) If $a_n$ is an interior point of $P$, then $|P| < |T| - |\sigma_T|$.

We now apply condition (B) to PHASE 2 of our algorithm.

PHASE 2

The arc of smallest value in $T$ is (6 13): 23. We thus rewrite $T$ with 6 as its last point.

$$T = (13\ 11\ 15\ 3\ 1\ 5\ 2\ 4\ 9\ 10\ 8\ 12\ 7\ 14\ 6).$$

From $T$, we obtain two $APM$'s (almost perfect matchings) and a fixed point 6.

$(13\ 11)\quad (15\ 3)\quad (1\ 5)\quad (2\ 4)\quad (9\ 10)\quad (8\ 12)\quad (7\ 14): 266$

$\quad\quad (11\ 15)\quad (3\ 1)\quad (5\ 2)\quad (4\ 9)\quad (10\ 8)\quad (12\ 7)\quad (14\ 6): 273$

$\sigma_T = (13\ 11)(15\ 3)(1\ 5)(2\ 4)(9\ 10)(8\ 12)(7\ 14): 532$. We also must add specific bounds for arcs out of $6$. On $T_{UPPERBOUND}$, $|(6\ 13)| = 23$. Thus, in applying the modified F-W algorithm to $\sigma_T^{-1}M^-$, the following hold.:

(1) the maximum value of each arc *not containing* 6 is 562 - 23 - 532 - 1 = 6;

(2) the maximum value of an arc whose *initial point* is 6 is $|(6\ 13)| + 6 = 29$.

(1) and (2) imply the following:

(a) a path *not containing* the point 6 as an *interior point* can have a value no greater than 6.

(b) a path *containing* 6 as an *interior point* cannot have a value greater than 29.

Furthermore, before we replace a path $(a\ b)$ by a smaller-valued path with the same initial and terminal points, we place each such path into a list that will be used later to obtain *all* possible acceptable cycles.

We now construct $\sigma_T^{-1}M^-$.

M

|   | 1 | 2 | 3 | 4 | 5 | 6 | 7 | 8 | 9 | 10 | 11 | 12 | 13 | 14 | 15 |   |
|---|---|---|---|---|---|---|---|---|---|----|----|----|----|----|----|---|
| 1 | ∞ | 53 | 52 | 51 | 50 | 49 | 48 | 48 | 48 | 48 | 50 | 51 | 52 | 53 | 54 | 1 |
| 2 | 53 | ∞ | 66 | 66 | 50 | 49 | 48 | 47 | 48 | 49 | 50 | 51 | 52 | 53 | 54 | 2 |
| 3 | 52 | 66 | ∞ | 62 | 65 | 64 | 48 | 48 | 48 | 49 | 50 | 75 | 69 | 77 | 36 | 3 |
| 4 | 51 | 66 | 62 | ∞ | 63 | 62 | 65 | 49 | 49 | 49 | 50 | 77 | 73 | 76 | 71 | 4 |
| 5 | 50 | 50 | 65 | 63 | ∞ | 66 | 65 | 50 | 73 | 76 | 74 | 71 | 77 | 75 | 76 | 5 |
| 6 | 49 | 49 | 64 | 62 | 66 | ∞ | 65 | 51 | 70 | 77 | 73 | 37 | 23 | 30 | 69 | 6 |
| 7 | 48 | 48 | 48 | 65 | 65 | 65 | ∞ | 73 | 74 | 71 | 75 | 26 | 39 | 27 | 72 | 7 |
| 8 | 48 | 47 | 48 | 49 | 50 | 51 | 73 | ∞ | 27 | 31 | 73 | 31 | 30 | 35 | 73 | 8 |
| 9 | 48 | 48 | 48 | 49 | 73 | 70 | 74 | 27 | ∞ | 30 | 38 | 29 | 26 | 36 | 77 | 9 |
| 10 | 48 | 49 | 49 | 49 | 76 | 77 | 71 | 31 | 30 | ∞ | 36 | 35 | 73 | 71 | 74 | 10 |
| 11 | 50 | 50 | 50 | 50 | 74 | 73 | 76 | 73 | 38 | 36 | ∞ | 30 | 26 | 37 | 35 | 11 |
| 12 | 51 | 51 | 75 | 77 | 71 | 37 | 26 | 31 | 29 | 35 | 30 | ∞ | 29 | 36 | 30 | 12 |
| 13 | 52 | 52 | 69 | 73 | 77 | 23 | 39 | 30 | 26 | 73 | 26 | 29 | ∞ | 30 | 36 | 13 |
| 14 | 53 | 63 | 77 | 76 | 75 | 30 | 27 | 35 | 36 | 71 | 37 | 36 | 30 | ∞ | 32 | 14 |
| 15 | 54 | 54 | 36 | 71 | 76 | 69 | 72 | 73 | 77 | 74 | 35 | 30 | 36 | 32 | ∞ | 15 |
|   | 1 | 2 | 3 | 4 | 5 | 6 | 7 | 8 | 9 | 10 | 11 | 12 | 13 | 14 | 15 |   |

$$\sigma_T^{-1} M^-$$

|   | 5 | 4 | 15 | 2 | 1 | 6 | 14 | 12 | 10 | 9 | 13 | 8 | 11 | 7 | 3 |   |
|---|---|---|---|---|---|---|---|---|---|---|---|---|---|---|---|---|
|   | 1 | 2 | 3 | 4 | 5 | 6 | 7 | 8 | 9 | 10 | 11 | 12 | 13 | 14 | 15 |   |
| 1 | 0 | 1 | 4 | 3 | ∞ | -1 | 3 | 1 | -2 | -2 | 2 | -2 | 0 | -2 | 2 | 1 |
| 2 | -16 | 0 | -12 | ∞ | -13 | -17 | -13 | -15 | -17 | -18 | -14 | -19 | -16 | -18 | 0 | 2 |
| 3 | 29 | 26 | 0 | 30 | 16 | 28 | 41 | 39 | 13 | 12 | 33 | 12 | 14 | 12 | ∞ | 3 |
| 4 | -3 | ∞ | 5 | 0 | -15 | -4 | 10 | 11 | -17 | -17 | 6 | -17 | -16 | -1 | -4 | 4 |
| 5 | ∞ | 13 | 26 | 0 | 0 | 16 | 25 | 21 | 26 | 23 | 27 | 0 | 24 | 15 | 15 | 5 |
| 6 | 66 | 62 | 69 | 49 | 49 | ∞ | 30 | 37 | 77 | 70 | 23 | 51 | 73 | 65 | 64 | 6 |
| 7 | 38 | 38 | 45 | 21 | 21 | 39 | 0 | -1 | 44 | 47 | 12 | 46 | 48 | ∞ | 21 | 7 |
| 8 | 19 | 18 | 42 | 16 | 17 | 20 | 4 | 0 | 0 | -4 | -1 | ∞ | 42 | 42 | 17 | 8 |
| 9 | 43 | 19 | 47 | 18 | 18 | 40 | 6 | -1 | 0 | ∞ | -4 | -3 | 8 | 44 | 18 | 9 |
| 10 | 46 | 19 | 44 | 19 | 18 | 47 | 41 | 5 | ∞ | 0 | 43 | 1 | 6 | 41 | 19 | 10 |
| 11 | 48 | 24 | 9 | 24 | 24 | 47 | 11 | 4 | 10 | 12 | 0 | 47 | ∞ | 50 | 24 | 11 |
| 12 | 40 | 46 | -1 | 20 | 20 | 6 | 5 | ∞ | 4 | -2 | -2 | 0 | -1 | -5 | 44 | 12 |
| 13 | 51 | 47 | 10 | 26 | 26 | -3 | 4 | 3 | 47 | 0 | ∞ | -6 | 0 | 13 | 43 | 13 |
| 14 | 48 | 49 | 5 | 36 | 26 | 3 | ∞ | 9 | 44 | 9 | 3 | 8 | 10 | 0 | 50 | 14 |
| 15 | 40 | 35 | ∞ | 18 | 18 | 33 | -4 | -6 | 38 | 41 | 0 | 37 | -1 | 36 | 0 | 15 |
|   | 1 | 2 | 3 | 4 | 5 | 6 | 7 | 8 | 9 | 10 | 11 | 12 | 13 | 14 | 15 |   |

j = 1

(2 1)(1 2) = (2 2): -15

CYCLE (2 1): -15

(2 1)(1 3) = (2 3): -12; (2 1)(1 6) = (2 6): -17; (2 1)(1 7) = (2 7): -13;

(2 1)(1 8) = (2 8): -15; (2 1)(1 9) = (2 9): -18; (2 1)(1 10) = (2 10): -18;

(2 1)(1 11) = (2 11): -14; (2 1)(1 12) = (2 12): -18; (2 1)(1 13) = (2 13): -16;

(2 1)(1 14) = (2 14): -18; (2 1)(1 15) = (2 15): -14; (4 1)(1 3) = (4 3): 0;

(4 1)(1 4) = (4 4): 0

CYCLE (4 1): 0

(4 1)(1 3) = (4 3): 0;

(4 1)(1 4) = (4 4): 0

CYCLE (4 1): 0

(4 1)(1 7) = (4 7): 0; (4 1)(1 8) = (4 8): -2; (4 1)(1 11) = (4 11): -1; (4 1)(1 14) = (4 14): -5;

*(2 1)(1 4) = (2 4): -13; (4 1)(1 2) = (4 2): -2*

j = 2

(1 2)(2 1) = (1 1): -15

CYCLE (1 2): -15

(1 2)(2 3) = (1 3): -11; (1 2)(2 6) = (1 6): -16; (1 2)(2 7) = (1 7): -12; (1 2)(2 8) = (1 8): -14;

(1 2)(2 9) = (1 9): -17; (1 2)(2 10) = (1 10): -17; (1 2)(2 11) = (1 11): -13;

(1 2)(2 12)= (1 12): -18; (1 2)(2 13) = (1 13): -15; (1 2)(2 14) = (1 14): -17;

(5 2)(2 3) = (5 3): 1;

(5 2)(2 5) = (5 5): 0

CYCLE (5 2): 0

*(1 2)(2 5) = (1 5): -12; (5 2)(2 1): -3; (8 2)(2 12) = (8 12): -1; (9 2)(2 10) = (9 10): 1;*

j = 4

(1 4)(4 1) = (1 1): 0

CYCLE (1 4): 0

(5 4)(4 5) = (5 5): -15

CYCLE (5 4): -15

(5 4)(4 9) = (5 9): -17; (5 4)(4 10) = (5 10): -17; (5 4)(4 12) = (5 12): -17;

(5 4)(4 13) = (5 13): -16; (5 4)(4 15) = (5 15): -4;

j = 5

(2 5)(5 2) = (2 2): 0

CYCLE (2 5): 0

(12 5)(5 12) = (12 12): 5

CYCLE (12 4 5): 5

*(2 5)(5 4) = (2 4): -13*

j = 7

(8 7)(7 8) = (8 8): 3

CYCLE (8 7): 3

j = 8

(1 8)(8 1) = (1 1): 5

CYCLE (1 2 8): 5

(1 8)(8 10) = (1 10): -18; (1 8)(8 11) = (1 11): -15;

(2 8)(8 2) = (2 2): 3

CYCLE (2 8): 3

(4 8)(8 11) = (4 11): -3;

(7 8)(8 7) = (7 7): 3

CYCLE (7 8): 3

(7 8)(8 9) = (7 9): -1; (7 8)(8 10) = (7 10): -5; (9 8)(8 7) = (9 7): <u>3</u>;

(9 8)(8 9) = (9 9): -1

CYCLE (9 8): -1

(10 8)(8 10) = (10 10): 1

CYCLE (10 8): 1

(10 8)(8 11) = (10 11): 4; (11 8)(8 9) = (11 9): 4; (11 8)(8 10) = (11 10): 0;

(11 8)(8 11) = (11 11): 3

CYCLE (11 8): 3

(13 8)(8 9) = (13 9): 3; (13 8)(8 10) = (13 10): -1;

(15 8)(8 9) = (15 9): -6; (15 8)(8 10) = (15 10): -10;

(15 8)(8 11) = (15 11): -7;

*(1 8)(8 5) = (1 5): 3; (13 8)(8 11) = (13 11): 2*

j = 9

(1 9)(9 8) = (1 8): -17; (1 9)(9 11) = (1 11): -20; (1 9)(9 12) = (1 12): -19;

(2 9)(9 2) = (2 2): 1

CYCLE (2 1 9): 1

(2 9)(9 8) = (2 8): -19; (2 9)(9 11) = (2 11): -22;

(4 9)(9 4) = (4 4): 1

CYCLE (4 9): 1

(4 9)(9 7): -11; (4 9)(9 8) = (4 8): -18; (4 9)(9 11) = (4 11): -21; (4 9)(9 12) = (4 12): -20;

(5 9)(9 5) = (5 5): 1

CYCLE (5 4 9): 1

(5 9)(9 7) = (5 7): -11; (5 9)(9 11) = (5 11): -21; (5 9)(9 12) = (5 12): -20;

(7 9)(9 7) = (7 7): 5

CYCLE (7 8 9): 5

(7 9)(9 11) = (7 11): -5;

(8 9)(9 8) = (8 8): -1

CYCLE (8 9): -1

(8 9)(9 11) = (8 11): -4;

(11 9)(9 11) = (11 11): 0

CYCLE (11 8 9): 0

(12 9)(9 12) = (12 12): 0

CYCLE (12 4 9): 0

(15 9)(9 11) = (15 11): -10

*(8 9)(9 12) = (8 12): -3; (13 9)(9 11) = (13 11): -1*

j = 10

(2 10)(10 2) = (2 2): 1

CYCLE (2 10): 1

(4 10)(10 4) = (4 4): 2

CYCLE (4 10): 2

(5 10)(10 5) = (5 5): 1

CYCLE (5 4 10): 1

(5 10)(10 8) = (5 8): <u>-12</u>;

(8 10)(10 8) = (8 8): 1

CYCLE (8 10): 1

(8 10)(10 13) = (8 13): -2

(12 10)(10 12) = (12 12): -1

CYCLE (12 10): -1

(13 10)(10 13) = (13 13): 5

CYCLE (13 8 10): 5

j = 11

(2 11)(11 2) = (2 2): 2

CYCLE (2 1 9 11): 2

(4 11)(11 4) = (4 4): 3

CYCLE (4 9 11): 3

(7 11)(11 3) = (7 3): 4;

(7 11)(11 7) = (7 7): 6

CYCLE (7 8 9 11): 6

(8 11)(11 3) = (8 3): 5; (9 11)(11 3) = (9 3): 5

*(15 11)(11 3) = (15 3): -1*

j = 12

(1 12)(12 14) = (1 14): -24; (2 12)(12 3) = (2 3): -22; (2 12)(12 7) = (2 7): -16;

(2 12)(12 10) = (2 10): -23; (2 12)(12 13) = (2 13): -20; (2 12)(12 14) = (2 14): -26;

(4 12)(12 3) = (4 3): -21;

(4 12)(12 4) = (4 4): 0

CYCLE (4 12 9): 0

(4 12)(12 6) = (4 6): -14; (4 12)(12 7) = (4 7): -15; (4 12)(12 11) = (4 11): -22;

(4 12)(12 13) = (4 13): -21; (4 12)(12 14) = (4 14): -25; (5 12)(12 3) = (6 3): -21;

(5 12)(12 5) = (5 5): 0

CYCLE (5 4 9 12): 0

(5 12)(12 6) = (5 6): -14; (5 12)(12 11) = (5 11): -22; (5 12)(12 13) = (5 13): -21;

(5 12)(12 14) = (5 14): -25; (9 12)(12 3) = (9 3): -4; (9 12)(12 7) = (9 7): 2;

(9 12)(12 9) = (9 9): 1

CYCLE (9 12): 1

(9 12)(12 11) = (9 11): -5; (9 12)(12 13) = (9 13): -4; (9 12)(12 14) = (9 14): -8;

(10 12)(12 3) = (10 3): 0; (10 12)(12 7) = (10 7): 6;

(10 12)(12 10) = (10 10): -1

CYCLE (10 12): -1

(10 12)(12 11) = (10 11): -1; (13 12)((12 3) = (13 3): -7; (13 12)(12 6) = (13 6): 0;

(13 12)(12 7) = (13 7): -1; (13 12)(12 9) = (13 9): -2; (13 12)(12 10) = (13 10): -8;

(13 12)(12 13) =  (13 13): -7

CYCLE (13 12): -7

*(9 12)(12 10) = (9 10): -4; (13 12)(12 11) = (13 11): -8*

j = 13

(1 13)(13 6) = (1 6): -18; (1 13)(13 12) = (1 12): -21; (2 13)(13 6) = (2 6): -23; (2 13)(13 7): -16;

(4 13)(13 4) = (4 4): 5

CYCLE (4 9 12 13): 5

(4 13)(13 7) = (4 7): <u>-17</u>; (5 13)(13 6) = (5 6): <u>-24</u>; (5 13)(13 7) = (5 7): <u>-17</u>;

(7 13)(13 6) = (7 6): <u>-2</u>;

(7 13)(13 7) = (7 7): 5

CYCLE (7 8 4 13): 5

(8 13)(13 6) = (8 6): <u>-5</u>;

(8 13)(13 8) = (8 8): 3

CYCLE (8 4 13): 3

(9 13)(13 6) = (9 6): <u>-7</u>; (9 13)(13 7) = (9 7): <u>0</u>; (10 13)(13 6) = (10 6): <u>-3</u>;

(10 13)(13 7) = (10 7): <u>4</u>

j = 14

(1 14)(14 3) = (1 3): <u>-21</u>;

(10 14)(14 10) = (10 10): 5

CYCLE (10 12 14): 5

(12 14)(14 12) = (12 12): 3

CYCLE (12 14): 3

j = 15

(2 15)(15 7) = (2 7): <u>-18</u>; (2 15)(15 8) = (2 8): <u>-20</u>

$P_{15}$

|    | 1 | 2 | 3  | 4 | 5 | 6  | 7  | 8  | 9  | 10 | 11 | 12 | 13 | 14 | 15 |    |
|----|---|---|----|---|---|----|----|----|----|----|----|----|----|----|----|----|
| 1  | 2 | 1 | 2  |   |   | 13 | 15 | 9  | 2  | 2  | 9  | 13 | 2  | 12 | 1  | 1  |
| 2  | 2 |   | 12 |   |   | 13 | 13 | 15 | 12 | 12 | 12 | 2  | 12 | 12 | 1  | 2  |
| 3  |   |   |    |   |   |    |    |    |    |    |    |    |    |    |    | 3  |
| 4  | 4 |   | 12 |   |   | 12 | 13 | 9  | 4  |    | 12 | 9  | 12 | 12 |    | 4  |
| 5  |   | 5 | 12 | 5 |   | 13 | 13 | 10 | 4  | 4  | 12 | 9  | 12 | 12 | 4  | 5  |
| 6  |   |   |    |   |   |    |    |    |    |    |    |    |    |    |    | 6  |
| 7  |   |   | 11 |   |   | 13 |    | 7  | 8  | 8  | 9  |    | 10 |    |    | 7  |
| 8  |   |   | 11 |   |   | 13 |    | 7  | 8  | 8  | 9  |    | 10 |    |    | 8  |
| 9  |   | 8 | 12 |   |   | 13 | 13 | 9  |    |    | 12 | 9  | 12 | 12 |    | 9  |
| 10 |   |   | 12 |   |   | 13 | 13 | 10 |    |    | 12 | 10 | 12 | 12 |    | 10 |
| 11 |   |   |    |   |   |    |    | 11 | 8  | 8  |    |    |    |    |    | 11 |
| 12 |   |   | 12 |   |   | 12 | 12 |    | 12 |    |    | 13 |    |    |    | 12 |
| 13 |   |   | 12 |   |   | 12 | 12 | 13 | 12 | 12 |    | 13 |    |    |    | 13 |
| 14 |   |   |    |   |   |    |    |    |    |    |    |    |    |    |    | 14 |
| 15 | 8 |   |    |   |   |    |    | 15 | 8  | 8  | 9  |    |    |    |    | 15 |
|    | 1 | 2 | 3  | 4 | 5 | 6  | 7  | 8  | 9  | 10 | 11 | 12 | 13 | 14 | 15 |    |

$$\sigma_T^{-1} M^-(15)$$

|  | 5 | 4 | 15 | 2 | 1 | 6 | 14 | 12 | 10 | 9 | 13 | 8 | 11 | 7 | 3 |  |
|---|---|---|---|---|---|---|---|---|---|---|---|---|---|---|---|---|
|  | 1 | 2 | 3 | 4 | 5 | 6 | 7 | 8 | 9 | 10 | 11 | 12 | 13 | 14 | 15 |  |
| 1 |  |  | -21 |  |  | -18 | -12 | -17 | -17 | -18 | -20 | -21 | -15 | -24 |  | 1 |
| 2 |  |  | -22 |  |  | -23 | -16 | -18 | -18 | -23 | -22 | -18 | -20 | -26 | -14 | 2 |
| 3 |  |  |  |  |  |  |  |  |  |  |  |  |  |  |  | 3 |
| 4 |  |  | -21 |  |  | -14 | -17 | -18 |  |  | -22 | -20 | -21 | -25 |  | 4 |
| 5 |  |  | -21 |  |  | -24 | -17 | -12 | -17 | -17 | -22 | -20 | -21 | -25 | -4 | 5 |
| 6 |  |  |  |  |  |  |  |  |  |  |  |  |  |  |  | 6 |
| 7 |  |  | 4 |  |  | -2 |  |  | -1 | -5 | -5 |  |  |  |  | 7 |
| 8 |  |  | 5 |  |  | -5 |  |  |  |  | -4 |  | -2 |  |  | 8 |
| 9 |  |  | -4 |  |  | -7 | 0 |  |  |  | -5 |  |  |  |  | 9 |
| 10 |  |  | 0 |  |  | -3 | 4 |  |  |  | -1 |  | -4 | -8 |  | 10 |
| 11 |  |  |  |  |  |  |  | 4 | 0 |  |  |  |  |  |  | 11 |
| 12 |  |  |  |  |  |  |  |  |  |  |  |  |  |  |  | 12 |
| 13 |  |  | -7 |  |  | 0 | -1 |  | -2 | -8 |  |  |  |  |  | 13 |
| 14 |  |  |  |  |  |  |  |  |  |  |  |  |  |  |  | 14 |
| 15 |  |  |  |  |  |  |  |  | -6 | -10 | -10 |  |  |  |  | 15 |
|  | 1 | 2 | 3 | 4 | 5 | 6 | 7 | 8 | 9 | 10 | 11 | 12 | 13 | 14 | 15 |  |

We now once again note the following rules:

Suppose that we must consider an entry of the form $|(a \ ... \ a_n \ ... \ x)|$ vis-a-vis $|(a \ x)|$. We use the following rule: If $|(a \ x)| + |(a_n \ a_1)|$ is greater than $|(a \ ... \ a_n \ ... \ x)|$, we choose $|(a \ ... \ a_n \ ... \ x)|$. Otherwise, $|(a \ x)|$ remains the entry.

j = 3

(2 3)(3 2) = (2 2): 4

CYCLE (2 12 3): 4

(2 3)(3 5) = (2 5): -6;

(5 3)(3 5) = (5 5): -5

CYCLE (5 4 9 12 3): -5

(13 3)(3 14) = (3 14): 5

*(1 3)(3 5) = (1 5): -5;*

j = 5

*(2 5)(5 4) = (2 4): -6  (13 12 6 11): 23**

j = 6

(8 6)(6 7) = (8 7): 25*;

*(13 6)(6 11) = (13 11): 23**

j = 7

(2 7)(7 5) = (2 5): <u>3</u>;

(4 7)(7 4) = (4 4): 4

CYCLE (4 9 12 13 7): 4

(4 7)(7 5) = (4 5): <u>4</u>;  (4 7)(7 15) = (4 15): 4;

(8 7)(7 8) = 28*

CYCLE (8 10 13 6 7): 28*

 (13 7)(7 8) = (13 8): -2

*(2 7)(7 4) = (2 4): 5*

j = 8

(1 8)(8 7) = (1 7): -13; (1 8)(8 15) = (1 15): 0;

(2 8)(8 2) = (2 2): -2

CYCLE (2 9 8): -2

(4 8)(8 1) = (4 1): 1;

(4 8)(8 4) = (4 4): -2

CYCLE (4 9 8): -2

(4 8)(8 15) = (4 15): -1;

(5 8)(8 5) = (5 5): 5

CYCLE (5 4 10 8): 5

*(1 8)(8 5): 0; (2 8)(8 4) = (2 4): -4; (4 8)(8 2) + (4 2): 0¿ (13 8)(8 11): -3*

j = 9

(13 9)(9 8) = (13 8): -3;

(13 9)(9 13) = (13 13): 6

CYCLE (13 8 9): 6

*(13 9)(9 11) = (13 11): -6*

j = 10

(2 10)(10 2) = (2 2): -4

CYCLE (2 12 10): -4

(13 10)(10 13) = (13 13): -2

CYCLE (13 8 10): -2

j = 11

(4 11)(11 4) = (4 4): 2

CYCLE (4 9 12 11): 2

(5 11)(11 5) = (5 5): 2

CYCLE (5 4 9 12 11): 2

(9 11)(11 9) = (9 9): 5

CYCLE (9 12 11): 5

*(4 11)(11 2) = (4 2): 2*

j = 12

(1 12)(12 3) = (1 3): -22; (1 12)(12 7) = (1 7): -16; (1 12)(12 10) = (1 10): -23;

(1 12)(12 14) = (1 14): -26

*(1 12)(12 5) = (1 5): -2*

j = 14

(1 14)(14 6) = (1 6): -23;

*(1 14)(14 5) = (1 5): 0*

$P_{30}$

|    | 1 | 2  | 3  | 4 | 5 | 6  | 7  | 8  | 9  | 10 | 11 | 12 | 13 | 14 | 15 |    |
|----|---|----|----|---|---|----|----|----|----|----|----|----|----|----|----|----|
| 1  | 2 | 1  | 12 |   | 3 | 14 | 12 | 9  | 2  | 12 | 9  | 13 | 2  | 12 | 8  | 1  |
| 2  | 2 |    | 12 |   | 7 | 13 | 13 | 9  | 1  | 12 | 12 | 2  | 12 | 12 |    | 2  |
| 3  |   |    |    |   |   |    |    |    |    |    |    |    |    |    |    | 3  |
| 4  | 8 |    | 12 |   | 7 | 12 | 13 | 9  | 4  |    | 12 | 9  | 12 | 12 | 8  | 4  |
| 5  |   | 5  | 12 | 5 |   | 13 | 13 | 10 | 4  | 4  | 12 | 9  | 12 | 12 | 4  | 5  |
| 6  |   |    |    |   |   |    |    |    |    |    |    |    |    |    |    | 6  |
| 7  |   |    | 11 |   |   | 13 |    | 7  | 8  | 8  | 9  |    | 10 |    |    | 7  |
| 8  |   |    | 11 |   |   | 13 | 6  | 7  | 8  | 8  | 9  |    | 10 |    |    | 8  |
| 9  |   | 8  | 12 |   |   | 13 | 13 | 9  |    |    | 12 | 9  | 12 | 12 |    | 9  |
| 10 |   | 11 | 12 |   |   | 13 | 13 | 10 |    |    | 12 | 10 | 12 | 12 |    | 10 |
| 11 |   |    |    |   |   |    |    | 11 | 8  | 8  |    |    |    |    |    | 11 |
| 12 |   |    | 12 |   |   | 12 | 12 |    | 12 |    |    | 13 |    |    |    | 12 |
| 13 |   |    | 12 |   |   | 12 | 12 | 13 | 12 | 12 |    | 13 |    | 3  |    | 13 |
| 14 |   |    |    |   |   |    |    |    |    |    |    |    |    |    |    | 14 |
| 15 | 8 |    |    |   |   |    |    | 15 | 8  | 8  | 9  |    |    |    |    | 15 |
|    | 1 | 2  | 3  | 4 | 5 | 6  | 7  | 8  | 9  | 10 | 11 | 12 | 13 | 14 | 15 |    |

$$\sigma_T^{-1} M^-(30)$$

|  | 5 | 4 | 15 | 2 | 1 | 6 | 14 | 12 | 10 | 9 | 13 | 8 | 11 | 7 | 3 |  |
|---|---|---|---|---|---|---|---|---|---|---|---|---|---|---|---|---|
|  | 1 | 2 | 3 | 4 | 5 | 6 | 7 | 8 | 9 | 10 | 11 | 12 | 13 | 14 | 15 |  |
| 1 |  |  | *-22* |  |  | *-23* | *-16* | *-18* | *-17* | *-23* | *-20* | *-21* | *-15* | *-26* | *0* | 1 |
| 2 |  |  | *-22* |  | *-6* | *-23* | *-18* | *-20* | *-18* | *-23* | *-22* | *-18* | *-20* | *-26* | *-14* | 2 |
| 3 |  |  |  |  |  |  |  |  |  |  |  |  |  |  |  | 3 |
| 4 | *1* |  | *-21* |  | *-5* | *-14* | *-17* | *-18* |  |  | *-22* | *-20* | *-21* | *-25* | *-1* | 4 |
| 5 |  |  | *-21* |  |  | *-24* | *-17* | *-12* | *-17* | *-17* | *-22* | *-20* | *-21* | *-25* | *-4* | 5 |
| 6 |  |  |  |  |  |  |  |  |  |  |  |  |  |  |  | 6 |
| 7 |  |  | *4* |  |  | *-2* |  |  | *-1* | *-5* | *-5* |  |  |  |  | 7 |
| 8 |  |  | *5* |  |  | *-5* | *25\** |  |  |  | *-4* |  | *-2* |  |  | 8 |
| 9 |  |  | *-4* |  |  | *-7* | *0* |  |  |  | *-5* |  |  |  |  | 9 |
| 10 |  |  | *0* |  |  | *-3* | *4* |  |  |  | *-1* |  | *-4* | *-8* |  | 10 |
| 11 |  |  |  |  |  |  |  |  | *4* | *0* |  |  |  |  |  | 11 |
| 12 |  |  |  |  |  |  |  |  |  |  |  |  |  |  |  | 12 |
| 13 |  |  | *-7* |  |  | *0* | *-1* | *-3* | *-2* | *-8* |  |  |  | *5* |  | 13 |
| 14 |  |  |  |  |  |  |  |  |  |  |  |  |  |  |  | 14 |
| 15 |  |  |  |  |  |  |  |  | *-6* | *-10* | *-10* |  |  |  |  | 15 |
|  | 1 | 2 | 3 | 4 | 5 | 6 | 7 | 8 | 9 | 10 | 11 | 12 | 13 | 14 | 15 |  |

*(1 3)(3 5) = (1 5): -6*

The only path that can be extended is (1 10): -21. (1 10)(10 15) = (1 15): -2. This cannot be extended. Thus, we cannot obtain any more acceptable cycles. We now list acceptable cycles according to the number of points contained in a cycle.

V. (i) (5 4 9 12 3): -5; (ii) (5 4 9 12 11): 2; (iii) (4 9 12 13 7): 4; (iv) (8 10 13 6 7): 28*

IV. (i) (5 4 9 12):0; (ii) (2 1 9 11): 2; (iii) (4 9 12 11): 2; (iv) (4 9 12 13): 5; (v) (7 8 9 11): 6; (vi) (7 8 4 13): 5

III. (i) (2 12 10): -4; (ii) (2 9 8): -2; (iii) (4 9 8): -2; (iv) (13 8 10): -2; (v) (4 12 9): 0;

(vi) (11 8 9): 0; (vii) (12 4 9): 0; (viii) (2 1 9): 1; (ix) (5 4 9): 1; (x) (5 4 10): 1

(xi) (4 9 11): 3; (xii) (8 4 13): 3; (xiii) (2 12 3): 4; (xiv) (1 2 8): 5; (xv) (7 8 9): 5;

(xvi) (12 4 5): 5; (xvii) (13 8 10): 5; (xviii) (13 8 9): 6

II. (i) (2 1): -15; (ii) (5 4): -15; (iii) (3 12): -7; (iv) (8 9): -1; (v) (9 8): -1; (vi) (4 1): 0;

(vii) (5 2): 0; (viii) (2 10): 1; (ix) (4 9): 1; (x) (9 12): 1; (xi) (10 8): 1; (xii) (4 10): 2;

(xiii) (2 8): 3; (xiv) (7 8): 3; (xv) (8 7): 3; (xvi): (11 8): 3

We now check all unlinked 2-circuit paths to initiate the construction of 2-circuit cycles by finding each path of smallest value having a fixed number as its initial point..

1. [1 2 5]: -12; [1 2 13 12 3 5]: -6; [1 2 9 8 5]: 0; [1 2 13 12 5]: -1; [1 2 13 12 14 5]: -1

2. [2 1 4]: -13; [2 1 15 8 4]: -4; [2 12 3 5 4]: -4; [2 12 13 7 4]: 5

4. [4 1 2]: -2; [4 9 12 11 2]: 2; [4 9 8 2]: 0

8. [8 9 12]: -3

9. [9 12 10]: -5

13. [13 7 8 11]: 2; [13 12 9 11]: -6; [13 12 11]: -8; [13 12 6 11]: 23*

15. [15 8 9 11 3]: -1

We thus choose the following:

UNLINKED 2-CIRCUIT $P_{45}$

|   | 5 | 4 | 15 | 2 | 1 | 6 | 14 | 12 | 10 | 9 | 13 | 8 | 11 | 7 | 3 |   |
|---|---|---|----|---|---|---|----|----|----|---|----|---|----|---|---|---|
|   | 1 | 2 | 3  | 4 | 5 | 6 | 7  | 8  | 9  | 10| 11 | 12| 13 | 14| 15|   |
| 1 |   | 1 | 12 |   | 3!|   |    |    |    |   |    | 13| 2  |   |   | 1 |
| 2 | 2 |   |    | 1!|   |   |    |    |    |   |    |   |    |   |   | 2 |
| 3 |   |   |    |   |   |   |    |    |    |   |    |   |    |   |   | 3 |
| 4 | 4 | 1!|    |   |   |   |    |    |    |   |    |   |    |   |   | 4 |
| 5 | 2!| 5 |    |   |   |   |    |    |    |   |    |   |    |   |   | 5 |
| 6 |   |   |    |   |   |   |    |    |    |   |    |   |    |   |   | 6 |
| 7 |   |   |    |   |   |   |    |    |    |   |    |   |    |   |   | 7 |
| 8 |   |   |    |   |   |   |    |    | 8  |   |    | 9!|    |   |   | 8 |
| 9 |   |   |    |   |   |   |    |    |    |12!|    | 9 |    |   |   | 9 |
| 10|   |   |    |   |   |   |    |    |    |   |    |   |    |   |   | 10|
| 11|   |   |    |   |   |   |    |    |    |   |    |   |    |   |   | 11|
| 12|   |   |    |   |   |   |    |    |    |   |    |   |    |   |   | 12|
| 13|   |   |    |   |   |   |    |    |    |12!| 13 |   |    |   |   | 13|
| 14|   |   |    |   |   |   |    |    |    |   |    |   |    |   |   | 14|
| 15|   |   | 11!|   |   |   |    | 15 | 8  |   | 9  |   |    |   |   | 15|
|   | 1 | 2 | 3  | 4 | 5 | 6 | 7  | 8  | 9  | 10| 11 | 12| 13 | 14| 15|   |

## UNLINKED 2-CIRCUIT $\sigma_T^{-1} M^-(45)$

|    | 1   | 2   | 3   | 4 | 5   | 6 | 7 | 8  | 9   | 10  | 11  | 12  | 13  | 14 | 15 |    |
|----|-----|-----|-----|---|-----|---|---|----|-----|-----|-----|-----|-----|----|----|----|
| 1  |     | 1   | -22 |   | -6! |   |   |    |     |     |     | -21 | -15 |    |    | 1  |
| 2  | -16 |     |     | - |     |   |   |    |     |     |     |     |     |    |    | 2  |
| 3  |     |     |     |   |     |   |   |    |     |     |     |     |     |    |    | 3  |
| 4  | -3  | -2! |     |   |     |   |   |    |     |     |     |     |     |    |    | 4  |
| 5  |     |     |     |   |     |   |   |    |     |     |     |     |     |    |    | 5  |
| 6  |     |     |     |   |     |   |   |    |     |     |     |     |     |    |    | 6  |
| 7  |     |     |     |   |     |   |   |    |     |     |     |     |     |    |    | 7  |
| 8  |     |     |     |   |     |   |   |    | 0   |     |     | -3! |     |    |    | 8  |
| 9  |     |     |     |   |     |   |   |    |     | -5! |     | -3  |     |    |    | 9  |
| 10 |     |     |     |   |     |   |   |    |     |     |     |     |     |    |    | 10 |
| 11 |     |     |     |   |     |   |   |    |     |     |     |     |     |    |    | 11 |
| 12 |     |     |     |   |     |   |   |    |     |     |     |     |     |    |    | 12 |
| 13 |     |     |     |   |     |   |   |    |     |     | -8! | -6  |     |    |    | 13 |
| 14 |     |     |     |   |     |   |   |    |     |     |     |     |     |    |    | 14 |
| 15 |     |     | -1! |   |     |   |   | -6 | -6  |     | -10 |     |     |    |    | 15 |
|    | 1   | 2   | 3   | 4 | 5   | 6 | 7 | 8  | 9   | 10  | 11  | 12  | 13  | 14 | 15 |    |

j = 2

(4 2)(2 3) = (4 3): -14; (4 2)(2 6) = (4 6): -19; (4 2)(2 7) = (4 7): -15; (4 2)(2 8) = (4 8): -17;

(4 2)(2 9) = (4 9): -19; (4 2)(2 10) = (4 10): -20; (4 2)(2 11) = (4 11): -16;

(4 2)(2 12) = (4 12): -21; (4 2)(2 13) = (4 13): -18; (4 2)(2 14) = (4 14): -20;

(4 2)(2 15) = (4 15): -2

*(4 2)(2 5) = (4 5): -15*

j = 4

(2 4)(4 3) = (2 3): <u>-8</u>;  (2 4)(4 6) = (2 6): -17; (2 4)(4 7) = (2 7): -3; (2 4)(4 8) = (2 8): -2;

(2 4)(4 9) = (2 9): -30; (2 4)(4 10) = (2 10): -30; (2 4)(4 11) = (2 11): -7;

(2 4)(4 12) = (2 12): -30;  (2 4)(4 13) = (2 13): -29; (2 4)(4 14) = (2 14): -14;

(2 4)(4 15) = (2 15): -17

*(2 4)(4 5) = (2 5): -28*

j = 5

(1 5)(5 6) = (1 6): 4;

(4 5)(5 4) = (4 4): -15

CYCLE (4 1 2 5): -15

*(1 5)(5 4) = (1 4): <u>-6</u>; (1 5)(5 11) = (1 11): 27\*; (1 5)(5 15) = (1 15): 3*

j = 6

(2 6)(6 11) = (2 11): 6\*;  (4 6)(6 11) = (4 11): 4\*

j = 8

(4 8)(8 10) = (4 10): -21

j = 9

(2 9)(9 2) = (2 2): -11

CYCLE (2 1 4 9): -11

(2 9)(9 7) = (2 7): <u>-24</u>; (2 9)(9 8) = (2 8): -31; (2 9)(9 11) = (2 11): -34;

(2 9)(9 12) = (2 12): -33; (4 9)(9 8) = (4 8): <u>-20</u>; (4 9)(9 11) = (4 11): -23;

(4 9)(9 12) = (4 12): -22

*(2 9)(9 5) = (2 9): -12*

j = 10

(4 10)(10 4) = (4 4): -2

CYCLE (4 1 2 8 10): -2

*(2 10)(10 5) = (2 5): <u>-12</u>; (9 10)(10 8) = (9 8): <u>0</u>; (9 10)(10 13) = (9 13): 1*

j = 11

(2 11)(11 2) = (2 2): -10

CYCLE (2 1 4 9 11): -10

(2 11)11 3) = (2 3): <u>-25</u>; (13 11)(11 3) = (13 3): <u>1</u>; (13 11)(11 7) = (13 7): <u>3</u>;

(13 11)(11 8) = (13 8): <u>-4</u>; (13 11)(11 9) = (13 9): <u>2</u>; (13 11)(11 10): <u>4</u>

*(2 11)(11 5) = (2 5): <u>-10</u>*

j = 12

(2 12)(12 3) = (2 3): <u>-34</u>; (2 12)(12 6) = (2 6): <u>-27</u>; (2 12)(12 7) = (2 7): <u>-28</u>;

(2 12)(12 11) = (2 11): <u>-35</u>; (2 12)(12 13) = (2 13): -34; (2 12)(12 14) = (2 14): -38;

(4 12)(12 3) = (4 3): <u>-23</u>;

(4 12)(12 4) = (4 4): -2

CYCLE (4 1 2 9 12): -2

(4 12)(12 7) = (4 7): <u>-17</u>; (4 12)(12 11) = (4 11): <u>-24</u>; (4 12)(12 13) = (4 13): -23;

(4 12)(12 14) = (4 14): -27; (8 12)(12 3) = (8 3): <u>-4</u>; (8 12)(12 6) = (8 6): <u>3</u>;

(8 12)(12 7) = (8 7): <u>2</u>; (8 12)(12 11) = (8 11): <u>-5</u>; (8 12)(12 13) = (8 13): -4; (8 12)(12 14): -8

*(8 12)(12 10) = (8 10): -5*

j = 13

(2 13)(13 6) = (2 6): <u>-34</u>; (2 13)(13 7) = (2 7): <u>-27</u>; (2 13)(13 10) = (2 10): <u>-31</u>;

(4 13)(13 6) = (4 6): <u>-26</u>; (4 13)(13 7) = (4 7): <u>-19</u>; (8 13)(13 6) = (8 6): <u>-7</u>;

(8 13)(13 7) = (8 7): <u>0</u>;

(8 13)(13 8) = (8 8): -1

CYCLE (8 9 12 13): -1

(9 13)(13 6) = (9 6): <u>-2</u>; (9 13)(13 7) = (9 7): <u>5</u>

*(9 13)(13 8) = (9 8): 4*

j = 14

(4 14)(14 6) = (4 6): <u>-24</u>;

(8 14)(14 8) = (8 8): 1

CYCLE (8 9 12 14): 1

*(8 14)(14 10) = (8 10): 1*

UNLINKED 2-CIRCUIT $P_{60}$

|    | 1  | 2  | 3   | 4  | 5  | 6  | 7  | 8  | 9  | 10  | 11 | 12 | 13 | 14 | 15 |    |
|----|----|----|-----|----|----|----|----|----|----|-----|----|----|----|----|----|----|
| 1  |    | 1  | 12  |    | 3! |    |    |    |    |     |    | 13 | 2  |    |    | 1  |
| 2  | 2  |    | 12  | 1! |    | 13 | 13 | 4  | 4  | 13  | 12 | 9  | 12 | 12 | 4  | 2  |
| 3  |    |    |     |    |    |    |    |    |    |     |    |    |    |    |    | 3  |
| 4  | 4  | 1! | 12  |    | 2  | 13 | 13 | 2  | 2  | 8   | 12 | 9  | 12 | 12 | 2  | 4  |
| 5  | 2! | 5  |     |    |    |    |    |    |    |     |    |    |    |    |    | 5  |
| 6  |    |    |     |    |    |    |    |    |    |     |    |    |    |    |    | 6  |
| 7  |    |    |     |    |    |    |    |    |    |     |    |    |    |    |    | 7  |
| 8  |    |    | 12  |    |    | 13 | 13 |    | 8  |     | 12 | 9! | 12 | 12 |    | 8  |
| 9  |    |    |     |    |    | 13 | 13 |    |    | 12! |    | 9  | 10 |    |    | 9  |
| 10 |    |    |     |    |    |    |    |    |    |     |    |    |    |    |    | 10 |
| 11 |    |    |     |    |    |    |    |    |    |     |    |    |    |    |    | 11 |
| 12 |    |    |     |    |    |    |    |    |    |     |    |    |    |    |    | 12 |
| 13 |    |    | 11  |    |    | 11 |    | 11 | 11 | 12! | 13 |    |    |    |    | 13 |
| 14 |    |    |     |    |    |    |    |    |    |     |    |    |    |    |    | 14 |
| 15 |    |    | 11! |    |    |    |    | 15 | 8  |     | 9  |    |    |    |    | 15 |
|    | 1  | 2  | 3   | 4  | 5  | 6  | 7  | 8  | 9  | 10  | 11 | 12 | 13 | 14 | 15 |    |

# UNLINKED 2-CIRCUIT $\sigma_T^{-1} M^-(60)$

|    | 1   | 2   | 3   | 4 | 5   | 6   | 7   | 8   | 9   | 10  | 11  | 12  | 13  | 14  | 15  |    |
|----|-----|-----|-----|---|-----|-----|-----|-----|-----|-----|-----|-----|-----|-----|-----|----|
| 1  |     | 1   | -22 |   | -6! | *4* |     |     |     |     |     | -21 | -15 | *3* |     | 1  |
| 2  | -16 |     | -34 | - |     | -34 | -30 | -31 | *-30* | -31 | -35 | *-33* | *-34* | *-38* | *-17* | 2  |
| 3  |     |     |     |   |     |     |     |     |     |     |     |     |     |     |     | 3  |
| 4  | -3  | -2! | -23 |   |     | -26 | -19 | -20 | *-19* | *-21* | -24 | *-22* | *-23* | *-27* | *-2* | 4  |
| 5  |     |     |     |   |     |     |     |     |     |     |     |     |     |     |     | 5  |
| 6  |     |     |     |   |     |     |     |     |     |     |     |     |     |     |     | 6  |
| 7  |     |     |     |   |     |     |     |     |     |     |     |     |     |     |     | 7  |
| 8  |     |     | -4  |   |     | -7  | 0   |     | 0   |     | -5  | -3! | *-4* | *-8* |     | 8  |
| 9  |     |     |     |   |     | *-2* | 5   | *-2* |     | *-5!* |     | *-3* | *1* |     |     | 9  |
| 10 |     |     |     |   |     |     |     |     |     |     |     |     |     |     |     | 10 |
| 11 |     |     |     |   |     |     |     |     |     |     |     |     |     |     |     | 11 |
| 12 |     |     |     |   |     |     |     |     |     |     |     |     |     |     |     | 12 |
| 13 |     |     | 1   |   |     |     | 3   |     | 2   | 4   | -8! | -6  |     |     |     | 13 |
| 14 |     |     |     |   |     |     |     |     |     |     |     |     |     |     |     | 14 |
| 15 |     |     | -1! |   |     |     |     | -6  | -6  |     | -10 |     |     |     |     | 15 |
|    | 1   | 2   | 3   | 4 | 5   | 6   | 7   | 8   | 9   | 10  | 11  | 12  | 13  | 14  | 15  |    |

j = 3

(2 3)(3 2) = (2 2): -8

CYCLE (2 1 4 12 3): -8

*(2 3)(3 5) = (2 5): -18; (4 3)(3 5) = (4 5): -7*

j = 7

(4 7)(7 4) = (4 4): 2

CYCLE (4 1 2 9 12 3 7): 2

(8 7)(7 8) = (8 8): -1

CYCLE (8 9 12 13 7): -1

*(4 7)(7 5) = (4 5): 2; (9 7)(7 8) = ( 8): 4*

j = 8

(2 8)(8 2) = (2 2): -13

CYCLE (2 1 4 8): -13

(4 8)(8 4) = (4 4): -4

CYCLE (4 1 2 8): -4

*(4 8)(8 5) = (4 5): -3*

j = 9

(4 9)(9 4) = (4 4): -1

CYCLE (4 1 2 9): -1

*(13 9)(9 8) = (13 8): 1*

j = 10

(2 10)(10 2) = (2 2): -31

CYCLE (2 1 4 12 13 10): -31

*(2 10)(10 5) = (2 5): -13*

j = 11

(2 11)(11 2) = (2 2): -10

CYCLE (2 1 4 9 12 11): -10

(4 11)(11 4) = (4 4): 0

CYCLE (4 1 2 9 12 11): 0

(8 11)(11 8) = (8 8): -1

CYCLE (8 9 12 11): -1

*(13 11)(11 8) = (13 8): 1*

We now list those linked 2-circuit paths that have the smallest value of all of such paths that have the same determining point:

1. (1 2 13 12 3 5 4): -6

2. (2 1 4 5): -28

4. (4 1 2 5): -15

8. (8 9 12 10): -5

9. (9 12 10 8): 0

13. (13 12 11 7 8): 2

We use these paths as initial paths for possible linked 2-circuit cycles.

# LINKED 2-CIRCUIT $P_{75}$

|   | 5 | 4 | 15 | 2 | 1 | 6 | 14 | 12 | 10 | 9 | 13 | 8 | 11 | 7 | 3 |   |
|---|---|---|----|---|---|---|----|----|----|---|----|---|----|---|---|---|
|   | 1 | 2 | 3  | 4 | 5 | 6 | 7  | 8  | 9  | 10| 11 | 12| 13 | 14| 15|   |
| 1 |   | 1 | 12 | 5!| 3 |   |    |    |    |   |    | 13| 2  |   |   | 1 |
| 2 | 2 |   |    | 1 | 4!|   |    |    |    |   |    |   |    |   |   | 2 |
| 3 |   |   |    |   |   |   |    |    |    |   |    |   |    |   |   | 3 |
| 4 | 4 | 1 |    |   | 2!|   |    |    |    |   |    |   |    |   |   | 4 |
| 5 |   |   |    |   |   |   |    |    |    |   |    |   |    |   |   | 5 |
| 6 |   |   |    |   |   |   |    |    |    |   |    |   |    |   |   | 6 |
| 7 |   |   |    |   |   |   |    |    |    |   |    |   |    |   |   | 7 |
| 8 |   |   |    |   |   |   |    |    | 8  |12!|    | 9 |    |   |   | 8 |
| 9 |   |   |    |   |   |   |    |10! |    |12 |    | 9 |    |   |   | 9 |
| 10|   |   |    |   |   |   |    |    |    |   |    |   |    |   |   | 10|
| 11|   |   |    |   |   |   |    |    |    |   |    |   |    |   |   | 11|
| 12|   |   |    |   |   |   |    |    |    |   |    |   |    |   |   | 12|
| 13|   |   |    |   |   |   | 11 | 7! |    |   | 12 | 13|    |   |   | 13|
| 14|   |   |    |   |   |   |    |    |    |   |    |   |    |   |   | 14|
| 15|   |   |    |   |   |   |    |    |    |   |    |   |    |   |   | 15|
|   | 1 | 2 | 3  | 4 | 5 | 6 | 7  | 8  | 9  | 10| 11 | 12| 13 | 14| 15|   |

LINKED 2-CIRCUIT $\sigma_T^{-1} M^-(75)$

|   | 5 | 4 | 15 | 2 | 1 | 6 | 14 | 12 | 10 | 9 | 13 | 8 | 11 | 7 | 3 |   |
|---|---|---|---|---|---|---|---|---|---|---|---|---|---|---|---|---|
|   | 1 | 2 | 3 | 4 | 5 | 6 | 7 | 8 | 9 | 10 | 11 | 12 | 13 | 14 | 15 |   |
| 1 |   | 1 | -22 | -6! | -6 |   |   |   |   |   |   | -21 | -15 |   |   | 1 |
| 2 | -16 |   |   | -13 | - |   |   |   |   |   |   |   |   |   |   | 2 |
| 3 |   |   |   |   |   |   |   |   |   |   |   |   |   |   |   | 3 |
| 4 | -3 | -2 |   |   | - |   |   |   |   |   |   |   |   |   |   | 4 |
| 5 |   |   |   |   |   |   |   |   |   |   |   |   |   |   |   | 5 |
| 6 |   |   |   |   |   |   |   |   |   |   |   |   |   |   |   | 6 |
| 7 |   |   |   |   |   |   |   |   |   |   |   |   |   |   |   | 7 |
| 8 |   |   |   |   |   |   |   |   | 0 | -5! |   | -3 |   |   |   | 8 |
| 9 |   |   |   |   |   |   |   | 0! |   | -5 |   | -3 |   |   |   | 9 |
| 10 |   |   |   |   |   |   |   |   |   |   |   |   |   |   |   | 10 |
| 11 |   |   |   |   |   |   |   |   |   |   |   |   |   |   |   | 11 |
| 12 |   |   |   |   |   |   |   |   |   |   |   |   |   |   |   | 12 |
| 13 |   |   |   |   |   |   | 3 | 2! |   |   | -8 | -6 |   |   |   | 13 |
| 14 |   |   |   |   |   |   |   |   |   |   |   |   |   |   |   | 14 |
| 15 |   |   |   |   |   |   |   |   |   |   |   |   |   |   |   | 15 |
|   | 1 | 2 | 3 | 4 | 5 | 6 | 7 | 8 | 9 | 10 | 11 | 12 | 13 | 14 | 15 |   |

j = 4

(1 4)(4 6) = (1 6): -13; (1 4)(4 7) = (1 7): -3; (1 4)(4 9) = (1 9): -23; (1 4)(4 10) = (1 10): -23

j = 5

(2 5)(5 2) = (2 2): -15

CYCLE (2 1 4 5): -15

(2 5)(5 6) = (2 6): -12; (2 5)(5 8) = (2 8): -7; (2 5)(5 12) = (2 12): -28;

(2 5)(5 14) = (2 14): -13; (2 5)(5 15) = (2 15): -13;

(4 5)(5 4) = (4 4): -15

CYCLE (4 1 2 5): -15

(4 5)(5 6) = (4 6): 1; (4 5)(5 12) = (4 12): -15; (4 5)(5 14) = (4 14): 0; (4 5)(5 15) = (4 15): 0

j = 6

(1 6)(6 7) = (1 7): 17*; (2 6)(6 11) = (2 11): 24*; (4 6)(6 11) = (4 11): 24*

j = 8

(2 8)(8 9) = (2 9): 6; (2 8)(8 10) = (2 10): 2; (9 8)(8 7) = (9 7): <u>4</u>;

(9 8)(8 9) = (9 9): 0

CYCLE (9 12 10 8): 0

(9 8)(8 11) = (9 11): -1; (13 8)(8 9) = (13 9): 2; (13 8)(8 10) = (13 10): -2

j = 9

(1 9)(9 7) = (1 7): <u>-17</u>; (2 9)(9 7) = <u>-1</u>; (2 9)(9 11) = (2 11): -28; (2 9)(9 13) = (2 13): -16

j = 10

(8 10)(10 8) = (8 8): 0

CYCLE (8 9 12 10): 0

(8 10)(10 13) = (8 13): 1;

(13 10)(10 13) = (13 13): 4

CYCLE (13 12 11 7 8 10): 4

j = 11

(2 11)(11 3) + (2 3): <u>1</u>

j = 12

(2 12)(12 3) = (2 3): <u>-16</u>; (2 12)(12 6) = (2 6): <u>-9</u>; (2 12)(12 9) = (2 9): <u>-11</u>;

(2 12)(12 10) = (2 10): <u>-17</u>; (2 12)(12 11) = (2 11): <u>-17</u>; (2 12)(12 13) = (2 13): -12;

(4 12)(12 3) = (4 3): <u>-16</u>; (4 12)(12 6) = (4 6): <u>-9</u>; (4 12)(12 7) = (4 7): <u>-10</u>;

(4 12)(12 14) = (4 14): -20; (4 12)(12 11) = (4 11): <u>-17</u>; (4 12)(12 13) = (4 13): -16

j = 13

(2 13)(13 6) = (2 6): <u>-19</u>; (2 13)(13 7) = (2 7): <u>-12</u>; (2 13)(13 14) = (2 14): -3;

(4 13)(13 6) = (4 6): <u>-19</u>; (4 13)(13 7) = (4 7): <u>-12</u>; (4 13)(13 10) = (4 10): <u>-16</u>;

(8 13)(13 6) = (8 6): <u>-2</u>; (8 13)(13 7) = (8 7): <u>5</u>

j = 15

(2 15)(15 7) = (2 7): <u>-17</u>; (2 15)(15 8) = (2 8): <u>-19</u>

# LINKED 2-CIRCUIT $P_{90}$

| | 5 | 4 | 15 | 2 | 1 | 6 | 14 | 12 | 10 | 9 | 13 | 8 | 11 | 7 | 3 | |
|---|---|---|---|---|---|---|---|---|---|---|---|---|---|---|---|---|
| | 1 | 2 | 3 | 4 | 5 | 6 | 7 | 8 | 9 | 10 | 11 | 12 | 13 | 14 | 15 | |
| 1 | | 1 | 12 | 5! | 3 | 4 | 9 | | 4 | 4 | | 13 | 2 | | | 1 |
| 2 | 2 | | 11 | 1 | 4! | 13 | 15 | 15 | 12 | 12 | 9 | 5 | 9 | 5 | 5 | 2 |
| 3 | | | | | | | | | | | | | | | | 3 |
| 4 | 4 | 1 | | | 2! | 12 | 13 | | | 13 | 12 | 5 | 12 | 12 | | 4 |
| 5 | | | | | | | | | | | | | | | | 5 |
| 6 | | | | | | | | | | | | | | | | 6 |
| 7 | | | | | | | | | | | | | | | | 7 |
| 8 | | | | | | 13 | 13 | | 8 | 12! | | 9 | 10 | | | 8 |
| 9 | | | | | | | 8 | 10! | | 12 | 8 | 9 | | | | 9 |
| 10 | | | | | | | | | | | | | | | | 10 |
| 11 | | | | | | | | | | | | | | | | 11 |
| 12 | | | | | | | | | | | | | | | | 12 |
| 13 | | | | | | | 11 | 7! | 8 | 8 | 12 | 13 | | | | 13 |
| 14 | | | | | | | | | | | | | | | | 14 |
| 15 | | | | | | | | | | | | | | | | 15 |
| | 1 | 2 | 3 | 4 | 5 | 6 | 7 | 8 | 9 | 10 | 11 | 12 | 13 | 14 | 15 | |

LINKED 2-CIRCUIT $\sigma_T^{-1} M^-(90)$

|   | 5 | 4 | 15 | 2 | 1 | 6 | 14 | 12 | 10 | 9 | 13 | 8 | 11 | 7 | 3 |   |
|---|---|---|---|---|---|---|---|---|---|---|---|---|---|---|---|---|
|   | 1 | 2 | 3 | 4 | 5 | 6 | 7 | 8 | 9 | 10 | 11 | 12 | 13 | 14 | 15 |   |
| 1 |   | 1 | -22 | -6! | -6 | *-13* | -17 |   | *-23* | -23 |   | -21 | -15 |   |   | 1 |
| 2 | -16 |   | -19 | -13 | - | -19 | -17 | -19 | -24 | -30 | -30 | *-28* | -29 | *-13* | *-13* | 2 |
| 3 |   |   |   |   |   |   |   |   |   |   |   |   |   |   |   | 3 |
| 4 | -3 | -2 | -16 |   | - | -19 | -12 |   |   | -16 | -17 | *-15* | -16 | *-20* | 0 | 4 |
| 5 |   |   |   |   |   |   |   |   |   |   |   |   |   |   |   | 5 |
| 6 |   |   |   |   |   |   |   |   |   |   |   |   |   |   |   | 6 |
| 7 |   |   |   |   |   |   |   |   |   |   |   |   |   |   |   | 7 |
| 8 |   |   |   |   |   | -2 | 5 |   | 0 | -5! |   | -3 | *1* |   |   | 8 |
| 9 |   |   |   |   |   |   | 4 | 0! |   | -5 | *-1* | -3 |   |   |   | 9 |
| 10 |   |   |   |   |   |   |   |   |   |   |   |   |   |   |   | 10 |
| 11 |   |   |   |   |   |   |   |   |   |   |   |   |   |   |   | 11 |
| 12 |   |   |   |   |   |   |   |   |   |   |   |   |   |   |   | 12 |
| 13 |   |   |   |   |   |   | 3 | 2! | *2* | -2 | -8 | -6 |   |   |   | 13 |
| 14 |   |   |   |   |   |   |   |   |   |   |   |   |   |   |   | 14 |
| 15 |   |   |   |   |   |   |   |   |   |   |   |   |   |   |   | 15 |
|   | 1 | 2 | 3 | 4 | 5 | 6 | 7 | 8 | 9 | 10 | 11 | 12 | 13 | 14 | 15 |   |

j = 7

(8 7)(7 8) = (8 8): 4

CYCLE (8 9 12 10 13 7): 4

j = 9

(2 9)(9 7) = (2 7): -18

j = 11

(2 11)(11 3) = (2 3): <u>-21</u>;  (2 11)(11 7) = (2 7): <u>-19</u>;  (4 11)(11 9) = (4 9): <u>-7</u>

LINKED 2-CIRCUIT $P_{105}$

|   | 5 | 4 | 15 | 2 | 1 | 6 | 14 | 12 | 10 | 9 | 13 | 8 | 11 | 7 | 3 |   |
|---|---|---|----|---|---|---|----|----|----|---|----|---|----|---|---|---|
|   | 1 | 2 | 3 | 4 | 5 | 6 | 7 | 8 | 9 | 10 | 11 | 12 | 13 | 14 | 15 |   |
| 1 |   | 1 | 12 | 5! | 3 | 4 | 9 |   | 4 | 4 |    | 13 | 2 |    |    | 1 |
| 2 | 2 |   | 11 | 1  | 4! | 13 | 11 | 15 | 12 | 12 | 9 | 5 | 9 | 5 | 5 | 2 |
| 3 |   |   |   |   |   |   |   |   |   |   |   |   |   |   |   | 3 |
| 4 | 4 | 1 | 12 |   | 2! | 12 | 9 |   |   | 13 | 12 | 5 | 12 | 12 |   | 4 |
| 5 |   |   |   |   |   |   |   |   |   |   |   |   |   |   |   | 5 |
| 6 |   |   |   |   |   |   |   |   |   |   |   |   |   |   |   | 6 |
| 7 |   |   |   |   |   |   |   |   |   |   |   |   |   |   |   | 7 |
| 8 |   |   |   |   |   | 13 | 13 |   | 8 | 12! |   | 9 | 10 |   |   | 8 |
| 9 |   |   |   |   |   |   | 8 | 10! |   | 12 | 8 | 9 |   |   |   | 9 |
| 10 |   |   |   |   |   |   |   |   |   |   |   |   |   |   |   | 10 |
| 11 |   |   |   |   |   |   |   |   |   |   |   |   |   |   |   | 11 |
| 12 |   |   |   |   |   |   |   |   |   |   |   |   |   |   |   | 12 |
| 13 |   |   |   |   |   |   | 11 | 7! | 8 | 8 | 12 | 13 |   |   |   | 13 |
| 14 |   |   |   |   |   |   |   |   |   |   |   |   |   |   |   | 14 |
| 15 |   |   |   |   |   |   |   |   |   |   |   |   |   |   |   | 15 |
|   | 1 | 2 | 3 | 4 | 5 | 6 | 7 | 8 | 9 | 10 | 11 | 12 | 13 | 14 | 15 |   |

## LINKED 2-CIRCUIT $\sigma_T^{-1} M^-$ (105)

|   | 5 | 4 | 15 | 2 | 1 | 6 | 14 | 12 | 10 | 9 | 13 | 8 | 11 | 7 | 3 |   |
|---|---|---|----|---|---|---|----|----|----|---|----|---|----|---|---|---|
|   | 1 | 2 | 3  | 4 | 5 | 6 | 7  | 8  | 9  | 10| 11 | 12| 13 | 14| 15|   |
| 1 |   | 1 | -22 | -6! | -6 | *-13* | *-17* |   | *-23* | *-23* |   | *-21* | -15 |   |   | 1 |
| 2 | -16 |   | <u>-21</u> | -13 | - | *-19* | <u>*-19*</u> | *-19* | -24 | -30 | -30 | -28 | -16 | -13 | -13 | 2 |
| 3 |   |   |    |   |   |   |    |    |    |   |    |   |    |   |   | 3 |
| 4 | -3 | -2 | *-16* |   | - | *-19* | -12 |   | <u>*-7*</u> | *-16* | -17 | -15 | -16 | -20 | 0 | 4 |
| 5 |   |   |    |   |   |   |    |    |    |   |    |   |    |   |   | 5 |
| 6 |   |   |    |   |   |   |    |    |    |   |    |   |    |   |   | 6 |
| 7 |   |   |    |   |   |   |    |    |    |   |    |   |    |   |   | 7 |
| 8 |   |   |    |   |   | *-2* | 5 |   | 0 | -5! |   | -3 | *1* |   |   | 8 |
| 9 |   |   |    |   |   |   | *4* | 0! |   | -5 | *-1* | -3 |    |   |   | 9 |
| 10|   |   |    |   |   |   |    |    |    |   |    |   |    |   |   | 10|
| 11|   |   |    |   |   |   |    |    |    |   |    |   |    |   |   | 11|
| 12|   |   |    |   |   |   |    |    |    |   |    |   |    |   |   | 12|
| 13|   |   |    |   |   |   | 3 | 2! | *2* | -2 | -8 | -6 |    |   |   | 13|
| 14|   |   |    |   |   |   |    |    |    |   |    |   |    |   |   | 14|
| 15|   |   |    |   |   |   |    |    |    |   |    |   |    |   |   | 15|
|   | 1 | 2 | 3  | 4 | 5 | 6 | 7  | 8  | 9  | 10| 11 | 12| 13 | 14| 15|   |

None of the following paths can be extended: (2 ... 3): -21, (2 ... 7): -19, (4 ... 9): -7.

We next construct the set of all cycles obtained thus far. As previously, they are grouped according to the number of 2-cycles containing a point on the cycle. Each such subset is listed in order of increasing value. Each point – followed by a colon – indicates the 2-cycle in which it is contained. A *capitalized Roman numeral* denotes the number of 2-cycles touched by each cycles following it. C denotes the 2-cycles containing a point of a cycle. DC denotes the 2-cycles not contained in the cycle. An *italicized point* is a non-linking point. A value of cycle with a superscripted asterik indicates a cycle containing the point 6.

1. (1 5)  2. (2 4)  3. (3 15)  4. 6  5. (7 14)  6. (8 12)  7. (9 10)  8. (11 13)

1: 1.  2: 2.  3: 3.  4: 2.  5: 1.  6: 4.  7: 5.  8: 6.  9: 7.  10: 7.  11: 8.  12: 6.  13: 8.

14: 5.  15: 3.

VI. *1.* DC *4*, 8; 2; (*4* 1 *2* 9 12 3 7).

V. *2.* DC 3, *4*, 5; -31; (*2* 1 *4* 12 13 10).  *3.* DC 3, *4*, 5; -10; (*2* 1 *4* 9 12 11).

*4.* DC *4*, 5, 8; -5; (5 *4* 9 12 3).  *5.* DC 3, *4*, 5; 2; (5 *4* 9 12 1).

*6.* DC 1, 3, *4*; 4; (*4* 9 12 13 7).  *7.* DC 1, 2, 3; 28*; (8 10 13 6 7).

IV. *8.* DC 3, *4*, 5, 6; -10; (*2* 1 *4* 9 10).  *9.* DC *4*, 5, 7, 8; -8; (*2* 1 *4* 12 3).

*10.* DC 3, *4*, 5, 8; -2; (*4* 1 *2* 8 10).  *11.* DC 3, *4*, 5, 8; -2; (*4* 1 *2* 9 12).

*12.* DC 3, *4*, 5, 8; 0; (5 *4* 9 12).  *13.* DC 3, *4*, 5, 6; 2; (*2* 1 9 11).  *14.* DC 1, 3, *4*, 5; 5; (*4* 9 12 13).

*15.* DC 1, 3, *4*, 7; 5; (7 8 *4* 13).  *16.* DC 1, 2, 3, *4*; 6; (7 8 9 11).

III. *17.* C 2, 6, 7; -4; ; (*2* 12 10).  *18.* C 2, 6, 7; -2; (*2* 9 8).  *19.* C 2, 6, 7; -2; (*4* 9 8).

*20.* C 6, 7, 8; 0; (13 8 9).  *21.* C 2, 6, 7; 0; (*4* 12 9.  *22.* C 6, 7, 8; 0; (11 8 9).

*23.* C 2, 6, 7; 0; (12 *4* 9).  *24.* C 1, 2 7; 1; (*2* 1 9).  *25.* C 1, 2, 7; 1; (5 *4* 9).

*26.* C 1, 2, 7; 1; (5 4 10). *27.* C 2, 7, 8; 3; (4 9 11). *28.* C 2, 6, 8; 3; 3; (8 4 13).

*29.* C 2, 3, 6; 6; (2 12 3). *30.* C 1, 2, 6; 5; (1 2 8). *31.* C 5, 6, 7; 5; (7 8 9).

*32.* C 1, 2, 6; 5; (12 4 5). *33.* C 6, 7, 8; 5; (13 8 10). *34.* C 6, 7, 8; 6; (13 8 9).

II. *35.* C 1, 2; -15; (2 1). *36.* C 1, 2; -15; (5 4). *37.* C 3, 6; -7; (3 12). *38.* C 6, 7; -1; (8 9).

*39.* C 6, 7; -1; (9 8). *40.* C 1, 2; 0; (4 1). *41.* C 1, 2; 0; (5 2) . *42.* C 2, 7; 1; (2 10).

*43.* C 2, 7; 1; (4 9). *44.* C 6, 7; 1; (9 12). *45.* C 6, 7; 1; (10 8). *46.* C 2, 7; 2; (4 10).

*47.* C 2, 6; 3; (2 8). *48.* C 5, 6; 3; (7 8). *49.* C 5, 6; 3; (8 7). *50.* C 6, 8; 3; (11 8).

In what follows, we say that two cycles *touch* if they have respective linking points in the same 2-cycle.

We now search for the point that is contained in the fewest cycles. It is 6. It occurs only in cycle *7.* .We thus construct tables containing all cycles that touch at most one 2-cycle in 7. . Since 7. has points in all cycles except 1., 2., 3., we only search for cycles that have points in at most four 2-cycles. At most one of these can be in a cycle from 5. to 8. .Alternately, DC cycles must include at least three of the cycles from 5. to 8. . We will use the tables to search for linkages to cycle *7.* .

$$M_{6,7.}$$

| | | | (1  5) | (2  4) | (3  15) | 6 | (7  14) | (8  12) | (9  10) | (11  13) |
|---|---|---|---|---|---|---|---|---|---|---|
| VALUE | 2-CYCLES | | 1 | 2 | 3 | 4 | 5 | 6 | 7 | 8 |
| 28* | 4,5,6,7,8 | 7. | | | | 6 | 7 | 8 | 10 | 13 |
| -8 | 1,2,3,6 | 9. | 1 | 2,4 | 3 | | | 12 | | |
| 1 | 1,2,7 | 24. | 1 | 2 | | | | | 9 | |
| | | | 1 | 2 | 3 | 4 | 5 | 6 | 7 | 8 |

| | | | 1 | 2 | 3 | 4 | 5 | 6 | 7 | 8 |
|---|---|---|---|---|---|---|---|---|---|---|
| 1 | 1,2,7 | *25.* | 5 | 4 | | | | | 9 | |

| | | | 1 | 2 | 3 | 4 | 5 | 6 | 7 | 8 | |
|---|---|---|---|---|---|---|---|---|---|---|---|
| 1 | 1,2,7 | *26.* | 5 | 4 | | | | | 10 | | |
| 4 | 2,3,6 | *29.* | | 2 | 3 | | | 12 | | | |
| 5 | 1,2,6 | *30.* | 1 | 2 | | | | 8 | | | |
| 5 | 1,2,6 | *32.* | 5 | 4 | | | | 12 | | | |
| -15 | 1,2 | 35. | 1 | 2 | | | | | | | |
| -15 | 1,2 | 36. | 5 | 4 | | | | | | | |
| -7 | 3,6 | 37. | | | 3 | | | | | 12 | |
| 0 | 1,2 | 40. | 1 | 4 | | | | | | | |
| 0 | 1,2 | 41. | 5 | 2 | | | | | | | |
| 1 | 2,7 | 42. | | 2 | | | | | 10 | | |
| 1 | 2,7 | 43. | | 4 | | | | | 9 | | |
| 2 | 2,7 | 46. | 4 | | | | | | 10 | | |
| 3 | 2,6 | 47. | 2 | | | | | 8 | | | |
| | | | 1 | 2 | 3 | 4 | 5 | 6 | 7 | 8 | |

Before proceeding, we eliminate all but one of a set of cycles that have (i) the same set of unlinked points in a 2-cycle, (ii) the same set of cycles containing points, the same value. We thus obtain

$$M_{6,7.}$$

| VALUE | IN 2-CYCLES | | (1 5) | (2 4) | (3 15) | 6 | (7 14) | (8 12) | (9 10) | (11 13) |
|---|---|---|---|---|---|---|---|---|---|---|
| | | | 1 | 2 | 3 | 4 | 5 | 6 | 7 | 8 |
| 28* | 4,5,6,7,8 | 7. | | | | 6 | 7 | 8 | 10 | 13 |
| -8 | 1,2,3,6 | 9. | 1 | 2,4 | 3 | | | 12 | | |
| 1 | 1,2,7 | 24. | 1 | 2 | | | | | 9 | |
| 6 | 2,3,6 | 29. | | 2 | 3 | | | 12 | | |
| 5 | 1,2,6 | 30. | 1 | 2 | | | | 8 | | |
| -15 | 1,2 | 35. | 1 | 2 | | | | | | |
| -7 | 3,6 | 37. | | | 3 | | | | | 12 |
| 0 | 1,2 | 40. | 1 | 4 | | | | | | |
| 1 | 2,7 | 42. | | 2 | | | | | 10 | |
| 2 | 2,7 | 46. | 4 | | | | | | 10 | |
| 3 | 2,6 | 47. | 2 | | | | | 8 | | |
| | | | 1 | 2 | 3 | 4 | 5 | 6 | 7 | 8 |

We now attempt to construct tours by linking cycles. If necessary, we include one phantom cycle that is removed once an existing cycle can take its place. As with the F-W algorithm applied to paths, we start with a determining cycle. Once we have obtained all linkages possible with a determining cycle, we no longer can use it for linkage in future cases. We start with the cycle having the smallest value as the first determining cycle.

Before proceeding further, we note the formula for the number of points in a set of cycles that can be linked to form a tour when $n$ is odd: $p = \dfrac{n+1}{2} + 3t + a - 1$ . As previously, $t$ is the number of 2-circuit cycles and $a$ the number of acceptable cycles. Linking points are printed in boldface.

*35.* ⊕ *29.* ⊕ *7.* = (1 **2**) ⊕ (**2** **12** 3) ⊕ (**8** 10 13 6 7)  yields  (1 4 3 15 12 7 14 6 13 11 10 9 8 2 5): 551
The value of the tour in $\sigma_{T_{UPPERBOUND}}^{-1} M^-$ is 19*.

No other tour can be obtained using *35.* as an initial determining cycle.

No tour can be obtained using the unlinked 2-cycle circuit cycle $9.$ as a determining cycle.

$37. \oplus 7. \oplus 24 = (3\ \mathbf{12}) \oplus (\mathbf{8\ 10}\ 13\ 6\ 7) \oplus (2\ 1\ \mathbf{9})$. The sum of the values of the cycles equals the value of the tour in $\sigma^{-1}_{T_{UPPERBOUND}} M^{-}$. Its value is $-7 + 28^{*} + 1 = 22^{*}$. Thus it is larger in value than the previous case.

There exist no more negatively-valued cycles that can be used for determining cycles. Thus, using the modified F-W algorithm, we can't obtain a tour of value less than 551. Our new upper bound is denoted by $T_1$. We now construct $\sigma^{-1}_{T_1} M^{-}$ in the following way: We search for the set of fourteen alternating arcs such that the difference between $T_1$ minus twice the value of the smaller of the two, minus the value of the missing arc, minus 1, is smallest possible. The set of alternating arcs whose sum of values is smaller yields the edges (2-cycles) comprising $\sigma_{T_1}$. Our fixed point is the point that is not included in the set chosen.

We wish to obtain paths that satisfy the following rule:

Assuming that $a_1$ is the first point of the smaller-valued of the two sets of alternating arcs, we wish to obtain $|u|$, an upper bound for the values of paths that don't contain the point $a_n$, as well as $u'$, an upper bound for the values of all paths, including those that contain $a_n$.

Using all possible sets of alternating arcs, assume that the initial point of the first arc of the smaller-valued set begins with $a_1$. For convenience sake, assume that the missing fixed point is $a_x$ while $\min(a_x)$ is the smallest value of all of the arcs in row $a_x$

$$|u| = |T| - 2(\sum_{i=1}^{\frac{n}{2}} |a_{2i-1}\ a_{2i}|) - |(a_x\ a_{x+1})| - 1,$$

.

$$|u'| = |T| - 2(\sum_{i=1}^{\frac{n}{2}} |a_{2i-1}\ a_{2i}|) - 1$$

where the right-hand side runs through all $n$ possible ways of obtaining alternating sets of arcs from $T$ as well as all $n$ possible ways of obtaining alternating sets of arcs from $T^{-1}$.

We comment on the above formulas:

The smaller the difference between the values of the two sets of alternating arcs, as well as the smaller the difference between the fixed point arc's value and the minimally valued entry in the fixed point row, the smaller the upper bounds will be for searching for cycles. In the case of clustered cost matrices, it is often preferable to go through the modified F-w algorithm again using the newly obtained upper bound tour to try to obtain a still smaller-valued tour. In this particular case,
$\sigma_{T_1} = (1\ 4)(3\ 15)(12\ 7)(14\ 6)(13\ 11)(10\ 9)(8\ 2)$. The fixed point arc is (5 1). We thus obtain:

$u = 551 - 492 - 50 - 1 = 8$,

$u' = 551 - 492 - 1 = 58*$.

M

|    | 1  | 2  | 3  | 4  | 5  | 6  | 7  | 8  | 9  | 10 | 11 | 12 | 13 | 14 | 15 |    |
|----|----|----|----|----|----|----|----|----|----|----|----|----|----|----|----|----|
| 1  | ∞  | 53 | 52 | 51 | 50 | 49 | 48 | 48 | 48 | 48 | 50 | 51 | 52 | 53 | 54 | 1  |
| 2  | 53 | ∞  | 66 | 66 | 50 | 49 | 48 | 47 | 48 | 49 | 50 | 51 | 52 | 53 | 54 | 2  |
| 3  | 52 | 66 | ∞  | 62 | 65 | 64 | 48 | 48 | 48 | 49 | 50 | 75 | 69 | 77 | 36 | 3  |
| 4  | 51 | 66 | 62 | ∞  | 63 | 62 | 65 | 49 | 49 | 49 | 50 | 77 | 73 | 76 | 71 | 4  |
| 5  | 50 | 50 | 65 | 63 | ∞  | 66 | 65 | 50 | 73 | 76 | 74 | 71 | 77 | 75 | 76 | 5  |
| 6  | 49 | 49 | 64 | 62 | 66 | ∞  | 65 | 51 | 70 | 77 | 73 | 37 | 23 | 30 | 69 | 6  |
| 7  | 48 | 48 | 48 | 65 | 65 | 65 | ∞  | 73 | 74 | 71 | 75 | 26 | 39 | 27 | 72 | 7  |
| 8  | 48 | 47 | 48 | 49 | 50 | 51 | 73 | ∞  | 27 | 31 | 73 | 31 | 30 | 35 | 73 | 8  |
| 9  | 48 | 48 | 48 | 49 | 73 | 70 | 74 | 27 | ∞  | 30 | 38 | 29 | 26 | 36 | 77 | 9  |
| 10 | 48 | 49 | 49 | 49 | 76 | 77 | 71 | 31 | 30 | ∞  | 36 | 35 | 73 | 71 | 74 | 10 |
| 11 | 50 | 50 | 50 | 50 | 74 | 73 | 76 | 73 | 38 | 36 | ∞  | 30 | 26 | 37 | 35 | 11 |
| 12 | 51 | 51 | 75 | 77 | 71 | 37 | 26 | 31 | 29 | 35 | 30 | ∞  | 29 | 36 | 30 | 12 |
| 13 | 52 | 52 | 69 | 73 | 77 | 23 | 39 | 30 | 26 | 73 | 26 | 29 | ∞  | 30 | 36 | 13 |
| 14 | 53 | 63 | 77 | 76 | 75 | 30 | 27 | 35 | 36 | 71 | 37 | 36 | 30 | ∞  | 32 | 14 |
| 15 | 54 | 54 | 36 | 71 | 76 | 69 | 72 | 73 | 77 | 74 | 35 | 30 | 36 | 32 | ∞  | 15 |
|    | 1  | 2  | 3  | 4  | 5  | 6  | 7  | 8  | 9  | 10 | 11 | 12 | 13 | 14 | 15 |    |

$$\sigma_{T_1}^{-1} M^-$$

|    | 4  | 8  | 15 | 1  | 5  | 14 | 12 | 2  | 10 | 9  | 13 | 7  | 11 | 6  | 3  |    |
|----|----|----|----|----|----|----|----|----|----|----|----|----|----|----|----|----|
|    | 1  | 2  | 3  | 4  | 5  | 6  | 7  | 8  | 9  | 10 | 11 | 12 | 13 | 14 | 15 |    |
| 1  | 0  | -3 | 3  | ∞  | -1 | 2  | 0  | 2  | -3 | -3 | 1  | -3 | -1 | -2 | 1  | 1  |
| 2  | 19 | 0  | 7  | 6  | 3  | 6  | 4  | ∞  | 2  | 1  | 5  | 1  | 3  | 2  | 19 | 2  |
| 3  | 26 | 12 | 0  | 16 | 29 | 41 | 39 | 30 | 13 | 12 | 33 | 12 | 14 | 28 | ∞  | 3  |
| 4  | ∞  | -2 | 20 | 0  | 12 | 25 | 26 | 15 | -2 | -2 | 22 | 14 | -1 | 11 | 11 | 4  |
| 5  | 63 | 50 | 76 | 50 | ∞  | 75 | 71 | 50 | 76 | 73 | 77 | 65 | 74 | 66 | 65 | 5  |
| 6  | 32 | 21 | 39 | 19 | 36 | 0  | 7  | 19 | 47 | 40 | -7 | 35 | 43 | ∞  | 34 | 6  |
| 7  | 39 | 47 | 46 | 22 | 39 | 1  | 0  | 22 | 45 | 48 | 13 | ∞  | 49 | 39 | 22 | 7  |
| 8  | 2  | ∞  | 26 | 1  | 3  | -12| -16| 0  | -16| -20| -17| 26 | 26 | 4  | 1  | 8  |
| 9  | 19 | -3 | 47 | 18 | 43 | 6  | -1 | 18 | 0  | ∞  | -4 | 44 | 8  | 40 | 18 | 9  |
| 10 | 19 | 1  | 44 | 18 | 46 | 41 | 5  | 19 | ∞  | 0  | 43 | 41 | 6  | 47 | 19 | 10 |
| 11 | 24 | 47 | 9  | 24 | 48 | 11 | 4  | 24 | 10 | 12 | 0  | 50 | ∞  | 47 | 24 | 11 |
| 12 | 52 | 5  | 4  | 25 | 45 | 10 | ∞  | 25 | 9  | 3  | 3  | 0  | 4  | 11 | 49 | 12 |
| 13 | 47 | 4  | 10 | 26 | 51 | 4  | 3  | 26 | 47 | 0  | ∞  | 13 | 0  | -3 | 43 | 13 |
| 14 | 46 | 5  | 2  | 23 | 45 | ∞  | 6  | 33 | 41 | 6  | 0  | -3 | 7  | 0  | 47 | 14 |
| 15 | 35 | 37 | ∞  | 18 | 46 | -4 | -6 | 18 | 38 | 41 | 0  | 36 | -1 | 33 | 36 | 15 |
|    | 1  | 2  | 3  | 4  | 5  | 6  | 7  | 8  | 9  | 10 | 11 | 12 | 13 | 14 | 15 |    |

We now obtain the minimum of $(j\ i\ k)$ where $\min |(j\ i)|$ is the smallest-valued of all arcs terminating in column $i$ while $|(i\ k)|$ is the smallest-valued of the arcs emanating from row $i$. We now give the values for the various points from 1 through 15.

(1) -1  (2) -2  (3) 14  (4) -3  (5) 49*  (6) -11  (7) -16  (8) -18  (9) -7  (10) -19  (11) -13  (12) 0

(13) -4  (14) -5  (15) -5

It follows that 3 has the largest value. We first note that 3 can't be the determining point of an acceptable or 2-circuit cycle. We thus must use theorem 1.22 to find acceptable paths that reach 3. From that point on, we try to construct acceptable or 2-circuit cycles containing 3.

j = 2

(1 2)(2 9) = (1 9): 0; (1 2)(2 10) = (1 10): -1; (1 2)(2 12) = (1 12): -1; (1 2)(2 14) + (1 14): 0

j = 4

(1 4)(4 2) = (1 2): 50*; (1 4)(4 9) = (1 9): 50*; (1 4)(4 13) = (1 13): 51*

j = 5

(4 5)(5 8) = (4 8): 51*

*(1 4)(4 5) = (1 5): 52*; (8 5)(5 2) = (8 2): 48**

j = 6

(7 6)(6 11) = (7 11): -6; (8 6)(6 7) = (8 7): -5; (8 6)(6 11) = (8 11): -14;

(15 6)(6 7) = (15 7): -11; (15 6)(6 11) = (15 11): -11

j = 7

(8 7)(7 6) = (8 6): <u>-15</u>; (9 7)(7 6) = (9 6): <u>0</u>; (15 7)(7 6) = (15 6): <u>-5</u>

j = 8

(5 8)(8 6) = (5 6): <u>-12</u>; (5 8)(8 7) = (5 7): <u>-16</u>; (5 8)(8 9) = (5 9): -16; (5 8)(8 10) = (5 10): -20;

(5 8)(8 11) = (5 11): -17

*(8 9)(9 2) = (8 2): -19*

j = 9

(4 9)(9 2) = (4 2): <u>46\*</u>

j = 10

(5 10)(10 13) = (5 13): -14

j = 11

(7 11)(11 7) = (7 7): -2

CYCLE (7 6 11): -2

(8 11)(11 3) = (8 3): <u>-11</u>

j = 12

(1 12)(12 3) = (1 3): 52\*

j = 13

(1 13)(13 14) = (1 14): 48\*; (5 13)(13 14) = (5 14): -17

*(8 13)(13 2) = (8 2): -4*

j = 14

(1 14)(14 3) = (1 3): <u>50\*</u>; (1 14)(14 11) = (1 11): 48\*; (4 14)(14 12) = (4 12): <u>44\*</u>;

(5 14)(14 3) = (5 3): <u>-15</u>; (5 14)(14 12) = (5 12): <u>-20</u>

$P_{15}$

|    | 1 | 2 | 3  | 4 | 5 | 6 | 7  | 8 | 9 | 10 | 11 | 12 | 13 | 14 | 15 |    |
|----|---|---|----|---|---|---|----|---|---|----|----|----|----|----|----|----|
| 1  |   | 9 | 12 | 1 |   |   | 9  |   | 4 | 2  | 14 | 14 | 4  | 13 |    | 1  |
| 2  |   |   |    |   |   |   |    |   |   |    |    |    |    |    |    | 2  |
| 3  |   |   |    |   |   |   |    |   |   |    |    |    |    |    |    | 3  |
| 4  |   | 9 | 14 |   | 4 |   | 9  | 5 | 4 | 13 | 9  | 14 | 4  | 13 |    | 4  |
| 5  |   |   | 14 |   |   | 8 | 8  | 5 | 8 | 8  | 8  | 14 | 10 | 13 |    | 5  |
| 6  |   |   |    |   |   |   |    |   |   |    |    |    |    |    |    | 6  |
| 7  |   |   |    |   |   | 7 |    |   |   |    | 6  |    |    |    |    | 7  |
| 8  |   |   | 11 |   |   | 7 | 8  |   | 8 |    | 9  |    | 9  |    |    | 8  |
| 9  |   |   |    |   |   | 7 | 9  |   |   |    |    |    |    |    |    | 9  |
| 10 |   |   |    |   |   |   |    |   |   |    |    |    |    |    |    | 10 |
| 11 |   |   |    |   |   |   |    |   |   |    |    |    |    |    |    | 11 |
| 12 |   |   |    |   |   |   |    |   |   |    |    |    |    |    |    | 12 |
| 13 |   |   |    |   |   |   |    |   |   |    |    |    |    |    |    | 13 |
| 14 |   |   |    |   |   |   |    |   |   |    |    |    |    |    |    | 14 |
| 15 |   |   |    |   |   | 7 | 15 |   |   |    | 6  |    |    |    |    | 15 |
|    | 1 | 2 | 3  | 4 | 5 | 6 | 7  | 8 | 9 | 10 | 11 | 12 | 13 | 14 | 15 |    |

$$\sigma_{T_1}^{-1}M^-(15)$$

|    | 5 | 8 | 15 | 4 | 1 | 14 | 12 | 2 | 10 | 9 | 13 | 7 | 11 | 6 | 3 |    |
|----|---|---|----|---|---|----|----|---|----|---|----|---|----|---|---|----|
|    | 1 | 2 | 3  | 4 | 5 | 6  | 7  | 8 | 9  | 10| 11 | 12| 13 | 14| 15|    |
| 1  |   | <u>-5</u> | <u>50*</u> |  |  |  | <u>-3</u> |  | 50* | *-1* | *-6* | *-1* | <u>51*</u> | *48* |  | 1  |
| 2  |   |   |    |   |   |    |    |   |    |   |    |   |    |   |   | 2  |
| 3  |   |   |    |   |   |    |    |   |    |   |    |   |    |   |   | 3  |
| 4  |   | <u>46*</u> | <u>49*</u> |  |  |  | <u>48*</u> | *51* |  | <u>50*</u> | <u>45*</u> | <u>44*</u> |  | *47* |  | 4  |
| 5  |   |   | <u>-15</u> |  |  | <u>-12</u> | <u>-16</u> |  | *-16* | *-20* | *-17* | <u>-20</u> | *-14* | *-17* |  | 5  |
| 6  |   |   |    |   |   |    |    |   |    |   |    |   |    |   |   | 6  |
| 7  |   |   |    |   |   |    |    |   |    |   | *-6* |   |    |   |   | 7  |
| 8  |   |   | <u>-11</u> |  |  | <u>-15</u> | *-17* |   |    |   | *-20* |   | *-8* |   |   | 8  |
| 9  |   |   |    |   |   |    |    |   |    |   |    |   |    |   |   | 9  |
| 10 |   |   |    |   |   |    |    |   |    |   |    |   |    |   |   | 10 |
| 11 |   |   |    |   |   |    |    |   |    |   |    |   |    |   |   | 11 |
| 12 |   |   |    |   |   |    |    |   |    |   |    |   |    |   |   | 12 |
| 13 |   |   |    |   |   |    |    |   |    |   |    |   |    |   |   | 13 |
| 14 |   |   |    |   |   |    |    |   |    |   |    |   |    |   |   | 14 |
| 15 |   |   |    |   |   |    | *-11* |  |    |   | *-11* |   |    |   |   | 15 |
|    | 1 | 2 | 3  | 4 | 5 | 6  | 7  | 8 | 9  | 10| 11 | 12| 13 | 14| 15|    |

j = 2

(1 2)(2 12) = (1 12): -4; (1 2)(2 13) = (1 13): -2; (1 2)(2 12) = (1 12): -4;

(4 2)(2 13) = (4 13): 49*

j = 3

*(8 3)(3 2) = (8 2): 1*

j = 6

(5 6)(6 11) = (5 11): -19; (8 6)(6 11) = (8 11): -22

j = 7

(4 7)(7 6) = (4 6): 49*; (5 7)(7 6) = (5 6): -15

j = 11

(8 11)(11 3) = (8 3): -13; (8 11)(11 7): -18

j = 12

(4 12)(12 3) = (4 3): 48*; (4 12)(12 10) = (4 10): 47*; (5 12)(12 3) = (5 3): -16

.

$P_{30}$

|    | 1 | 2 | 3  | 4 | 5 | 6 | 7  | 8 | 9 | 10 | 11 | 12 | 13 | 14 | 15 |    |
|----|---|---|----|---|---|---|----|---|---|----|----|----|----|----|----|----|
| 1  |   | 9 | 12 | 1 |   |   | 9  |   | 4 | 2  | 14 | 2  | 2  | 13 |    | 1  |
| 2  |   |   |    |   |   |   |    |   |   |    |    |    |    |    |    | 2  |
| 3  |   |   |    |   |   |   |    |   |   |    |    |    |    |    |    | 3  |
| 4  |   | 9 | 12 |   | 4 | 7 | 9  | 5 | 4 | 13 | 9  | 14 | 2  | 13 |    | 4  |
| 5  |   |   | 12 |   |   | 7 | 8  | 5 | 8 | 8  | 6  | 14 | 10 | 13 |    | 5  |
| 6  |   |   |    |   |   |   |    |   |   |    |    |    |    |    |    | 6  |
| 7  |   |   |    |   |   | 7 |    |   |   |    | 6  |    |    |    |    | 7  |
| 8  |   |   | 11 |   |   | 7 | 11 |   | 8 |    | 9  |    | 9  |    |    | 8  |
| 9  |   |   |    |   |   | 7 | 9  |   |   |    |    |    |    |    |    | 9  |
| 10 |   |   |    |   |   |   |    |   |   |    |    |    |    |    |    | 10 |
| 11 |   |   |    |   |   |   |    |   |   |    |    |    |    |    |    | 11 |
| 12 |   |   |    |   |   |   |    |   |   |    |    |    |    |    |    | 12 |
| 13 |   |   |    |   |   |   |    |   |   |    |    |    |    |    |    | 13 |
| 14 |   |   |    |   |   |   |    |   |   |    |    |    |    |    |    | 14 |
| 15 |   |   |    |   |   | 7 | 15 |   |   |    | 6  |    |    |    |    | 15 |
|    | 1 | 2 | 3  | 4 | 5 | 6 | 7  | 8 | 9 | 10 | 11 | 12 | 13 | 14 | 15 |    |

$$\sigma_{T_1}^{-1} M^-(30)$$

|   | 5 | 8 | 15 | 4 | 1 | 14 | 12 | 2 | 10 | 9 | 13 | 7 | 11 | 6 | 3 |   |
|---|---|---|----|---|---|----|----|---|----|---|----|---|----|---|---|---|
|   | 1 | 2 | 3 | 4 | 5 | 6 | 7 | 8 | 9 | 10 | 11 | 12 | 13 | 14 | 15 |   |
| 1 |  | <u>47*</u> | *50* |  |  | <u>49*</u> |  | *50* | -1 | -6 | *48* | *51* | *48* |  |  | 1 |
| 2 |  |  |  |  |  |  |  |  |  |  |  |  |  |  |  | 2 |
| 3 |  |  |  |  |  |  |  |  |  |  |  |  |  |  |  | 3 |
| 4 |  | <u>46*</u> | <u>48*</u> |  |  | <u>49*</u> | <u>48*</u> | *51* |  | <u>47*</u> | <u>42*</u> | <u>44*</u> | <u>49*</u> | *47* |  | 4 |
| 5 |  |  | <u>-16</u> |  |  | <u>-15</u> | -16 |  | -16 | -20 | -19 | -20 | -14 | -17 |  | 5 |
| 6 |  |  |  |  |  |  |  |  |  |  |  |  |  |  |  | 6 |
| 7 |  |  |  |  |  |  |  |  |  |  | -6 |  |  |  |  | 7 |
| 8 |  |  | <u>-13</u> |  |  | -15 | <u>-18</u> |  |  |  | -22 |  | -8 |  |  | 8 |
| 9 |  |  |  |  |  |  |  |  |  |  |  |  |  |  |  | 9 |
| 10 |  |  |  |  |  |  |  |  |  |  |  |  |  |  |  | 10 |
| 11 |  |  |  |  |  |  |  |  |  |  |  |  |  |  |  | 11 |
| 12 |  |  |  |  |  |  |  |  |  |  |  |  |  |  |  | 12 |
| 13 |  |  |  |  |  |  |  |  |  |  |  |  |  |  |  | 13 |
| 14 |  |  |  |  |  |  |  |  |  |  |  |  |  |  |  | 14 |
| 15 |  |  |  |  |  |  | *-11* |  |  |  | *-11* |  |  |  |  | 15 |
|   | 1 | 2 | 3 | 4 | 5 | 6 | 7 | 8 | 9 | 10 | 11 | 12 | 13 | 14 | 15 |   |

j = 2

(1 2)(2 13) = (1 13): 50*

j = 3

(8 3)(3 12) = (8 12): -1

*(8 9 11 3)(3 2) = (8 2): -1*

j = 6

(4 6)(6 11) = (4 11): 42*; (5 6)(6 11) = (5 11): -22

j = 7

(1 7)(7 6) = (1 6): <u>50*</u>; (8 7)(7 6) = (8 6): <u>-17</u>

j = 12

(1 12)(12 3) = (1 3): <u>49*</u>; (1 12)(12 13) = (1 13): 49*

$P_{45}$

|    | 1 | 2 | 3  | 4 | 5 | 6  | 7  | 8 | 9 | 10 | 11 | 12 | 13 | 14 | 15 |    |
|----|---|---|----|---|---|----|----|---|---|----|----|----|----|----|----|----|
| 1  |   | 9 | 12 | 1 |   |    | 9  |   | 4 | 2  | 14 | 2  | 12 | 13 |    | 1  |
| 2  |   |   |    |   |   |    |    |   |   |    |    |    |    |    |    | 2  |
| 3  |   |   |    |   |   |    |    |   |   |    |    |    |    |    |    | 3  |
| 4  |   | 9 | 12 |   | 4 | 13 | 9  | 5 | 4 | 13 | 6  | 14 | 2  | 13 |    | 4  |
| 5  |   |   | 12 |   |   | 7  | 8  | 5 | 8 | 8  | 6  | 14 | 10 | 13 |    | 5  |
| 6  |   |   |    |   |   |    |    |   |   |    |    |    |    |    |    | 6  |
| 7  |   |   |    |   |   | 7  |    |   |   |    | 6  |    |    |    |    | 7  |
| 8  |   |   | 11 |   |   | 7  | 11 |   | 8 |    | 9  | 3  | 9  |    |    | 8  |
| 9  |   |   |    |   |   | 7  | 9  |   |   |    |    |    |    |    |    | 9  |
| 10 |   |   |    |   |   |    |    |   |   |    |    |    |    |    |    | 10 |
| 11 |   |   |    |   |   |    |    |   |   |    |    |    |    |    |    | 11 |
| 12 |   |   |    |   |   |    |    |   |   |    |    |    |    |    |    | 12 |
| 13 |   |   |    |   |   |    |    |   |   |    |    |    |    |    |    | 13 |
| 14 |   |   |    |   |   |    |    |   |   |    |    |    |    |    |    | 14 |
| 15 |   |   |    |   |   | 7  | 15 |   |   |    | 6  |    |    |    |    | 15 |
|    | 1 | 2 | 3  | 4 | 5 | 6  | 7  | 8 | 9 | 10 | 11 | 12 | 13 | 14 | 15 |    |

$$\sigma_{T_1}^{-1} M^-(45)$$

| | 5 | 8 | 15 | 4 | 1 | 14 | 12 | 2 | 10 | 9 | 13 | 7 | 11 | 6 | 3 | |
|---|---|---|---|---|---|---|---|---|---|---|---|---|---|---|---|---|
| | 1 | 2 | 3 | 4 | 5 | 6 | 7 | 8 | 9 | 10 | 11 | 12 | 13 | 14 | 15 | |
| 1 | | 47* | <u>49*</u> | | | <u>50*</u> | 49* | | 50* | -1 | -6 | 45* | 49* | 48* | | 1 |
| 2 | | | | | | | | | | | | | | | | 2 |
| 3 | | | | | | | | | | | | | | | | 3 |
| 4 | | 46* | 48* | | | <u>46*</u> | 48* | 51* | | 47* | 45* | 44* | 49* | 47* | | 4 |
| 5 | | | -16 | | | -15 | -16 | | -16 | -20 | -22 | -20 | -14 | -17 | | 5 |
| 6 | | | | | | | | | | | | | | | | 6 |
| 7 | | | | | | | | | | | -6 | | | | | 7 |
| 8 | | | -13 | | | <u>-17</u> | -18 | | | | -22 | -1 | -8 | | | 8 |
| 9 | | | | | | | | | | | | | | | | 9 |
| 10 | | | | | | | | | | | | | | | | 10 |
| 11 | | | | | | | | | | | | | | | | 11 |
| 12 | | | | | | | | | | | | | | | | 12 |
| 13 | | | | | | | | | | | | | | | | 13 |
| 14 | | | | | | | | | | | | | | | | 14 |
| 15 | | | | | | -11 | | | | | -11 | | | | | 15 |
| | 1 | 2 | 3 | 4 | 5 | 6 | 7 | 8 | 9 | 10 | 11 | 12 | 13 | 14 | 15 | |

We now choose 2-cycle (3 15) as one that is difficult to belong to a cycle that, in turn, can be linked to a set of cycles that yield a tour. Since every cycle that contains 15 corresponds to a companion cycle containing 3, we need only try to obtain acceptable or 2-circuit cycles containing 3. We first note that 3 cannot be a determining point of an acceptable or 2-circuit cycle. We therefore use theorem 1.22 to try to obtain a cycle in which 3 is not a determining point. As we see from the following trees, we are unable to obtain a cycle containing 3. Thus, $T_{OPT} = T_{FWK}$.

Chapter 3

The Asymmetric Traveling Salesman Problem

I. Introduction. Let M be an $n \times n$ asymmetric cost matrix. Using PHASE 2 of the modified F-W algorithm, [1], we obtain negative cycles in $T_0^{-1}M^-$ where $T_0$ is an arbitrary $n$-cycle in M. Using products of disjoint negative cycles, we obtain $D_1, D_2, \ldots, D_r$, a sequence of derangements in M each of which has a smaller value than the previous one. When we no longer can obtain a derangement of smaller value, we have obtained $D_{ASSIGNMENT}$, an optimal solution to the Assignment Problem. Using patching, we obtain an $n$-cycle which we name $T_{UPPERBOUND}$. An "average arc-value of paths" procedure in the Floyd-Warshall algorithm to M, is called the FWK algorithm. Using it, we construct acceptable average arc-value paths in M. Let $C$ be a cycle containing $r$ points in M whose average arc-value value is no greater than $u_C = \dfrac{|C|}{r}$. In theorem 3.1, we prove that there always exists a sequence of subpaths of $C$, say $P_1, P_2, \ldots, P_r$, each of which has an average arc-value no greater than $u_C$. Using the FWK algorithm, we obtain an $n$-cycle, $T_{FWKOPT}$. Let $u_{FWK}$ be the average arc-value of $T_{FWKOPT}$. During the algorithm, we place in a file **REPLACED & CAN'T REPLACE** all acceptable paths that have either been replaced or had larger average arc values than the path currently represented by an entry. (All such paths have an average arc-value value no greater than $\dfrac{|T_{UPPERBOUND}|}{n}$.) We now eliminate all paths in **REPLACED & CAN'T REPLACE** that have an average arc value greater than $\dfrac{|T_{FWKOPT}|}{n}$. To obtain $T_{OPT+}$, we construct trees whose initial paths are the paths still in **REPLACED & CAN'T REPLACE**. Either we obtain an $n$-cycle whose average arc-value is less than $u_{FWKOPT}$ or else $T_{OPT} = T_{FWKOPT}$. In the former case, the $n$-cycle having the smallest average arc-value is $T_{OPT}$. branches are acceptable "average arc value" paths whose average arc value is no greater than $\dfrac{|T_{FWOPT+}|}{n}$. In the next chapter, we discuss replacing trees by a sequence M, $\mathbf{M_2}, \mathbf{M_3}, \ldots, \mathbf{M_r}$ **(r < n)** where each $\mathbf{M_i}$, **(i ≥ 2)** consists of rows of paths obtained from $\mathbf{M_{i-1}}$'s set of **REPLACED & CAN'T REPLACE**. We will apply the FWK algorithm to each $M_i$ until no paths remain in the last **REPLACED & CAN'T REPLACE**.

II. Theorems.

Let $C$ be an $n$-cycle each of whose arcs has a non-negative weight. A subpath of $C$ is *acceptable* if its average arc-value is no greater than $u_C = \frac{|C|}{n}$.

Theorem 3.1. *Let $C$ be a cycle whose arcs have non-negative weights. Then there always exists a determining point $a_C$ such that all subpaths of $C$ whose initial point is $a_C$ have an average arc value no greater than $u_C = \frac{|C|}{n}$.*

Proof. Let $C = (a_1\ a_2\ a_3\ \ldots\ a_{n-1}\ a_n)$. For simplicity, let the arcs of $C$ be partitioned in the following way:

$[a_1\ a_2\ \ldots\ a_{m_1}],\ [a_{m_1}\ a_{m_1+1}\ a_{m_1+2}\ \ldots\ a_{m_1+l_2}],\ [a_{m_1+l_2+1}\ a_{m_1+l_2+1}\ a_{m_1+l_2+2}\ \ldots\ a_{m_1+l_2+m_2}],\ [a_{m_1+l_2+m_2}\ a_{m_1+l_2+m_2+1}\ \ldots\ a_{m_1+l_2+m_2+l}$

Our purpose here is to partition the arcs of $C$ into sets in which each arc in a partition has the same property: its value when divided by the number of arcs in the partition is less than [greater than] $\frac{|C|}{n}$.

We might use colors to illustrate this. Let those whose average value is less than $\frac{|C|}{n}$ be colored red, while those whose value is greater than $\frac{|C|}{n}$ are colored black. We now go clockwise around the cycle.

Suppose there exists no case where when the values in a red colored partition when added to the values of the black valued partition always has a weighted average that is greater than $\frac{|C|}{n}$. Then the weighted average of all of the partitions, (i.e., $\frac{|C|}{n}$) is greater than $\frac{|C|}{n}$. This cannot occur. Thus, there must exist at least a pair of successive partitions that have a weighted average no greater than $\frac{|C|}{n}$. We add these two partitions together to form a new acceptable partition.

We then repeat the operation until we have obtained $C$ as an acceptable partition.

We record all acceptable paths that have been replaced or are not smaller in value than corresponding entries to form a set $S$. We then use each path obtained in $S$ to construct a tree whose branches are all acceptable paths. If each such tree contains at most polynomial-in-$n$ branches or yields an acceptable tour before we have constructed all such trees, then we can obtain an optimal tour in polynomial time.

**Theorem 3.1a** *Let M be an $n \times n$ cost matrix. Denote $T_{UPPERBOUND}$ as an upper bound tour for $T_{OPT}$. Suppose $C$ satisfies $|C| < |T_{UPPERBOUND}|$. Then there exists a determining point, $d_C$ for $C$ such that every subpath along $C$ whose initial point is $d_C$ has an average arc-value no greater than*

$$\frac{|C|}{n} < \frac{|T_{UPPERBOUND}|}{n} = u$$

Proof. Let $u_C$ be the average arc-value of $C$. If an arc of $C$ has a value no greater than $u_C$, color it red. Otherwise, color it black. W.l.o.g., assume that if $C$ consists of directed arcs, then they are oriented in a clockwise manner. Construct subpaths of $C$ by taking the union of consecutive red arcs to form red subpaths, $R_i$, $i = 1, 2, \ldots$, consecutive black arcs to form subpaths $B_j$, $j = 1, 2, \ldots$. Call a subpath acceptable if it has an average arc-value no greater than $u_C$. Thus, every subpath consisting of only red arcs is always acceptable. Suppose only one colored path exists on $C$. It can't be black since then its average arc-value would be greater than $u_C$. Thus, it must be red, proving the theorem for this case. Now assume that $C$ contains two subpaths - the first red, the second black. Suppose their union $R_I \cup B_I = P_I$ contains a black subpath. Then it must be that $P_I$ is also black. The reason is that once we start adding black-colored arcs to $R_I$ the average arc-value of each new subpath becomes larger in a strictly monotonic sense. It follows that the average arc-value of $P_I = C$ must be greater than $u_C$ which is impossible. It follows that the initial point of $R_I$ is $d_C$. Assume that if $C$ consists of $2s$ subpaths, each of which consists of the union of an acceptable path and a black subpath, then each subpath always has an average arc-value no greater than $u_C$. Now consider the case when the number of subpaths is $2s + 2$. Taking the union of each alternating pair, suppose that each new subpath had an average arc-value greater than $u_C$. But then $C$ would have an average arc-value greater than $u_C$ which is impossible. It follows that at least one pair, say $R_i \cup B_i$, has an average arc-value no greater than $u_C$. But therefore $R_i \cup B_i \cup R_{i+1}$ has the same property, i.e., it is acceptable. By replacing $R_{i+1} \cup B_{i+1}$ by $R_i \cup B_i \cup R_{i+1} \cup B_{i+1}$, we now have $2s$ subpaths each pair of which consists of the union of an acceptable subpath and a black subpath. It follows by induction that $C$ always contains a determining point $d_C$.

**Theorem 3.2** $d_C$ is unique if none of its subpaths other than its last one has an average arc-value equal to $aav_C$.

Proof. Suppose $C_i = (R_1\ B_1\ R_2\ B_2\ \ldots\ R_i\ B_i)$, a subpath of $C$ has an average arc-value less than $aav_C$ for all values of i except for the last $RB^i$ that completes the cycle $C$. Without loss of generality, assume that the initial point of $R_{i+1}$, $d_{i+1}$, is a determining point of $C$. Then, by definition, every subpath of $C' = (RB^{i+1}\ RB^{i+2}\ \ldots\ RB^s\ RB^1\ \ldots\ RB^{i-1}\ RB^i)$ has a subpath whose average arc-value is no greater than $aav_C$. But $C' = (RB^{i+1}\ RB^{i+2}\ \ldots\ RB^s) \oplus C_i$ where the average arc-value of $C_i$ is less than $aav_C$. It follows that the average arc-value of $C'$ is less than $aav_C$ which is impossible. It follows that $d_{i+1}$ can't be a determining point of $C$.

In order to obtain it in the general case, starting with $2s$ dubpaths, following the procedure used to reduce subpaths from $2s+2$ to $2s$, we eventually end up with one union of an acceptable path and a black subpath. The initial point of this subpath is $d_C$.

**Theorem 3.3** *Let M be an $n \times n$ cost matrix all of whose entries are positive. Assume $T_{UPPERBOUND}$ has an average-arc value of $u$. Suppose that $P = [a_1\ a_2\ \ldots\ a_r\ a_{r+1}\ \ldots\ a_{s-1}]$ is an acceptable average arc-value path in M that has an average arc-value, $u_p$, where $u_p < u$. Let $P' = P \cup \{(a_{s-1}\ a_r)\}$.*

*Then the following is true:*

*(1) $P' = P \cup [a_r\ a_{r+1}\ \ldots\ a_{s-1}\ a_r] = P \cup C$.*

*(2) If the average arc-value of $C$ is $u_C$, and $u_C \geq u$, then $u_C$ can never be considered as a replacement of $u_p$ in the entry $(a_1\ a_r)$ of M when applying the FWK algorithm to M.*

Proof. Suppose we wish to replace $P$ by $P'$. Then the average arc-value of $P'$, say $u'$, is less than $u$.
$u' = \dfrac{ru_p + (s-r)u_C}{s} < u$. Let us now replace $u_C$ by $u_p$ in the inequality. We then obtain
$\dfrac{ru_p + (s-r)u_p}{s} = \dfrac{su_p}{s} = u_p$. $u_p < u \leq u_C$. Since $u_C > u_p$, $P'$ can't be considered to replace $P$ in entry $(a_1\ a_r)$ unless $u_C < u$.

*Note 1.* Suppose $v_i$ is a point on an acceptable path $Q$ and $w_j$ has as its last arc a proposed new arc to extend $Q$. If $w_j$ completes a subpath of $Q$ that is a cycle, then the cycle must have $j - i$ arcs and must terminate in the same entry of $P_{in}$ containing $v_i$. Furthermore, the length of the cycle must be precisely $w - v$. We thus need to backtrack at must $j - i$ arcs to see if a cycle has been formed by the addition of a new arc.

*Note 2. Finding $T_{OPT}$.* We can use the acceptable paths in **REPLACED & CAN'T REPLACE** to obtain $T_{FWKOPT}$ to form a new $m_2 \ X \ n$ matrix, say $\mathbf{M_2}$ to which we apply FWK. We then obtain a new set of acceptable paths to form a new **REPLACED & CAN'T REPLACE** whose entries form the $m_3$ rows of an $m_3 \ X \ n$ matrix, $\mathbf{M_3}$, to which we again apply the FWK. If we obtain an $n$-cycle whose average arc-value is less than $T_{UPPERBOUND}$, we use this value instead of that obtained from $T_{UPPERBOUND}$. We continue in this manner until there are no longer acceptable paths obtainable from **REPLACED & CAN'T REPLACE**. In that case, the smallest $n$-cycle obtained is $T_{OPT}$.

*Note 3.* Checking to see if a new arc makes $P$ a non-simple path, i.e., the new arc completes a non-acceptable cycle.

Suppose $P = (a_1 \ a_2 \ ... \ a_n \ b) \cup (b \ c)$. Let the entry $(a_1 \ c)$ be $v_i$ while $P$ has the value $w_j$. Then the value of the cycle formed would be $w - v$ while its number of arcs is $j - i$. Therefore, if we backtrack a value $w-v$ or $j-1$ arcs (whichever occurs first) and we don't reach, $c$ we know that $P$ is an acceptable path. If $j < i$, then replacing $v_i$ by $w_j$ always yields a simple path. Furthermore, if we use the $v_i$ notation when applying the modified F-W algorithm to M as well as to M $_i$ $(i = 1, 2, \ ... \ r)$, where the rows of $\mathbf{M}_i$ consists of the acceptable paths in REPLACED and CAN'T REPLACE, each row is independent of every other row when obtaining negative cycles and suffixes.

Example 8

M

|   | 1 | 2 | 3 | 4 | 5 | 6 | 7 |   |
|---|---|---|---|---|---|---|---|---|
| 1 | ∞ | 26 | 4 | 30 | 74 | 5 | 4 | 1 |
| 2 | 38 | ∞ | 28 | 78 | 81 | 7 | 97 | 2 |
| 3 | 10 | 94 | ∞ | 40 | 98 | 49 | 40 | 3 |
| 4 | 70 | 67 | 69 | ∞ | 30 | 41 | 80 | 4 |
| 5 | 30 | 74 | 1 | 60 | ∞ | 9 | 9 | 5 |
| 6 | 31 | 87 | 89 | 91 | 6 | ∞ | 82 | 6 |
| 7 | 23 | 85 | 23 | 7 | 61 | 60 | ∞ | 7 |
|   | 1 | 2 | 3 | 4 | 5 | 6 | 7 |   |

.

MIN(M)

|   | 1 | 2 | 3 | 4 | 5 | 6 | 7 |   |
|---|---|---|---|---|---|---|---|---|
| 1 | 3 | 7 | 6 | 4 | 74 | 1 |   | 1 |
| 2 | 6 | 3 | 1 | 4 | 5 | 7 |   | 2 |
| 3 | 1 | 4 | 7 | 6 | 2 | 5 |   | 3 |
| 4 | 6 | 6 | 2 | 3 | 1 | 7 |   | 4 |
| 5 | 3 | 6 | 7 | 1 | 4 | 6 |   | 5 |
| 6 | 5 | 1 | 7 | 2 | 5 | 4 |   | 6 |
| 7 | 4 | 1 | 3 | 6 | 5 | 2 |   | 7 |
|   | 1 | 2 | 3 | 4 | 5 | 6 | 7 |   |

Let $T = (1\ 2\ 3\ 4\ 5\ 6\ 7): 238$

Our next step is to obtain an upper bound tour for M. We will use PHASE 2 of the modified F-W algorithm. This always yields a solution to the Assignment Problem. We then patch cycles to obtain a tour that is used as an upper bound for both $T_{FWOPT}$ and $T_{OPT}$.

$$T^{-1}M^-$$

|   | 2 | 3 | 4 | 5 | 6 | 7 | 1 |   |
|---|---|---|---|---|---|---|---|---|
|   | 1 | 2 | 3 | 4 | 5 | 6 | 7 |   |
| 1 | 0 | -22 | 4 | 48 | -21 | -22 | ∞ | 1 |
| 2 | ∞ | 0 | 50 | 53 | -21 | 69 | 10 | 2 |
| 3 | 54 | ∞ | 0 | 58 | 9 | 0 | -3 | 3 |
| 4 | 37 | 39 | ∞ | 0 | 11 | 50 | 40 | 4 |
| 5 | 65 | -8 | 51 | ∞ | 0 | 0 | 21 | 5 |
| 6 | 5 | 7 | 9 | -76 | ∞ | 0 | -51 | 6 |
| 7 | 62 | 0 | -16 | 38 | 37 | ∞ | 0 | 7 |
|   | 1 | 2 | 3 | 4 | 5 | 6 | 7 |   |

j = 2

(1 2)(2 5) = (1 5): -65;  (1 2)(2 7) = (1 7): -12;

(5 2)(2 5) = (5 5): -29

CYCLE (5 2): -29

j = 3

(7 3)(3 7) = (7 7): -19

CYCLE (7 3): -19

j = 4

(6 4)(4 1) = (6 1): -39; (6 4)(4 2) = (6 2): -38; (6 4)(4 5) = (6 5): -65;

(6 4)(4 6) = (6 6): -26

CYCLE (6 4): -26

j = 5

(1 5)(5 2) = (1 2): -29; (1 5)(5 6) = (1 6): -21

j = 6

(1 6)(6 1) = (1 1): -17

CYCLE (1 6): -17

(1 6)(6 2) = (1 2): -15; (1 6)(6 3) = (1 3): -13; (1 6)(6 4) = (1 4): -98; (1 6)(6 7) = (1 7): -73

j = 7

(3 7)(7 2) = (3 2): -3;

(3 7)(7 3) = (3 3): -19

CYCLE (3 7): -19

(6 7)(7 2) = (6 2): <u>-51</u>; (6 7)(7 3) = (6 3): <u>-67</u>; (6 7)(7 4) = (6 4): <u>-13</u>;

(6 7)(7 5) = (6 5): <u>-14</u>

Perusing the negative cycles obtained in order to obtain a product that is a permutation, we obtain:

$s_1 = (1\ 6)(2\ 5)(7\ 3) : -92$.

$Ts_1 = (1\ 7\ 4\ 5\ 3)(2\ 6) = D_1$. We next apply $s_1^{-1}$ to $T^{-1}M^-$ in order to obtain $D_1^{-1}M^-$.

$$D_1 = \begin{matrix} 1 & 2 & 3 & 4 & 5 & 6 & 7 \\ 7 & 6 & 1 & 5 & 3 & 2 & 4 \end{matrix}$$

$$D_1^{-1} = \begin{matrix} 1 & 2 & 3 & 4 & 5 & 6 & 7 \\ 3 & 6 & 5 & 7 & 4 & 2 & 1 \end{matrix}$$

$$D_1^{-1}M^-$$

|   | 7 | 6 | 1 | 5 | 3 | 2 | 4 |   |
|---|---|---|---|---|---|---|---|---|
|   | 1 | 2 | 3 | 4 | 5 | 6 | 7 |   |
| 1 | 0 | 1 | ∞ | 70 | 0 | 22 | 26 | 1 |
| 2 | 90 | 0 | 31 | 74 | 21 | ∞ | 71 | 2 |
| 3 | 30 | 39 | 0 | 88 | ∞ | 84 | 30 | 3 |
| 4 | 50 | 11 | 40 | 0 | 39 | 37 | ∞ | 4 |
| 5 | 8 | 8 | 29 | ∞ | 0 | 73 | 59 | 5 |
| 6 | -5 | ∞ | -56 | -81 | 2 | 0 | 4 | 6 |
| 7 | ∞ | 53 | 16 | 54 | 16 | 77 | 0 | 7 |
|   | 1 | 2 | 3 | 4 | 5 | 6 | 7 |   |

We now attempt to obtain a smaller-valued derangement.

j = 1

(6 1)(1 2) = (6 2): -4;  (6 1)(1 5) = (6 5): -5

j = 3

(6 3)(3 1) = (6 1): <u>-26</u>;  (6 3)(3 2) = (6 2): -17;  (6 3)(3 7) = (6 7): -26

j = 4

(6 4)(4 1) = (6 1): <u>-31</u>;  (6 4)(4 2) = (6 2): <u>-71</u>;  (6 4)(4 3) = (6 3): <u>-41</u>;

(6 4)(4 5) = (6 5): -42;

(6 4)(4 6) = (6 6): -44

CYCLE (6 4): -44

j = 7

(6 7)(7 3) = (6 3): <u>-10</u>; (6 7)(7 5) = (6 5): <u>-10</u>

Let $s_2 = (6\ 4)$. Then $D_2 = D_1 s_2 = (1\ 7\ 4\ 2\ 6\ 5\ 3): 102$. We now construct $D_2^{-1} M^-$.

(6 4) → (6 5); (4 6) → (4 2).

$$D_1^{-1} M^-$$

| | 7 | 6 | 1 | 2 | 3 | 5 | 4 | |
|---|---|---|---|---|---|---|---|---|
| | 1 | 2 | 3 | 4 | 5 | 6 | 7 | |
| 1 | 0 | 1 | ∞ | 26 | 0 | 70 | 26 | 1 |
| 2 | 90 | 0 | 31 | ∞ | 21 | 74 | 71 | 2 |
| 3 | 30 | 39 | 0 | 94 | ∞ | 88 | 30 | 3 |
| 4 | 13 | -26 | 3 | 0 | 2 | -37 | ∞ | 4 |
| 5 | 8 | 8 | 29 | 74 | 0 | ∞ | 59 | 5 |
| 6 | 76 | ∞ | 4 | 81 | 92 | 0 | 85 | 6 |
| 7 | ∞ | 53 | 16 | 85 | 16 | 54 | 0 | 7 |
| | 1 | 2 | 3 | 4 | 5 | 6 | 7 | |

j = 2

(4 2)(2 5) = (4 5): -5

j = 6

(4 6)(6 3) = (4 3): <u>-33</u>

We now try to extend the path [4 6 3] into another negative path.

[4 6 3 1]: -3; [4 6 3 7]: -3.

[4 6 3 1 5]: -3

Neither [4 6 3 7] nor [4 6 3 1 5] can be extended. Thus, $D_{ASSIGNMENT} = (1\ 7\ 4\ 2\ 6\ 5\ 3) : 102$.

Since $D_{ASSIGNMENT}$ is a 7-cycle, we rename it $T_{UPPERBOUND}$. Theorem 3.1 allows us to apply FWK directly to the matrix $M$. However, because $T_{UPPERBOUND}$ was obtained by using the modified F-W algorithm, we can go directly to the last step that we use to obtain $T_{OPT}$. Let $u = \frac{|T_{UPPERBOUND}|}{n}$. In theorem 3.1a, we proved that we can obtain any $r$-cycle, $C$, using a sequence of paths each of whose average arc value is less than $\frac{|C|}{r}$. We now define the FWK algorithm more generally. Instead of using modified F-W, we define an acceptable (or – as the case may require) 2-circuit cycle if its average arc value is less than $u$. Let **REPLACED & CAN'T REPLACE** be the set of all paths replaced during our application of the modified F-W algorithm to $T_{UPPERBOUND}^{-1} M^{-}$ or paths which – though acceptable – have a larger average value than the current entry. Also, if two paths have the same value, we choose the one having the larger number of arcs. If both paths have the same value, we keep the current entry and place the contender in **REPLACED & CAN'T REPLACE**  We then use each path in **REPLACED & CAN'T REPLACE** as an initial path of a tree containing all possible acceptable paths. If $|T_{UPPERBOUND}| > |T_{OPT}|$, then $T_{OPT}$ must be a branch of one of the trees obtained. In general, FWK is most useful when the matrix or tree branch to which it is applied contains no negative cycles. If negative paths exist, we use the following rule:

If a given path, $P$, is negative, it can be replaced by *any* negative path of smaller value regardless of the respective number of arcs in the two paths. It is also worth noting that we can apply our algorithm directly to the matrix M. The reason is that if an arc is $(a\ b)$, by using our usual transposition, it becomes

$(a\ T_{UPPERBOUND}(b)\ b)$ where $|(T_{UPPERBOUND}(b)\ b)| = 0$. In this case, $u = \frac{102}{7} = 14.571429$

|   | 1 | 2 | 3 | 4 | 5 | 6 | 7 |   |
|---|---|---|---|---|---|---|---|---|
|   |   |   |   | M |   |   |   |   |
|   | 1 | 2 | 3 | 4 | 5 | 6 | 7 |   |
| 1 | ∞ | 26 | 4 | 30 | 74 | 5 | 4 | 1 |
| 2 | 38 | ∞ | 28 | 78 | 81 | 7 | 97 | 2 |
| 3 | 10 | 94 | ∞ | 40 | 98 | 49 | 40 | 3 |
| 4 | 70 | 67 | 69 | ∞ | 30 | 41 | 80 | 4 |
| 5 | 30 | 74 | 1 | 60 | ∞ | 9 | 9 | 5 |
| 6 | 31 | 87 | 89 | 91 | 6 | ∞ | 82 | 6 |
| 7 | 23 | 85 | 23 | 7 | 61 | 60 | ∞ | 7 |
|   | 1 | 2 | 3 | 4 | 5 | 6 | 7 |   |

j = 1

(3 1)(1 6) = (3 6): $15_2$; (3 1)(1 7) = (3 7): $14_2$

j = 3

(5 3)(3 1) = (5 1): $11_2$

j = 5

(6 5)(5 3) = (6 3): $7_2$; (6 5)(5 7) = (6 7): $15_2$

j = 6

(1 6)(6 5) = (1 5): $11_2$; (2 6)(6 5) = (2 5): $13_2$; (3 6)(6 5) = (3 5): $21_3$

j = 7

(1 7)(7 3) = (1 3): $27_2$; (1 7)(7 4) = (1 4): $11_2$; (3 7)(7 4) = (3 4): $21_3$; (5 7)(7 4) = $16_2$;

(6 7)(7 4) = (6 4): $22_3$

$P_7$

|   | 1 | 2 | 3 | 4 | 5 | 6 | 7 |   |
|---|---|---|---|---|---|---|---|---|
| 1 |   |   | 7 | 7 | 6 | 1 | 1 | 1 |
| 2 |   |   |   |   | 6 | 2 |   | 2 |
| 3 | 3 |   |   | 7 |   | 1 | 1 | 3 |
| 4 |   |   |   |   |   |   |   | 4 |
| 5 | 3 |   | 5 |   |   |   |   | 5 |
| 6 |   |   | 5 | 7 | 6 |   | 5 | 6 |
| 7 |   |   |   |   |   |   |   | 7 |
|   | 1 | 2 | 3 | 4 | 5 | 6 | 7 |   |

$T_1(7)$

|   | 1 | 2 | 3 | 4 | 5 | 6 | 7 |   |
|---|---|---|---|---|---|---|---|---|
| 1 |   |   | $\underline{27_2}$ | $\underline{11_2}$ | $\underline{11_2}$ |   |   | 1 |
| 2 |   |   |   |   | $\underline{13_2}$ |   |   | 2 |
| 3 |   |   |   | $\underline{21_3}$ | $\underline{21_3}$ | $15_2$ | $14_2$ | 3 |
| 4 |   |   |   |   |   |   |   | 4 |
| 5 | $\underline{11_2}$ |   |   | $\underline{16_2}$ |   |   |   | 5 |
| 6 |   |   | $\underline{7_2}$ | $\underline{22_3}$ |   |   | $15_2$ | 6 |
| 7 |   |   |   |   |   |   |   | 7 |
|   | 1 | 2 | 3 | 4 | 5 | 6 | 7 |   |

j = 1

(5 1)(1 2) = (5 2): $\underline{37_3}$; (5 1)(1 6) = (5 6): $\underline{16_3}$

j = 5

(1 5)(5 3) = (1 3): $12_3$;  (1 5)(5 7) = (1 7): $20_3$;  (2 5)(5 3) = (2 3): $14_3$

j = 7

(6 7)(7 1) = (6 1): $38_3$

## REPLACED & CAN'T REPLACE

1. (5 3 1 4): $41_3$  2. (1 7 4 5): $41_3$  3. (3 1 7 4 5): $51_4$  4. (5 3 1 2 6): $44_4$

5. (3 1 6 5 7): $30_4$  6. (1 6 5 7): $20_3$  7. (6 5 7 3): $38_3$  8. (1 7 3): $27_2$

$P_{14}$

|   | 1 | 2 | 3 | 4 | 5 | 6 | 7 |   |
|---|---|---|---|---|---|---|---|---|
| 1 |   |   | 5 | 7 | 6 | 1 | 1 | 1 |
| 2 |   |   | 5 |   | 6 | 2 |   | 2 |
| 3 | 3 |   |   | 7 | 6 | 1 | 1 | 3 |
| 4 |   |   |   |   |   |   |   | 4 |
| 5 | 3 | 1 | 5 | 7 | 6 | 1 | 5 | 5 |
| 6 | 7 |   | 5 | 7 | 6 |   | 5 | 6 |
| 7 |   |   |   |   |   |   |   | 7 |
|   | 1 | 2 | 3 | 4 | 5 | 6 | 7 |   |

$T_1(14)$

|   | 1 | 2 | 3 | 4 | 5 | 6 | 7 |   |
|---|---|---|---|---|---|---|---|---|
| 1 |   |   | $12_3$ | $11_2$ | $11_2$ |   |   | 1 |
| 2 |   |   | $14_3$ |   | $13_2$ |   |   | 2 |
| 3 |   |   |   | $21_3$ | $21_3$ | $15_2$ | $14_2$ | 3 |
| 4 |   |   |   |   |   |   |   | 4 |
| 5 | $11_2$ | $37_3$ |   | $16_2$ |   | $16_3$ |   | 5 |
| 6 | $38_3$ |   | $7_2$ | $22_3$ |   |   | $15_2$ | 6 |
| 7 |   |   |   |   |   |   |   | 7 |

|   | 1 | 2 | 3 | 4 | 5 | 6 | 7 |   |
|---|---|---|---|---|---|---|---|---|

j = 3

(2 3)(3 1) = (2 1): $\underline{24}_4$

Since we have only one path that possibly might be extended to a 7-cycle, we now try to extend it.

(2 6 5 3 1)(1 7) = (2 6 5 3 1 7): $28_5$. This, in turn, can be extended to (2 6 5 3 1 7 4): $35_6$.

Interestingly, the 7-cycle (2 6 5 3 1 7 4) = $T_{UPPERBOUND}$. Since the number of entries in

**REPLACED & CAN'T REPLACE** is small, we construct trees using its paths as initial branches

of trees consisting of acceptable branches. Either one such branch yields a 7-cycle of smaller value than

102 or else $T_{OPT} = T_{UPPERBOUND}$.

**1.** $5 \to 3 \to 1 \to \underline{4}:41_3$  **2.** $1 \to 7 \to 4 \to 5:41_3$  **3.** $3 \to 1 \to 7 \to 4 \to 5$  **4.** $5 \to 3 \to 1 \to 2 \to \underline{6}:44$
$\swarrow \quad \swarrow$ $\swarrow$
$\underline{3}:42_4 \quad \underline{6}:50_4$ $\underline{6}:60_5$

**5.** $3 \to 1 \to 6 \to 5 \to 7:30_4$  **6.** $1 \to 6 \to 5 \to 7:20_3$  **7.** $6 \to 5 \to 7 \to 3:38_3$  **8.** $1 \to 7 \to \underline{3}:27_2$
$\swarrow$ $\swarrow \quad \swarrow$ $\swarrow$
$\underline{4}:37_5$ $\underline{3}:43_4 \quad \underline{4}:27_4$ $\underline{1}:48_4$

Since no branch is a 7-cycle, $T_{OPT} = T_{UPPERBOUND}$.

Chapter 4

I. INTRODUCTION. In this chapter, we discuss the application of the FWK algorithm to symmetric cost matrices. (As shown in chapter 3, it works with asymmetric matrices too. However, our examples will all be symmetric).We first use a negative cycle procedure that yields a derangement. We then use a special patching procedure that yields a tour or an $n$-cycle, depending on whether or not our cost matrix is symmetric or asymmetric. W.l.o.g., call this $n$-cycle $T_{UPPERBOUND}$. We next use both the modified Floyd-Warshall algorithm as well as the average arc-value of paths to obtain paths each of whose average arc-value is less than $/\frac{T_{UPPERBOUND}}{n}/$. This procedure allows us to use parallel processors. We either obtain an $n$-cycle, $T_{UPPERBOUND2}$, such that $|T_{FWKOPT}|=|T_{UPPERBOUND2}|<|T_{UPPERBOUND}|$ or else $|T_{FWKOPT}|=|T_{UPPERBOUND}|$. In either case, $T_{FWKOPT}$ denotes the solution to a heuristic. In order for our algorithm to be exact in all cases, we go through our algorithm a second time, using $T_{FWKOPT}$ as our upper bound. All that is required to obtain an exact solution is that each subpath whose aav is less than $/\frac{T_{FWKOPT}}{n}|$ be obtained. To obtain a heuristic, we delete subpaths that can't be extended as well as constructing new subpaths by copying existing subpaths at branch points and pasting them onto new rows inserted into the matrix We then add any arc that keeps such a new path acceptable. To retain the independence of the parallel processors, we use the first $k$ rows of a matrix to create possible small subpaths that might be initial paths of tours (n-cycles). As we proceed, the great majority of subpaths have greater aavs (average arc-values) than even the aav in the $k$-th row. In that case, it - together with its row - is deleted. Otherwise, an ordered pair of form (name, its aav) is placed in its proper place among the first $k$ rows. The algorithm concludes when we can obtain no new $T_{UPPERBOUND}$. If the total number of paths obtained when we have reached that point is not exponential in $n$, then we can generally obtain $T_{OPT}$ in polynomial time. Also, if $|T_{FWKOPT}| = HK$, the Held-Karp lower bound, then $| T_{FWKOPT} | = | T_{OPT} |$. Next, if we assume that no exact algorithm has a running time that is less than that of FWK by a polynomial-in-$n$ factor and there exists a lower bound $L$ larger than $HK$ such that $|T_{FWKOPT}|=L$, then again we can obtain $T_{OPT}$ in polynomial time. However, given these conditions, if there exists a cost matrix M such that none of these conditions holds, then $P \neq NP$.

We now introduce Theorem 3.1a again.

Theorem 3.1a *Let M be an $n$ X $n$ cost matrix containing an n-cycle $C$. Then there always exists a determining node in $C$, say $d$, such that every subpath emanating from d that traverses $C$ has an average arc-value no greater than $\frac{|C|}{n}$.*

**Theorem 3.2** *Let M be a symmetric $n \times n$ cost matrix. Let $T$ be a tour. Suppose that there exists a permutation of the points of $T$ that yields a smaller-valued tour $T'$ such that the points from $a$ through $b$ of $T$ contain all points that have been permuted while the subpath $b$ through $a$ of $T$ remains unchanged. Then we can always use the subpath $(a \ldots b)$ of $T$ given in $T^{-1}M^-$ or those of the subpath $(b \ldots a)$ of $T^{-1}$ given in $TM^-$ to obtain a negative path consisting of arcs of $T$ to obtain $T'$.*

**Theorem 3.3** *Let (i k), (k j), (i j) be entries in a symmetric $m \times n$ cost matrix. Denote $aav_{ik}$, $aav_{kj}$, $aav_{ij}$ as their respective average arc-values. Then if $aav_{ik} + aav_{kj} < aav_{ij}$ where the interior point sets of the subpaths (I k) and (kj) are disjoint, the subpath (i k) $\cup$ (k j) has a smaller aav then that of the subpath (i j).*

Proof.

$$aav_{ik} + aav_{kj} = \frac{value(i\ k)}{no.\ of\ arcs(i\ k)} + \frac{value(k\ j)}{no.\ of\ arcs(k\ j)} = \frac{value(i\ k) + value(k\ j)}{no.\ of\ arcs(i\ k) + no.\ of\ arcs(k\ j)} < \frac{value(i\ j)}{no.\ of\ arcs(i\ j)}$$

(a) PROLOGUE. M is either a symmetric or an asymmetric $n \times n$ cost matrix. As we construct M, we also construct the corresponding matrix MIN(M) in which the column numbers of M of each row are sorted in ascending order of their respective values. $D_0 = (1\ 2\ 3\ \ldots\ n)$ is a permutation on $n$ points. We write the permutation as a set of two rows:

$$\begin{matrix} 1 & 2 & 3 & \ldots & n \\ D_0(1) & D_0(2) & D_0(3) & \ldots & D_0(n) \end{matrix}$$. If $P$ is a path, $|P|$ is the value of $P$.

(b) OBTAINING $T_{UPPERBOUND}$. We apply $D_0^{-1}$ to the columns of M obtain the matrix $D_0^{-1}M$. Given each row $i$, we subtract the value of the entry $(i, D_0^{-1}(i))$ from each of the entries in row $i$ to obtain the matrix $D_0^{-1}M^-$. Each time we subtract a diagonal entry from each entry in a row, if we obtain a negative value, we first place the entry in an ordered list, NEGATIVE, containing the value of the negative entry as well as its location. The list is ordered in increasing value of negative numbers. On the other hand, if we subtract a negative diagonal entry from the other entries in a row, we sometimes change a negative entry into a positive one. In that case, we delete the original negative entry from NEGATIVE. We use entries from NEGATIVE to start creating negative cycles. Such a cycle is constructed by choosing smallest-valued entries in each row until we obtain at least one negative cycle. In what follows, we use the two-row representation of a permutation together with MIN(M) to change $D_k$ into $D_{k+1}$. Our aim is to obtain negative cycles in each row of $D_k^{-1}M^-$ ($k = 1, 2, \ldots, r$). We first dedicate a fixed processor $FP$ to change the rows and columns of each new derangement arc that is derived from a negative cycle obtained by a processor. It then subtracts the diagonal entries of each row and column from all other entries in the row. This yields $D_{k+1}$. This is useful because it means that the other processors need only change a row that it is working on if that row's diagonal entry is a new derangement arc of $D_{k+1}$. If $(i\ j)$ is an entry in MIN(M), then $(i\ D_k(j))$ tells us where to find the $j-th$ smallest-valued entry in $D_k^{-1}M^-$. Suppose we have $p$ parallel processors at our disposal. If $p \geq n$, then we assign a parallel processor to each row of $D_k^{-1}M^-$. If $p < n$, one parallel processor is assigned to $\left[\frac{n}{p}\right] + 1$ rows. In each row $i$, we choose the $[\sqrt{n}]$ smallest-valued negative entries in the row. Using a row of form $[1\ 2\ 3\ ...\ n]$ with a blank row underneath it, we try to construct a negative cycle of smallest possible value. As we construct our negative path, if M is symmetric, we cannot choose an arc that is symmetric to an arc of $D_k$. If the initial node of our path is $q$, after a new arc $(a\ b)$ extends a path, we check to see whether $(b\ q)$ yields a negatively-valued cycle. All negative cycles obtained are placed in a separate list. When we can no longer obtain a negative cycle, we choose the one, say $C$, with the smallest value. Each arc of $C$ yields an edge that replaces an edge of $D_k$. When all of the new arcs have replaced corresponding arcs in $D_k$, we have obtained $D_{k+1}$. We obtain the matrix $D_{k+1}^{-1}M^-$ by permuting columns using the new arcs obtained from $C$. Thus, if $(a\ b)$ was an arc of $D_k$ that has been replaced by $(a\ b')$, we delete column $b$ of $D_k^{-1}M^-$ and replace it by column $b'$ of the same matrix. We then subtract the value of entry $(a\ b')$ from every entry in row $a$. We continue this process until we no longer can obtain a negative cycle. When this occurs, we have obtained a symmetric minimally-valued derangement on $n$ points, i.e., a permutation that moves every one of its $n$ points, i.e., a derangement containing no 2-cycles.. We then use the patching method employed in chapter 1 to obtain an $n$-cycle, $T_{UPPERBOUND}$.

(c) FWK. Before applying FWK, we first transform M into M', the matrix obtained by subtracting the smallest-valued entry in each row from the remaining entries in the row. We then obtain the value that $T_{UPPERBOUND}$ has in M' and use its average arc-value (aav) for the remainder of the algorithm. For an exact algorithm, we need only to obtain all paths each of whose aav is less than that of $T_{UPPERBOUND}$. We now note a special feature of our selection process. We first obtain acceptable paths each of whose aav has a value less than $/\frac{T_{UPPERBOUND}}{n}/$. If this yields an $n$-cycle, $T_0$, we substitute $/\frac{T_0}{n}/$ for $/\frac{T_{UPPERBOUND}}{n}/$. Otherwise, we try to extend each path obtained earlier by allowing precisely *one* arc whose value is greater than $/\frac{T_0}{n}/$ to be included in a new acceptable path. If we obtain an acceptable $n$-cycle, $T_1$, whose value is less than $/\frac{T_0}{n}/$, we substitute $/\frac{T_1}{n}/$ for $/\frac{T_0}{n}/$ as a new upper bound. Eventually, we either obtain $T_{OPT}$ or else the last $T_i$ we obtained is $T_{OPT}$. Using the Floyd-Warshall method, we proceed systematically from column to column. If we are using a parallel processor in a row, we place an exclamation point after the terminal node to indicate where we should continue our extension of a subpath. It is worth noting that if we use only parallel processors, then each subpath can be extended independently. Thus, some subpaths may many more arcs than others. This means that we might be able to obtain smaller-valued n-cycles. This would require all of the subpaths extended henceforth to have smaller aavs. If there is at most a polynomial-in-$n$ number of acceptable paths, then we can obtain $T_{OPT}$ in polynomial-in-$n$ running time. In one other case, we can also obtain $T_{OPT}$ polynomial time. If $|T_{FWKOPT}| = \text{HK}$ where HK is the Held-Karp lower bound, $|T_{FWKOPT}| = |T_{OPT}|$. Assume now that our procedure obtains an exponential number of 0acceptable paths. Given the assumption that no other exact algorithm has a running time that is more than a polynomial-in-$n$ different factor in running time than our algorithm, we cannot obtain an optimal tour for every $n \, X \, n$ cost matrix in polynomial time. Therefore, $P \neq NP$.

(d). Exact Algorithm. If $C$ is a cycle, define $|C|$ as its value. We now discuss our method of transforming the procedure of obtaining $T_{FWKOPT}$ into one which can obtain $T_{OPT}$. As mentioned earlier, if we are willing to spend the time and use the storage, we can always obtain an optimal tour by obtaining all subpaths that have an aav smaller than $\frac{|T_{UPPERBOUND}|}{n}$. If each subpath lies in a unique parallel processor, we don't require an exclamation point at the end of each iteration. In general, we won't even need to keep track of iterations at all. While applying the FWK algorithm, we generally obtain a large number of acceptable paths that are periodically replaced by smaller-valued paths that have the same initial and terminal nodes. If an acceptable smaller-valued path exists that eventually yields a smaller-valued $n$-cycle, it may have as an initial subpath one of the acceptable paths that we didn't use to obtain the smallest-valued path obtainable using the FWK algorithm. We thus arrange these paths in order of the number of arcs in each set. As we obtain these paths, we sort them in terms of increasing average arc-value. We do these operations while applying the FWK algorithm. We call the sets obtained REPLACED AND CAN'T REPLACE. Let $T_{FWKOPT}$ be the smallest-valued $n$-cycle obtained using FWK. If $|T_{UPPERBOUND}| > |T_{FWKOPT}|$, we delete all paths in REPLACED AND CAN'T REPLACE whose average arc-value is greater than or equal to $|T_{FWKUPPERBOUND}|$. We then replace $T_{UPPERBOUND}$ by $T_{FWKOPT}$. Otherwise, $|T_{FWKOPT}| = |T_{UPPERBOUND}|$. Suppose we have $r$ parallel processors. Let $a_i, i = 1,2,...,r$ be the full set of numbers of arcs in the paths in REPLACED AND CAN'T REPLACE. Pick the smallest of the arc-valued paths in each of the sets in REPLACED AND CAN'T REPLACE. Assign a parallel processor to each such set and in each path print in boldface each node of the path. We then try to extend each path. We go through all iterations. As we obtain new acceptable paths while doing operations, we place them according to their average arc values in the sets of REPLACED AND CAN'T REPLACE. For such paths. there is only one REPLACED AND CAN'T REPLACE in which they can be placed. This applies to all iterations and all paths obtained until we either obtain an $n$-cycle with smaller arc-value than the current $T_{UPPERBOUND}$ or else $T_{OPT} = T_{UPPERBOUND}$. If there is at most a polynomial-in-$n$ number of acceptable paths, then we can obtain $T_{OPT}$ in polynomial-in-$n$ running time. In one other case, we can also obtain $T_{OPT}$ polynomial time. If $|T_{FWKOPT}| = \text{HK}$ where $\text{HK}$ is the Held-Karp lower bound, $|T_{FWKOPT}| = |T_{OPT}|$. Assume now that our procedure obtains exponential-in-n acceptable paths. Given the assumption that no other exact algorithm has a running time that is more than a fixed polynomial-in-$n$ different factor in running time than our algorithm, we cannot obtain an optimal tour for every $n \times n$ cost matrix in polynomial time. Therefore, $P \neq NP$.

*Comment*. If we have an unlimited number of parallel processors, we can assign a parallel processor to each subpath whose average aav is less than $|\frac{T_{UPPERBOUND}}{n}|$. Each subpath which can't be extended has its row deleted. At each branch point of a subpath, we check to see which arcs extending the subpath are acceptable. We then insert blank rows into M' in which we paste the main body of new subpaths derived from existing ones at their branchpoints. We then add an arc obtained from those emanating from the branchpoint to complete each subpath. We will always be able to obtain $T_{OPT}$ using this procedure. How often its running time will be polynomial (if at all) is another question.

(e) Heuristic Algorithm (1).

As we have mentioned before, we use a matrix each row of which represents a subpath that may eventually lead to a tour (n-cycle). We first obtain $T_{UPPERBOUND}$. For sake of discussion, assume that we wish to obtain $k$ tours (n-cycles). We then leave the first $k$ rows of our matrix blank. Column $i$ of the first $k$ rows is defined to be the column in which subpaths containing $i$ arcs are placed. As we construct subpaths in rows $k+1, k+2, ...$, we place in ascending value of aavs, the at most $k$ smallest-valued subpaths in the correct column of rows $1, 2, 3, ..., k$. If a column $i$ in the first $k$ rows has fewer than $k$ subpaths in it, we insert a subpath in it in its correct place with respect to its aav. We continue this procedure until we have at most $k$ subpaths in each column of rows $1, 2, 3, ..., k$. Any subpath whose aav is at least as great as the aav of the correct column of the subpath in row $k$ has its row deleted. A subpath whose aav is less than that of correct column of the $k-th$ row is represented by an ordered pair of form *( name of subpath, its aav )* placed in its proper place in the first $k$ rows. The set of these k ordered paths is denoted by BEST. The subpaths in BEST are kept unchanged until the end of each iteration. At that point, starting the right-most column of BEST containing a subpath, we we place the extension (if one0 occurs) of each subpath in the proper column to which its extension belongs. We do the same procedure going from right to left. Since each column is run on a parallel processor, we begin sorting as soon as new additions or deletions are made in any column. We continue this procedure always going to the left of a column we have finished working on. Concerning the construction of subpaths, whenever a subpath reaches a branch-point, before adding new rows, check to see if the addition of any of the possible new arcs would not yield an acceptable new subpath. Such arcs should be deleted from consideration before new rows are added. If the column $n$ of BEST contains a subpath, then it is a tour (n-cycle). The first subpath in column $n$ of BEST is a tour (n-cycle). Thus, we either obtain a tour (n-cycle) with smaller aav than that of $T_{UPPERBOUND}$ or else $T_{OPT-APPROXZ} = T_{UPPERBOUND}$.

Heuristic Algorithm (2)

We obtain $T_{UPPERBOUND}$ as in the exact algorithm. In the process of obtaining $T_{FWKOPT}$, we save and extend all subpaths thar were replaced, but delete all acceptable subpaths that can't replace the current best subpath.

Heuristic Algorithm (3)

We find the [log(n)] + 1 rows in M that have the largest number of entries each of whose values is less than the aav of $T_{UPPERBOUND}$. From these rows, we obtain all acceptable subpaths in M'.

Heuristic Algorithm (4)

We obtain the determining point of $T_{UPPERBOUND}$. Writing it with its determining point as its first point, we find the first arc that satisfies the following conditions:

(1) Its aav is greater than that of $T_{UPPERBOUND}$.

(2) The majority of the arcs following it have aavs that are greater than the aav of $T_{UPPERBOUND}$.

Call this point $p$. We mark the ordinal number of its initial point. We now construct all subpaths whose aav is less than that of $T_{UPPERBOUND}$. At the end of each iteration, we choose the [log(n)] + 1 subpaths with the smallest aavs. We do this until one or more subpaths have $p$ arcs. From that point on, we always count subpaths having $p$ or more arcs as part of the [log(n)] + 1 subpaths chosen. The portion of [log(n) + 1 remaining is used to choose those subpaths having the smallest aavs.

.

Example 9.

M

| | 1 | 2 | 3 | 4 | 5 | 6 | 7 | 8 | 9 | 10 | 11 | 12 | 13 | 14 | 15 | |
|---|---|---|---|---|---|---|---|---|---|---|---|---|---|---|---|---|
| 1 | ∞ | 53 | 52 | 51 | 50 | 49 | 48 | 48 | 48 | 48 | 50 | 51 | 52 | 53 | 54 | 1 |
| 2 | 53 | ∞ | 66 | 66 | 50 | 49 | 48 | 47 | 48 | 49 | 50 | 51 | 52 | 53 | 54 | 2 |
| 3 | 52 | 66 | ∞ | 62 | 65 | 64 | 48 | 48 | 48 | 49 | 50 | 75 | 69 | 77 | 76 | 3 |
| 4 | 51 | 66 | 62 | ∞ | 63 | 62 | 65 | 49 | 49 | 49 | 50 | 77 | 73 | 76 | 71 | 4 |
| 5 | 50 | 50 | 65 | 63 | ∞ | 66 | 65 | 50 | 73 | 76 | 74 | 71 | 77 | 75 | 76 | 5 |
| 6 | 49 | 49 | 64 | 62 | 66 | ∞ | 65 | 51 | 70 | 77 | 73 | 37 | 23 | 30 | 69 | 6 |
| 7 | 48 | 48 | 48 | 65 | 65 | 65 | ∞ | 73 | 74 | 71 | 75 | 26 | 39 | 27 | 72 | 7 |
| 8 | 48 | 47 | 48 | 49 | 50 | 51 | 73 | ∞ | 27 | 31 | 73 | 31 | 30 | 35 | 73 | 8 |
| 9 | 48 | 48 | 48 | 49 | 73 | 70 | 74 | 27 | ∞ | 30 | 38 | 29 | 26 | 36 | 77 | 9 |
| 10 | 48 | 49 | 49 | 49 | 76 | 77 | 71 | 31 | 30 | ∞ | 36 | 35 | 73 | 71 | 74 | 10 |
| 11 | 50 | 50 | 50 | 50 | 74 | 73 | 76 | 73 | 38 | 36 | ∞ | 30 | 26 | 37 | 35 | 11 |
| 12 | 51 | 51 | 75 | 77 | 71 | 37 | 26 | 31 | 29 | 35 | 30 | ∞ | 29 | 36 | 30 | 12 |
| 13 | 52 | 52 | 69 | 73 | 77 | 23 | 39 | 30 | 26 | 73 | 26 | 29 | ∞ | 30 | 36 | 13 |
| 14 | 53 | 63 | 77 | 76 | 75 | 30 | 27 | 35 | 36 | 71 | 37 | 36 | 30 | ∞ | 32 | 14 |
| 15 | 54 | 54 | 36 | 71 | 76 | 69 | 72 | 73 | 77 | 74 | 35 | 30 | 36 | 32 | ∞ | 15 |
| | 1 | 2 | 3 | 4 | 5 | 6 | 7 | 8 | 9 | 10 | 11 | 12 | 13 | 14 | 15 | |

MIN(M)

|    | 1   | 2    | 3    | 4    | 5  | 6  | 7  | 8  | 9  | 10 | 11 | 12 | 13 | 14 |    |    |
|----|-----|------|------|------|----|----|----|----|----|----|----|----|----|----|----|----|
| 1  | 7*  | 8*   | 9*   | 10*  | 6  | 5  | 11 | 4  | 12 | 3  | 13 | 2  | 14 | 15 |    | 1  |
| 2  | 8   | 7*   | 9*   | 6    | 10 | 5  | 11 | 12 | 13 | 1  | 14 | 15 | 3  | 4  |    | 2  |
| 3  | 7*  | 8*   | 9*   | 10   | 11 | 1  | 2  | 5  | 6  | 4  | 12 | 15 | 14 | 13 |    | 3  |
| 4  | 8*  | 9*   | 10*  | 11   | 1  | 6  | 7  | 2  | 5  | 13 | 3  | 14 | 15 | 12 |    | 4  |
| 5  | 1*  | 2*   | 8*   | 3    | 6  | 7  | 4  | 11 | 12 | 9  | 15 | 14 | 13 | 10 |    | 5  |
| 6  | 13  | 14   | 12   | 1    | 2  | 8  | 4  | 7  | 5  | 3  | 11 | 9  | 10 | 15 |    | 6  |
| 7  | 12  | 14   | 13   | 1    | 2  | 3  | 4  | 6  | 5  | 8  | 15 | 11 | 10 | 9  |    | 7  |
| 8  | 9   | 13   | 10   | 12   | 14 | 2  | 1  | 3  | 4  | 5  | 11 | 6  | 7  | 15 |    | 8  |
| 9  | 13  | 8    | 10   | 12   | 14 | 11 | 1  | 2  | 3  | 4  | 5  | 15 | 6  | 7  |    | 9  |
| 10 | 9   | 8    | 12   | 11   | 1  | 2  | 3  | 4  | 15 | 6  | 7  | 14 | 13 | 5  |    | 10 |
| 11 | 13  | 120  | 15   | 10   | 14 | 9  | 1  | 2  | 3  | 4  | 8  | 5  | 6  | 7  |    | 11 |
| 12 | 7   | 9*   | 13*  | 11   | 15 | 8  | 10 | 14 | 6  | 1  | 2  | 3  | 5  | 4  |    | 12 |
| 13 | 6   | 9*   | 11*  | 12   | 8  | 14 | 15 | 7  | 1  | 2  | 4  | 5  | 10 | 3  |    | 13 |
| 14 | 7   | 6*   | 13*  | 15   | 8  | 9  | 12 | 11 | 1  | 2  | 4  | 3  | 5  | 10 |    | 14 |
| 15 | 12  | 14   | 11   | 13   | 1  | 2  | 6  | 7  | 10 | 5  | 3  | 8  | 9  | 4  |    | 15 |
|    | 1   | 2    | 3    | 4    | 5  | 6  | 7  | 8  | 9  | 10 | 11 | 12 | 13 | 14 | 15 |    |

$D_0^{-1}$

|    | 2  | 3  | 4  | 5  | 6  | 7  | 8  | 9  | 10 | 11 | 12 | 13 | 14 | 15 | 1  |    |
|----|----|----|----|----|----|----|----|----|----|----|----|----|----|----|----|----|
|    | 1  | 2  | 3  | 4  | 5  | 6  | 7  | 8  | 9  | 10 | 11 | 12 | 13 | 14 | 15 |    |
| 1  | 0  | -1 | -2 | -3 | -4 | -5 | -5 | -5 | -5 | -3 | -2 | -1 | 0  | 1  | ∞  | 1  |
| 2  | ∞  | 0  | 0  | -16| -17| -18| -19| -18| -17| -16| -15| -14| -13| -12| -13| 2  |
| 3  | 4  | ∞  | 0  | 3  | 2  | -14| -14| -14| -13| -12| 13 | 7  | 15 | 14 | -10| 3  |
| 4  | 3  | -1 | ∞  | 0  | -1 | 2  | -14| -14| -14| -13| 14 | 10 | 13 | 8  | -12| 4  |
| 5  | -16| -1 | -3 | ∞  | 0  | -1 | -16| 7  | 10 | 8  | 5  | 11 | 9  | 10 | -16| 5  |
| 6  | -16| -1 | -3 | 1  | ∞  | 0  | -14| 5  | 12 | 8  | -28| -42| -35| 4  | -16| 6  |
| 7  | -25| 21 | -8 | -8 | -8 | ∞  | 0  | 1  | -2 | 2  | -47| -34| -46| -1 | -25| 7  |
| 8  | 20 | 21 | 22 | 23 | 24 | 46 | ∞  | 0  | 4  | 46 | 4  | 3  | 8  | 46 | 18 | 8  |
| 9  | 18 | 18 | 19 | 43 | 40 | 44 | -3 | ∞  | 0  | 8  | -1 | -4 | 6  | 47 | 18 | 9  |
| 10 | 13 | 13 | 13 | 40 | 42 | 35 | -5 | -6 | ∞  | 0  | -1 | 37 | 35 | 38 | 12 | 10 |
| 11 | 20 | 20 | 20 | 44 | 43 | 46 | 43 | 8  | 6  | ∞  | 0  | -4 | 7  | 5  | 20 | 11 |
| 12 | 22 | 46 | 48 | 42 | 8  | -3 | 2  | 0  | 6  | 1  | ∞  | 0  | 7  | 1  | 22 | 12 |
| 13 | 22 | 39 | 43 | 47 | -7 | 9  | 0  | -4 | 43 | -4 | -1 | ∞  | 0  | 6  | 22 | 13 |
| 14 | 31 | 45 | 44 | 43 | -2 | -5 | 3  | 4  | 39 | 5  | 4  | -2 | ∞  | 0  | 21 | 14 |
| 15 | 0  | 22 | 17 | 22 | 15 | 18 | 19 | 23 | 20 | -19| -24| -18| -22| ∞  | 54 | 15 |
|    | 1  | 2  | 3  | 4  | 5  | 6  | 7  | 8  | 9  | 10 | 11 | 12 | 13 | 14 | 15 |    |

We first obtain the [log(n)] + 1 smallest-valued negative entries in our matrix.

NEGATIVE CYCLES

|         | 2 | 3 | 4 | 5 | 6  | 7  | 8    | 9   | 10  | 11 | 12   | 13  | 14  | 15  | 1   |
|---------|---|---|---|---|----|----|------|-----|-----|----|------|-----|-----|-----|-----|
|         | 1 | 2 | 3 | 4 | 5  | 6  | 7    | 8   | 9   | 10 | 11   | 12  | 13  | 14  | 15  |
| (7 11) -47 |   |   |   |   |    |    | <u>10</u> <u>-44</u> | 12 -51 | 8 -47 | 9 -39 | 11 -51 |     |     |     |     |
| (7 13) -46 | 5 -69 |   |   |   | 13 -53 | 1 -74 | <u>10</u> <u>-101</u> |   |   | 12 -96 | 6 -102 | 14 -99 | 7 -46 | 11 -97 |     |
| (6 12) -45 | 5 -59 |   |   |   | 14 -43 | <u>12</u> <u>-71</u> | 10 -62 | 1 -64 |   |   | 13 -57 | 8 -60 | 6 -42 | 11 -53 | 12 -41 | 7 -87 |
| (6 13) -35 |   |   |   |   | 13 -42 | <u>12</u> <u>-85</u> |      |     |     | 15 -82 | 11 -86 | 6 -35 |     | 5 -58 |     |
| (7 12) -34 |   |   |   |   |    | 12 -37 | <u>11</u> <u>-88</u> |     |     |    | 6 -84 | 7 -34 |     |     |     |
|         | 1 | 2 | 3 | 4 | 5  | 6  | 7    | 8   | 9   | 10 | 11   | 12  | 13  | 14  | 15  |

Our smallest-valued negative cycle: (7 13): (7 13 5 1 6 11 14 12 10): -101

We obtain the following new arcs: $(7\ 13) \to (7\ 14); (13\ 5) \to (13\ 6); (5\ 1) \to (5\ 2);$
$(1\ 6) \to (1\ 7); (6\ 11) \to (6\ 12); (11\ 14) \to (11\ 15);$
$(14\ 12) \to (14\ 13); (12\ 10) \to (12\ 11); (10\ 7) \to (10\ 8).$

$$D_1^{-1}$$

| | 7 | 3 | 4 | 5 | 2 | 12 | 14 | 9 | 10 | 8 | 15 | 11 | 6 | 13 | 1 | |
|---|---|---|---|---|---|---|---|---|---|---|---|---|---|---|---|---|
| | 1 | 2 | 3 | 4 | 5 | 6 | 7 | 8 | 9 | 10 | 11 | 12 | 13 | 14 | 15 | |
| 1 | 0 | 4 | 3 | 2 | 5 | 3 | 5 | 0 | 0 | 0 | 6 | 2 | 1 | 4 | ∞ | 1 |
| 2 | -18 | 0 | 0 | -16 | ∞ | -15 | -13 | -18 | -17 | -19 | -12 | -16 | -17 | -14 | -13 | 2 |
| 3 | -14 | ∞ | 0 | 3 | 4 | 13 | 15 | -14 | -13 | -14 | 14 | -12 | 2 | 7 | -10 | 3 |
| 4 | 2 | -1 | ∞ | 0 | 3 | 14 | 13 | -14 | -14 | -14 | 8 | -13 | -1 | 10 | -12 | 4 |
| 5 | 15 | 15 | 13 | ∞ | 0 | 21 | 25 | 23 | 26 | 0 | 26 | 24 | 16 | 27 | -16 | 5 |
| 6 | 0 | -1 | 24 | 29 | 12 | 0 | -7 | 33 | 40 | 14 | 32 | 36 | ∞ | 14 | -16 | 6 |
| 7 | ∞ | 68 | 38 | 38 | 21 | -1 | 0 | 47 | 44 | 46 | 45 | 48 | 38 | 12 | -25 | 7 |
| 8 | 46 | 21 | 22 | 23 | 20 | 4 | 8 | 0 | 4 | ∞ | 46 | 46 | 24 | 3 | 18 | 8 |
| 9 | 44 | 18 | 19 | 43 | 18 | -1 | 6 | ∞ | 0 | -3 | 47 | 8 | 40 | -4 | 18 | 9 |
| 10 | 40 | 18 | 18 | 45 | 18 | 4 | 40 | -1 | ∞ | 0 | 43 | 5 | 47 | 42 | 12 | 10 |
| 11 | 41 | 15 | 15 | 39 | 15 | -5 | 2 | 3 | 1 | 38 | 0 | ∞ | 38 | -9 | 20 | 11 |
| 12 | -4 | 45 | 47 | 41 | 21 | ∞ | 6 | -1 | 5 | 1 | 0 | 0 | 7 | -1 | 22 | 12 |
| 13 | 16 | 46 | 50 | 54 | 29 | 6 | 7 | 3 | 50 | 7 | 13 | 3 | 0 | ∞ | 22 | 13 |
| 14 | -3 | 47 | 46 | 45 | 33 | 6 | ∞ | 6 | 41 | 5 | 2 | 7 | 0 | 0 | 21 | 14 |
| 15 | 37 | 41 | 36 | 41 | 19 | -5 | -3 | 42 | 39 | 38 | ∞ | 0 | 34 | 1 | 54 | 15 |
| | 1 | 2 | 3 | 4 | 5 | 6 | 7 | 8 | 9 | 10 | 11 | 12 | 13 | 14 | 15 | |

NEGATIVE CYCLES

| | 7 | 3 | 4 | 5 | 2 | 12 | 14 | 9 | 10 | 8 | 15 | 11 | 6 | 13 | 1 |
|---|---|---|---|---|---|---|---|---|---|---|---|---|---|---|---|
| | 1 | 2 | 3 | 4 | 5 | 6 | 7 | 8 | 9 | 10 | 11 | 12 | 13 | 14 | 15 |
| (2 9) -19 | | 6 -16 | | | | 10 -15 | | | 2 -19 | | | | | | |
| (2 1) -18 | 2 -18 | 6 -20 | | | | 9 -19 | | | 1 -18 | | | | | | |
| (2 8) -18 | 12 -16 | 6 -18 | | | | 9 -17 | | 2 -18 | 1 -16 | | | 13 -12 | 14 -15 | 8 -15 | |
| (2 9) -17 | | 6 -24 * | | | | 15 -23 | | | 2 -17 | | | | 14 -21 | 9 -21 | 13 -18 |
| (2 13) -17 | | 6 -20 | | | | 15 -19 | | | | | | | 2 -17 | | 13 -14 |
| | 1 | 2 | 3 | 4 | 5 | 6 | 7 | 8 | 9 | 10 | 11 | 12 | 13 | 14 | 15 |

Our smallest-valued negative cycle is (2  9  14  13  15  6): -24. We obtain the following new arcs:

$( 2\ 9 ) \rightarrow ( 2\ 10 ); ( 9\ 14 ) \rightarrow ( 9\ 13 ); ( 14\ 13 ) \rightarrow ( 14\ 6 );$
$( 13\ 15 ) \rightarrow ( 13\ 1 ); ( 15\ 6 ) \rightarrow ( 15\ 12 ); ( 6\ 2 ) \rightarrow ( 6\ 3 ).$

Our new derangement is (1  7  14  6  3  4  5  2  10  8  9  13)(11  15  12).

$$D_2^{-1}$$

| | 7 | 10 | 4 | 5 | 2 | 3 | 14 | 9 | 13 | 8 | 15 | 11 | 1 | 6 | 12 | |
|---|---|---|---|---|---|---|---|---|---|---|---|---|---|---|---|---|
| | 1 | 2 | 3 | 4 | 5 | 6 | 7 | 8 | 9 | 10 | 11 | 12 | 13 | 14 | 15 | |
| 1 | 0 | 0 | 3 | 2 | 5 | 4 | 5 | 0 | 4 | 0 | 6 | 2 | ∞ | 1 | 3 | 1 |
| 2 | -1 | 0 | 17 | 1 | ∞ | 17 | 4 | -1 | 3 | -2 | 5 | 1 | 4 | 0 | 2 | 2 |
| 3 | -14 | -13 | 0 | 3 | 4 | ∞ | 15 | -14 | 7 | -14 | 14 | -12 | -9 | 2 | 13 | 3 |
| 4 | 2 | -14 | ∞ | 0 | 3 | -1 | 13 | -14 | 10 | -14 | 8 | -13 | -12 | -1 | 14 | 4 |
| 5 | 15 | 26 | 13 | ∞ | 0 | 15 | 25 | 23 | 27 | 0 | 26 | 24 | 0 | 16 | 21 | 5 |
| 6 | 1 | 41 | 25 | 30 | 13 | 0 | -6 | 34 | 15 | 15 | 33 | 37 | -15 | ∞ | 1 | 6 |
| 7 | ∞ | 44 | 38 | 38 | 21 | 68 | 0 | 47 | 12 | 46 | 45 | 48 | 21 | 38 | -1 | 7 |
| 8 | 46 | 4 | 22 | 23 | 20 | 21 | 8 | 0 | 3 | ∞ | 46 | 46 | 21 | 24 | 4 | 8 |
| 9 | 48 | 4 | 23 | 47 | 22 | 22 | 10 | ∞ | 0 | 1 | 51 | 12 | 22 | 44 | 3 | 9 |
| 10 | 40 | ∞ | 18 | 45 | 18 | 18 | 40 | -1 | 42 | 0 | 43 | 5 | 17 | 47 | 4 | 10 |
| 11 | 41 | 1 | 15 | 39 | 15 | 15 | 2 | 3 | -9 | 38 | 0 | ∞ | 15 | 38 | -5 | 11 |
| 12 | -4 | 5 | 47 | 41 | 21 | 45 | 6 | -1 | -1 | 1 | 0 | 0 | 21 | 7 | ∞ | 12 |
| 13 | 13 | 47 | 47 | 51 | 26 | 43 | 4 | 0 | ∞ | 4 | 10 | 0 | 0 | -3 | 3 | 13 |
| 14 | -3 | 41 | 46 | 45 | 33 | 47 | ∞ | 6 | 0 | 5 | 2 | 7 | 23 | 0 | 6 | 14 |
| 15 | 42 | 44 | 41 | 46 | 24 | 46 | 2 | 47 | 6 | 43 | ∞ | 5 | 24 | 39 | 0 | 15 |
| | 1 | 2 | 3 | 4 | 5 | 6 | 7 | 8 | 9 | 10 | 11 | 12 | 13 | 14 | 15 | |

NEGATIVE CYCLES

| | 7 | 10 | 4 | 5 | 2 | 3 | 14 | 9 | 13 | 8 | 15 | 11 | 1 | 6 | 12 |
|---|---|---|---|---|---|---|---|---|---|---|---|---|---|---|---|
| | 1 | 2 | 3 | 4 | 5 | 6 | 7 | 8 | 9 | 10 | 11 | 12 | 13 | 14 | 15 |
| (3 1) -14 | 3 -14 | 1 -14 | <u>1</u> <u>-11</u> | | | | 15 -9! | 10 -17 | 8 -14 | 2 -16 | | | | | 9 -11 |
| (3 8) -14 | | | | | | | 15 -4! | 3 -14 | 8 -11 | 9 -10 | | | | | 10 -6 |
| (3 10) -14 | | | | | | | 15 -7! | 10 -15 | 8 -12 | 3 -14 | | | | | 9 -9 |
| (4 2) -14 | | 4 -14 | | | | | 15 -9! | 10 -17 | 8 -14 | 2 -16 | | | | | 9 -11 |
| (4 8) -14 | | | | | | | 15 -7! | 4 -14 | 8 -11 | 9 -13 | | | | | 10 -8 |
| (4 10) -14 | | | | | | | 15 -7! | 10 -15 | 8 -12 | 4 -14 | | | | | 9 -7 |
| | 1 | 2 | 3 | 4 | 5 | 6 | 7 | 8 | 9 | 10 | 11 | 12 | 13 | 14 | 15 |

We are unable to obtain any cycles by choosing the smallest-valued entry in each successive row. However, we did obtain a cycle - (1 3) - by checking to see if the arc from the endpoint of a path to the its initial point yielded a negatively-valued valued cycle. We now obtain new arcs from (1 3).

$(1\ 3) \rightarrow (1\ 4); (3\ 1) \rightarrow (3\ 7).$

$$D_3^{-1}$$

|    | 4 | 10 | 7 | 5 | 2 | 3 | 14 | 9 | 13 | 8 | 15 | 11 | 1 | 6 | 12 |    |
|----|---|----|---|---|---|---|----|---|----|---|----|----|---|---|----|----|
|    | 1 | 2  | 3 | 4 | 5 | 6 | 7  | 8 | 9  | 10 | 11 | 12 | 13 | 14 | 15 |    |
| 1  | 0 | 0 | -3 | -3 | 2 | 1 | 2 | -3 | 1 | -3 | 3 | -1 | ∞ | -2 | 0 | 1 |
| 2  | 17 | 0 | -1 | 1 | ∞ | 17 | 4 | -1 | 3 | -2 | 5 | 1 | 4 | 0 | 2 | 2 |
| 3  | 14 | 17 | 0 | 1 | 18 | ∞ | 29 | 0 | 21 | 0 | 28 | 2 | 5 | 16 | 29 | 3 |
| 4  | ∞ | 0 | 2 | 0 | 3 | -1 | 13 | -14 | 10 | -14 | 8 | -13 | -12 | -1 | 14 | 4 |
| 5  | 13 | 26 | 15 | ∞ | 0 | 15 | 25 | 23 | 27 | 0 | 26 | 24 | 0 | 16 | 21 | 5 |
| 6  | 25 | 41 | 1 | 30 | 13 | 0 | -6 | 34 | 15 | 15 | 33 | 37 | -15 | ∞ | 1 | 6 |
| 7  | 38 | 44 | ∞ | 38 | 21 | 68 | 0 | 47 | 12 | 46 | 45 | 48 | 21 | 38 | -1 | 7 |
| 8  | 22 | 4 | 46 | 23 | 20 | 21 | 8 | 0 | 3 | ∞ | 46 | 46 | 21 | 24 | 4 | 8 |
| 9  | 23 | 4 | 48 | 47 | 22 | 22 | 10 | ∞ | 0 | 1 | 51 | 12 | 22 | 44 | 3 | 9 |
| 10 | 18 | ∞ | 40 | 45 | 18 | 18 | 40 | -1 | 42 | 0 | 43 | 5 | 17 | 47 | 4 | 10 |
| 11 | 15 | 1 | 41 | 39 | 15 | 15 | 2 | 3 | -9 | 38 | 0 | ∞ | 15 | 38 | -5 | 11 |
| 12 | 47 | 5 | -4 | 41 | 21 | 45 | 6 | -1 | -1 | 1 | 0 | 0 | 21 | 7 | ∞ | 12 |
| 13 | 47 | 47 | 13 | 51 | 26 | 43 | 4 | 0 | ∞ | 4 | 10 | 0 | 0 | -3 | 3 | 13 |
| 14 | 46 | 41 | -3 | 45 | 33 | 47 | ∞ | 6 | 0 | 5 | 2 | 7 | 23 | 0 | 6 | 14 |
| 15 | 41 | 44 | 42 | 46 | 24 | 46 | 2 | 47 | 6 | 43 | ∞ | 5 | 24 | 39 | 0 | 15 |
|    | 1 | 2 | 3 | 4 | 5 | 6 | 7 | 8 | 9 | 10 | 11 | 12 | 13 | 14 | 15 |    |

NEGATIVE CYCLES

| | 4 | 5 | 7 | 10 | 2 | 3 | 14 | 9 | 13 | 8 | 15 | 11 | 1 | 6 | 12 |
|---|---|---|---|---|---|---|---|---|---|---|---|---|---|---|---|
| | 1 | 2 | 3 | 4 | 5 | 6 | 7 | 8 | 9 | 10 | 11 | 12 | 13 | 14 | 15 |
| (6 13) -15 | | | | | | 9 -3 | 15 -20! | | 11 -25 | | 14 -16 | | 6 -15 | 13 -18 | 9 -22 |
| (4 8) -14 | | | | | | 15 -4! | 4 -14 | 8 -11 | 9 -10 | | | | | | 10 -6 |
| (4 10) -14 | | 9 -8 | | 2 -7 | | | 10 -15 | 8 -12 | 4 -14 | | | | | | |
| (4 12) -13 | | | 12 -16 | 3 -15 | | | | | | | | 4 -13 | | | |
| (11 9) -9 | | 9 -5 | | 2 -4 | | | 11 -9 | | | | | | | | |
| | | | | | | | | | | | | | | | |
| | | | | | | | | | | | | | | | |
| | | | | | | | | | | | | | | | |
| | 1 | 2 | 3 | 4 | 5 | 6 | 7 | 8 | 9 | 10 | 11 | 12 | 13 | 14 | 15 |

   The cycle of smallest value is (4 12 2). We note that it was obtained by first checking the smallest-valued entry in column 4 and testing and noting their values. This was done because the negative-valued initial arcs had large values. Thus, there were only a small number of rows that could yield cycles. There were only a small number of possibilities. Otherwise, for 6, the only practical values were (1 6), (2 6) and (3 6). For 4, there were (1 4) and (2 4). For 11, since there were a reasonable of possibilities, we didn't do this.

   Our smallest negatively valued cycle was (4 12 3): -15. It yields the following arcs:

$( 4\ 12 ) \to ( 4\ 11 ); ( 12\ 3 ) \to ( 12\ 7 ); ( 3\ 4 ) \to ( 3\ 5 ).$

$$D_4^{-1}$$

|   | 4 | 10 | 5 | 11 | 2 | 3 | 14 | 9 | 13 | 8 | 15 | 7 | 1 | 6 | 12 |   |
|---|---|---|---|---|---|---|---|---|---|---|---|---|---|---|---|---|
|   | 1 | 2 | 3 | 4 | 5 | 6 | 7 | 8 | 9 | 10 | 11 | 12 | 13 | 14 | 15 |   |
| 1 | 0 | 0 | -3 | -1 | 2 | 1 | 2 | -3 | 1 | -3 | 3 | -3 | ∞ | -2 | 0 | 1 |
| 2 | 17 | 0 | 1 | 1 | ∞ | 17 | 4 | -1 | 3 | -2 | 5 | -1 | 4 | 0 | 2 | 2 |
| 3 | 13 | 16 | 0 | 1 | 17 | ∞ | 28 | -1 | 20 | -1 | 27 | -1 | 4 | 15 | 28 | 3 |
| 4 | ∞ | 13 | 13 | 0 | 6 | 12 | 26 | -1 | 23 | -1 | 21 | 15 | 1 | 12 | 27 | 4 |
| 5 | 13 | 26 | ∞ | 24 | 0 | 15 | 25 | 23 | 27 | 0 | 26 | 15 | 0 | 16 | 21 | 5 |
| 6 | 25 | 41 | 30 | 37 | 13 | 0 | -6 | 34 | 15 | 15 | 33 | 1 | -15 | ∞ | 1 | 6 |
| 7 | 38 | 44 | 38 | 48 | 21 | 68 | 0 | 47 | 12 | 46 | 45 | ∞ | 21 | 38 | -1 | 7 |
| 8 | 22 | 4 | 23 | 46 | 20 | 21 | 8 | 0 | 3 | ∞ | 46 | 46 | 21 | 24 | 4 | 8 |
| 9 | 23 | 4 | 47 | 12 | 22 | 22 | 10 | ∞ | 0 | 1 | 51 | 48 | 22 | 44 | 3 | 9 |
| 10 | 18 | ∞ | 45 | 5 | 18 | 18 | 40 | -1 | 42 | 0 | 43 | 40 | 17 | 47 | 4 | 10 |
| 11 | 15 | 1 | 39 | ∞ | 15 | 15 | 2 | 3 | -9 | 38 | 0 | 41 | 15 | 38 | -5 | 11 |
| 12 | 51 | 9 | 45 | 4 | 25 | 49 | 10 | 3 | 3 | 5 | 5 | 0 | 25 | 11 | ∞ | 12 |
| 13 | 47 | 47 | 51 | 0 | 26 | 43 | 4 | 0 | ∞ | 4 | 10 | 13 | 0 | -3 | 3 | 13 |
| 14 | 46 | 41 | 45 | 7 | 33 | 47 | ∞ | 6 | 0 | 5 | 2 | -3 | 23 | 0 | 6 | 14 |
| 15 | 41 | 44 | 46 | 5 | 24 | 46 | 2 | 47 | 6 | 43 | ∞ | 42 | 24 | 39 | 0 | 15 |
|   | 1 | 2 | 3 | 4 | 5 | 6 | 7 | 8 | 9 | 10 | 11 | 12 | 13 | 14 | 15 |   |

## NEGATIVE CYCLES

|        | 4 | 10 | 5 | 11 | 2 | 3 | 14 | 9 | 13 | 8 | 15 | 7 | 1 | 6 | 12 |
|---|---|---|---|---|---|---|---|---|---|---|---|---|---|---|---|
|        | 1 | 2 | 3 | 4 | 5 | 6 | 7 | 8 | 9 | 10 | 11 | 12 | 13 | 14 | 15 |
| (6 13) -15 |  | 9 -21 |  |  |  | <u>2</u> <u>-4</u> |  |  | 11 -25 |  | 14 -16 |  | 6 -15 | 13 -18 |  |
| (11 9) -9 |  | 8 -5! |  |  |  |  |  | 10 -9 | 11 -9 | 9 -8 |  |  |  |  |  |
| (11 15) -5 |  |  |  |  |  |  | 15 -3! |  |  |  |  |  |  |  | 11 -5 |
| (1 3) -3 |  |  | 1 -3 |  |  |  |  | 10 -5 | 8 -1! | 3 -4 |  |  |  |  |  |
| (1 8) -3 |  |  |  |  |  |  |  | 1 -3! |  |  |  |  |  |  |  |
| (1 10) -3 |  |  |  |  |  |  |  | 10 -4 | 8 -1! | 1 -3 |  |  |  |  |  |
| (1 12) -3 |  |  |  |  |  |  |  |  |  |  |  | 1 -3! |  |  |  |
|        | 1 | 2 | 3 | 4 | 5 | 6 | 7 | 8 | 9 | 10 | 11 | 12 | 13 | 14 | 15 |

The only cycle we can obtain is (6 13 14 11 9 10): -6. The new arcs obtained are

*( 6 13 ) → ( 6 1 ); ( 13 14 ) → ( 13 6 ); ( 14 11 ) → ( 14 15 );*
*( 11 9 ) → ( 11 13 ); ( 9 2 ) → ( 9 10 ); ( 2 6 ) → ( 2 3 ).*

$$D_5^{-1}$$

| | 4 | 3 | 5 | 11 | 2 | 1 | 14 | 9 | 10 | 8 | 13 | 7 | 6 | 15 | 12 | |
|---|---|---|---|---|---|---|---|---|---|---|---|---|---|---|---|---|
| | 1 | 2 | 3 | 4 | 5 | 6 | 7 | 8 | 9 | 10 | 11 | 12 | 13 | 14 | 15 | |
| 1 | 0 | 1 | -3 | -1 | 2 | ∞ | 2 | -3 | 0 | -3 | 1 | -3 | -2 | 3 | 0 | 1 |
| 2 | 0 | 0 | -16 | -16 | ∞ | -13 | -13 | -18 | -17 | -19 | -14 | -18 | -17 | -12 | -15 | 2 |
| 3 | 13 | ∞ | 0 | 1 | 17 | 4 | 28 | -1 | 16 | -1 | 20 | -1 | 15 | 27 | 28 | 3 |
| 4 | ∞ | 12 | 13 | 0 | 6 | 1 | 26 | -1 | 13 | -1 | 23 | 15 | 12 | 21 | 27 | 4 |
| 5 | 13 | 15 | ∞ | 24 | 0 | 0 | 25 | 23 | 26 | 0 | 27 | 15 | 16 | 26 | 21 | 5 |
| 6 | 40 | 15 | 45 | 52 | 28 | 0 | 9 | 49 | 56 | 30 | 30 | 16 | ∞ | 49 | 16 | 6 |
| 7 | 38 | 68 | 38 | 48 | 21 | 21 | 0 | 47 | 44 | 46 | 12 | ∞ | 38 | 45 | -1 | 7 |
| 8 | 22 | 21 | 23 | 46 | 20 | 21 | 8 | 0 | 4 | ∞ | 3 | 46 | 24 | 46 | 4 | 8 |
| 9 | 19 | 18 | 43 | 8 | 18 | 18 | 6 | ∞ | 0 | -3 | -4 | 44 | 40 | 47 | -1 | 9 |
| 10 | 18 | 18 | 45 | 5 | 18 | 17 | 40 | -1 | ∞ | 0 | 42 | 40 | 47 | 43 | 4 | 10 |
| 11 | 24 | 24 | 48 | ∞ | 24 | 24 | 11 | 12 | 10 | 47 | 0 | 50 | 47 | 9 | 4 | 11 |
| 12 | 51 | 49 | 45 | 4 | 25 | 25 | 10 | 3 | 9 | 5 | 3 | 0 | 11 | 5 | ∞ | 12 |
| 13 | 50 | 46 | 54 | 3 | 29 | 3 | 7 | 3 | 50 | 7 | ∞ | 16 | 0 | 13 | 6 | 13 |
| 14 | 44 | 45 | 43 | 5 | 31 | 21 | ∞ | 4 | 39 | 3 | -2 | -5 | -2 | 0 | 4 | 14 |
| 15 | 41 | 46 | 46 | 5 | 24 | 24 | 2 | 47 | 44 | 43 | 6 | 42 | 39 | ∞ | 0 | 15 |
| | 1 | 2 | 3 | 4 | 5 | 6 | 7 | 8 | 9 | 10 | 11 | 12 | 13 | 14 | 15 | |

NEGATIVE CYCLES

| | 4 | 3 | 5 | 11 | 2 | 1 | 14 | 9 | 10 | 8 | 13 | 7 | 6 | 15 | 12 |
|---|---|---|---|---|---|---|---|---|---|---|---|---|---|---|---|
| | 1 | 2 | 3 | 4 | 5 | 6 | 7 | 8 | 9 | 10 | 11 | 12 | 13 | 14 | 15 |
| (2 10) -19 | | <u>4</u> <u>-2</u> | | 10 -14 | | | | | | 2 -19 | | | | | |
| (2 8) -18 | | | | 10 -12! | | | | 2 -18 | 8 -14 | 9 -17 | | | | | |
| (2 12) -18 | | <u>4</u> <u>-2</u> | | 12 -14 | | | | | | | | 2 -18 | | | |
| (2 9) -17 | | <u>4</u> <u>-1</u> | | 15 -13 | | | | 2 -17 | | | | | | | 9 -18 |
| (2 13) -17 | | <u>4</u> <u>-2</u> | | 13 -14 | | | | | | | | | 2 -17 | | |
| * (2 40) -16 | | <u>4</u> <u>-4</u> | | 2 -16 | | | | | | | | | | | |
| (2 15) -15 | | | | 15 -10! | | | | | | | | | | | 2 -15 |
| (2 11) -14 | | | | 15 -5! | | | | | | | 2 -14 | | | | 11 -10 |
| (2 6) -13 | | | | | 2 -13! | | | | | | | | | | |
| (2 7) -13 | | | | 15 -9! | | 2 -13 | | | | | | | | | 7 -14 |
| (2 14) -12 | | <u>4</u> <u>-1</u> | | 12 -13 | | | | | | | | 14 -17 | | 2 -12 | |
| | 1 | 2 | 3 | 4 | 5 | 6 | 7 | 8 | 9 | 10 | 11 | 12 | 13 | 14 | 15 |

Our smallest-valued cycle is (2 4): -4. It yields the following arcs: *( 2 4 )* → *( 2 11 ); ( 4 2 )* → *( 4 3 )*.

$$D_6^{-1}$$

| | 4 | 11 | 5 | 3 | 2 | 1 | 14 | 9 | 10 | 8 | 13 | 7 | 6 | 15 | 12 | |
|---|---|---|---|---|---|---|---|---|---|---|---|---|---|---|---|---|
| | 1 | 2 | 3 | 4 | 5 | 6 | 7 | 8 | 9 | 10 | 11 | 12 | 13 | 14 | 15 | |
| 1 | 0 | -1 | -3 | 1 | 2 | ∞ | 2 | -3 | 0 | -3 | 1 | -3 | -2 | 3 | 0 | 1 |
| 2 | 0 | 0 | 0 | 16 | ∞ | 3 | 3 | -2 | -1 | -3 | 2 | -2 | -1 | 4 | 1 | 2 |
| 3 | 13 | 1 | 0 | ∞ | 17 | 4 | 28 | -1 | 16 | -1 | 20 | -1 | 15 | 27 | 28 | 3 |
| 4 | ∞ | -12 | 1 | 0 | -6 | -11 | 14 | -13 | 1 | -13 | 11 | 3 | 0 | 9 | 15 | 4 |
| 5 | 13 | 24 | ∞ | 15 | 0 | 0 | 25 | 23 | 26 | 0 | 27 | 15 | 16 | 26 | 21 | 5 |
| 6 | 40 | 52 | 45 | 15 | 28 | 0 | 9 | 49 | 56 | 30 | 30 | 16 | ∞ | 49 | 16 | 6 |
| 7 | 38 | 48 | 38 | 68 | 21 | 21 | 0 | 47 | 44 | 46 | 12 | ∞ | 38 | 45 | -1 | 7 |
| 8 | 22 | 46 | 23 | 21 | 20 | 21 | 8 | 0 | 4 | ∞ | 3 | 46 | 24 | 46 | 4 | 8 |
| 9 | 19 | 8 | 43 | 18 | 18 | 18 | 6 | ∞ | 0 | -3 | -4 | 44 | 40 | 47 | -1 | 9 |
| 10 | 18 | 5 | 45 | 18 | 18 | 17 | 40 | -1 | ∞ | 0 | 42 | 40 | 47 | 43 | 4 | 10 |
| 11 | 24 | ∞ | 48 | 24 | 24 | 24 | 11 | 12 | 10 | 47 | 0 | 50 | 47 | 9 | 4 | 11 |
| 12 | 51 | 4 | 45 | 49 | 25 | 25 | 10 | 3 | 9 | 5 | 3 | 0 | 11 | 5 | ∞ | 12 |
| 13 | 50 | 3 | 54 | 46 | 29 | 3 | 7 | 3 | 50 | 7 | ∞ | 16 | 0 | 13 | 6 | 13 |
| 14 | 44 | 5 | 43 | 45 | 31 | 21 | ∞ | 4 | 39 | 3 | -2 | -5 | -2 | 0 | 4 | 14 |
| 15 | 41 | 5 | 46 | 46 | 24 | 24 | 2 | 47 | 44 | 43 | 6 | 42 | 39 | ∞ | 0 | 15 |
| | 1 | 2 | 3 | 4 | 5 | 6 | 7 | 8 | 9 | 10 | 11 | 12 | 13 | 14 | 15 | |

.

NEGATIVE CYCLES

|  | 4 | 11 | 5 | 3 | 2 | 1 | 14 | 9 | 10 | 8 | 13 | 7 | 6 | 15 | 12 |
|---|---|---|---|---|---|---|---|---|---|---|---|---|---|---|---|
|  | 1 | 2 | 3 | 4 | 5 | 6 | 7 | 8 | 9 | 10 | 11 | 12 | 13 | 14 | 15 |
| (4 2) -12 | 2 -12 | 4 -12 | 1 -15 |  |  |  |  |  |  | 3 -16 | 15 -6! |  |  |  | 10 -12 |
| (4 5) -6 |  |  |  |  | 4 -6 | 5 -6! |  |  |  |  |  |  |  |  |  |
| (4 6) -11 |  |  |  |  | 4 -11 | 6 -2 |  |  |  |  |  |  |  |  | 7 -3! |
| (4 8) -13 | 2 -1! | 13 -1 |  | 1 0? |  |  |  | 4 -14 |  | 8 -11 |  | 14 -4 | 11 -2 |  |  |
| (4 10) -13 | 2 -8 | 10 -8 |  | <u>1</u> <u>-7</u> |  |  |  |  |  | 4 -13 |  |  |  |  |  |
| (14 12) -5 |  |  |  |  |  |  |  |  |  |  |  | 14 -5! |  |  |  |
|  | 1 | 2 | 3 | 4 | 5 | 6 | 7 | 8 | 9 | 10 | 11 | 12 | 13 | 14 | 15 |

We've obtained the cycle (4 10 2 1): -7. The corresponding arcs are:

*( 4 10 ) →( 4 8 ); ( 10 2 ) →( 10 11 ); ( 2 1 ) →( 2 4 ); ( 1 4 ) →( 1 3 ).*

$$D_7^{-1}$$

| | 3 | 4 | 5 | 8 | 2 | 1 | 14 | 9 | 10 | 11 | 13 | 7 | 6 | 15 | 12 | |
|---|---|---|---|---|---|---|---|---|---|---|---|---|---|---|---|---|
| | 1 | 2 | 3 | 4 | 5 | 6 | 7 | 8 | 9 | 10 | 11 | 12 | 13 | 14 | 15 | |
| 1 | 0 | 0 | -4 | -4 | 1 | ∞ | 1 | -4 | -1 | -2 | 0 | -4 | -3 | 2 | -1 | 1 |
| 2 | 16 | 0 | 0 | -3 | ∞ | 3 | 3 | -2 | -1 | 0 | 2 | -2 | -1 | 4 | 1 | 2 |
| 3 | ∞ | 13 | 0 | -1 | 17 | 4 | 28 | -1 | 16 | 1 | 20 | -1 | 15 | 27 | 28 | 3 |
| 4 | 13 | ∞ | 14 | 0 | 7 | 2 | 27 | 0 | 14 | 1 | 24 | 16 | 13 | 22 | 28 | 4 |
| 5 | 15 | 13 | ∞ | 0 | 0 | 0 | 25 | 23 | 26 | 24 | 27 | 15 | 16 | 26 | 21 | 5 |
| 6 | 15 | 40 | 45 | 30 | 28 | 0 | 9 | 49 | 56 | 52 | 30 | 16 | ∞ | 49 | 16 | 6 |
| 7 | 68 | 38 | 38 | 46 | 21 | 21 | 0 | 47 | 44 | 48 | 12 | ∞ | 38 | 45 | -1 | 7 |
| 8 | 21 | 22 | 23 | ∞ | 20 | 21 | 8 | 0 | 4 | 46 | 3 | 46 | 24 | 46 | 4 | 8 |
| 9 | 18 | 19 | 43 | -3 | 18 | 18 | 6 | ∞ | 0 | 8 | -4 | 44 | 40 | 47 | -1 | 9 |
| 10 | 13 | 13 | 40 | -5 | 13 | 12 | 35 | -6 | ∞ | 0 | 37 | 35 | 42 | 38 | -1 | 10 |
| 11 | 24 | 24 | 48 | 47 | 24 | 24 | 11 | 12 | 10 | ∞ | 0 | 50 | 47 | 9 | 4 | 11 |
| 12 | 49 | 51 | 45 | 5 | 25 | 25 | 10 | 3 | 9 | 4 | 3 | 0 | 11 | 5 | ∞ | 12 |
| 13 | 46 | 50 | 54 | 7 | 29 | 3 | 7 | 3 | 50 | 3 | ∞ | 16 | 0 | 13 | 6 | 13 |
| 14 | 45 | 44 | 43 | 3 | 31 | 21 | ∞ | 4 | 39 | 5 | -2 | -5 | -2 | 0 | 4 | 14 |
| 15 | 46 | 41 | 46 | 43 | 24 | 24 | 2 | 47 | 44 | 5 | 6 | 42 | 39 | ∞ | 0 | 15 |
| | 1 | 2 | 3 | 4 | 5 | 6 | 7 | 8 | 9 | 10 | 11 | 12 | 13 | 14 | 15 | |

.

NEGATIVE CYCLES

| | 3 | 4 | 5 | 8 | 2 | 1 | 14 | 9 | 10 | 11 | 13 | 7 | 6 | 15 | 12 |
|---|---|---|---|---|---|---|---|---|---|---|---|---|---|---|---|
| | 1 | 2 | 3 | 4 | 5 | 6 | 7 | 8 | 9 | 10 | 11 | 12 | 13 | 14 | 15 |
| (10 4) -5 | | | | 10 -5 | | | | | | <u>4</u> <u>-4</u> | | | | | |
| (1 4) -4 | | | | 1 -4 | | | | | | 4 -3 | | | | | 10 -4! |
| (1 3) -4 | | | 1 -4 | | | | | 3 -5 | | | 8 -2! | | | | |
| (1 8) -4 | | | | | | | | 1 -4 | | | 8 -1! | | | | |
| (1 12) -4 | | | | | | | | 12 -1! | | | | 1 -4 | | | |
| (9 13) -4 | | | | | | | | 13 -1! | | | | | 9 -4 | | |
| | 1 | 2 | 3 | 4 | 5 | 6 | 7 | 8 | 9 | 10 | 11 | 12 | 13 | 14 | 15 |

The only cycle we obtain is (10 4): -4. This yields the following arcs:

$(4\ 10) \rightarrow (4\ 11); (10\ 4) \rightarrow (10\ 8).$

$D_8^{-1}$

| | 3 | 4 | 5 | 11 | 2 | 1 | 14 | 9 | 10 | 8 | 13 | 7 | 6 | 15 | 12 |
|---|---|---|---|---|---|---|---|---|---|---|---|---|---|---|---|
| | 1 | 2 | 3 | 4 | 5 | 6 | 7 | 8 | 9 | 10 | 11 | 12 | 13 | 14 | 15 |
| 1 | 0 | 0 | -4 | -2 | 1 | ∞ | 1 | -4 | -1 | -4 | 0 | -4 | -3 | 2 | -1 |
| 2 | 16 | 0 | 0 | 0 | ∞ | 3 | 3 | -2 | -1 | -3 | 2 | -2 | -1 | 4 | 1 |
| 3 | ∞ | 13 | 0 | 2 | 17 | 4 | 28 | -1 | 16 | -1 | 20 | -1 | 13 | 27 | 28 |
| 4 | 12 | ∞ | 13 | 0 | 6 | 1 | 26 | -1 | 13 | -1 | 23 | 15 | 12 | 21 | 27 |
| 5 | 15 | 13 | ∞ | 24 | 0 | 0 | 25 | 23 | 26 | 0 | 27 | 15 | 16 | 26 | 21 |
| 6 | 15 | 40 | 45 | 52 | 28 | 0 | 9 | 49 | 56 | 30 | 30 | 16 | ∞ | 49 | 16 |
| 7 | 68 | 38 | 38 | 48 | 21 | 21 | 0 | 47 | 44 | 46 | 12 | ∞ | 38 | 45 | -1 |
| 8 | 21 | 22 | 23 | 46 | 20 | 21 | 8 | 0 | 4 | ∞ | 3 | 46 | 24 | 46 | 4 |
| 9 | 18 | 19 | 43 | 8 | 18 | 18 | 6 | ∞ | 0 | -3 | -4 | 44 | 40 | 47 | -1 |
| 10 | 18 | 18 | 45 | 5 | 18 | 17 | 40 | -1 | ∞ | 0 | 42 | 40 | 47 | 43 | 4 |
| 11 | 24 | 24 | 48 | ∞ | 24 | 24 | 11 | 12 | 10 | 47 | 0 | 50 | 47 | 9 | 4 |
| 12 | 49 | 51 | 45 | 4 | 25 | 25 | 10 | 3 | 9 | 5 | 3 | 0 | 11 | 5 | ∞ |
| 13 | 46 | 50 | 54 | 3 | 29 | 3 | 7 | 3 | 50 | 7 | ∞ | 16 | 0 | 13 | 6 |
| 14 | 45 | 44 | 43 | 5 | 31 | 21 | ∞ | 4 | 39 | 3 | -2 | -5 | -2 | 0 | 4 |
| 15 | 46 | 41 | 46 | 5 | 24 | 24 | 2 | 47 | 44 | 43 | 6 | 42 | 39 | ∞ | 0 |
| | 1 | 2 | 3 | 4 | 5 | 6 | 7 | 8 | 9 | 10 | 11 | 12 | 13 | 14 | 15 |

We first note that it is very unlikely that we can obtain a cycle whose initial node is 1. The acceptable negative values (i.e., those that don't yield a 2-cycle) are such that we would have to obtain a path whose value is -13 or smaller in order to obtain a negatively valued cycle. The same is true of row 2. We thus use as our initial nodes the following: (9 11): -4, (14 13): -2.

## NEGATIVE CYCLES

|         | 3 | 4 | 5 | 11 | 2 | 1 | 14 | 9 | 10 | 8 | 13 | 7 | 6 | 15 | 12 |
|---------|---|---|---|----|---|---|----|---|----|---|----|---|---|----|----|
|         | 1 | 2 | 3 | 4  | 5 | 6 | 7  | 8 | 9  | 10| 11 | 12| 13| 14 | 15 |
| (9 11) -4 |   |   |   |    |   |   |    |   |    |   | 9 -4! |   |   |    |    |
| (14 13) -2 |   |   |   |    |   |   |    |   |    |   |    |   | 14 -2! |    |    |
|         | 1 | 2 | 3 | 4  | 5 | 6 | 7  | 8 | 9  | 10| 11 | 12| 13| 14 | 15 |

Since we have obtained no further negative cycles, we now patch the cycles of the derangement $D_8$:

$(1\ 3\ 5\ 2\ 4\ 11\ 13\ 6)(8\ 9\ 10)(7\ 14\ 15\ 12) : 584.$ Using the arcs (4 10), (9 12), (15 11), we obtain the tour $T_{UPPERBOUND} = (4\ 10\ 8\ 9\ 12\ 7\ 14\ 15\ 11\ 13\ 6\ 1\ 3\ 5\ 2) : 587$. The average arc-value of $T_{UPPERBOUND}$ is $\frac{587}{15} = 39.133...$ .We now employ our FWK algorithm to obtain cycles each of which has an average arc-value less than that of $T_{UPPERBOUND}$.

We now construct paths using FWK.

$j = 6$
$(12\ 6)(6\ 14) = (12\ 14) : 67_2; (13\ 6)(6\ 14) = (13\ 14) : 53_2; (14\ 6)(6\ 12) = (14\ 12) : 67_2;$
$(14\ 6)(6\ 13) = (14\ 13).$
$j = 7$
$(12\ 7)(7\ 14) = (12\ 14) : 53_2; (13\ 7)(7\ 14) = (13\ 14) : 62_2.$
$j = 8$
$(9\ 8)(8\ 6) = (9\ 6) : \underline{78_2}; (9\ 8)(8\ 10) = (9\ 10) : 58_2; (10\ 8)(8\ 9) = (10\ 9) : 58_2;$
$(10\ 8)(8\ 12) = (10\ 12) : 62_2; (10\ 8)(8\ 13) : (10\ 13) : 61_2; (10\ 8)(8\ 14) = (10\ 14) : 66_2;$
$(12\ 8)(8\ 9) = (12\ 9) : 58_2; (12\ 8)(8\ 10) = (12\ 10) : 62_2; (13\ 8)(8\ 3) = (13\ 3) : \underline{78_2};$
$(13\ 8)(8\ 10) = (13\ 10).$

*j = 9*

$( 8\ 9 )( 9\ 10 ) = ( 8\ 10 ): 57_2; ( 8\ 9 )( 9\ 11 ) = ( 8\ 11 ): 65_2; ( 8\ 9 )( 9\ 12 ) = ( 8\ 12 ): 58_2;$

$( 8\ 9 )( 9\ 13 ) = ( 8\ 13 ): 53_2; ( 8\ 9 )( 9\ 14 ) = ( 8\ 14 ): 63_2; ( 11\ 9 )( 9\ 10 ) = ( 11\ 10 ): 68_2;$

$( 12\ 9 )( 9\ 11 ) = ( 12\ 11 ): 96_3; ( 12\ 9 )( 9\ 13 ) = ( 12\ 13 ): 84_3; ( 12\ 9 )( 9\ 14 ) = ( 12\ 14 ): 94_3;$

$( 13\ 9 )( 9\ 8 ) = ( 13\ 8 ): \underline{53_2}; ( 13\ 9 )( 9\ 10 ) = ( 13\ 10 ): 56_2; ( 13\ 9 )( 9\ 12 ) = ( 13\ 12 ): 55_2;$

$( 14\ 9 )( 9\ 8 ) = ( 14\ 8 ): \underline{53_2}; ( 14\ 9 )( 9\ 10 ) = ( 14\ 10 ): 66_2; ( 14\ 9 )( 9\ 12 ) = ( 14\ 12 ): 65_2.$

*j = 10*

$( 8\ 10 )( 10\ 11 ) = ( 8\ 11 ): 93_3; ( 11\ 10 )( 10\ 12 ) = ( 11\ 12 ): 103_3; ( 13\ 10 )( 10\ 11 ) = ( 13\ 11 ): 92_3;$

$( 14\ 10 )( 10\ 11 ) = ( 14\ 11 ): 102_3.$

*j = 12*

$( 6\ 12 )( 12\ 7 ) = ( 6\ 7 ): \underline{63_2}; ( 6\ 12 )( 12\ 9 ) = ( 6\ 9 ): \underline{66_2}; ( 6\ 12 )( 12\ 13 ) = ( 6\ 13 ): 66_2;$

$( 6\ 12 )( 12\ 11 ) = ( 6\ 11 ): \underline{67_2}; ( 6\ 12 )( 12\ 15 ) = ( 6\ 15 ): 67_2; ( 6\ 12 )( 12\ 8 ) = ( 6\ 8 ): \underline{68_2};$

$( 6\ 12 )( 12\ 10 ) = ( 6\ 10 ): \underline{72_2}; ( 6\ 12 )( 12\ 14 ) = ( 6\ 14 ): 73_2; ( 7\ 12 )( 12\ 9 ) = ( 7\ 9 ): \underline{55_2};$

$( 7\ 12 )( 12\ 13 ) = ( 7\ 13 ): 55_2; ( 7\ 12 )( 12\ 11 ) = ( 7\ 11 ): \underline{56_2}; ( 7\ 12 )( 12\ 15 ) = ( 7\ 15 ): 56_2;$

$( 8\ 12 )( 12\ 7 ) = ( 8\ 7 ): \underline{84_3}; ( 8\ 12 )( 12\ 11 ) = ( 8\ 11 ): \underline{88_3}; ( 8\ 12 )( 12\ 15 ) = ( 8\ 15 ): 88_3;$

$( 8\ 12 )( 12\ 14 ) = ( 8\ 14 ): 92_3; ( 9\ 12 )( 12\ 7 ) = ( 9\ 7 ): \underline{55_2}; ( 9\ 12 )( 12\ 11 ) = ( 9\ 11 ): \underline{59_2};$

$( 9\ 12 )( 12\ 15 ) = ( 9\ 15 ): 59_2; ( 9\ 12 )( 12\ 14 ) = ( 9\ 14 ): 65_2; ( 9\ 12 )( 12\ 6 ) = ( 9\ 6 ): \underline{66_2};$

$( 10\ 12 )( 12\ 7 ) = ( 10\ 7 ): \underline{88_3}; ( 10\ 12 )( 12\ 11 ) = ( 10\ 11 ): \underline{92_3}; ( 10\ 12 )( 12\ 6 ) = ( 10\ 6 ): \underline{99_3};$

$( 11\ 12 )( 12\ 7 ) = ( 11\ 7 ): \underline{56_2}; ( 11\ 12 )( 12\ 9 ) = ( 11\ 9 ): \underline{59_2}; ( 11\ 12 )( 12\ 15 ) = ( 11\ 15 ): 60_2;$

$( 11\ 12 )( 12\ 8 ) = ( 11\ 8 ): \underline{61_2}; ( 11\ 12 )( 12\ 10 ) = ( 11\ 10 ): \underline{65_2}; ( 11\ 12 )( 12\ 14 ) = ( 11\ 14 ): 66_2;$

$( 11\ 12 )( 12\ 6 ) = ( 11\ 6 ): \underline{67_2}; ( 13\ 12 )( 12\ 7 ) = ( 13\ 7 ): \underline{81_3}; ( 14\ 12 )( 12\ 9 ) = ( 14\ 9 ): \underline{82_3};$

$( 14\ 12 )( 12\ 15 ) = ( 14\ 15 ): 83_3; ( 14\ 12 )( 12\ 10 ) = ( 14\ 10 ): \underline{88_3}.$

*j = 13*

$( 6\ 13 )( 13\ 7 ) = ( 6\ 7 ): \underline{62_2}; ( 6\ 13 )( 13\ 9 ) = ( 6\ 9 ): \underline{49_2}; ( 6\ 13 )( 13\ 11 ) = ( 6\ 11 ): \underline{49_2};$

$( 6\ 13 )( 13\ 12 ) = ( 6\ 12 ): \underline{52_2}; ( 6\ 13 )( 13\ 14 ) = ( 6\ 14 ): 53_2; ( 6\ 13 )( 13\ 15 ) = ( 6\ 15 ): 59_2;$

$( 7\ 13 )( 13\ 11 ) = ( 7\ 11 ): \underline{81_3}; ( 7\ 13 )( 13\ 14 ) = ( 7\ 14 ): 85_3; ( 8\ 13 )( 13\ 6 ) = ( 8\ 6 ): \underline{79_3};$

$( 8\ 13 )( 13\ 11 ): \underline{79_3}; ( 8\ 13 )( 13\ 14 ) = ( 8\ 14 ): 83_3; ( 8\ 13 )( 13\ 15 ) = ( 8\ 15 ): 89_3;$

$( 9\ 13 )( 13\ 6 ) = ( 9\ 6 ): \underline{49_2}; ( 9\ 13 )( 13\ 11 ) = ( 9\ 11 ): \underline{52_2}; ( 9\ 13 )( 13\ 12 ) = ( 9\ 12 ): \underline{55_2};$

$( 9\ 13 )( 13\ 8 ) = ( 9\ 8 ): \underline{56_2}; ( 9\ 13 )( 13\ 14 ) = ( 9\ 14 ): 56_2; ( 10\ 13 )( 13\ 6 ) = ( 10\ 6 ): \underline{84_3};$

$( 10\ 13 )( 13\ 9 ) = ( 10\ 9 ): \underline{87_3}; ( 10\ 13 )( 13\ 11 ) = ( 10\ 11 ): \underline{87_3}; ( 10\ 13 )( 13\ 12 ) = ( 10\ 12 ): \underline{90_3};$

$( 10\ 13 )( 13\ 14 ) = ( 10\ 14 ): 91_3; ( 10\ 13 )( 13\ 15 ) = ( 10\ 15 ): 97_3.$

$(11\ 13)(13\ 6) = (11\ 6): \underline{49}_2; (11\ 13)(13\ 9) = (11\ 9): \underline{52}_2; (11\ 13)(13\ 12) = (11\ 12): \underline{55}_2;$

$(14\ 13)(13\ 9) = (14\ 9): \underline{79}_3; (14\ 13)(13\ 11) = (14\ 11): \underline{79}_3; (14\ 13)(13\ 12) = (14\ 12): \underline{82}_3;$
$(14\ 13)(13\ 8) = (14\ 8): \underline{83}_3; (14\ 13)(13\ 15) = (14\ 15): \underline{89}_3; (15\ 13)(13\ 6) = (15\ 6): \underline{59}_2;$
$(15\ 13)(13\ 9) = (15\ 9): \underline{62}_2; (15\ 13)(13\ 11) = (15\ 11): \underline{62}_2; (15\ 13)(13\ 12) = (15\ 12): \underline{66}_2;$
$(15\ 13)(13\ 7) = (15\ 7): \underline{75}_2.$

*j = 14*
$(6\ 14)(14\ 7) = (6\ 7): \underline{57}_2; (6\ 14)(14\ 13) = (6\ 13): \underline{60}_2; (6\ 14)(14\ 8) = (6\ 8): \underline{65}_2;$
$(6\ 14)(14\ 12) = (6\ 12): \underline{66}_2; (7\ 14)(14\ 6) = (7\ 6): \underline{57}_2; (8\ 14)(14\ 7) = (8\ 7): \underline{119}_4;$
$(9\ 14)(14\ 15) = (9\ 15): \underline{88}_3; (10\ 14)(14\ 7) = (10\ 7): \underline{93}_3; (10\ 14)(14\ 6) = (10\ 6): \underline{96}_3;$
$(10\ 14)(14\ 15) = (10\ 15): \underline{98}_3; (11\ 14)(14\ 6) = (11\ 6): \underline{86}_3; (11\ 14)(14\ 15) = (11\ 15): \underline{88}_3;$
$(11\ 14)(14\ 1) = (11\ 1): \underline{109}_3; (12\ 14)(14\ 6) = (12\ 6): \underline{83}_3; (12\ 14)(14\ 13) = (12\ 13): \underline{83}_3;$
$(12\ 14)(14\ 15) = (12\ 15): \underline{85}_3; (15\ 14)(14\ 13) = (15\ 13): \underline{62}_2.$

*j = 15*
$(7\ 15)(15\ 11) = (7\ 11): \underline{71}_3; (8\ 15)(15\ 11) = (8\ 11): \underline{124}_4.$

*P₁₅*

|    | 6  | 7  | 8  | 9   | 10 | 11    | 12    | 13      | 14      | 15 |     |
|----|----|----|----|-----|----|-------|-------|---------|---------|----|-----|
| 6  |    | 14 | 14 | 13  |    | 13    | 14    | 6       | 6       | 13 | 6   |
| P  |    | 3  | 4  | 5   |    | 6     | 7     | 1,2,5,7 | 3,4,6,8 |    |     |
| NP |    |    |    |     |    |       |       |         |         |    |     |
|    |    |    |    |     |    |       |       |         |         |    |     |
| 7  | 14 |    | 13 | 12  |    | 12    | 7     | 12      | 7       | 12 | 7   |
| P  | 3  |    | 4  | 5   |    | 6     | 1,2,4,5 | 1,2,4 | 3       |    |     |
| NP |    |    |    |     |    |       |       |         |         |    |     |
|    |    |    |    |     |    |       |       |         |         |    |     |
| 8  | 14 | 12 |    | 8   | 9  | 10    | 9     | 9       | 8       | 13 | 8   |
| P  | 5  | 6  |    | a-5 |    |       | 6     | 1       | 5       |    | a=6 |
| NP |    |    |    |     |    |       |       |         |         |    |     |

| | | | | | | 0 | | | | | |
|---|---|---|---|---|---|---|---|---|---|---|---|
| 9 P NP | 13 6 | 12 7 | 13 a-6.7 | | 8 9 | 13 7.10 | 13 a. | 9 | 9 | 14 | 9 a=10 |
| 10 P NP | 14 3 | 14 4 | 10 A | 8 1.2 | | 13 5 | 13 4.6 | 8 5.6 | 8 3.4 | 14 | 10 a=6 |
| 11 P NP | 14 3 | 12 4 | 13 5 | 13 | 12 | | 13 4.6 | 11 A | 13 1.3 | 13 2 | 11 a=6 |
| 12 P NP | 14 5 | 12 5.6 | 12 1.2.3.4 | 8 1.2.3.4 | 9 | 10 | | 14 6 | 7 5.6 | 14 | 12 |
| 13 P NP | 13 2 | 12 5 | 9 | 13 1.3.4.5 | 9 | 9 | 9 1.5 | | 6 | 11 | 13 |
| 14 P NP | 14 5 | 14 a- | 9 4 | 12 3.4.6 | 9 6 | 13 5 | 7 a-5 | 6 5 | | 12 | 14 a=6 |
| 15 P NP | 13 1 | 12 2 | 14 | 13 3 | | 13 4 | 13 2 | 14 A | 15 a | | 15 a=5 |
| | 6 | 7 | 8 | 9 | 10 | 11 | 12 | 13 | 14 | 15 | |

INITIAL ARCS

6. (6 13), (6 14). 7. (7 12), (7 14). 8. (8 9), (8 14). 9. (9 13), (9 14). 10. (10 8). 11. (11 13). 12. (12 6), (12 7), (12 8). 13. (13 6), (13 9). 14. (14 6), (14 7). 15. (15 14).

M(15)

| | 6 | 7 | 8 | 9 | 10 | 11 | 12 | 13 | 14 | 15 | |
|---|---|---|---|---|---|---|---|---|---|---|---|
| 6 | $\infty$ | $\underline{62_2}$ | $\underline{53_2}$ | $\underline{49_2}$ | 48 | $\underline{49_2}$ | $\underline{52_2}$ | 23 | 30 | $\underline{62_2}$ | 6 |
| 7 | $\underline{57_2}$ | $\infty$ | $\underline{85_3}$ | $\underline{81_3}$ | $\underline{61_2}$ | $\underline{81_3}$ | 26 | $55_2$ | 27 | $\underline{56_2}$ | 7 |
| 8 | $\underline{108_4}$ | $\underline{82_3}$ | $\infty$ | 27 | $57_2$ | $\underline{86_3}$ | $\underline{56_2}$ | $53_2$ | $\underline{83_3}$ | $\underline{89_3}$ | 8 |
| 9 | $\underline{49_2}$ | $\underline{81_3}$ | 27 | $\infty$ | $\underline{87_3}$ | $\underline{52_2}$ | $\underline{55_2}$ | 26 | $56_2$ | $\underline{62_2}$ | 9 |

| | 6 | 7 | 8 | 9 | 10 | 11 | 12 | 13 | 14 | 15 | |
|---|---|---|---|---|---|---|---|---|---|---|---|
| 10 | $\underline{99_3}$ | $\underline{93_3}$ | 31 | 30 | ∞ | $\underline{87_3}$ | $\underline{90_3}$ | $61_2$ | $93_3$ | $97_3$ | 10 |
| 11 | $\underline{86_3}$ | $\underline{81_3}$ | $\underline{56_2}$ | $\underline{52_2}$ | $68_2$ | ∞ | $\underline{56_2}$ | 26 | $56_2$ | $88_3$ | 11 |
| 12 | $\underline{83_3}$ | 26 | 31 | 29 | $88_3$ | 30 | ∞ | $\underline{83_3}$ | $53_2$ | $85_3$ | 12 |
| 13 | 23 | $\underline{81_3}$ | $\underline{53_2}$ | 26 | $56_2$ | 26 | $55_2$ | ∞ | $53_2$ | $91_3$ | 13 |
| 14 | 30 | 27 | $\underline{109_4}$ | $\underline{82_3}$ | $\underline{112_4}$ | $\underline{79_3}$ | $53_2$ | $53_2$ | ∞ | $83_3$ | 14 |
| 15 | $\underline{88_3}$ | $\underline{117_4}$ | 73 | $\underline{88_3}$ | 74 | $\underline{88_3}$ | 30 | $\underline{62_2}$ | 32 | ∞ | 15 |
| | 6 | 7 | 8 | 9 | 10 | 11 | 12 | 13 | 14 | 15 | |

We now give underlined entries as well as initial arcs of paths in a row:

6. *(6 13), (6 14).* 1, 2, 7, 8, 9, 10, 11, 12. 7. *(7 12), (7 14).* 1, 2, 6, 8, 9, 10, 11. 8. *(8 9), (8 14).* 1, 2, 3, 4, 6, 7, 1, 12.

9. *(9 13), (9 14).* 1, 2, 3, 4, 5, 6, 10, 11, 12. 10. *(10 8).* 1, 2, 6, 7, 11, 12. 11. *(11 13).* 1, 2, 6, 7, 8, 9, 12.

12. *(12 6), (12 7), (12 8).* 1, 2, 3, 4, 6, 13. 13. *(13 6), (13 9).* 1, 2, 3, 4, 7, 8. 14. *(14 6), (14 7).* 1, 2, 8, 9, 10, 11.

15. *(15 14).* 6, 7, 9, 11, 13.

We place the respective path numbers in $P_{15}$.

Next, we give cycles whose average arc values have been arranged in ascending order.

<div align="center">CYCLES</div>

(6 13 8 12): 25.25; (12 7 14): 26.3…; (14 6 13 11 12 7): 27; (14 7 12 1 6): 27;

(13 6 14): 27.6…; (8 9 13): 27.6…; (12 8 9 13 11): 27.6…; (6 13 14): 27.6…; (14 6 13): 27.6…; (13 9 12 11):

27.75; (9 13 11 12): 27.75; (13 11 12 9): 27.75; (12 7 14 13 9 8): 27.83…;

(13 9 12 7 14 6): 27.83…; (12 8 9 13 11): 28; (7 12 13 14): 28; (13 9 12): 28;

(11 13 12): 28.3…; (7 12 13): 28.3…; (13 6 14 9): 28.75; (14 13 12 7): 28.75;

(7 14 15 12): 28.75; (13 6 12 9): 28.75; (7 12 15 14): 28.75; (12 7 14 15): 28.75;

(12 8 9 10 11): 28.8; (13 6 14 15 11): 28.8; (13 6 14 11): 29; (13 9 12 8): 29;

(14 6 13 11): 29; (6 13 11 12): 29; (11 13 6 12): 29; (8 9 12): 29; (8 7 14 12 9): 29.2;
(14 6 13 8 9): 29.2; (13 9 12 15 11): 29.2; (13 6 14 15 11): 29.2; (13 9 10 8): 29.25;
(9 8 10): 29.3...; (12 8 9 13 6): 29.4; (14 6 12 8 9): 29.4; (13 9 8 14 6): 29.4;
(14 7 12 9 8): 29.4; (13 9 10 11): 29.5; (13 9 10 11): 29.5; (14 6 12 9): 29.5;
(12 7 *14* 9): 29.5; (9 8 14 13): 29.5; (12 9 14 7): 29.5; (12 7 14): 29.6...; (13 6 14): 29.6...;
(6 13 12): 29.6...; (8 12 13): 29.6...; (14 6 12): 29.6...; (13 6 14 7): 29.75; (14 7 13 6): 29.75;
(14 15 12 7): 29.75; (7 12 8 14): 29.75; (12 7 14 8): 29.75; (12 7 14 13 15): 29.8;
(12 7 14 6): 30; (15 14 7 12 13): 30; (14 7 12 6): 30; (13 9 12 14 6): 30; (13 9 12 7): 30;
(12 7 14 11): 30; (9 13 11): 30; (9 12 7 13): 30; (6 14 7 12): 30; (12 7 14 11): 30;
(14 6 13 11 10 9): 30.16...; (12 8 9 14 6): 30.2; (6 13 11 15 12): 30.2; (12 9 10 8): 30.25;
(15 13 6 14): 30.25; (9 13 14 12): 30.25; (14 6 13 15): 30.25; (9 12 14 13): 30.25;
(13 9 11 15 14 6): 30.3...; (8 13 14 12 9): 30.4; (11 13 14 12): 30.5; (9 13 14): 30.6...;
(12 11 15): 30.6...; (6 14 8 12): 30.75; (11 13 14 15): 30.75; (13 6 14 12 9): 30.8;
(14 6 12 10 9): 30.8; (6 14 15 13 12): 30.8; (13 9 12 *15*): 31; (11 13 14): 31;
(12 8 9 10 11 15): 31;  (6 13 7 12): 31.25; (9 10 12): 31.3...; (7 12 13): 31.3...;
(6 13 15 12): 1.5; (11 12 15): 31.6...; (8 9 12 14): 31.75; (10 8 12 11): 32; (8 12 14): 32;
(8 9 12 14): 32; (14 6 12 15): 32.25; (9 12 11): 32.3...; (10 8 12): 32.3...;
(11 13 15): 32.3...; (11 13 15): 32.3...; (11 13 14 9): 32.5; (15 13 9 14): 32.5; (15 14 13): 32.6...;
(15 13 11 14): 32.75; (6 14 9): 33; (15 13 8 14): 33.25; (11 12 10 9): 33.25;
(15 13 12 14): 33.25; (6 12 11 14): 33.5; (12 11 15): 34.3...; (14 7 12): 34.3...; (6 14 12): 34.3...;
(11 12 14): 34.3...; (6 14 12): 34.3...; (10 8 14 11): 34.75; (13 9 10 *1* 6): 35.2; (9 10 11): 35.3...;
(13 9 12 1 14 6): 35.3...; (10 8 9 11): 35.5;  (11 9 10): 35.6...; (9 15 11): 36; (9 12 6 8): 36;
(13 6 1 9): 36.5; (13 6 2 9): 36.5; (9 11 *14*): 37; (10 9 2 8): 37.25; (12 8 9 1 7): 38;
(12 8 9 2 7): 38; (12 8 9 3 7): 38; (9 8 2 13): 38; (9 8 1 13): 38.25; (8 9 13 1): 38.25;
 (14 6 13 9): 38.3...; (10 8 9 1): 38.5; (7 12 13 1): 38.75; (10 8 9 2): 38.75; (7 12 13 2): 38.75.

PATHS

6.
$P_1 = (6\ 13\ 1)$; $P_2 = (6\ 13\ 2)$; $P_3 = (6\ 14\ 7)$; $P_4 = (6\ 14\ 8)$; $P_5 = (6\ 13\ 9)$; $P_6 = (6\ 14\ 12\ 10)$; $P_7 = (6\ 13\ 11)$; $P_8 = (6\ 14\ 12)$.

7.
$P_1 = (7\ 12\ 13\ 1)$; $P_2 = (7\ 12\ 13\ 2)$; $P_3 = (7\ 14\ 6)$; $P_4 = (7\ 12\ 13\ 8)$; $P_5 = (7\ 12\ 9)$; $P_6 = (7\ 12\ 10)$; $P_7 = (7\ 12\ 11)$.

8.
$P_1 = (8\ 9\ 13\ 1)$; $P_2 = (8\ 9\ 2)$; $P_3 = (8\ 9\ 3)$; $P_4 = (8\ 9\ 4)$; $P_5 = (8\ 14\ 6)$; $P_6 = (8\ 9\ 12\ 7)$.

9.
$P_1 = (9\ 13\ 8\ 1)$; $P_2 = (9\ 13\ 8\ 2)$; $P_3 = (9\ 13\ 8\ 3)$; $P_4 = (9\ 13\ 8\ 4)$; $P_5 = (9\ 13\ 8\ 5)$; $P_6 = (9\ 13\ 6)$; $P_7 = (9\ 13\ 12\ 7)$; $P_8 = (9\ 13\ 8)$; $P_9 = (9\ 13\ 11)$; $P_{10} = (9\ 13\ 12)$.

10.
$P_1 = (10\ 8\ 9\ 1)$; $P_2 = (10\ 8\ 9\ 2)$; $P_3 = (10\ 8\ 14\ 6)$; $P_4 = (10\ 8\ 13\ 12\ 7)$; $P_5 = (10\ 8\ 13\ 11)$; $P_6 = (10\ 8\ 13\ 12)$.

11.
$P_1 = (11\ 13\ 14\ 1)$; $P_2 = (11\ 13\ 15\ 2)$; $P_3 = (11\ 13\ 14\ 6)$; $P_4 = (11\ 13\ 12\ 7)$; $P_5 = (11\ 13\ 8)$; $P_6 = (11\ 13\ 12)$.

12.
$P_1 = (12\ 8\ 9\ 1)$; $P_2 = (12\ 8\ 9\ 2)$; $P_3 = (12\ 8\ 9\ 3)$; $P_4 = (12\ 8\ 9\ 4)$; $P_5 = (12\ 7\ 14\ 6)$; $P_6 = (12\ 7\ 14\ 13)$.

13. $P_1 = (13\ 9\ 12\ 1)$; $P_2 = (13\ 6\ 2)$; $P_3 = (13\ 9\ 3)$; $P_4 = (13\ 9\ 4)$; $P_5 = (13\ 9\ 12\ 7)$.

14.

$P_1 = ( 14\ 7\ 12\ 1 ); P_2 = ( 14\ 7\ 12\ 2 ); P_3 = ( 14\ 7\ 12\ 9 ); P_4 = ( 14\ 7\ 12\ 9\ 8 ); P_5 = ( 14\ 6\ 13\ 11 );$
$P_6 = ( 14\ 7\ 12\ 9\ 10 ).$

15.

$P_1 = ( 15\ 14\ 13\ 6 ); P_2 = ( 15\ 14\ 13\ 12\ 7 ); P_3 = ( 15\ 14\ 13\ 9 ); P_4 = ( 15\ 14\ 13\ 11 ); P_5 = ( 15\ 14\ 13 ).$

$j = 1$
$( 9\ 1 )( 1\ 3 ) = ( 9\ 13 ) : \underline{156}_4; ( 9\ 1 )( 1\ 4 ) = ( 9\ 4 ) : \underline{155}_4; ( 10\ 1 )( 1\ 5 ) = ( 10\ 5 ) : \underline{156}_4;$
$( 14\ 1 )( 1\ 5 ) = ( 14\ 5 ) : \underline{154}_4; ( 14\ 1 )( 1\ 4 ) = ( 14\ 4 ) : \underline{155}_4; ( 14\ 1 )( 1\ 3 ) = ( 14\ 3 ) : \underline{156}_4.$

$j = 6$
$( 7\ 6 )( 6\ 1 ) = ( 7\ 1 ) : \underline{106}_3; ( 7\ 6 )( 6\ 2 ) = ( 7\ 2 ) : \underline{106}_3; ( 8\ 6 )( 6\ 2 ) : \underline{171}_5;$
$( 8\ 6 )( 6\ 4 ) = ( 8\ 4 ) : \underline{184}_5; ( 8\ 6 )( 6\ 5 ) = ( 8\ 5 ) : \underline{188}_5; ( 8\ 6 )( 6\ 3 ) = ( 8\ 3 ) : \underline{186}_5;$
$( 9\ 6 )( 6\ 14 ) : \underline{79}_3; ( 9\ 6 )( 6\ 1 ) = ( 9\ 1 ) : \underline{98}_3; ( 9\ 6 )( 6\ 2 ) = ( 9\ 2 ) : \underline{98}_3;$
$( 11\ 6 )( 6\ 1 ) = ( 11\ 1 ) : \underline{127}_4; ( 11\ 6 )( 6\ 2 ) = ( 11\ 2 ) : \underline{127}_4; ( 11\ 6 )( 6\ 3 ) = ( 11\ 3 ) : \underline{150}_4;$
$( 11\ 6 )( 6\ 4 ) = ( 11\ 4 ) : \underline{148}_4; ( 11\ 6 )( 6\ 5 ) = ( 11\ 5 ) : \underline{152}_4; ( 12\ 6 )( 6\ 1 ) = ( 12\ 1 ) : \underline{132}_4;$
$( 12\ 6 )( 6\ 5 ) = ( 12\ 5 ) : \underline{149}_4; ( 12\ 6 )( 6\ 13 ) = ( 12\ 13 ) : \underline{106}_4; ( 15\ 6 )( 6\ 1 ) = ( 15\ 1 ) : \underline{137}_4;$
$( 15\ 6 )( 6\ 2 ) = ( 15\ 2 ) : \underline{137}_4; ( 15\ 6 )( 6\ 4 ) = ( 15\ 4 ) : \underline{150}_4; ( 15\ 6 )( 6\ 5 ) = ( 15\ 5 ) : \underline{154}_4;$
$( 15\ 6 )( 6\ 8 ) = ( 15\ 8 ) : \underline{139}_4.$

$j = 7$
$( 6\ 7 )( 7\ 1 ) = ( 6\ 1 ) : \underline{110}_3; ( 6\ 7 )( 7\ 2 ) = ( 6\ 2 ) : \underline{110}_3; ( 9\ 7 )( 7\ 1 ) : \underline{129}_4;$
$( 9\ 7 )( 7\ 2 ) : \underline{129}_4; ( 9\ 7 )( 7\ 3 ) = ( 9\ 3 ) : \underline{129}_4; ( 10\ 7 )( 7\ 1 ) = ( 10\ 1 ) : \underline{166}_5;$
$( 10\ 7 )( 7\ 2 ) = ( 10\ 2 ) : \underline{166}_5; ( 10\ 7 )( 7\ 3 ) = ( 10\ 3 ) : \underline{166}_5; ( 10\ 7 )( 7\ 4 ) = ( 10\ 4 ) : \underline{183}_5;$
$( 10\ 7 )( 7\ 5 ) = ( 10\ 5 ) : \underline{183}_5; ( 10\ 7 )( 7\ 12 ) = ( 10\ 12 ) : \underline{144}_5; ( 11\ 7 )( 7\ 1 ) = ( 11\ 1 ) : \underline{129}_4;$
$( 11\ 7 )( 7\ 2 ) = ( 11\ 2 ) : \underline{129}_4; ( 11\ 7 )( 7\ 3 ) = ( 11\ 3 ) : \underline{129}_4; ( 11\ 7 )( 7\ 4 ) = ( 11\ 4 ) : \underline{146}_4;$
$( 11\ 7 )( 7\ 5 ) = ( 11\ 5 ) : \underline{146}_4; ( 11\ 7 )( 7\ 14 ) = ( 11\ 14 ) : \underline{108}_4; ( 11\ 7 )( 7\ 15 ) = ( 11\ 15 ) : \underline{116}_4;$
$( 13\ 7 )( 7\ 1 ) = ( 13\ 1 ) : \underline{129}_4; ( 13\ 7 )( 7\ 2 ) = ( 13\ 2 ) : \underline{129}_4; ( 13\ 7 )( 7\ 3 ) = ( 13\ 3 ) : \underline{129}_4;$
$( 13\ 7 )( 7\ 4 ) = ( 13\ 4 ) : \underline{146}_4; ( 13\ 7 )( 7\ 5 ) = ( 13\ 5 ) : \underline{146}_4; ( 15\ 7 )( 7\ 1 ) = ( 15\ 1 ) : \underline{165}_5;$
$( 15\ 7 )( 7\ 2 ) = ( 15\ 2 ) : \underline{165}_5; ( 15\ 7 )( 7\ 3 ) = ( 15\ 3 ) : \underline{165}_5; ( 15\ 7 )( 7\ 4 ) = ( 15\ 4 ) : \underline{182}_5;$
$( 15\ 7 )( 7\ 5 ) = ( 15\ 5 ) : \underline{182}_5.$

$j = 8$

$(6\ 8)(8\ 1) = (6\ 1): \underline{101}_3;\ (6\ 8)(8\ 2) = (6\ 2): \underline{100}_3;\ (6\ 8)(8\ 3) = (6\ 3): \underline{101}_3;$
$(6\ 8)(8\ 4) = (6\ 4): \underline{102}_3;\ (6\ 8)(8\ 5) = (6\ 5): \underline{103}_3;\ (6\ 8)(8\ 9) = (6\ 9): 80_3;$
$(7\ 8)(8\ 3) = (7\ 3): \underline{133}_4;\ (7\ 8)(8\ 4) = (7\ 4): \underline{134}_4;\ (7\ 8)(8\ 5) = (7\ 5): \underline{135}_4;$
$(7\ 8)(8\ 10) = (7\ 10): \underline{116}_4;\ (11\ 8)(8\ 4) = (11\ 4): \underline{105}_3;\ (11\ 8)(8\ 5) = (11\ 5): \underline{106}_3;$
$(11\ 8)(8\ 10): 87_3;\ (13\ 8)(8\ 4) = (13\ 4): \underline{102}_3;\ (13\ 8)(8\ 5) = (13\ 5): \underline{106}_3;$
$(14\ 8)(8\ 1) = (14\ 1): \underline{157}_5;\ (14\ 8)(8\ 2) = (14\ 2): \underline{156}_5;\ (14\ 8)(8\ 3) = (14\ 3): \underline{157}_5;$
$(14\ 8)(8\ 4) = (14\ 4): \underline{158}_5;\ (14\ 8)(8\ 5) = (14\ 5): \underline{159}_5;\ (15\ 8)(8\ 10) = (15\ 10): 170_5.$

$j = 9$

$(6\ 9)(9\ 1) = (6\ 1): \underline{97}_3;\ (6\ 9)(9\ 2) = (6\ 2): \underline{97}_3;\ (6\ 9)(9\ 3) = (6\ 3): \underline{97}_3;$
$(6\ 9)(9\ 4) = (6\ 4): \underline{97}_3;\ (6\ 9)(9\ 10) = (6\ 10): 99_3;\ (7\ 9)(9\ 1) = (7\ 1): \underline{129}_4;$
$(7\ 9)(9\ 2) = (7\ 2): \underline{129}_4;\ (7\ 9)(9\ 3) = (7\ 3): \underline{129}_4;\ (7\ 9)(9\ 4) = (7\ 4): \underline{130}_4;$
$(7\ 9)(9\ 5) = (7\ 5): \underline{154}_4;\ (7\ 9)(9\ 8) = (7\ 8): \underline{108}_4;\ (7\ 9)(9\ 10) = (7\ 10): \underline{111}_4;$
$(11\ 9)(9\ 4) = (11\ 4): \underline{101}_3;\ (11\ 9)(9\ 8) = (11\ 8): \underline{79}_3;\ (11\ 9)(9\ 10) = (11\ 10): 82_3;$
$(11\ 9)(9\ 12) = (11\ 12): 81_3;\ (15\ 9)(9\ 4) = (15\ 4): \underline{137}_4;\ (15\ 9)(9\ 8) = (15\ 8): \underline{115}_4;$
$(15\ 9)(9\ 10) = (15\ 10): 118_4;\ (15\ 9)(9\ 12) = (15\ 12): \underline{117}_4.$

$j = 11$

$(6\ 11)(11\ 15) = (6\ 15): 84_3;\ (7\ 11)(11\ 13) = (7\ 13): 82_3;\ (9\ 11)(11\ 12) = (9\ 12): 82_3;$
$(9\ 11)(11\ 15) = (9\ 15): 87_3;\ (10\ 11)(11\ 15) = (10\ 15): 122_4;\ (15\ 11)(11\ 4) = (15\ 4): 138_4;$
$(15\ 11)(11\ 12) = (15\ 12): 118_4.$

$j = 12$

$(6\ 12)(12\ 7) = (6\ 7): \underline{147}_5;\ (8\ 12)(12\ 1) = (8\ 1): \underline{159}_5;\ (8\ 12)(12\ 2) = (8\ 2): \underline{159}_5;$
$(8\ 12)(12\ 5) = (8\ 5): \underline{180}_5;\ (8\ 12)(12\ 6) = (8\ 6): \underline{146}_5;\ (8\ 12)(12\ 7) = (8\ 7): \underline{135}_5;$
$(10\ 12)(12\ 15) = (10\ 15): 174_6;\ (11\ 12)(12\ 7) = (11\ 7): \underline{107}_4;\ (11\ 12)(12\ 15) = (11\ 15): 111_4.$

$j = 13$

$(7\ 13)(13\ 6) = (7\ 6): \underline{105}_4.$

$j = 14$

$(6\ 14)(14\ 7) = (6\ 7): \underline{112}_4;\ (9\ 14)(14\ 7) = (9\ 7): \underline{106}_4;\ (11\ 14)(14\ 1) = (11\ 1): \underline{161}_5;$
$(11\ 14)(14\ 6) = (11\ 6): \underline{138}_5.$

*j = 15*

( 6 15 )( 15 12 ) = ( 6 12 ) : $\underline{114}_4$; ( 6 15 )( 15 14 ) = ( 6 14 ) : $\underline{116}_4$; ( 9 15 )( 15 11 ) = ( 9 11 ) : $\underline{146}_5$;

( 10 15 )( 15 5 ) = ( 10 5 ) : $\underline{250}_7$; ( 11 15 )( 15 3 ) = ( 11 3 ) : $\underline{147}_5$; ( 11 15 )( 15 14 ) = ( 11 14 ) : $\underline{143}_5$.

$P_{30}$

|    | 6   | 7     | 8   | 9       | 10 | 11 | 12  | 13  | 14 | 15  |     |
|----|-----|-------|-----|---------|----|----|-----|-----|----|-----|-----|
| 6  |     |       | 9   | *13*    |    | 13 |     | 6   | *15* | 11 | 6   |
| P  |     |       | 5   | a-6     |    | 6  |     | a   | 6  | 6   | a=6 |
| NP |     |       |     |         |    |    |     |     |    |     |     |
|    |     |       |     |         |    |    |     |     |    |     |     |
| 7  | *13* |      | 13  | *13*    |    | 12 | 7   | 11  |    |     | 7   |
| P  | 6   |       | 5   | a-4 4,5,6 |    | a  | a   | a   |    |     | a=6 |
| NP |     |       |     |         |    |    |     |     |    |     |     |
|    |     |       |     |         |    |    |     |     |    |     |     |
| 8  | 12  | 12    |     | 8       |    | 13 | 11  | 9   |    |     | 8   |
| P  | 6   | 3,4   |     | a       |    | a  | a   | a   |    |     | a=6 |
| NP |     |       |     |         |    |    |     |     |    |     |     |
|    |     |       |     |         |    |    |     |     |    |     |     |
| 9  | 13  | 14    | 13  |         | 8  | *15* |   | 9   | 6  | 14  | 9   |
| P  | a-4 | a-4,5 | 4   |         | 4  | 5  |     | a,  | a-4 | 5  | a=5 |
| NP |     |       |     |         |    |    |     |     |    |     |     |
|    |     |       |     |         |    |    |     |     |    |     |     |
| 10 |     | 14    | 10  |         |    |    | 7   | 8   | 13 | 12  | 10  |
| P  |     | a     | a   |         |    |    | 2,5 | a   | a  | 2,5 | a=5 |
| NP |     |       |     |         |    |    |     |     |    |     |     |
|    | 11   | *14* | 12  | 13  | 13  |     |        | 9     | 11    | 15    | 12    | 11    |
|    | P    | 6    | 2   | 5   | a-5 |     |        | a-4,5 | a     | 1     | 1,3,6 | a=6   |
|    | NP   |      |     |     |     |     |        |       |       |       |       |       |
|    | 12   | 14   | 12  |     |     |     |        |       |       | 7     |       | 12    |
|    | P    | a    | a   |     |     |     |        |       |       | a     |       | a=2   |
|    | NP   |      |     |     |     |     |        |       |       |       |       |       |
|    | 13   |      | 12  | 9   | 13  |     |        | 9     |       |       |       | 13    |
|    | P    |      | a-4,5| 4  | a   |     |        | a-4,5 |       |       |       | a=5   |
|    | NP   |      |     |     |     |     |        |       |       |       |       |       |
|    | 14   |      | 14  | 9   | 12  |     |        | 7     |       |       |       | 14    |
|    | P    |      | a   | a   | a   |     |        | a     |       |       |       | a=5   |
|    | NP   |      |     |     |     |     |        |       |       |       |       |       |
|    | 15   |      | 12  | *9* | 13  |     | 13     | 11    | 14    | 15    |       | 15    |
|    | P    |      | a-3,4| 4  | 4   |     | a-4    | a-3,4 | a     | a     |       | a=5   |
|    | NP   |      |     |     |     |     |        |       |       |       |       |       |
|    |      | 6    | 7   | 8   | 9   | 10  | 11     | 12    | 13    | 14    | 15    |       |

M(30)

|    | 6 | 7 | 8 | 9 | 10 | 11 | 12 | 13 | 14 | 15 |    |
|----|---|---|---|---|----|----|----|----|----|----|----|
| 6  | ∞ |   |   |   |    |    |    | 23 | <u>*116*<sub>4</sub></u> | *84*<sub>3</sub> | 6 |
| 7  | <u>*105*<sub>4</sub></u> | ∞ | *112*<sub>4</sub> | *108*<sub>4</sub> |    | *56*<sub>2</sub> | 26 | *82*<sub>3</sub> | 27 |    | 7 |
| 8  | <u>*146*<sub>5</sub></u> | *135*<sub>5</sub> | ∞ | 27 |    | *79*<sub>3</sub> | *109*<sub>4</sub> | *53*<sub>2</sub> |    |    | 8 |
| 9  | *49*<sub>2</sub> | *129*<sub>4</sub> | 27 | ∞ |    | <u>*146*<sub>5</sub></u> |    | 26 | *79*<sub>3</sub> | *111*<sub>4</sub> | 9 |
| 10 |   | *118*<sub>4</sub> | 31 | 30 | ∞ |    | *144*<sub>5</sub> | *61*<sub>2</sub> | *91*<sub>3</sub> | *174*<sub>6</sub> | 10 |
| 11 | <u>*173*<sub>6</sub></u> | *107*<sub>4</sub> |   | *52*<sub>2</sub> |    | ∞ | *81*<sub>3</sub> | 26 | *143*<sub>5</sub> | *111*<sub>4</sub> | 11 |
| 12 | *83*<sub>3</sub> | 26 | 31 | 29 |    | 30 | ∞ |   | *53*<sub>2</sub> |    | 12 |
| 13 | 23 | *81*<sub>3</sub> |   | 26 |    | 26 | *55*<sub>2</sub> | ∞ |   |    | 13 |
| 14 | 30 | 27 | *169*<sub>4</sub> | *82*<sub>3</sub> |    |    | *53*<sub>2</sub> |   | ∞ |    | 14 |
| 15 |   | *144*<sub>5</sub> | <u>*115*<sub>4</sub></u> | *88*<sub>3</sub> | 74 | *88*<sub>3</sub> | *118*<sub>4</sub> | *62*<sub>2</sub> | 32 | ∞ | 15 |
|    | 6 | 7 | 8 | 9 | 10 | 11 | 12 | 13 | 14 | 15 |    |

*j = 6*

( 8 6 )( 6 14 ) = ( 8 14 ) : $176_6$.

*j = 11*

( 9 11 )( 11 12 ) = ( 9 12 ) : $176_6$.

*j = 12*

( 9 12 )( 12 7 ) = $\underline{202_7}$; ( 15 12 )( 12 6 ) = ( 15 6 ) : $\underline{183_6}$; ( 15 12 )( 12 7 ) = ( 15 7 ) : $\underline{172_6}$.

*j = 14*

( 6 14 )( 14 7 ) = ( 6 7 ) : $\underline{114_5}$; ( 6 14 )( 14 12 ) = ( 6 12 ) : $\underline{152_5}$; ( 11 14 )( 14 6 ) : $\underline{173_6}$; ( 11 14 )( 14 7 ) = ( 11 7 ) : $\underline{170_6}$.

$P_{45}$

|    | 6  | 7  | 8 | 9 | 10 | 11 | 12 | 13 | 14 | 15 |     |
|----|----|----|---|---|----|----|----|----|----|----|-----|
| 6  | ∞  | *14* |   |   |    | 13 | *14* | 6  | *15* | 11 | 6   |
| P  |    | 1  |   |   |    | a  | 2  | a  | a  | a  | a=2 |
| NP |    |    |   |   |    |    |    |    |    |    |     |
|    |    |    |   |   |    |    |    |    |    |    |     |
| 9  | 13 | *12* |   |   | *11* | 15 | 11 | 9  | 6  | 14 | 9   |
| P  | a  | 1  |   |   | 2  | a  | a  | a  | a  | a  | a=2 |
| NP |    |    |   |   |    |    |    |    |    |    |     |
|    |    |    |   |   |    |    |    |    |    |    |     |
| 11 | *14* | *14* |   | 13 |    |    | 9  | 11 | 15 | 12 | 11  |
| P  | 1  | 2  |   | a |    |    | a  | a  | a  | a  | a=2 |
| NP |    |    |   |   |    |    |    |    |    |    |     |

| | | | | | | | | | | |
|---|---|---|---|---|---|---|---|---|---|---|
| 15 | *12* | *12* | 9 | 13 | | | 8 | 14 | 15 | 15 |
| P | 1 | 2 | a | a | | | a | a | a | a=2 |
| NP | | | | | | | | | | |
| | | | | | | | | | | |
| | 6 | 7 | 8 | 9 | 10 | 11 | 12 | 13 | 14 | 15 |

M(45)

| | 6 | 7 | 8 | 9 | 10 | 11 | 12 | 13 | 14 | 15 | |
|---|---|---|---|---|---|---|---|---|---|---|---|
| 6 | ∞ | $\underline{143_5}$ | | | | $49_2$ | $\underline{152_5}$ | 23 | $116_4$ | $84_3$ | 6 |
| 9 | $49_2$ | $\underline{202_7}$ | 27 | ∞ | $\underline{182_6}$ | $146_5$ | $176_6$ | 26 | $79_3$ | $111_4$ | 9 |
| 11 | $\underline{173_6}$ | $\underline{170_6}$ | | $52_2$ | | ∞ | $81_3$ | 26 | $143_5$ | $111_4$ | 11 |
| 15 | $\underline{183_6}$ | $\underline{172_6}$ | $115_4$ | $88_3$ | | $88_3$ | $146_5$ | $62_2$ | 32 | ∞ | 15 |
| | 6 | 7 | 8 | 9 | 10 | 11 | 12 | 13 | 14 | 15 | |

6. $P_1 = (6\ 13\ 11\ 15\ 14\ 7): 143_5;\ P_2 = (6\ 13\ 11\ 15\ 14\ 12): 152_5.$

9. $P_1 = (9\ 13\ 6\ 14\ 15\ 11\ 12\ 7): 202_7;\ P_2 = (9\ 13\ 6\ 14\ 15\ 11\ 10): 182_6.$

11. $P_1 = (11\ 13\ 9\ 12\ 15\ 14\ 6): 173_6;\ P_2 = (11\ 13\ 9\ 12\ 15\ 14\ 7): 170_6.$

15. $P_1 = (15\ 14\ 13\ 9\ 8\ 12\ 6): 183_6;\ P_2 = (15\ 14\ 13\ 9\ 8\ 12\ 7): 172_6.$

$j = 7$
$(6\ 7)(7\ 12) = (6\ 12): 169_6.$
$j = 10$
$(9\ 10)(10\ 8) = (9\ 8): \underline{213_7}.$
$j = 12$
$(6\ 12)(12\ 8) = (6\ 8): \underline{200_7};\ (6\ 12)(12\ 9) = (6\ 9): \underline{198_7};\ (6\ 12)(12\ 10) = (6\ 10): \underline{204_7}.$

$P_{45}$

| | 6 | 7 | 8 | 9 | 10 | 11 | 12 | 13 | 14 | 15 | |
|---|---|---|---|---|---|---|---|---|---|---|---|

| 6 | ∞ | 14 | *12* | *12* | *12* | 13 | 7 | 6 | *15* | 11 | 6 |
|---|---|---|---|---|---|---|---|---|---|---|---|
| P |   | a | 1 | 2 | 3 | a | a | a | a | a | a=3 |
| NP |   |   |   |   |   |   |   |   |   |   |   |
|   |   |   |   |   |   |   |   |   |   |   |   |
| 9 | 13 |   | *10* |   | 11 | 15 | 11 | 9 | 6 | 14 | 9 |
| P | a |   | 1 |   | A | a | a | a | a | a | a=1 |
| NP |   |   |   |   |   |   |   |   |   |   |   |
|   |   |   |   |   |   |   |   |   |   |   |   |
|   | 6 | 7 | 8 | 9 | 10 | 11 | 12 | 13 | 14 | 15 |   |

M(45)

|   | 6 | 7 | 8 | 9 | 10 | 11 | 12 | 13 | 14 | 15 |   |
|---|---|---|---|---|---|---|---|---|---|---|---|
| 6 | ∞ | $143_5$ | $\underline{200_7}$ | $\underline{198_7}$ | $\underline{204_7}$ | $49_2$ | $169_6$ | 23 | $116_4$ | $84_3$ | 6 |
| 9 | $49_2$ |   | $\underline{213_7}$ | ∞ | $182_6$ | $146_5$ |   | 26 | $79_3$ | $111_4$ | 9 |
|   | 6 | 7 | 8 | 9 | 10 | 11 | 12 | 13 | 14 | 15 |   |

6. $P_1 = (6\ 13\ 11\ 15\ 14\ 12\ 9\ 8) : 225_8$.

9. $P_1 = (9\ 13\ 6\ 14\ 15\ 11\ 10\ 8\ 12\ 7) : 270_9$.

$P_{45}$

|   | 6 | 7 | 8 | 9 | 10 | 11 | 12 | 13 | 14 | 15 |   |
|---|---|---|---|---|---|---|---|---|---|---|---|
| 6 | ∞ | 14 | *9* | 12 |   | 13 | 7 | 6 | *15* | 11 | 6 |
| P |   | a | 2 | 2 |   | a | a | a | a | a | a=3 |
| NP |   |   |   |   |   |   |   |   |   |   |   |
|   |   |   |   |   |   |   |   |   |   |   |   |
| 9 | 13 | *12* | 10 |   | 11 | 15 | 8 | 9 | 6 | 14 | 9 |
| P | a | 1 | a |   | a | *a* | *a* | *a* | *a* | a | a=1 |
| NP |   |   |   |   |   |   |   |   |   |   |   |
|   |   |   |   |   |   |   |   |   |   |   |   |

|   | 6 | 7 | 8 | 9 | 10 | 11 | 12 | 13 | 14 | 15 |   |
|---|---|---|---|---|----|----|----|----|----|----|---|

M(45)

|   | 6 | 7 | 8 | 9 | 10 | 11 | 12 | 13 | 14 | 15 |   |
|---|---|---|---|---|----|----|----|----|----|----|---|
| 6 | ∞ | $143_5$ | $\underline{225_8}$ | $198_7$ |  | $49_2$ | $169_6$ | 23 | $116_4$ | $84_3$ | 6 |
| 9 | $49_2$ | $\underline{270_9}$ | $213_7$ | ∞ | $182_6$ | $146_5$ | $244_8$ | 26 | $79_3$ | $111_4$ | 9 |
|   | 6 | 7 | 8 | 9 | 10 | 11 | 12 | 13 | 14 | 15 |   |

$j = 8$
$(9\ 8)(8\ 12) = (9\ 12) : 244_8$.
$j = 9$
$(6\ 9)(9\ 8) = (6\ 8) : \underline{225_8}$; $(6\ 9)(9\ 10) = (6\ 10) : 228_8$.
$j = 12$
$(9\ 12)(12\ 7) = (9\ 7) : 270_9$.

$P_{60}$

|   | 6 | 7 | 8 | 9 | 10 | 11 | 12 | 13 | 14 | 15 |   |
|---|---|---|---|---|----|----|----|----|----|----|---|
| 6 | ∞ | 14 | *9* | 12 |   | 13 | 7 | 6 | *15* | 11 | 6 |
| P |   | a | 2 | 2 |   | A | a | a | a | a | a=1 |
| NP |   |   |   |   |   |   |   |   |   |   |   |
|   |   |   |   |   |   |   |   |   |   |   |   |
| 9 | 13 | *12* | 10 |   | 11 | 15 | 8 | 9 | 6 | 14 | 9 |
| P | a | 1 | a |   | a |   | a | a | a | a | a=1 |
| NP |   |   |   |   |   |   |   |   |   |   |   |
|   |   |   |   |   |   |   |   |   |   |   |   |
|   | 6 | 7 | 8 | 9 | 10 | 11 | 12 | 13 | 14 | 15 |   |

|   | 6 | 7 | 8 | 9 | 10 | 11 | 12 | 13 | 14 | 15 |   |
|---|---|---|---|---|---|---|---|---|---|---|---|
| 6 | $\infty$ | $143_5$ | $\underline{225_8}$ | $198_7$ |  | $49_2$ | $169_6$ | 23 | $116_4$ | $84_3$ | 6 |
| 9 | $49_2$ | $\underline{270_9}$ | $213_7$ | $\infty$ | $182_6$ | $146_5$ | $244_8$ | 26 | $79_3$ | $111_4$ | 9 |
|   | 6 | 7 | 8 | 9 | 10 | 11 | 12 | 13 | 14 | 15 |   |

<div align="center">M(60)</div>

6. $P_1 = ($ 6 13 11 15 14 7 12 9 8 $) : 225_8$.

9. $P_1 = ($ 9 13 6 14 15 11 10 8 12 7 $) : 270_9$.

$j = 8$

$(6\ 8)(8\ 10) = (6\ 10) : 256_9$.

6. $P_{61} = ($ 6 13 11 15 14 7 12 9 8 10 $) : 256_9$.

9. $P_{91} = ($ 9 13 6 14 15 11 10 8 12 7 $) : 270_9$.

From this point on, we must extend the paths so that we obtain a 15-cycle. Checking column 6, we note that the entries in (1 6) and (2 6) are considerably smaller than (i 6), i = 3,4,5. Thus, using $P_{61}$ as an initial path, we must search for the smallest-valued path from 10 to 6. Each interior node must belong to {1,2,3,4,5}. Checking $T_{UPPERBOUND}$, we note that the following subpath occurs:
(6 1 3 5 2 4 10) = (10 4 2 5 3 1 6). Thus, $T_{OPT-APPROX} = ($ 6 13 11 15 14 7 12 9 8 10 4 2 5 3 1 6 $)$.
NOT NECESSARILY CORRECT!! For one thing, we must obtain the smallest-valued path from 10 to 6.

Example 10.

|   | | | | | M | | | | | |   |
|---|---|---|---|---|---|---|---|---|---|---|---|
|   | 1 | 2 | 3 | 4 | 5 | 6 | 7 | 8 | 9 | 10 |   |
| 1 | ∞ | 53 | 52 | 51 | 50 | 49 | 48 | 48 | 48 | 48 | 1 |
| 2 | 53 | ∞ | 66 | 66 | 50 | 49 | 48 | 47 | 48 | 49 | 2 |
| 3 | 52 | 66 | ∞ | 62 | 65 | 64 | 48 | 48 | 48 | 49 | 3 |
| 4 | 51 | 66 | 62 | ∞ | 63 | 62 | 65 | 49 | 49 | 49 | 4 |
| 5 | 50 | 50 | 65 | 63 | ∞ | 66 | 65 | 50 | 73 | 76 | 5 |
| 6 | 49 | 49 | 64 | 62 | 66 | ∞ | 65 | 51 | 70 | 77 | 6 |
| 7 | 48 | 48 | 48 | 65 | 65 | 65 | ∞ | 73 | 74 | 71 | 7 |
| 8 | 48 | 47 | 48 | 49 | 50 | 51 | 73 | ∞ | 27 | 31 | 8 |
| 9 | 48 | 48 | 48 | 49 | 73 | 70 | 74 | 27 | ∞ | 30 | 9 |
| 10 | 48 | 49 | 49 | 49 | 76 | 77 | 71 | 31 | 30 | ∞ | 10 |
|   | 1 | 2 | 3 | 4 | 5 | 6 | 7 | 8 | 9 | 10 |   |

MIN(M)

|   | 1 | 2 | 3 | 4 | 5 | 6 | 7 | 8 | 9 | 10 |   |
|---|---|---|---|---|---|---|---|---|---|----|---|
| 1 | 7 | 8 | 9 | 10 | 6 | 5 | 4 | 3 | 2 |   | 1 |
| 2 | 8 | 7 | 9 | 6 | 10 | 5 | 1 | 3 | 4 |   | 2 |
| 3 | 7 | 8 | 9 | 10 | 1 | 4 | 6 | 5 | 2 |   | 3 |
| 4 | 8 | 9 | 10 | 1 | 6 | 3 | 5 | 7 | 2 |   | 4 |
| 5 | 1 | 2 | 8 | 4 | 3 | 7 | 6 | 9 | 10 |   | 5 |
| 6 | 1 | 2 | 8 | 4 | 3 | 7 | 5 | 9 | 10 |   | 6 |
| 7 | 1 | 2 | 3 | 4 | 5 | 6 | 10 | 8 | 9 |   | 7 |
| 8 | 9 | 10 | 2 | 1 | 3 | 4 | 5 | 6 | 7 |   | 8 |
| 9 | 8 | 10 | 1 | 2 | 3 | 4 | 6 | 5 | 7 |   | 9 |
| 10 | 9 | 8 | 1 | 2 | 3 | 4 | 7 | 5 | 6 |   | 10 |
|   | 1 | 2 | 3 | 4 | 5 | 6 | 7 | 8 | 9 | 10 |   |

.

There is a weakness if we apply FWK directly to M. In obtaining an $n$-cycle, we wish to obtain the smallest possible value of each entry with respect to each entry in the row. If we use M itself, as the number of arcs in a path increases, generally the average arc-value of the path increases because we have eliminated the column entries that are available for extending the path. One way to address this problem is to subtract the smallest entry in each row from the remaining entries to form the matrix M'. This helps reveal the relative value of the remaining entries compared to an entry chosen for the $n$-cycle. Assume that the total of the values subtracted is T. Since an $n$-cycle contains precisely one entry from each row and column, every $n$-cycle in M' corresponds to a unique $n$-cycle in M. Furthermore, this cycle, C', obtained in M', has the property that |C'| + T = |C| where C is its corresponding cycle in M. Thus, an optimal $n$-cycle in M' corresponds to one in M and vice-versa. Since we are subtracting the same amount from each entry of a row in M, MIN(M) = MIN(M').

M'

|    | 1  | 2  | 3  | 4  | 5  | 6  | 7  | 8  | 9  | 10 |    |
|----|----|----|----|----|----|----|----|----|----|----|----|
| 1  | ∞  | 5  | 4  | 3  | 2  | 1  | 0  | 0  | 0  | 0  | 1  |
| 2  | 6  | ∞  | 19 | 19 | 3  | 2  | 1  | 0  | 1  | 2  | 2  |
| 3  | 4  | 18 | ∞  | 14 | 17 | 16 | 0  | 0  | 0  | 1  | 3  |
| 4  | 2  | 17 | 13 | ∞  | 14 | 13 | 16 | 0  | 0  | 0  | 4  |
| 5  | 0  | 0  | 15 | 13 | ∞  | 16 | 15 | 0  | 23 | 26 | 5  |
| 6  | 0  | 0  | 15 | 13 | 17 | ∞  | 16 | 2  | 21 | 28 | 6  |
| 7  | 0  | 0  | 0  | 17 | 17 | 17 | ∞  | 25 | 26 | 23 | 7  |
| 8  | 21 | 20 | 21 | 22 | 23 | 24 | 46 | ∞  | 0  | 4  | 8  |
| 9  | 21 | 21 | 21 | 22 | 46 | 43 | 47 | 0  | ∞  | 3  | 9  |
| 10 | 18 | 19 | 19 | 19 | 46 | 47 | 41 | 1  | 0  | ∞  | 10 |
|    | 1  | 2  | 3  | 4  | 5  | 6  | 7  | 8  | 9  | 10 |    |

$D_0^{-1}$

|    | 2   | 3   | 4   | 5   | 6   | 7   | 8   | 9   | 10  | 1   |    |
|----|-----|-----|-----|-----|-----|-----|-----|-----|-----|-----|----|
|    | 1   | 2   | 3   | 4   | 5   | 6   | 7   | 8   | 9   | 10  |    |
| 1  | 0   | -1  | 3   | -3  | -4  | -5  | -5  | -5  | -5  | ∞   | 1  |
| 2  | ∞   | 0   | 0   | -16 | -17 | -18 | -19 | -18 | -17 | -13 | 2  |
| 3  | 4   | ∞   | 0   | 3   | 2   | -14 | -14 | -14 | -13 | -10 | 3  |
| 4  | 3   | -1  | ∞   | 0   | -1  | 2   | -14 | -14 | -14 | -12 | 4  |
| 5  | -16 | -1  | -3  | ∞   | 0   | -1  | -16 | 7   | 10  | -16 | 5  |
| 6  | -16 | -1  | -3  | 1   | ∞   | 0   | -14 | 5   | 12  | -16 | 6  |
| 7  | -25 | -25 | -8  | -8  | -8  | ∞   | 0   | 1   | -2  | -25 | 7  |
| 8  | 20  | 21  | 22  | 23  | 24  | 46  | ∞   | 0   | 4   | 21  | 8  |
| 9  | 18  | 18  | 19  | 43  | 40  | 44  | -3  | ∞   | 0   | 18  | 9  |
| 10 | 1   | 1   | 1   | 28  | 29  | 23  | -17 | -18 | ∞   | 0   | 10 |
|    | 1   | 2   | 3   | 4   | 5   | 6   | 7   | 8   | 9   | 10  |    |

Each parallel processor obtains the [log(n)] + 1 smallest-valued negative entries in each row. As soon as a parallel processor obtains a negative cycle, it signals the derangement arcs obtained from these cycle arcs. Each parallel processor then changes the columns of the rows it is working on if it is necessary to do so, i.e., it permutes the columns of the rows it is currently working on, replacing columns that the new arcs indicate must be replaced. For simplicity, we use the [log(n)] + 1 smallest negative values of the matrices in our example until we can no longer obtain a negative cycle. Then we must work on each row individually.

NEGATIVE CYCLES

|  | 2 | 3 | 4 | 5 | 6 | 7 | 8 | 9 | 10 | 1 |
|---|---|---|---|---|---|---|---|---|---|---|
|  | 1 | 2 | 3 | 4 | 5 | 6 | 7 | 8 | 9 | 10 |
| (7 1):<br>-25 | 7<br>-25 |  |  |  |  | 1<br>-30 | 9!<br>-62 | 10<br>-64 | 8<br>-60 | 6<br>-46 |
|  | 1 | 2 | 3 | 4 | 5 | 6 | 7 | 8 | 9 | 10 |

Our smallest-valued negative cycle: (7 1 6 10 8 9): -62.

We obtain the following new arcs: (7 2), (1 7), (9 8), (8 10), (10 9), (6 1) .

$$D_1^{-1}$$

|   | 7 | 3 | 4 | 5 | 6 | 1 | 2 | 10 | 8 | 9 |   |
|---|---|---|---|---|---|---|---|----|---|---|---|
|   | 1 | 2 | 3 | 4 | 5 | 6 | 7 | 8  | 9 | 10|   |
| 1 | 0 | -1 | 13 | 7 | 6 | ∞ | 0 | 0 | 5 | 0 | 1 |
| 2 | -18 | 0 | 18 | 2 | 1 | 5 | ∞ | -17 | -1 | -18 | 2 |
| 3 | -14 | ∞ | 0 | 3 | 2 | -10 | 4 | -13 | -14 | -14 | 3 |
| 4 | 2 | -1 | ∞ | 0 | -1 | -12 | 3 | -14 | -14 | -14 | 4 |
| 5 | -1 | -1 | -3 | ∞ | 0 | -16 | -16 | 10 | -16 | 7 | 5 |
| 6 | 0 | -1 | 13 | 17 | ∞ | 0 | -16 | 12 | 2 | 5 | 6 |
| 7 | ∞ | 0 | 42 | 42 | 42 | 25 | 0 | 23 | 50 | 26 | 7 |
| 8 | 42 | 21 | 18 | 19 | 20 | 17 | 20 | 0 | ∞ | -4 | 8 |
| 9 | 44 | 18 | 22 | 46 | 43 | 21 | 18 | 3 | 0 | ∞ | 9 |
| 10 | 41 | 1 | 18 | 45 | 46 | 17 | 1 | ∞ | 0 | 0 | 10 |
|   | 1 | 2 | 3 | 4 | 5 | 6 | 7 | 8 | 9 | 10 |   |

NEGATIVE CYCLES

|   | 7 | 3 | 4 | 5 | 6 | 1 | 2 | 10 | 8 | 9 |
|---|---|---|---|---|---|---|---|----|---|---|
|   | 1 | 2 | 3 | 4 | 5 | 6 | 7 | 8 | 9 | 10 |
| (2 9): -18 |  |  |  |  |  |  |  |  | 2! -18 |  |
| (2 8): -17 |  |  |  |  |  |  |  | 2! -17 |  |  |
| (5 6): -16 |  | 7 -32 | 10 -32 |  | <u>3! -30</u> | 5 -16 | 6 -32 |  |  | 2 -50 |
|   | 1 | 2 | 3 | 4 | 5 | 6 | 7 | 8 | 9 | 10 |

Our smallest-valued negative cycle: (5 6 7 2 10 3): -30

We obtain the following new arcs: (7 3), (10 4), (3 6), (5 1), (6 2), (2 9).

$$D_2^{-1}$$

|   | 7 | 9 | 6 | 5 | 1 | 2 | 3 | 10 | 8 | 4 |   |
|---|---|---|---|---|---|---|---|----|---|---|---|
|   | 1 | 2 | 3 | 4 | 5 | 6 | 7 | 8  | 9 | 10|   |
| 1 | 0 | 0 | 6 | 7 | ∞ | 0 | -1| 0  | 5 | 13| 1 |
| 2 | 0 | 0 | 19| 20| 23| ∞ | 18| 1  | 17| 36| 2 |
| 3 | -16| -16 | 0 | 1 | -12 | 2 | ∞ | -15 | -16 | -2 | 3 |
| 4 | 2 | -14 | -1 | 0 | -12 | 3 | -1 | -14 | -14 | ∞ | 4 |
| 5 | 15 | 23 | 16 | ∞ | 0 | 0 | 15 | 26 | 0 | 13 | 5 |
| 6 | 16 | 21 | ∞ | 33 | 16 | 0 | 15 | 28 | 18 | 29 | 6 |
| 7 | ∞ | 26 | 42 | 42 | 25 | 0 | 0 | 23 | 50 | 42 | 7 |
| 8 |   | -4 | 20 | 19 | 17 | 20 | 21 | 4 | ∞ | 18 | 8 |
| 9 | 44 | ∞ | 43 | 46 | 21 | 18 | 18 | 0 | 0 | 22 | 9 |
| 10 | 23 | -18 | 28 | 27 | -1 | -17 | -17 | ∞ | -18 | 0 | 10 |
|   | 1 | 2 | 3 | 4 | 5 | 6 | 7 | 8 | 9 | 10 |   |

NEGATIVE CYCLES

|   | 7 | 9 | 6 | 5 | 1 | 2 | 3 | 10 | 8 | 4 |
|---|---|---|---|---|---|---|---|----|---|---|
|   | 1 | 2 | 3 | 4 | 5 | 6 | 7 | 8 | 9 | 10 |
| (10 2): -18 | 2 -18 | 10 -18 |   |   | 6 -3 | 7 -19 | 1 -19 | 9 -3 | 5! -3 |   |
| (10 6): -17 |   |   |   |   |   | 10 -17 | 6! -2 |   |   |   |
| (10 7): -17 | 6 -1 |   |   |   |   | 7 -17 | 10 -17 | 1! -1 |   |   |
| (3 2): -16 | 2 -16 | 3 -16 | <u>1</u> <u>-10</u> |   |   |   |   |   |   |   |
|   | 1 | 2 | 3 | 4 | 5 | 6 | 7 | 8 | 9 | 10 |

A negatively-valued cycle: (3 2 1): -10

We obtain the following new arcs: (2 7), (3 9), (1 6).

$$D_3^{-1}$$

|   | 6 | 7 | 9 | 5 | 1 | 2 | 3 | 10 | 8 | 4 |   |
|---|---|---|---|---|---|---|---|----|---|---|---|
|   | 1 | 2 | 3 | 4 | 5 | 6 | 7 | 8  | 9 | 10|   |
| 1 | 0 | -6 | 0 | 1 | ∞ | -6 | -7 | -6 | -1 | 7 | 1 |
| 2 | 19 | 0 | 0 | 20 | 23 | ∞ | 18 | 1 | 17 | 36 | 2 |
| 3 | 16 | 0 | -16 | 17 | 4 | 18 | ∞ | 1 | 0 | 14 | 3 |
| 4 | -1 | 2 | -14 | 0 | -12 | 3 | -1 | -14 | -14 | ∞ | 4 |
| 5 | 16 | 15 | 23 | ∞ | 0 | 0 | 15 | 26 | 0 | 13 | 5 |
| 6 | ∞ | 16 | 21 | 33 | 16 | 0 | 15 | 28 | 18 | 29 | 6 |
| 7 | 42 | ∞ | 26 | 42 | 25 | 0 | 0 | 23 | 50 | 42 | 7 |
| 8 | 20 | 42 | -4 | 19 | 17 | 20 | 21 | 0 | ∞ | 18 | 8 |
| 9 | 43 | 44 | ∞ | 46 | 21 | 18 | 18 | 0 | 0 | 22 | 9 |
| 10 | 28 | 23 | -18 | 27 | -1 | -17 | -17 | ∞ | -18 | 0 | 10 |
|   | 1 | 2 | 3 | 4 | 5 | 6 | 7 | 8 | 9 | 10 |   |

NEGATIVE CYCLES

|   | 7 | 9 | 6 | 5 | 1 | 2 | 3 | 10 | 8 | 4 |
|---|---|---|---|---|---|---|---|----|---|---|
|   | 1 | 2 | 3 | 4 | 5 | 6 | 7 | 8  | 9 | 10 |
| (1 7): -7 |   |   |   |   |   | 7! -7 | 1 -7 |   |   |   |
| (1 2): -6 |   | 1 -6 |   |   |   |   |   | 2! -5 |   |   |
| (4 3): -14 |   | 3 -14 | 4 -14 |   |   |   |   | 2! -14 |   |   |
| (10 3): -18 |   |   | 10 -18 |   |   |   |   | 9! -18 | 3 -18 |   |
|   | 1 | 2 | 3 | 4 | 5 | 6 | 7 | 8 | 9 | 10 |

|        | 7 | 9 | 6 | 5 | 1 | 2 | 3 | 10 | 8 | 4 |
|--------|---|---|---|---|---|---|---|----|---|---|
|        | 1 | 2 | 3 | 4 | 5 | 6 | 7 | 8  | 9 | 10 |
| (10 9): -18 |   | 3! -1 | 5 -1 |   | 8 -1 |   |   | 9 -18 | 10 -18 |   |
| (10 6): -17 |   |   |   |   |   | 10 -17 | 6! -2 |   |   |   |
|        | 1 | 2 | 3 | 4 | 5 | 6 | 7 | 8  | 9 | 10 |

We cannot proceed further with Prologue. Since $D_3$ is a 10-cycle, let
$T_{UPPERBOUND} = D_3 = (1\ 6\ 2\ 7\ 3\ 9\ 8\ 10\ 4\ 5):39$. Each of the arcs (8 10), (10 4), (4 5) has a value greater than 3.9. Let us now represent $T_{UPPERBOUND}$ in terms of entries in M.
$(1^{49}\ 6^{49}\ 2^{48}\ 7^{49}\ 3^{48}\ 9^{27}\ 8^{31}\ 10^{49}\ 4^{63}\ 5^{50}) = (9\ 8\ 10\ 4\ 5\ 1\ 6\ 2\ 7\ 3)$. Its value is 463; its average arc-value is 46.3 . Thus, all of the rows of M that have no value less than the average arc-value of $T_{UPPERBOUND}$ occur after 10. Since M is symmetric, there generally exists a set of arcs at the end of $T_{UPPERBOUND}$ each of which has a value greater than or equal to its average arc-value. One reason why this often occurs is that as we progress in constructing subpaths of a possible $T_{OPT}$, a subpath with a considerably smaller average arc-value than that of $T_{UPPERBOUND}$ is constructed, leaving latitude for the values of the remaining arcs. Now suppose M contains $r$ rows, each of whose entries has no value less than the average arc-value of $T_{UPPERBOUND}$. Generally, $T_{OPT}$ is the same whether we apply FWK to M or M'. Consider the case if we apply FWK to M. Suppose $(a\ b\ c)$ is a subpath of $T_{OPT}$ in which no entry of row $b$ of M has a value less than $|\frac{T_{UPPERBOUND}}{n}|$. Since M is symmetric, $|(a\ b)| = |(b\ a)|$. Thus, each arc has a value greater than or equal to the average arc-value of $T_{UPPERBOUND}$. Let the initial node of $T_{OPT}$ be a determining node. If entries from each of these rows occur at the end of $T_{OPT}$, it would have at least $r+1$ entries each of whose arc-values isn't less than that of $T_{UPPERBOUND}$. Now suppose one such arc occurs separately from the remaining $r-1$. As we demonstrated using $(a\ b\ c)$, we would then have at least $r+2$ entries of $T_{OPT}$ each of whose arc-values isn't less than that of $T_{UPPERBOUND}$. Now consider the general case. Assume that we start FWK at a determining node of an $n$-cycle $T$. From our previous remarks, most (almost all?) of arcs $\{(a\ b)\}$ satisfying $|(a\ b)| \geq \frac{|T_{UPPERBOUND}|}{n}$ occur at the end of our construction of $T$. Although this is not an iron-clad rule, never-the-less, it appears to be a helpful pragmatic tool to use when applying FWK. Only two arcs of $T_{UPPERBOUND}$ with respect to M have values less than its average arc-value. For this reason and since NEGCYCLE will be discussed more completely in the new book, we won't use it in this case.

P(10a)

|      | 1    | 2   | 3   | 4     | 5     | 6 | 7 | 8   | 9   | 10 |
|------|------|-----|-----|-------|-------|---|---|-----|-----|----|
| 1(1) | $1_2$ | 6!  |     |       |       | 1 |   |     |     |    |
| 1(2) | $0_2$ |     | 7!  |       |       |   | 1 |     |     |    |
| 1(3) | $0_2$ |     |     |       |       |   |   | 10! |     | 1  |
| 1(4) | $0_2$ |     |     |       |       |   |   |     | 10! | 1  |
| 2(1) | 7!   | $1_2$ |   |       |       |   | 2 |     |     |    |
| 2(2) |      | $1_2$ | 7! |       |       |   | 2 |     |     |    |
| 3(1) | 7!   |     | $1_2$ |     |       |   | 3 |     |     |    |
| 3(2) |      | 7!  | $1_2$ |     |       |   | 3 |     |     |    |
| 3(3) |      |     | $1_2$ |     |       |   |   | 10! |     | 3  |
| 3(4) |      |     | $1_2$ |     |       |   |   |     | 10! | 3  |
| 4(1) | 4    | 5!  |     | $4_3$ | 1     |   |   |     |     |    |
| 4(2) | 4    | 7!  |     | $2_3$ |       |   | 1 |     |     |    |
| 4(3) | 4    |     | 7!  | $2_3$ |       |   | 1 |     |     |    |
| 4(4) |      |     |     | $0_2$ |       |   |   |     | 10! | 4  |
| 5(1) | 5    | 6!  |     |       | $1_3$ | 1 |   |     |     |    |
| 5(2) | 5    | 7!  |     |       | $0_3$ |   | 1 |     |     |    |
| 5(3) | 5    | 7!  | 1   |       | $4_4$ |   | 3 |     |     |    |
| 5(4) | 5    |     |     |       | $0_3$ |   |   |     | 10! | 1  |
| 5(5) | 7!   | 5   |     |       | $1_3$ |   | 2 |     |     |    |
| 5(6) |      | 5   | 7!  |       | $1_3$ |   | 2 |     |     |    |
|      | 1    | 2   | 3   | 4     | 5     | 6 | 7 | 8   | 9   | 10 |

|       | 1  | 2  | 3 | 4 | 5 | 6     | 7     | 8   | 9   | 10 |
|-------|----|----|---|---|---|-------|-------|-----|-----|----|
| 6(1)  | 6  |    |   | 1 |   | $4_4$ |       | 10! |     | 4  |
| 6(2)  | 6  |    |   | 1 |   | $3_4$ |       |     | 10! | 4  |
| 6(3)  | 6  | 5! |   |   | 1 | $2_3$ |       |     |     |    |
| 6(4)  | 6  | 7! |   |   |   | $0_3$ | 1     |     |     |    |
| 6(5)  | 6  |    |   |   |   | $1_3$ |       | 10! |     | 1  |
| 6(6)  | 6  |    |   |   |   | $0_3$ |       |     | 10! | 1  |
| 6(7)  | 7! | 6  |   |   |   | $1_3$ | 2     |     |     |    |
| 7(1)  | 7  | 6! |   |   |   | 1     | $1_3$ |     |     |    |
| 7(2)  | 7  |    |   |   |   |       | $1_3$ | 10! |     | 1  |
| 7(3)  | 7  |    |   |   |   |       | $0_3$ |     | 10! | 1  |
| 10(1) |    |    |   |   |   |       | $0_2$ | 9!  | 10  |    |
|       | 1  | 2  | 3 | 4 | 5 | 6     | 7     | 8   | 9   | 10 |

P(10)(b)

|      | 1    | 2    | 3   | 4    | 5    | 6    | 7    | 8   | 9   | 10 |
|------|------|------|-----|------|------|------|------|-----|-----|----|
| 1(1) | $3_2$ |      |     | 1    |      |      |      | 4   | 10! | 8  |
| 1(2) | $3_3$ |      |     | 1    |      |      |      | 9!  | 4   |    |
| 1(3) | $7_4$ |      |     | 1    |      |      |      | 10! | 4   | 9  |
| 1(4) | $3_3$ |      |     | 1    |      |      |      |     | 10! | 4  |
| 1(5) | $4_3$ |      |     | 1    |      |      |      | 10! |     | 4  |
| 1(6) | $4_3$ |      |     |      |      |      |      | 10! | 1   | 9  |
| 1(7) | $6_4$ |      |     |      | 1    |      |      | 5   | 10! | 8  |
| 1(8) | $2_2$ | 5    |     |      | 1    |      |      |     |     |    |
| 2(1) | 6!   | $1_2$ |     |      |      | 2    |      |     |     |    |
|      |      | $2_2$ |     |      |      |      |      |     | 10! | 2  |
|      |      | $7_4$ |     |      | 2    |      |      | 5   | 10! | 8  |
|      | 5!   | $3_2$ |     |      | 2    |      |      |     |     |    |
|      |      | $5_3$ |     |      |      |      |      | 10! | 2   | 9  |
|      |      | $8_4$ |     |      | 2    |      |      | 6   | 10! | 8  |
|      | 7!   | $1_2$ |     |      |      |      | 2    |     |     |    |
| 3(1) | 7!   |      | $0_3$ |      |      |      | 3    |     |     |    |
| 4(1) | 4    |      |     | $6_4$ |      |      |      | 1   | 10! | 8  |
|      | 4    |      |     | $2_3$ |      |      |      | 9!  | 1   |    |
|      | 4    |      |     | $2_3$ |      |      |      |     | 10! | 1  |
|      | 4    |      |     | $3_3$ |      |      |      | 10! |     | 1  |
|      |      |      |     | $4_3$ |      |      |      | 4   | 10! | 8  |
|      | 4    | 7!   | 1   | $6_4$ |      |      | 3    |     |     |    |
|      | 4    | 7!   |     | $2_3$ |      |      | 1    |     |     |    |
|      | 4    |      | 7!  | $2_3$ |      |      | 1    |     |     |    |
| 5(1) | 2!   | 5    |     |      | $6_2$ |      |      |     |     |    |
| 5()  | 6!   | 5    |     |      | $2_3$ | 2    |      |     |     |    |
|      |      | 5    |     |      | $3_3$ |      |      | 10! |     | 2  |
|      |      | 5    |     |      | $2_3$ |      |      |     | 10! | 2  |
|      | 5    | 1    | 7!  |      | $6_4$ |      | 2    |     |     |    |
|      | 5    | 1    |     |      | $6_4$ |      |      | 9!  | 2   |    |
|      | 5    | 1    |     |      | $7_4$ |      |      |     | 10! | 2  |
|      | 6    | 1    | 7!  |      |      | $6_4$ | 2    |     |     |    |
|      | 6    |      | 7!  |      |      | $0_3$ | 1    |     |     |    |
|      | 5!   | 6    |     |      | 2    | $2_3$ |      |     |     |    |
|      |      | 6    |     |      |      | $2_3$ |      |     | 10! | 2  |
|      | 7    | 1    |     |      |      |      | $7_4$ |     | 10! | 2  |
|      | 7    | 6!   |     |      |      | 1    | $1_3$ |     |     |    |
|      | 2!   | 7    |     |      |      | $6_2$ |      |     |     |    |
|      | 5!   | 7    |     |      | 2    |      | $3_3$ |     |     |    |
|      | 1    | 2    | 3   | 4    | 5    | 6    | 7    | 8   | 9   | 10 |

| | 1 | 2 | 3 | 4 | 5 | 6 | 7 | 8 | 9 | 10 |
|---|---|---|---|---|---|---|---|---|---|---|
| | 6! | 7 | | | | 2 | $2_3$ | | | |
| | | 7 | | | | | $1_3$ | 9! | 2 | |
| | | 7 | | | | | $2_3$ | | 10! | 2 |
| | 1 | 2 | 3 | 4 | 5 | 6 | 7 | 8 | 9 | 10 |

P(20)(b)

| | 1 | 2 | 3 | 4 | 5 | 6 | 7 | 8 | 9 | 10 |
|---|---|---|---|---|---|---|---|---|---|---|
| 1(1) | $3_4$ | 5 | 7! | | 1 | | 2 | | | |
| 1(2) | $6_5$ | 5 | | | 1 | | | 2 | 10! | 8 |
| 1(3) | $3_4$ | 5 | | | 1 | | | 9! | 2 | |
| 1(4) | $7_5$ | 5 | | | 1 | | | 10! | 2 | 9 |
| 1(5) | $5_4$ | 5 | | | 1 | | | 10! | | 2 |
| 1(6) | $4_4$ | 5 | | | 1 | | | | 10! | 2 |
| 2(1) | 6 | $9_6$ | | 1 | | 2 | | 4 | 10! | 8 |
| 2(2) | 6 | $5_5$ | | 1 | | 2 | | 9! | 4 | |
| 2(3) | 6 | $6_5$ | | 1 | | 2 | | 10! | 4 | 9 |
| 2(4) | 6 | $5_5$ | | 1 | | 2 | | | 10! | 4 |
| 2(5) | 6 | $6_5$ | | 1 | | 2 | | 10! | | 4 |
| 2(6) | 6 | $2_4$ | 7! | | | 2 | 1 | | | |
| 2(7) | 6 | $6_5$ | | | | 2 | | 1 | 10! | 8 |
| 2(8) | 6 | $2_4$ | | | | 2 | | 9! | 1 | |
| 2(9) | 6 | $6_5$ | | | | 2 | | 10! | 1 | 9 |
| 2(10) | 6 | $2_4$ | | | | 2 | | | 10! | 1 |
| 2(11) | 6 | $3_4$ | | | | 2 | | 10! | | 1 |
| 2(12) | | $2_3$ | | | | | | 9! | 10 | 2 |
| 2(14) | 5 | $11_6$ | 1 | | 2 | | | 3 | 10! | 8 |
| 2(15) | 5 | $7_5$ | 1 | | 2 | | | 9! | 3 | |
| 2(16) | 5 | $11_6$ | 1 | | 2 | | | 10! | 3 | 9 |
| 2(17) | 5 | $8_5$ | 1 | | 2 | | | | 10! | 3 |
| 2(18) | 5 | $9_5$ | 1 | | 2 | | | 10! | | 3 |
| 2(19) | 5 | $10_6$ | | 1 | 2 | | | 4 | 10! | 8 |
| 2(20) | 5 | $6_5$ | | 1 | 2 | | | 9! | 4 | |
| 2(21) | 5 | $10_6$ | | 1 | 2 | | | 10! | 4 | 9 |
| 2(22) | 5 | $6_5$ | | 1 | 2 | | | | 10! | 4 |
| 2(23) | 5 | $10_6$ | | | 2 | 1 | | 6 | 10! | 8 |
| 2(24) | 5 | $3_4$ | 7! | | 2 | | 1 | | | |
| | 1 | 2 | 3 | 4 | 5 | 6 | 7 | 8 | 9 | 10 |

|       | 1 | 2    | 3   | 4    | 5 | 6 | 7 | 8   | 9   | 10 |
|-------|---|------|-----|------|---|---|---|-----|-----|----|
| 2(25) | 5 | $7_5$ |     |      | 2 |   |   | 1   | 10! | 8  |
| 2(26) | 5 | $3_4$ |     |      | 2 |   |   | 9!  | 1   |    |
| 2(27) | 5 | $7_5$ |     |      | 2 |   |   | 10! | 1   | 9  |
| 2(28) | 5 | $3_4$ |     |      | 2 |   |   |     | 10! | 1  |
| 2(29) | 5 | $4_4$ |     |      | 2 |   |   | 10! |     | 1  |
| 2(30) |   | $5_3$ |     |      |   |   |   | 10! | 2   | 9  |
| 2(31) |   | $8_4$ |     |      |   | 2 |   | 6   | 10! | 8  |
| 2(32) | 7 | $2_4$ |     |      |   |   | 2 | 10! |     | 1  |
| 2(33) | 7 | $25_7$ | 10! |     | 1 |   | 2 | 5   | 8   | 9  |
| 2(34) | 7 | $25_7$ |     | 10!  | 1 |   | 2 | 5   | 8   | 9  |
| 2(35) | 7 | $7_6$ |     |      | 1 |   | 2 | 5   | 10! | 8  |
| 2(36) | 7 | $26_7$ | 10! | 1   |   |   | 2 | 4   | 8   | 9  |
| 2(37) | 7 | $7_6$ |     | 1    |   |   | 2 | 4   | 10! | 8  |
| 2(38) | 7 | $4_5$ |     | 1    |   |   | 2 | 9!  | 4   |    |
| 2(39) | 7 | $8_6$ |     | 1    |   |   | 2 | 10! | 4   | 9  |
| 2(40) | 7 | $4_5$ |     | 1    |   |   | 2 |     | 10! | 4  |
| 2(41) | 7 | $5_5$ |     | 1    |   |   | 2 | 10! |     | 4  |
| 2(42) | 7 | $8_6$ |     |      |   | 1 | 2 | 6   | 10! | 8  |
| 2(43) | 7 | $4_5$ |     | 1    |   |   | 2 | 9!  | 4   |    |
| 2(44) | 7 | $8_6$ |     | 1    |   |   | 2 | 10! | 4   | 9  |
| 2(45) | 7 | $5_5$ |     | 1    |   |   | 2 | 10! |     | 4  |
| 2(46) | 7 | $4_5$ |     | 1    |   |   | 2 |     | 10! | 4  |
| 2(47) | 7 | $23_7$ | 10! |     | 1 |   | 2 | 5   | 8   | 9  |
| 3(1)  | 7 |      | $7_6$ | 1  |   |   | 3 | 4   | 10! | 8  |
| 3(2)  | 7 |      | $3_5$ | 1  |   |   | 3 | 9!  | 4   |    |
| 3(3)  | 7 |      | $7_6$ | 1  |   |   | 3 | 10! | 4   | 9  |
| 3(4)  | 7 |      | $3_5$ | 1  |   |   | 3 |     | 10! | 4  |
| 3(5)  | 7 |      | $4_5$ | 1  |   |   | 3 | 10! |     | 4  |
| 3(6)  | 7 |      | $4_5$ |    |   |   | 3 | 1   | 10! | 8  |
| 3(7)  | 7 |      | $18_7$ | 6! |   | 1 | 3 |     |     |    |
| 3(8)  | 7 |      | $6_6$ |    | 1 |   | 3 | 5   | 10! | 8  |
| 4(1)  | 4 | 7    | 1   | $13_8$ | 2 |   | 3 | 5   | 10! | 8  |
| 4(2)  | 4 | 7    | 1   | $14_8$ |   | 2 | 3 | 6   | 10! | 8  |
| 4(3)  | 4 | 7    | 1   | $10_7$ |   |   | 3 | 2   | 10! | 8  |
| 4(4)  | 4 | 7    | 1   | $7_6$  |   |   | 3 | 9!  | 2   |    |
| 4(5)  | 4 | 7    | 1   | $11_7$ |   |   | 3 | 10! | 2   | 9  |
| 4(6)  | 4 | 7    | 1   | $9_6$  |   |   | 3 | 10! |     | 2  |

|  | 1 | 2 | 3 | 4 | 5 | 6 | 7 | 8 | 9 | 10 |
|---|---|---|---|---|---|---|---|---|---|---|
| 4(7) | 4 | 7 | 1 | $8_6$ |  |  | 3 |  | 10! | 2 |
| 4(8) | 4 |  | 7 | $10_6$ |  |  | 1 | 3 | 10! | 8 |
| 4(9) | 4 |  | 7 | $6_5$ |  |  | 1 | 9! | 3 |  |
| 4(10) | 4 |  | 7 | $6_6$ |  |  | 1 | 10! | 3 | 9 |
| 4(11) | 4 | 7 |  | $3_5$ |  |  | 1 | 9! | 2 |  |
| 4(12) | 4 | 7 |  | $7_6$ |  |  | 1 | 10! | 2 | 9 |
| 4(13) | 4 | 7 |  | $4_5$ |  |  | 1 |  | 10! | 2 |
| 4(14) | 4 | 7 |  | $5_5$ |  |  | 1 | 10! |  | 2 |
| 5(1) | 6 | 5 | 1 | 10! | $28_8$ | 2 |  | 3 | 8 | 9 |
| 5(2) | 6 | 5 | 1 |  | $10_7$ | 2 |  | 3 | 10! | 8 |
| 5(3) | 6 | 5 | 1 |  | $6_6$ | 2 |  | 9! | 3 |  |
| 5(4) | 6 | 5 | 1 |  | $10_7$ | 2 |  | 10! | 3 | 9 |
| 5(5) | 6 | 5 | 1 |  | $8_6$ | 2 |  | 10! |  | 3 |
| 5(6) | 6 | 5 | 1 |  | $7_6$ | 2 |  |  | 10! | 3 |
| 5(7) | 6 | 5 | 4! | 1 | $18_5$ | 2 |  |  |  |  |
| 5(8) | 6 | 5 | 10! | 1 | $27_8$ | 2 |  | 4 | 8 | 9 |
| 5(9) | 6 | 5 |  | 1 | $9_7$ | 2 |  | 4 | 10! | 8 |
| 5(10) | 6 | 5 |  | 1 | $5_6$ | 2 |  | 9! | 4 |  |
| 5(11) | 6 | 5 |  | 1 | $9_7$ | 2 |  | 10! | 4 | 9 |
| 5(12) | 6 | 5 |  | 1 | $6_6$ | 2 |  | 10! |  | 4 |
| 5(13) | 6 | 5 |  | 1 | $5_6$ | 2 |  |  | 10! | 4 |
| 5(14) | 6 | 5 |  |  | $6_6$ | 2 |  | 1 | 10! | 8 |
| 5(15) |  | 5 |  |  | $2_4$ |  |  | 9! | 10 | 2 |
| 5(16) | 5 | 1 | 7 |  | $10_7$ |  | 2 | 3 | 10! | 8 |
| 5(17) | 5 | 1 | 7 |  | $6_6$ |  | 2 | 9! | 3 |  |
| 5(18) | 5 | 1 | 7 |  | $10_7$ |  | 2 | 10! | 3 | 9 |
| 5(19) | 5 | 1 |  |  | $7_5$ |  |  | 9! | 10 | 2 |
| 5(20) | 2 | 5 | 1 |  | $14_6$ |  |  | 3 | 10! | 8 |
| 5(21) | 2 | 5 |  | 1 | $13_6$ |  |  | 4 | 10! | 8 |
| 5(22) | 2 | 5 |  | 1 | $9_5$ |  |  | 9! | 4 |  |
| 5(23) | 2 | 5 |  | 1 | $12_6$ |  |  | 10! | 4 | 9 |
| 5(24) | 2 | 5 |  | 1 | $9_5$ |  |  |  | 10! | 4 |
| 5(25) | 2 | 5 |  | 1 | $10_5$ |  |  | 10! |  | 4 |
| 5(26) | 2 | 5 |  |  | $13_6$ | 1 |  | 6 | 10! | 8 |
| 5(27) | 2 | 5 | 7! |  | $6_4$ |  | 1 |  |  |  |
| 5(28) | 2 | 5 |  |  | $10_5$ |  |  | 1 | 10! | 8 |
|  | 1 | 2 | 3 | 4 | 5 | 6 | 7 | 8 | 9 | 10 |

|       | 1 | 2 | 3   | 4   | 5 | 6       | 7 | 8   | 9   | 10 |
|-------|---|---|-----|-----|---|---------|---|-----|-----|----|
| 5(29) | 2 | 5 |     |     | $6_4$ |   |   | 9!  | 1   |    |
| 5(30) | 2 | 5 |     |     | $10_5$ |  |   | 10! | 1   | 9  |
| 5(31) | 2 | 5 |     |     | $6_4$ |   |   |     | 10! | 1  |
| 5(32) | 2 | 5 |     |     | $7_4$ |   |   | 10! |     | 1  |
| 6(1)  |   | 6 |     |     |   | $2_4$   |   | 9!  | 10  | 2  |
| 6(2)  | 5 | 6 | 1   | 10! | 2 | $29_8$  |   | 3   | 8   | 9  |
| 6(3)  | 5 | 6 | 1   |     | 2 | $11_7$  |   | 3   | 10! | 8  |
| 6(4)  | 5 | 6 | 1   |     | 2 | $7_6$   |   | 9!  | 3   |    |
| 6(5)  | 5 | 6 | 1   |     | 2 | $11_7$  |   | 10! | 3   | 9  |
| 6(6)  | 5 | 6 | 1   |     | 2 | $9_6$   |   | 10! |     | 3  |
| 6(7)  | 5 | 6 | 1   |     | 2 | $8_6$   |   |     | 10! | 3  |
| 6(8)  | 5 | 6 | 7!  |     | 2 | $3_5$   | 1 |     |     |    |
| 6(9)  | 5 | 6 | 10! |     | 2 | $25_7$  |   | 1   | 8   | 9  |
| 6(10) | 5 | 6 |     | 10! | 2 | $25_7$  |   | 1   | 8   | 9  |
| 6(11) | 5 | 6 |     |     | 2 | $7_6$   |   | 1   | 10! | 8  |
| 6(12) | 5 | 6 |     |     | 2 | $3_5$   |   | 9!  | 1   |    |
| 6(13) | 5 | 6 |     |     | 2 | $7_6$   |   | 10! | 1   | 9  |
| 6(14) | 5 | 6 |     |     | 2 | $4_5$   |   | 10! |     | 1  |
| 6(15) | 5 | 6 |     |     | 2 | $3_5$   |   |     | 10! | 1  |
| 6(16) | 5 | 6 | 10! | 1   | 2 | $28_8$  |   | 4   | 8   | 9  |
| 6(17) | 5 | 6 |     | 1   | 2 | $10_7$  |   | 4   | 10! | 8  |
| 6(18) | 5 | 6 |     | 1   | 2 | $6_6$   |   | 9!  | 4   |    |
| 6(19) | 5 | 6 |     | 1   | 2 | $10_7$  |   | 10! | 4   | 9  |
| 6(20) | 5 | 6 |     | 1   | 2 | $7_6$   |   | 10! |     | 4  |
| 6(21) | 5 | 6 |     | 1   | 2 | $6_6$   |   |     | 10! | 4  |
| 6(22) | 6 | 1 | 7   | 10! |   | $28_8$  | 2 | 3   | 8   | 9  |
| 6(23) | 6 | 1 | 7   |     |   | $10_7$  | 2 | 3   | 10! | 8  |
| 6(24) | 6 | 1 | 7   |     |   | $6_6$   | 2 | 9!  | 3   |    |
| 6(25) | 6 | 1 | 7   |     |   | $10_7$  | 2 | 10! | 3   | 9  |
| 6(26) | 6 | 1 | 7   |     |   | $8_6$   | 2 | 10! |     | 3  |
| 6(27) | 6 | 1 | 7   |     |   | $7_6$   | 2 |     | 10! | 3  |
| 6(28) | 2 | 6 | 1   |     |   | $10_6$  |   | 3   | 10! | 8  |
| 6(29) | 2 | 6 | 1   |     |   | $10_5$  |   | 9!  | 3   |    |
| 6(30) | 2 | 6 | 1   |     |   | $14_6$  |   | 10! | 3   | 9  |
|       | 1 | 2 | 3   | 4   | 5 | 6       | 7 | 8   | 9   | 10 |

|       | 1 | 2   | 3 | 4   | 5 | 6       | 7       | 8   | 9   | 10 |
|-------|---|-----|---|-----|---|---------|---------|-----|-----|----|
| 6(31) | 2 | 6   | 1 |     |   | $11_5$  |         |     | 10! | 3  |
| 6(32) | 2 | 6   | 1 |     |   | $12_5$  |         | 10! |     | 3  |
| 6(33) | 2 | 6   |   |     | 1 |         |         | 5   | 10! | 8  |
| 6(34) | 2 | 6   |   |     |   | $10_5$  |         | 1   | 10! | 8  |
| 6(35) | 2 | 6   |   |     |   | $6_4$   |         | 9!  | 1   |    |
| 6(36) | 2 | 6   |   |     |   | $10_5$  |         | 10! | 1   | 9  |
| 6(37) | 2 | 6   |   |     |   | $6_4$   |         |     | 10! | 1  |
| 6(38) | 2 | 6   |   |     |   | $7_4$   |         | 10! |     | 1  |
| 6(39) | 6 | 7   |   |     |   | $2_5$   | 1       |     | 10! | 2  |
| 6(40) | 6 | 7   |   |     |   | $3_5$   | 1       | 10! |     | 2  |
| 6(41) | 6 | 7   |   |     |   | $4_6$   | 1       | 2   | 10! | 8  |
| 6(42) | 6 | 7   |   | 5!  | 2 | $16_5$  | 1       |     |     |    |
| 6(43) | 6 | 10! | 7 |     |   | $22_7$  | 1       | 3   | 8   | 9  |
| 6(44) | 6 |     | 7 | 10! |   | $22_7$  | 1       | 3   | 8   | 9  |
| 6(45) | 6 | 10! | 7 |     |   | $23_6$  | 1       | 3   |     | 8  |
| 6(46) | 6 |     | 7 | 10! |   | $23_6$  | 1       | 3   |     | 8  |
| 6(47) | 6 |     | 7 |     |   | $3_6$   | 1       | 3   | 10! | 8  |
| 7(1)  |   | 7   |   |     |   |         | $2_4$   | 9!  | 10  | 2  |
| 7(2)  | 6 | 7   | 1 | 10! |   | 2       | $28_8$  | 3   | 8   | 9  |
| 7(3)  | 6 | 7   | 1 |     |   | 2       | $10_7$  | 3   | 10! | 8  |
| 7(4)  | 6 | 7   | 1 |     |   | 2       | $6_6$   | 9!  | 3   |    |
| 7(5)  | 6 | 7   | 1 |     |   | 2       | $10_7$  | 10! | 3   | 9  |
| 7(6)  | 6 | 7   | 1 |     |   | 2       | $8_6$   | 10! |     | 3  |
| 7(7)  | 6 | 7   | 1 |     |   | 2       | $10_7$  | 10! | 3   | 9  |
| 7(9)  | 6 | 7   | 1 |     |   | 2       | $7_6$   |     | 10! | 3  |
| 7(10) | 5 | 7   | 1 | 10! | 2 |         | $10_8$  | 3   | 8   | 9  |
| 7(11) | 7 | 1   |   |     |   |         | $7_5$   | 9!  | 10  | 2  |
| 7(12) | 2 | 7   | 1 |     |   |         | $13_6$  | 3   | 10! | 8  |
| 7(13) | 2 | 7   | 1 |     |   |         | $9_5$   | 9!  | 3   |    |
| 7(14) | 2 | 7   | 1 |     |   |         | $13_6$  | 10! | 3   | 9  |
| 7(15) | 2 | 7   | 1 |     |   |         | $10_5$  |     | 10! | 3  |
| 7(16) | 2 | 7   |   | 1   |   |         | $8_5$   | 9!  | 4   |    |
| 7(17) | 2 | 7   |   | 1   |   |         | $12_6$  | 10! | 4   | 9  |
| 7(18) | 2 | 7   |   | 1   |   |         | $8_5$   |     | 10! | 4  |
| 7(19) | 2 | 7   |   | 1   |   |         | $9_5$   | 10! |     | 4  |
| 7(20) | 2 | 7   |   |     | 1 |         | $11_6$  | 5   | 10! | 8  |
|       | 1 | 2   | 3 | 4   | 5 | 6       | 7       | 8   | 9   | 10 |

We now construct $M_2$. It generally contains a much larger number of paths than $n$. Also, the average value of paths generally increases as the number of arcs in a path gets larger. The reason for this is that the number of rows and columns from which to choose entries decreases as each new arc is chosen. The smallest acceptable entries in a row are generally chosen first. Perhaps an approximation formula may be found for what occurs when entries are randomly chosen. In what follows, we use a crude method that allows us to deal with a fractional number of cases at a time. In what follows, we obtain the following numbers for each set of arcs: $(\frac{number\ of\ arcs}{n})(average\ arc\ value\ of\ T_{UPPERBOUND})$.

These values are used as upper bounds to obtain the first set of paths that we will try to extend to $n$-cycles. However, for a large value of $n$ we would have to use the fractional representation of average value. From it, we could obtain a decimal approximation that would allow us to place the latter fraction in its proper place in the table.

P(20)(a)

|       | 1    | 2     | 3      | 4  | 5 | 6 | 7 | 8   | 9   | 10 |
|-------|------|-------|--------|----|---|---|---|-----|-----|----|
| 1(1)  | $2_4$ | 6    | 7!     |    |   | 1 | 2 |     |     |    |
| 1(2)  | $2_4$ | 6    |        |    |   | 1 |   | 9!  | 2   |    |
| 1(3)  | $4_4$ | 6    |        |    |   | 1 |   | 10! |     | 2  |
| 1(4)  | $3_4$ | 6    |        |    |   | 1 |   |     | 10! | 2  |
| 1(5)  | $4_5$ |      | 7      |    |   |   | 1 | 3   | 10! | 8  |
| 1(6)  | $4_5$ |      | 7      |    |   |   | 1 | 9!  | 3   |    |
| 1(7)  | $3_4$ |      | 7      |    |   |   | 1 | 10! | 3   | 9  |
| 1(8)  | $2_4$ |      | 7      |    |   |   | 1 | 10! |     | 3  |
| 1(9)  | $1_4$ |      | 7      |    |   |   | 1 |     | 10! | 3  |
| 2(1)  | 7    | $1_4$ |        |    |   |   | 2 | 9!  | 1   |    |
| 2(2)  | 7    | $5_5$ |        |    |   |   | 2 | 10! | 1   | 9  |
| 2(3)  | 7    | $2_4$ |        |    |   |   | 2 | 10! |     | 1  |
| 2(4)  |      | $5_5$ | 7      |    |   |   | 2 | 3   | 10! | 8  |
| 2(5)  |      | $1_4$ | 7      |    |   |   | 2 | 9!  | 3   |    |
| 2(6)  |      | $2_4$ | 7      |    |   |   | 2 | 10! |     | 3  |
| 2(7)  | 7    | $5_5$ |        |    |   |   | 2 | 1   | 10! | 8  |
| 2(8)  | 3    | $9_6$ | 7      |    |   |   | 2 | 1   | 10! | 8  |
| 2(9)  | 7    | $9_6$ | 1      |    |   |   | 2 | 3   | 10! | 8  |
| 2(10) | 7    | $1_4$ |        |    |   |   | 2 |     | 10! | 1  |
| 3(1)  |      |       | $1_3$  |    |   |   |   | 9!  | 10  | 3  |
| 3(2)  |      | 7     | $4_5$  |    |   |   | 3 | 2   | 10! | 8  |
| 3(3)  |      | 7     | $1_4$  |    |   |   | 3 | 9!  | 2   |    |
| 3(4)  |      | 7     | $5_5$  |    |   |   | 3 | 10! | 2   | 9  |
| 3(5)  |      | 7     | $2_4$  |    |   |   | 3 |     | 10! | 2  |
| 3(6)  |      | 7     | $2_4$  |    |   |   | 3 | 10! |     | 2  |
| 3(7)  | 6!   | 7     | $2_4$  |    |   | 2 | 3 |     |     |    |
| 3(8)  |      | 7     | $15_4$ | 6! |   | 2 | 3 |     |     |    |
| 3(9)  |      | 7     | $10_6$ |    |   | 2 | 3 | 6   | 10! | 8  |
| 3(7)  | 5!   | 7     | $3_4$  |    | 2 |   | 3 |     |     |    |
| 3(8)  |      | 7     | $7_6$  |    | 2 |   | 3 | 5   | 10! | 8  |
| 3(9)  |      | 7     | $10_6$ |    |   | 2 | 3 | 6   | 10! | 8  |
| 3(10) | 5!   | 7     | $3_4$  |    | 2 |   | 3 |     |     |    |
| 3(11) |      | 7     | $7_6$  |    | 2 |   | 3 | 5   | 10! | 8  |
|       | 1    | 2     | 3      | 4  | 5 | 6 | 7 | 8   | 9   | 10 |

|       | 1  | 2 | 3     | 4    | 5 | 6 | 7 | 8   | 9   | 10 |
|-------|----|---|-------|------|---|---|---|-----|-----|----|
| 3(1)  |    |   | $1_3$ |      |   |   |   | 9!  | 10  | 3  |
| 3(2)  |    | 7 | $4_5$ |      |   |   | 3 | 2   | 10! | 8  |
| 3(3)  |    | 7 | $1_4$ |      |   |   | 3 | 9!  | 2   |    |
| 3(4)  |    | 7 | $5_5$ |      |   |   | 3 | 10! | 2   | 9  |
| 3(5)  |    | 7 | $2_4$ |      |   |   | 3 |     | 10! | 2  |
| 3(6)  |    | 7 | $2_4$ |      |   |   | 3 | 10! |     | 2  |
| 3(7)  | 6! | 7 | $2_4$ |      |   | 2 | 3 |     |     |    |
| 3(8)  |    | 7 | $15_4$| 6!   |   | 2 | 3 |     |     |    |
| 3(9)  |    | 7 | $10_6$|      |   | 2 | 3 | 6   | 10! | 8  |
| 3(10) | 5! | 7 | $3_4$ |      | 2 |   | 3 |     |     |    |
| 3(11) |    | 7 | $7_6$ |      | 2 |   | 3 | 5   | 10! | 8  |
| 3(12) | 7  |   | $0_5$ |      |   |   | 3 | 9!  | 10  | 1  |
| 3(13) | 7  | 6 | $17_6$| 5!   | 2 | 1 | 3 |     |     |    |
| 3(14) | 7  | 6 | $8_8$ |      | 2 | 1 | 3 | 5   | 10! | 8  |
| 3(15) |    | 7 | $4_5$ |      |   |   | 3 | 2   | 10! | 8  |
| 4(1)  | 4  | 7 | 10!   | $27_8$| 2 |   | 1 | 5   | 8   | 9  |
| 4(2)  | 4  | 7 |       | $9_7$ | 2 |   | 1 | 5   | 10! | 8  |
| 4(3)  | 4  | 7 | 6!    | $19_5$|   | 2 | 1 |     |     |    |
| 4(4)  | 4  | 7 | 9!    | $27_7$|   | 2 | 1 | 6   | 8   |    |
| 4(5)  | 4  | 7 |       | $10_7$|   | 2 | 1 | 6   | 10! | 8  |
| 4(6)  | 4  | 7 | 9!    | $23_6$|   |   | 1 | 2   | 8   |    |
| 4(7)  | 4  | 7 | 10!   | $24_7$|   |   | 1 | 2   | 8   | 9  |
| 4(8)  | 4  | 7 |       | $6_6$ |   |   | 1 | 2   | 10! | 8  |
| 4(9)  | 4  | 7 |       | $3_5$ |   |   | 1 | 9!  | 2   |    |
| 4(10) | 4  | 7 |       | $7_6$ |   |   | 1 | 10! | 2   | 9  |
| 4(11) | 4  | 7 |       | $5_5$ |   |   | 1 | 10! |     | 2  |
| 4(12) | 4  | 5 | 10!   | $30_8$| 1 | 2 |   | 6   | 8   | 9  |
| 4(13) | 4  | 5 |       | $12_7$| 1 | 2 |   | 6   | 10! | 8  |
| 4(14) | 4  | 5 | 7!    | $5_5$ | 1 |   | 2 |     |     |    |
| 4(15) | 4  | 5 | 10!   | $26_7$| 1 |   |   | 2   | 8   | 9  |
| 4(16) | 4  | 5 |       | $8_5$ | 1 |   |   | 2   | 10! | 8  |
| 4(17) | 4  | 5 |       | $5_5$ | 1 |   |   | 9!  | 2   |    |
| 4(18) | 4  | 5 |       | $9_6$ | 1 |   |   | 10! | 2   | 9  |
| 4(19) | 4  | 5 |       | $7_5$ | 1 |   |   | 10! |     | 2  |
| 4(20) | 4  | 5 |       | $6_5$ | 1 |   |   |     | 10! | 2  |
|       | 1  | 2 | 3     | 4    | 5 | 6 | 7 | 8   | 9   | 10 |

|  | 1 | 2 | 3 | 4 | 5 | 6 | 7 | 8 | 9 | 10 |
|---|---|---|---|---|---|---|---|---|---|---|
| 5(1) | 7 | 5 | 1 |  | $9_7$ |  | 2 | 3 | 10! | 8 |
| 5(2) | 7 | 5 | 1 |  | $9_7$ |  | 2 | 10! | 3 | 9 |
| 5(3) | 7 | 5 | 1 |  | $3_5$ |  | 2 | 10! |  | 3 |
| 5(4) | 7 | 5 | 1 |  | $2_5$ |  | 2 |  | 10! | 3 |
| 5(5)* | 3 | 5 | 7 |  | $7_4$ |  | 2 |  |  |  |
| 5(6) | 4! | 5 | 7 | 3 | $17_5$ |  | 2 |  |  |  |
| 5(7) |  | 5 | 7 | 3 | $19_7$ |  | 2 | 4 | 10! | 8 |
| 5(8) | 5 | 6 | 7! |  | $2_5$ | 1 | 2 |  |  |  |
| 5(9) | 5 | 6 |  | 7! | $19_5$ | 1 | 2 |  |  |  |
| 5(10) | 5 | 7 | 1 | 6! | $19_6$ | 2 | 3 |  |  |  |
| 5(11) | 5 | 7 | 1 |  | $12_8$ | 2 | 3 | 6 | 10! | 8 |
| 5(12) | 3! | 5 | 7 |  | $5_4$ |  | 2 |  |  |  |
| 5(13) | 4! | 5 | 7 | 3 | $17_5$ |  | 2 |  |  |  |
| 5(14) | 5 | 6 | 7! |  | $1_5$ | 1 | 2 |  |  |  |
| 5(15) | 5 | 6 |  | 7! | $23_8$ | 1 | 2 | 4 | 10! | 8 |
| 5(16) | 5 | 7 | 1 | 6 | $23_9$ | 2 | 3 | 4 | 10! | 8 |
| 6(1) | 7 | 6 | 1 |  |  | $9_7$ | 2 | 3 | 10! | 8 |
| 6(2) | 7 | 6 | 1 | 10! |  | $27_8$ | 2 | 3 | 8 | 9 |
| 6(3) | 7 | 6 | 1 |  |  | $6_6$ | 2 |  | 10! | 3 |
| 6(4) | 6 | 5 | 7! |  | 1 | $3_5$ | 2 |  |  |  |
| 6(5) | 6 | 5 | 10! |  | 1 | $25_7$ |  | 2 | 8 | 9 |
| 6(6) | 6 | 5 |  |  | 1 | $3_5$ |  | 9! | 2 |  |
| 6(7) | 6 | 5 |  |  | 1 | $4_5$ |  |  | 10! | 2 |
| 6(8) | 6 | 5 |  |  | 1 | $5_5$ |  | 10! |  | 2 |
| 6(9) | 6 |  |  |  |  | $0_4$ |  | 9! | 10 | 1 |
| 6(10) | 6 |  |  | 1 |  | $3_5$ |  | 9! | 10 | 4 |
| 6(11) | 6 | 7 |  |  |  | $1_5$ | 1 | 9! | 2 |  |
| 7(1) | 7 | 5 |  |  | 1 |  | $3_5$ | 9! | 2 |  |
| 7(2) | 7 | 5 |  |  | 1 |  | $7_6$ | 10! | 2 | 9 |
| 7(3) | 7 | 5 |  |  | 1 |  | $5_5$ | 10! |  | 2 |
| 7(4) | 7 | 5 |  |  | 1 |  | $4_5$ |  | 10! | 2 |
| 7(5) | 7 |  |  |  |  |  | $0_4$ | 9! | 10 | 1 |
| 7(6) | 7 |  |  | 1 |  |  | $3_5$ | 9! | 10 | 4 |
|  | 1 | 2 | 3 | 4 | 5 | 6 | 7 | 8 | 9 | 10 |

|      | 1 | 2 | 3   | 4   | 5 | 6 | 7      | 8 | 9 | 10 |
|------|---|---|-----|-----|---|---|--------|---|---|----|
| 7(7) | 7 | 6 | 10! |     |   | 1 | $23_7$ | 2 | 8 | 9  |
| 7(8) | 7 | 6 |     | 10! |   | 1 | $23_7$ | 2 | 8 | 9  |
|      | 1 | 2 | 3   | 4   | 5 | 6 | 7      | 8 | 9 | 10 |

P(20)(b)

|       | 1 | 2      | 3   | 4 | 5 | 6 | 7 | 8   | 9   | 10 |
|-------|---|--------|-----|---|---|---|---|-----|-----|----|
| 1(1)  | $3_4$ | 5 | 7! |   | 1 |   | 2 |     |     |    |
| 1(2)  | $6_5$ | 5 |    |   | 1 |   |   | 2   | 10! | 8  |
| 1(3)  | $3_4$ | 5 |    |   | 1 |   |   | 9!  | 2   |    |
| 1(4)  | $7_5$ | 5 |    |   | 1 |   |   | 10! | 2   | 9  |
| 1(5)  | $5_4$ | 5 |    |   | 1 |   |   | 10! |     | 2  |
| 1(6)  | $4_4$ | 5 |    |   | 1 |   |   |     | 10! | 2  |
| 2(1)  | 6 | $9_6$  |    | 1 |   | 2 |   | 4   | 10! | 8  |
| 2(2)  | 6 | $5_5$  |    | 1 |   | 2 |   | 9!  | 4   |    |
| 2(3)  | 6 | $6_5$  |    | 1 |   | 2 |   | 10! | 4   | 9  |
| 2(4)  | 6 | $5_5$  |    | 1 |   | 2 |   |     | 10! | 4  |
| 2(5)  | 6 | $6_5$  |    | 1 |   | 2 |   | 10! |     | 4  |
| 2(6)  | 6 | $2_4$  | 7! |   |   | 2 | 1 |     |     |    |
| 2(7)  | 6 | $6_5$  |    |   |   | 2 |   | 1   | 10! | 8  |
| 2(8)  | 6 | $2_4$  |    |   |   | 2 |   | 9!  | 1   |    |
| 2(9)  | 6 | $6_5$  |    |   |   | 2 |   | 10! | 1   | 9  |
| 2(10) | 6 | $2_4$  |    |   |   | 2 |   |     | 10! | 1  |
| 2(11) | 6 | $3_4$  |    |   |   | 2 |   | 10! |     | 1  |
| 2(12) |   | $2_3$  |    |   |   |   |   | 9!  | 10  | 2  |
| 2(14) | 5 | $11_6$ | 1  |   | 2 |   |   | 3   | 10! | 8  |
| 2(15) | 5 | $7_5$  | 1  |   | 2 |   |   | 9!  | 3   |    |
| 2(16) | 5 | $11_6$ | 1  |   | 2 |   |   | 10! | 3   | 9  |
| 2(17) | 5 | $8_5$  | 1  |   | 2 |   |   |     | 10! | 3  |
| 2(18) | 5 | $9_5$  | 1  |   | 2 |   |   | 10! |     | 3  |
| 2(19) | 5 | $10_6$ |    | 1 | 2 |   |   | 4   | 10! | 8  |
| 2(20) | 5 | $6_5$  |    | 1 | 2 |   |   | 9!  | 4   |    |
| 2(21) | 5 | $10_6$ |    | 1 | 2 |   |   | 10! | 4   | 9  |
| 2(22) | 5 | $6_5$  |    | 1 | 2 |   |   |     | 10! | 4  |
| 2(23) | 5 | $10_6$ |    |   | 2 | 1 |   | 6   | 10! | 8  |
| 2(24) | 5 | $3_4$  | 7! |   | 2 |   | 1 |     |     |    |
|       | 1 | 2      | 3  | 4 | 5 | 6 | 7 | 8   | 9   | 10 |

|       | 1 | 2    | 3   | 4     | 5 | 6 | 7 | 8   | 9   | 10 |
|-------|---|------|-----|-------|---|---|---|-----|-----|-----|
| 2(25) | 5 | $7_5$ |     |       | 2 |   |   | 1   | 10! | 8 |
| 2(26) | 5 | $3_4$ |     |       | 2 |   |   | 9!  | 1   |   |
| 2(27) | 5 | $7_5$ |     |       | 2 |   |   | 10! | 1   | 9 |
| 2(28) | 5 | $3_4$ |     |       | 2 |   |   |     | 10! | 1 |
| 2(29) | 5 | $4_4$ |     |       | 2 |   |   | 10! |     | 1 |
| 2(30) |   | $5_3$ |     |       |   |   |   | 10! | 2   | 9 |
| 2(31) |   | $8_4$ |     |       |   | 2 |   | 6   | 10! | 8 |
| 2(32) | 7 | $2_4$ |     |       |   |   | 2 | 10! |     | 1 |
| 2(33) | 7 | $25_7$ | 10! |      | 1 |   | 2 | 5   | 8   | 9 |
| 2(34) | 7 | $25_7$ |     | 10!  | 1 |   | 2 | 5   | 8   | 9 |
| 2(35) | 7 | $7_6$ |     |       | 1 |   | 2 | 5   | 10! | 8 |
| 2(36) | 7 | $26_7$ | 10! | 1    |   |   | 2 | 4   | 8   | 9 |
| 2(37) | 7 | $7_6$ |     | 1     |   |   | 2 | 4   | 10! | 8 |
| 2(38) | 7 | $4_5$ |     | 1     |   |   | 2 | 9!  | 4   |   |
| 2(39) | 7 | $8_6$ |     | 1     |   |   | 2 | 10! | 4   | 9 |
| 2(40) | 7 | $4_5$ |     | 1     |   |   | 2 |     | 10! | 4 |
| 2(41) | 7 | $5_5$ |     | 1     |   |   | 2 | 10! |     | 4 |
| 2(42) | 7 | $8_6$ |     |       |   | 1 | 2 | 6   | 10! | 8 |
| 2(43) | 7 | $4_5$ |     | 1     |   |   | 2 | 9!  | 4   |   |
| 2(44) | 7 | $8_6$ |     | 1     |   |   | 2 | 10! | 4   | 9 |
| 2(45) | 7 | $5_5$ |     | 1     |   |   | 2 | 10! |     | 4 |
| 2(46) | 7 | $4_5$ |     | 1     |   |   | 2 |     | 10! | 4 |
| 2(47) | 7 | $23_7$ | 10! |      | 1 |   | 2 | 5   | 8   | 9 |
| 3(1)  | 7 |      | $7_6$ | 1   |   |   | 3 | 4   | 10! | 8 |
| 3(2)  | 7 |      | $3_5$ | 1   |   |   | 3 | 9!  | 4   |   |
| 3(3)  | 7 |      | $7_6$ | 1   |   |   | 3 | 10! | 4   | 9 |
| 3(4)  | 7 |      | $3_5$ | 1   |   |   | 3 |     | 10! | 4 |
| 3(5)  | 7 |      | $4_5$ | 1   |   |   | 3 | 10! |     | 4 |
| 3(6)  | 7 |      | $4_5$ |     |   |   | 3 | 1   | 10! | 8 |
| 3(7)  | 7 |      | $18_7$ | 6! |   | 1 | 3 |     |     |   |
| 3(8)  | 7 |      | $6_6$ |     | 1 |   | 3 | 5   | 10! | 8 |
| 4(1)  | 4 | 7    | 1   | $13_8$ | 2 |   | 3 | 5   | 10! | 8 |
| 4(2)  | 4 | 7    | 1   | $14_8$ |   | 2 | 3 | 6   | 10! | 8 |
| 4(3)  | 4 | 7    | 1   | $10_7$ |   |   | 3 | 2   | 10! | 8 |
| 4(4)  | 4 | 7    | 1   | $7_6$  |   |   | 3 | 9!  | 2   |   |
| 4(5)  | 4 | 7    | 1   | $11_7$ |   |   | 3 | 10! | 2   | 9 |
| 4(6)  | 4 | 7    | 1   | $9_6$  |   |   | 3 | 10! |     | 2 |

|       | 1 | 2 | 3  | 4    | 5      | 6 | 7 | 8   | 9   | 10 |
|-------|---|---|----|------|--------|---|---|-----|-----|----|
| 4(7)  | 4 | 7 | 1  | $8_6$ |        |   | 3 |     | 10! | 2  |
| 4(8)  | 4 |   | 7  | $10_6$ |        |   | 1 | 3   | 10! | 8  |
| 4(9)  | 4 |   | 7  | $6_5$ |        |   | 1 | 9!  | 3   |    |
| 4(10) | 4 |   | 7  | $6_6$ |        |   | 1 | 10! | 3   | 9  |
| 4(11) | 4 | 7 |    | $3_5$ |        |   | 1 | 9!  | 2   |    |
| 4(12) | 4 | 7 |    | $7_6$ |        |   | 1 | 10! | 2   | 9  |
| 4(13) | 4 | 7 |    | $4_5$ |        |   | 1 |     | 10! | 2  |
| 4(14) | 4 | 7 |    | $5_5$ |        |   | 1 | 10! |     | 2  |
| 5(1)  | 6 | 5 | 1  | 10!  | $28_8$ | 2 |   | 3   | 8   | 9  |
| 5(2)  | 6 | 5 | 1  |      | $10_7$ | 2 |   | 3   | 10! | 8  |
| 5(3)  | 6 | 5 | 1  |      | $6_6$  | 2 |   | 9!  | 3   |    |
| 5(4)  | 6 | 5 | 1  |      | $10_7$ | 2 |   | 10! | 3   | 9  |
| 5(5)  | 6 | 5 | 1  |      | $8_6$  | 2 |   | 10! |     | 3  |
| 5(6)  | 6 | 5 | 1  |      | $7_6$  | 2 |   |     | 10! | 3  |
| 5(7)  | 6 | 5 | 4! | 1    | $18_5$ | 2 |   |     |     |    |
| 5(8)  | 6 | 5 | 10!| 1    | $27_8$ | 2 |   | 4   | 8   | 9  |
| 5(9)  | 6 | 5 |    | 1    | $9_7$  | 2 |   | 4   | 10! | 8  |
| 5(10) | 6 | 5 |    | 1    | $5_6$  | 2 |   | 9!  | 4   |    |
| 5(11) | 6 | 5 |    | 1    | $9_7$  | 2 |   | 10! | 4   | 9  |
| 5(12) | 6 | 5 |    | 1    | $6_6$  | 2 |   | 10! |     | 4  |
| 5(13) | 6 | 5 |    | 1    | $5_6$  | 2 |   |     | 10! | 4  |
| 5(14) | 6 | 5 |    |      | $6_6$  | 2 |   | 1   | 10! | 8  |
| 5(15) |   | 5 |    |      | $2_4$  |   |   | 9!  | 10  | 2  |
| 5(16) | 5 | 1 | 7  |      | $10_7$ |   | 2 | 3   | 10! | 8  |
| 5(17) | 5 | 1 | 7  |      | $6_6$  |   | 2 | 9!  | 3   |    |
| 5(18) | 5 | 1 | 7  |      | $10_7$ |   | 2 | 10! | 3   | 9  |
| 5(19) | 5 | 1 |    |      | $7_5$  |   |   | 9!  | 10  | 2  |
| 5(20) | 2 | 5 | 1  |      | $14_6$ |   |   | 3   | 10! | 8  |
| 5(21) | 2 | 5 |    | 1    | $13_6$ |   |   | 4   | 10! | 8  |
| 5(22) | 2 | 5 |    | 1    | $9_5$  |   |   | 9!  | 4   |    |
| 5(23) | 2 | 5 |    | 1    | $12_6$ |   |   | 10! | 4   | 9  |
| 5(24) | 2 | 5 |    | 1    | $9_5$  |   |   |     | 10! | 4  |
| 5(25) | 2 | 5 |    | 1    | $10_5$ |   |   | 10! |     | 4  |
| 5(26) | 2 | 5 |    |      | $13_6$ | 1 |   | 6   | 10! | 8  |
| 5(27) | 2 | 5 | 7! |      | $6_4$  |   | 1 |     |     |    |
| 5(28) | 2 | 5 |    |      | $10_5$ |   |   | 1   | 10! | 8  |
|       | 1 | 2 | 3  | 4    | 5      | 6 | 7 | 8   | 9   | 10 |

|  | 1 | 2 | 3 | 4 | 5 | 6 | 7 | 8 | 9 | 10 |
|---|---|---|---|---|---|---|---|---|---|---|
| 5(29) | 2 | 5 |  |  | $6_4$ |  |  | 9! | 1 |  |
| 5(30) | 2 | 5 |  |  | $10_5$ |  |  | 10! | 1 | 9 |
| 5(31) | 2 | 5 |  |  | $6_4$ |  |  |  | 10! | 1 |
| 5(32) | 2 | 5 |  |  | $7_4$ |  |  | 10! |  | 1 |
| 6(1) |  | 6 |  |  |  | $2_4$ |  | 9! | 10 | 2 |
| 6(2) | 5 | 6 | 1 | 10! | 2 | $29_8$ |  | 3 | 8 | 9 |
| 6(3) | 5 | 6 | 1 |  | 2 | $11_7$ |  | 3 | 10! | 8 |
| 6(4) | 5 | 6 | 1 |  | 2 | $7_6$ |  | 9! | 3 |  |
| 6(5) | 5 | 6 | 1 |  | 2 | $11_7$ |  | 10! | 3 | 9 |
| 6(6) | 5 | 6 | 1 |  | 2 | $9_6$ |  | 10! |  | 3 |
| 6(7) | 5 | 6 | 1 |  | 2 | $8_6$ |  |  | 10! | 3 |
| 6(8) | 5 | 6 | 7! |  | 2 | $3_5$ | 1 |  |  |  |
| 6(9) | 5 | 6 | 10! |  | 2 | $25_7$ |  | 1 | 8 | 9 |
| 6(10) | 5 | 6 |  | 10! | 2 | $25_7$ |  | 1 | 8 | 9 |
| 6(11) | 5 | 6 |  |  | 2 | $7_6$ |  | 1 | 10! | 8 |
| 6(12) | 5 | 6 |  |  | 2 | $3_5$ |  | 9! | 1 |  |
| 6(13) | 5 | 6 |  |  | 2 | $7_6$ |  | 10! | 1 | 9 |
| 6(14) | 5 | 6 |  |  | 2 | $4_5$ |  | 10! |  | 1 |
| 6(15) | 5 | 6 |  |  | 2 | $3_5$ |  |  | 10! | 1 |
| 6(16) | 5 | 6 | 10! | 1 | 2 | $28_8$ |  | 4 | 8 | 9 |
| 6(17) | 5 | 6 |  | 1 | 2 | $10_7$ |  | 4 | 10! | 8 |
| 6(18) | 5 | 6 |  | 1 | 2 | $6_6$ |  | 9! | 4 |  |
| 6(19) | 5 | 6 |  | 1 | 2 | $10_7$ |  | 10! | 4 | 9 |
| 6(20) | 5 | 6 |  | 1 | 2 | $7_6$ |  | 10! |  | 4 |
| 6(21) | 5 | 6 |  | 1 | 2 | $6_6$ |  |  | 10! | 4 |
| 6(22) | 6 | 1 | 7 | 10! |  | $28_8$ | 2 | 3 | 8 | 9 |
| 6(23) | 6 | 1 | 7 |  |  | $10_7$ | 2 | 3 | 10! | 8 |
| 6(24) | 6 | 1 | 7 |  |  | $6_6$ | 2 | 9! | 3 |  |
| 6(25) | 6 | 1 | 7 |  |  | $10_7$ | 2 | 10! | 3 | 9 |
| 6(26) | 6 | 1 | 7 |  |  | $8_6$ | 2 | 10! |  | 3 |
| 6(27) | 6 | 1 | 7 |  |  | $7_6$ | 2 |  | 10! | 3 |
| 6(28) | 2 | 6 | 1 |  |  | $10_6$ |  | 3 | 10! | 8 |
| 6(29) | 2 | 6 | 1 |  |  | $10_5$ |  | 9! | 3 |  |
| 6(30) | 2 | 6 | 1 |  |  | $14_6$ |  | 10! | 3 | 9 |
|  | 1 | 2 | 3 | 4 | 5 | 6 | 7 | 8 | 9 | 10 |

|       | 1 | 2   | 3 | 4   | 5 | 6       | 7       | 8   | 9   | 10 |
|-------|---|-----|---|-----|---|---------|---------|-----|-----|----|
| 6(31) | 2 | 6   | 1 |     |   | $11_5$  |         |     | 10! | 3  |
| 6(32) | 2 | 6   | 1 |     |   | $12_5$  |         | 10! |     | 3  |
| 6(33) | 2 | 6   |   |     | 1 |         |         | 5   | 10! | 8  |
| 6(34) | 2 | 6   |   |     |   | $10_5$  |         | 1   | 10! | 8  |
| 6(35) | 2 | 6   |   |     |   | $6_4$   |         | 9!  | 1   |    |
| 6(36) | 2 | 6   |   |     |   | $10_5$  |         | 10! | 1   | 9  |
| 6(37) | 2 | 6   |   |     |   | $6_4$   |         |     | 10! | 1  |
| 6(38) | 2 | 6   |   |     |   | $7_4$   |         | 10! |     | 1  |
| 6(39) | 6 | 7   |   |     |   | $2_5$   | 1       |     | 10! | 2  |
| 6(40) | 6 | 7   |   |     |   | $3_5$   | 1       | 10! |     | 2  |
| 6(41) | 6 | 7   |   |     |   | $4_6$   | 1       | 2   | 10! | 8  |
| 6(42) | 6 | 7   |   | 5!  | 2 | $16_5$  | 1       |     |     |    |
| 6(43) | 6 | 10! | 7 |     |   | $22_7$  | 1       | 3   | 8   | 9  |
| 6(44) | 6 |     | 7 | 10! |   | $22_7$  | 1       | 3   | 8   | 9  |
| 6(45) | 6 | 10! | 7 |     |   | $23_6$  | 1       | 3   |     | 8  |
| 6(46) | 6 |     | 7 | 10! |   | $23_6$  | 1       | 3   |     | 8  |
| 6(47) | 6 |     | 7 |     |   | $3_6$   | 1       | 3   | 10! | 8  |
| 7(1)  |   | 7   |   |     |   |         | $2_4$   | 9!  | 10  | 2  |
| 7(2)  | 6 | 7   | 1 | 10! |   | 2       | $28_8$  | 3   | 8   | 9  |
| 7(3)  | 6 | 7   | 1 |     |   | 2       | $10_7$  | 3   | 10! | 8  |
| 7(4)  | 6 | 7   | 1 |     |   | 2       | $6_6$   | 9!  | 3   |    |
| 7(5)  | 6 | 7   | 1 |     |   | 2       | $10_7$  | 10! | 3   | 9  |
| 7(6)  | 6 | 7   | 1 |     |   | 2       | $8_6$   | 10! |     | 3  |
| 7(7)  | 6 | 7   | 1 |     |   | 2       | $10_7$  | 10! | 3   | 9  |
| 7(9)  | 6 | 7   | 1 |     |   | 2       | $7_6$   |     | 10! | 3  |
| 7(10) | 5 | 7   | 1 | 10! | 2 |         | $10_8$  | 3   | 8   | 9  |
| 7(11) | 7 | 1   |   |     |   |         | $7_5$   | 9!  | 10  | 2  |
| 7(12) | 2 | 7   | 1 |     |   |         | $13_6$  | 3   | 10! | 8  |
| 7(13) | 2 | 7   | 1 |     |   |         | $9_5$   | 9!  | 3   |    |
| 7(14) | 2 | 7   | 1 |     |   |         | $13_6$  | 10! | 3   | 9  |
| 7(15) | 2 | 7   | 1 |     |   |         | $10_5$  |     | 10! | 3  |
| 7(16) | 2 | 7   |   | 1   |   |         | $8_5$   | 9!  | 4   |    |
| 7(17) | 2 | 7   |   | 1   |   |         | $12_6$  | 10! | 4   | 9  |
| 7(18) | 2 | 7   |   | 1   |   |         | $8_5$   |     | 10! | 4  |
| 7(19) | 2 | 7   |   | 1   |   |         | $9_5$   | 10! |     | 4  |
| 7(20) | 2 | 7   |   |     | 1 |         | $11_6$  | 5   | 10! | 8  |
|       | 1 | 2   | 3 | 4   | 5 | 6       | 7       | 8   | 9   | 10 |

|       | 1 | 2 | 3 | 4  | 5 | 6 | 7      | 8   | 9   | 10 |
|-------|---|---|---|----|---|---|--------|-----|-----|----|
| 7(12) | 7 | 1 |   |    |   |   | $7_5$  | 9!  | 10  | 2  |
| 7(13) | 2 | 7 | 1 |    |   |   | $13_6$ | 3   | 10! | 8  |
| 7(14) | 2 | 7 | 1 |    |   |   | $9_5$  | 9!  | 3   |    |
| 7(15) | 2 | 7 | 1 |    |   |   | $13_6$ | 10! | 3   | 9  |
| 7(16) | 2 | 7 | 1 |    |   |   | $10_5$ |     | 10! | 3  |
| 7(17) | 2 | 7 |   | 1  |   |   | $8_5$  | 9!  | 4   |    |
| 7(18) | 2 | 7 |   | 1  |   |   | $12_6$ | 10! | 4   | 9  |
| 7(19) | 2 | 7 |   | 1  |   |   | $8_5$  |     | 10! | 4  |
| 7(20) | 2 | 7 |   | 1  |   |   | $9_5$  | 10! |     | 4  |
| 7(21) | 2 | 7 |   |    | 1 |   | $11_6$ | 5   | 10! | 8  |
| 7(22) | 2 | 7 |   |    |   | 1 | $12_6$ | 6   | 10! | 8  |
| 7(23) | 2 | 7 |   |    |   |   | $9_5$  | 1   | 10! | 8  |
| 7(24) | 2 | 7 |   |    |   |   | $5_4$  | 9!  | 1   |    |
| 7(25) | 2 | 7 |   |    |   |   | $9_5$  | 10! | 1   | 9  |
| 7(26) | 2 | 7 |   |    |   |   | $5_4$  |     | 10! | 1  |
| 7(27) | 7 | 6 |   |    |   | 1 | $2_5$  | 9!  | 2   |    |
| 7(28) | 7 | 6 |   |    |   | 1 | $6_6$  | 10! | 2   | 9  |
| 7(29) | 7 | 6 |   |    | 2 | 1 | $7_7$  | 5   | 10! | 8  |
| 7(30) | 7 | 6 |   |    |   | 1 | $5_6$  | 2   | 10! | 8  |
| 7(31) | 7 | 6 |   | 5! | 2 | 1 | $17_5$ |     |     |    |
|       | 1 | 2 | 3 | 4  | 5 | 6 | 7      | 8   | 9   | 10 |

P(30)(a)

|        | 1       | 2       | 3       | 4  | 5 | 6 | 7 | 8   | 9   | 10 |
|--------|---------|---------|---------|----|---|---|---|-----|-----|----|
| 1(2)   | $16_7$  | 6       | 7       | 3  |   | 1 | 2 | 9!  | 4   |    |
| 1(4)   | $16_8$  | 6       | 7       | 3  |   | 1 | 2 | 9!  | 10  | 4  |
| 1(5)   | $17_7$  | 6       | 7       | 3  |   | 1 | 2 | 10! |     | 4  |
| 1(6)   | $23_8$  | 6       | 7       |    | 3 | 1 | 2 | 5   | 10! | 8  |
| 1(8)   | $6_7$   | 6       | 7       |    |   | 1 | 2 | 3   | 10! | 8  |
| 1(9)   | $24_7$  | 6       | 7       |    |   | 1 | 2 | 9!  | 3   |    |
| 1(10)  | $25_7$  | 6       | 7       |    |   | 1 | 2 | 9!  | 3   |    |
| 1(11)  | $6_7$   | 6       | 7       |    |   | 1 | 2 | 10! | 3   | 9  |
| 1(12)  | $2_6$   | 6       | 7       |    |   | 1 | 2 |     | 10! | 3  |
| 1(13)  | $2_6$   | 6       | 7       |    |   | 1 | 2 | 10! |     | 3  |
| 1(14)  | $3_5$   | 6       |         |    |   | 1 |   | 9!  | 10  | 2  |
| 1(15)  | $23_6$  | 9!      | 7       |    |   |   | 1 | 10  | 8   | 3  |
| 1(16)  | $1_5$   |         | 7       |    |   |   | 1 | 9!  | 10  | 3  |
| 2(1)   | 7       | $23_6$  | 9!      |    |   |   | 2 | 10  | 8   | 1  |
| 2(2)   | 10!     | $23_6$  | 7       |    |   |   | 2 | 9   | 3   | 8  |
| 2(3)   | 9!      | $23_6$  | 7       |    |   |   | 2 | 10  | 8   | 3  |
| 2(4)   | 7       | $1_5$   |         |    |   |   | 2 | 9!  | 10  | 1  |
| 3(1)   | 10!     | 7       | $23_6$  |    |   |   | 3 | 9   | 2   | 8  |
| 3(2)   |         | 7       | $2_5$   |    |   |   | 3 | 9!  | 10  | 2  |
| 3(3)   | 9!      | 7       | $23_6$  |    |   |   | 3 | 10  | 8   | 2  |
| 3(4)   | 6       | 7       | $23_9$  | 1  | 4 | 2 | 3 | 5   | 10! | 8  |
| 3(5)   | 6       | 7       | $8_8$   |    | 1 | 2 | 3 | 5   | 10! | 8  |
| 3(6)   | 6       | 7       | $2_6$   |    |   | 2 | 3 | 9!  | 1   |    |
| 3(7)   | 6       | 7       | $8_7$   |    |   | 2 | 3 | 10! | 1   | 9  |
| 3(8)   | 6       | 7       | $5_6$   |    |   | 2 | 3 | 10! |     | 1  |
| 3(9)   | 6       | 7       | $2_6$   |    |   | 2 | 3 |     | 10! | 1  |
| 3(10)  | 6       | 7       | $8_8$   | 1  |   | 2 | 3 | 4   | 10! | 8  |
| 3(11)  | 6       | 7       | $5_7$   | 1  |   | 2 | 3 | 9!  | 4   |    |
| 3(12)  | 6       | 7       | $9_8$   | 1  |   | 2 | 3 | 10! | 4   | 9  |
| 3(13)  | 6       | 7       | $6_7$   | 1  |   | 2 | 3 | 10! |     | 4  |
| 3(14)  | 6       | 7       | $5_7$   | 1  |   | 2 | 3 |     | 10! | 4  |
| 3(15)  |         | 7       | $32_7$  | 9! |   | 2 | 3 | 6   | 10  | 8  |
|        | 1       | 2       | 3       | 4  | 5 | 6 | 7 | 8   | 9   | 10 |

|       | 1 | 2 | 3      | 4      | 5      | 6 | 7 | 8   | 9   | 10 |
|-------|---|---|--------|--------|--------|---|---|-----|-----|----|
| 3(16) | 5 | 7 | $25_9$ | 1      | 2      | 4 | 3 | 6   | 10! | 8  |
| 3(17) | 7 |   | $0_5$  |        |        |   | 3 | 9!  | 10  | 1  |
| 3(18) | 7 | 6 | $17_6$ | 5      | 2      | 1 | 3 | 4   | 10! | 8  |
| 3(19) | 7 | 6 | $17_6$ | 5      | 2      | 1 | 3 | 9!  | 4   |    |
| 3(20) | 7 | 6 | $17_6$ | 5      | 2      | 1 | 3 | 10! | 4   | 9  |
| 3(21) | 7 | 6 | $17_6$ | 5      | 2      | 1 | 3 | 10! |     | 4  |
| 3(22) | 7 | 6 | $17_6$ | 5      | 2      | 1 | 3 |     | 10! | 4  |
| 3(23) | 7 | 6 | $8_8$  |        | 2      | 1 | 3 | 5   | 10! | 8  |
| 3(24) | 7 | 8 | $20_6$ |        |        |   |   | 9   | 10  | 1  |
| 3(25) | 7 | 6 | $8_8$  |        | 2      | 1 | 3 | 5   | 10! | 8  |
| 4(1)  | 4 | 7 | 9!     | $30_8$ | 2      |   | 1 | 5   | 10  | 8  |
| 4(2)  | 4 | 7 | 6      | $23_8$ |        | 2 | 1 | 3   | 10! | 8  |
| 4(3)  | 4 | 7 | 6      | $23_8$ |        | 2 | 1 | 10! | 3   | 9  |
| 4(4)  | 4 | 7 | 9!     | $31_8$ |        | 2 | 1 | 6   | 10  | 8  |
| 4(5)  | 4 | 7 | 9!     | $27_7$ |        |   | 1 | 2   | 10  | 8  |
| 4(6)  | 4 | 7 | 10!    | $26_7$ |        |   | 1 | 9   | 2   | 8  |
| 4(7)  | 4 | 7 | 10!    | $26_7$ |        |   | 1 | 10  | 2   | 9  |
| 4(8)  | 4 | 7 | 8!     | $26_6$ |        |   | 1 | 10  |     | 2  |
| 4(9)  | 4 | 7 | 9!     | $26_7$ |        |   | 1 | 10  | 8   | 2  |
| 4(10) | 4 | 5 | 7      | $5_7$  | 1      |   | 2 | 9!  | 3   |    |
| 4(11) | 4 | 5 | 7      | $9_8$  | 1      |   | 2 | 10! | 3   | 9  |
| 4(12) | 4 | 5 | 10!    | $27_7$ | 1      |   |   | 9   | 2   | 8  |
| 4(13) | 4 | 5 |        | $7_6$  | 1      |   |   | 9!  | 10  | 2  |
| 5(1)  | 7 | 5 | 1      |        | $2_6$  |   | 2 | 9!  | 10  | 3  |
| 5(2)  | 3 | 5 | 7      | 1      | $14_8$ |   | 2 | 4   | 10! | 8  |
| 5(3)  | 3 | 5 | 7      | 1      | $10_7$ |   | 2 | 9!  | 4   |    |
| 5(4)  | 3 | 5 | 7      | 1      | $14_8$ |   | 2 | 10! | 4   | 9  |
| 5(5)  | 3 | 5 | 7      | 1      | $11_7$ |   | 2 | 10! |     | 4  |
| 5(6)  | 3 | 5 | 7      |        | $11_7$ |   | 2 | 1   | 10! | 8  |
| 5(7)  | 3 | 5 | 7      |        | $7_6$  |   | 2 | 9!  | 1   |    |
| 5(8)  | 3 | 5 | 7      |        | $11_7$ |   | 2 | 10! | 1   | 9  |
| 5(9)  | 3 | 5 | 7      |        | $8_6$  |   | 2 | 10! |     | 1  |
| 5(10) | 4 | 5 | 7      | 3      | $24_9$ | 1 | 2 | 6   | 10! | 8  |
| 5(11) | 4 | 5 | 7      | 3      | $21_8$ |   | 2 | 1   | 10! | 8  |
| 5(12) | 4 | 5 | 7      | 3      | $17_7$ |   | 2 | 9!  | 1   |    |
| 5(13) | 4 | 5 | 7      | 3      | $21_8$ |   | 2 | 10! | 1   | 9  |
| 5(14) | 4 | 5 | 7      | 3      | $18_7$ |   | 2 | 10! |     | 1  |
|       | 1 | 2 | 3      | 4      | 5      | 6 | 7 | 8   | 9   | 10 |

|       | 1 | 2 | 3 | 4 | 5 | 6 | 7 | 8 | 9 | 10 |
|-------|---|---|---|---|---|---|---|---|---|----|
| 5(15) | 4 | 5 | 7 | 3 | $17_7$ |   | 2 |   | 10! | 1 |
| 5(16) | 5 | 6 | 7 | 3 | $14_8$ | 1 | 2 | 9! | 4 |   |
| 5(17) | 5 | 6 | 7 | 3 | $15_8$ | 1 | 2 | 10! |   | 4 |
| 5(18) | 5 | 6 | 7 | 3 | $14_8$ | 1 | 2 |   | 10! | 4 |
| 5(19) | 5 | 6 | 7 | 8 | $24_9$ | 1 | 2 | 3 | 10! | 4 |
| 5(20) | 5 | 6 | 7 | 9 | $25_9$ | 1 | 2 | 10! | 3 | 4 |
| 5(21) | 5 | 6 | 7 |   | $2_7$ | 1 | 2 | 9! | 3 |   |
| 5(22) | 5 | 6 | 7 |   | $3_7$ | 1 | 2 | 10! |   | 3 |
| 5(23) | 5 | 6 | 7 |   | $2_7$ | 1 | 2 |   | 10! | 3 |
| 5(24) | 5 | 6 |   | 7 | $23_8$ | 1 | 2 | 4 | 10! | 8 |
| 5(25) | 5 | 6 |   | 7 | $19_7$ | 1 | 2 | 9! | 4 |   |
| 5(26) | 5 | 6 |   | 7 | $23_8$ | 1 | 2 | 10! | 4 | 9 |
| 5(27) | 5 | 6 |   | 7 | $20_7$ | 1 | 2 | 10! |   | 4 |
| 5(28) | 5 | 6 |   | 7 | $19_7$ | 1 | 2 |   | 10! | 4 |
| 5(29) | 5 | 7 | 1 | 6 | $23_9$ | 2 | 3 | 4 | 10! | 8 |
| 5(30) | 5 | 7 | 1 | 6 | $20_8$ | 2 | 3 | 9! | 4 |   |
| 5(31) | 5 | 7 | 1 | 6 | $23_9$ | 2 | 3 | 10! | 4 | 9 |
| 5(32) | 5 | 7 | 1 | 6 | $20_8$ | 2 | 3 | 10! |   | 4 |
| 6(1)  | 7 | 6 | 1 |   |   | $6_7$ | 2 | 9! | 10 | 3 |
| 6(2)  | 6 | 5 | 7 | 3 | 1 | $21_9$ | 2 | 4 | 10! | 8 |
| 6(3)  | 6 | 5 | 7 | 3 | 1 | $17_8$ | 2 | 9! | 4 |   |
| 6(4)  | 6 | 5 | 7 | 3 | 1 | $18_8$ | 2 | 10! |   | 4 |
| 6(5)  | 6 | 5 | 7 | 3 | 1 | $17_8$ | 2 |   | 10! | 4 |
| 6(6)  | 6 | 5 | 7 |   | 1 | $3_7$ | 2 | 9! | 3 |   |
| 6(7)  | 6 | 5 | 7 | 9! | 1 | $25_7$ | 2 |   | 3 |   |
| 6(8)  | 6 | 5 | 7 |   | 1 | $7_8$ | 2 | 10! | 3 | 9 |
| 6(9)  | 6 | 5 | 7 |   | 1 | $4_7$ | 2 | 10! |   | 3 |
| 6(10) | 6 | 5 | 7 |   | 1 | $3_7$ | 2 |   | 10! | 3 |
| 6(11) | 6 | 5 | 10! |   | 1 | $26_7$ |   | 9 | 2 | 8 |
| 6(12) | 6 | 5 |   | 10! | 1 | $26_7$ |   | 9 | 2 | 8 |
| 6(13) | 6 | 5 |   |   | 1 | $4_6$ |   | 9! | 10 | 2 |
| 6(14) | 6 | 5 | 9! |   | 1 | $26_7$ |   | 10 | 8 | 2 |
| 6(15) | 6 | 5 |   | 9! | 1 | $26_7$ |   | 10 | 8 | 2 |
| 6(16) | 6 | 7 | 10! |   |   | $24_7$ | 1 | 9 | 2 | 8 |
| 6(17) | 6 | 7 |   | 10! |   | $24_7$ | 1 | 9 | 2 | 8 |
| 7(1)  | 7 | 5 | 10! |   | 1 |   | $26_7$ | 9 | 2 | 8 |
| 7(2)  | 7 | 5 |   | 10! | 1 |   | $26_7$ | 9 | 2 | 8 |
| 7(3)  | 7 | 5 | 9! |   | 1 |   | $26_7$ | 10 | 8 | 2 |
| 7(4)  | 7 | 5 |   | 9! | 1 |   | $26_7$ | 10 | 8 | 2 |
| 7(5)  | 7 | 5 |   |   | 1 |   | $5_6$ | 9! | 10 | 2 |
| 7(6)  | 7 | 8! |   | 1 |   |   | $23_6$ | 9 | 10 | 4 |
|       | 1 | 2 | 3 | 4 | 5 | 6 | 7 | 8 | 9 | 10 |

| | 1 | 2 | 3 | 4 | 5 | 6 | 7 | 8 | 9 | 10 |
|---|---|---|---|---|---|---|---|---|---|---|
| 1(1) | $17_7$ | 5 | 7 | 3 | 1 | | 2 | 9! | 4 | |
| 1(2) | $18_7$ | 5 | 7 | 3 | 1 | | 2 | 10! | | 4 |
| 1(3) | $17_7$ | 5 | 7 | 3 | 1 | | 2 | | 10! | 4 |
| 1(4) | $25_7$ | 5 | 7 | 9! | 1 | | 2 | 3 | 8 | |
| 1(5) | $3_7$ | 5 | 7 | | 1 | | 2 | 3 | 10! | 8 |
| 1(6) | $3_6$ | 5 | 7 | | 1 | | 2 | 9! | 3 | |
| 1(7) | $5_6$ | 5 | 7 | | 1 | | 2 | 10! | | 3 |
| 1(8) | $4_6$ | 5 | 7 | | 1 | | 2 | | 10! | 3 |
| 1(9) | $4_5$ | 5 | | | 1 | | | 9! | 10 | 2 |
| 2(1) | 6 | $5_6$ | | 1 | | 2 | | 9! | 10 | 4 |
| 2(2) | 6 | $20_8$ | 7 | 3 | | 2 | 1 | 4 | 10! | 8 |
| 2(3) | 6 | $24_7$ | 7 | 9! | | 2 | 1 | $32_5$ | $82_6$ | |
| 2(4) | 6 | $24_8$ | 7 | 10! | | 2 | 1 | 3 | 8 | 9 |
| 2(5) | 6 | $25_7$ | 7 | 10! | | 2 | 1 | $32_5$ | | $86_6$ |
| 2(6) | 6 | $6_7$ | 7 | | | 2 | 1 | 3 | 10! | 8 |
| 2(7) | 6 | $2_6$ | 7 | | | 2 | 1 | 9! | 3 | |
| 2(8) | 6 | $24_7$ | 7 | 10! | | 2 | 1 | | 3 | 9 |
| 2(9) | 6 | $3_7$ | 7 | | | 2 | 1 | 10! | 3 | 9 |
| 2(10) | 6 | $4_6$ | 7 | | | 2 | 1 | 10! | | 3 |
| 2(11) | 6 | $3_6$ | 7 | | | 2 | 1 | | 10! | 3 |
| 2(12) | 6 | $2_5$ | | | | 2 | | 9! | 10 | 1 |
| 2(13) | 5 | $8_6$ | 1 | | 2 | | | 9! | 10 | 3 |
| 2(14) | 5 | $6_6$ | | 1 | 2 | | | 9! | 10 | 4 |
| 2(15) | 5 | $17_7$ | 7 | 3 | 2 | | 1 | 9! | 4 | |
| 2(16) | 5 | $18_7$ | 7 | 3 | 2 | | 1 | 10! | | 4 |
| 2(17) | 5 | $17_7$ | 7 | 3 | 2 | | 1 | | 10! | 4 |
| 2(18) | 5 | $25_7$ | 7 | 9! | 2 | | 1 | 3 | 8 | |
| 2(19) | 5 | $25_8$ | 7 | 10! | 2 | | 1 | 3 | 8 | 9 |
| 2(20) | 5 | $26_7$ | 7 | 10! | 2 | | 1 | 3 | | 8 |
| 2(21) | 5 | $7_7$ | 7 | | 2 | | 1 | 3 | 10! | 8 |
| 2(22) | 5 | $3_6$ | 7 | | 2 | | 1 | 9! | 3 | |
| 2(23) | 5 | $25_7$ | 7 | 10! | 2 | | 1 | | 3 | 9 |
| 2(24) | 5 | $4_7$ | 7 | | 2 | | 1 | 10! | 3 | 9 |
| 2(25) | 5 | $5_6$ | 7 | | 2 | | 1 | 10! | | 3 |
| 2(26) | 5 | $4_6$ | 7 | | 2 | | 1 | | 10! | 3 |
| 2(27) | 7 | $4_6$ | | 1 | | | 2 | 9! | 10 | 4 |

|       | 1 | 2   | 3    | 4    | 5       | 6 | 7 | 8   | 9   | 10 |
|-------|---|-----|------|------|---------|---|---|-----|-----|----|
| 2(28) | 7 | $26_7$ | 9!   | 1    |         |   | 2 | `0  | 8   | 4  |
| 2(29) | 7 | $4_6$ |      | 1    |         |   | 2 | 9!  | 10  | 4  |
| 3(1)  | 7 | 10! | $26_7$ | 1    |         |   | 3 | 9   | 4   | 8  |
| 3(2)  | 7 |     | $3_6$ | 1    |         |   | 3 | 9!  | 10  | 4  |
| 3(3)  | 7 | 9!  | $25_7$ | 1    |         |   | 3 | 10  | 8   | 4  |
| 3(4)  | 7 |     | $14_6$ | 6    |         | 1 | 3 |     | 10! | 4  |
| 4(1)  | 4 |     |      | $2_4$ |         |   |   | 9!  | 10  | 1  |
| 4(2)  | 4 | 7   | 1    | $8_7$ |         |   | 3 | 9!  | 10  | 2  |
| 4(3)  | 4 | 8!  | 7    | $26_7$ |         |   | 1 | 10  | 3   | 9  |
| 4(4)  | 4 | 7   | 10!  | $26_7$ |         |   | 1 | 9   | 2   | 8  |
| 4(5)  | 4 | 7   |      | $4_6$ |         |   | 1 | 9!  | 10  | 2  |
| 5(1)  | 6 | 5   | 1    | 10!  | $29_8$  | 2 |   | 9   | 3   | 8  |
| 5(2)  | 6 | 5   | 1    | 9!   | $30_8$  | 2 |   | 10  | 8   | 3  |
| 5(3)  | 6 | 5   | 1    |      | $7_7$   | 2 |   | 9!  | 10  | 3  |
| 5(4)  | 6 | 5   | 4    | 1    | $22_8$  | 2 |   | 3   | 10! | 8  |
| 5(5)  | 6 | 5   | 4    | 1    | $18_7$  | 2 |   | 9!  | 3   |    |
| 5(6)  | 6 | 5   | 4    | 1    | $22_8$  | 2 |   | 10! | 3   | 9  |
| 5(7)  | 6 | 5   | 4    | 1    | $20_7$  | 2 |   | 10! |     | 3  |
| 5(8)  | 6 | 5   | 4    | 1    | $19_7$  | 2 |   |     | 10! | 3  |
| 5(9)  | 6 | 5   | 10!  | 1    | $28_8$  | 2 |   | 9   | 4   | 8  |
| 5(10) | 6 | 5   | 8!   | 1    | $30_8$  | 2 |   | 10  | 4   | 9  |
| 5(11) | 6 | 5   | 9!   | 1    | $29_8$  | 2 |   | 10  | 8   | 4  |
| 5(12) | 6 | 5   |      | 1    | $5_7$   | 2 |   | 9!  | 10  | 4  |
| 5(13) | 6 | 5   | 8!   | 1    | $30_8$  | 2 |   | 10  | 4   | 9  |
| 5(14) | 5 | 1   | 7    | 10!  | $29_8$  |   | 2 | 9   | 3   | 8  |
| 5(15) | 2 | 5   | 7    |      | $6_6$   |   | 1 | 9!  | 3   |    |
| 5(16) | 2 | 5   | 7    |      | $10_7$  |   | 1 | 10! | 3   | 9  |
| 5(17) | 2 | 5   | 7    |      | $7_6$   |   | 1 | 10! |     | 3  |
| 5(18) | 2 | 5   |      |      | $6_5$   |   |   | 9!  | 10  | 1  |
| 5(19) | 2 | 5   | 7    | 10!  | $28_8$  |   | 1 | 3   | 8   | 9  |
| 5(20) | 2 | 5   | 7    |      | $10_7$  |   | 1 | 3   | 10! | 8  |
| 5(21) | 2 | 5   | 7    |      | $6_6$   |   | 1 | 9!  | 3   |    |
| 5(22) | 2 | 5   | 7    |      | $8_6$   |   | 1 | 10! |     | 3  |
| 5(23) | 2 | 5   | 7    |      | $8_7$   |   | 1 |     | 10! | 3  |
| 5(24) | 2 | 5   |      |      | $6_5$   |   |   | 9!  | 10  | 1  |
| 5(25) | 2 | 5   | 7    | 10!  | $28_8$  |   | 1 | 3   | 8   | 9  |
|       | 1 | 2   | 3    | 4    | 5       | 6 | 7 | 8   | 9   | 10 |

|       | 1 | 2 | 3  | 4   | 5 | 6        | 7 | 8   | 9   | 10 |
|-------|---|---|----|-----|---|----------|---|-----|-----|----|
| 5(26) | 2 | 5 | 7  |     | $6_6$ |      | 1 | 9!  | 3   |    |
| 5(27) | 2 | 5 | 7  |     | $10_7$ |     | 1 | 10! | 3   | 9  |
| 5(28) | 2 | 5 | 7  |     | $8_6$ |      | 1 | 10! |     | 3  |
| 5(29) | 2 | 5 | 7  |     | $7_6$ |      | 1 |     | 10! | 3  |
| 5(30) | 2 | 5 |    |     | $6_5$ |      |   | 9!  | 10  | 1  |
| 6(1)  | 5 | 6 | 1  |     | 2 | $8_7$    |   | 9!  | 10  | 3  |
| 6(2)  | 5 | 6 | 7  | 3   | 2 | $17_8$   | 1 | 9!  | 4   |    |
| 6(3)  | 5 | 6 | 7  | 3   | 2 | $18_8$   | 1 | 10! |     | 4  |
| 6(4)  | 5 | 6 | 7  | 3   | 2 | $17_8$   | 1 |     | 10! | 4  |
| 6(5)  | 5 | 6 | 7  | 8!  | 2 | $25_7$   | 1 | $3\,3_6$ |   |    |
| 6(6)  | 5 | 6 | 7  | 9!  | 2 | $25_8$   | 1 | 3   | 8   |    |
| 6(7)  | 5 | 6 | 7  | 10! | 2 | $25_9$   | 1 | 3   | 8   | 9  |
| 6(8)  | 5 | 6 | 7  | 10! | 2 | $26_8$   | 1 | 3   |     | 8  |
| 6(9)  | 5 | 6 | 7  |     | 2 | $7_8$    | 1 | 3   | 10! | 8  |
| 6(10) | 5 | 6 | 7  |     | 2 | $3_7$    | 1 | 9!  | 3   |    |
| 6(11) | 5 | 6 | 7  | 10! | 2 | $25_8$   | 1 |     | 3   | 9  |
| 6(12) | 5 | 6 | 7  |     | 2 | $7_8$    | 1 | 10! | 3   | 9  |
| 6(13) | 5 | 6 | 7  | 10! | 2 | $23_7$   | 1 |     |     | 3  |
| 6(14) | 5 | 6 | 7  |     | 2 | $5_7$    | 1 | 10! |     | 3  |
| 6(15) | 5 | 6 | 7  |     | 2 | $4_7$    | 1 |     | 10! | 3  |
| 6(16) | 5 | 6 | 10!|     | 2 | $26_7$   |   | 9   | 1   | 8  |
| 6(17) | 5 | 6 |    | 10! | 2 | $26_7$   |   | 9   | 1   | 8  |
| 6(18) | 5 | 6 | 9! |     | 2 | $25_7$   |   | 10  | 8   | 1  |
| 6(19) | 5 | 6 |    | 9!  | 2 | $26_7$   |   | 10  | 8   | 1  |
| 6(20) | 5 | 6 |    |     | 2 | $3_6$    |   | 9!  | 10  | 1  |
| 6(21) | 5 | 6 | 10!| 1   | 2 | $29_8$   |   | 9   | 4   | 8  |
| 6(22) | 5 | 6 | 9! | 1   | 2 | $28_8$   |   | 10  | 8   | 4  |
| 6(23) | 5 | 6 |    | 1   | 2 | $6_7$    |   | 9!  | 10  | 4  |
| 6(24) | 6 | 1 | 7  | 10! |   | $29_8$   | 2 | 9   | 3   | 8  |
| 6(25) | 6 | 1 | 7  | 9!  |   | $30_8$   | 2 | 10  | 8   | 3  |
| 6(26) | 6 | 1 | 7  |     |   | $7_7$    | 2 | 9!  | 10  | 3  |
| 6(27) | 2 | 6 | 1  |     |   | $11_6$   |   | 9!  | 10  | 3  |
| 6(28) | 2 | 6 |    |     |   | $6_5$    |   | 9!  | 10  | 1  |
| 6(29) | 6 | 7 |    |     |   | $2_6$    | 1 | 9!  | 10  | 2  |
| 6(30) | 6 | 7 | 9! |     |   | $24_7$   | 1 | 10  | 8   | 2  |
| 6(31) | 6 | 7 |    | 9!  |   | $25_7$   | 1 | 10  | 8   | 2  |
|       | 1 | 2 | 3  | 4   | 5 | 6        | 7 | 8   | 9   | 10 |

|        | 1 | 2 | 3   | 4   | 5 | 6      | 7        | 8  | 9  | 10 |
|--------|---|---|-----|-----|---|--------|----------|----|----|----|
| 6(32)) | 6 | 7 | 9!  |     |   | $25_7$ | 1        | 2  | 10 | 8  |
| 6(33)  | 6 | 7 |     | 9!  |   | $26_7$ | 1        | 2  | 10 | 8  |
| 6(34)  | 6 | 7 |     | 5   | 2 | $16_8$ | 1        | 9! | 10 | 4  |
| 7(1)   | 6 | 7 | 1   | 10! |   | 2      | $29_8$   | 9  | 3  | 8  |
| 7(2)   | 6 | 7 | 1   |     |   | 2      | $7_7$    | 9! | 10 | 3  |
| 7(3)   | 2 | 7 | 1   |     |   |        | $10_6$   | 9! | 10 | 3  |
| 7(4)   | 2 | 7 |     | 1   |   |        | $8_6$    | 9! | 10 | 4  |
| 7(5)   | 2 | 7 |     |     |   |        | $5_5$    | 9! | 10 | 1  |
| 7(6)   | 7 | 6 | 10! |     |   | 1      | $25_7$   | 9  | 2  | 8  |
| 7(7)   | 7 | 6 | 9!  |     | 2 | 1      | $28_8$   | 5  | 10 | 8  |
| 7(8)   | 7 | 6 |     | 9!  | 2 | 1      | $29_8$   | 5  | 10 | 8  |
| 7(9)   | 7 | 6 | 9!  |     |   | 1      | $26_7$   | 2  | 10 | 8  |
| 7(10)  | 7 | 6 |     | 5   | 2 | 1      | $17_8$   | 9! | 10 | 4  |
|        | 1 | 2 | 3   | 4   | 5 | 6      | 7        | 8  | 9  | 10 |

P(30)(b)

|       | 1      | 2       | 3 | 4   | 5 | 6 | 7 | 8     | 9    | 10   |
|-------|--------|---------|---|-----|---|---|---|-------|------|------|
| 1(1)  | $17_7$ | 5       | 7 | 3   | 1 |   | 2 | 9!    | 4    |      |
| 1(2)  | $18_7$ | 5       | 7 | 3   | 1 |   | 2 | 10!   |      | 4    |
| 1(3)  | $17_7$ | 5       | 7 | 3   | 1 |   | 2 |       | 10!  | 4    |
| 1(4)  | $25_7$ | 5       | 7 | 9!  | 1 |   | 2 | 3     | 8    |      |
| 1(5)  | $3_7$  | 5       | 7 |     | 1 |   | 2 | 3     | 10!  | 8    |
| 1(6)  | $3_6$  | 5       | 7 |     | 1 |   | 2 | 9!    | 3    |      |
| 1(7)  | $5_6$  | 5       | 7 |     | 1 |   | 2 | 10!   |      | 3    |
| 1(8)  | $4_6$  | 5       | 7 |     | 1 |   | 2 |       | 10!  | 3    |
| 1(9)  | $4_5$  | 5       |   |     | 1 |   |   | 9!    | 10   | 2    |
| 2(1)  | 6      | $5_6$   |   | 1   |   | 2 |   | 9!    | 10   | 4    |
| 2(2)  | 6      | $20_8$  | 7 | 3   |   | 2 | 1 | 4     | 10!  | 8    |
| 2(3)  | 6      | $24_7$  | 7 | 9!  |   | 2 | 1 | $3 2_5$ | $8 2_6$ |      |
| 2(4)  | 6      | $24_8$  | 7 | 10! |   | 2 | 1 | 3     | 8    | 9    |
| 2(5)  | 6      | $25_7$  | 7 | 10! |   | 2 | 1 | $3 2_5$ |      | $8 6_6$ |
| 2(6)  | 6      | $6_7$   | 7 |     |   | 2 | 1 | 3     | 10!  | 8    |
| 2(7)  | 6      | $2_6$   | 7 |     |   | 2 | 1 | 9!    | 3    |      |
| 2(8)  | 6      | $24_7$  | 7 | 10! |   | 2 | 1 |       | 3    | 9    |
| 2(9)  | 6      | $3_7$   | 7 |     |   | 2 | 1 | 10!   | 3    | 9    |
| 2(10) | 6      | $4_6$   | 7 |     |   | 2 | 1 | 10!   |      | 3    |
| 2(11) | 6      | $3_6$   | 7 |     |   | 2 | 1 |       | 10!  | 3    |
| 2(12) | 6      | $2_5$   |   |     |   | 2 |   | 9!    | 10   | 1    |
| 2(13) | 5      | $8_6$   | 1 |     | 2 |   |   | 9!    | 10   | 3    |
| 2(14) | 5      | $6_6$   |   | 1   | 2 |   |   | 9!    | 10   | 4    |
| 2(15) | 5      | $17_7$  | 7 | 3   | 2 |   | 1 | 9!    | 4    |      |
| 2(16) | 5      | $18_7$  | 7 | 3   | 2 |   | 1 | 10!   |      | 4    |
| 2(17) | 5      | $17_7$  | 7 | 3   | 2 |   | 1 |       | 10!  | 4    |
| 2(18) | 5      | $25_7$  | 7 | 9!  | 2 |   | 1 | 3     | 8    |      |
| 2(19) | 5      | $25_8$  | 7 | 10! | 2 |   | 1 | 3     | 8    | 9    |
| 2(20) | 5      | $26_7$  | 7 | 10! | 2 |   | 1 | 3     |      | 8    |
| 2(21) | 5      | $7_7$   | 7 |     | 2 |   | 1 | 3     | 10!  | 8    |
| 2(22) | 5      | $3_6$   | 7 |     | 2 |   | 1 | 9!    | 3    |      |
| 2(23) | 5      | $25_7$  | 7 | 10! | 2 |   | 1 |       | 3    | 9    |
| 2(24) | 5      | $4_7$   | 7 |     | 2 |   | 1 | 10!   | 3    | 9    |
| 2(25) | 5      | $5_6$   | 7 |     | 2 |   | 1 | 10!   |      | 3    |
| 2(26) | 5      | $4_6$   | 7 |     | 2 |   | 1 |       | 10!  | 3    |
| 2(27) | 7      | $4_6$   |   | 1   |   |   | 2 | 9!    | 10   | 4    |

|       | 1 | 2 | 3 | 4 | 5 | 6 | 7 | 8 | 9 | 10 |
|-------|---|---|---|---|---|---|---|---|---|----|
| 2(28) | 7 | $26_7$ | 9! | 1 |   |   | 2 | 10 | 8 | 4 |
| 2(29) | 7 | $4_6$ |   | 1 |   |   | 2 | 9! | 10 | 4 |
| 3(1)  | 7 | 10! | $26_7$ | 1 |   |   | 3 | 9 | 4 | 8 |
| 3(2)  | 7 |   | $3_6$ | 1 |   |   | 3 | 9! | 10 | 4 |
| 3(3)  | 7 | 9! | $25_7$ | 1 |   |   | 3 | 10 | 8 | 4 |
| 3(4)  | 7 |   | $14_6$ | 6 |   | 1 | 3 |   | 10! | 4 |
| 4(1)  | 4 |   |   | $2_4$ |   |   |   | 9! | 10 | 1 |
| 4(2)  | 4 | 7 | 1 | $8_7$ |   |   | 3 | 9! | 10 | 2 |
| 4(3)  | 4 | 8! | 7 | $26_7$ |   |   | 1 | 10 | 3 | 9 |
| 4(4)  | 4 | 7 | 10! | $26_7$ |   |   | 1 | 9 | 2 | 8 |
| 4(5)  | 4 | 7 |   | $4_6$ |   |   | 1 | 9! | 10 | 2 |
| 5(1)  | 6 | 5 | 1 | 10! | $29_8$ | 2 |   | 9 | 3 | 8 |
| 5(2)  | 6 | 5 | 1 | 9! | $30_8$ | 2 |   | 10 | 8 | 3 |
| 5(3)  | 6 | 5 | 1 |   | $7_7$ | 2 |   | 9! | 10 | 3 |
| 5(4)  | 6 | 5 | 4 | 1 | $22_8$ | 2 |   | 3 | 10! | 8 |
| 5(5)  | 6 | 5 | 4 | 1 | $18_7$ | 2 |   | 9! | 3 |   |
| 5(6)  | 6 | 5 | 4 | 1 | $22_8$ | 2 |   | 10! | 3 | 9 |
| 5(7)  | 6 | 5 | 4 | 1 | $20_7$ | 2 |   | 10! |   | 3 |
| 5(8)  | 6 | 5 | 4 | 1 | $19_7$ | 2 |   |   | 10! | 3 |
| 5(9)  | 6 | 5 | 10! | 1 | $28_8$ | 2 |   | 9 | 4 | 8 |
| 5(10) | 6 | 5 | 8! | 1 | $30_8$ | 2 |   | 10 | 4 | 9 |
| 5(11) | 6 | 5 | 9! | 1 | $29_8$ | 2 |   | 10 | 8 | 4 |
| 5(12) | 6 | 5 |   | 1 | $5_7$ | 2 |   | 9! | 10 | 4 |
| 5(13) | 6 | 5 | 8! | 1 | $30_8$ | 2 |   | 10 | 4 | 9 |
| 5(14) | 5 | 1 | 7 | 10! | $29_8$ |   | 2 | 9 | 3 | 8 |
| 5(15) | 2 | 5 | 7 |   | $6_6$ |   | 1 | 9! | 3 |   |
| 5(16) | 2 | 5 | 7 |   | $10_7$ |   | 1 | 10! | 3 | 9 |
| 5(17) | 2 | 5 | 7 |   | $7_6$ |   | 1 | 10! |   | 3 |
| 5(18) | 2 | 5 |   |   | $6_5$ |   |   | 9! | 10 | 1 |
| 5(19) | 2 | 5 | 7 | 10! | $28_8$ |   | 1 | 3 | 8 | 9 |
| 5(20) | 2 | 5 | 7 |   | $10_7$ |   | 1 | 3 | 10! | 8 |
| 5(21) | 2 | 5 | 7 |   | $6_6$ |   | 1 | 9! | 3 |   |
| 5(22) | 2 | 5 | 7 |   | $8_6$ |   | 1 | 10! |   | 3 |
| 5(23) | 2 | 5 | 7 |   | $8_7$ |   | 1 |   | 10! | 3 |
| 5(24) | 2 | 5 |   |   | $6_5$ |   |   | 9! | 10 | 1 |
| 5(25) | 2 | 5 | 7 | 10! | $28_8$ |   | 1 | 3 | 8 | 9 |
|       | 1 | 2 | 3 | 4 | 5 | 6 | 7 | 8 | 9 | 10 |

|       | 1 | 2 | 3   | 4   | 5 | 6      | 7 | 8    | 9   | 10 |
|-------|---|---|-----|-----|---|--------|---|------|-----|----|
| 5(26) | 2 | 5 | 7   |     | $6_6$ |      | 1 | 9!   | 3   |    |
| 5(27) | 2 | 5 | 7   |     | $10_7$ |     | 1 | 10!  | 3   | 9  |
| 5(28) | 2 | 5 | 7   |     | $8_6$ |      | 1 | 10!  |     | 3  |
| 5(29) | 2 | 5 | 7   |     | $7_6$ |      | 1 |      | 10! | 3  |
| 5(30) | 2 | 5 |     |     | $6_5$ |      |   | 9!   | 10  | 1  |
| 6(1)  | 5 | 6 | 1   |     | 2 | $8_7$  |   | 9!   | 10  | 3  |
| 6(2)  | 5 | 6 | 7   | 3   | 2 | $17_8$ | 1 | 9!   | 4   |    |
| 6(3)  | 5 | 6 | 7   | 3   | 2 | $18_8$ | 1 | 10!  |     | 4  |
| 6(4)  | 5 | 6 | 7   | 3   | 2 | $17_8$ | 1 |      | 10! | 4  |
| 6(5)  | 5 | 6 | 7   | 8!  | 2 | $25_7$ | 1 | $33_6$ |   |    |
| 6(6)  | 5 | 6 | 7   | 9!  | 2 | $25_8$ | 1 | 3    | 8   |    |
| 6(7)  | 5 | 6 | 7   | 10! | 2 | $25_9$ | 1 | 3    | 8   | 9  |
| 6(8)  | 5 | 6 | 7   | 10! | 2 | $26_8$ | 1 | 3    |     | 8  |
| 6(9)  | 5 | 6 | 7   |     | 2 | $7_8$  | 1 | 3    | 10! | 8  |
| 6(10) | 5 | 6 | 7   |     | 2 | $3_7$  | 1 | 9!   | 3   |    |
| 6(11) | 5 | 6 | 7   | 10! | 2 | $25_8$ | 1 |      | 3   | 9  |
| 6(12) | 5 | 6 | 7   |     | 2 | $7_8$  | 1 | 10!  | 3   | 9  |
| 6(13) | 5 | 6 | 7   | 10! | 2 | $23_7$ | 1 |      |     | 3  |
| 6(14) | 5 | 6 | 7   |     | 2 | $5_7$  | 1 | 10!  |     | 3  |
| 6(15) | 5 | 6 | 7   |     | 2 | $4_7$  | 1 |      | 10! | 3  |
| 6(16) | 5 | 6 | 10! |     | 2 | $26_7$ |   | 9    | 1   | 8  |
| 6(17) | 5 | 6 |     | 10! | 2 | $26_7$ |   | 9    | 1   | 8  |
| 6(18) | 5 | 6 | 9!  |     | 2 | $25_7$ |   | 10   | 8   | 1  |
| 6(19) | 5 | 6 |     | 9!  | 2 | $26_7$ |   | 10   | 8   | 1  |
| 6(20) | 5 | 6 |     |     | 2 | $3_6$  |   | 9!   | 10  | 1  |
| 6(21) | 5 | 6 | 10! | 1   | 2 | $29_8$ |   | 9    | 4   | 8  |
| 6(22) | 5 | 6 | 9!  | 1   | 2 | $28_8$ |   | 10   | 8   | 4  |
| 6(23) | 5 | 6 |     | 1   | 2 | $6_7$  |   | 9!   | 10  | 4  |
| 6(24) | 6 | 1 | 7   | 10! |   | $29_8$ | 2 | 9    | 3   | 8  |
| 6(25) | 6 | 1 | 7   | 9!  |   | $30_8$ | 2 | 10   | 8   | 3  |
| 6(26) | 6 | 1 | 7   |     |   | $7_7$  | 2 | 9!   | 10  | 3  |
| 6(27) | 2 | 6 | 1   |     |   | $11_6$ |   | 9!   | 10  | 3  |
| 6(28) | 2 | 6 |     |     |   | $6_5$  |   | 9!   | 10  | 1  |
| 6(29) | 6 | 7 |     |     |   | $2_6$  | 1 | 9!   | 10  | 2  |
| 6(30) | 6 | 7 | 9!  |     |   | $24_7$ | 1 | 10   | 8   | 2  |
| 6(31) | 6 | 7 |     | 9!  |   | $25_7$ | 1 | 10   | 8   | 2  |
|       | 1 | 2 | 3   | 4   | 5 | 6      | 7 | 8    | 9   | 10 |

|       | 1 | 2 | 3   | 4   | 5 | 6   | 7    | 8  | 9  | 10 |
|-------|---|---|-----|-----|---|-----|------|----|----|----|
| 6(32))| 6 | 7 | 9!  |     |   | $25_7$ | 1   | 2  | 10 | 8  |
| 6(33) | 6 | 7 |     | 9!  |   | $26_7$ | 1   | 2  | 10 | 8  |
| 6(34) | 6 | 7 |     | 5   | 2 | $16_8$ | 1   | 9! | 10 | 4  |
| 7(1)  | 6 | 7 | 1   | 10! |   | 2   | $29_8$ | 9  | 3  | 8  |
| 7(2)  | 6 | 7 | 1   |     |   | 2   | $7_7$  | 9! | 10 | 3  |
| 7(3)  | 2 | 7 | 1   |     |   |     | $10_6$ | 9! | 10 | 3  |
| 7(4)  | 2 | 7 |     | 1   |   |     | $8_6$  | 9! | 10 | 4  |
| 7(5)  | 2 | 7 |     |     |   |     | $5_5$  | 9! | 10 | 1  |
| 7(6)  | 7 | 6 | 10! |     |   | 1   | $25_7$ | 9  | 2  | 8  |
| 7(7)  | 7 | 6 | 9!  |     | 2 | 1   | $28_8$ | 5  | 10 | 8  |
| 7(8)  | 7 | 6 |     | 9!  | 2 | 1   | $29_8$ | 5  | 10 | 8  |
| 7(9)  | 7 | 6 | 9!  |     |   | 1   | $26_7$ | 2  | 10 | 8  |
| 7(10) | 7 | 6 |     | 5   | 2 | 1   | $17_8$ | 9! | 10 | 4  |
|       | 1 | 2 | 3   | 4   | 5 | 6   | 7    | 8  | 9  | 10 |

P(40)(a)

|       | 1     | 2   | 3      | 4   | 5  | 6 | 7 | 8  | 9    | 10 |
|-------|-------|-----|--------|-----|----|---|---|----|------|----|
| 1(4)  | $16_8$ | 6   | 7      | 3   |    | 1 | 2 | 9! | 10   | 4  |
| 1(8)  | $28_8$ | 6   | 7      | 9!  |    | 1 | 2 | 3  | 10   | 8  |
| 1(9)  | $24_7$ | 6   | 7      | 8!  |    | 1 | 2 | 9  | 3    |    |
|       | $25_7$ | 6   | 7      |     | 8! | 1 | 2 | 9  | 3    |    |
|       | $25_8$ | 6   | 7      | 10! |    | 1 | 2 | 9  | 3    | 8  |
| 1(13) | $2_6$  | 6   | 7      |     |    | 1 | 2 |    | 10!  | 3  |
| 1(14) | $26_8$ | 6   | 7      | 9!  |    | 1 | 2 | 10 | 8    | 3  |
| 1(17) | $21_6$ | 8!  | 7      |     |    |   | 1 | 9  | 10   | 3  |
| 2(4)  | 7     | $22_6$ | 8!  |     |    |   | 2 | 9  | 10   | 1  |
| 3(4)  | 6     | 7   | $23_9$  | 1   | 4  | 2 | 3 | 5  | 10!  | 8  |
| 3(5)  | 6     | 7   | $8_8$   |     | 1  | 2 | 3 | 5  | 10!  | 8  |
| 3(6)  | 6     | 7   | $24_7$  | 8!  |    | 2 | 3 | 9  | 1    |    |
|       | 6     | 7   | $25_8$  | 10! |    | 2 | 3 | 9  | 1    | 8  |
| 3(7)  | 6     | 7   | $30_8$  | 8!  |    | 2 | 3 | 10 | 1    | 9  |
| 3(8)  | 6     | 7   | $27_8$  | 9!  |    | 2 | 3 | 10 | $85_7$ | 1  |
| 3(9)  | 6     | 7   | $2_6$   |     |    | 2 | 3 |    | 10!  | 1  |
| 3(10) | 6     | 7   | $8_8$   | 1   |    | 2 | 3 | 4  | 10!  | 8  |
| 3(12) | 6     | 7   | $32_9$  | 1   | 8! | 2 | 3 | 10 | 4    | 9  |
| 3(14) | 6     | 7   | $5_7$   | 1   |    | 2 | 3 |    | 10!  | 4  |
|       | 1     | 2   | 3      | 4   | 5  | 6 | 7 | 8  | 9    | 10 |

|       | 1 | 2 | 3    | 4   | 5    | 6   | 7   | 8  | 9   | 10 |
|-------|---|---|------|-----|------|-----|-----|----|-----|----|
| 3(16) | 5 | 7 | $25_9$ | 1 | 2 | 4 | 3 | 6 | 10! | 8 |
| 3(17) | 7 | 8! | $20_6$ |   |   |   | 3 | 9 | 10 | 1 |
|       | 7 |   | $22_6$ | 8! |   |   | 3 | 9 | 10 | 1 |
|       | 7 |   | $23_6$ |   | 8! |   | 3 | 9 | 10 | 1 |
| 3(18) | 7 | 6 | $21_9$ | 5 | 2 | 1 | 3 | 4 | 10! | 8 |
| 3(22) | 7 | 6 | $17_6$ | 5 | 2 | 1 | 3 |   | 10! | 4 |
| 3(23) | 7 | 6 | $8_8$ |   | 2 | 1 | 3 | 5 | 10! | 8 |
| 3(24) | 7 | 8 | $20_6$ |   |   |   |   | 9 | 10 | 1 |
| 3(25) | 7 | 6 | $8_8$ |   | 2 | 1 | 3 | 5 | 10! | 8 |
| 4(2)  | 4 | 7 | 6 | $23_8$ |   | 2 | 1 | 3 | 10! | 8 |
| 4(11) | 4 | 5 | 7 | $33_9$ | 1 | 8! | 2 | 10 | 3 | 9 |
| 5(1)  | 7 | 5 | 1 | 8! | $24_7$ |   | 2 | 9 | 10 | 3 |
| 5(2)  | 3 | 5 | 7 | 1 | $14_8$ |   | 2 | 4 | 10! | 8 |
| 5(6)  | 3 | 5 | 7 |   | $11_7$ |   | 2 | 1 | 10! | 8 |
| 5(7)  | 3 | 5 | 7 | 10! | $30_8$ |   | 2 | 9 | 1 | 8 |
| 5(9)  | 3 | 5 | 7 | 9! | $30_8$ |   | 2 | 10 | 8 | 1 |
| 5(10) | 4 | 5 | 7 | 3 | $24_9$ | 1 | 2 | 6 | 10! | 8 |
| 5(11) | 4 | 5 | 7 | 3 | $21_8$ |   | 2 | 1 | 10! | 8 |
|       | 1 | 2 | 3 | 4 | 5 | 6 | 7 | 8 | 9 | 10 |

|       | 1 | 2 | 3 | 4 | 5 | 6 | 7 | 8 | 9 | 10 |
|-------|---|---|---|---|---|---|---|---|---|----|
| 5(15) | 4 | 5 | 7 | 3 | $17_7$ |   | 2 |   | 10! | 1 |
| 5(18) | 5 | 6 | 7 | 3 | $14_8$ | 1 | 2 |   | 10! | 4 |
| 5(19) | 5 | 6 | 7 | 8 | $24_9$ | 1 | 2 | 3 | 10! | 4 |
| 5(22) | 5 | 6 | 7 | 8! | $25_8$ | 1 | 2 | 10 |   | 3 |
|       | 5 | 6 | 7 | 9! | $25_9$ | 1 | 2 | 10 | 8 | 3 |
| 5(23) | 5 | 6 | 7 | 8! | $25_9$ | 1 | 2 | 9 | 10 | 3 |
| 5(28) | 5 | 6 |   | 7 | $19_7$ | 1 | 2 |   | 10! | 4 |
| 5(29) | 5 | 7 | 1 | 6 | $23_9$ | 2 | 3 | 4 | 10! | 8 |
| 6(1)  | 7 | 6 | 1 | 8! |   | $28_8$ | 2 | 9 | 10 | 3 |
| 6(2)  | 6 | 5 | 7 | 3 | 1 | $21_9$ | 2 | 4 | 10! | 8 |
| 6(5)  | 6 | 5 | 7 | 3 | 1 | $17_8$ | 2 |   | 10! | 4 |
| 6(6)  | 6 | 5 | 7 | 10! | 1 | $26_9$ | 2 | 9 | 3 | 8 |
|       | 6 | 5 | 7 | 9 | 1 | $25_7$ | 2 |   | 3 | 4 $25_8$ |
| 6(8)  | 6 | 5 | 7 | 8! | 1 | $28_9$ | 2 | 10 | 3 | 9 |
| 6(9)  | 6 | 5 | 7 | 9! | 1 | $26_9$ | 2 | 10 | 8 | 3 |
| 6(10) | 6 | 5 | 7 |   | 1 | $3_7$ | 2 |   | 10! | 3 |
| 6(13) | 6 | 5 | 8! |   | 1 | $25_7$ |   | 9 | 10 | 2 |
|       | 6 | 5 |   | 8! | 1 | $26_7$ |   | 9 | 10 | 2 |
| 7(5)  | 7 | 5 | 8! |   | 1 |   | $26_7$ | 9 | 10 | 2 |
|       | 7 | 5 |   | 8! | 1 |   | $27_7$ | 9 | 10 | 2 |

*Comment.* Suppose we ranked arcs in each row by their row rank. This will be the same as in MIN(M') except that if two entries have the same value then they have the *same* row rank. Then we could judge paths roughly by the average row rank of their entries. Perhaps we could change MIN(M') by printing S as a suffix if a number of entries in a row have the same value. This would make it easier to choose paths from R & CR. We could judge them by their *average value* instead of by their *average row rank*. It might take more effort to add row rank to each average value but it would make it easier to choose paths that might yield smaller $n$ - cycles.  Average row rank would be most useful when the cost matrix has randomly chosen entries since then it might be a good indicator - along with average arc-value - as *to which paths are most likely to yield an optimal $n$ - cycle or tour.*

*Comment.* We can eliminate PATHS as well as M'(R & CR i)(j) by placing the aav of each subpath in the column number that is identical to the row number. Further, as we extend paths, we place the new value of the path in the same box as the path number of the new arc used to extend the path. In that way, we may even be able to eliminate the writing down of the new path numbers. We need only use P(R & CR i)(j) for the entire algorithm. However, we still need to construct the initial sorted M'(R & CR i), i = 2, 3, …   Henceforth, we shall use this new procedure. It will save running time as well as storage space.

*Comment 1.* Henceforth, we apply the following rules:

Our method expresses a rooted tree of paths such that each path is generally displayed in a sequence of rows. One great advantage of this procedure is that we don't have to backtrack to see if a point is repeated. In what follows, the term "original path" denotes a path that was obtained in the previous iteration. Those points which haven't been used in a path lie in blank entries of the row. Branching is shown in rows underneath the previous path. All entries that belong to an original path are printed in boldface. Since we are using parallel processors, we need only consider the paths obtained using a given processor. In order to identify each path separately, every time a path reaches $i$ acceptable branches, we construct $i - 1$ rows underneath it so that each path is unique. It thus may take several rows to represent a new path. Upon reaching the last point of a successful acceptable path, it must go into a blank entry of the row in which it lies.  In what follows, the "name" of a path is its original name followed by the ordered pairs denoting the points in which its branch points lie. Before going on, we note that (1) the set of ordered pairs containing its branch points uniquely identifies a path.; (2) using (1), by obtaining the point whose column value is closest on the left to that of a proposed last point of a path during the current iteration, we can peruse that row to see if the proposed final point leads to an acceptable path (There may be an entry that repeats a number of the path.). We now discuss columns where the last point of an acceptable path may not be placed.  It can't lie in a column containing a boldface entry of the original path. It can't lie in a column containing one of the branch points in its name or a non-blank entry in the same row of such a branch point. However, given an entry that doesn't occur before the last branch point in the current path, then we can place an entry in any column whose number is greater than that of the branch point's. If at the end of an iteration a path can't be extended, we delete it before proceeding further. Finally, it may occur that two different paths have the same name. This because the respective branch points of the paths -although not identical - may lie in the same columns. However, this doesn't cause difficulties since when we renumber the paths they will then have different numbers.

(2) We use only P(R & CR i) as expressed in matrix form.

(3) Each row contains the points of an acceptable path. The average arc-value of the path is placed in the column whose number is that of the first number of the path.

4) The endpoint of each path has an exclamation point following it. These points indicate from where we should try to extend the path. Once we have worked with an entry having an exclamation point following it, we remove the exclamation point.

(5) We print in boldface each point that extends a path. This allows us to distinguish points of the original path from those that extend it.

(6) If more than one point emanates from a column of the path, we add a row underneath the row containing the path and place the second point in its correct column. No matter how many points emanate from a column of a path, we need only place points that extend the path in new paths underneath the row of the original path. Thus, we may have 100(23) followed by 100(23,1), 100(23,2),

…, etc. .

(7) We place an exclamation point after any point - in any new row - that completes an iteration of a path.

(8) We print in boldface any point in a new row that has an exclamation point after it.

(9) In order to obtain all extended paths as easily as possible, we do the following:

   (a) Let $P$ be an original path while $P_E$ is an extension of $P$. Let $B_i$ be a block of $P$ of maximum length that has no point of $P_E$ underneath it. We then copy $B_i$ to $P_E$.

   (b) After completing this for all subscripts $i$, we delete the exclamation point after the endpoint of $P$ that was contained in a block that has been copied and change it from boldface to regular print.

(10) At the end of an iteration, we place an asterisk in front of any row that has extended an original path.

(11) We delete all rows that have not been extended including original paths.

(12) We print all paths that have been extended.

(13) We renumber the paths obtained in (11). The matrix obtained becomes P(R & CR (i+1)).

*Comment 2.* Suppose that we obtain an acceptable path containing $n$ arcs, say $C = (a_1 \ldots a_r f_1 f_2 \ldots f_s)$. Now assume that $F = (f_1 f_2 \ldots f_s a_1)$ has the property that every one of its arcs has an arc-value less than the average arc-value of $T_{UPPERBOUND}$. From this assumption, it follows that $F$ is a red block. This implies that $A = (a_1 a_2 \ldots a_r f_1)$ is a set of alternating red and black blocks. Therefore, $f_1$ is also a determining point of $C$. Thus, we may expect the final arcs of an $n$-cycle obtained by applying FWK to often be greater than or equal to the average arc-value of $T_{UPPERBOUND}$.

In the following list, we have deleted all paths that could not be extended during the next iteration.

P(40)(a)

|  | 1 | 2 | 3 | 4 | 5 | 6 | 7 | 8 | 9 | 10 |
|---|---|---|---|---|---|---|---|---|---|---|
| 1(1) | $16_8$ | 6 | 7 | 3 |  | 1 | 2 | 9! | 10 | 4 |
| 1(2) | 5-- <u>38</u> | 6 | 7 | 10 | 4 | 1 | 2 | 3 | 8 | 9 |
| 1(3) | $28_8$ | 6 | 7 | 8! |  | 1 | 2 | 10 | 3 | 9 |
| 1(4) | $24_8$ | 6 | 7 | 8! |  | 1 | 2 | 9 | 10 | 3 |
| 1(5) | $26_7$ | 6 | 7 | 8! |  | 1 | 2 | 10 |  | 3 |
| 1(6) | $22_8$ | 6 | 7 | 10 |  | 1 | 2 | 9! | 4 | 3 |
| 1(7) | $21_6$ | 8! | 7 |  |  |  | 1 | 9 | 10 | 3 |
| 1(8) | $23_6$ |  | 7 | 8! |  |  | 1 | 9 | 10 | 3 |
| 3(1) | 6 | 7 | $30_9$ | 9! | 1 | 2 | 3 | 5 | 10 | 8 |
| 3(2) | 6 | 7 | $24_7$ | 8! |  | 2 | 3 | 9 | 1 |  |
| 3(3) | 6 | 7 | $25_7$ |  | 8! | 2 | 3 | 9 | 1 |  |
| 3(4) | 6 | 7 | $30_8$ | 8! |  | 2 | 3 | 10 | 1 | 9 |
| 3(5) | 6 | 7 | $27_8$ | 9! |  | 2 | 3 | 10 | 8 | 1 |
| 3(6) | 6 | 7 | $2_7$ |  |  | 2 | 3 | 9! | 10 | 1 |
| 3(7) | 6 | 7 | $24_7$ | 9! |  | 2 | 3 |  | 10 | 1 |
| 3(8) | 6 | 7 | $32_9$ | 1 | 8! | 2 | 3 | 10 | 4 | 9 |
| 3(9) | 6 | 7 | $29_8$ | 1 | 8! | 2 | 3 | 10 |  | 4 |
| 3(10) | 6 | 7 | $28_9$ | 1 | 8! | 2 | 3 | 9 | 10 | 4 |
| 3(11) | 7 | 6 | $21_9$ | 5 | 2 | 1 | 3 | 4 | 10! | 8 |
| 3(12) | 7 | 6 | $17_8$ | 5 | 2 | 1 | 3 | 9! | 4 |  |
| 3(13) | 7 | 6 | $23_9$ | 5 | 2 | 1 | 3 | 10! | 4 | 9 |
| 3(14) | 7 | 6 | $17_8$ | 5 | 2 | 1 | 3 |  | 10! | 4 |
| 3(15) | 7 | 6 | $18_8$ | 5 | 2 | 1 | 3 | 10! |  | 4 |
| 3(16) | 7 | 6 | $30_9$ | 9! | 2 | 1 | 3 | 5 | 10 | 8 |
| 4(1) | 4 | 5 | 7 | $29_8$ | 1 | 8! | 2 | 9 | 3 |  |
| 5(1) | 7 | 5 | 1 | 8! | $24_7$ |  | 2 | 9 | 10 | 3 |
| 5(2) | 3 | 5 | 7 | 10! | $30_8$ |  | 2 | 9 | 1 | 8 |
| 5(3) | 5 | 6 | 7 | 8! | $24_8$ | 1 | 2 | 9 | 3 |  |
| 5(4) | 5 | 6 | 7 | 9! | $24_8$ | 1 | 2 |  | 10 | 3 |
| 5(5) | 5 | 6 | 7 | 8! | $24_9$ | 1 | 2 | 9 | 10 | 3 |
| 5(6) | 5 | 6 | 7 | 9! | $25_8$ | 1 | 2 |  | 10 | 3 |
|  | 1 | 2 | 3 | 4 | 5 | 6 | 7 | 8 | 9 | 10 |

|      | 1 | 2 | 3 | 4   | 5      | 6      | 7 | 8   | 9  | 10 |
|------|---|---|---|-----|--------|--------|---|-----|----|----|
| 5(7) | 5 | 6 |   | 7   | $19_8$ | 1      | 2 | 9!  | 10 | 4  |
| 6(1) | 7 | 6 | 1 | 8!  |        | $28_8$ | 2 | 9   | 10 | 3  |
| 6(2) | 6 | 5 | 7 | 3   | 1      | $17_9$ | 2 | 9!  | 10 | 4  |
| 6(3) | 6 | 5 | 7 | 10! | 1      | $26_9$ | 2 | 9   | 3  | 8  |
| 6(4) | 6 | 5 | 7 | 9   | 1      | $28_9$ | 2 | 10! | 3  | 4  |
| 6(5) | 6 | 5 | 7 | 8!  | 1      | $29_9$ | 2 | 10  | 3  | 9  |
| 6(6) | 6 | 5 | 7 | 8!  | 1      | $26_8$ | 2 | 10  |    | 3  |
| 6(7) | 6 | 5 | 7 | 9!  | 1      | $26_9$ | 2 | 10  | 8  | 3  |
| 6(8) | 6 | 5 | 7 |     | 1      | $3_8$  | 2 | 9!  | 10 | 3  |
|      | 1 | 2 | 3 | 4   | 5      | 6      | 7 | 8   | 9  | 10 |

P(40)(b)

|       | 1      | 2      | 3      | 4      | 5 | 6 | 7 | 8   | 9   | 10 |
|-------|--------|--------|--------|--------|---|---|---|-----|-----|----|
| 1(1)  | $17_8$ | 5      | 7      | 3      | 1 |   | 2 | 9!  | 10  | 4  |
| 1(2)  | $25_8$ | 5      | 7      | 9!     | 1 |   | 2 | 3   | 10  | 8  |
| 1(3)  | $26_8$ | 5      | 7      | 10!    | 1 |   | 2 | 9   | 3   | 8  |
| 1(4)  | $27_8$ | 5      | 7      | 9!     | 1 |   | 2 | 10  | 8   | 3  |
| 1(5)  | $26_7$ | 5      | 7      | 9!     | 1 |   | 2 |     | 10  | 3  |
| 1(6)  | $4_7$  | 5      | 7      |        | 1 |   | 2 | 9!  | 10  | 3  |
| 2(1)  | 6      | $26_7$ | 8!     | 1      |   | 2 |   | 9   | 10  | 4  |
| 2(2)  | 6      | $28_8$ | 7      | 9!     |   | 2 | 1 | 3   | 10  | 8  |
| 2(3)  | 6      | $25_8$ | 7      | 10!    |   | 2 | 1 | 9   | 3   | 8  |
| 2(4)  | 6      | $25_8$ | 7      | 8!     |   | 2 | 1 | 10  | 3   | 9  |
| 2(5)  | 6      | $26_8$ | 7      | 8!     |   | 2 | 1 | 10  | 8   | 3  |
| 2(6)  | 6      | $25_7$ | 7      | 9!     |   | 2 | 1 |     | 10  | 3  |
| 2(7)  | 6      | $3_7$  | 7      |        |   | 2 | 1 | 9!  | 10  | 3  |
| 2(8)  | 5      | $17_8$ | 7      | 3      | 2 |   | 1 | 9!  | 10  | 4  |
| 2(9)  | 5      | $7_7$  | 7      |        | 2 |   | 1 | 3   | 10! | 8  |
| 2(10) | 5      | $25_7$ | 7      | 8!     | 2 |   | 1 | 9   | 3   |    |
| 2(11) | 5      | $26_8$ | 7      | 10!    | 2 |   | 1 | 9   | 3   | 8  |
| 2(12) | 5      | $26_8$ | 7      | 8!     | 2 |   | 1 | 10  | 3   | 9  |
| 2(13) | 5      | $27_8$ | 7      | 9!     | 2 |   | 1 | 10  | 8   | 3  |
| 2(14) | 5      | $26_7$ | 7      | 9!     | 2 |   | 1 |     | 10  | 3  |
| 2(15) | 5      | $4_7$  | 7      |        | 2 |   | 1 | 9!  | 10  | 3  |
| 2(16) | 7      | $25_7$ | 8!     | 1      |   |   | 2 | 9   | 10  | 4  |
| 2(17) | 7      | $25_7$ | 8!     | 1      |   |   | 2 | 9   | 10  | 4  |
| 3(1)  | 7      | 8!     | $25_7$ | 1      |   |   | 3 | 9   | 10  | 4  |
| 3(2)  | 7      |        | $14_7$ | 6      |   | 1 | 3 | 9!  | 10  | 4  |
| 4(1)  | 4      | 7      | 8!     | $25_7$ |   |   | 1 | 9   | 10  | 2  |

|       | 1 | 2 | 3  | 4   | 5      | 6     | 7 | 8  | 9  | 10 |
|-------|---|---|----|-----|--------|-------|---|----|----|----|
| 5(1)  | 6 | 5 | 1  | 8!  | $29_8$ | 2     |   | 9! | 10 | 3  |
| 5(2)  | 6 | 5 | 4  | 1   | $19_8$ | 2     |   | 9! | 10 | 3  |
| 5(3)  | 6 | 5 | 8! | 1   | $26_8$ | 2     |   | 9  | 10 | 4  |
| 5(4)  | 2 | 5 | 7  | 10! | $29_8$ |       | 1 | 9  | 3  | 8  |
| 5(5)  | 2 | 5 | 7  | 9!  | $29_8$ |       | 1 | 10 | 8  | 3  |
| 5(6)  | 2 | 5 | 7  | 10! | $26_8$ |       | 1 | 9  | 3  | 8  |
| 5(7)  | 2 | 5 | 7  | 9!  | $30_8$ |       | 1 | 10 | 8  | 3  |
| 5(8)  | 2 | 5 | 7  |     | $8_7$  |       | 1 | 9! | 10 | 3  |
| 5(9)  | 2 | 5 | 7  | 10! | $29_8$ |       | 1 | 9  | 3  | 8  |
| 5(10) | 2 | 5 | 7  | 9!  | $30_8$ |       | 1 | 10 | 8  | 3  |
| 5(11) | 2 | 5 | 7  |     | $7_7$  |       | 1 | 9! | 10 | 3  |
| 6(1)  | 5 | 6 | 1  | 8!  | 2      | $30_8$ |  | 9  | 10 | 3  |
| 6(2)  | 5 | 6 | 7  | 3   | 2      | $17_9$ | 1 | 9! | 10 | 4  |
| 6(3)  | 5 | 6 | 7  | 9!  | 2      | $29_9$ | 1 | 3  | 10 | 8  |
| 6(4)  | 5 | 6 | 7  | 10! | 2      | $29_9$ | 1 | 9  | 3  | 8  |
| 6(5)  | 5 | 6 | 7  | 8!  | 2      | $29_9$ | 1 | 10 | 3  | 9  |
| 6(4)  | 5 | 6 | 7  | 10! | 2      | $29_9$ | 1 | 9  | 3  | 8  |
| 6(5)  | 5 | 6 | 7  | 8!  | 2      | $29_9$ | 1 | 10 | 3  | 9  |
| 6(6)  | 5 | 6 | 7  | 10  | 2      | $23_9$ | 1 | 9! | 4  | 3  |
| 6(7)  | 5 | 6 | 7  | 8!  | 2      | $27_8$ | 1 | 10 |    | 3  |
| 6(8)  | 5 | 6 | 7  | 9!  | 2      | $27_9$ | 1 | 10 | 8  | 3  |
| 6(9)  | 5 | 6 | 7  | 9!  | 2      | $26_8$ | 1 |    | 10 | 3  |
| 6(10) | 5 | 6 | 10!|     | 2      | $26_7$ |   | 9  | 1  | 8  |
| 6(11) | 5 | 6 |    | 10! | 2      | $26_7$ |   | 9  | 1  | 8  |
| 6(12) | 5 | 6 | 8! |     | 2      | $24_7$ |   | 9  | 10 | 1  |
| 6(13) | 5 | 6 |    | 8!  | 2      | $25_7$ |   | 9  | 10 | 1  |
| 6(14) | 5 | 6 | 8! | 1   | 2      | $27_8$ |   | 9  | 10 | 4  |
| 6(15) | 6 | 1 | 7  | 8!  |        | $29_8$ | 2 | 9  | 10 | 3  |
| 6(16) | 6 | 7 |    | 8!  |        | $24_7$ | 1 | 9  | 10 | 2  |
|       | 1 | 2 | 3  | 4   | 5      | 6     | 7 | 8  | 9  | 10 |

P(50)(a)

|       | 1      | 2 | 3      | 4  | 5 | 6 | 7 | 8  | 9  | 10 |
|-------|--------|---|--------|----|---|---|---|----|----|----|
| 1(1)  | 5-- 38 | 6 | 7      | 10 | 4 | 1 | 2 | 3  | 8  | 9  |
| 3(15) | 7      | 6 | $17_9$ | 5  | 2 | 1 | 3 | 9! | 10 | 4  |
|       | 1      | 2 | 3      | 4  | 5 | 6 | 7 | 8  | 9  | 10 |

P(50)(b)

|  | 1 | 2 | 3 | 4 | 5 | 6 | 7 | 8 | 9 | 10 |
|---|---|---|---|---|---|---|---|---|---|---|
| 1(6) | $26_8$ | 5 | 7 | 8! | 1 |  | 2 | 9 | 10 | 3 |
| 2(7) | 6 | $25_8$ | 7 | 8! |  | 2 | 1 | 9 | 10 | 3 |
| 2(9) | 5 | $29_8$ | 7 | 9! | 2 |  | 1 | 3 | 10 | 8 |
| 2(15) | 5 | $26_8$ | 7 | 8! | 2 |  | 1 | 9 | 10 | 3 |
| 5(8) | 2 | 5 | 7 |  | $8_7$ |  | 1 | 9 | 10 | 3 |
| 5(9) | 2 | 5 | 7 | 10! | $29_8$ |  | 1 | 9 | 3 | 8 |
| 5(10) | 2 | 5 | 7 | 9! | $30_8$ |  | 1 | 10 | 8 | 3 |
| 5(11) | 2 | 5 | 7 |  | $7_7$ |  | 1 | 9! | 10 | 3 |
| 6(1) | 5 | 6 | 1 | 8! | 2 | $30_8$ |  | 9 | 10 | 3 |
| 6(2) | 5 | 6 | 7 | 3 | 2 | $17_9$ | 1 | 9! | 10 | 4 |
| 6(3) | 5 | 6 | 7 | 9! | 2 | $29_9$ | 1 | 3 | 10 | 8 |
| 6(4) | 5 | 6 | 7 | 10! | 2 | $29_9$ | 1 | 9 | 3 | 8 |
| 6(5) | 5 | 6 | 7 | 8! | 2 | $29_9$ | 1 | 10 | 3 | 9 |
| 6(4) | 5 | 6 | 7 | 10! | 2 | $29_9$ | 1 | 9 | 3 | 8 |
| 6(5) | 5 | 6 | 7 | 8! | 2 | $29_9$ | 1 | 10 | 3 | 9 |
| 6(6) | 5 | 6 | 7 | 10 | 2 | $23_9$ | 1 | 9! | 4 | 3 |
| 6(7) | 5 | 6 | 7 | 8! | 2 | $27_8$ | 1 | 10 |  | 3 |
| 6(8) | 5 | 6 | 7 | 9! | 2 | $27_9$ | 1 | 10 | 8 | 3 |
| 6(9) | 5 | 6 | 7 | 9! | 2 | $26_8$ | 1 |  | 10 | 3 |
| 6(10) | 5 | 6 | 10! |  | 2 | $26_7$ |  | 9 | 1 | 8 |
| 6(11) | 5 | 6 |  | 10! | 2 | $26_7$ |  | 9 | 1 | 8 |
| 6(12) | 5 | 6 | 8! |  | 2 | $24_7$ |  | 9 | 10 | 1 |
| 6(13) | 5 | 6 |  | 8! | 2 | $25_7$ |  | 9 | 10 | 1 |
| 6(14) | 5 | 6 | 8! | 1 | 2 | $27_8$ |  | 9 | 10 | 4 |
| 6(15) | 6 | 1 | 7 | 8! |  | $29_8$ | 2 | 9 | 10 | 3 |
| 6(16) | 6 | 7 |  | 8! |  | $24_7$ | 1 | 9 | 10 | 2 |
|  | 1 | 2 | 3 | 4 | 5 | 6 | 7 | 8 | 9 | 10 |

3(15) yields a cycle whose aav is 3.8 .

P(10)(b)

|  | 1 | 2 | 3 | 4 | 5 | 6 | 7 | 8 | 9 | 10 |
|---|---|---|---|---|---|---|---|---|---|---|
| 1(1) | $3_2$ |  |  | 1 |  |  |  | 4 | 10! | 8 |
| 1(2) | $3_3$ |  |  | 1 |  |  |  | 9! | 4 |  |
| 1(3) | $7_4$ |  |  | 1 |  |  |  | 10! | 4 | 9 |
| 1(4) | $3_3$ |  |  | 1 |  |  |  |  | 10! | 4 |
| 1(5) | $4_3$ |  |  | 1 |  |  |  | 10! |  | 4 |
| 1(6) | $4_3$ |  |  |  |  |  |  | 10! | 1 | 9 |
| 1(7) | $6_4$ |  |  |  | 1 |  |  | 5 | 10! | 8 |
| 1(8) | $2_2$ | 5 |  |  | 1 |  |  |  |  |  |
| 2(1) | 6! | $1_2$ |  |  |  | 2 |  |  |  |  |
|  |  | $2_2$ |  |  |  |  |  |  | 10! | 2 |
|  |  | $7_4$ |  |  | 2 |  |  | 5 | 10! | 8 |
|  | 5! | $3_2$ |  |  | 2 |  |  |  |  |  |
|  |  | $5_3$ |  |  |  |  |  | 10! | 2 | 9 |
|  |  | $8_4$ |  |  |  | 2 |  | 6 | 10! | 8 |
|  | 7! | $1_2$ |  |  |  |  | 2 |  |  |  |
| 3(1) | 7! |  | $0_3$ |  |  |  | 3 |  |  |  |
| 4(1) | 4 |  |  | $6_4$ |  |  |  | 1 | 10! | 8 |
|  | 4 |  |  | $2_3$ |  |  |  | 9! | 1 |  |
|  | 4 |  |  | $2_3$ |  |  |  |  | 10! | 1 |
|  | 4 |  |  | $3_3$ |  |  |  | 10! |  | 1 |
|  |  |  |  | $4_3$ |  |  |  | 4 | 10! | 8 |
|  | 4 | 7! | 1 | $6_4$ |  |  | 3 |  |  |  |
|  | 4 | 7! |  | $2_3$ |  |  | 1 |  |  |  |
|  | 4 |  | 7! | $2_3$ |  |  | 1 |  |  |  |
| 5(1) | 2! | 5 |  |  | $6_2$ |  |  |  |  |  |
| 5() | 6! | 5 |  |  | $2_3$ | 2 |  |  |  |  |
|  |  | 5 |  |  | $3_3$ |  |  | 10! |  | 2 |
|  |  | 5 |  |  | $2_3$ |  |  |  | 10! | 2 |
|  | 5 | 1 | 7! |  | $6_4$ |  | 2 |  |  |  |
|  | 5 | 1 |  |  | $6_4$ |  |  | 9! | 2 |  |
|  | 5 | 1 |  |  | $7_4$ |  |  |  | 10! | 2 |
|  | 6 | 1 | 7! |  |  | $6_4$ | 2 |  |  |  |
|  | 6 |  | 7! |  |  | $0_3$ | 1 |  |  |  |
|  | 5! | 6 |  |  | 2 | $2_3$ |  |  |  |  |
|  |  | 6 |  |  |  | $2_3$ |  |  | 10! | 2 |
|  | 7 | 1 |  |  |  | $7_4$ |  |  | 10! | 2 |
|  | 7 | 6! |  |  |  | 1 | $1_3$ |  |  |  |
|  | 2! | 7 |  |  |  | $6_2$ |  |  |  |  |
|  | 5! | 7 |  |  | 2 |  | $3_3$ |  |  |  |
|  | 1 | 2 | 3 | 4 | 5 | 6 | 7 | 8 | 9 | 10 |

|   | 1  | 2 | 3 | 4 | 5 | 6 | 7     | 8  | 9   | 10 |
|---|----|---|---|---|---|---|-------|----|-----|----|
|   | 6! | 7 |   |   |   | 2 | $2_3$ |    |     |    |
|   |    | 7 |   |   |   |   | $1_3$ | 9! | 2   |    |
|   |    | 7 |   |   |   |   | $2_3$ |    | 10! | 2  |
|   | 1  | 2 | 3 | 4 | 5 | 6 | 7     | 8  | 9   | 10 |

P(20)(b)

|        | 1      | 2        | 3  | 4 | 5 | 6 | 7 | 8   | 9   | 10 |
|--------|--------|----------|----|---|---|---|---|-----|-----|----|
| 1(1)   | $3_4$  | 5        | 7! |   | 1 |   | 2 |     |     |    |
| 1(2)   | $6_5$  | 5        |    |   | 1 |   |   | 2   | 10! | 8  |
| 1(3)   | $3_4$  | 5        |    |   | 1 |   |   | 9!  | 2   |    |
| 1(4)   | $7_5$  | 5        |    |   | 1 |   |   | 10! | 2   | 9  |
| 1(5)   | $5_4$  | 5        |    |   | 1 |   |   | 10! |     | 2  |
| 1(6)   | $4_4$  | 5        |    |   | 1 |   |   |     | 10! | 2  |
| 2(1)   | 6      | $9_6$    |    | 1 |   | 2 |   | 4   | 10! | 8  |
| 2(2)   | 6      | $5_5$    |    | 1 |   | 2 |   | 9!  | 4   |    |
| 2(3)   | 6      | $6_5$    |    | 1 |   | 2 |   | 10! | 4   | 9  |
| 2(4)   | 6      | $5_5$    |    | 1 |   | 2 |   |     | 10! | 4  |
| 2(5)   | 6      | $6_5$    |    | 1 |   | 2 |   | 10! |     | 4  |
| 2(6)   | 6      | $2_4$    | 7! |   |   | 2 | 1 |     |     |    |
| 2(7)   | 6      | $6_5$    |    |   |   | 2 |   | 1   | 10! | 8  |
| 2(8)   | 6      | $2_4$    |    |   |   | 2 |   | 9!  | 1   |    |
| 2(9)   | 6      | $6_5$    |    |   |   | 2 |   | 10! | 1   | 9  |
| 2(10)  | 6      | $2_4$    |    |   |   | 2 |   |     | 10! | 1  |
| 2(11)  | 6      | $3_4$    |    |   |   | 2 |   | 10! |     | 1  |
| 2(12)  |        | $2_3$    |    |   |   |   |   | 9!  | 10  | 2  |
| 2(14)  | 5      | $11_6$   | 1  |   | 2 |   |   | 3   | 10! | 8  |
| 2(15)  | 5      | $7_5$    | 1  |   | 2 |   |   | 9!  | 3   |    |
| 2(16)  | 5      | $11_6$   | 1  |   | 2 |   |   | 10! | 3   | 9  |
| 2(17)  | 5      | $8_5$    | 1  |   | 2 |   |   |     | 10! | 3  |
| 2(18)  | 5      | $9_5$    | 1  |   | 2 |   |   | 10! |     | 3  |
| 2(19)  | 5      | $10_6$   |    | 1 | 2 |   |   | 4   | 10! | 8  |
| 2(20)  | 5      | $6_5$    |    | 1 | 2 |   |   | 9!  | 4   |    |
| 2(21)  | 5      | $10_6$   |    | 1 | 2 |   |   | 10! | 4   | 9  |
| 2(22)  | 5      | $6_5$    |    | 1 | 2 |   |   |     | 10! | 4  |
| 2(23)  | 5      | $10_6$   |    |   | 2 | 1 |   | 6   | 10! | 8  |
| 2(24)  | 5      | $3_4$    | 7! |   | 2 |   | 1 |     |     |    |
|        | 1      | 2        | 3  | 4 | 5 | 6 | 7 | 8   | 9   | 10 |

|       | 1 | 2    | 3   | 4    | 5 | 6 | 7 | 8   | 9   | 10 |
|-------|---|------|-----|------|---|---|---|-----|-----|----|
| 2(25) | 5 | $7_5$ |     |      | 2 |   |   | 1   | 10! | 8  |
| 2(26) | 5 | $3_4$ |     |      | 2 |   |   | 9!  | 1   |    |
| 2(27) | 5 | $7_5$ |     |      | 2 |   |   | 10! | 1   | 9  |
| 2(28) | 5 | $3_4$ |     |      | 2 |   |   |     | 10! | 1  |
| 2(29) | 5 | $4_4$ |     |      | 2 |   |   | 10! |     | 1  |
| 2(30) |   | $5_3$ |     |      |   |   |   | 10! | 2   | 9  |
| 2(31) |   | $8_4$ |     |      |   | 2 |   | 6   | 10! | 8  |
| 2(32) | 7 | $2_4$ |     |      |   |   | 2 | 10! |     | 1  |
| 2(33) | 7 | $25_7$ | 10! |      | 1 |   | 2 | 5   | 8   | 9  |
| 2(34) | 7 | $25_7$ |     | 10!  | 1 |   | 2 | 5   | 8   | 9  |
| 2(35) | 7 | $7_6$ |     |      | 1 |   | 2 | 5   | 10! | 8  |
| 2(36) | 7 | $26_7$ | 10! | 1    |   |   | 2 | 4   | 8   | 9  |
| 2(37) | 7 | $7_6$ |     | 1    |   |   | 2 | 4   | 10! | 8  |
| 2(38) | 7 | $4_5$ |     | 1    |   |   | 2 | 9!  | 4   |    |
| 2(39) | 7 | $8_6$ |     | 1    |   |   | 2 | 10! | 4   | 9  |
| 2(40) | 7 | $4_5$ |     | 1    |   |   | 2 |     | 10! | 4  |
| 2(41) | 7 | $5_5$ |     | 1    |   |   | 2 | 10! |     | 4  |
| 2(42) | 7 | $8_6$ |     |      |   | 1 | 2 | 6   | 10! | 8  |
| 2(43) | 7 | $4_5$ |     | 1    |   |   | 2 | 9!  | 4   |    |
| 2(44) | 7 | $8_6$ |     | 1    |   |   | 2 | 10! | 4   | 9  |
| 2(45) | 7 | $5_5$ |     | 1    |   |   | 2 | 10! |     | 4  |
| 2(46) | 7 | $4_5$ |     | 1    |   |   | 2 |     | 10! | 4  |
| 2(47) | 7 | $23_7$ | 10! |      | 1 |   | 2 | 5   | 8   | 9  |
| 3(1)  | 7 |      | $7_6$ | 1    |   |   | 3 | 4   | 10! | 8  |
| 3(2)  | 7 |      | $3_5$ | 1    |   |   | 3 | 9!  | 4   |    |
| 3(3)  | 7 |      | $7_6$ | 1    |   |   | 3 | 10! | 4   | 9  |
| 3(4)  | 7 |      | $3_5$ | 1    |   |   | 3 |     | 10! | 4  |
| 3(5)  | 7 |      | $4_5$ | 1    |   |   | 3 | 10! |     | 4  |
| 3(6)  | 7 |      | $4_5$ |      |   |   | 3 | 1   | 10! | 8  |
| 3(7)  | 7 |      | $18_7$ | 6!   |   | 1 | 3 |     |     |    |
| 3(8)  | 7 |      | $6_6$ |      | 1 |   | 3 | 5   | 10! | 8  |
| 4(1)  | 4 | 7    | 1   | $13_8$ | 2 |   | 3 | 5   | 10! | 8  |
| 4(2)  | 4 | 7    | 1   | $14_8$ |   | 2 | 3 | 6   | 10! | 8  |
| 4(3)  | 4 | 7    | 1   | $10_7$ |   |   | 3 | 2   | 10! | 8  |
| 4(4)  | 4 | 7    | 1   | $7_6$  |   |   | 3 | 9!  | 2   |    |
| 4(5)  | 4 | 7    | 1   | $11_7$ |   |   | 3 | 10! | 2   | 9  |
| 4(6)  | 4 | 7    | 1   | $9_6$  |   |   | 3 | 10! |     | 2  |
|       | 1 | 2    | 3   | 4    | 5 | 6 | 7 | 8   | 9   | 10 |

|  | 1 | 2 | 3 | 4 | 5 | 6 | 7 | 8 | 9 | 10 |
|---|---|---|---|---|---|---|---|---|---|---|
| 4(7) | 4 | 7 | 1 | $8_6$ |  |  | 3 |  | 10! | 2 |
| 4(8) | 4 |  | 7 | $10_6$ |  |  | 1 | 3 | 10! | 8 |
| 4(9) | 4 |  | 7 | $6_5$ |  |  | 1 | 9! | 3 |  |
| 4(10) | 4 |  | 7 | $6_6$ |  |  | 1 | 10! | 3 | 9 |
| 4(11) | 4 | 7 |  | $3_5$ |  |  | 1 | 9! | 2 |  |
| 4(12) | 4 | 7 |  | $7_6$ |  |  | 1 | 10! | 2 | 9 |
| 4(13) | 4 | 7 |  | $4_5$ |  |  | 1 |  | 10! | 2 |
| 4(14) | 4 | 7 |  | $5_5$ |  |  | 1 | 10! |  | 2 |
| 5(1) | 6 | 5 | 1 | 10! | $28_8$ | 2 |  | 3 | 8 | 9 |
| 5(2) | 6 | 5 | 1 |  | $10_7$ | 2 |  | 3 | 10! | 8 |
| 5(3) | 6 | 5 | 1 |  | $6_6$ | 2 |  | 9! | 3 |  |
| 5(4) | 6 | 5 | 1 |  | $10_7$ | 2 |  | 10! | 3 | 9 |
| 5(5) | 6 | 5 | 1 |  | $8_6$ | 2 |  | 10! |  | 3 |
| 5(6) | 6 | 5 | 1 |  | $7_6$ | 2 |  |  | 10! | 3 |
| 5(7) | 6 | 5 | 4! | 1 | $18_5$ | 2 |  |  |  |  |
| 5(8) | 6 | 5 | 10! | 1 | $27_8$ | 2 |  | 4 | 8 | 9 |
| 5(9) | 6 | 5 |  | 1 | $9_7$ | 2 |  | 4 | 10! | 8 |
| 5(10) | 6 | 5 |  | 1 | $5_6$ | 2 |  | 9! | 4 |  |
| 5(11) | 6 | 5 |  | 1 | $9_7$ | 2 |  | 10! | 4 | 9 |
| 5(12) | 6 | 5 |  | 1 | $6_6$ | 2 |  | 10! |  | 4 |
| 5(13) | 6 | 5 |  | 1 | $5_6$ | 2 |  |  | 10! | 4 |
| 5(14) | 6 | 5 |  |  | $6_6$ | 2 |  | 1 | 10! | 8 |
| 5(15) |  | 5 |  |  | $2_4$ |  |  | 9! | 10 | 2 |
| 5(16) | 5 | 1 | 7 |  | $10_7$ |  | 2 | 3 | 10! | 8 |
| 5(17) | 5 | 1 | 7 |  | $6_6$ |  | 2 | 9! | 3 |  |
| 5(18) | 5 | 1 | 7 |  | $10_7$ |  | 2 | 10! | 3 | 9 |
| 5(19) | 5 | 1 |  |  | $7_5$ |  |  | 9! | 10 | 2 |
| 5(20) | 2 | 5 | 1 |  | $14_6$ |  |  | 3 | 10! | 8 |
| 5(21) | 2 | 5 |  | 1 | $13_6$ |  |  | 4 | 10! | 8 |
| 5(22) | 2 | 5 |  | 1 | $9_5$ |  |  | 9! | 4 |  |
| 5(23) | 2 | 5 |  | 1 | $12_6$ |  |  | 10! | 4 | 9 |
| 5(24) | 2 | 5 |  | 1 | $9_5$ |  |  |  | 10! | 4 |
| 5(25) | 2 | 5 |  | 1 | $10_5$ |  |  | 10! |  | 4 |
| 5(26) | 2 | 5 |  |  | $13_6$ | 1 |  | 6 | 10! | 8 |
| 5(27) | 2 | 5 | 7! |  | $6_4$ |  | 1 |  |  |  |
| 5(28) | 2 | 5 |  |  | $10_5$ |  |  | 1 | 10! | 8 |
|  | 1 | 2 | 3 | 4 | 5 | 6 | 7 | 8 | 9 | 10 |

|  | 1 | 2 | 3 | 4 | 5 | 6 | 7 | 8 | 9 | 10 |
|---|---|---|---|---|---|---|---|---|---|---|
| 5(29) | 2 | 5 |  |  | $6_4$ |  |  | 9! | 1 |  |
| 5(30) | 2 | 5 |  |  | $10_5$ |  |  | 10! | 1 | 9 |
| 5(31) | 2 | 5 |  |  | $6_4$ |  |  |  | 10! | 1 |
| 5(32) | 2 | 5 |  |  | $7_4$ |  |  | 10! |  | 1 |
| 6(1) |  | 6 |  |  |  | $2_4$ |  | 9! | 10 | 2 |
| 6(2) | 5 | 6 | 1 | 10! | 2 | $29_8$ |  | 3 | 8 | 9 |
| 6(3) | 5 | 6 | 1 |  | 2 | $11_7$ |  | 3 | 10! | 8 |
| 6(4) | 5 | 6 | 1 |  | 2 | $7_6$ |  | 9! | 3 |  |
| 6(5) | 5 | 6 | 1 |  | 2 | $11_7$ |  | 10! | 3 | 9 |
| 6(6) | 5 | 6 | 1 |  | 2 | $9_6$ |  | 10! |  | 3 |
| 6(7) | 5 | 6 | 1 |  | 2 | $8_6$ |  |  | 10! | 3 |
| 6(8) | 5 | 6 | 7! |  | 2 | $3_5$ | 1 |  |  |  |
| 6(9) | 5 | 6 | 10! |  | 2 | $25_7$ |  | 1 | 8 | 9 |
| 6(10) | 5 | 6 |  | 10! | 2 | $25_7$ |  | 1 | 8 | 9 |
| 6(11) | 5 | 6 |  |  | 2 | $7_6$ |  | 1 | 10! | 8 |
| 6(12) | 5 | 6 |  |  | 2 | $3_5$ |  | 9! | 1 |  |
| 6(13) | 5 | 6 |  |  | 2 | $7_6$ |  | 10! | 1 | 9 |
| 6(14) | 5 | 6 |  |  | 2 | $4_5$ |  | 10! |  | 1 |
| 6(15) | 5 | 6 |  |  | 2 | $3_5$ |  |  | 10! | 1 |
| 6(16) | 5 | 6 | 10! | 1 | 2 | $28_8$ |  | 4 | 8 | 9 |
| 6(17) | 5 | 6 |  | 1 | 2 | $10_7$ |  | 4 | 10! | 8 |
| 6(18) | 5 | 6 |  | 1 | 2 | $6_6$ |  | 9! | 4 |  |
| 6(19) | 5 | 6 |  | 1 | 2 | $10_7$ |  | 10! | 4 | 9 |
| 6(20) | 5 | 6 |  | 1 | 2 | $7_6$ |  | 10! |  | 4 |
| 6(21) | 5 | 6 |  | 1 | 2 | $6_6$ |  |  | 10! | 4 |
| 6(22) | 6 | 1 | 7 | 10! |  | $28_8$ | 2 | 3 | 8 | 9 |
| 6(23) | 6 | 1 | 7 |  |  | $10_7$ | 2 | 3 | 10! | 8 |
| 6(24) | 6 | 1 | 7 |  |  | $6_6$ | 2 | 9! | 3 |  |
| 6(25) | 6 | 1 | 7 |  |  | $10_7$ | 2 | 10! | 3 | 9 |
| 6(26) | 6 | 1 | 7 |  |  | $8_6$ | 2 | 10! |  | 3 |
| 6(27) | 6 | 1 | 7 |  |  | $7_6$ | 2 |  | 10! | 3 |
| 6(28) | 2 | 6 | 1 |  |  | $10_6$ |  | 3 | 10! | 8 |
| 6(29) | 2 | 6 | 1 |  |  | $10_5$ |  | 9! | 3 |  |
| 6(30) | 2 | 6 | 1 |  |  | $14_6$ |  | 10! | 3 | 9 |
|  | 1 | 2 | 3 | 4 | 5 | 6 | 7 | 8 | 9 | 10 |

|  | 1 | 2 | 3 | 4 | 5 | 6 | 7 | 8 | 9 | 10 |
|---|---|---|---|---|---|---|---|---|---|---|
| 6(31) | 2 | 6 | 1 |  |  | $11_5$ |  |  | 10! | 3 |
| 6(32) | 2 | 6 | 1 |  |  | $12_5$ |  | 10! |  | 3 |
| 6(33) | 2 | 6 |  |  | 1 |  |  | 5 | 10! | 8 |
| 6(34) | 2 | 6 |  |  |  | $10_5$ |  | 1 | 10! | 8 |
| 6(35) | 2 | 6 |  |  |  | $6_4$ |  | 9! | 1 |  |
| 6(36) | 2 | 6 |  |  |  | $10_5$ |  | 10! | 1 | 9 |
| 6(37) | 2 | 6 |  |  |  | $6_4$ |  |  | 10! | 1 |
| 6(38) | 2 | 6 |  |  |  | $7_4$ |  | 10! |  | 1 |
| 6(39) | 6 | 7 |  |  |  | $2_5$ | 1 |  | 10! | 2 |
| 6(40) | 6 | 7 |  |  |  | $3_5$ | 1 | 10! |  | 2 |
| 6(41) | 6 | 7 |  |  |  | $4_6$ | 1 | 2 | 10! | 8 |
| 6(42) | 6 | 7 |  | 5! | 2 | $16_5$ | 1 |  |  |  |
| 6(43) | 6 | 10! | 7 |  |  | $22_7$ | 1 | 3 | 8 | 9 |
| 6(44) | 6 |  | 7 | 10! |  | $22_7$ | 1 | 3 | 8 | 9 |
| 6(45) | 6 | 10! | 7 |  |  | $23_6$ | 1 | 3 |  | 8 |
| 6(46) | 6 |  | 7 | 10! |  | $23_6$ | 1 | 3 |  | 8 |
| 6(47) | 6 |  | 7 |  |  | $3_6$ | 1 | 3 | 10! | 8 |
| 7(1) |  | 7 |  |  |  |  | $2_4$ | 9! | 10 | 2 |
| 7(2) | 6 | 7 | 1 | 10! |  | 2 | $28_8$ | 3 | 8 | 9 |
| 7(3) | 6 | 7 | 1 |  |  | 2 | $10_7$ | 3 | 10! | 8 |
| 7(4) | 6 | 7 | 1 |  |  | 2 | $6_6$ | 9! | 3 |  |
| 7(5) | 6 | 7 | 1 |  |  | 2 | $10_7$ | 10! | 3 | 9 |
| 7(6) | 6 | 7 | 1 |  |  | 2 | $8_6$ | 10! |  | 3 |
| 7(7) | 6 | 7 | 1 |  |  | 2 | $10_7$ | 10! | 3 | 9 |
| 7(9) | 6 | 7 | 1 |  |  | 2 | $7_6$ |  | 10! | 3 |
| 7(10) | 5 | 7 | 1 | 10! | 2 |  | $10_8$ | 3 | 8 | 9 |
| 7(11) | 7 | 1 |  |  |  |  | $7_5$ | 9! | 10 | 2 |
| 7(12) | 2 | 7 | 1 |  |  |  | $13_6$ | 3 | 10! | 8 |
| 7(13) | 2 | 7 | 1 |  |  |  | $9_5$ | 9! | 3 |  |
| 7(14) | 2 | 7 | 1 |  |  |  | $13_6$ | 10! | 3 | 9 |
| 7(15) | 2 | 7 | 1 |  |  |  | $10_5$ |  | 10! | 3 |
| 7(16) | 2 | 7 |  | 1 |  |  | $8_5$ | 9! | 4 |  |
| 7(17) | 2 | 7 |  | 1 |  |  | $12_6$ | 10! | 4 | 9 |
| 7(18) | 2 | 7 |  | 1 |  |  | $8_5$ |  | 10! | 4 |
| 7(19) | 2 | 7 |  | 1 |  |  | $9_5$ | 10! |  | 4 |
| 7(20) | 2 | 7 |  |  | 1 |  | $11_6$ | 5 | 10! | 8 |
|  | 1 | 2 | 3 | 4 | 5 | 6 | 7 | 8 | 9 | 10 |

|       | 1 | 2 | 3 | 4  | 5 | 6 | 7      | 8   | 9   | 10 |
|-------|---|---|---|----|---|---|--------|-----|-----|----|
| 7(12) | 7 | 1 |   |    |   |   | $7_5$  | 9!  | 10  | 2  |
| 7(13) | 2 | 7 | 1 |    |   |   | $13_6$ | 3   | 10! | 8  |
| 7(14) | 2 | 7 | 1 |    |   |   | $9_5$  | 9!  | 3   |    |
| 7(15) | 2 | 7 | 1 |    |   |   | $13_6$ | 10! | 3   | 9  |
| 7(16) | 2 | 7 | 1 |    |   |   | $10_5$ |     | 10! | 3  |
| 7(17) | 2 | 7 |   | 1  |   |   | $8_5$  | 9!  | 4   |    |
| 7(18) | 2 | 7 |   | 1  |   |   | $12_6$ | 10! | 4   | 9  |
| 7(19) | 2 | 7 |   | 1  |   |   | $8_5$  |     | 10! | 4  |
| 7(20) | 2 | 7 |   | 1  |   |   | $9_5$  | 10! |     | 4  |
| 7(21) | 2 | 7 |   |    | 1 |   | $11_6$ | 5   | 10! | 8  |
| 7(22) | 2 | 7 |   |    |   | 1 | $12_6$ | 6   | 10! | 8  |
| 7(23) | 2 | 7 |   |    |   |   | $9_5$  | 1   | 10! | 8  |
| 7(24) | 2 | 7 |   |    |   |   | $5_4$  | 9!  | 1   |    |
| 7(25) | 2 | 7 |   |    |   |   | $9_5$  | 10! | 1   | 9  |
| 7(26) | 2 | 7 |   |    |   |   | $5_4$  |     | 10! | 1  |
| 7(27) | 7 | 6 |   |    |   | 1 | $2_5$  | 9!  | 2   |    |
| 7(28) | 7 | 6 |   |    |   | 1 | $6_6$  | 10! | 2   | 9  |
| 7(29) | 7 | 6 |   |    | 2 | 1 | $7_7$  | 5   | 10! | 8  |
| 7(30) | 7 | 6 |   |    |   | 1 | $5_6$  | 2   | 10! | 8  |
| 7(31) | 7 | 6 |   | 5! | 2 | 1 | $17_5$ |     |     |    |
|       | 1 | 2 | 3 | 4  | 5 | 6 | 7      | 8   | 9   | 10 |

P(30)(b)

|      | 1 | 2 | 3 | 4 | 5 | 6 | 7 | 8 | 9 | 10 |
|------|---|---|---|---|---|---|---|---|---|----|
| 1(1) | $17_7$ | 5 | 7 | 3 | 1 |   | 2 | 9! | 4 |   |
| 1(2) | $18_7$ | 5 | 7 | 3 | 1 |   | 2 | 10! |   | 4 |
| 1(3) | $17_7$ | 5 | 7 | 3 | 1 |   | 2 |   | 10! | 4 |
| 1(4) | $25_7$ | 5 | 7 | 9! | 1 |   | 2 | 3 | 8 |   |
| 1(5) | $3_7$ | 5 | 7 |   | 1 |   | 2 | 3 | 10! | 8 |
| 1(6) | $3_6$ | 5 | 7 |   | 1 |   | 2 | 9! | 3 |   |
| 1(7) | $5_6$ | 5 | 7 |   | 1 |   | 2 | 10! |   | 3 |
| 1(8) | $4_6$ | 5 | 7 |   | 1 |   | 2 |   | 10! | 3 |
| 1(9) | $4_5$ | 5 |   |   | 1 |   |   | 9! | 10 | 2 |
| 2(1) | 6 | $5_6$ |   | 1 |   | 2 |   | 9! | 10 | 4 |
| 2(2) | 6 | $20_8$ | 7 | 3 |   | 2 | 1 | 4 | 10! | 8 |
| 2(3) | 6 | $24_7$ | 7 | 9! |   | 2 | 1 | $32_5$ | $82_6$ |   |
| 2(4) | 6 | $24_8$ | 7 | 10! |   | 2 | 1 | 3 | 8 | 9 |
| 2(5) | 6 | $25_7$ | 7 | 10! |   | 2 | 1 | $32_5$ |   | $86_6$ |
| 2(6) | 6 | $6_7$ | 7 |   |   | 2 | 1 | 3 | 10! | 8 |
| 2(7) | 6 | $2_6$ | 7 |   |   | 2 | 1 | 9! | 3 |   |
| 2(8) | 6 | $24_7$ | 7 | 10! |   | 2 | 1 |   | 3 | 9 |
| 2(9) | 6 | $3_7$ | 7 |   |   | 2 | 1 | 10! | 3 | 9 |
| 2(10) | 6 | $4_6$ | 7 |   |   | 2 | 1 | 10! |   | 3 |
| 2(11) | 6 | $3_6$ | 7 |   |   | 2 | 1 |   | 10! | 3 |
| 2(12) | 6 | $2_5$ |   |   |   | 2 |   | 9! | 10 | 1 |
| 2(13) | 5 | $8_6$ | 1 |   | 2 |   |   | 9! | 10 | 3 |
| 2(14) | 5 | $6_6$ |   | 1 | 2 |   |   | 9! | 10 | 4 |
| 2(15) | 5 | $17_7$ | 7 | 3 | 2 |   | 1 | 9! | 4 |   |
| 2(16) | 5 | $18_7$ | 7 | 3 | 2 |   | 1 | 10! |   | 4 |
| 2(17) | 5 | $17_7$ | 7 | 3 | 2 |   | 1 |   | 10! | 4 |
| 2(18) | 5 | $25_7$ | 7 | 9! | 2 |   | 1 | 3 | 8 |   |
| 2(19) | 5 | $25_8$ | 7 | 10! | 2 |   | 1 | 3 | 8 | 9 |
| 2(20) | 5 | $26_7$ | 7 | 10! | 2 |   | 1 | 3 |   | 8 |
| 2(21) | 5 | $7_7$ | 7 |   | 2 |   | 1 | 3 | 10! | 8 |
| 2(22) | 5 | $3_6$ | 7 |   | 2 |   | 1 | 9! | 3 |   |
| 2(23) | 5 | $25_7$ | 7 | 10! | 2 |   | 1 |   | 3 | 9 |
| 2(24) | 5 | $4_7$ | 7 |   | 2 |   | 1 | 10! | 3 | 9 |
| 2(25) | 5 | $5_6$ | 7 |   | 2 |   | 1 | 10! |   | 3 |
| 2(26) | 5 | $4_6$ | 7 |   | 2 |   | 1 |   | 10! | 3 |
| 2(27) | 7 | $4_6$ |   | 1 |   |   | 2 | 9! | 10 | 4 |
| 2(28) | 7 | $26_7$ | 9! | 1 |   |   | 2 | 10 | 8 | 4 |

|       | 1 | 2   | 3               | 4    | 5          | 6 | 7 | 8   | 9   | 10 |
|-------|---|-----|-----------------|------|------------|---|---|-----|-----|----|
| 2(29) | 7 | $4_6$ |                 | 1    |            |   | 2 | 9!  | 10  | 4  |
| 3(1)  | 7 | 10! | $26_7$          | 1    |            |   | 3 | 9   | 4   | 8  |
| 3(2)  | 7 |     | $3_6$           | 1    |            |   | 3 | 9!  | 10  | 4  |
| 3(3)  | 7 | 9!  | $25_7$          | 1    |            |   | 3 | 10  | 8   | 4  |
| 3(4)  | 7 |     | $14_6$          | 6    |            | 1 | 3 |     | 10! | 4  |
| 4(1)  | 4 |     |                 | $2_4$ |            |   |   | 9!  | 10  | 1  |
| 4(2)  | 4 | 7   | 1               | $8_7$ |            |   | 3 | 9!  | 10  | 2  |
| 4(3)  | 4 | 8!  | 7               | $26_7$ |           |   | 1 | 10  | 3   | 9  |
| 4(4)  | 4 | 7   | 10!             | $26_7$ |           |   | 1 | 9   | 2   | 8  |
| 4(5)  | 4 | 7   |                 | $4_6$ |            |   | 1 | 9!  | 10  | 2  |
| 5(1)  | 6 | 5   | 1               | 10!  | $29_8$     | 2 |   | 9   | 3   | 8  |
| 5(2)  | 6 | 5   | 1               | 9!   | $30_8$     | 2 |   | 10  | 8   | 3  |
| 5(3)  | 6 | 5   | 1               |      | $7_7$      | 2 |   | 9!  | 10  | 3  |
| 5(4)  | 6 | 5   | 4               | 1    | $22_8$     | 2 |   | 3   | 10! | 8  |
| 5(5)  | 6 | 5   | 4               | 1    | $18_7$     | 2 |   | 9!  | 3   |    |
| 5(6)  | 6 | 5   | 4               | 1    | $22_8$     | 2 |   | 10! | 3   | 9  |
| 5(7)  | 6 | 5   | 4               | 1    | $20_7$     | 2 |   | 10! |     | 3  |
| 5(8)  | 6 | 5   | 4               | 1    | $19_7$     | 2 |   |     | 10! | 3  |
| 5(9)  | 6 | 5   | 10!             | 1    | $28_8$     | 2 |   | 9   | 4   | 8  |
| 5(10) | 6 | 5   | 8!              | 1    | $30_8$     | 2 |   | 10  | 4   | 9  |
| 5(11) | 6 | 5   | 9!              | 1    | $29_8$     | 2 |   | 10  | 8   | 4  |
| 5(12) | 6 | 5   |                 | 1    | $5_7$      | 2 |   | 9!  | 10  | 4  |
| 5(13) | 6 | 5   | 8!              | 1    | $30_8$     | 2 |   | 10  | 4   | 9  |
| 5(14) | 5 | 1   | 7               | 10!  | $29_8$     |   | 2 | 9   | 3   | 8  |
| 5(15) | 2 | 5   | 7               |      | $6_6$      |   | 1 | 9!  | 3   |    |
| 5(16) | 2 | 5   | 7               |      | $10_7$     |   | 1 | 10! | 3   | 9  |
| 5(17) | 2 | 5   | 7               |      | $7_6$      |   | 1 | 10! |     | 3  |
| 5(18) | 2 | 5   |                 |      | $6_5$      |   |   | 9!  | 10  | 1  |
| 5(19) | 2 | 5   | 7               | 10!  | $28_8$     |   | 1 | 3   | 8   | 9  |
| 5(20) | 2 | 5   | 7               |      | $10_7$     |   | 1 | 3   | 10! | 8  |
| 5(21) | 2 | 5   | 7               |      | $6_6$      |   | 1 | 9!  | 3   |    |
| 5(22) | 2 | 5   | 7               |      | $8_6$      |   | 1 | 10! |     | 3  |
| 5(23) | 2 | 5   | 7               |      | $8_7$      |   | 1 |     | 10! | 3  |
| 5(24) | 2 | 5   |                 |      | $6_5$      |   |   | 9!  | 10  | 1  |
| 5(25) | 2 | 5   | 7               | 10!  | $28_8$     |   | 1 | 3   | 8   | 9  |
|       | 1 | 2   | 3               | 4    | 5          | 6 | 7 | 8   | 9   | 10 |

|       | 1 | 2 | 3   | 4   | 5 | 6       | 7 | 8   | 9   | 10 |
|-------|---|---|-----|-----|---|---------|---|-----|-----|----|
| 5(26) | 2 | 5 | 7   |     | $6_6$ |    | 1 | 9!  | 3   |    |
| 5(27) | 2 | 5 | 7   |     | $10_7$ |   | 1 | 10! | 3   | 9  |
| 5(28) | 2 | 5 | 7   |     | $8_6$ |    | 1 | 10! |     | 3  |
| 5(29) | 2 | 5 | 7   |     | $7_6$ |    | 1 |     | 10! | 3  |
| 5(30) | 2 | 5 |     |     | $6_5$ |    |   | 9!  | 10  | 1  |
| 6(1)  | 5 | 6 | 1   |     | 2 | $8_7$   |   | 9!  | 10  | 3  |
| 6(2)  | 5 | 6 | 7   | 3   | 2 | $17_8$  | 1 | 9!  | 4   |    |
| 6(3)  | 5 | 6 | 7   | 3   | 2 | $18_8$  | 1 | 10! |     | 4  |
| 6(4)  | 5 | 6 | 7   | 3   | 2 | $17_8$  | 1 |     | 10! | 4  |
| 6(5)  | 5 | 6 | 7   | 8!  | 2 | $25_7$  | 1 | $33_6$ |  |    |
| 6(6)  | 5 | 6 | 7   | 9!  | 2 | $25_8$  | 1 | 3   | 8   |    |
| 6(7)  | 5 | 6 | 7   | 10! | 2 | $25_9$  | 1 | 3   | 8   | 9  |
| 6(8)  | 5 | 6 | 7   | 10! | 2 | $26_8$  | 1 | 3   |     | 8  |
| 6(9)  | 5 | 6 | 7   |     | 2 | $7_8$   | 1 | 3   | 10! | 8  |
| 6(10) | 5 | 6 | 7   |     | 2 | $3_7$   | 1 | 9!  | 3   |    |
| 6(11) | 5 | 6 | 7   | 10! | 2 | $25_8$  | 1 |     | 3   | 9  |
| 6(12) | 5 | 6 | 7   |     | 2 | $7_8$   | 1 | 10! | 3   | 9  |
| 6(13) | 5 | 6 | 7   | 10! | 2 | $23_7$  | 1 |     |     | 3  |
| 6(14) | 5 | 6 | 7   |     | 2 | $5_7$   | 1 | 10! |     | 3  |
| 6(15) | 5 | 6 | 7   |     | 2 | $4_7$   | 1 |     | 10! | 3  |
| 6(16) | 5 | 6 | 10! |     | 2 | $26_7$  |   | 9   | 1   | 8  |
| 6(17) | 5 | 6 |     | 10! | 2 | $26_7$  |   | 9   | 1   | 8  |
| 6(18) | 5 | 6 | 9!  |     | 2 | $25_7$  |   | 10  | 8   | 1  |
| 6(19) | 5 | 6 |     | 9!  | 2 | $26_7$  |   | 10  | 8   | 1  |
| 6(20) | 5 | 6 |     |     | 2 | $3_6$   |   | 9!  | 10  | 1  |
| 6(21) | 5 | 6 | 10! | 1   | 2 | $29_8$  |   | 9   | 4   | 8  |
| 6(22) | 5 | 6 | 9!  | 1   | 2 | $28_8$  |   | 10  | 8   | 4  |
| 6(23) | 5 | 6 |     | 1   | 2 | $6_7$   |   | 9!  | 10  | 4  |
| 6(24) | 6 | 1 | 7   | 10! |   | $29_8$  | 2 | 9   | 3   | 8  |
| 6(25) | 6 | 1 | 7   | 9!  |   | $30_8$  | 2 | 10  | 8   | 3  |
| 6(26) | 6 | 1 | 7   |     |   | $7_7$   | 2 | 9!  | 10  | 3  |
| 6(27) | 2 | 6 | 1   |     |   | $11_6$  |   | 9!  | 10  | 3  |
| 6(28) | 2 | 6 |     |     |   | $6_5$   |   | 9!  | 10  | 1  |
| 6(29) | 6 | 7 |     |     |   | $2_6$   | 1 | 9!  | 10  | 2  |
| 6(30) | 6 | 7 | 9!  |     |   | $24_7$  | 1 | 10  | 8   | 2  |
| 6(31) | 6 | 7 |     | 9!  |   | $25_7$  | 1 | 10  | 8   | 2  |
|       | 1 | 2 | 3   | 4   | 5 | 6       | 7 | 8   | 9   | 10 |

|  | 1 | 2 | 3 | 4 | 5 | 6 | 7 | 8 | 9 | 10 |
|---|---|---|---|---|---|---|---|---|---|---|
| 6(32)) | 6 | 7 | 9! |  |  | $25_7$ | 1 | 2 | 10 | 8 |
| 6(33) | 6 | 7 |  | 9! |  | $26_7$ | 1 | 2 | 10 | 8 |
| 6(34) | 6 | 7 |  | 5 | 2 | $16_8$ | 1 | 9! | 10 | 4 |
| 7(1) | 6 | 7 | 1 | 10! |  | 2 | $29_8$ | 9 | 3 | 8 |
| 7(2) | 6 | 7 | 1 |  |  | 2 | $7_7$ | 9! | 10 | 3 |
| 7(3) | 2 | 7 | 1 |  |  |  | $10_6$ | 9! | 10 | 3 |
| 7(4) | 2 | 7 |  | 1 |  |  | $8_6$ | 9! | 10 | 4 |
| 7(5) | 2 | 7 |  |  |  |  | $5_5$ | 9! | 10 | 1 |
| 7(6) | 7 | 6 | 10! |  |  | 1 | $25_7$ | 9 | 2 | 8 |
| 7(7) | 7 | 6 | 9! |  | 2 | 1 | $28_8$ | 5 | 10 | 8 |
| 7(8) | 7 | 6 |  | 9! | 2 | 1 | $29_8$ | 5 | 10 | 8 |
| 7(9) | 7 | 6 | 9! |  |  | 1 | $26_7$ | 2 | 10 | 8 |
| 7(10) | 7 | 6 |  | 5 | 2 | 1 | $17_8$ | 9! | 10 | 4 |
|  | 1 | 2 | 3 | 4 | 5 | 6 | 7 | 8 | 9 | 10 |

P(40)(b)

|  | 1 | 2 | 3 | 4 | 5 | 6 | 7 | 8 | 9 | 10 |
|---|---|---|---|---|---|---|---|---|---|---|
| 1(1) | $17_8$ | 5 | 7 | 3 | 1 |  | 2 | 9! | 10 | 4 |
| 1(2) | $25_8$ | 5 | 7 | 9! | 1 |  | 2 | 3 | 10 | 8 |
| 1(3) | $26_8$ | 5 | 7 | 10! | 1 |  | 2 | 9 | 3 | 8 |
| 1(4) | $27_8$ | 5 | 7 | 9! | 1 |  | 2 | 10 | 8 | 3 |
| 1(5) | $26_7$ | 5 | 7 | 9! | 1 |  | 2 |  | 10 | 3 |
| 1(6) | $4_7$ | 5 | 7 |  | 1 |  | 2 | 9! | 10 | 3 |
| 2(1) | 6 | $26_7$ | 8! | 1 |  | 2 |  | 9 | 10 | 4 |
| 2(2) | 6 | $28_8$ | 7 | 9! |  | 2 | 1 | 3 | 10 | 8 |
| 2(3) | 6 | $25_8$ | 7 | 10! |  | 2 | 1 | 9 | 3 | 8 |
| 2(4) | 6 | $25_8$ | 7 | 8! |  | 2 | 1 | 10 | 3 | 9 |
| 2(5) | 6 | $26_8$ | 7 | 8! |  | 2 | 1 | 10 | 8 | 3 |
| 2(6) | 6 | $25_7$ | 7 | 9! |  | 2 | 1 |  | 10 | 3 |
| 2(7) | 6 | $3_7$ | 7 |  |  | 2 | 1 | 9! | 10 | 3 |
| 2(8) | 5 | $17_8$ | 7 | 3 | 2 |  | 1 | 9! | 10 | 4 |
| 2(9) | 5 | $7_7$ | 7 |  | 2 |  | 1 | 3 | 10! | 8 |
| 2(10) | 5 | $25_7$ | 7 | 8! | 2 |  | 1 | 9 | 3 |  |
| 2(11) | 5 | $26_8$ | 7 | 10! | 2 |  | 1 | 9 | 3 | 8 |
| 2(12) | 5 | $26_8$ | 7 | 8! | 2 |  | 1 | 10 | 3 | 9 |
| 2(13) | 5 | $27_8$ | 7 | 9! | 2 |  | 1 | 10 | 8 | 3 |
| 2(14) | 5 | $26_7$ | 7 | 9! | 2 |  | 1 |  | 10 | 3 |
|  | 1 | 2 | 3 | 4 | 5 | 6 | 7 | 8 | 9 | 10 |

|        | 1 | 2      | 3      | 4    | 5      | 6      | 7 | 8   | 9  | 10 |
|--------|---|--------|--------|------|--------|--------|---|-----|----|----|
| 2(15)  | 5 | $4_7$  | 7      |      | 2      |        | 1 | 9!  | 10 | 3  |
| 2(16)  | 7 | $25_7$ | 8!     | 1    |        |        | 2 | 9   | 10 | 4  |
| 2(17)  | 7 | $25_7$ | 8!     | 1    |        |        | 2 | 9   | 10 | 4  |
| 3(1)   | 7 | 8!     | $25_7$ | 1    |        |        | 3 | 9   | 10 | 4  |
| 3(2)   | 7 |        | $14_7$ | 6    |        | 1      | 3 | 9!  | 10 | 4  |
| 4(1)   | 4 | 7      | 8!     | $25_7$ |      |        | 1 | 9   | 10 | 2  |
| 5(1)   | 6 | 5      | 1      | 8!   | $29_8$ | 2      |   | 9!  | 10 | 3  |
| 5(2)   | 6 | 5      | 4      | 1    | $19_8$ | 2      |   | 9!  | 10 | 3  |
| 5(3)   | 6 | 5      | 8!     | 1    | $26_8$ | 2      |   | 9   | 10 | 4  |
| 5(4)   | 2 | 5      | 7      | 10!  | $29_8$ |        | 1 | 9   | 3  | 8  |
| 5(5)   | 2 | 5      | 7      | 9!   | $29_8$ |        | 1 | 10  | 8  | 3  |
| 5(6)   | 2 | 5      | 7      | 10!  | $26_8$ |        | 1 | 9   | 3  | 8  |
| 5(7)   | 2 | 5      | 7      | 9!   | $30_8$ |        | 1 | 10  | 8  | 3  |
| 5(8)   | 2 | 5      | 7      |      | $8_7$  |        | 1 | 9!  | 10 | 3  |
| 5(9)   | 2 | 5      | 7      | 10!  | $29_8$ |        | 1 | 9   | 3  | 8  |
| 5(10)  | 2 | 5      | 7      | 9!   | $30_8$ |        | 1 | 10  | 8  | 3  |
| 5(11)  | 2 | 5      | 7      |      | $7_7$  |        | 1 | 9!  | 10 | 3  |
| 6(1)   | 5 | 6      | 1      | 8!   | 2      | $30_8$ |   | 9   | 10 | 3  |
| 6(2)   | 5 | 6      | 7      | 3    | 2      | $17_9$ | 1 | 9!  | 10 | 4  |
| 6(3)   | 5 | 6      | 7      | 9!   | 2      | $29_9$ | 1 | 3   | 10 | 8  |
| 6(4)   | 5 | 6      | 7      | 10!  | 2      | $29_9$ | 1 | 9   | 3  | 8  |
| 6(5)   | 5 | 6      | 7      | 8!   | 2      | $29_9$ | 1 | 10  | 3  | 9  |
| 6(4)   | 5 | 6      | 7      | 10!  | 2      | $29_9$ | 1 | 9   | 3  | 8  |
| 6(5)   | 5 | 6      | 7      | 8!   | 2      | $29_9$ | 1 | 10  | 3  | 9  |
| 6(6)   | 5 | 6      | 7      | 10   | 2      | $23_9$ | 1 | 9!  | 4  | 3  |
| 6(7)   | 5 | 6      | 7      | 8!   | 2      | $27_8$ | 1 | 10  |    | 3  |
| 6(8)   | 5 | 6      | 7      | 9!   | 2      | $27_9$ | 1 | 10  | 8  | 3  |
| 6(9)   | 5 | 6      | 7      | 9!   | 2      | $26_8$ | 1 |     | 10 | 3  |
| 6(10)  | 5 | 6      | 10!    |      | 2      | $26_7$ |   | 9   | 1  | 8  |
| 6(11)  | 5 | 6      |        | 10!  | 2      | $26_7$ |   | 9   | 1  | 8  |
| 6(12)  | 5 | 6      | 8!     |      | 2      | $24_7$ |   | 9   | 10 | 1  |
| 6(13)  | 5 | 6      |        | 8!   | 2      | $25_7$ |   | 9   | 10 | 1  |
| 6(14)  | 5 | 6      | 8!     | 1    | 2      | $27_8$ |   | 9   | 10 | 4  |
| 6(15)  | 6 | 1      | 7      | 8!   |        | $29_8$ | 2 | 9   | 10 | 3  |
| 6(16)  | 6 | 7      |        | 8!   |        | $24_7$ | 1 | 9   | 10 | 2  |
|        | 1 | 2      | 3      | 4    | 5      | 6      | 7 | 8   | 9  | 10 |

P(50)(b)

|  | 1 | 2 | 3 | 4 | 5 | 6 | 7 | 8 | 9 | 10 |
|---|---|---|---|---|---|---|---|---|---|---|
| 1(6) | $26_8$ | 5 | 7 | 8! | 1 |  | 2 | 9 | 10 | 3 |
| 2(7) | 6 | $25_8$ | 7 | 8! |  | 2 | 1 | 9 | 10 | 3 |
| 2(9) | 5 | $29_8$ | 7 | 9! | 2 |  | 1 | 3 | 10 | 8 |
| 2(15) | 5 | $26_8$ | 7 | 8! | 2 |  | 1 | 9 | 10 | 3 |
| 5(8) | 2 | 5 | 7 |  | $8_7$ |  | 1 | 9 | 10 | 3 |
| 5(9) | 2 | 5 | 7 | 10! | $29_8$ |  | 1 | 9 | 3 | 8 |
| 5(10) | 2 | 5 | 7 | 9! | $30_8$ |  | 1 | 10 | 8 | 3 |
| 5(11) | 2 | 5 | 7 |  | $7_7$ |  | 1 | 9! | 10 | 3 |
| 6(1) | 5 | 6 | 1 | 8! | 2 | $30_8$ |  | 9 | 10 | 3 |
| 6(2) | 5 | 6 | 7 | 3 | 2 | $17_9$ | 1 | 9! | 10 | 4 |
| 6(3) | 5 | 6 | 7 | 9! | 2 | $29_9$ | 1 | 3 | 10 | 8 |
| 6(4) | 5 | 6 | 7 | 10! | 2 | $29_9$ | 1 | 9 | 3 | 8 |
| 6(5) | 5 | 6 | 7 | 8! | 2 | $29_9$ | 1 | 10 | 3 | 9 |
| 6(4) | 5 | 6 | 7 | 10! | 2 | $29_9$ | 1 | 9 | 3 | 8 |
| 6(5) | 5 | 6 | 7 | 8! | 2 | $29_9$ | 1 | 10 | 3 | 9 |
| 6(6) | 5 | 6 | 7 | 10 | 2 | $23_9$ | 1 | 9! | 4 | 3 |
| 6(7) | 5 | 6 | 7 | 8! | 2 | $27_8$ | 1 | 10 |  | 3 |
| 6(8) | 5 | 6 | 7 | 9! | 2 | $27_9$ | 1 | 10 | 8 | 3 |
| 6(9) | 5 | 6 | 7 | 9! | 2 | $26_8$ | 1 |  | 10 | 3 |
| 6(10) | 5 | 6 | 10! |  | 2 | $26_7$ |  | 9 | 1 | 8 |
| 6(11) | 5 | 6 |  | 10! | 2 | $26_7$ |  | 9 | 1 | 8 |
| 6(12) | 5 | 6 | 8! |  | 2 | $24_7$ |  | 9 | 10 | 1 |
| 6(13) | 5 | 6 |  | 8! | 2 | $25_7$ |  | 9 | 10 | 1 |
| 6(14) | 5 | 6 | 8! | 1 | 2 | $27_8$ |  | 9 | 10 | 4 |
| 6(15) | 6 | 1 | 7 | 8! |  | $29_8$ | 2 | 9 | 10 | 3 |
| 6(16) | 6 | 7 |  | 8! |  | $24_7$ | 1 | 9 | 10 | 2 |
|  | 1 | 2 | 3 | 4 | 5 | 6 | 7 | 8 | 9 | 10 |

|        | 1        | 2        | 3 | 4    | 5 | 6 | 7 | 8    | 9    | 10 |
|--------|----------|----------|---|------|---|---|---|------|------|----|
| 1(1)   | $25_8$   | 5        | 7 | 10!  | 1 |   | 2 | 3    | 8    | 9  |
| 1(2)   | $7_7$    | 5        | 7 |      | 1 |   | 2 | 3    | 10!  | 8  |
| 1(3)   | $26_7$   | 5        | 7 | 10!  | 1 |   | 2 | 3    |      | 8  |
| 1(4)   | $25_8$   | 5        | 7 |      | 1 |   | 2 | 9!   | 3    |    |
| 1(5)   | $4_6$    | 5        | 7 |      | 1 |   | 2 |      | 10!  | 3  |
| 1(6)   | $5_6$    | 5        | 7 |      | 1 |   | 2 | 10!  |      | 3  |
| 1(10)  | $3_3$    |          |   | 1    |   |   |   |      | 10!  | 4  |
| 1(11)  | $4_3$    |          |   | 1    |   |   |   | 10!  |      | 4  |
| 1(12)  | $4_3$    |          |   |      |   |   |   | 10!  | 1    | 9  |
| 2(1)   | 6        | $9_6$    |   | 1    |   | 2 |   | 4    | 10!  | 8  |
| 2(2)   | 6        | $2_4$    |   | 1    |   | 2 |   | 9!   | 4    |    |
| 2(3)   | 6        | $6_5$    |   | 1    |   | 2 |   | 10!  | 4    | 9  |
| 2(4)   | 6        | $5_5$    |   | 1    |   | 2 |   |      | 10!  | 4  |
| 2(5)   | 6        | $3_4$    |   | 1    |   | 2 |   | 10!  |      | 4  |
| 2(6)   | 6        | $24_8$   | 7 | 10!  |   | 2 | 1 | 3    | 8    | 9  |
| 2(7)   | 6        | $6_7$    | 7 |      |   | 2 | 1 | 3    | 10!  | 8  |
| 2(8)   | 6        | $2_6$    | 7 |      |   | 2 | 1 | 9!   | 3    |    |
| 2(9)   | 6        | $6_7$    | 7 |      |   | 2 | 1 | 10!  | 3    | 9  |
| 2(10)  | 6        | $3_6$    | 7 |      |   | 2 | 1 |      | 10!  | 3  |
| 2(11)  |          | $4_6$    |   |      |   |   |   | 10!  |      | 3  |
| 2(12)  | 6        | $6_5$    |   |      |   | 2 |   | 1    | 10!  | 8  |
| 2(13)  | 6        | $2_4$    |   |      |   | 2 |   | 9!   | 1    |    |
| 2(14)  | 6        | $6_5$    |   |      |   | 2 |   | 10!  | 1    | 9  |
| 2(15)  | 6        | $2_4$    |   |      |   | 2 |   |      | 10!  | 1  |
| 2(16)  | 6        | $3_4$    |   |      |   | 2 |   | 10!  |      | 1  |
| 2(17)  |          | $2_3$    |   |      |   |   |   | 9!   | 10   | 2  |
| 2(18)  |          | $1_3$    |   |      |   |   |   | 10   | 8!   | 2  |
| 2(19)  |          | $7_4$    |   |      | 2 |   |   | 5    | 10!  | 8  |
| 2(20)  | 5        | $11_6$   | 1 |      | 2 |   |   | 3    | 10!  | 8  |
|        | 1        | 2        | 3 | 4    | 5 | 6 | 7 | 8    | 9    | 10 |

| | 1 | 2 | 3 | 4 | 5 | 6 | 7 | 8 | 9 | 10 |
|---|---|---|---|---|---|---|---|---|---|---|
| 2(21) | 5 | $7_5$ | 1 | | 2 | | | | 9! | 3 | |
| 2(22) | 5 | $11_6$ | 1 | | 2 | | | 10! | 3 | 9 |
| 2(23) | 5 | $8_5$ | 1 | | 2 | | | | 10! | 3 |
| 2(24) | 5 | $9_5$ | 1 | | 2 | | | 10! | | 3 |
| 2(25) | 5 | $10_6$ | | 1 | 2 | | | 4 | 10! | 8 |
| 2(26) | 5 | $6_5$ | | 1 | 2 | | | 9! | 4 | |
| 2(27) | 5 | $10_6$ | | 1 | 2 | | | 10! | 4 | 9 |
| 2(28) | 5 | $6_5$ | | 1 | 2 | | | | 10! | 4 |
| 2(29) | 5 | $7_5$ | | 1 | 2 | | | 10! | | 4 |
| 2(30) | 5 | $10_6$ | | | 2 | 1 | | 6 | 10! | 8 |
| 2(31) | 5 | $7_5$ | | | 2 | | | 1 | 10! | 8 |
| 2(32) | 5 | $3_4$ | | | 2 | | | 9! | 1 | |
| 2(33) | 5 | $7_5$ | | | 2 | | | 10! | 1 | 9 |
| 2(34) | 5 | $3_4$ | | | 2 | | | | 10! | 1 |
| 2(35) | 5 | $4_4$ | | | 2 | | | 10! | | 1 |
| 2(36) | | $8_4$ | | | | 2 | | 6 | 10! | 8 |
| 2(37) | 7 | $7_6$ | | | 1 | | 2 | 5 | 10! | 8 |
| 2(38) | 7 | $8_6$ | | 1 | | | 2 | 4 | 10! | 8 |
| 2(39) | 7 | $4_5$ | | 1 | | | 2 | 9! | 4 | |
| 2(40) | 7 | $8_6$ | | 1 | | | 2 | 10! | 4 | 9 |
| 2(41) | 7 | $4_5$ | | 1 | | | 2 | | 10! | 4 |
| 2(42) | 7 | $5_5$ | | 1 | | | 2 | 10! | | 4 |
| 2(43) | 7 | $19_7$ | | 6 | | 1 | 2 | 4 | 10! | 8 |
| 2(44) | 7 | $15_6$ | | 6 | | 1 | 2 | 9! | 4 | |
| 2(45) | 7 | $19_7$ | | 6 | | 1 | 2 | 10! | 4 | 9 |
| 2(46) | 7 | $15_6$ | | 6 | | 1 | 2 | | 10! | 4 |
| | 1 | 2 | 3 | 4 | 5 | 6 | 7 | 8 | 9 | 10 |

|       | 1 | 2   | 3               | 4   | 5               | 6 | 7 | 8   | 9   | 10 |
|-------|---|-----|-----------------|-----|-----------------|---|---|-----|-----|----|
| 3(1)  | 7 |     | $7_6$           | 1   |                 |   | 3 | 4   | 10! | 8  |
| 3(2)  | 7 |     | $3_5$           | 1   |                 |   | 3 | 9!  | 4   |    |
| 3(3)  | 7 |     | $7_6$           | 1   |                 |   | 3 | 10! | 4   | 9  |
| 3(4)  | 7 |     | $3_5$           | 1   |                 |   | 3 |     | 10! | 4  |
| 3(5)  | 7 |     | $4_5$           | 1   |                 |   | 3 | 10! |     | 4  |
| 3(6)  | 7 |     | $4_5$           |     |                 |   | 3 | 1   | 10! | 8  |
| 3(7)  | 7 |     | $18_7$          | 6   |                 | 1 | 3 | 4   | 10! | 8  |
| 3(8)  | 7 |     | $14_6$          | 6   |                 | 1 | 3 |     | 10! | 4  |
| 3(9)  | 7 |     | $15_6$          | 6   |                 | 1 | 3 | 10! |     | 4  |
| 3(10) | 7 |     | $4_5$           |     |                 |   | 3 | 1   | 10! | 8  |
| 3(11) | 7 |     | $18_7$          | 6   |                 | 1 | 3 | 4   | 10! | 8  |
| 3(12) | 7 |     | $14_6$          | 6   |                 | 1 | 3 |     | 10! | 4  |
| 3(13) | 7 |     | $15_6$          | 6   |                 | 1 | 3 | 10! |     | 4  |
| 3(14) | 7 | 10! | $24_7$          |     | 1               |   | 3 | 5   | 8   | 9  |
| 3(15) | 7 |     | $24_7$          | 10! | 1               |   | 3 | 5   | 8   | 9  |
| 3(16) | 7 |     | $6_6$           |     | 1               |   | 3 | 5   | 10! | 8  |
| 4(1)  | 4 |     |                 | $6_4$ |               |   |   | 1   | 10! | 8  |
| 4(2)  | 4 |     |                 | $2_3$ |               |   |   |     | 10! | 1  |
| 4(3)  | 4 |     |                 | $3_3$ |               |   |   | 10! |     | 1  |
| 4(4)  | 4 |     | 7               | $2_5$ |               |   | 1 | 9!  | 3   |    |
| 4(5)  | 4 |     | 7               | $9_6$ |               |   | 1 | 10! | 3   | 9  |
| 4(6)  | 4 | 7   |                 | $4_5$ |               |   | 1 |     | 10! | 2  |
| 4(7)  | 4 | 7   |                 | $5_5$ |               |   | 1 | 10! |     | 2  |
| 5(1)  | 6 | 5   | 1               |     | $10_7$          | 2 |   | 3   | 10! | 8  |
| 5(2)  | 6 | 5   | 1               |     | $6_6$           | 2 |   | 9!  | 3   |    |
| 5(3)  | 6 | 5   | 1               |     | $7_6$           | 2 |   |     | 10! | 3  |
| 5(4)  | 6 | 5   | 1               |     | $8_6$           | 2 |   | 10! |     | 3  |
| 5(5)  | 6 | 5   | 10!             | 1   | $27_8$          | 2 |   | 4   | 8   | 9  |
| 5(6)  | 6 | 5   | 7!              |     | $2_5$           | 2 | 1 |     |     |    |
| 5(7)  | 6 | 5   | 10!             |     | $24_7$          | 2 |   | 1   | 8   | 9  |
| 5(8)  | 6 | 5   |                 |     | $6_6$           | 2 |   | 1   | 10! | 8  |
| 5(9)  | 6 | 5   |                 |     | $2_5$           | 2 |   | 9!  | 1   |    |
|       | 1 | 2   | 3               | 4   | 5               | 6 | 7 | 8   | 9   | 10 |

|       | 1 | 2 | 3 | 4 | 5 | 6 | 7 | 8 | 9 | 10 |
|-------|---|---|---|---|---|---|---|---|---|----|
| 5(10) | 6 | 5 |   |   | $2_5$ | 2 |   |   | 10! | 1 |
| 5(11) |   | 5 |   |   | $10_5$ | 2 |   | 6 | 10! | 8 |
| 5(12) |   | 5 |   |   | $2_3$ |   |   |   | 10! | 2 |
| 5(13) |   | 5 |   |   | $3_3$ |   |   | 10! |   | 2 |
| 5(14) | 5 | 1 | 7 | 10! | $28_8$ |   | 2 | 3 | 8 | 9 |
| 5(15) | 5 | 1 | 7 |   | $10_7$ |   | 2 | 3 | 10! | 8 |
| 5(16) | 5 | 1 | 7 |   | $6_6$ |   | 2 | 9! | 3 |   |
| 5(17) | 5 | 1 | 7 |   | $10_6$ |   |   | 10! | 3 | 9 |
| 5(18) | 5 | 1 | 7 |   | $7_6$ |   | 2 |   | 10! | 3 |
| 5(19) | 5 | 1 | 7 |   | $8_6$ |   | 2 | 10! |   | 3 |
| 5(20) | 5 | 1 |   |   | $6_4$ |   |   | 9! | 2 |   |
| 5(21) | 5 | 1 |   |   | $10_5$ |   |   | 10! | 2 | 9 |
| 5(22) | 5 | 1 |   |   | $7_5$ |   |   | 9! | 10 | 2 |
| 5(23) | 5 | 1 |   |   | $8_4$ |   |   | 10! |   | 2 |
| 5(24) | 2 | 5 |   | 1 | $13_6$ |   |   | 4 | 10! | 8 |
| 5(25) | 2 | 5 |   | 1 | $9_5$ |   |   | 9! | 4 |   |
| 5(26) | 2 | 5 |   | 1 | $12_6$ |   |   | 10! | 4 | 9 |
| 5(27) | 2 | 5 |   | 1 | $9_5$ |   |   |   | 10! | 4 |
| 5(28) | 2 | 5 |   | 1 | $10_5$ |   |   | 10! |   | 4 |
| 5(29) | 2 | 5 |   |   | $13_6$ | 1 |   | 6 | 10! | 8 |
| 5(30) | 2 | 5 | 7 | 10! | $28_8$ |   | 1 | 3 | 8 | 9 |
| 5(31) | 2 | 5 | 7 |   | $10_7$ |   | 1 | 3 | 10! | 8 |
| 5(32) | 2 | 5 | 7 |   | $6_6$ |   | 1 | 9! | 3 |   |
| 5(33) | 2 | 5 | 7 |   | $10_7$ |   | 1 | 10! | 3 | 9 |
| 5(34) | 2 | 5 | 7 |   | $7_6$ |   | 1 |   | 10! | 3 |
| 5(35) | 2 | 5 | 7 |   | $8_6$ |   | 1 | 10! |   | 3 |
| 5(36) | 2 | 5 |   |   | $10_5$ |   |   | 1 | 10! | 8 |
| 5(37) | 2 | 5 |   |   | $6_4$ |   |   | 9! | 1 |   |
| 5(38) | 2 | 5 |   |   | $6_4$ |   |   | 9! | 1 |   |
| 5(39) | 2 | 5 |   |   | $10_5$ |   |   | 10! | 1 | 9 |
| 5(40) | 2 | 5 |   |   | $6_4$ |   |   |   | 10! | 1 |
| 5(41) | 2 | 5 |   |   | $7_4$ |   |   | 10! |   | 1 |
| 6(1)  |   | 6 |   |   |   | $2_4$ |   | 9! | 10 | 2 |
| 6(2)  | 5 | 6 | 1 |   | 2 | $7_5$ |   |   | 10! | 3 |
| 6(3)  | 5 | 6 | 1 |   | 2 | $8_5$ |   | 10! |   | 3 |
| 6(4)  | 5 | 6 | 1 |   | 2 | $7_6$ |   | 9! | 3 |   |
| 6(5)  | 5 | 6 | 1 |   | 2 | $11_7$ |   | 10! | 3 | 9 |
| 6(6)  | 5 | 6 | 1 |   | 2 | $11_7$ |   | 3 | 10! | 8 |
| 6(7)  | 5 | 6 |   | 1 | 2 | $6_6$ |   | 9! | 4 |   |
| 6(8)  | 5 | 6 |   | 1 | 2 | $10_7$ |   | 10! | 4 | 9 |
| 6(9)  | 5 | 6 |   | 1 | 2 | $10_7$ |   | 4 | 10! | 8 |
|       | 1 | 2 | 3 | 4 | 5 | 6 | 7 | 8 | 9 | 10 |

|  | 1 | 2 | 3 | 4 | 5 | 6 | 7 | 8 | 9 | 10 |
|---|---|---|---|---|---|---|---|---|---|---|
| 6(12) |  | 6 |  |  |  | $2_4$ |  | 9! | 10 | 2 |
| 6(13) | 5 | 6 | 1 |  | 2 | $7_5$ |  |  | 10! | 3 |
| 6(14) | 5 | 6 | 1 |  | 2 | $8_5$ |  | 10! |  | 3 |
| 6(15) | 5 | 6 | 1 |  | 2 | $7_6$ |  | 9! | 3 |  |
| 6(16) | 5 | 6 | 1 |  | 2 | $11_7$ |  | 10! | 3 | 9 |
| 6(17) | 5 | 6 | 1 |  | 2 | $11_7$ |  | 3 | 10! | 8 |
| 6(18) | 5 | 6 |  | 1 | 2 | $6_6$ |  | 9! | 4 |  |
| 6(19) | 5 | 6 |  | 1 | 2 | $10_7$ |  | 10! | 4 | 9 |
| 6(21) | 5 | 6 |  | 1 | 2 | $10_7$ |  | 4 | 10! | 8 |
| 6(22) | 6 | 1 |  |  | 2 | $12_6$ |  | 5 | 10! | 8 |
| 6(23) | 6 | 1 | 7 |  |  | $10_7$ | 2 | 3 | 10! | 8 |
| 6(24) | 6 | 1 | 7 |  |  | $6_6$ | 2 | 9! | 3 |  |
| 6(25) | 6 | 1 | 7 |  |  | $7_6$ | 2 |  | 10! | 3 |
| 6(26) | 6 | 1 |  |  |  | $6_4$ |  | 9! | 2 |  |
| 6(27) | 6 | 1 |  |  |  | $10_5$ |  | 10! | 2 | 9 |
| 6(28) | 6 | 1 |  |  |  | $8_4$ |  | 10! |  | 2 |
| 6(29) | 2 | 6 | 1 |  |  | $14_6$ |  | 3 | 10! | 8 |
| 6(30) | 2 | 6 | 1 |  |  | $10_5$ |  | 9! | 3 |  |
| 6(31) | 2 | 6 | 1 |  |  | $14_6$ |  | 10! | 3 | 9 |
| 6(32) | 2 | 6 | 1 |  |  | $11_5$ |  |  | 10! | 3 |
| 6(33) | 2 | 6 | 1 |  |  | $12_5$ |  | 10! |  | 3 |
| 6(34) | 2 | 6 |  |  | 1 | $12_6$ |  | 5 | 10! | 8 |
| 6(35) | 2 | 6 |  |  |  | $10_5$ |  | 1 | 10! | 8 |
| 6(36) | 2 | 6 |  |  |  | $6_4$ |  | 9! | 1 |  |
| 6(37) | 2 | 6 |  |  |  | $10_5$ |  | 10! | 1 | 9 |
| 6(38) | 2 | 6 |  |  |  | $6_4$ |  |  | 10! | 1 |
| 6(39) | 2 | 6 |  |  |  | $7_4$ |  | 10! |  | 1 |
| 6(40) | 6 | 7 |  |  |  | $2_5$ | 1 |  | 10! | 2 |
| 6(41) | 6 | 7 |  |  |  | $3_5$ | 1 | 10! |  | 2 |
| 6(42) | 6 | 7 |  |  |  | $4_6$ | 1 | 2 | 10! | 8 |
|  | 1 | 2 | 3 | 4 | 5 | 6 | 7 | 8 | 9 | 10 |

|  | 1 | 2 | 3 | 4 | 5 | 6 | 7 | 8 | 9 | 10 |
|---|---|---|---|---|---|---|---|---|---|---|
| 6(43) | 6 | 7 |  | 5 | 2 | $20_8$ | 1 | 4 | 10! | 8 |
| 6(44) | 6 | 7 |  | 5 | 2 | $16_7$ | 1 | 9! | 4 |  |
| 6(45) | 6 | 7 |  | 5 | 2 | $16_7$ | 1 |  | 10! | 4 |
| 6(46) | 6 | 7 |  | 5 | 2 | $17_7$ | 1 | 10! |  | 4 |
| 7(1) |  | 7 |  |  |  |  | $2_3$ |  | 10! | 2 |
| 7(2) |  | 7 |  |  |  |  | $3_3$ | 10! |  | 2 |
| 7(3) | 6 | 7 | 1 |  |  | 2 | $10_7$ | 3 | 10! | 8 |
| 7(4) | 6 | 7 |  | 1 |  | 2 | $10_7$ | 4 | 10! | 8 |
| 7(5) | 6 | 7 |  |  | 1 | 2 | $8_7$ | 5 | 10! | 8 |
| 7(6) | 6 | 7 |  |  |  | 2 | $6_6$ | 1 | 10! | 8 |
| 7(7) | 6 | 7 |  |  |  | 2 | $2_5$ | 9! | 1 |  |
| 7(8) | 6 | 7 |  |  |  | 2 | $6_6$ | 10! | 1 | 9 |
| 7(9) | 6 | 7 |  |  |  | 2 | $2_5$ |  | 10! | 1 |
| 7(10) | 6 | 7 |  |  |  | 2 | $3_5$ | 10! |  | 1 |
| 7(11) |  | 7 |  |  |  | 2 | $8_5$ | 6 | 10! | 8 |
| 7(12) | 5 | 7 | 1 |  | 2 |  | $11_7$ | 3 | 10! | 8 |
| 7(13) | 5 | 7 | 1 |  | 2 |  | $7_6$ | 9! | 3 |  |
| 7(14) | 5 | 7 | 1 |  | 2 |  | $8_6$ |  | 10! | 3 |
| 7(15) | 5 | 7 | 1 |  | 2 |  | $9_6$ | 10! |  | 3 |
| 7(16) | 5 | 7 |  | 1 | 2 |  | $10_7$ | 4 | 10! | 8 |
| 7(17) | 5 | 7 |  | 1 | 2 |  | $6_6$ | 9! | 4 |  |
| 7(18) | 5 | 7 |  | 1 | 2 |  | $10_7$ | 10! | 4 | 9 |
| 7(19) | 5 | 7 |  | 1 | 2 |  | $6_6$ |  | 10! | 4 |
| 7(20) | 5 | 7 |  |  | 2 | 1 | $10_7$ | 6 | 10! | 8 |
| 7(21) | 5 | 7 |  |  | 2 |  | $7_6$ | 1 | 10! | 8 |
| 7(22) | 5 | 7 |  |  | 2 |  | $3_5$ | 9! | 1 |  |
| 7(23) | 5 | 7 |  |  | 2 |  | $7_6$ | 10! | 1 | 9 |
| 7(24) | 5 | 7 |  |  | 2 |  | $3_5$ |  | 10! | 1 |
| 7(25) | 5 | 7 |  |  | 2 |  | $6_5$ | 10! |  | 1 |
| 7(26) |  | 7 |  |  | 2 |  | $7_5$ | 5 | 10! | 8 |
| 7(27) | 7 | 1 |  |  | 2 |  | $11_6$ | 5 | 10! | 8 |
| 7(28) | 7 | 1 |  |  |  | 2 | $12_6$ | 6 | 10! | 8 |
| 7(29) | 7 | 1 |  |  |  |  | $8_5$ | 2 | 10! | 8 |
| 7(30) | 7 | 1 |  |  |  |  | $5_4$ | 9! | 2 |  |
| 7(31) | 7 | 1 |  |  |  |  | $9_5$ | 10! | 2 | 9 |
| 7(32) | 7 | 1 |  |  |  |  | $6_4$ |  | 10! | 2 |
| 7(33) | 7 | 1 |  |  |  |  | $7_4$ | 10! |  | 2 |
| 7(34) | 2 | 7 | 1 |  |  |  |  | 3 | 10! | 8 |
| 7(35) | 2 | 7 | 1 |  |  |  | $14_7$ | 10! | 3 | 9 |
| 7(36) | 2 | 7 | 1 |  |  |  | $10_5$ |  | 10! | 3 |
|  | 1 | 2 | 3 | 4 | 5 | 6 | 7 | 8 | 9 | 10 |

|       | 1 | 2 | 3 | 4 | 5 | 6 | 7      | 8   | 9   | 10 |
|-------|---|---|---|---|---|---|--------|-----|-----|----|
| 7(37) | 2 | 7 |   | 1 |   |   | $8_5$  | 9!  | 4   |    |
| 7(38) | 2 | 7 |   | 1 |   |   | $12_6$ | 10! | 4   | 9  |
| 7(39) | 2 | 7 |   | 1 |   |   | $8_5$  |     | 10! | 4  |
| 7(40) | 2 | 7 |   | 1 |   |   | $9_5$  | 10! |     | 4  |
| 7(41) | 2 | 7 |   |   |   | 1 | $12_6$ | 6   | 10! | 8  |
| 7(42) | 2 | 7 |   |   |   |   | $9_5$  | 1   | 10! | 8  |
| 7(43) | 2 | 7 |   |   |   |   | $5_4$  | 9!  | 1   |    |
| 7(44) | 2 | 7 |   |   |   |   | $9_5$  | 10! | 1   | 9  |
| 7(45) | 2 | 7 |   |   |   |   | $5_4$  |     | 10! | 1  |
| 7(46) | 7 | 6 |   |   |   | 1 | $2_5$  | 9!  | 2   |    |
| 7(47) | 7 | 6 |   |   |   | 1 | $6_6$  | 10! | 2   | 9  |
| 7(48) | 7 | 6 |   |   |   | 1 | $3_5$  |     | 10! | 2  |
| 7(49) | 7 | 6 |   |   |   | 1 | $5_6$  | 2   | 10! | 8  |
| 7(50) | 7 | 6 |   | 5 | 2 | 1 | $21_8$ | 4   | 10! | 8  |
| 7(51) | 7 | 6 |   | 5 | 2 | 1 | $17_7$ | 9!  | 4   |    |
| 7(52) | 7 | 6 |   | 5 | 2 | 1 | $21_8$ | 10! | 4   | 9  |
| 7(52) | 7 | 6 |   | 5 | 2 | 1 | $17_7$ |     | 10! | 4  |
|       | 1 | 2 | 3 | 4 | 5 | 6 | 7      | 8   | 9   | 10 |

(R & CR 3)(20 b)

|       | 1    | 2    | 3    | 4    | 5    | 6 | 7 | 8   | 9   | 10 |
|-------|------|------|------|------|------|---|---|-----|-----|----|
| 1(1)  | $29_8$ | 5 | 7 | 9! | 1 |   | 2 | 3 | 10 | 8 |
| 1(2)  | $4_7$ | 5 | 7 |   | 1 |   | 2 | 9! | 10 | 3 |
| 1(3)  | $26_8$ | 5 | 7 | 9! | 1 |   | 2 |   | 10 | 3 |
| 1(4)  | $27_8$ | 5 | 7 | 9! | 1 |   | 2 | 10 | 8 | 3 |
| 1(5)  | $3_4$ |   |   | 1 |   |   |   | 9! | 10 | 4 |
| 2(1)  | 6 | $5_6$ |   | 1 |   | 2 |   | 9! | 10 | 4 |
| 2(2)  | 6 | $28_8$ | 7 | 9! |   | 2 | 1 | 3 | 10 | 8 |
| 2(3)  | 6 | $24_8$ | 7 | 10! |   | 2 | 1 | 9 | 3 | 8 |
| 2(4)  | 6 | $28_8$ | 7 | 8! |   | 2 | 1 | 10 | 3 | 9 |
| 2(5)  | 6 | $3_7$ | 7 |   |   | 2 | 1 | 9! | 10 | 3 |
| 2(6)  | 6 | $26_8$ | 7 | 9! |   | 2 | 1 | 10 | 8 | 3 |
| 2(7)  | 6 | $2_5$ |   |   |   | 2 |   | 9! | 10 | 1 |
| 2(8)  | 5 | $8_6$ | 1 |   | 2 |   |   | 9! | 10 | 3 |
| 2(9)  | 5 | $6_6$ |   | 1 | 2 |   |   | 9! | 10 | 4 |
| 2(10) | 5 | $3_5$ |   |   | 2 |   |   | 9! | 10 | 1 |
| 2(11) | 7 | $4_6$ |   | 1 |   |   | 2 | 9! | 10 | 4 |
| 2(12) | 7 | $15_7$ |   | 6 |   | 1 | 2 | 9! | 10 | 4 |
| 3(1)  | 7 | 10! | $22_7$ | 1 |   |   | 3 | 9 | 4 | 8 |
| 3(2)  | 7 |   | $3_6$ | 1 |   |   | 3 | 9! | 10 | 4 |
| 3(3)  | 7 |   | $14_7$ | 6 |   | 1 | 3 | 9! | 10 | 4 |
| 3(4)  | 7 |   | $14_7$ | 6 |   | 1 | 3 | 9! | 10 | 4 |
| 4(1)  | 4 |   |   | $2_4$ |   |   |   | 9! | 10 | 1 |
| 4(2)  | 4 | 10! | 7 | $24_7$ |   |   | 1 | 9 | 3 | 8 |
| 4(3)  | 4 | 7 |   | $4_6$ |   |   | 1 | 9! | 10 | 2 |
| 5(1)  | 6 | 5 | 1 | 10! | $29_8$ | 2 |   | 9 | 3 | 8 |
| 5(2)  | 6 | 5 | 1 | 9! | $30_8$ | 2 |   | 10 | 8 | 3 |
| 5(3)  | 6 | 5 | 7 | 10! | $24_9$ | 2 | 1 | 3 | 8 | 9 |
| 5(4)  | 6 | 5 | 7 |   | $2_7$ | 2 | 1 | 9! | 3 |   |
| 5(5)  | 6 | 5 | 7 |   | $6_8$ | 2 | 1 | 10! | 3 | 9 |
| 5(6)  | 6 | 5 | 7 |   | $2_7$ | 2 | 1 |   | 10! | 3 |
| 5(7)  | 6 | 5 | 7 |   | $3_7$ | 2 | 1 | 10! |   | 3 |
| 5(8)  | 6 | 5 | 10! |   | $25_7$ | 2 |   | 9 | 1 | 8 |
| 5(9)  | 6 | 5 |   | 10! | $25_7$ | 2 |   | 9 | 1 | 8 |
| 5(10) | 1 | 2 | 3 | 4 | 5 | 6 | 7 | 8 | 9 | 10 |

|       | 1 | 2 | 3   | 4   | 5       | 6      | 7        | 8   | 9   | 10 |
|-------|---|---|-----|-----|---------|--------|----------|-----|-----|----|
| 5(11) | 6 | 5 |     |     | $2_6$   | 2      |          | 9!  | 10  | 1  |
| 5(12) | 5 | 1 | 7   |     | $7_7$   |        | 2        | 9!  | 10  | 3  |
| 5(13) | 5 | 1 | 7   | 9!  | $30_8$  |        | 2        | 10  | 8   | 3  |
| 5(14) | 2 | 5 |     | 1   | $9_6$   |        |          | 9!  | 10  | 4  |
| 5(15) | 2 | 5 | 7   | 10! | $29_8$  |        | 1        | 9   | 3   | 8  |
| 5(16) | 2 | 5 | 7   |     | $7_7$   |        | 1        | 9!  | 10  | 3  |
| 5(17) | 2 | 5 | 7   | 9!  | $30_8$  |        | 1        | 10  | 8   | 3  |
| 5(18) | 2 | 5 |     |     | $6_5$   |        |          | 9!  | 10  | 1  |
| 6(1)  | 5 | 6 | 1   |     | 2       | $7_6$  |          | 9!  | 10  | 3  |
| 6(2)  | 5 | 6 | 1   | 10! | 2       | $30_8$ |          | 9   | 3   | 8  |
| 6(3)  | 5 | 6 | 10! | 1   | 2       | $29_8$ |          | 9   | 4   | 8  |
| 6(4)  | 6 | 1 | 7   |     |         | $7_7$  | 2        | 9!  | 10  | 3  |
| 6(5)  | 2 | 6 | 1   |     |         | $11_6$ |          | 9!  | 10  | 3  |
| 6(6)  | 2 | 6 |     |     |         | $6_5$  |          | 9!  | 10  | 1  |
| 6(7)  | 6 | 7 |     |     |         | $2_6$  | 1        | 9!  | 10  | 2  |
| 6(8)  | 6 | 7 | 9!  |     |         | $24_7$ | 1        | 10  | 8   | 2  |
| 6(9)  | 6 | 7 | 9!  |     |         | $25_7$ | 1        | 2   | 10  | 8  |
| 6(10) | 6 | 7 |     | 9!  |         | $26_7$ | 1        | 2   | 10  | 8  |
| 6(11) | 6 | 7 |     | 5   | 2       | $16_8$ | 1        | 9!  | 10  | 4  |
| 7(1)  |   | 7 |     |     |         |        | $2_4$    | 9!  | 10  | 2  |
| 7(2)  | 6 | 7 | 9!  |     | 1       | 2      | $29_8$   | 5   | 10  | 8  |
| 7(3)  | 6 | 7 |     | 9!  | 1       | 2      | $30_8$   | 5   | 10  | 8  |
| 7(4)  | 6 | 7 |     |     |         | 2      | $2_6$    | 9!  | 10  | 1  |
| 7(5)  | 6 | 7 | 9!  |     |         | 2      | $24_7$   | 10  | 8   | 1  |
| 7(6)  | 6 | 7 |     | 9!  |         | 2      | $25_7$   | 10  | 8   | 1  |
| 7(7)  | 5 | 7 | 1   |     | 2       |        | $8_7$    | 9!  | 10  | 3  |
| 7(8)  | 5 | 7 |     | 1   | 2       |        | $6_7$    | 9!  | 10  | 4  |
| 7(9)  | 5 | 7 | 10! |     | 2       |        | $3_5$    | 9   | 1   | 8  |
| 7(10) | 5 | 7 |     |     | 2       |        | $3_6$    | 9!  | 10  | 1  |
| 7(11) | 7 | 1 |     |     |         |        | $6_4$    |     | 10! | 2  |
| 7(12) | 2 | 7 | 1   |     |         |        | $10_6$   | 9!  | 10  | 3  |
| 7(13) | 2 | 7 |     | 1   |         |        | $8_6$    | 9!  | 10  | 4  |
| 7(14) | 2 | 7 |     |     |         |        | $9_5$    | 10! | 1   | 9  |
| 7(15) | 2 | 7 |     |     |         |        | $5_5$    | 9!  | 10  | 1  |
| 7(16) | 7 | 6 |     |     |         | 1      | $2_5$    | 9!  | 2   |    |
| 7(17) | 7 | 6 |     |     |         | 1      | $3_6$    | 9!  | 10  | 2  |
| 7(18) | 7 | 6 | 9!  |     |         | 1      | $26_7$   | 2   | 10  | 8  |
| 7(19) | 7 | 6 |     | 5   | 2       | 1      | $17_8$   | 9!  | 10  | 4  |
|       | 1 | 2 | 3   | 4   | 5       | 6      | 7        | 8   | 9   | 10 |

(R & CR 3)(30 b)

|  | 1 | 2 | 3 | 4 | 5 | 6 | 7 | 8 | 9 | 10 |
|---|---|---|---|---|---|---|---|---|---|---|
| 3(1) | 7 | 8! | $23_7$ | 1 |  |  | 3 | 9 | 10 | 4 |
| 4(1) | 4 | 7 | 8! | $25_7$ |  |  | 1 | 9 | 10 | 2 |
| 5(1) | 6 | 5 | 7 | 10! | $25_9$ | 2 | 1 | 9 | 3 | 8 |
| 5(2) | 6 | 5 | 7 | 8! | $28_9$ | 2 | 1 | 10 | 3 | 9 |
| 5(3) | 6 | 5 | 7 |  | $2_8$ | 2 | 1 | 9! | 10 | 3 |
| 5(4) | 6 | 5 | 7 | 9! | $25_9$ | 2 | 1 | 10 | 8 | 3 |
| 5(5) | 6 | 5 | 8! |  | $23_7$ | 2 |  | 9 | 10 | 1 |
| 5(6) | 6 | 5 |  | 8! | $24_7$ | 2 |  | 9 | 10 | 1 |
| 5(7) | 5 | 1 | 7 | 8! | $29_8$ |  | 2 | 9 | 10 | 3 |
| 5(8) | 2 | 5 | 7 | 8! | $29_8$ |  | 1 | 9 | 10 | 3 |
| 6(1) | 5 | 6 | 1 | 8! | 2 | $30_8$ |  | 9 | 10 | 3 |
| 6(2) | 5 | 6 | 1 | 10! | 2 | $30_8$ |  | 9 | 3 | 8 |
| 6(3) | 6 | 1 | 7 | 8! |  | $29_8$ | 2 | 9 | 10 | 3 |
| 6(4) | 6 | 7 | 8! |  |  | $23_7$ | 1 | 9 | 10 | 2 |
| 7(1) | 6 | 7 | 8! |  | 1 | 2 | $23_7$ | 9 | 10 | 1 |
| 7(2) | 5 | 7 | 1 | 8! | 2 |  | $30_8$ | 9 | 10 | 3 |
| 7(3) | 5 | 7 | 8! | 1 | 2 |  | $27_8$ | 9 | 10 | 4 |
| 7(4) | 5 | 7 | 8! |  | 2 |  | $24_7$ | 9 | 10 | 1 |
| 7(5) | 5 | 7 |  | 8! | 2 |  | $25_7$ | 9 | 10 | 1 |
| 7(6) | 7 | 1 |  |  |  |  | $6_5$ | 9! | 10 | 2 |
| 7(7) | 7 | 6 | 8! |  |  | 1 | $24_7$ | 9 | 10 | 2 |
| 7(8) | 7 | 6 |  | 8! |  | 1 | $25_7$ | 9 | 10 | 2 |
|  | 1 | 2 | 3 | 4 | 5 | 6 | 7 | 8 | 9 | 10 |

(R & CR 3)(40 b)

|      | 1 | 2 | 3 | 4  | 5      | 6 | 7 | 8  | 9  | 10 |
|------|---|---|---|----|--------|---|---|----|----|----|
| 5(3) | 6 | 5 | 7 | 8! | $24_9$ | 2 | 1 | 9! | 10 | 3  |
|      | 1 | 2 | 3 | 4  | 5      | 6 | 7 | 8  | 9  | 10 |

Path 5(3) can't be extended. Thus, $T_{OPT} = T_{UPPERBOUND}$.

Example 11.

M

|    | 1  | 2  | 3  | 4  | 5  | 6  | 7  | 8  | 9  | 10 | 11 | 12 | 13 | 14 | 15 | 16 | 17 | 18 | 19 | 20 |    |
|----|----|----|----|----|----|----|----|----|----|----|----|----|----|----|----|----|----|----|----|----|----|
| 1  | ∞  | 47 | 19 | 49 | 38 | 43 | 59 | 99 | 51 | 55 | 59 | 84 | 80 | 16 | 58 | 95 | 50 | 57 | 48 | 31 | 1  |
| 2  | 47 | ∞  | 14 | 90 | 92 | 38 | 5  | 1  | 32 | 39 | 68 | 50 | 59 | 53 | 63 | 83 | 11 | 60 | 5  | 13 | 2  |
| 3  | 19 | 14 | ∞  | 58 | 49 | 33 | 18 | 93 | 34 | 68 | 57 | 66 | 37 | 5  | 91 | 17 | 62 | 99 | 99 | 87 | 3  |
| 4  | 49 | 90 | 58 | ∞  | 20 | 81 | 38 | 70 | 66 | 30 | 11 | 88 | 47 | 36 | 21 | 46 | 94 | 30 | 84 | 71 | 4  |
| 5  | 38 | 92 | 49 | 20 | ∞  | 6  | 39 | 81 | 88 | 81 | 61 | 12 | 15 | 84 | 25 | 37 | 35 | 96 | 64 | 10 | 5  |
| 6  | 43 | 38 | 33 | 81 | 6  | ∞  | 62 | 73 | 82 | 55 | 79 | 69 | 55 | 62 | 92 | 70 | 62 | 49 | 99 | 92 | 6  |
| 7  | 59 | 5  | 18 | 38 | 21 | 62 | ∞  | 55 | 23 | 19 | 30 | 80 | 57 | 18 | 62 | 79 | 16 | 68 | 75 | 74 | 7  |
| 8  | 99 | 1  | 93 | 70 | 81 | 73 | 55 | ∞  | 27 | 58 | 20 | 15 | 72 | 78 | 17 | 88 | 75 | 24 | 91 | 20 | 8  |
| 9  | 51 | 32 | 34 | 66 | 88 | 82 | 23 | 27 | ∞  | 22 | 62 | 48 | 33 | 6  | 22 | 4  | 5  | 18 | 39 | 62 | 9  |
| 10 | 55 | 39 | 68 | 30 | 81 | 55 | 19 | 58 | 22 | ∞  | 62 | 6  | 41 | 47 | 43 | 4  | 51 | 81 | 59 | 52 | 10 |
| 11 | 59 | 68 | 57 | 11 | 61 | 79 | 30 | 20 | 62 | 62 | ∞  | 20 | 40 | 9  | 31 | 32 | 95 | 30 | 78 | 17 | 11 |
| 12 | 84 | 50 | 66 | 88 | 12 | 69 | 80 | 15 | 48 | 6  | 20 | ∞  | 48 | 63 | 99 | 20 | 19 | 98 | 61 | 23 | 12 |
| 13 | 80 | 59 | 37 | 47 | 15 | 55 | 57 | 72 | 33 | 41 | 40 | 48 | ∞  | 92 | 86 | 49 | 36 | 11 | 42 | 99 | 13 |
| 14 | 16 | 53 | 5  | 36 | 84 | 62 | 18 | 78 | 6  | 47 | 9  | 63 | 92 | ∞  | 44 | 11 | 56 | 63 | 43 | 59 | 14 |
| 15 | 58 | 63 | 91 | 21 | 25 | 92 | 62 | 17 | 22 | 43 | 31 | 99 | 86 | 44 | ∞  | 45 | 67 | 99 | 64 | 56 | 15 |
| 16 | 95 | 83 | 17 | 46 | 37 | 70 | 79 | 88 | 4  | 4  | 32 | 20 | 49 | 11 | 45 | ∞  | 21 | 11 | 27 | 79 | 16 |
| 17 | 50 | 11 | 62 | 94 | 35 | 62 | 16 | 75 | 5  | 51 | 95 | 19 | 36 | 56 | 67 | 21 | ∞  | 17 | 63 | 46 | 17 |
| 18 | 57 | 60 | 99 | 30 | 96 | 49 | 68 | 24 | 18 | 81 | 30 | 98 | 11 | 63 | 99 | 11 | 17 | ∞  | 96 | 47 | 18 |
| 19 | 48 | 5  | 99 | 84 | 64 | 99 | 75 | 91 | 39 | 59 | 78 | 61 | 42 | 43 | 64 | 17 | 63 | 96 | ∞  | 36 | 19 |
| 20 | 31 | 13 | 87 | 71 | 10 | 92 | 74 | 20 | 62 | 52 | 17 | 23 | 99 | 59 | 56 | 79 | 46 | 47 | 36 | ∞  | 20 |
|    | 1  | 2  | 3  | 4  | 5  | 6  | 7  | 8  | 9  | 10 | 11 | 12 | 13 | 14 | 15 | 16 | 17 | 18 | 19 | 20 |    |

MIN(M)

|    | 1  | 2  | 3  | 4  | 5  | 6  | 7  | 8  | 9  | 10 | 11 | 12 | 13 | 14 | 15 | 16 | 17 | 18 | 19 | 20 |
|----|----|----|----|----|----|----|----|----|----|----|----|----|----|----|----|----|----|----|----|----|
| 1  | 14 | 3  | 20 | 5  | 6  | 2  | 19 | 17 | 4  | 9  | 10 | 18 | 16 | 7  | 11 | 13 | 12 | 16 | 8  |    |
| 2  | 8  | 7  | 19 | 17 | 20 | 3  | 9  | 6  | 10 | 1  | 12 | 14 | 13 | 18 | 15 | 11 | 16 | 4  | 5  |    |
| 3  | 14 | 2  | 16 | 7  | 1  | 6  | 9  | 13 | 5  | 11 | 4  | 17 | 12 | 10 | 20 | 15 | 8  | 18 | 19 |    |
| 4  | 11 | 5  | 15 | 10 | 18 | 14 | 7  | 16 | 13 | 1  | 3  | 9  | 8  | 20 | 6  | 19 | 12 | 2  | 17 |    |
| 5  | 6  | 20 | 12 | 13 | 4  | 15 | 17 | 16 | 1  | 7  | 3  | 11 | 19 | 8  | 10 | 14 | 9  | 2  | 18 |    |
| 6  | 5  | 3  | 2  | 1  | 18 | 10 | 13 | 7  | 14 | 17 | 12 | 16 | 8  | 11 | 4  | 9  | 15 | 20 | 19 |    |
| 7  | 2  | 17 | 3  | 14 | 10 | 5  | 9  | 11 | 4  | 8  | 13 | 1  | 15 | 6  | 18 | 20 | 19 | 16 | 12 |    |
| 8  | 2  | 12 | 15 | 20 | 11 | 18 | 9  | 7  | 10 | 4  | 13 | 6  | 17 | 14 | 5  | 16 | 19 | 3  | 1  |    |
| 9  | 16 | 17 | 14 | 18 | 10 | 15 | 7  | 8  | 2  | 13 | 3  | 19 | 12 | 1  | 11 | 20 | 4  | 6  | 5  |    |
| 10 | 16 | 12 | 7  | 9  | 4  | 2  | 13 | 15 | 14 | 17 | 20 | 1  | 6  | 8  | 19 | 11 | 3  | 5  | 18 |    |
| 11 | 14 | 4  | 20 | 8  | 12 | 7  | 17 | 18 | 15 | 16 | 3  | 1  | 5  | 9  | 10 | 2  | 19 | 6  | 17 |    |
| 12 | 10 | 5  | 8  | 17 | 11 | 16 | 20 | 9  | 13 | 2  | 19 | 14 | 3  | 6  | 7  | 1  | 4  | 18 | 15 |    |
| 13 | 18 | 5  | 9  | 17 | 3  | 11 | 10 | 19 | 4  | 12 | 16 | 6  | 7  | 2  | 8  | 1  | 15 | 14 | 20 |    |
| 14 | 3  | 9  | 11 | 16 | 1  | 7  | 4  | 19 | 15 | 10 | 2  | 17 | 20 | 6  | 12 | 18 | 8  | 5  | 13 |    |
| 15 | 8  | 4  | 9  | 5  | 11 | 10 | 14 | 16 | 20 | 1  | 7  | 2  | 19 | 17 | 13 | 3  | 6  | 12 | 18 |    |
| 16 | 9  | 10 | 14 | 18 | 3  | 12 | 17 | 19 | 11 | 5  | 15 | 4  | 13 | 6  | 20 | 7  | 2  | 8  | 1  |    |
| 17 | 9  | 2  | 7  | 18 | 12 | 16 | 5  | 13 | 20 | 1  | 10 | 14 | 3  | 6  | 19 | 15 | 8  | 4  | 11 |    |
| 18 | 13 | 16 | 17 | 9  | 8  | 4  | 11 | 20 | 6  | 1  | 2  | 14 | 7  | 10 | 5  | 19 | 12 | 3  | 15 |    |
| 19 | 2  | 16 | 20 | 9  | 13 | 14 | 1  | 10 | 12 | 17 | 5  | 15 | 7  | 11 | 4  | 8  | 18 | 3  | 6  |    |
| 20 | 5  | 2  | 11 | 8  | 12 | 1  | 19 | 17 | 18 | 10 | 15 | 14 | 9  | 4  | 7  | 16 | 3  | 6  | 13 |    |
|    | 1  | 2  | 3  | 4  | 5  | 6  | 7  | 8  | 9  | 10 | 11 | 12 | 13 | 14 | 15 | 16 | 17 | 18 | 19 | 20 |

*Comment 1*. Let $U = \{1,2,...,M\}$, $M$ *fixed* be the set $R$ from which elements are randomly chosen to form $n \times n$ random matrices whose diagonal elements are all $\infty$. When applying FWK to a random matrix, we should generally expect the ratio of the average arc-value of $T_{OPT}$ to $n$ to decrease substantially as $n$ increases. The reasoning behind this is that we would expect small natural numbers to occur in each row no matter how large $n$ becomes. Perhaps this implies that at least for random matrices, we may generally be able to obtain an optimal tour ($n$-cycle) for large values of $n$ in polynomial time.

*Comment 2.* Suppose that $m$ parallel processors are employed in FWK. Then - if possible - when constructing negative cycles in a heuristic obtained from FWK, we construct up to $m$ of them. We choose the one with smallest value to reduce the value of the last derangement. Since we don't have parallel processors in this example, we henceforth use the [ln n] + 1 smallest-valued cycles obtained. In general, we continue constructing negative cycles in the following manner: (a) we check each subpath of a negative path to see what its value is and then keep in memory whichever subpath together with the arc to the initial point has the smallest value; (b) since our matrix is symmetric, we can't choose the arc symmetric to one already chosen. A 2-cycle in a tour constitutes *one* edge. Thus, *we would no longer have a derangement*. In particular, if the last arc leads to a 2-cycle, we backtrack to the next to last arc. We then choose a new terminal point for the arc. (c) In constructing a negative path, if no acceptable non-positive arc can be obtained to extend the path, we find the smallest positive value such that it leads to a negative entry where the sum of the two values is negative.. (d) if - using (d) - no negative cycle can be found, we continue extending our negative path until it becomes positive. We then check each subpath to see which cycle obtained from them is the smallest-valued negative cycle. (e) If - using (a)-(e), no negative cycle can be found, we use the last derangement obtained to construct an $n$-cycle.

$$D_0^{-1}M^-$$

| | 2 | 3 | 4 | 5 | 6 | 7 | 8 | 9 | 10 | 11 | 12 | 13 | 14 | 15 | 16 | 17 | 18 | 19 | 20 | 1 | |
|---|---|---|---|---|---|---|---|---|---|---|---|---|---|---|---|---|---|---|---|---|---|
| | 1 | 2 | 3 | 4 | 5 | 6 | 7 | 8 | 9 | 10 | 11 | 12 | 13 | 14 | 15 | 16 | 17 | 18 | 19 | 20 | |
| 1 | 0 | -28 | 2 | 38 | -9 | 12 | 52 | 4 | 8 | 12 | 37 | 33 | -31 | 11 | 48 | 3 | 10 | 1 | 16 | ∞ | 1 |
| 2 | ∞ | 0 | 76 | 78 | 24 | -9 | -13 | 18 | 25 | 54 | 36 | 45 | 39 | 49 | 69 | -3 | 46 | -9 | -1 | 33 | 2 |
| 3 | -44 | ∞ | 0 | -9 | -25 | -40 | 35 | -24 | 18 | -1 | 8 | -21 | -53 | 33 | -41 | 4 | 41 | 41 | 29 | -39 | 3 |
| 4 | 70 | 38 | ∞ | 0 | 61 | 18 | 50 | 46 | 10 | -9 | 68 | 27 | 16 | 1 | 26 | 74 | 10 | 64 | 51 | 29 | 4 |
| 5 | 86 | 43 | 14 | ∞ | 0 | 33 | 75 | 82 | 75 | 55 | 6 | 9 | 78 | 19 | 31 | 29 | 90 | 58 | 4 | 32 | 5 |
| 6 | -24 | -29 | 19 | -56 | ∞ | 0 | 11 | 20 | -7 | 17 | 7 | -7 | 0 | 30 | 8 | 0 | -13 | 37 | 30 | -19 | 6 |
| 7 | -50 | -37 | -17 | -34 | 7 | ∞ | 0 | -32 | -36 | -25 | 25 | 2 | -37 | 7 | 24 | -39 | 13 | 20 | 19 | 4 | 7 |
| 8 | -26 | 66 | 43 | 54 | 46 | 28 | ∞ | 0 | 31 | -7 | -12 | 45 | 51 | -10 | 61 | 48 | -3 | 64 | -7 | 72 | 8 |
| 9 | 10 | 12 | 44 | 66 | 60 | 1 | 5 | ∞ | 0 | 40 | 26 | 11 | -16 | 0 | -18 | -17 | -4 | 17 | 40 | 29 | 9 |
| 10 | -23 | 6 | -32 | 19 | -7 | -43 | -4 | -40 | ∞ | 0 | -56 | -21 | -15 | -19 | -58 | -11 | 19 | -3 | -10 | -7 | 10 |
| 11 | 48 | 37 | -9 | 41 | 59 | 10 | 0 | 42 | 42 | ∞ | 0 | 20 | -11 | 11 | 12 | 75 | 10 | 58 | 7 | 39 | 11 |
| 12 | 2 | 18 | 40 | -36 | 21 | 32 | -33 | 0 | -42 | -28 | ∞ | 0 | 15 | 51 | -28 | -29 | 50 | 15 | -25 | 36 | 12 |
| 13 | -33 | -55 | -45 | -77 | -37 | -35 | -20 | -59 | -51 | -52 | -44 | ∞ | 0 | -6 | -43 | -56 | -81 | -50 | 7 | -12 | 13 |
| 14 | 9 | -39 | -8 | 40 | 18 | -26 | 34 | -38 | 3 | -35 | 19 | 48 | ∞ | 0 | -33 | 12 | 19 | -1 | 15 | -28 | 14 |
| 15 | 18 | 46 | -24 | -20 | 47 | 17 | -28 | 23 | -2 | -14 | 54 | 41 | -1 | ∞ | 0 | 22 | 54 | 19 | 11 | 13 | 15 |
| 16 | 62 | -4 | 25 | 16 | 49 | 58 | 67 | -17 | -17 | 11 | -1 | 28 | -10 | 45 | ∞ | 0 | -10 | 6 | 58 | 74 | 16 |
| 17 | -6 | 45 | 77 | 18 | 45 | -1 | 58 | -12 | 34 | 78 | 2 | 19 | 39 | 50 | 4 | ∞ | 0 | 46 | 29 | 33 | 17 |
| 18 | -36 | 3 | -66 | 0 | -47 | -28 | -72 | -78 | -15 | -66 | 2 | -85 | -33 | 3 | -85 | -79 | ∞ | 0 | -49 | -39 | 18 |
| 19 | 31 | 63 | 48 | 28 | 63 | 39 | 55 | 3 | 23 | 42 | 25 | 6 | 7 | 28 | -19 | 27 | 60 | ∞ | 0 | 12 | 19 |
| 20 | -18 | 56 | 40 | -21 | 61 | 43 | -11 | 31 | 21 | -14 | -8 | 68 | 28 | 25 | 48 | 15 | 16 | 5 | ∞ | 0 | 20 |
| | 1 | 2 | 3 | 4 | 5 | 6 | 7 | 8 | 9 | 10 | 11 | 12 | 13 | 14 | 15 | 16 | 17 | 18 | 19 | 20 | |

In the following table we omit the "minus signs" when giving values of paths.

*Comment.* At the left-hand side of the bottom row of each mini-matrix that follows, we print ARCS. We place 1,2, … , n in the remainder of the columns of the row. Each time we obtain a new arc belonging to a possible *derangement*, we place the terminal point of the arc underneath its initial point. As we proceed to construct a negative cycle, we can use these arcs to check if we already have chosen an arc symmetric to one already chosen. This would yield a 2-cycle which is prohibited.

|        | 2   | 3 | 4   | 5   | 6 | 7   | 8   | 9   | 10 | 11 | 12  | 13 | 14  | 15 | 16 | 17 | 18 | 19   | 20 | 1 |
|--------|-----|---|-----|-----|---|-----|-----|-----|----|----|-----|----|-----|----|----|----|----|------|----|---|
|        | 1   | 2 | 3   | 4   | 5 | 6   | 7   | 8   | 9  | 10 | 11  | 12 | 13  | 14 | 15 | 16 | 17 | 18   | 19 | 20 |
| (18 15) | 7  |   | 11  | 6   |   | 3   | 15  | 13  |    |    | 8   |    | 1   |    | 18 |    |    | 4    |    |   |
| -307   | 163 |   | 275 | 371 |   | 315 | 113 | 253 |    |    | 265 |    | 194 |    | 85 |    |    | 307! |    |   |
| ARCS   | 1   | 2 | 3   | 4   | 5 | 6   | 7   | 8   | 9  | 10 | 11  | 12 | 13  | 14 | 15 | 16 | 17 | 18   | 19 | 20 |
|        | 14  | 3 | 7   | 19  | 6 | 5   | 2   | 12  | 10 | 11 | 4   | 13 | 9   | 15 | 8  | 17 | 18 | 16   | 20 | 1 |

Checking the possible cycles obtainable from the above table, the smallest-valued cycle is:

C(1) = (18 15 7 1 13 8 11 3 6 4): -307

|        | 2   | 3 | 4   | 5   | 6 | 7   | 8   | 9   | 10  | 11  | 12  | 13 | 14  | 15 | 16  | 17 | 18 | 19   | 20 | 1   |
|--------|-----|---|-----|-----|---|-----|-----|-----|-----|-----|-----|----|-----|----|-----|----|----|------|----|-----|
|        | 1   | 2 | 3   | 4   | 5 | 6   | 7   | 8   | 9   | 10  | 11  | 12 | 13  | 14 | 15  | 16 | 17 | 18   | 19 | 20  |
| (18 12) | 7  |   | 11  | 20  |   |     | 15  | 13  | 12  |     | 8   | 18 | 1   |    | 9   |    |    | 4    |    | 3   |
| -330   | 223 |   | 334 | 394 |   | 305 | 173 | 313 | 127 | 403 | 325 | 85 | 254 |    | 145 |    |    | 330! |    | 373 |
| ARCS   | 1   | 2 | 3   | 4   | 5 | 6   | 7   | 8   | 9   | 10  | 11  | 12 | 13  | 14 | 15  | 16 | 17 | 18   | 19 | 20  |
|        | 14  | 3 | 1   | 19  | 6 | 7   | 2   | 12  | 16  | 11  | 4   | 10 | 9   | 15 | 8   | 17 | 18 | 13   | 20 | 5   |

C(2) = (18 12 9 15 7 1 13 8 11 3 20 4) : -330

|        | 2   | 3   | 4 | 5   | 6 | 7   | 8  | 9   | 10  | 11 | 12   | 13 | 14 | 15 | 16 | 17 | 18 | 19 | 20 | 1   |
|--------|-----|-----|---|-----|---|-----|----|-----|-----|----|------|----|----|----|----|----|----|----|----|-----|
|        | 1   | 2   | 3 | 4   | 5 | 6   | 7  | 8   | 9   | 10 | 11   | 12 | 13 | 14 | 15 | 16 | 17 | 18 | 19 | 20  |
| (13 17) | 8  | 1   |   | 20  |   | 2   |    | 17  |     | 4  | 10   |    | 11 |    |    |    |    | 13 |    | 6   |
| -281   | 119 | 157 |   | 205 |   | 166 |    | 93  |     | 214| 270  |    | 281!|   |    |    |    | 81 |    | 185 |
| ARCS   | 1   | 2   | 3 | 4   | 5 | 6   | 7  | 8   | 9   | 10 | 11   | 12 | 13 | 14 | 15 | 16 | 17 | 18 | 19 | 20  |
|        | 3   | 7   | 4 | 11  | 6 | 1   | 8  | 2   | 10  | 12 | 14   | 13 | 18 | 15 | 16 | 17 | 9  | 19 | 20 | 5   |

C(3) = (13 17 8 1 2 6 20 4 10 11): -281

Thus, C(2) is the smallest-valued cycle. We use the following arcs obtained from it to construct $D_1^{-1}$ :

(18 13), (12 10), (9 16), (15 8), (7 2), (1 14), (13 9), (8 12), (11 4), (3 1), (20 5), (4 19)

$$D_1^{-1}M^-$$

|  | 14 | 3 | 1 | 19 | 6 | 7 | 2 | 12 | 16 | 11 | 4 | 10 | 9 | 15 | 8 | 17 | 18 | 13 | 20 | 5 |  |
|---|---|---|---|---|---|---|---|---|---|---|---|---|---|---|---|---|---|---|---|---|---|
|  | 1 | 2 | 3 | 4 | 5 | 6 | 7 | 8 | 9 | 10 | 11 | 12 | 13 | 14 | 15 | 16 | 17 | 18 | 19 | 20 |  |
| 1 | 0 | 3 | ∞ | 32 | 22 | 43 | 31 | 68 | 79 | 43 | 33 | 39 | 35 | 42 | 83 | 34 | 41 | 64 | 47 | 69 | 1 |
| 2 | 39 | 0 | 33 | -9 | 24 | -9 | ∞ | 36 | 69 | 54 | 76 | 25 | 18 | 49 | -13 | -3 | 46 | 45 | -1 | 78 | 2 |
| 3 | -14 | ∞ | 0 | 80 | 14 | -1 | -5 | 47 | -2 | 38 | 39 | 57 | 15 | 72 | 74 | 43 | 80 | 18 | 68 | 30 | 3 |
| 4 | -48 | -26 | -35 | 0 | -3 | -46 | 6 | 4 | -38 | -73 | ∞ | -54 | -18 | -63 | -14 | 10 | -54 | -37 | -13 | -64 | 4 |
| 5 | 78 | 43 | 32 | 58 | 0 | 33 | 86 | 6 | 31 | 55 | 14 | 75 | 82 | 19 | 75 | 29 | 90 | 9 | 4 | ∞ | 5 |
| 6 | 0 | -29 | -19 | 37 | ∞ | 0 | -24 | 7 | 8 | 17 | 19 | -7 | 20 | 30 | 11 | 0 | -13 | -7 | 30 | -56 | 6 |
| 7 | 13 | 13 | 54 | 70 | 57 | ∞ | 0 | 75 | 74 | 25 | 33 | 14 | 18 | 57 | 50 | 11 | 63 | 52 | 69 | 16 | 7 |
| 8 | 63 | 78 | 84 | 76 | 58 | 40 | -14 | 0 | 73 | 5 | 55 | 43 | 12 | 2 | ∞ | 60 | 9 | 57 | 5 | 66 | 8 |
| 9 | 2 | 30 | 47 | 35 | 78 | 19 | 28 | 44 | 0 | 58 | 62 | 18 | ∞ | 18 | 23 | 1 | 14 | 29 | 58 | 84 | 9 |
| 10 | -15 | 6 | -7 | -3 | -7 | -43 | -23 | -56 | -58 | 0 | -32 | ∞ | -40 | -19 | -4 | -11 | 19 | -21 | -10 | 19 | 10 |
| 11 | -2 | 46 | 48 | 67 | 68 | 19 | 57 | 9 | 21 | ∞ | 0 | 51 | 51 | 20 | 9 | 84 | 19 | 29 | 16 | 50 | 11 |
| 12 | 57 | 60 | 78 | 67 | 53 | 74 | 44 | ∞ | 14 | 14 | 82 | 0 | 42 | 93 | 9 | 13 | 92 | 42 | 17 | 6 | 12 |
| 13 | 59 | 4 | 47 | 9 | 22 | 24 | 26 | 15 | 16 | 7 | 14 | 8 | 0 | 53 | 39 | 3 | -22 | ∞ | 66 | -18 | 13 |
| 14 | ∞ | -39 | -28 | -1 | 18 | -26 | 9 | 19 | -33 | -35 | -8 | 3 | -38 | 0 | 34 | 12 | 19 | 48 | 15 | 40 | 14 |
| 15 | 27 | 74 | 41 | 47 | 75 | 45 | 46 | 72 | 28 | 14 | 4 | 26 | 51 | ∞ | 0 | 50 | 82 | 69 | 39 | 8 | 15 |
| 16 | -10 | -4 | 74 | 6 | 49 | 58 | 62 | -1 | ∞ | 11 | 25 | -17 | -17 | 45 | 67 | 0 | -10 | 28 | 58 | 16 | 16 |
| 17 | 39 | 45 | 33 | 46 | 45 | -1 | -6 | 2 | 4 | 78 | 77 | 34 | -12 | 50 | 58 | ∞ | 0 | 19 | 29 | 18 | 17 |
| 18 | 52 | 88 | 46 | 85 | 38 | 57 | 49 | 87 | 0 | 19 | 19 | 70 | 7 | 99 | 13 | 6 | ∞ | 0 | 36 | 85 | 18 |
| 19 | 7 | 63 | 12 | ∞ | 63 | 39 | 31 | 25 | -19 | 42 | 48 | 23 | 3 | 28 | 55 | 27 | 60 | 6 | 0 | 28 | 19 |
| 20 | 49 | 77 | 21 | 26 | 82 | 64 | 3 | 13 | 69 | 7 | 61 | 42 | 52 | 46 | 10 | 36 | 37 | 89 | ∞ | 0 | 20 |
|  | 1 | 2 | 3 | 4 | 5 | 6 | 7 | 8 | 9 | 10 | 11 | 12 | 13 | 14 | 15 | 16 | 17 | 18 | 19 | 20 |  |

|  | 14 | 3 | 1 | 19 | 6 | 7 | 2 | 12 | 16 | 11 | 4 | 10 | 9 | 15 | 8 | 17 | 18 | 13 | 20 | 5 |
|---|---|---|---|---|---|---|---|---|---|---|---|---|---|---|---|---|---|---|---|---|
|  | 1 | 2 | 3 | 4 | 5 | 6 | 7 | 8 | 9 | 10 | 11 | 12 | 13 | 14 | 15 | 16 | 17 | 18 | 19 | 20 |
| (4 20) |  | 6 |  | 2 |  | 17 |  |  | 10 | 20 |  |  |  |  |  | 9 | 16 |  |  | 4 |
| -163 |  | 144 |  | 153! |  | 115 |  |  | 115 | -57 |  |  |  |  |  | 114 | 124 |  |  | 64 |
| ARCS | 1 | 2 | 3 | 4 | 5 | 6 | 7 | 8 | 9 | 10 | 11 | 12 | 13 | 14 | 15 | 16 | 17 | 18 | 19 | 20 |
|  | 14 | 19 | 1 | 5 | 6 | 3 | 2 | 12 | 17 | 16 | 4 | 10 | 9 | 15 | 8 | 18 | 7 | 13 | 20 | 11 |

C(1) = (4 20 10 9 16 17 6 2): -163

|  | 14 | 3 | 1 | 19 | 6 | 7 | 2 | 12 | 16 | 11 | 4 | 10 | 9 | 15 | 8 | 17 | 18 | 13 | 20 | 5 |
|---|---|---|---|---|---|---|---|---|---|---|---|---|---|---|---|---|---|---|---|---|
|  | 1 | 2 | 3 | 4 | 5 | 6 | 7 | 8 | 9 | 10 | 11 | 12 | 13 | 14 | 15 | 16 | 17 | 18 | 19 | 20 |
| (4 14) |  | 6 |  | 2! |  | 17 |  |  | 10 | 20 |  |  | 14 | 4 |  | 9 | 16 |  |  | 13 |
| -217 |  | 209 |  | 218 |  | 180 |  |  | 170 | 112 |  |  | 101 | 63 |  | 169 | 179 |  |  | 119 |
| ARCS | 1 | 2 | 3 | 4 | 5 | 6 | 7 | 8 | 9 | 10 | 11 | 12 | 13 | 14 | 15 | 16 | 17 | 18 | 19 | 20 |
|  | 14 | 19 | 1 | 15 | 6 | 7 | 2 | 12 | 17 | 16 | 4 | 10 | 5 | 9 | 8 | 18 | 7 | 13 | 20 | 11 |

C(2) = (4 14 13 20 10 9 16 17 6 2): -217

|  | 14 | 3 | 1 | 19 | 6 | 7 | 2 | 12 | 16 | 11 | 4 | 10 | 9 | 15 | 8 | 17 | 18 | 13 | 20 | 5 |
|---|---|---|---|---|---|---|---|---|---|---|---|---|---|---|---|---|---|---|---|---|
|  | 1 | 2 | 3 | 4 | 5 | 6 | 7 | 8 | 9 | 10 | 11 | 12 | 13 | 14 | 15 | 16 | 17 | 18 | 19 | 20 |
| (10 9) |  |  |  | 13 |  |  |  |  | 10 | 14! |  |  | 17 | 4 |  | 9 | 16 |  |  |  |
| -168 |  |  |  | 70 |  |  |  |  | 58 | 168 |  |  | 79 | 133 |  | 57 | 67 |  |  |  |
| ARCS | 1 | 2 | 3 | 4 | 5 | 6 | 7 | 8 | 9 | 10 | 11 | 12 | 13 | 14 | 15 | 16 | 17 | 18 | 19 | 20 |
|  | 14 | 3 | 1 | 15 | 6 | 7 | 2 | 12 | 17 | 16 | 4 | 10 | 19 | 11 | 8 | 18 | 9 | 13 | 20 | 5 |

C(3) = (10 9 16 17 13 4 14): -168

We choose C(2). The arcs obtained are;

(4 15), (14 9), (13 5), (20 11), (10 16), (9 17), (16 18), (17 7), (6 3), (2 19).

$$D_2^{-1}M^-$$

| | 14 | 19 | 1 | 15 | 6 | 3 | 2 | 12 | 17 | 16 | 4 | 10 | 5 | 9 | 8 | 18 | 7 | 13 | 20 | 11 | |
|---|---|---|---|---|---|---|---|---|---|---|---|---|---|---|---|---|---|---|---|---|---|
| | 1 | 2 | 3 | 4 | 5 | 6 | 7 | 8 | 9 | 10 | 11 | 12 | 13 | 14 | 15 | 16 | 17 | 18 | 19 | 20 | |
| 1 | 0 | 32 | ∞ | 42 | 22 | 3 | 31 | 68 | 34 | 79 | 33 | 39 | 69 | 35 | 83 | 41 | 43 | 64 | 47 | 43 | 1 |
| 2 | 48 | 0 | 42 | 58 | 33 | 9 | ∞ | 45 | 6 | 78 | 85 | 34 | 87 | 27 | -4 | 55 | 0 | 54 | 8 | 63 | 2 |
| 3 | -14 | 80 | 0 | 72 | 14 | ∞ | -5 | 47 | 43 | -2 | 39 | 57 | 30 | 15 | 74 | 80 | -1 | 18 | 68 | 38 | 3 |
| 4 | 15 | 0 | 28 | 0 | 60 | 34 | 69 | 67 | 73 | 25 | ∞ | 9 | -1 | 45 | 49 | 9 | 17 | 26 | 50 | -10 | 4 |
| 5 | 78 | 58 | 32 | 19 | 0 | 43 | 86 | 6 | 29 | 31 | 14 | 75 | ∞ | 82 | 75 | 90 | 33 | 9 | 4 | 55 | 5 |
| 6 | 29 | 66 | 10 | 59 | ∞ | 0 | 5 | 36 | 29 | 37 | 48 | 22 | -27 | 49 | 40 | 16 | 29 | 22 | 59 | 46 | 6 |
| 7 | 13 | 70 | 54 | 57 | 57 | 13 | 0 | 75 | 11 | 74 | 33 | 14 | 16 | 18 | 50 | 63 | ∞ | 52 | 69 | 25 | 7 |
| 8 | 63 | 76 | 84 | 2 | 58 | 78 | -14 | 0 | 60 | 73 | 55 | 43 | 66 | 12 | ∞ | 9 | 40 | 57 | 5 | 5 | 8 |
| 9 | 1 | 34 | 46 | 17 | 77 | 29 | 27 | 43 | 0 | -1 | 60 | 16 | 82 | ∞ | 22 | 13 | 18 | 28 | 57 | 57 | 9 |
| 10 | 43 | 55 | 51 | 39 | 51 | 64 | 35 | 2 | 47 | 0 | 26 | ∞ | 77 | 18 | 54 | 77 | 15 | 37 | 48 | 58 | 10 |
| 11 | -2 | 67 | 48 | 20 | 68 | 46 | 57 | 9 | 84 | 21 | 0 | 51 | 50 | 51 | 9 | 19 | 19 | 29 | 16 | ∞ | 11 |
| 12 | 57 | 67 | 78 | 93 | 53 | 60 | 44 | ∞ | 13 | 14 | 82 | 0 | 6 | 42 | 9 | 92 | 74 | 42 | 17 | 14 | 12 |
| 13 | 77 | 27 | 65 | 71 | 40 | 22 | 44 | 33 | 21 | 34 | 32 | 26 | 0 | 18 | 57 | 14 | 60 | ∞ | 84 | 25 | 13 |
| 14 | ∞ | 37 | 10 | 38 | 56 | -1 | 47 | 57 | 50 | 5 | 30 | 41 | 78 | 0 | 72 | 57 | 64 | 86 | 53 | 3 | 14 |
| 15 | 27 | 47 | 41 | ∞ | 75 | 74 | 46 | 72 | 50 | 28 | 4 | 26 | 8 | 51 | 0 | 82 | 45 | 69 | 39 | 14 | 15 |
| 16 | 0 | 16 | 84 | 55 | 59 | 6 | 72 | 9 | 10 | ∞ | 35 | -7 | 26 | -7 | 77 | 0 | 68 | 38 | 68 | 21 | 16 |
| 17 | 40 | 47 | 34 | 51 | 46 | 46 | -5 | 3 | ∞ | 5 | 78 | 35 | 19 | -11 | 59 | 1 | 0 | 20 | 30 | 79 | 17 |
| 18 | 52 | 85 | 46 | 99 | 38 | 88 | 49 | 87 | 6 | 0 | 19 | 70 | 85 | 7 | 13 | ∞ | 57 | 0 | 36 | 19 | 18 |
| 19 | 7 | ∞ | 12 | 28 | 63 | 63 | 31 | 25 | 27 | -19 | 48 | 23 | 28 | 3 | 55 | 60 | 39 | 6 | 0 | 42 | 19 |
| 20 | 42 | 19 | 14 | 39 | 75 | 70 | -4 | 6 | 29 | 62 | 54 | 35 | -7 | 45 | 3 | 30 | 57 | 82 | ∞ | 0 | 20 |
| | 1 | 2 | 3 | 4 | 5 | 6 | 7 | 8 | 9 | 10 | 11 | 12 | 13 | 14 | 15 | 16 | 17 | 18 | 19 | 20 | |

A *symmetric derangement* is a derangement containing no 2-cycles. Thus, if each arc is converted into an edge, the number of edges equals the number of arcs. We now prove that $D_2$ is a minimal symmetric derangement. We can no longer obtain a negative cycle or permutation that yields a symmetric derangement. It follows that $D_2$ is a minimally-valued symmetric derangement. Since it is also a 20-cycle, it follows that $T_{OPT} = D_2$.

*Comment.* We first note that we continue until we have obtained m(n) acceptable paths. In most cases, each path has a much smaller average arc-value than $T_{OPT}$ until a path is close to becoming a tour ($n$-cycle). Thus, a moderately-sized heuristic generally can be used to obtain $T_{OPT}$. The key thing is that the parallel processors generally run independently of each other. This reduces the running time of both the exact algorithm as well as the heuristic. To be more precise, when constructing a heuristic, we construct paths each of whose aav's is less than the minimally-valued derangement obtained earlier in the algorithm or the Held-Karp lower bound for $T_{OPT}$. (It may be that $HK$ is larger in value than than the value of a minimally-valued derangement.) We accept all such paths obtained by parallel processors. We use one of these upper bounds until we have obtained $\lfloor \frac{n}{2} \rfloor$ arcs in a path. Once we have finished with that procedure, each parallel processor tries to extend the path with the smallest arc available. This procedure continues to the end of the algorithm.

```
Chapter 5
```

The General Bottleneck Traveling Salesman Problem

I. INTRODUCTION. The bottleneck traveling salesman problem asks us to find a Hamilton circuit in a weighted graph which minimizes the weight of the weightiest edge of the circuit. Here we give a heuristic for finding such a circuit.
II. PRELIMINARIES. In what follows, M is an *n X n matrix*, each of whose arcs has been assigned a positive weight. If H is an n-cycle in M, an *H-admissible permutation*, s, is a permutation such that Hs = H', where H' is also an n-cycle. An *even permutation* has the property that the number of its disjoint cycles containing an even number of points is even. Thus, $s = H^{-1} H'$ is even. Finally, every even permutation can be expressed as a product of 3-cycles. Our idea is to use equences of 3-cycles to try eventually minimize the weightiest arc of an n-tour.
III. THE ALGORITHM. In our algorithm, using the algorithm given in chapter 4 of [1], we construct an n-circuit, H, in M that is approximately minimal in weight. We then use a sequence of H-admissible 3-cycles, $s_i$, (I = 1, 2, …, m) such that the first point of each $s_i$ has as its initial point the initial point of the weightiest edge of H. Each of the first two permutation arcs of any 3-cycle used can never have a value as great or greater than the weightiest arc in each $H_i$. With this caveat, we hopefully can obtain a set of n-circuits $H_1$, $H_2$, …. $H_m$ each of which has a smaller weightiest edge than the previous one. In this procedure, we only permit at most (log n) 3-cycles in a sequence  The following matrix is on page 126, [1].

EXAMPLE 1.

M

| | 1 | 2 | 3 | 4 | 5 | 6 | 7 | 8 | 9 | 10 | 11 | 12 | 13 | 14 | 15 | 16 | 17 | 18 | 19 | 20 | |
|---|---|---|---|---|---|---|---|---|---|---|---|---|---|---|---|---|---|---|---|---|---|
| 1 | ∞ | 26 | 1 | 30 | 74 | 5 | 3 | 38 | 28 | 78 | 81 | 7 | 97 | 10 | 94 | 40 | 98 | 49 | 40 | 70 | 1 |
| 2 | 26 | ∞ | 69 | 30 | 1 | 80 | 50 | 74 | 1 | 60 | 3 | 9 | 31 | 87 | 89 | 91 | 6 | 82 | 23 | 85 | 2 |
| 3 | 1 | 69 | ∞ | 23 | 7 | 61 | 2 | 98 | 99 | 90 | 84 | 57 | 4 | 56 | 66 | 30 | 51 | 3 | 25 | 47 | 3 |
| 4 | 30 | 30 | 23 | ∞ | 33 | 59 | 5 | 3 | 26 | 48 | 84 | 18 | 57 | 28 | 47 | 1 | 81 | 48 | 70 | 17 | 4 |
| 5 | 74 | 1 | 7 | 33 | ∞ | 82 | 29 | 80 | 5 | 87 | 2 | 97 | 3 | 45 | 72 | 94 | 20 | 9 | 90 | 20 | 5 |
| 6 | 5 | 80 | 61 | 82 | 82 | ∞ | 1 | 6 | 43 | 9 | 39 | 41 | 3 | 45 | 62 | 38 | 50 | 1 | 41 | 50 | 6 |
| 7 | 3 | 50 | 2 | 29 | 29 | 1 | ∞ | 34 | 78 | 49 | 73 | 10 | 56 | 36 | 87 | 31 | 45 | 59 | 88 | 42 | 7 |
| 8 | 38 | 74 | 98 | 80 | 80 | 6 | 34 | ∞ | 14 | 55 | 43 | 91 | 85 | 93 | 75 | 2 | 64 | 78 | 60 | 1 | 8 |
| 9 | 28 | 1 | 99 | 5 | 5 | 43 | 78 | 14 | ∞ | 50 | 28 | 81 | 98 | 95 | 3 | 31 | 73 | 63 | 87 | 2 | 9 |
| 10 | 78 | 60 | 90 | 87 | 87 | 9 | 49 | 55 | 50 | ∞ | 37 | 1 | 95 | 59 | 30 | 25 | 3 | 90 | 64 | 36 | 10 |
| 11 | 81 | 3 | 84 | 2 | 2 | 39 | 73 | 43 | 28 | 37 | ∞ | 3 | 61 | 14 | 11 | 3 | 3 | 74 | 22 | 26 | 11 |
| 12 | 7 | 9 | 57 | 97 | 97 | 41 | 10 | 91 | 81 | 1 | 3 | ∞ | 84 | 62 | 56 | 34 | 2 | 17 | 71 | 30 | 12 |
| 13 | 97 | 31 | 4 | 3 | 3 | 3 | 56 | 85 | 98 | 95 | 61 | 84 | ∞ | 66 | 66 | 30 | 49 | 2 | 23 | 86 | 13 |
| 14 | 10 | 87 | 56 | 45 | 45 | 45 | 36 | 93 | 95 | 59 | 14 | 62 | 66 | ∞ | 25 | 14 | 47 | 57 | 2 | 5 | 14 |
| 15 | 94 | 89 | 66 | 72 | 72 | 62 | 87 | 75 | 3 | 30 | 11 | 56 | 66 | 25 | ∞ | 89 | 85 | 87 | 8 | 2 | 15 |
| 16 | 40 | 91 | 30 | 94 | 94 | 38 | 31 | 2 | 31 | 25 | 3 | 34 | 30 | 14 | 89 | ∞ | 54 | 18 | 92 | 90 | 16 |
| 17 | 98 | 6 | 51 | 20 | 20 | 50 | 45 | 64 | 73 | 3 | 3 | 2 | 49 | 47 | 85 | 54 | ∞ | 66 | 48 | 90 | 17 |
| 18 | 49 | 82 | 3 | 9 | 9 | 1 | 59 | 78 | 63 | 90 | 74 | 17 | 2 | 57 | 87 | 18 | 66 | ∞ | 87 | 16 | 18 |
| 19 | 40 | 23 | 25 | 90 | 30 | 41 | 88 | 60 | 87 | 64 | 22 | 71 | 23 | 2 | 8 | 92 | 48 | 87 | ∞ | 22 | 19 |
| 20 | 70 | 85 | 47 | 20 | 20 | 50 | 42 | 1 | 2 | 36 | 26 | 30 | 86 | 5 | 2 | 90 | 90 | 16 | 22 | ∞ | 20 |
| | 1 | 2 | 3 | 4 | 5 | 6 | 7 | 8 | 9 | 10 | 11 | 12 | 13 | 14 | 15 | 16 | 17 | 18 | 19 | 20 | |

For simplicity, we use $T_{FWOPT}$, page 126, as an approximate optimal solution of the traveling salesman problem for 3-cycle M

$$T_{FWKOPT} = (17^3\ 10^1\ 12^7\ 1^1\ 3^4\ 13^2\ 18^1\ 6^2\ 7^5\ 4^1\ 16^2\ 8^1\ 20^5\ 14^2\ 19^8\ 15^3\ 9^1\ 2^1\ 5^2\ 11^3) : 54.$$

The weightiest edge is (19 15): 8. (Although we're discussing edges, we are treating $T_{FWKOPT}$ as a cycle oriented in a clockwise direction.) The edge of smallest weight out of 19 is (19 14): 2. Since (14 19) is an arc out of $T_{FWKOPT}$. All of the other arcs out of 19 (other than (19 15)) have a value greater than 8. Thus, $T_{FWKOPT}$ is our approximate solution to example 1.

The following matrix is given in example 9, page 319.

EXAMPLE 2.

M

|    | 1  | 2  | 3  | 4  | 5  | 6  | 7  | 8  | 9  | 10 | 11 | 12 | 13 | 14 | 15 |     |
|----|----|----|----|----|----|----|----|----|----|----|----|----|----|----|----|-----|
| 1  | ∞  | 53 | 52 | 51 | 50 | 49 | 48 | 48 | 48 | 48 | 50 | 51 | 52 | 53 | 54 | 1   |
| 2  | 53 | ∞  | 66 | 66 | 50 | 49 | 48 | 47 | 48 | 49 | 50 | 51 | 52 | 53 | 54 | 2   |
| 3  | 52 | 66 | ∞  | 62 | 65 | 64 | 48 | 48 | 48 | 49 | 50 | 75 | 69 | 77 | 76 | 3   |
| 4  | 51 | 66 | 62 | ∞  | 63 | 62 | 65 | 49 | 49 | 49 | 50 | 77 | 73 | 76 | 71 | 4   |
| 5  | 50 | 50 | 65 | 63 | ∞  | 66 | 65 | 50 | 73 | 76 | 74 | 71 | 77 | 75 | 76 | 5   |
| 6  | 49 | 49 | 64 | 62 | 66 | ∞  | 65 | 51 | 70 | 77 | 73 | 37 | 23 | 30 | 69 | 6   |
| 7  | 48 | 48 | 48 | 65 | 65 | 54 | ∞  | 73 | 74 | 71 | 75 | 26 | 39 | 27 | 72 | 7   |
| 8  | 48 | 47 | 48 | 49 | 50 | 51 | 73 | ∞  | 27 | 31 | 73 | 31 | 30 | 35 | 73 | 8   |
| 9  | 48 | 48 | 48 | 49 | 73 | 70 | 74 | 27 | ∞  | 30 | 38 | 29 | 26 | 36 | 77 | 9   |
| 10 | 48 | 49 | 49 | 49 | 76 | 77 | 71 | 31 | 30 | ∞  | 36 | 35 | 73 | 71 | 74 | 100 |
| 11 | 50 | 50 | 50 | 50 | 74 | 73 | 75 | 73 | 38 | 36 | ∞  | 30 | 26 | 37 | 35 | 11  |
| 12 | 51 | 51 | 75 | 77 | 71 | 37 | 26 | 31 | 29 | 35 | 30 | ∞  | 29 | 36 | 30 | 12  |
| 13 | 52 | 52 | 69 | 73 | 77 | 23 | 39 | 30 | 26 | 73 | 26 | 29 | ∞  | 30 | 36 | 13  |
| 14 | 53 | 53 | 77 | 76 | 75 | 30 | 27 | 35 | 36 | 71 | 37 | 36 | 30 | ∞  | 32 | 14  |
| 15 | 54 | 54 | 76 | 71 | 75 | 69 | 72 | 73 | 77 | 74 | 35 | 30 | 36 | 32 | ∞  | 15  |
|    | 1  | 2  | 3  | 4  | 5  | 6  | 7  | 8  | 9  | 10 | 11 | 12 | 13 | 14 | 15 |     |

The optimal solution for the above matrix was given on page 338 of [1]:

$T_{OPT} = (4^{49}\ 10^{31}\ 8^{27}\ 9^{29}\ 12^{26}\ 7^{27}\ 14^{32}\ 15^{35}\ 11^{26}\ 13^{26}\ 6^{49}\ 1^{52}\ 3^{65}\ 5^{50}\ 2^{66})$: 587

The weightiest arc is (2 4):66. For simplicity, we place the weight of the edge obtained as a superscript of the initial node of the corresponding arc of a 3-cycle..

$(2^{47}\ 10^{30}\ 8^{49})$. Applied to $T_{OPT}$, this yields $W_1 = (1^{52}3^{65}5^{50}2^{47}8^{27}4^{49}10^{30}9^{29}12^{26}7^{27}14^{32}15^{35}11^{26}13^{52})$.

Our weightiest edge in $W_1$ is [3 5]: 65.

By testing the possibilities for a $W_1$-admissible 3-cycle, we obtain $(3^{49}10^{48}1^{50})$ which yields

$W_2 = (1^{50}5^{50}2^{47}8^{49}4^{49}10^{49}3^{48}9^{29}12^{26}7^{27}14^{32}15^{35}11^{26}13^{23}6^{49})$

As we go along, we must always check to see if the weightiest edge in any case is the arc of smallest weight in a row. In this case, we cannot obtain an n-cycle with a weightiest edge smaller than that edge. In this case, two edges have the same weightiest value. Interestingly, the total weight of $W_2$ is 589, just two units greater than $T_{OPT}$. Whether this procedure is really effective for large examples would require extensive testing.

Comment. Let $s_{GBTSPOPT}$ be an even permutation such that $Ts_{GBTSPOPT} = T_{GBTDPOPT}$ is an optimal solution to the bottleneck traveling salesman problem. If we eliminate the condition on the weight of the third arc in 3-cycles of a sequence in the set of sequences then we can always obtain an optimal solution. However, the running time will generally not be polynomial in at least some cases unless P = NP. It is also worth noting that a similar sort of algorithm could be used in the general traveling salesman case.

EXAMPLE 3.

M

|   | 1 | 2 | 3 | 4 | 5 | 6 | 7 | 8 | 9 | 10 | 11 | 12 | 13 | 14 | 15 |   |
|---|---|---|---|---|---|---|---|---|---|----|----|----|----|----|----|---|
| 1 | ∞ | 49 | 72 | 48 | 36 | 28 | 15 | 4 | 36 | 26 | 7 | 76 | 98 | 37 | 4 | 1 |
| 2 | 57 | ∞ | 47 | 87 | 54 | 35 | 28 | 71 | 99 | 1 | 47 | 34 | 96 | 53 | 4 | 2 |
| 3 | 69 | 78 | ∞ | 40 | 35 | 7 | 50 | 87 | 23 | 53 | 21 | 96 | 74 | 94 | 69 | 3 |
| 4 | 81 | 91 | 90 | ∞ | 93 | 54 | 91 | 99 | 9 | 77 | 19 | 85 | 29 | 3 | 9 | 4 |
| 5 | 94 | 19 | 48 | 11 | ∞ | 9 | 11 | 3 | 48 | 71 | 28 | 26 | 78 | 60 | 32 | 5 |
| 6 | 11 | 94 | 65 | 79 | 26 | ∞ | 79 | 33 | 89 | 88 | 12 | 38 | 79 | 55 | 36 | 6 |
| 7 | 90 | 42 | 26 | 86 | 84 | 98 | ∞ | 37 | 80 | 64 | 88 | 19 | 83 | 10 | 48 | 7 |
| 8 | 53 | 9 | 46 | 86 | 97 | 25 | 66 | ∞ | 77 | 23 | 55 | 84 | 95 | 30 | 63 | 8 |
| 9 | 5 | 57 | 86 | 67 | 98 | 1 | 40 | 53 | ∞ | 35 | 74 | 33 | 84 | 49 | 27 | 9 |
| 10 | 91 | 22 | 86 | 27 | 45 | 80 | 94 | 54 | 71 | ∞ | 72 | 52 | 49 | 28 | 12 | 10 |
| 11 | 49 | 49 | 74 | 43 | 79 | 26 | 21 | 13 | 56 | 92 | ∞ | 70 | 79 | 60 | 82 | 11 |
| 12 | 3 | 68 | 64 | 12 | 12 | 7 | 27 | 74 | 65 | 62 | 92 | ∞ | 54 | 77 | 11 | 12 |
| 13 | 21 | 51 | 38 | 65 | 30 | 90 | 51 | 23 | 94 | 49 | 35 | 14 | ∞ | 97 | 29 | 13 |
| 14 | 14 | 7 | 51 | 70 | 7 | 41 | 20 | 66 | 92 | 58 | 13 | 32 | 97 | ∞ | 53 | 14 |
| 15 | 59 | 29 | 91 | 98 | 74 | 54 | 25 | 18 | 75 | 94 | 42 | 34 | 16 | 36 | ∞ | 15 |
|   | 1 | 2 | 3 | 4 | 5 | 6 | 7 | 8 | 9 | 10 | 11 | 12 | 13 | 14 | 15 | 15 |

MIN(M)

|   | 1 | 2 | 3 | 4 | 5 | 6 | 7 | 8 | 9 | 10 | 11 | 12 | 13 | 14 | 15 |   |
|---|---|---|---|---|---|---|---|---|---|----|----|----|----|----|----|---|
| 1 | 8 | 15 | 11 | 7 | 10 | 6 | 5 | 9 | 14 | 4 | 2 | 3 | 12 | 13 |   | 1 |
| 2 | 10 | 15 | 7 | 12 | 6 | 3 | 11 | 14 | 5 | 1 | 8 | 4 | 13 | 9 |   | 2 |
| 3 | 6 | 11 | 9 | 5 | 4 | 7 | 10 | 1 | 15 | 2 | 13 | 8 | 14 | 12 |   | 3 |
| 4 | 14 | 9 | 15 | 11 | 13 | 6 | 10 | 1 | 12 | 3 | 2 | 7 | 5 | 8 |   | 4 |
| 5 | 8 | 9 | 4 | 7 | 2 | 12 | 11 | 15 | 3 | 9 | 14 | 10 | 13 | 1 |   | 5 |
| 6 | 1 | 11 | 5 | 8 | 15 | 11 | 14 | 3 | 4 | 7 | 13 | 10 | 9 | 2 |   | 6 |
| 7 | 14 | 12 | 3 | 8 | 2 | 15 | 10 | 9 | 12 | 5 | 4 | 11 | 1 | 6 |   | 7 |
| 8 | 2 | 6 | 10 | 14 | 3 | 1 | 11 | 15 | 7 | 9 | 12 | 4 | 13 | 5 |   | 8 |
| 9 | 6 | 1 | 5 | 12 | 10 | 7 | 14 | 8 | 2 | 4 | 11 | 13 | 3 | 5 |   | 9 |
| 10 | 15 | 2 | 4 | 14 | 5 | 13 | 12 | 8 | 9 | 11 | 6 | 3 | 1 | 7 |   | 10 |
| 11 | 8 | 7 | 6 | 4 | 1 | 2 | 9 | 14 | 12 | 3 | 5 | 13 | 15 | 10 |   | 11 |
| 12 | 1 | 6 | 15 | 4 | 5 | 7 | 13 | 10 | 3 | 9 | 2 | 8 | 14 | 11 |   | 12 |
| 13 | 12 | 1 | 8 | 13 | 5 | 11 | 3 | 10 | 2 | 7 | 4 | 6 | 9 | 14 |   | 13 |
| 14 | 2 | 5 | 11 | 1 | 7 | 14 | 6 | 3 | 15 | 10 | 8 | 4 | 9 | 13 |   | 14 |
| 15 | 13 | 8 | 7 | 2 | 12 | 14 | 11 | 6 | 1 | 5 | 9 | 3 | 10 | 4 |   | 15 |
|   | 1 | 2 | 3 | 4 | 5 | 6 | 7 | 8 | 9 | 10 | 11 | 12 | 13 | 14 | 15 |   |

$T_0 = (1^{43}2^{47}3^{40}4^{93}5^{9}6^{79}7^{37}8^{77}9^{35}10^{72}11^{70}12^{54}13^{97}14^{53}15^{59})$.

$T_0^{-1}M$

|     |     |     | 2   | 3   | 4   | 5   | 6   | 7   | 8   | 9   | 10  | 11  | 12  | 13  | 14  | 15  | 1   |     |
| --- | --- | --- | --- | --- | --- | --- | --- | --- | --- | --- | --- | --- | --- | --- | --- | --- | --- | --- |
|     |     |     | 1   | 2   | 3   | 4   | 5   | 6   | 7   | 8   | 9   | 10  | 11  | 12  | 13  | 14  | 15  |     |
| -14 | -45 | 1   | 0   | 23  | -1  | -13 | -21 | -34 | -45 | -13 | -23 | -42 | 27  | 59  | -12 | -45 | ∞   | 1   |
| -32 | -46 | 2   | ∞   | 0   | 40  | 7   | -12 | -19 | 24  | 52  | -46 | 0   | -13 | 49  | 6   | -43 | 10  | 2   |
| -17 | -33 | 3   | 38  | ∞   | 0   | -5  | -33 | 10  | 47  | -17 | 13  | -19 | 56  | 34  | 54  | 29  | 29  | 3   |
| -84 | -90 | 4   | -2  | -3  | ∞   | 0   | -39 | -2  | 6   | -84 | -16 | -74 | -8  | -64 | -90 | -84 | -12 | 4   |
| -6  | -85 | 5   | 10  | 39  | 2   | ∞   | 0   | 2   | -6  | 38  | 62  | 19  | 17  | 69  | 51  | 23  | 85  | 5   |
| -67 | -68 | 6   | 15  | -14 | 0   | -53 | ∞   | 0   | -46 | 10  | 9   | -67 | -41 | 0   | -24 | -43 | -68 | 6   |
| -18 | -27 | 7   | 5   | -11 | 49  | 47  | 61  | ∞   | 0   | 43  | 27  | 51  | -18 | 46  | -27 | 11  | 53  | 7   |
| -52 | -54 | 8   | -68 | -31 | 9   | 20  | -52 | -1  | ∞   | 0   | -54 | -22 | 7   | 18  | -47 | -14 | -24 | 8   |
| -34 | -42 | 9   | 22  | 51  | 32  | 63  | -34 | 5   | 18  | ∞   | 0   | 39  | -2  | 49  | 14  | -8  | -30 | 9   |
| -50 | -60 | 10  | -50 | 14  | -45 | -27 | 8   | 22  | -18 | -1  | ∞   | 0   | -20 | -23 | -44 | -60 | 19  | 10  |
| -49 | -57 | 11  | -21 | 4   | -27 | 9   | -44 | -49 | -57 | -14 | 22  | ∞   | 0   | 9   | -10 | 12  | -21 | 11  |
| -47 | -51 | 12  | 14  | 10  | -42 | -42 | -47 | -27 | 20  | 11  | 8   | 28  | ∞   | 0   | 23  | -43 | -51 | 12  |
| -76 | -83 | 13  | -46 | -59 | -32 | -67 | -7  | -46 | -74 | -3  | -48 | -62 | -83 | ∞   | 0   | -68 | -76 | 13  |
| -46 | -46 | 14  | -46 | -2  | 17  | -46 | -12 | -33 | 13  | -39 | 5   | -40 | -21 | 44  | ∞   | 0   | -39 | 14  |
| -41 | -43 | 15  | -30 | 32  | 39  | 15  | -5  | -34 | -41 | 16  | 35  | -17 | -25 | -43 | -23 | ∞   | 0   | 15  |
|     |     |     | 1   | 2   | 3   | 4   | 5   | 6   | 7   | 8   | 9   | 10  | 11  | 12  | 13  | 14  | 15  | 15  |

$D_1 = (\, 4^{-90} 13^{-83} 11^{-57} 7^{-11} 2^{-46} 9^{-42} 15^{-41} 12^{-47} 5^2 6^{-53} \,)$

$T_1 = (\, 1\ 2\ 10\ 11\ 8\ 9\,)(\, 3\ 4\ 14\ 15\ 13\ 12\ 6\ 5\ 7\,)$

$T_1^{-1}M$

|     |     |     | 2   | 10  | 4   | 14  | 7   | 5   | 3   | 9   | 1   | 11  | 8   | 6   | 12  | 15  | 13  |     |
| --- | --- | --- | --- | --- | --- | --- | --- | --- | --- | --- | --- | --- | --- | --- | --- | --- | --- | --- |
|     |     |     | 1   | 2   | 3   | 4   | 5   | 6   | 7   | 8   | 9   | 10  | 11  | 12  | 13  | 14  | 15  |     |
| -14 | -45 | 1   | 0   | -23 | -1  | -12 | -34 | -13 | 23  | -13 | ∞   | -42 | -45 | -21 | 27  | -45 | 59  | 1   |
|     |     | 2   | ∞   | 0   | 86  | 52  | 27  | 53  | 46  | 98  | 56  | 46  | 70  | 34  | 33  | 3   | 95  | 2   |
| -17 | -33 | 3   | 38  | 13  | 0   | 54  | -10 | -5  | ∞   | -17 | 29  | -19 | 47  | -33 | 56  | 29  | 34  | 3   |
|     |     | 4   | 88  | 74  | ∞   | 0   | 88  | 90  | 87  | 6   | 78  | 16  | 96  | 51  | 82  | 6   | 26  | 4   |
| -2  | -8  | 5   | 8   | 60  | 0   | 49  | 0   | ∞   | 37  | 36  | 15  | 17  | -8  | -2  | 15  | 21  | 67  | 5   |
|     | -14 | 6   | 68  | 62  | 53  | 29  | 53  | 0   | 39  | 63  | -15 | -14 | 7   | ∞   | 12  | 10  | 53  | 6   |
|     | -7  | 7   | 16  | 38  | 60  | -16 | ∞   | 58  | 0   | 54  | 64  | 62  | 11  | 72  | -7  | 22  | 57  | 7   |
| -52 | -54 | 8   | -68 | -54 | 9   | -47 | -1  | 20  | -31 | 0   | -24 | -22 | ∞   | -52 | 7   | -14 | 18  | 8   |
|     |     | 9   | 17  | 30  | 62  | 44  | 35  | 93  | 81  | ∞   | 0   | 69  | 48  | -4  | 28  | 27  | 79  | 9   |
| -50 | -60 | 10  | -50 | ∞   | -45 | -44 | 22  | -27 | 14  | -1  | 19  | 0   | -18 | 8   | -20 | -60 | -23 | 10  |
|     |     | 11  | 36  | 79  | 30  | 47  | 8   | 66  | 61  | 43  | 36  | ∞   | 0   | 13  | 57  | 69  | 66  | 11  |
|     | -4  | 12  | 61  | 55  | 5   | 70  | 24  | 9   | 61  | 62  | -4  | 79  | 71  | 0   | ∞   | 8   | 47  | 12  |
|     |     | 13  | 37  | 35  | 51  | 83  | 37  | 16  | 24  | 80  | 7   | 21  | 9   | 76  | 0   | 15  | ∞   | 13  |
| -46 | -46 | 14  | -46 | 5   | 17  | ∞   | -33 | -46 | -2  | -39 | -39 | -40 | 13  | -12 | -21 | 0   | 44  | 14  |
|     |     | 15  | 13  | 78  | 82  | 20  | 9   | 58  | 75  | 59  | 43  | 26  | 2   | 38  | 18  | ∞   | 0   | 15  |
|     |     |     | 1   | 2   | 3   | 4   | 5   | 6   | 7   | 8   | 9   | 10  | 11  | 12  | 13  | 14  | 15  |     |

$D_2 = (\, 10^{-60} 14^{-46} 1^{-42} \,),\ T_2 = (\, 1\ 11\ 8\ 9\,)(\, 2\ 10\ 15\ 12\ 6\ 5\ 7\ 3\ 4\ 14\,)$

$$T_2^{-1}M$$

|   |   |   | 11 | 10 | 4 | 14 | 7 | 5 | 3 | 9 | 1 | 15 | 8 | 6 | 12 | 2 | 13 |   |
|---|---|---|----|----|---|----|---|---|---|---|---|----|---|---|----|---|----|---|
|   |   |   | 1 | 2 | 3 | 4 | 5 | 6 | 7 | 8 | 9 | 10 | 11 | 12 | 13 | 14 | 15 |   |
| -3 | -3 | 1 | 0 | 19 | 41 | 30 | 8 | 29 | 65 | 29 | ∞ | -3 | -3 | 21 | 69 | 42 | 91 | 1 |
|   |   | 2 | 46 | 0 | 86 | 52 | 27 | 53 | 46 | 98 | 56 | 3 | 70 | 34 | 33 | ∞ | 95 | 2 |
| -17 | -33 | 3 | -19 | 13 | 0 | 54 | -10 | -5 | ∞ | -17 | 29 | 29 | 47 | -33 | 56 | 38 | 34 | 3 |
|   |   | 4 | 16 | 74 | ∞ | 0 | 88 | 90 | 87 | 6 | 78 | 6 | 96 | 51 | 82 | 88 | 26 | 4 |
| -2 | -8 | 5 | 17 | 60 | 0 | 49 | 0 | ∞ | 37 | 36 | 15 | 21 | -8 | -2 | 15 | 8 | 67 | 5 |
| -14 | -15 | 6 | -14 | 62 | 53 | 29 | 53 | 0 | 39 | 63 | -15 | 10 | 7 | ∞ | 12 | 68 | 53 | 6 |
| -7 | -16 | 7 | 62 | 38 | 60 | -16 | ∞ | 58 | 0 | 54 | 64 | 22 | 11 | 72 | -7 | 16 | 57 | 7 |
| -54 | -68 | 8 | -22 | -54 | 9 | -47 | -1 | 20 | -31 | 0 | -24 | -14 | ∞ | -52 | 7 | -68 | 18 | 8 |
|   |   | 9 | 69 | 30 | 62 | 44 | 35 | 93 | 81 | ∞ | 0 | 22 | 48 | -4 | 28 | 17 | 22 | 9 |
|   |   | 10 | 60 | ∞ | 15 | 16 | 82 | 33 | 74 | 59 | 79 | 0 | 42 | 68 | 40 | 10 | 37 | 10 |
|   |   | 11 | ∞ | 79 | 30 | 47 | 8 | 66 | 61 | 43 | 36 | 69 | 0 | 13 | 57 | 36 | 66 | 11 |
|   | -4 | 12 | 79 | 55 | 5 | 70 | 24 | 9 | 61 | 62 | -4 | 8 | 71 | 0 | ∞ | 61 | 47 | 12 |
|   |   | 13 | 21 | 35 | 51 | 83 | 37 | 16 | 24 | 80 | 7 | 15 | 9 | 76 | 0 | 37 | ∞ | 13 |
|   |   | 14 | 6 | 51 | 63 | ∞ | 13 | 0 | 44 | 7 | 7 | 46 | 59 | 34 | 25 | 0 | 90 | 14 |
|   |   | 15 | 26 | 78 | 82 | 20 | 9 | 58 | 75 | 59 | 43 | ∞ | 2 | 38 | 18 | 13 | 0 | 15 |
|   |   |   | 1 | 2 | 3 | 4 | 5 | 6 | 7 | 8 | 9 | 10 | 11 | 12 | 13 | 14 | 15 |   |

$$D_3 = (\,8^{-66}\ 14^7\,),\ T_3 = (\,1\ 11\ 8\ 2\ 10\ 15\ 13\ 12\ 6\ 5\ 7\ 3\ 4\ 14\ 9\,)$$

$$T_3^{-1}M$$

|   |   |   | 11 | 10 | 4 | 14 | 7 | 5 | 3 | 2 | 1 | 15 | 8 | 6 | 12 | 9 | 13 |   |
|---|---|---|----|----|---|----|---|---|---|---|---|----|---|---|----|---|----|---|
|   |   |   | 1 | 2 | 3 | 4 | 5 | 6 | 7 | 8 | 9 | 10 | 11 | 12 | 13 | 14 | 15 |   |
| -3 | -3 | 1 | 0 | 19 | 41 | 30 | 8 | 29 | 65 | 42 | ∞ | -3 | -3 | 21 | 69 | 29 | 91 | 1 |
|   |   | 2 | 46 | 0 | 86 | 52 | 27 | 53 | 46 | ∞ | 56 | 3 | 70 | 34 | 33 | 98 | 95 | 2 |
| -19 | -33 | 3 | -19 | 13 | 0 | 54 | -10 | -5 | ∞ | 38 | 29 | 29 | 47 | -33 | 56 | -17 | 34 | 3 |
|   |   | 4 | 16 | 74 | ∞ | 0 | 88 | 90 | 87 | 88 | 78 | 6 | 96 | 51 | 82 | 6 | 26 | 4 |
| -2 | -8 | 5 | 17 | 60 | 0 | 49 | 0 | ∞ | 37 | 8 | 15 | 21 | -8 | -2 | 15 | 36 | 67 | 5 |
| -14 | -15 | 6 | -14 | 62 | 53 | 29 | 53 | 0 | 39 | 68 | -15 | 10 | 7 | ∞ | 12 | 63 | 53 | 6 |
| -7 | -16 | 7 | 62 | 38 | 60 | -16 | ∞ | 58 | 0 | 16 | 64 | 22 | 11 | 72 | -7 | 54 | 57 | 7 |
|   |   | 8 | 46 | 14 | 77 | 21 | 67 | 88 | 37 | 0 | 44 | 54 | ∞ | 16 | 61 | 68 | 86 | 8 |
|   |   | 9 | 69 | 30 | 62 | 44 | 35 | 93 | 81 | 17 | 0 | 22 | 48 | -4 | 28 | ∞ | 79 | 9 |
|   |   | 10 | 60 | ∞ | 15 | 16 | 82 | 33 | 74 | 10 | 79 | 0 | 42 | 68 | 40 | 59 | 37 | 10 |
|   |   | 11 | ∞ | 79 | 30 | 47 | 8 | 66 | 61 | 36 | 36 | 69 | 0 | 13 | 57 | 43 | 66 | 11 |
|   | -4 | 12 | 79 | 55 | 5 | 70 | 24 | 9 | 61 | 61 | -4 | 8 | 71 | 0 | ∞ | 62 | 47 | 12 |
|   |   | 13 | 21 | 35 | 51 | 83 | 37 | 16 | 24 | 37 | 7 | 15 | 9 | 76 | 0 | 80 | ∞ | 13 |
| -7 | -7 | 14 | -1 | 44 | 56 | ∞ | 6 | -7 | 44 | -7 | 0 | 39 | 52 | 27 | 18 | 0 | 83 | 14 |
|   |   | 15 | 26 | 78 | 82 | 20 | 9 | 58 | 75 | 13 | 43 | ∞ | 2 | 38 | 18 | 59 | 0 | 15 |
|   |   |   | 1 | 2 | 3 | 4 | 5 | 6 | 7 | 8 | 9 | 10 | 11 | 12 | 13 | 14 | 15 |   |

We note that $T_3$ is a 15-cycle. We thus will try to obtain a $T_3$ – admissible 3-cycle that has a negative value.

$D_4 = (\,3^{-33}14^{27}12^5\,)$ has a value of -1. We apply it to $T_3$ to obtain

$T_4 = (\,1^7 11^{13}\,8^9\,2^1 10^{12} 15^{16} 13^{14} 12^{12} 4^3 14^{41} 6^{26} 5^{11} 7^{26} 3^{23} 9^5\,)$

|     |     |    | 11 | 10 | 4  | 14 | 7   | 5  | 3  | 2  | 1  | 15 | 8  | 6  | 12 | 9  | 13 |    |
|     |     |    | 1  | 2  | 3  | 4  | 5   | 6  | 7  | 8  | 9  | 10 | 11 | 12 | 13 | 14 | 15 |    |
| -3  | -3  | 1  | 0  | 19 | 41 | 30 | 8   | 29 | 65 | 42 | ∞  | -3 | -3 | 21 | 69 | 29 | 91 | 1  |
|     |     | 2  | 46 | 0  | 86 | 52 | 27  | 53 | 46 | ∞  | 56 | 3  | 70 | 34 | 33 | 98 | 95 | 2  |
| -17 | -33 | 3  | -19| 13 | 0  | 54 | -10 | -5 | ∞  | 38 | 29 | 29 | 47 | -33| 56 | -17| 34 | 3  |
|     |     | 4  | 16 | 74 | ∞  | 0  | 88  | 90 | 87 | 88 | 78 | 6  | 96 | 51 | 82 | 6  | 26 | 4  |
| -2  | -8  | 5  | 17 | 60 | 0  | 49 | 0   | ∞  | 37 | 8  | 15 | 21 | -8 | -2 | 15 | 36 | 67 | 5  |
| -14 | -15 | 6  | -14| 62 | 53 | 29 | 53  | 0  | 39 | 68 | -15| 10 | 7  | ∞  | 12 | 63 | 53 | 6  |
| -7  | -16 | 7  | 62 | 38 | 60 | -16| ∞   | 58 | 0  | 16 | 64 | 22 | 11 | 72 | -7 | 54 | 57 | 7  |
|     |     | 8  | 46 | 14 | 77 | 21 | 67  | 88 | 37 | 0  | 44 | 54 | ∞  | 16 | 61 | 68 | 86 | 8  |
|     |     | 9  | 69 | 30 | 62 | 44 | 35  | 93 | 81 | 17 | 0  | 22 | 48 | -4 | 28 | ∞  | 22 | 9  |
|     |     | 10 | 60 | ∞  | 15 | 16 | 82  | 33 | 74 | 10 | 79 | 0  | 42 | 68 | 40 | 59 | 37 | 10 |
|     |     | 11 | ∞  | 79 | 30 | 47 | 8   | 66 | 61 | 36 | 36 | 69 | 0  | 13 | 57 | 43 | 66 | 11 |
|     | -4  | 12 | 79 | 55 | 5  | 70 | 24  | 9  | 61 | 61 | -4 | 8  | 71 | 0  | ∞  | 62 | 47 | 12 |
|     |     | 13 | 21 | 35 | 51 | 83 | 37  | 16 | 24 | 37 | 7  | 15 | 9  | 76 | 0  | 80 | ∞  | 13 |
| -7  | -7  | 14 | -1 | 44 | 56 | ∞  | 6   | -7 | 44 | -7 | 0  | 39 | 52 | 27 | 18 | 0  | 83 | 14 |
|     |     | 15 | 26 | 78 | 82 | 20 | 9   | 58 | 75 | 13 | 43 | ∞  | 2  | 38 | 18 | 59 | 0  | 15 |
|     |     |    | 1  | 2  | 3  | 4  | 5   | 6  | 7  | 8  | 9  | 10 | 11 | 12 | 13 | 14 | 15 |    |

We now rename $T_4$, $H_0$. Our next xtep is to use $H_0$ - admissible 3-cycles to obtain an approximate solution to the bottleneck traveling salesman problem. In what follows, the values given are the actual values of the arcs obtained after applying $s_i$ to $H_i$. Let $s_0 = (14^7 6^{12} 1^{28})$.

$(1^{28}\ 6^{12}\ 11^{13} 8^9\ 2^1\ 10^{12} 15^{16} 13^{14} 12^{12} 4^3 14^7 5^{11} 7^{26} 3^{23} 9^5)$.

We next try to reduce the value of the arc emanating from 1. $s_1 = (1^4\ 11^{21} 5^9)$ lowers the value of the weightiest arc to 26. $H_2 = (1^4 8^9 2^1 10^{12} 15^{16} 13^{14} 12^{12} 4^3 14^7 5^9 6^{12} 11^{21} 7^{26} 3^{23} 9^5)$. We cannot construct an

$H_2$ - admissible 3-cycle whose first number is 7, because the last number of the 3-cycle must yield an arc whose value lies in column 3 of M. But every element of that column in M has a value at least as great as 26. Thus, $H_2$ is the best approximate solution that this algorithm can obtain for the asymmetric bottleneck traveling salesman problem.

COMMENT. In our method of constructing a near optimal or optimal n-cycle, after using negative cycles to obtain a near minimally-valued derangement, we then connect the cycles obtained to obtain a near minimal n-cycle. As $n \to \infty$, the number of cycles in a random permutation approaches log(n). Thus, we have a comparatively small number of arcs that are generally larger than arcs in the derangement. By using our sequences of 3-cycles, we can often reduce the value of our n-cycle to one that is near optimal or optimal.

Example 4.

Introduction. In this paper, we use the fact that every even permutation can be written as the product of (not necessarily disjoint) 3-cycles. Another property of even permutations is that if a permutation can be written as a product that contains an even number of cycles of even length, then it is even. Now suppose there exists a permutation, s, such that $Ts = T'$ where both T and T' are n-cycles. Since $s = T^{-1}T'$, s is always an even permutation even if T has an even number of points. Assume that H is an arbitrary n-cycle. In what follows, let M be an n X n matrix containing only positive entries. Then the following are true: (a) Every cycle containing an odd number of numbers is even. (b) Let $H_0$ be a randomly chosen n-cycle. Then there exists a sequence of 3-cycles which, when sequentially applied to $H_0$, yields an optimal tour. At the beginning of the algorithm, since our first n-cycle is chosen at random, we generally can choose s as a cycle of odd length that has a negative value. This helps speed up the algorithm. However, when we get closer to an optimal tour, we must rely on $H_i$-admissible 3-cycles alone. We define a *chain* of 3-cycles. A chain consists of no more than $[\ln(n)] + 1$ $H_i$-admissible 3-cycles. Some of the values of the chain are positive but the *sum of the values obtained must be negative*. This last statement is important because otherwise we might go into a loop. The algorithm ends when we can no longer find a chain of 3-cycles the sum of whose values is negative.

II. Preliminaries. We first obtain the 380 smallest odd prime numbers. We then obtain the remainder of each one modulo 100. Using subsets of consecutive remainders, we construct subsets each containing 10 or 9 remainders. We then randomly place them in an n X n matrix, M, containing no entries except along the diagonal where the symbol $\infty$ has been placed. In what follows the expression $a < b$ is defined to mean that b is further along going clockwise than a..

Theorem 1 Let H be a n-cycle, while $S = \{a_1, a_2, \ldots, a_{2r+1}; r > 0\}$ where

$a_1 < a_2 < \ldots < a_{2r+1} \leq n$. Then the cycle $(a_1\ a_2\ \ldots\ a_{2r+1})$ is always H-admissible.

Proof. H = (1 2 3 4 5 6 7 8 9 10), while s = (1 2 3 4 5).
$(1\ 2) \to (1\ 3), (2\ 3) \to (2\ 4), (3\ 4) \to (3\ 5), (4\ 5) \to (4\ 6), (5\ 1) \to (5\ 2)$.
*These yield the subpath (1 3 5 2 4 6). Thus our original path becomes (1 3 5 2 4 6 7 8 9 10).*

To save space, in the example that follows, negative entries in matrices are printed in boldface.

$H_0 = (1^3\ 2^9\ 3^{81}\ 4^{43}\ 5^{49}\ 6^1\ 7^{29}\ 8^{21}\ 9^{19}\ 10^{31}\ 11^{51}\ 12^{11}\ 13^{91}\ 14^{81}\ 15^{33}\ 16^{53}\ 17^{63}\ 18^{69}\ 19^{57}\ 20^{43})$: 838

$$H_0^{-1}M$$

|   | 2 | 3 | 4 | 5 | 6 | 7 | 8 | 9 | 10 | 11 | 12 | 13 | 14 | 15 | 16 | 17 | 18 | 19 | 20 | 1 |   |
|---|---|---|---|---|---|---|---|---|----|----|----|----|----|----|----|----|----|----|----|---|---|
|   | 1 | 2 | 3 | 4 | 5 | 6 | 7 | 8 | 9 | 10 | 11 | 12 | 13 | 14 | 15 | 16 | 17 | 18 | 19 | 20 |   |
| 1 | 0 | 2 | 4 | 2 | 10 | 14 | 16 | 20 | 26 | 60 | 64 | 76 | 84 | 88 | 96 | 0 | 4 | 10 | 14 | ∞ | 1 |
| 2 | ∞ | 0 | 4 | 10 | 12 | 22 | 24 | 30 | 40 | 54 | 31 | 38 | 52 | 58 | 62 | 64 | 68 | 70 | 80 | 88 | 2 |
| 3 | 83 | ∞ | 0 | 10 | 12 | 16 | 18 | 70 | 58 | 52 | 48 | 42 | 40 | 30 | 24 | 18 | 12 | 10 | 4 | 8 | 3 |
| 4 | 12 | 2 | ∞ | 0 | 4 | 10 | 16 | 18 | 30 | 40 | 32 | 26 | 16 | 14 | 4 | 10 | 20 | 26 | 38 | 24 | 4 |
| 5 | 20 | 18 | 12 | ∞ | 0 | 10 | 28 | 30 | 34 | 32 | 38 | 44 | 48 | 40 | 38 | 16 | 10 | 8 | 2 | 26 | 5 |
| 6 | 78 | 82 | 88 | 96 | ∞ | 0 | 2 | 6 | 8 | 96 | 8 | 12 | 18 | 20 | 30 | 32 | 38 | 48 | 62 | 72 | 6 |
| 7 | 54 | 60 | 70 | 18 | 16 | ∞ | 0 | 10 | 20 | 24 | 30 | 48 | 44 | 50 | 54 | 60 | 68 | 28 | 22 | 54 | 7 |
| 8 | 70 | 76 | 20 | 17 | 14 | 2 | ∞ | 0 | 6 | 20 | 26 | 32 | 38 | 56 | 62 | 66 | 68 | 20 | 10 | 68 | 8 |
| 9 | 68 | 72 | 74 | 78 | 16 | 10 | 6 | ∞ | 0 | 68 | 70 | 74 | 80 | 8 | 4 | 12 | 24 | 30 | 34 | 50 | 9 |
| 10 | 56 | 58 | 62 | 68 | 24 | 20 | 18 | 12 | ∞ | 0 | 6 | 10 | 12 | 16 | 22 | 28 | 30 | 36 | 40 | 52 | 10 |
| 11 | 30 | 20 | 14 | 12 | 2 | 50 | 6 | 12 | 16 | ∞ | 0 | 6 | 20 | 26 | 30 | 32 | 38 | 42 | 48 | 38 | 11 |
| 12 | 10 | 20 | 22 | 28 | 38 | 40 | 50 | 52 | 58 | 8 | ∞ | 0 | 6 | 16 | 18 | 28 | 42 | 52 | 58 | 8 | 12 |
| 13 | 78 | 70 | 54 | 52 | 48 | 40 | 24 | 22 | 18 | 14 | 8 | ∞ | 0 | 90 | 82 | 72 | 64 | 58 | 52 | 84 | 13 |
| 14 | 6 | 12 | 16 | 78 | 72 | 64 | 57 | 52 | 30 | 28 | 18 | 10 | ∞ | 0 | 6 | 12 | 80 | 68 | 64 | 6 | 14 |
| 15 | 34 | 38 | 46 | 50 | 64 | 24 | 12 | 10 | 0 | 32 | 26 | 20 | 2 | ∞ | 0 | 16 | 18 | 40 | 46 | 26 | 15 |
| 16 | 30 | 40 | 46 | 42 | 40 | 36 | 22 | 16 | 6 | 24 | 22 | 16 | 12 | 10 | ∞ | 0 | 8 | 26 | 50 | 28 | 16 |
| 17 | 26 | 22 | 16 | 10 | 4 | 8 | 14 | 20 | 28 | 24 | 20 | 14 | 12 | 6 | 4 | ∞ | 0 | 8 | 14 | 34 | 17 |
| 18 | 36 | 30 | 22 | 18 | 16 | 10 | 2 | 12 | 14 | 20 | 60 | 50 | 48 | 32 | 6 | 2 | ∞ | 0 | 12 | 40 | 18 |
| 19 | 34 | 54 | 48 | 54 | 36 | 28 | 6 | 4 | 0 | 48 | 38 | 36 | 26 | 24 | 18 | 14 | 8 | ∞ | 0 | 30 | 19 |
| 20 | 8 | 14 | 18 | 26 | 30 | 44 | 36 | 32 | 24 | 24 | 14 | 12 | 2 | 6 | 18 | 24 | 36 | 40 | ∞ | 0 | 20 |
|   | 1 | 2 | 3 | 4 | 5 | 6 | 7 | 8 | 9 | 10 | 11 | 12 | 13 | 14 | 15 | 16 | 17 | 18 | 19 | 20 |   |

We now construct $s_1 = (13^{-90}\ 14^{-80}\ 17^{-16}\ 3^{-42}\ 12^6)$. The 3-cycle (13 14 17) has a value of -198. The 5-cycle $s_1$ has a value of -222. Therefore, we use $s_1$ as our $H_0$ - admissible permutation.

$$H_1 = H_0 s_1 = (1^3\ 2^9\ 3^{39}\ 13^1\ 15^{33}\ 16^{49}\ 17^{47}\ 4^{43}\ 5^{49}\ 6^1\ 7^{29}\ 8^{21}\ 9^{19}\ 10^{31}\ 11^{51}\ 12^{17}\ 14^1\ 18^{69}\ 19^{57}\ 20^{43}): 612$$

Our new arcs are

(13 15), (14 18), (17 4), (3 13), (12 14).

$$H_1^{-1}M$$

|    | 2 | 3 | 13 | 5 | 6 | 7 | 8 | 9 | 10 | 11 | 12 | 14 | 15 | 18 | 16 | 17 | 4 | 19 | 20 | 1 |    |
|----|---|---|----|---|---|---|---|---|----|----|----|----|----|----|----|----|---|----|----|---|----|
|    | 1 | 2 | 3  | 4 | 5 | 6 | 7 | 8 | 9  | 10 | 11 | 12 | 13 | 14 | 15 | 16 | 17| 18 | 19 | 20|    |
| 1  | 0 | 2 | 76 | 2 | 10 | 14 | 16 | 20 | 26 | 60 | 64 | 84 | 88 | 4 | 96 | 0 | 4 | 10 | 14 | ∞ | 1 |
| 2  | ∞ | 0 | 38 | 10 | 12 | 22 | 24 | 30 | 40 | 54 | 31 | 52 | 58 | 68 | 62 | 64 | 4 | 70 | 80 | 88 | 2 |
| 3  | 40 | ∞ | 0 | 52 | 54 | 58 | 60 | 28 | 16 | 10 | 6 | 2 | 12 | 30 | 11 | 24 | 42 | 32 | 38 | 34 | 3 |
| 4  | 12 | 2 | 26 | 0 | 4 | 10 | 16 | 18 | 30 | 40 | 32 | 16 | 14 | 20 | 4 | 10 | ∞ | 26 | 38 | 24 | 4 |
| 5  | 20 | 18 | 44 | ∞ | 0 | 10 | 28 | 30 | 34 | 32 | 38 | 48 | 40 | 10 | 38 | 16 | 12 | 8 | 2 | 26 | 5 |
| 6  | 78 | 82 | 12 | 96 | ∞ | 0 | 2 | 6 | 8 | 96 | 8 | 18 | 20 | 38 | 30 | 32 | 88 | 48 | 62 | 72 | 6 |
| 7  | 54 | 60 | 48 | 18 | 16 | ∞ | 0 | 10 | 20 | 24 | 30 | 44 | 50 | 68 | 54 | 60 | 70 | 28 | 22 | 54 | 7 |
| 8  | 70 | 76 | 32 | 17 | 14 | 2 | ∞ | 0 | 6 | 20 | 26 | 38 | 56 | 68 | 62 | 66 | 20 | 20 | 10 | 68 | 8 |
| 9  | 68 | 72 | 74 | 78 | 16 | 10 | 6 | ∞ | 0 | 68 | 70 | 80 | 8 | 24 | 4 | 12 | 74 | 30 | 34 | 50 | 9 |
| 10 | 56 | 58 | 10 | 68 | 24 | 20 | 18 | 12 | ∞ | 0 | 6 | 12 | 16 | 30 | 22 | 28 | 62 | 36 | 40 | 52 | 10 |
| 11 | 30 | 20 | 6 | 12 | 2 | 50 | 6 | 12 | 16 | ∞ | 0 | 20 | 26 | 38 | 30 | 32 | 14 | 42 | 48 | 38 | 11 |
| 12 | 4 | 14 | 6 | 22 | 32 | 34 | 44 | 48 | 52 | 14 | ∞ | 0 | 10 | 36 | 12 | 22 | 16 | 46 | 52 | 2 | 12 |
| 13 | 12 | 20 | ∞ | 38 | 42 | 50 | 66 | 68 | 72 | 76 | 82 | 0 | 0 | 26 | 8 | 18 | 36 | 32 | 38 | 6 | 13 |
| 14 | 86 | 92 | 70 | 2 | 8 | 16 | 23 | 28 | 50 | 52 | 62 | ∞ | 0 | 0 | 86 | 92 | 96 | 12 | 16 | 86 | 14 |
| 15 | 34 | 38 | 20 | 50 | 64 | 24 | 12 | 10 | 0 | 32 | 26 | 2 | ∞ | 18 | 0 | 16 | 46 | 40 | 46 | 26 | 15 |
| 16 | 30 | 40 | 16 | 42 | 40 | 36 | 22 | 16 | 6 | 24 | 22 | 12 | 10 | 8 | ∞ | 0 | 46 | 26 | 50 | 28 | 16 |
| 17 | 10 | 6 | 0 | 10 | 20 | 24 | 30 | 36 | 44 | 8 | 4 | 4 | 10 | 16 | 12 | ∞ | 0 | 24 | 30 | 18 | 17 |
| 18 | 36 | 30 | 50 | 18 | 16 | 10 | 2 | 12 | 14 | 20 | 60 | 48 | 32 | ∞ | 6 | 2 | 22 | 0 | 12 | 40 | 18 |
| 19 | 34 | 54 | 36 | 54 | 36 | 28 | 6 | 4 | 0 | 48 | 38 | 26 | 24 | 8 | 18 | 14 | 48 | ∞ | 0 | 30 | 19 |
| 20 | 8 | 14 | 12 | 26 | 30 | 44 | 36 | 32 | 24 | 24 | 14 | 2 | 6 | 36 | 18 | 24 | 18 | 40 | ∞ | 0 | 20 |
|    | 1 | 2 | 3 | 4 | 5 | 6 | 7 | 8 | 9 | 10 | 11 | 12 | 13 | 14 | 15 | 16 | 17 | 18 | 19 | 20 |    |

Using $s_2 = (18^{-50}\ 3^{-38}\ 8^{-20})$: -108: , let $H_2 = H_1 s_2$ ..

Our new arcs are $(18^{-50}\ 13)$, $(3^{-28}\ 9)$, $(8^{-20}\ 19)$.

$H_2 = (1^3\ 2^9\ 3^{11}\ 9^{19}\ 10^{31}\ 11^{51}\ 12^{17}\ 14^1\ 18^{19}\ 13^1\ 15^{33}\ 16^{53}\ 17^{47}\ 4^{43}\ 5^{49}\ 6^1\ 7^{29}\ 8^1\ 19^{57}\ 20^{43})$: 518.

$H_2^{-1}M$

|    | 2  | 3  | 9  | 5  | 6  | 7  | 8  | 19 | 10 | 11 | 12 | 14 | 15 | 18 | 16 | 17 | 4  | 13 | 20 | 1  |    |
|----|----|----|----|----|----|----|----|----|----|----|----|----|----|----|----|----|----|----|----|----|----|
|    | 1  | 2  | 3  | 4  | 5  | 6  | 7  | 8  | 9  | 10 | 11 | 12 | 13 | 14 | 15 | 16 | 17 | 18 | 19 | 20 |    |
| 1  | 0  | 2  | 20 | 2  | 10 | 14 | 16 | 10 | 26 | 60 | 64 | 84 | 88 | 4  | 96 | 0  | 4  | 76 | 14 | ∞  | 1  |
| 2  | ∞  | 0  | 30 | 10 | 12 | 22 | 24 | 70 | 40 | 54 | 31 | 52 | 58 | 68 | 62 | 64 | 4  | 38 | 80 | 88 | 2  |
| 3  | 68 | ∞  | 0  | 80 | 82 | 86 | 88 | 60 | 12 | 18 | 22 | 30 | 40 | 58 | 39 | 52 | 70 | 28 | 66 | 62 | 3  |
| 4  | 12 | 2  | 18 | 0  | 4  | 10 | 16 | 26 | 30 | 40 | 32 | 16 | 14 | 20 | 4  | 10 | ∞  | 26 | 38 | 24 | 4  |
| 5  | 20 | 18 | 30 | ∞  | 0  | 10 | 28 | 8  | 34 | 32 | 38 | 48 | 40 | 10 | 38 | 16 | 12 | 44 | 2  | 26 | 5  |
| 6  | 78 | 82 | 6  | 96 | ∞  | 0  | 2  | 48 | 8  | 96 | 8  | 18 | 20 | 38 | 30 | 32 | 88 | 12 | 62 | 72 | 6  |
| 7  | 54 | 60 | 10 | 18 | 16 | ∞  | 0  | 28 | 20 | 24 | 30 | 44 | 50 | 68 | 54 | 60 | 70 | 48 | 22 | 54 | 7  |
| 8  | 90 | 96 | 0  | 3  | 6  | 18 | ∞  | 0  | 26 | 40 | 46 | 58 | 76 | 88 | 82 | 86 | 0  | 52 | 10 | 88 | 8  |
| 9  | 68 | 72 | ∞  | 78 | 16 | 10 | 6  | 30 | 0  | 68 | 70 | 80 | 8  | 24 | 4  | 12 | 74 | 74 | 34 | 50 | 9  |
| 10 | 56 | 58 | 12 | 68 | 24 | 20 | 18 | 36 | ∞  | 0  | 6  | 12 | 16 | 30 | 22 | 28 | 62 | 10 | 40 | 52 | 10 |
| 11 | 30 | 20 | 12 | 12 | 2  | 50 | 6  | 42 | 16 | ∞  | 0  | 20 | 26 | 38 | 30 | 32 | 14 | 6  | 48 | 38 | 11 |
| 12 | 4  | 14 | 48 | 22 | 32 | 34 | 44 | 46 | 52 | 14 | ∞  | 0  | 10 | 36 | 12 | 22 | 16 | 6  | 52 | 2  | 12 |
| 13 | 12 | 20 | 68 | 38 | 42 | 50 | 66 | 32 | 72 | 76 | 82 | 0  | 0  | 26 | 8  | 18 | 36 | ∞  | 38 | 6  | 13 |
| 14 | 86 | 92 | 28 | 2  | 8  | 16 | 23 | 12 | 50 | 52 | 62 | ∞  | 0  | 0  | 86 | 92 | 96 | 70 | 16 | 86 | 14 |
| 15 | 34 | 38 | 10 | 50 | 64 | 24 | 12 | 40 | 0  | 32 | 26 | 2  | ∞  | 18 | 0  | 16 | 46 | 20 | 46 | 26 | 15 |
| 16 | 30 | 40 | 16 | 42 | 40 | 36 | 22 | 26 | 6  | 24 | 22 | 12 | 10 | 8  | ∞  | 0  | 46 | 16 | 50 | 28 | 16 |
| 17 | 10 | 6  | 36 | 10 | 20 | 24 | 30 | 24 | 44 | 8  | 4  | 4  | 10 | 16 | 12 | ∞  | 0  | 0  | 30 | 18 | 17 |
| 18 | 14 | 20 | 62 | 32 | 34 | 40 | 52 | 50 | 64 | 70 | 10 | 2  | 18 | ∞  | 44 | 48 | 28 | 0  | 62 | 10 | 18 |
| 19 | 34 | 54 | 4  | 54 | 36 | 28 | 6  | ∞  | 0  | 48 | 38 | 26 | 24 | 8  | 18 | 14 | 48 | 36 | 0  | 30 | 19 |
| 20 | 8  | 14 | 32 | 26 | 30 | 44 | 36 | 40 | 24 | 24 | 14 | 2  | 6  | 36 | 18 | 24 | 18 | 12 | ∞  | 0  | 20 |
|    | 1  | 2  | 3  | 4  | 5  | 6  | 7  | 8  | 9  | 10 | 11 | 12 | 13 | 14 | 15 | 16 | 17 | 18 | 19 | 20 |    |

Let $s_3 = (19^{-48}\ 10^{-24}\ 5^{-2})$: -74.

$H_3 = (1^3\ 2^9\ 3^{11}\ 9^{19}\ 10^7\ 6^1\ 7^{29}\ 8^1\ 19^9\ 11^{51}\ 12^{17}\ 14^1\ 18^{37}\ 15^{33}\ 16^{53}\ 17^{47}\ 4^{43}\ 5^{47}\ 20^{43})$: 491

Our new arcs are (19 11), (10 6), (5 20).

$$H_3^{-1}M$$

|   | 2 | 3 | 9 | 5 | 20 | 7 | 8 | 19 | 10 | 6 | 12 | 14 | 15 | 18 | 16 | 17 | 4 | 13 | 11 | 1 |   |
|---|---|---|---|---|---|---|---|---|---|---|---|---|---|---|---|---|---|---|---|---|---|
|   | 1 | 2 | 3 | 4 | 5 | 6 | 7 | 8 | 9 | 10 | 11 | 12 | 13 | 14 | 15 | 16 | 17 | 18 | 19 | 20 |   |
| 1 | 0 | 2 | 20 | 2 | 14 | 14 | 16 | 10 | 26 | 10 | 64 | 84 | 88 | 4 | 96 | 0 | 4 | 76 | 60 | $\infty$ | 1 |
| 2 | $\infty$ | 0 | 30 | 10 | 80 | 22 | 24 | 70 | 40 | 12 | 31 | 52 | 58 | 68 | 62 | 64 | 4 | 38 | 54 | 88 | 2 |
| 3 | 68 | $\infty$ | 0 | 80 | 66 | 86 | 88 | 60 | 12 | 82 | 22 | 30 | 40 | 58 | 39 | 52 | 70 | 28 | 18 | 62 | 3 |
| 4 | 12 | 2 | 18 | 0 | 38 | 10 | 16 | 26 | 30 | 4 | 32 | 16 | 14 | 20 | 4 | 10 | $\infty$ | 26 | 40 | 24 | 4 |
| 5 | 20 | 18 | 12 | $\infty$ | 2 | 10 | 28 | 30 | 34 | 32 | 38 | 44 | 48 | 40 | 28 | 16 | 10 | 8 | 2 | 26 | 5 |
| 6 | 78 | 82 | 6 | 96 | 62 | 0 | 2 | 48 | 8 | $\infty$ | 8 | 18 | 20 | 38 | 30 | 32 | 88 | 12 | 96 | 72 | 6 |
| 7 | 54 | 60 | 10 | 18 | 22 | $\infty$ | 0 | 28 | 20 | 16 | 30 | 44 | 50 | 68 | 54 | 60 | 70 | 48 | 24 | 54 | 7 |
| 8 | 90 | 96 | 0 | 3 | 10 | 18 | $\infty$ | 0 | 26 | 6 | 46 | 58 | 76 | 88 | 82 | 86 | 0 | 52 | 40 | 88 | 8 |
| 9 | 68 | 72 | $\infty$ | 78 | 34 | 10 | 6 | 30 | 0 | 16 | 70 | 80 | 8 | 24 | 4 | 12 | 74 | 74 | 68 | 50 | 9 |
| 10 | 80 | 82 | 12 | 92 | 40 | 4 | 6 | 60 | $\infty$ | 0 | 30 | 36 | 40 | 54 | 46 | 52 | 86 | 34 | 24 | 76 | 10 |
| 11 | 30 | 20 | 12 | 12 | 48 | 50 | 6 | 42 | 16 | 2 | 0 | 20 | 26 | 38 | 30 | 32 | 14 | 6 | $\infty$ | 38 | 11 |
| 12 | 4 | 14 | 48 | 22 | 52 | 34 | 44 | 46 | 52 | 32 | $\infty$ | 0 | 10 | 36 | 12 | 22 | 16 | 6 | 14 | 2 | 12 |
| 13 | 12 | 20 | 68 | 38 | 38 | 50 | 66 | 32 | 72 | 42 | 82 | 0 | 0 | 26 | 8 | 18 | 36 | $\infty$ | 76 | 6 | 13 |
| 14 | 86 | 92 | 28 | 2 | 16 | 16 | 23 | 12 | 50 | 8 | 62 | $\infty$ | 0 | 0 | 86 | 92 | 96 | 70 | 52 | 86 | 14 |
| 15 | 34 | 38 | 10 | 50 | 46 | 24 | 12 | 40 | 0 | 64 | 26 | 2 | $\infty$ | 18 | 0 | 16 | 46 | 20 | 32 | 26 | 15 |
| 16 | 30 | 40 | 16 | 42 | 50 | 36 | 22 | 26 | 6 | 40 | 22 | 12 | 10 | 8 | $\infty$ | 0 | 46 | 16 | 24 | 28 | 16 |
| 17 | 10 | 6 | 36 | 10 | 30 | 24 | 30 | 24 | 44 | 20 | 4 | 4 | 10 | 16 | 12 | $\infty$ | 0 | 0 | 8 | 18 | 17 |
| 18 | 14 | 20 | 62 | 32 | 62 | 40 | 52 | 50 | 64 | 34 | 10 | 2 | 18 | $\infty$ | 44 | 48 | 28 | 0 | 70 | 10 | 18 |
| 19 | 78 | 6 | 44 | 6 | 0 | 20 | 42 | $\infty$ | 48 | 12 | 10 | 22 | 24 | 40 | 30 | 34 | 0 | 12 | 0 | 78 | 19 |
| 20 | 8 | 14 | 32 | 26 | $\infty$ | 44 | 36 | 40 | 24 | 30 | 14 | 2 | 6 | 36 | 18 | 24 | 18 | 12 | 24 | 0 | 20 |
|   | 1 | 2 | 3 | 4 | 5 | 6 | 7 | 8 | 9 | 10 | 11 | 12 | 13 | 14 | 15 | 16 | 17 | 18 | 19 | 20 |   |

$s_4 = (16^{-50} 5^{-20} 1^0)$: -70. Our new arcs are: (16 20), (5 2), (1 17).

$H_4 = (1^3\ 17^{47}\ 4^{43}\ 5^{29}\ 2^9\ 3^{11}\ 9^{19}\ 10^7\ 6^1\ 7^{29}\ 8^1\ 19^9\ 11^{51}\ 12^{17}\ 14^1\ 18^{19}\ 13^9\ 16^3\ 20^{13})$: 321.

$$H_4^{-1}M$$

|  | 17 | 3 | 9 | 5 | 2 | 7 | 8 | 19 | 10 | 6 | 12 | 14 | 15 | 18 | 16 | 20 | 4 | 13 | 11 | 1 |  |
|---|---|---|---|---|---|---|---|---|---|---|---|---|---|---|---|---|---|---|---|---|---|
|  | 1 | 2 | 3 | 4 | 5 | 6 | 7 | 8 | 9 | 10 | 11 | 12 | 13 | 14 | 15 | 16 | 17 | 18 | 19 | 20 |  |
| 1 | 0 | 2 | 20 | 2 | 0 | 14 | 16 | 10 | 26 | 10 | 64 | 84 | 88 | 4 | 96 | 14 | 4 | 76 | 60 | ∞ | 1 |
| 2 | 64 | 0 | 30 | 10 | ∞ | 22 | 24 | 70 | 40 | 12 | 31 | 52 | 58 | 68 | 62 | 80 | 4 | 38 | 54 | 88 | 2 |
| 3 | 52 | ∞ | 0 | 80 | 68 | 86 | 88 | 60 | 12 | 82 | 22 | 30 | 40 | 58 | 39 | 66 | 70 | 28 | 18 | 62 | 3 |
| 4 | 10 | 2 | 18 | 0 | 12 | 10 | 16 | 26 | 30 | 4 | 32 | 16 | 14 | 20 | 4 | 38 | ∞ | 26 | 40 | 24 | 4 |
| 5 | 4 | 2 | 8 | ∞ | 0 | 30 | 48 | 50 | 54 | 52 | 58 | 64 | 68 | 20 | 0 | 2 | 10 | 12 | 18 | 6 | 5 |
| 6 | 32 | 82 | 6 | 96 | 78 | 0 | 2 | 48 | 8 | ∞ | 8 | 18 | 20 | 38 | 30 | 62 | 88 | 12 | 96 | 72 | 6 |
| 7 | 60 | 60 | 10 | 18 | 54 | ∞ | 0 | 28 | 20 | 16 | 30 | 44 | 50 | 68 | 54 | 22 | 70 | 48 | 24 | 54 | 7 |
| 8 | 86 | 96 | 0 | 3 | 90 | 18 | ∞ | 0 | 26 | 6 | 46 | 58 | 76 | 88 | 82 | 10 | 0 | 52 | 40 | 88 | 8 |
| 9 | 12 | 72 | ∞ | 78 | 68 | 10 | 6 | 30 | 0 | 16 | 70 | 80 | 8 | 24 | 4 | 34 | 74 | 74 | 68 | 50 | 9 |
| 10 | 52 | 82 | 12 | 92 | 80 | 4 | 6 | 60 | ∞ | 0 | 30 | 36 | 40 | 54 | 46 | 40 | 86 | 34 | 24 | 76 | 10 |
| 11 | 32 | 20 | 12 | 12 | 30 | 50 | 6 | 42 | 16 | 2 | 0 | 20 | 26 | 38 | 30 | 48 | 14 | 6 | ∞ | 38 | 11 |
| 12 | 22 | 14 | 48 | 22 | 4 | 34 | 44 | 46 | 52 | 32 | ∞ | 0 | 10 | 36 | 12 | 52 | 16 | 6 | 14 | 2 | 12 |
| 13 | 18 | 20 | 68 | 38 | 12 | 50 | 66 | 32 | 72 | 42 | 82 | 0 | 0 | 26 | 8 | 38 | 36 | ∞ | 76 | 6 | 13 |
| 14 | 92 | 92 | 28 | 2 | 86 | 16 | 23 | 12 | 50 | 8 | 62 | ∞ | 0 | 0 | 86 | 16 | 96 | 70 | 52 | 86 | 14 |
| 15 | 16 | 38 | 10 | 50 | 34 | 24 | 12 | 40 | 0 | 64 | 26 | 2 | ∞ | 18 | 0 | 46 | 46 | 20 | 32 | 26 | 15 |
| 16 | 50 | 90 | 34 | 8 | 80 | 14 | 28 | 76 | 44 | 10 | 28 | 38 | 40 | 58 | ∞ | 0 | 4 | 34 | 26 | 78 | 16 |
| 17 | ∞ | 6 | 36 | 10 | 10 | 24 | 30 | 24 | 44 | 20 | 4 | 4 | 10 | 16 | 12 | 30 | 0 | 0 | 8 | 18 | 17 |
| 18 | 48 | 20 | 62 | 32 | 14 | 40 | 52 | 50 | 64 | 34 | 10 | 2 | 18 | ∞ | 44 | 62 | 28 | 0 | 70 | 10 | 18 |
| 19 | 34 | 6 | 44 | 6 | 78 | 20 | 42 | ∞ | 48 | 12 | 10 | 22 | 24 | 40 | 30 | 0 | 0 | 12 | 0 | 78 | 19 |
| 20 | 24 | 14 | 32 | 26 | 8 | 44 | 36 | 40 | 24 | 30 | 14 | 2 | 6 | 36 | 18 | ∞ | 18 | 12 | 24 | 0 | 20 |
|  | 1 | 2 | 3 | 4 | 5 | 6 | 7 | 8 | 9 | 10 | 11 | 12 | 13 | 14 | 15 | 16 | 17 | 18 | 19 | 20 |  |

We use the second smallest arc of the matrix: (11  20).  $S_5 = (11^{-38}\ 20^{-24}\ 19^{10})$: -52.  Our new arcs are:

(11  1), (20  11) and (19  12).

H₅ = (1³ 17⁴⁷ 4⁴³ 5²⁹ 2⁹ 3¹¹ 9¹⁹ 10⁷ 6¹ 7²⁹ 8¹ 19¹⁹ 12¹⁷ 14¹ 18¹⁹ 13¹ 15³³ 16³ 20¹⁹ 11¹³): 281.

$H_5^{-1}M$

|   | 17 | 3 | 9 | 5 | 2 | 7 | 8 | 19 | 10 | 6 | 1 | 14 | 15 | 18 | 16 | 20 | 4 | 13 | 12 | 11 |   |
|---|---|---|---|---|---|---|---|---|---|---|---|---|---|---|---|---|---|---|---|---|---|
|   | 1 | 2 | 3 | 4 | 5 | 6 | 7 | 8 | 9 | 10 | 11 | 12 | 13 | 14 | 15 | 16 | 17 | 18 | 19 | 20 |   |
| 1  | 0 | 2 | 20 | 2 | 0 | 14 | 16 | 10 | 26 | 10 | ∞ | 84 | 88 | 4 | 96 | 14 | 4 | 76 | 64 | 60 | 1 |
| 2  | 64 | 0 | 30 | 10 | ∞ | 22 | 24 | 70 | 40 | 12 | 88 | 52 | 58 | 68 | 62 | 80 | 4 | 38 | 31 | 54 | 2 |
| 3  | 52 | ∞ | 0 | 80 | 68 | 86 | 88 | 60 | 12 | 82 | 62 | 30 | 40 | 58 | 39 | 66 | 70 | 28 | 22 | 18 | 3 |
| 4  | 10 | 2 | 18 | 0 | 12 | 10 | 16 | 26 | 30 | 4 | 24 | 16 | 14 | 20 | 4 | 38 | ∞ | 26 | 32 | 40 | 4 |
| 5  | 4 | 2 | 8 | ∞ | 0 | 30 | 48 | 50 | 54 | 52 | 6 | 64 | 68 | 20 | 0 | 2 | 10 | 12 | 58 | 18 | 5 |
| 6  | 32 | 82 | 6 | 96 | 78 | 0 | 2 | 48 | 8 | ∞ | 72 | 18 | 20 | 38 | 30 | 62 | 88 | 12 | 8 | 96 | 6 |
| 7  | 60 | 60 | 10 | 18 | 54 | ∞ | 0 | 28 | 20 | 16 | 54 | 44 | 50 | 68 | 54 | 22 | 70 | 48 | 30 | 24 | 7 |
| 8  | 86 | 96 | 0 | 3 | 90 | 18 | ∞ | 0 | 26 | 6 | 88 | 58 | 76 | 88 | 82 | 10 | 0 | 52 | 46 | 40 | 8 |
| 9  | 12 | 72 | ∞ | 78 | 68 | 10 | 6 | 30 | 0 | 16 | 50 | 80 | 8 | 24 | 4 | 34 | 74 | 74 | 70 | 68 | 9 |
| 10 | 52 | 82 | 12 | 92 | 80 | 4 | 6 | 60 | ∞ | 0 | 76 | 36 | 40 | 54 | 46 | 40 | 86 | 34 | 30 | 24 | 10 |
| 11 | 70 | 18 | 50 | 26 | 8 | 12 | 44 | 80 | 54 | 36 | 0 | 58 | 64 | 76 | 68 | 86 | 26 | 44 | 38 | ∞ | 11 |
| 12 | 22 | 14 | 48 | 22 | 4 | 34 | 44 | 46 | 52 | 32 | 2 | 0 | 10 | 36 | 12 | 52 | 16 | 6 | ∞ | 14 | 12 |
| 13 | 18 | 20 | 68 | 38 | 12 | 50 | 66 | 32 | 72 | 42 | 6 | 0 | 0 | 26 | 8 | 38 | 36 | ∞ | 82 | 76 | 13 |
| 14 | 92 | 92 | 28 | 2 | 86 | 16 | 23 | 12 | 50 | 8 | 86 | ∞ | 0 | 0 | 86 | 16 | 96 | 70 | 62 | 52 | 14 |
| 15 | 16 | 38 | 10 | 50 | 34 | 24 | 12 | 40 | 0 | 64 | 26 | 2 | ∞ | 18 | 0 | 46 | 46 | 20 | 26 | 32 | 15 |
| 16 | 50 | 90 | 34 | 8 | 80 | 14 | 28 | 76 | 44 | 10 | 78 | 38 | 40 | 58 | ∞ | 0 | 4 | 34 | 28 | 26 | 16 |
| 17 | ∞ | 6 | 36 | 10 | 10 | 24 | 30 | 24 | 44 | 20 | 18 | 4 | 10 | 16 | 12 | 30 | 0 | 0 | 4 | 8 | 17 |
| 18 | 48 | 20 | 62 | 32 | 14 | 40 | 52 | 50 | 64 | 34 | 10 | 2 | 18 | ∞ | 44 | 62 | 28 | 0 | 10 | 70 | 18 |
| 19 | 24 | 16 | 24 | 16 | 68 | 10 | 32 | ∞ | 38 | 2 | 68 | 12 | 14 | 30 | 20 | 20 | 24 | 2 | 0 | 10 | 19 |
| 20 | 48 | 38 | 8 | 50 | 32 | 68 | 12 | 64 | 0 | 54 | 24 | 22 | 30 | 60 | 42 | ∞ | 42 | 12 | 10 | 0 | 20 |
|   | 1 | 2 | 3 | 4 | 5 | 6 | 7 | 8 | 9 | 10 | 11 | 12 | 13 | 14 | 15 | 16 | 17 | 18 | 19 | 20 |   |

s₆ = (4⁻³² 19² 18³²): 2   Our new arcs are: (4 12), (19 13), (18 5)..

$H_6 = (1^3\ 17^{47}\ 4^{11}\ 12^{17}\ 14^1\ 18^{51}\ 5^{29}\ 2^9\ 3^{11}\ 9^{19}\ 10^7\ 6^1\ 7^{29}\ 8^1\ 19^{21}\ 13^1\ 15^{33}\ 16^3\ 20^{19}\ 11^{13}\ )$: 326

$$H_6^{-1}M$$

| | 17 | 3 | 9 | 12 | 2 | 7 | 8 | 19 | 10 | 6 | 1 | 14 | 15 | 18 | 16 | 20 | 4 | 5 | 13 | 11 | |
|---|---|---|---|---|---|---|---|---|---|---|---|---|---|---|---|---|---|---|---|---|---|
| | 1 | 2 | 3 | 4 | 5 | 6 | 7 | 8 | 9 | 10 | 11 | 12 | 13 | 14 | 15 | 16 | 17 | 18 | 19 | 20 | |
| 1 | 0 | 2 | 20 | 64 | 0 | 14 | 16 | 10 | 26 | 10 | ∞ | 84 | 88 | 4 | 96 | 14 | 4 | 2 | 76 | 60 | 1 |
| 2 | 64 | 0 | 30 | 31 | ∞ | 22 | 24 | 70 | 40 | 12 | 88 | 52 | 58 | 68 | 62 | 80 | 4 | 10 | 38 | 54 | 2 |
| 3 | 52 | ∞ | 0 | 22 | 68 | 86 | 88 | 60 | 12 | 82 | 62 | 30 | 40 | 58 | 39 | 66 | 70 | 80 | 28 | 18 | 3 |
| 4 | 42 | 30 | 50 | 0 | 44 | 42 | 48 | 58 | 62 | 36 | 8 | 16 | 18 | 52 | 28 | 70 | ∞ | 32 | 58 | 8 | 4 |
| 5 | 4 | 2 | 8 | 58 | 0 | 30 | 48 | 50 | 54 | 52 | 6 | 64 | 68 | 20 | 0 | 2 | 10 | ∞ | 12 | 18 | 5 |
| 6 | 32 | 82 | 6 | 8 | 78 | 0 | 2 | 48 | 8 | ∞ | 72 | 18 | 20 | 38 | 30 | 62 | 88 | 96 | 12 | 96 | 6 |
| 7 | 60 | 60 | 10 | 30 | 54 | ∞ | 0 | 28 | 20 | 16 | 54 | 44 | 50 | 68 | 54 | 22 | 70 | 18 | 48 | 24 | 7 |
| 8 | 86 | 96 | 0 | 46 | 90 | 18 | ∞ | 0 | 26 | 6 | 88 | 58 | 76 | 88 | 82 | 10 | 0 | 3 | 52 | 40 | 8 |
| 9 | 12 | 72 | ∞ | 70 | 68 | 10 | 6 | 30 | 0 | 16 | 50 | 80 | 8 | 24 | 4 | 34 | 74 | 78 | 74 | 68 | 9 |
| 10 | 52 | 82 | 12 | 30 | 80 | 4 | 6 | 60 | ∞ | 0 | 76 | 36 | 40 | 54 | 46 | 40 | 86 | 92 | 34 | 24 | 10 |
| 11 | 70 | 18 | 50 | 38 | 8 | 12 | 44 | 80 | 54 | 36 | 0 | 58 | 64 | 76 | 68 | 86 | 26 | 26 | 44 | ∞ | 11 |
| 12 | 22 | 14 | 48 | ∞ | 4 | 34 | 44 | 46 | 52 | 32 | 2 | 0 | 10 | 36 | 12 | 52 | 16 | 22 | 6 | 14 | 12 |
| 13 | 18 | 20 | 68 | 82 | 12 | 50 | 66 | 32 | 72 | 42 | 6 | 0 | 0 | 26 | 8 | 38 | 36 | 38 | ∞ | 76 | 13 |
| 14 | 92 | 92 | 28 | 62 | 86 | 16 | 23 | 12 | 50 | 8 | 86 | ∞ | 0 | 0 | 86 | 16 | 96 | 2 | 70 | 52 | 14 |
| 15 | 16 | 38 | 10 | 26 | 34 | 24 | 12 | 40 | 0 | 64 | 26 | 2 | ∞ | 18 | 0 | 46 | 46 | 50 | 20 | 32 | 15 |
| 16 | 50 | 90 | 34 | 28 | 80 | 14 | 28 | 76 | 44 | 10 | 78 | 38 | 40 | 58 | ∞ | 0 | 4 | 8 | 34 | 26 | 16 |
| 17 | ∞ | 6 | 36 | 4 | 10 | 24 | 30 | 24 | 44 | 20 | 18 | 4 | 10 | 16 | 12 | 30 | 0 | 10 | 0 | 8 | 17 |
| 18 | 16 | 12 | 30 | 42 | 18 | 8 | 20 | 18 | 32 | 2 | 22 | 30 | 14 | ∞ | 12 | 30 | 4 | 0 | 32 | 38 | 18 |
| 19 | 22 | 18 | 22 | 2 | 66 | 8 | 30 | ∞ | 36 | 0 | 66 | 10 | 12 | 28 | 18 | 18 | 22 | 18 | 0 | 12 | 19 |
| 20 | 48 | 38 | 8 | 10 | 32 | 68 | 12 | 64 | 0 | 54 | 24 | 22 | 30 | 60 | 42 | ∞ | 42 | 50 | 12 | 0 | 20 |
| | 1 | 2 | 3 | 4 | 5 | 6 | 7 | 8 | 9 | 10 | 11 | 12 | 13 | 14 | 15 | 16 | 17 | 18 | 19 | 20 | |

Searching for a negative 3-cycle, we obtain $(18^{-42}\ 4^{16}\ 12^{22})$: -4. We now note the following: As long as we eventually obtain a negative $H_i$-adnmissible 3-cycle, our chain works O.K. By construction, the sum of the values of the first two arcs of a 3-cycle is always negative. The value of the third arc is not important unless we are unable to eventually obtain a negative 3-cycle. The chains that we construct always delete the third arc of each 3-cycle until we finally obtain a negative 3-cycle. If we are unable to do so, starting from any negatively-valued arc, then the algorithm can proceed no further.

Our new arcs are: (18 12), (4 14), (12 5).

$H_7 = (1^3 \ 17^{47} \ 4^{27} \ 14^1 \ 18^9 \ 12^{39} \ 5^{29} \ 2^9 \ 3^{11} \ 9^{19} \ 10^7 \ 6^1 \ 7^{29} \ 8^1 \ 19^{21} \ 13^1 \ 15^{33} \ 16^3 \ 20^{19} \ 11^{13}) : 322.$

$$H_7^{-1}M$$

| | 17 | 3 | 9 | 14 | 2 | 7 | 8 | 19 | 10 | 6 | 1 | 5 | 15 | 18 | 16 | 20 | 4 | 12 | 13 | 11 | |
|---|---|---|---|---|---|---|---|---|---|---|---|---|---|---|---|---|---|---|---|---|---|
| | 1 | 2 | 3 | 4 | 5 | 6 | 7 | 8 | 9 | 10 | 11 | 12 | 13 | 14 | 15 | 16 | 17 | 18 | 19 | 20 | |
| 1 | 0 | 2 | 20 | 84 | 0 | 14 | 16 | 10 | 26 | 10 | ∞ | 2 | 88 | 4 | 96 | 14 | 4 | 64 | 76 | 60 | 1 |
| 2 | 64 | 0 | 30 | 52 | ∞ | 22 | 24 | 70 | 40 | 12 | 88 | 10 | 58 | 68 | 62 | 80 | 4 | 31 | 38 | 54 | 2 |
| 3 | 52 | ∞ | 0 | 30 | 68 | 86 | 88 | 60 | 12 | 82 | 62 | 80 | 40 | 58 | 39 | 66 | 70 | 22 | 28 | 18 | 3 |
| 4 | 26 | 14 | 34 | 0 | 28 | 26 | 32 | 42 | 46 | 20 | 8 | 16 | 2 | 36 | 12 | 54 | ∞ | 16 | 42 | 24 | 4 |
| 5 | 4 | 2 | 8 | 64 | 0 | 30 | 48 | 50 | 54 | 52 | 6 | ∞ | 68 | 20 | 0 | 2 | 10 | 58 | 12 | 18 | 5 |
| 6 | 32 | 82 | 6 | 18 | 78 | 0 | 2 | 48 | 8 | ∞ | 72 | 96 | 20 | 38 | 30 | 62 | 88 | 8 | 12 | 96 | 6 |
| 7 | 60 | 60 | 10 | 44 | 54 | ∞ | 0 | 28 | 20 | 16 | 54 | 18 | 50 | 68 | 54 | 22 | 70 | 30 | 48 | 24 | 7 |
| 8 | 86 | 96 | 0 | 58 | 90 | 18 | ∞ | 0 | 26 | 6 | 88 | 3 | 76 | 88 | 82 | 10 | 0 | 46 | 52 | 40 | 8 |
| 9 | 12 | 72 | ∞ | 80 | 68 | 10 | 6 | 30 | 0 | 16 | 50 | 78 | 8 | 24 | 4 | 34 | 74 | 70 | 74 | 68 | 9 |
| 10 | 52 | 82 | 12 | 36 | 80 | 4 | 6 | 60 | ∞ | 0 | 76 | 92 | 40 | 54 | 46 | 40 | 86 | 30 | 34 | 24 | 10 |
| 11 | 70 | 18 | 50 | 58 | 8 | 12 | 44 | 80 | 54 | 36 | 0 | 26 | 64 | 76 | 68 | 86 | 26 | 38 | 44 | ∞ | 11 |
| 12 | 0 | 8 | 26 | 22 | 18 | 12 | 22 | 24 | 30 | 10 | 20 | 0 | 12 | 14 | 10 | 30 | 6 | ∞ | 28 | 36 | 12 |
| 13 | 18 | 20 | 68 | 0 | 12 | 50 | 66 | 32 | 72 | 42 | 6 | 38 | 0 | 26 | 8 | 38 | 36 | 82 | ∞ | 76 | 13 |
| 14 | 92 | 92 | 28 | ∞ | 86 | 16 | 23 | 12 | 50 | 8 | 86 | 2 | 0 | 0 | 86 | 16 | 96 | 62 | 70 | 52 | 14 |
| 15 | 16 | 38 | 10 | 2 | 34 | 24 | 12 | 40 | 0 | 64 | 26 | 50 | ∞ | 18 | 0 | 46 | 46 | 26 | 20 | 32 | 15 |
| 16 | 50 | 90 | 34 | 38 | 80 | 14 | 28 | 76 | 44 | 10 | 78 | 8 | 40 | 58 | ∞ | 0 | 4 | 28 | 34 | 26 | 16 |
| 17 | ∞ | 6 | 36 | 4 | 10 | 24 | 30 | 24 | 44 | 20 | 18 | 10 | 10 | 16 | 12 | 30 | 0 | 4 | 0 | 8 | 17 |
| 18 | 58 | 30 | 72 | 54 | 18 | 50 | 62 | 60 | 74 | 44 | 20 | 42 | 28 | ∞ | 54 | 72 | 38 | 0 | 10 | 80 | 18 |
| 19 | 22 | 18 | 22 | 10 | 66 | 8 | 30 | ∞ | 36 | 0 | 66 | 18 | 12 | 28 | 18 | 18 | 22 | 2 | 0 | 12 | 19 |
| 20 | 48 | 38 | 8 | 22 | 32 | 68 | 12 | 64 | 0 | 54 | 24 | 50 | 30 | 60 | 42 | ∞ | 42 | 10 | 12 | 0 | 20 |
| | 1 | 2 | 3 | 4 | 5 | 6 | 7 | 8 | 9 | 10 | 11 | 12 | 13 | 14 | 15 | 16 | 17 | 18 | 19 | 20 | |

We start with (12 20):-36. We have very few choices to start a new chain with. None of them are good. We thus see if can obtain a better option using the next smallest arc (15 20). Going in a clockwise direction, 15 is close to 20. Thus, we have many choices. In particular, (20 9) is an excellent choice because (9 15) has a value of 4. Thus, we can obtain a negatively-valued 3-cycle:

Let $s_8 = (15^{-32} \ 20^0 \ 9^4) : -28$. Our new arcs are (15 11), (20 10), (9 16)..

H₈ = (1³ 17⁴⁷ 4²⁷ 14¹ 18⁹ 12³⁹ 5²⁹ 2⁹ 3¹¹ 9²³ 16³ 20¹⁹ 10⁷ 6¹ 7²⁹ 8¹ 19²¹ 13¹ 15¹ 11¹³): 294

$$H_8^{-1}M$$

|    | 17 | 3  | 9  | 14 | 2  | 7  | 8  | 19 | 16 | 6  | 1  | 5  | 15 | 18 | 11 | 20 | 4  | 12 | 13 | 10 |    |
|----|----|----|----|----|----|----|----|----|----|----|----|----|----|----|----|----|----|----|----|----|----|
|    | 1  | 2  | 3  | 4  | 5  | 6  | 7  | 8  | 9  | 10 | 11 | 12 | 13 | 14 | 15 | 16 | 17 | 18 | 19 | 20 |    |
| 1  | 0  | 2  | 20 | 84 | 0  | 14 | 16 | 10 | 96 | 10 | ∞  | 2  | 88 | 4  | 60 | 14 | 4  | 64 | 76 | 26 | 1  |
| 2  | 64 | 0  | 30 | 52 | ∞  | 22 | 24 | 70 | 62 | 12 | 88 | 10 | 58 | 68 | 54 | 80 | 4  | 31 | 38 | 40 | 2  |
| 3  | 52 | ∞  | 0  | 30 | 68 | 86 | 88 | 60 | 39 | 82 | 62 | 80 | 40 | 58 | 18 | 66 | 70 | 22 | 28 | 12 | 3  |
| 4  | 26 | 14 | 34 | 0  | 28 | 26 | 32 | 42 | 12 | 20 | 8  | 16 | 2  | 36 | 24 | 54 | ∞  | 16 | 42 | 46 | 4  |
| 5  | 4  | 2  | 8  | 64 | 0  | 30 | 48 | 50 | 0  | 52 | 6  | ∞  | 68 | 20 | 18 | 2  | 10 | 58 | 12 | 54 | 5  |
| 6  | 32 | 82 | 6  | 18 | 78 | 0  | 2  | 48 | 30 | ∞  | 72 | 96 | 20 | 38 | 96 | 62 | 88 | 8  | 12 | 8  | 6  |
| 7  | 60 | 60 | 10 | 44 | 54 | ∞  | 0  | 28 | 54 | 16 | 54 | 18 | 50 | 68 | 24 | 22 | 70 | 30 | 48 | 20 | 7  |
| 8  | 86 | 96 | 0  | 58 | 90 | 18 | ∞  | 0  | 82 | 6  | 88 | 3  | 76 | 88 | 40 | 10 | 0  | 46 | 52 | 26 | 8  |
| 9  | 8  | 68 | ∞  | 76 | 64 | 14 | 10 | 26 | 0  | 20 | 46 | 74 | 12 | 20 | 64 | 30 | 70 | 66 | 70 | 4  | 9  |
| 10 | 52 | 82 | 12 | 36 | 80 | 4  | 6  | 60 | 46 | 0  | 76 | 92 | 40 | 54 | 24 | 40 | 86 | 30 | 34 | ∞  | 10 |
| 11 | 70 | 18 | 50 | 58 | 8  | 12 | 44 | 80 | 68 | 36 | 0  | 26 | 64 | 76 | ∞  | 86 | 26 | 38 | 44 | 54 | 11 |
| 12 | 0  | 8  | 26 | 22 | 18 | 12 | 22 | 24 | 10 | 10 | 20 | 0  | 12 | 14 | 36 | 30 | 6  | ∞  | 28 | 30 | 12 |
| 13 | 18 | 20 | 68 | 0  | 12 | 50 | 66 | 32 | 8  | 42 | 6  | 38 | 0  | 26 | 76 | 38 | 36 | 82 | ∞  | 72 | 13 |
| 14 | 92 | 92 | 28 | ∞  | 86 | 16 | 23 | 12 | 86 | 8  | 86 | 2  | 0  | 0  | 52 | 16 | 96 | 62 | 70 | 50 | 14 |
| 15 | 48 | 70 | 22 | 30 | 66 | 8  | 20 | 72 | 32 | 96 | 58 | 82 | ∞  | 50 | 0  | 78 | 78 | 6  | 12 | 32 | 15 |
| 16 | 50 | 90 | 34 | 38 | 80 | 14 | 28 | 76 | ∞  | 10 | 78 | 8  | 40 | 58 | 26 | 0  | 4  | 28 | 34 | 44 | 16 |
| 17 | ∞  | 6  | 36 | 4  | 10 | 24 | 30 | 24 | 12 | 20 | 18 | 10 | 10 | 16 | 8  | 30 | 0  | 4  | 0  | 44 | 17 |
| 18 | 58 | 30 | 72 | 54 | 18 | 50 | 62 | 60 | 54 | 44 | 20 | 42 | 28 | ∞  | 80 | 72 | 38 | 0  | 10 | 74 | 18 |
| 19 | 22 | 18 | 22 | 10 | 66 | 8  | 30 | ∞  | 18 | 0  | 66 | 18 | 12 | 28 | 12 | 18 | 22 | 2  | 0  | 36 | 19 |
| 20 | 48 | 38 | 8  | 22 | 32 | 68 | 12 | 64 | 42 | 54 | 24 | 50 | 30 | 60 | 0  | ∞  | 42 | 10 | 12 | 0  | 20 |
|    | 1  | 2  | 3  | 4  | 5  | 6  | 7  | 8  | 9  | 10 | 11 | 12 | 13 | 14 | 15 | 16 | 17 | 18 | 19 | 20 |    |

We now start a new chain (12⁻²⁸ 19⁻¹² 15⁸²): 42. Our new arcs are (12  13),  (19  11),  (15  5).

$H_9 = (1^3\ 17^{47}\ 4^{27}\ 14^1\ 18^9\ 12^{11}\ 13^1\ 15^{83}\ 5^{29}\ 2^9\ 3^{11}\ 9^{23}\ 16^3\ 20^{19}\ 10^7\ 6^1\ 7^{29}\ 8^1\ 19^9\ 11^{13})$: 336

$$H_9^{-1}M$$

|    | 17 | 3  | 9  | 14 | 2  | 7  | 8  | 19 | 16 | 6  | 1  | 13 | 15 | 18 | 5  | 20 | 4  | 12 | 11 | 10 |    |
|----|----|----|----|----|----|----|----|----|----|----|----|----|----|----|----|----|----|----|----|----|----|
|    | 1  | 2  | 3  | 4  | 5  | 6  | 7  | 8  | 9  | 10 | 11 | 12 | 13 | 14 | 15 | 16 | 17 | 18 | 19 | 20 |    |
| 1  | 0  | 2  | 20 | 84 | 0  | 14 | 16 | 10 | 96 | 10 | ∞  | 76 | 88 | 4  | 2  | 14 | 4  | 64 | 60 | 26 | 1  |
| 2  | 64 | 0  | 30 | 52 | ∞  | 22 | 24 | 70 | 62 | 12 | 88 | 38 | 58 | 68 | 10 | 80 | 4  | 31 | 54 | 40 | 2  |
| 3  | 52 | ∞  | 0  | 30 | 68 | 86 | 88 | 60 | 39 | 82 | 62 | 28 | 40 | 58 | 80 | 66 | 70 | 22 | 18 | 12 | 3  |
| 4  | 26 | 14 | 34 | 0  | 28 | 26 | 32 | 42 | 12 | 20 | 8  | 42 | 2  | 36 | 16 | 54 | ∞  | 16 | 24 | 46 | 4  |
| 5  | 4  | 2  | 8  | 64 | 0  | 30 | 48 | 50 | 0  | 52 | 6  | 12 | 68 | 20 | ∞  | 2  | 10 | 58 | 18 | 54 | 5  |
| 6  | 32 | 82 | 6  | 18 | 78 | 0  | 2  | 48 | 30 | ∞  | 72 | 12 | 20 | 38 | 96 | 62 | 88 | 8  | 96 | 8  | 6  |
| 7  | 60 | 60 | 10 | 44 | 54 | ∞  | 0  | 28 | 54 | 16 | 54 | 48 | 50 | 68 | 18 | 22 | 70 | 30 | 24 | 20 | 7  |
| 8  | 86 | 96 | 0  | 58 | 90 | 18 | ∞  | 0  | 82 | 6  | 88 | 52 | 76 | 88 | 3  | 10 | 0  | 46 | 40 | 26 | 8  |
| 9  | 8  | 68 | ∞  | 76 | 64 | 14 | 10 | 26 | 0  | 20 | 46 | 70 | 12 | 20 | 74 | 30 | 70 | 66 | 64 | 4  | 9  |
| 10 | 52 | 82 | 12 | 36 | 80 | 4  | 6  | 60 | 46 | 0  | 76 | 34 | 40 | 54 | 92 | 40 | 86 | 30 | 24 | ∞  | 10 |
| 11 | 70 | 18 | 50 | 58 | 8  | 12 | 44 | 80 | 68 | 36 | 0  | 44 | 64 | 76 | 26 | 86 | 26 | 38 | ∞  | 54 | 11 |
| 12 | 28 | 20 | 64 | 6  | 10 | 40 | 60 | 52 | 10 | 38 | 8  | 0  | 16 | 42 | 28 | 58 | 22 | ∞  | 8  | 58 | 12 |
| 13 | 18 | 20 | 68 | 0  | 12 | 50 | 66 | 32 | 8  | 42 | 6  | ∞  | 0  | 26 | 38 | 38 | 36 | 82 | 76 | 72 | 13 |
| 14 | 92 | 92 | 28 | ∞  | 86 | 16 | 23 | 12 | 86 | 8  | 86 | 70 | 0  | 0  | 2  | 16 | 96 | 62 | 52 | 50 | 14 |
| 15 | 34 | 12 | 60 | 52 | 16 | 74 | 62 | 10 | 50 | 14 | 24 | 70 | ∞  | 32 | 0  | 4  | 4  | 76 | 82 | 50 | 15 |
| 16 | 50 | 90 | 34 | 38 | 80 | 14 | 28 | 76 | ∞  | 10 | 78 | 34 | 40 | 58 | 8  | 0  | 4  | 28 | 26 | 44 | 16 |
| 17 | ∞  | 6  | 36 | 4  | 10 | 24 | 30 | 24 | 12 | 20 | 18 | 0  | 10 | 16 | 10 | 30 | 0  | 4  | 8  | 44 | 17 |
| 18 | 58 | 30 | 72 | 54 | 18 | 50 | 62 | 60 | 54 | 44 | 20 | 10 | 28 | ∞  | 42 | 72 | 38 | 0  | 80 | 74 | 18 |
| 19 | 34 | 6  | 34 | 22 | 78 | 20 | 42 | ∞  | 30 | 12 | 78 | 12 | 24 | 40 | 6  | 30 | 34 | 10 | 0  | 48 | 19 |
| 20 | 48 | 38 | 8  | 22 | 32 | 68 | 12 | 64 | 42 | 54 | 24 | 12 | 30 | 60 | 50 | ∞  | 42 | 10 | 0  | 0  | 20 |
|    | 1  | 2  | 3  | 4  | 5  | 6  | 7  | 8  | 9  | 10 | 11 | 12 | 13 | 14 | 15 | 16 | 17 | 18 | 19 | 20 |    |

$(15^{-74}\ 6^2\ 7^{-18})$: -90. Our new arcs are (15 7), (6 8), (7 5).

$H_{10} = (1^3\ 17^{47}\ 4^{27}\ 14^1\ 18^9\ 12^{11}\ 13^1\ 15^9\ 7^{11}\ 5^{29}\ 2^9\ 3^{11}\ 9^{23}\ 16^3\ 20^{19}\ 10^7\ 6^3\ 8^1\ 19^{11}\ 11^{13})$: 248

$$H_{10}^{-1}M$$

|    | 17 | 3  | 9  | 14 | 2  | 8  | 5  | 19 | 16 | 6  | 1  | 13 | 15 | 18 | 7  | 20 | 4  | 12 | 11 | 10 |    |
|----|----|----|----|----|----|----|----|----|----|----|----|----|----|----|----|----|----|----|----|----|----|
|    | 1  | 2  | 3  | 4  | 5  | 6  | 7  | 8  | 9  | 10 | 11 | 12 | 13 | 14 | 15 | 16 | 17 | 18 | 19 | 20 |    |
| 1  | 0  | 2  | 20 | 84 | 0  | 16 | 2  | 10 | 96 | 10 | ∞  | 76 | 88 | 4  | 14 | 14 | 4  | 64 | 60 | 26 | 1  |
| 2  | 64 | 0  | 30 | 52 | ∞  | 24 | 10 | 70 | 62 | 12 | 88 | 38 | 58 | 68 | 22 | 80 | 4  | 31 | 54 | 40 | 2  |
| 3  | 52 | ∞  | 0  | 30 | 68 | 88 | 80 | 60 | 39 | 82 | 62 | 28 | 40 | 58 | 86 | 66 | 70 | 22 | 18 | 12 | 3  |
| 4  | 26 | 14 | 34 | 0  | 28 | 32 | 16 | 42 | 12 | 20 | 8  | 42 | 2  | 36 | 26 | 54 | ∞  | 16 | 24 | 46 | 4  |
| 5  | 4  | 2  | 8  | 64 | 0  | 48 | ∞  | 50 | 0  | 52 | 6  | 12 | 68 | 10 | 30 | 2  | 10 | 58 | 18 | 54 | 5  |
| 6  | 28 | 78 | 2  | 14 | 74 | 0  | 92 | 44 | 26 | ∞  | 68 | 8  | 16 | 34 | 4  | 58 | 84 | 4  | 92 | 4  | 6  |
| 7  | 78 | 78 | 28 | 62 | 72 | 0  | 0  | 10 | 72 | 2  | 72 | 66 | 68 | 86 | ∞  | 4  | 88 | 48 | 42 | 38 | 7  |
| 8  | 86 | 96 | 0  | 58 | 90 | ∞  | 3  | 0  | 82 | 6  | 88 | 52 | 76 | 88 | 18 | 10 | 0  | 46 | 40 | 26 | 8  |
| 9  | 8  | 68 | ∞  | 76 | 64 | 10 | 74 | 26 | 0  | 20 | 46 | 70 | 12 | 20 | 14 | 30 | 70 | 66 | 64 | 4  | 9  |
| 10 | 52 | 82 | 12 | 36 | 80 | 6  | 92 | 60 | 46 | 0  | 76 | 34 | 40 | 54 | 4  | 40 | 86 | 30 | 24 | ∞  | 10 |
| 11 | 70 | 18 | 50 | 58 | 8  | 44 | 26 | 80 | 68 | 36 | 0  | 44 | 64 | 76 | 12 | 86 | 26 | 38 | ∞  | 54 | 11 |
| 12 | 28 | 20 | 64 | 6  | 10 | 60 | 28 | 52 | 10 | 38 | 8  | 0  | 16 | 42 | 40 | 58 | 22 | ∞  | 8  | 58 | 12 |
| 13 | 18 | 20 | 68 | 0  | 12 | 66 | 38 | 32 | 8  | 42 | 6  | ∞  | 0  | 26 | 50 | 38 | 36 | 82 | 76 | 72 | 13 |
| 14 | 92 | 92 | 28 | ∞  | 86 | 23 | 2  | 12 | 86 | 8  | 86 | 70 | 0  | 0  | 16 | 16 | 96 | 62 | 52 | 50 | 14 |
| 15 | 40 | 62 | 14 | 22 | 58 | 12 | 74 | 64 | 24 | 88 | 50 | 4  | ∞  | 42 | 0  | 70 | 70 | 2  | 8  | 24 | 15 |
| 16 | 50 | 90 | 34 | 38 | 80 | 28 | 8  | 76 | ∞  | 10 | 78 | 34 | 40 | 58 | 14 | 0  | 4  | 28 | 26 | 44 | 16 |
| 17 | ∞  | 6  | 36 | 4  | 10 | 30 | 10 | 24 | 12 | 20 | 18 | 0  | 10 | 16 | 24 | 30 | 0  | 4  | 8  | 44 | 17 |
| 18 | 58 | 30 | 72 | 54 | 18 | 62 | 42 | 60 | 54 | 44 | 20 | 10 | 28 | ∞  | 50 | 72 | 38 | 0  | 80 | 74 | 18 |
| 19 | 34 | 6  | 34 | 22 | 78 | 42 | 6  | ∞  | 30 | 12 | 78 | 12 | 24 | 40 | 20 | 30 | 34 | 10 | 0  | 48 | 19 |
| 20 | 48 | 38 | 8  | 22 | 32 | 12 | 50 | 64 | 42 | 54 | 24 | 12 | 30 | 60 | 68 | ∞  | 42 | 10 | 0  | 0  | 20 |
|    | 1  | 2  | 3  | 4  | 5  | 6  | 7  | 8  | 9  | 10 | 11 | 12 | 13 | 14 | 15 | 16 | 17 | 18 | 19 | 20 |    |

We know that three points on a cycle $H_i$ can form an $H_i$-admissible 3-cycle if each one can go clockwise to the following one without backtracking.

$s_{11} = (9^{-20}\ 10^4\ 15^{24})$: 8. Our new arcs are (9 6), (10 7), (15 16). .

H$_{11}$ = (1$^3$ 17$^{47}$ 4$^{27}$ 14$^1$ 18$^9$ 12$^{11}$ 13$^1$ 15$^{33}$ 16$^3$ 20$^{19}$ 10$^{11}$ 7$^{11}$ 5$^{29}$ 2$^9$ 3$^{11}$ 9$^3$ 6$^3$ 8$^1$ 19$^9$ 11$^{13}$) : 254.

$H_{11}^{-1}M$

|    | 17 | 3  | 9  | 14 | 2  | 8  | 5  | 19 | 6  | 7  | 1  | 13 | 15 | 18 | 16 | 20 | 4  | 12 | 11 | 10 |    |
|----|----|----|----|----|----|----|----|----|----|----|----|----|----|----|----|----|----|----|----|----|----|
|    | 1  | 2  | 3  | 4  | 5  | 6  | 7  | 8  | 9  | 10 | 11 | 12 | 13 | 14 | 15 | 16 | 17 | 18 | 19 | 20 |    |
| 1  | 0  | 2  | 20 | 84 | 0  | 16 | 2  | 10 | 10 | 14 | ∞  | 76 | 88 | 4  | 96 | 14 | 4  | 64 | 60 | 26 | 1  |
| 2  | 64 | 0  | 30 | 52 | ∞  | 24 | 10 | 70 | 12 | 22 | 88 | 38 | 58 | 68 | 62 | 80 | 4  | 31 | 54 | 40 | 2  |
| 3  | 52 | ∞  | 0  | 30 | 68 | 88 | 80 | 60 | 82 | 86 | 62 | 28 | 40 | 58 | 39 | 66 | 70 | 22 | 18 | 12 | 3  |
| 4  | 26 | 14 | 34 | 0  | 28 | 32 | 16 | 42 | 20 | 26 | 8  | 42 | 2  | 36 | 12 | 54 | ∞  | 16 | 24 | 46 | 4  |
| 5  | 4  | 2  | 8  | 64 | 0  | 48 | ∞  | 50 | 52 | 30 | 6  | 12 | 68 | 10 | 0  | 2  | 10 | 58 | 18 | 54 | 5  |
| 6  | 28 | 78 | 2  | 14 | 74 | 0  | 92 | 44 | ∞  | 4  | 68 | 8  | 16 | 34 | 26 | 58 | 84 | 4  | 92 | 4  | 6  |
| 7  | 78 | 78 | 28 | 62 | 72 | 0  | 0  | 10 | 2  | ∞  | 72 | 66 | 68 | 86 | 72 | 4  | 88 | 48 | 42 | 38 | 7  |
| 8  | 86 | 96 | 0  | 58 | 90 | ∞  | 3  | 0  | 6  | 18 | 88 | 52 | 76 | 88 | 82 | 10 | 0  | 46 | 40 | 26 | 8  |
| 9  | 28 | 88 | ∞  | 96 | 84 | 10 | 94 | 46 | 0  | 6  | 66 | 90 | 8  | 40 | 20 | 50 | 90 | 86 | 84 | 16 | 9  |
| 10 | 48 | 78 | 8  | 32 | 76 | 2  | 88 | 56 | 4  | 0  | 72 | 30 | 36 | 50 | 42 | 36 | 82 | 26 | 20 | ∞  | 10 |
| 11 | 70 | 18 | 50 | 58 | 8  | 44 | 26 | 80 | 36 | 12 | 0  | 44 | 64 | 76 | 68 | 86 | 26 | 38 | ∞  | 54 | 11 |
| 12 | 28 | 20 | 64 | 6  | 10 | 60 | 28 | 52 | 38 | 40 | 8  | 0  | 16 | 42 | 18 | 58 | 22 | ∞  | 8  | 58 | 12 |
| 13 | 18 | 20 | 68 | 0  | 12 | 66 | 38 | 32 | 42 | 50 | 6  | ∞  | 0  | 26 | 8  | 38 | 36 | 82 | 76 | 72 | 13 |
| 14 | 92 | 92 | 28 | ∞  | 86 | 23 | 2  | 12 | 8  | 16 | 86 | 70 | 0  | 0  | 86 | 16 | 96 | 62 | 52 | 50 | 14 |
| 15 | 16 | 38 | 10 | 2  | 34 | 3  | 50 | 40 | 64 | 24 | 26 | 20 | ∞  | 18 | 24 | 46 | 46 | 26 | 32 | 0  | 15 |
| 16 | 50 | 90 | 34 | 38 | 80 | 28 | 8  | 76 | 10 | 14 | 78 | 34 | 40 | 58 | ∞  | 0  | 4  | 28 | 26 | 44 | 16 |
| 17 | ∞  | 6  | 36 | 4  | 10 | 30 | 10 | 24 | 20 | 24 | 18 | 0  | 10 | 16 | 12 | 30 | 0  | 4  | 8  | 44 | 17 |
| 18 | 58 | 30 | 72 | 54 | 18 | 62 | 42 | 60 | 44 | 50 | 20 | 10 | 28 | ∞  | 54 | 72 | 38 | 0  | 80 | 74 | 18 |
| 19 | 34 | 6  | 34 | 22 | 78 | 42 | 6  | ∞  | 12 | 20 | 78 | 12 | 24 | 40 | 30 | 30 | 34 | 10 | 0  | 48 | 19 |
| 20 | 48 | 38 | 8  | 22 | 32 | 12 | 50 | 64 | 54 | 68 | 24 | 12 | 30 | 60 | 42 | ∞  | 42 | 10 | 0  | 0  | 20 |
|    | 1  | 2  | 3  | 4  | 5  | 6  | 7  | 8  | 9  | 10 | 11 | 12 | 13 | 14 | 15 | 16 | 17 | 18 | 19 | 20 |    |

Let s$_{12}$ = (12$^{-12}$ 4$^{-16}$ 15): - 2.

Our new arcs are: (12 14), (4 12), (15 16)..

$H_{12} = (1^3\ 17^{47}\ 4^{27}\ 14^1\ 18^{19}\ 13^1\ 15^7\ 12^{29}\ 16^3\ 20^{19}\ 10^{11}\ 7^{11}\ 5^{29}\ 2^9\ 3^{11}\ 9^3\ 6^3\ 8^1\ 19^9\ 11^{13}): .266$

$$H_{12}^{-1}M$$

|    | 17 | 3 | 9 | 14 | 2 | 8 | 5 | 19 | 6 | 7 | 1 | 16 | 15 | 18 | 12 | 20 | 4 | 13 | 11 | 10 |    |
|----|----|----|----|----|----|----|----|----|----|----|----|----|----|----|----|----|----|----|----|----|----|
|    | 1 | 2 | 3 | 4 | 5 | 6 | 7 | 8 | 9 | 10 | 11 | 12 | 13 | 14 | 15 | 16 | 17 | 18 | 19 | 20 |    |
| 1  | 0 | 2 | 20 | 84 | 0 | 16 | 2 | 10 | 10 | 14 | ∞ | 96 | 88 | 4 | 64 | 14 | 4 | 76 | 60 | 26 | 1 |
| 2  | 64 | 0 | 30 | 52 | ∞ | 24 | 10 | 70 | 12 | 22 | 88 | 62 | 58 | 68 | 31 | 80 | 4 | 38 | 54 | 40 | 2 |
| 3  | 52 | ∞ | 0 | 30 | 68 | 88 | 80 | 60 | 82 | 86 | 62 | 39 | 40 | 58 | 22 | 66 | 70 | 28 | 18 | 12 | 3 |
| 4  | 26 | 14 | 34 | 0 | 28 | 32 | 16 | 42 | 20 | 26 | 8 | 12 | 2 | 36 | 16 | 54 | ∞ | 42 | 24 | 46 | 4 |
| 5  | 4 | 2 | 8 | 64 | 0 | 48 | ∞ | 50 | 52 | 30 | 6 | 0 | 68 | 10 | 58 | 2 | 10 | 12 | 18 | 54 | 5 |
| 6  | 26 | 76 | 0 | 12 | 72 | 0 | 92 | 44 | ∞ | 4 | 68 | 26 | 16 | 34 | 4 | 58 | 84 | 8 | 92 | 4 | 6 |
| 7  | 78 | 78 | 28 | 62 | 72 | 0 | 0 | 10 | 2 | ∞ | 72 | 72 | 68 | 86 | 48 | 4 | 88 | 66 | 42 | 38 | 7 |
| 8  | 86 | 96 | 0 | 58 | 90 | ∞ | 3 | 0 | 6 | 18 | 88 | 82 | 76 | 88 | 46 | 10 | 0 | 52 | 40 | 26 | 8 |
| 9  | 28 | 88 | ∞ | 96 | 84 | 10 | 94 | 46 | 0 | 6 | 66 | 20 | 8 | 40 | 86 | 50 | 90 | 90 | 84 | 16 | 9 |
| 10 | 48 | 78 | 8 | 32 | 76 | 2 | 88 | 56 | 4 | 0 | 72 | 42 | 36 | 50 | 26 | 36 | 82 | 30 | 20 | ∞ | 10 |
| 11 | 70 | 18 | 50 | 58 | 8 | 44 | 26 | 80 | 36 | 12 | 0 | 68 | 64 | 76 | 38 | 86 | 26 | 44 | ∞ | 54 | 11 |
| 12 | 10 | 2 | 34 | 12 | 8 | 32 | 10 | 34 | 20 | 22 | 10 | 0 | 2 | 24 | ∞ | 40 | 4 | 18 | 26 | 48 | 12 |
| 13 | 18 | 20 | 68 | 0 | 12 | 66 | 38 | 32 | 42 | 50 | 6 | 8 | 0 | 26 | 82 | 38 | 36 | ∞ | 76 | 72 | 13 |
| 14 | 92 | 92 | 28 | ∞ | 86 | 23 | 2 | 12 | 8 | 16 | 86 | 86 | 0 | 0 | 62 | 16 | 96 | 70 | 52 | 50 | 14 |
| 15 | 42 | 64 | 16 | 24 | 60 | 14 | 76 | 66 | 90 | 2 | 52 | 26 | ∞ | 44 | 0 | 72 | 72 | 6 | 6 | 26 | 15 |
| 16 | 50 | 90 | 34 | 38 | 80 | 28 | 8 | 76 | 10 | 14 | 78 | ∞ | 40 | 58 | 28 | 0 | 4 | 34 | 26 | 44 | 16 |
| 17 | ∞ | 6 | 36 | 4 | 10 | 30 | 10 | 24 | 20 | 24 | 18 | 12 | 10 | 16 | 4 | 30 | 0 | 0 | 8 | 44 | 17 |
| 18 | 48 | 20 | 62 | 44 | 8 | 52 | 32 | 50 | 34 | 40 | 10 | 44 | 18 | ∞ | 10 | 62 | 28 | 0 | 70 | 64 | 18 |
| 19 | 34 | 6 | 34 | 22 | 78 | 42 | 6 | ∞ | 12 | 20 | 78 | 30 | 24 | 40 | 10 | 30 | 34 | 12 | 0 | 48 | 19 |
| 20 | 48 | 38 | 8 | 22 | 32 | 12 | 50 | 64 | 54 | 68 | 24 | 42 | 30 | 60 | 10 | ∞ | 42 | 12 | 0 | 0 | 20 |
|    | 1 | 2 | 3 | 4 | 5 | 6 | 7 | 8 | 9 | 10 | 11 | 12 | 13 | 14 | 15 | 16 | 17 | 18 | 19 | 20 |    |

The purpose of the chain is to continually eliminate the last arc of the previous 3-cycle when the sum of the values of the first two arcs is negative. In this way, if we can obtain a negatively-valued 3-cycle in which the last arc is comparatively small (or negative), we would reduce the value of the tour (n-cycle).

$(17^{-10}\ 5^2\ 2^4)$: -4 . Our new arcs are (17 2), (5 3), (2 4).

$H_{13} = (1^3\ 17^{37}\ 2^{13}\ 4^{27}\ 14^1\ 18^{19}\ 13^1\ 15^7\ 12^{29}\ 16^3\ 20^{19}\ 10^{11}\ 7^{11}\ 5^{31}\ 3^{11}\ 9^3\ 6^3\ 8^1\ 19^9\ 11^{13})$: 252

$$H_{13}^{-1}M$$

|  | 17 | 4 | 9 | 14 | 3 | 8 | 5 | 19 | 6 | 7 | 1 | 16 | 15 | 18 | 12 | 20 | 2 | 13 | 11 | 10 |  |
|---|---|---|---|---|---|---|---|---|---|---|---|---|---|---|---|---|---|---|---|---|---|
|  | 1 | 2 | 3 | 4 | 5 | 6 | 7 | 8 | 9 | 10 | 11 | 12 | 13 | 14 | 15 | 16 | 17 | 18 | 19 | 20 |  |
| 1 | 0 | 4 | 20 | 84 | 2 | 16 | 2 | 10 | 10 | 14 | ∞ | 96 | 88 | 4 | 64 | 14 | 0 | 76 | 60 | 26 | 1 |
| 2 | 56 | 0 | 26 | 48 | 18 | 20 | 6 | 66 | 8 | 18 | 84 | 58 | 54 | 64 | 27 | 76 | ∞ | 34 | 50 | 36 | 2 |
| 3 | 52 | 70 | 0 | 30 | ∞ | 88 | 80 | 60 | 82 | 86 | 62 | 39 | 40 | 58 | 22 | 66 | 68 | 28 | 18 | 12 | 3 |
| 4 | 26 | ∞ | 34 | 0 | 14 | 32 | 16 | 42 | 20 | 26 | 8 | 12 | 2 | 36 | 16 | 54 | 28 | 42 | 24 | 46 | 4 |
| 5 | 2 | 6 | 48 | 66 | 0 | 46 | ∞ | 10 | 18 | 28 | 8 | 20 | 22 | 8 | 56 | 16 | 2 | 84 | 72 | 74 | 5 |
| 6 | 26 | 84 | 0 | 12 | 76 | 0 | 92 | 44 | ∞ | 4 | 68 | 26 | 16 | 34 | 4 | 58 | 72 | 8 | 92 | 4 | 6 |
| 7 | 78 | 88 | 28 | 62 | 78 | 0 | 0 | 10 | 2 | ∞ | 72 | 72 | 68 | 86 | 48 | 4 | 72 | 66 | 42 | 38 | 7 |
| 8 | 86 | 0 | 0 | 58 | 96 | ∞ | 3 | 0 | 6 | 18 | 88 | 82 | 76 | 88 | 46 | 10 | 90 | 52 | 40 | 26 | 8 |
| 9 | 28 | 90 | ∞ | 96 | 88 | 10 | 94 | 46 | 0 | 6 | 66 | 20 | 8 | 40 | 86 | 50 | 84 | 90 | 84 | 16 | 9 |
| 10 | 48 | 82 | 8 | 32 | 78 | 2 | 88 | 56 | 4 | 0 | 72 | 42 | 36 | 50 | 26 | 36 | 76 | 30 | 20 | ∞ | 10 |
| 11 | 70 | 26 | 50 | 58 | 18 | 44 | 26 | 80 | 36 | 12 | 0 | 68 | 64 | 76 | 38 | 86 | 8 | 44 | ∞ | 54 | 11 |
| 12 | 10 | 4 | 34 | 12 | 2 | 32 | 10 | 34 | 20 | 22 | 10 | 0 | 2 | 24 | ∞ | 40 | 8 | 18 | 26 | 48 | 12 |
| 13 | 18 | 36 | 68 | 0 | 20 | 66 | 38 | 32 | 42 | 50 | 6 | 8 | 0 | 26 | 82 | 38 | 12 | ∞ | 76 | 72 | 13 |
| 14 | 92 | 96 | 28 | ∞ | 92 | 23 | 2 | 12 | 8 | 16 | 86 | 86 | 0 | 0 | 62 | 16 | 86 | 70 | 52 | 50 | 14 |
| 15 | 42 | 72 | 16 | 24 | 64 | 14 | 76 | 66 | 90 | 2 | 52 | 26 | ∞ | 44 | 0 | 72 | 60 | 6 | 6 | 26 | 15 |
| 16 | 50 | 4 | 34 | 38 | 90 | 28 | 8 | 76 | 10 | 14 | 78 | ∞ | 40 | 58 | 28 | 0 | 80 | 34 | 26 | 44 | 16 |
| 17 | ∞ | 0 | 46 | 14 | 6 | 40 | 20 | 34 | 30 | 34 | 0 | 22 | 20 | 26 | 19 | 40 | 0 | 10 | 2 | 54 | 17 |
| 18 | 48 | 28 | 62 | 44 | 20 | 52 | 32 | 50 | 34 | 40 | 10 | 44 | 18 | ∞ | 10 | 62 | 8 | 0 | 70 | 64 | 18 |
| 19 | 34 | 34 | 34 | 22 | 6 | 42 | 6 | ∞ | 12 | 20 | 78 | 30 | 24 | 40 | 10 | 30 | 78 | 12 | 0 | 48 | 19 |
| 20 | 48 | 42 | 8 | 22 | 38 | 12 | 50 | 64 | 54 | 68 | 24 | 42 | 30 | 60 | 10 | ∞ | 32 | 12 | 0 | 0 | 20 |
|  | 1 | 2 | 3 | 4 | 5 | 6 | 7 | 8 | 9 | 10 | 11 | 12 | 13 | 14 | 15 | 16 | 17 | 18 | 19 | 20 |  |

$s_{14} = (5^{-22}\ 13^8\ 12^2)$: -12 . Our new arcs are (5  15), (13  16), (12  3)

$H_{14} = (1^3 \ 17^{37} \ 2^{13} \ 4^{27} \ 14^1 \ 18^{19} \ 13^9 \ 16^3 \ 20^{19} \ 10^{11} \ 7^{11} \ 5^9 \ 15^7 \ 12^{31} \ 3^{11} \ 9^3 \ 6^3 \ 8^1 \ 19^9 \ 11^{13})$: 240

$$H_{14}^{-1}M$$

|   | 17 | 4 | 9 | 14 | 15 | 8 | 5 | 19 | 6 | 7 | 1 | 3 | 16 | 18 | 12 | 20 | 2 | 13 | 11 | 10 |   |
|---|---|---|---|---|---|---|---|---|---|---|---|---|---|---|---|---|---|---|---|---|---|
|   | 1 | 2 | 3 | 4 | 5 | 6 | 7 | 8 | 9 | 10 | 11 | 12 | 13 | 14 | 15 | 16 | 17 | 18 | 19 | 20 |   |
| 1 | 0 | 4 | 20 | 84 | 88 | 16 | 2 | 10 | 10 | 14 | ∞ | 2 | 96 | 4 | 64 | 14 | 0 | 76 | 60 | 26 | 1 |
| 2 | 56 | 0 | 26 | 48 | 54 | 20 | 6 | 66 | 8 | 18 | 84 | 18 | 58 | 64 | 27 | 76 | ∞ | 34 | 50 | 36 | 2 |
| 3 | 52 | 70 | 0 | 30 | 40 | 88 | 80 | 60 | 82 | 86 | 62 | ∞ | 39 | 58 | 22 | 66 | 68 | 28 | 18 | 12 | 3 |
| 4 | 26 | ∞ | 34 | 0 | 2 | 32 | 16 | 42 | 20 | 26 | 8 | 14 | 12 | 36 | 16 | 54 | 28 | 42 | 24 | 46 | 4 |
| 5 | 46 | 28 | 70 | 88 | 0 | 68 | ∞ | 32 | 40 | 50 | 14 | 22 | 2 | 30 | 78 | 38 | 20 | 84 | 72 | 38 | 5 |
| 6 | 26 | 84 | 0 | 12 | 16 | 0 | 92 | 44 | ∞ | 4 | 68 | 76 | 26 | 34 | 4 | 58 | 72 | 8 | 92 | 4 | 6 |
| 7 | 78 | 88 | 28 | 62 | 68 | 0 | 0 | 10 | 2 | ∞ | 72 | 78 | 72 | 86 | 48 | 4 | 72 | 66 | 42 | 38 | 7 |
| 8 | 86 | 0 | 0 | 58 | 76 | ∞ | 3 | 0 | 6 | 18 | 88 | 96 | 82 | 88 | 46 | 10 | 90 | 52 | 40 | 26 | 8 |
| 9 | 28 | 90 | ∞ | 96 | 8 | 10 | 94 | 46 | 0 | 6 | 66 | 88 | 20 | 40 | 86 | 50 | 84 | 90 | 84 | 16 | 9 |
| 10 | 48 | 82 | 8 | 32 | 36 | 2 | 88 | 56 | 4 | 0 | 72 | 78 | 42 | 50 | 26 | 36 | 76 | 30 | 20 | ∞ | 10 |
| 11 | 70 | 26 | 50 | 58 | 64 | 44 | 26 | 80 | 36 | 12 | 0 | 18 | 68 | 76 | 38 | 86 | 8 | 44 | ∞ | 54 | 11 |
| 12 | 8 | 2 | 32 | 14 | 4 | 30 | 8 | 32 | 18 | 20 | 12 | 0 | 2 | 22 | ∞ | 38 | 10 | 20 | 28 | 46 | 12 |
| 13 | 10 | 28 | 60 | 8 | 8 | 58 | 30 | 24 | 34 | 22 | 2 | 12 | 0 | 18 | 74 | 30 | 4 | ∞ | 68 | 64 | 13 |
| 14 | 92 | 96 | 28 | ∞ | 0 | 23 | 2 | 12 | 8 | 16 | 86 | 92 | 86 | 0 | 62 | 16 | 86 | 70 | 52 | 50 | 14 |
| 15 | 42 | 72 | 16 | 24 | ∞ | 14 | 76 | 66 | 90 | 2 | 52 | 64 | 26 | 44 | 0 | 72 | 60 | 6 | 6 | 26 | 15 |
| 16 | 50 | 4 | 34 | 38 | 40 | 28 | 8 | 76 | 10 | 14 | 78 | 90 | ∞ | 58 | 28 | 0 | 80 | 34 | 26 | 44 | 16 |
| 17 | ∞ | 0 | 46 | 14 | 20 | 40 | 20 | 34 | 30 | 34 | 0 | 6 | 22 | 26 | 19 | 40 | 0 | 10 | 2 | 54 | 17 |
| 18 | 48 | 28 | 62 | 44 | 18 | 52 | 32 | 50 | 34 | 40 | 10 | 20 | 44 | ∞ | 10 | 62 | 8 | 0 | 70 | 64 | 18 |
| 19 | 34 | 34 | 34 | 22 | 24 | 42 | 6 | ∞ | 12 | 20 | 78 | 6 | 30 | 40 | 10 | 30 | 78 | 12 | 0 | 48 | 19 |
| 20 | 48 | 42 | 8 | 22 | 30 | 12 | 50 | 64 | 54 | 68 | 24 | 38 | 42 | 60 | 10 | ∞ | 32 | 12 | 0 | 0 | 20 |
|   | 1 | 2 | 3 | 4 | 5 | 6 | 7 | 8 | 9 | 10 | 11 | 12 | 13 | 14 | 15 | 16 | 17 | 18 | 19 | 20 |   |

$s_{15} = (12^{-20} \ 18^{-10} \ 15^{64})$: 34 . Our new arcs are (12 13), (18 12), (15 3).

H$_{15}$ = (1³ 17³⁷ 2¹³ 4²⁷ 14¹ 18²⁰ 12¹¹ 13⁹ 16³ 20¹¹ 10¹¹ 7¹¹ 5⁹ 15⁷¹ 3¹¹ 9⁶ 6³ 8¹ 19⁹ 11¹³): 280

$$H_{15}^{-1}M$$

| I | 17 | 4 | 9 | 14 | 15 | 8 | 5 | 19 | 6 | 7 | 1 | 13 | 16 | 18 | 3 | 20 | 2 | 12 | 11 | 10 | |
|---|---|---|---|---|---|---|---|---|---|---|---|---|---|---|---|---|---|---|---|---|---|
| | 1 | 2 | 3 | 4 | 5 | 6 | 7 | 8 | 9 | 10 | 11 | 12 | 13 | 14 | 15 | 16 | 17 | 18 | 19 | 20 | |
| 1 | 0 | 4 | 20 | 84 | 88 | 16 | 2 | 10 | 10 | 14 | ∞ | 76 | 96 | 4 | 2 | 14 | 0 | 64 | 60 | 26 | 1 |
| 2 | 56 | 0 | 26 | 48 | 54 | 20 | 6 | 66 | 8 | 18 | 84 | 34 | 58 | 64 | 18 | 76 | ∞ | 27 | 50 | 36 | 2 |
| 3 | 52 | 70 | 0 | 30 | 40 | 88 | 80 | 60 | 82 | 86 | 62 | 28 | 39 | 58 | ∞ | 66 | 68 | 22 | 18 | 12 | 3 |
| 4 | 26 | ∞ | 34 | 0 | 2 | 32 | 16 | 42 | 20 | 26 | 8 | 42 | 12 | 36 | 14 | 54 | 28 | 16 | 24 | 46 | 4 |
| 5 | 46 | 28 | 70 | 88 | 0 | 68 | ∞ | 32 | 40 | 50 | 14 | 84 | 2 | 30 | 22 | 38 | 20 | 78 | 72 | 38 | 5 |
| 6 | 26 | 84 | 0 | 12 | 16 | 0 | 92 | 44 | ∞ | 4 | 68 | 8 | 26 | 34 | 76 | 58 | 72 | 4 | 92 | 4 | 6 |
| 7 | 78 | 88 | 28 | 62 | 68 | 0 | 0 | 10 | 2 | ∞ | 72 | 66 | 72 | 86 | 78 | 4 | 72 | 48 | 42 | 38 | 7 |
| 8 | 86 | 0 | 0 | 58 | 76 | ∞ | 3 | 0 | 6 | 18 | 88 | 52 | 82 | 88 | 96 | 10 | 90 | 46 | 40 | 26 | 8 |
| 9 | 28 | 90 | ∞ | 96 | 8 | 10 | 94 | 46 | 0 | 6 | 66 | 90 | 20 | 40 | 88 | 50 | 84 | 86 | 84 | 16 | 9 |
| 10 | 48 | 82 | 8 | 32 | 36 | 2 | 88 | 56 | 4 | 0 | 72 | 30 | 42 | 50 | 78 | 36 | 76 | 26 | 20 | ∞ | 10 |
| 11 | 70 | 26 | 50 | 58 | 64 | 44 | 26 | 80 | 36 | 12 | 0 | 44 | 68 | 76 | 18 | 86 | 8 | 38 | ∞ | 54 | 11 |
| 12 | 28 | 22 | 52 | 6 | 16 | 50 | 28 | 52 | 38 | 40 | 8 | 0 | 18 | 42 | 20 | 58 | 10 | ∞ | 8 | 66 | 12 |
| 13 | 10 | 28 | 60 | 8 | 8 | 58 | 30 | 24 | 34 | 22 | 2 | ∞ | 0 | 18 | 12 | 30 | 4 | 74 | 68 | 64 | 13 |
| 14 | 92 | 96 | 28 | ∞ | 0 | 23 | 2 | 12 | 8 | 16 | 86 | 70 | 86 | 0 | 92 | 16 | 86 | 62 | 52 | 50 | 14 |
| 15 | 22 | 8 | 48 | 40 | ∞ | 50 | 12 | 2 | 26 | 58 | 12 | 58 | 38 | 20 | 0 | 8 | 4 | 64 | 70 | 38 | 15 |
| 16 | 50 | 4 | 34 | 38 | 40 | 28 | 8 | 76 | 10 | 14 | 78 | 34 | ∞ | 58 | 90 | 0 | 80 | 28 | 26 | 44 | 16 |
| 17 | ∞ | 0 | 46 | 14 | 20 | 40 | 20 | 34 | 30 | 34 | 0 | 10 | 22 | 26 | 6 | 40 | 0 | 19 | 2 | 54 | 17 |
| 18 | 58 | 38 | 72 | 54 | 28 | 62 | 42 | 60 | 44 | 50 | 20 | 10 | 54 | ∞ | 30 | 72 | 18 | 0 | 80 | 74 | 18 |
| 19 | 34 | 34 | 34 | 22 | 24 | 42 | 6 | ∞ | 12 | 20 | 78 | 12 | 30 | 40 | 6 | 30 | 78 | 10 | 0 | 48 | 19 |
| 20 | 48 | 42 | 8 | 22 | 30 | 12 | 50 | 64 | 54 | 68 | 24 | 12 | 42 | 60 | 38 | ∞ | 32 | 10 | 0 | 0 | 20 |
| | 1 | 2 | 3 | 4 | 5 | 6 | 7 | 8 | 9 | 10 | 11 | 12 | 13 | 14 | 15 | 16 | 17 | 18 | 19 | 20 | |

s$_{16}$ = (15$^{-70}$ 19⁸ 7⁶⁸): 6.  Our new arcs are (15  11), (19  5), (7  3)

$H_{16}$ = ($1^3$ $17^{37}$ $2^{13}$ $4^{27}$ $14^1$ $18^9$ $12^{11}$ $13^9$ $16^3$ $20^{19}$ $10^{11}$ $7^{89}$ $3^{11}$ $9^3$ $6^3$ $8^1$ $19^{17}$ $5^9$ $15^1$ $11^{13}$ ): 290

$$H_{16}^{-1}M$$

| II | 17 | 4 | 9 | 14 | 15 | 8 | 3 | 19 | 6 | 7 | 1 | 13 | 16 | 18 | 11 | 20 | 2 | 12 | 5 | 10 | |
|---|---|---|---|---|---|---|---|---|---|---|---|---|---|---|---|---|---|---|---|---|---|
| | 1 | 2 | 3 | 4 | 5 | 6 | 7 | 8 | 9 | 10 | 11 | 12 | 13 | 14 | 15 | 16 | 17 | 18 | 19 | 20 | |
| 1 | 0 | 4 | 20 | 84 | 88 | 16 | 2 | 10 | 10 | 14 | ∞ | 76 | 96 | 4 | 60 | 14 | 0 | 64 | 2 | 26 | 1 |
| 2 | 56 | 0 | 26 | 48 | 54 | 20 | 18 | 66 | 8 | 18 | 84 | 34 | 58 | 64 | 50 | 76 | ∞ | 27 | 6 | 36 | 2 |
| 3 | 52 | 70 | 0 | 30 | 40 | 88 | ∞ | 60 | 82 | 86 | 62 | 28 | 39 | 58 | 18 | 66 | 68 | 22 | 80 | 12 | 3 |
| 4 | 26 | ∞ | 34 | 0 | 2 | 32 | 14 | 42 | 20 | 26 | 8 | 42 | 12 | 36 | 24 | 54 | 28 | 16 | 16 | 46 | 4 |
| 5 | 46 | 28 | 70 | 88 | 0 | 68 | 22 | 32 | 40 | 50 | 14 | 84 | 2 | 30 | 72 | 38 | 20 | 78 | ∞ | 38 | 5 |
| 6 | 26 | 84 | 0 | 12 | 16 | 0 | 76 | 44 | ∞ | 4 | 68 | 8 | 26 | 34 | 92 | 58 | 72 | 4 | 92 | 4 | 6 |
| 7 | 0 | 10 | 50 | 16 | 10 | 78 | 0 | 88 | 76 | ∞ | 6 | 12 | 6 | 8 | 36 | 82 | 16 | 30 | 78 | 40 | 7 |
| 8 | 86 | 0 | 0 | 58 | 76 | ∞ | 96 | 0 | 6 | 18 | 88 | 52 | 82 | 88 | 40 | 10 | 90 | 46 | 3 | 26 | 8 |
| 9 | 28 | 90 | ∞ | 96 | 8 | 10 | 88 | 46 | 0 | 6 | 66 | 90 | 20 | 40 | 84 | 50 | 84 | 86 | 94 | 16 | 9 |
| 10 | 48 | 82 | 8 | 32 | 36 | 2 | 78 | 56 | 4 | 0 | 72 | 30 | 42 | 50 | 20 | 36 | 76 | 26 | 88 | ∞ | 10 |
| 11 | 70 | 26 | 50 | 58 | 64 | 44 | 18 | 80 | 36 | 12 | 0 | 44 | 68 | 76 | ∞ | 86 | 8 | 38 | 26 | 54 | 11 |
| 12 | 28 | 22 | 52 | 6 | 16 | 50 | 20 | 52 | 38 | 40 | 8 | 0 | 18 | 42 | 8 | 58 | 10 | ∞ | 28 | 66 | 12 |
| 13 | 10 | 28 | 60 | 8 | 8 | 58 | 12 | 24 | 34 | 22 | 2 | ∞ | 0 | 18 | 68 | 30 | 4 | 74 | 30 | 64 | 13 |
| 14 | 92 | 96 | 28 | ∞ | 0 | 23 | 92 | 12 | 8 | 16 | 86 | 70 | 86 | 0 | 52 | 16 | 86 | 62 | 2 | 50 | 14 |
| 15 | 48 | 78 | 30 | 30 | ∞ | 20 | 70 | 72 | 96 | 20 | 58 | 12 | 32 | 50 | 0 | 78 | 66 | 6 | 82 | 78 | 15 |
| 16 | 50 | 4 | 34 | 38 | 40 | 28 | 90 | 76 | 10 | 14 | 78 | 34 | ∞ | 58 | 26 | 0 | 80 | 28 | 8 | 44 | 16 |
| 17 | ∞ | 0 | 46 | 14 | 20 | 40 | 6 | 34 | 30 | 34 | 0 | 10 | 22 | 26 | 2 | 40 | 0 | 19 | 20 | 54 | 17 |
| 18 | 58 | 38 | 72 | 54 | 28 | 62 | 30 | 60 | 44 | 50 | 20 | 10 | 54 | ∞ | 80 | 72 | 18 | 0 | 42 | 74 | 18 |
| 19 | 40 | 40 | 40 | 28 | 30 | 48 | 14 | ∞ | 18 | 26 | 84 | 18 | 36 | 46 | 6 | 36 | 84 | 16 | 0 | 54 | 19 |
| 20 | 48 | 42 | 8 | 22 | 30 | 12 | 38 | 64 | 54 | 68 | 24 | 12 | 42 | 60 | 0 | ∞ | 32 | 10 | 50 | 0 | 20 |
| | 1 | 2 | 3 | 4 | 5 | 6 | 7 | 8 | 9 | 10 | 11 | 12 | 13 | 14 | 15 | 16 | 17 | 18 | 19 | 20 | |

$s_{17}$ = ($7^{-88}$ $8^2$ $19^{-14}$): -100, This complets this chain since our last arc (19 7) is small. Our new arcs are

(7 19), (8 5), (19 3).

$H_{17} = (1^3\ 17^{37}\ 2^{13}\ 4^{27}\ 14^1\ 18^9\ 12^{11}\ 13^9\ 16^3\ 20^{19}\ 10^{11}\ 7^1\ 19^3\ 3^{11}\ 9^3\ 6^3\ 8^3\ 5^9\ 15^1\ 11^{13})$: 190

$$H_{17}^{-1}M$$

| III | 17 | 4 | 9 | 14 | 15 | 8 | 19 | 5 | 6 | 7 | 1 | 13 | 16 | 18 | 11 | 20 | 2 | 12 | 3 | 10 | |
|---|---|---|---|---|---|---|---|---|---|---|---|---|---|---|---|---|---|---|---|---|---|
| | 1 | 2 | 3 | 4 | 5 | 6 | 7 | 8 | 9 | 10 | 11 | 12 | 13 | 14 | 15 | 16 | 17 | 18 | 19 | 20 | |
| 1 | 0 | 4 | 20 | 84 | 88 | 16 | 10 | 2 | 10 | 14 | ∞ | 76 | 96 | 4 | 60 | 14 | 0 | 64 | 2 | 26 | 1 |
| 2 | 56 | 0 | 26 | 48 | 54 | 20 | 66 | 6 | 8 | 18 | 84 | 34 | 58 | 64 | 50 | 76 | ∞ | 27 | 18 | 36 | 2 |
| 3 | 52 | 70 | 0 | 30 | 40 | 88 | 60 | 80 | 82 | 86 | 62 | 28 | 39 | 58 | 18 | 66 | 68 | 22 | ∞ | 12 | 3 |
| 4 | 26 | ∞ | 34 | 0 | 2 | 32 | 42 | 16 | 20 | 26 | 8 | 42 | 12 | 36 | 24 | 54 | 28 | 16 | 14 | 46 | 4 |
| 5 | 46 | 28 | 70 | 88 | 0 | 68 | 32 | ∞ | 40 | 50 | 14 | 84 | 2 | 30 | 72 | 38 | 20 | 78 | 22 | 38 | 5 |
| 6 | 26 | 84 | 0 | 12 | 16 | 0 | 44 | 92 | ∞ | 4 | 68 | 8 | 26 | 34 | 92 | 58 | 72 | 4 | 76 | 4 | 6 |
| 7 | 88 | 98 | 38 | 72 | 78 | 28 | 0 | 10 | 12 | ∞ | 82 | 76 | 82 | 96 | 52 | 6 | 72 | 58 | 88 | 48 | 7 |
| 8 | 83 | 3 | 3 | 55 | 73 | ∞ | 2 | 0 | 3 | 15 | 85 | 49 | 79 | 85 | 37 | 7 | 87 | 43 | 93 | 23 | 8 |
| 9 | 28 | 90 | ∞ | 96 | 8 | 10 | 46 | 94 | 0 | 6 | 66 | 90 | 20 | 40 | 84 | 50 | 84 | 86 | 88 | 16 | 9 |
| 10 | 48 | 82 | 8 | 32 | 36 | 2 | 56 | 88 | 4 | 0 | 72 | 30 | 42 | 50 | 20 | 36 | 76 | 26 | 78 | ∞ | 10 |
| 11 | 70 | 26 | 50 | 58 | 64 | 44 | 80 | 26 | 36 | 12 | 0 | 44 | 68 | 76 | ∞ | 86 | 8 | 38 | 18 | 54 | 11 |
| 12 | 28 | 22 | 52 | 6 | 16 | 50 | 52 | 28 | 38 | 40 | 8 | 0 | 18 | 42 | 8 | 58 | 10 | ∞ | 20 | 66 | 12 |
| 13 | 10 | 28 | 60 | 8 | 8 | 58 | 24 | 30 | 34 | 22 | 2 | ∞ | 0 | 18 | 68 | 30 | 4 | 74 | 12 | 64 | 13 |
| 14 | 92 | 96 | 28 | ∞ | 0 | 23 | 12 | 2 | 8 | 16 | 86 | 70 | 86 | 0 | 52 | 16 | 86 | 62 | 92 | 50 | 14 |
| 15 | 48 | 78 | 30 | 30 | ∞ | 20 | 72 | 82 | 96 | 20 | 58 | 12 | 32 | 50 | 0 | 78 | 66 | 6 | 70 | 78 | 15 |
| 16 | 50 | 4 | 34 | 38 | 40 | 28 | 76 | 8 | 10 | 14 | 78 | 34 | ∞ | 58 | 26 | 0 | 80 | 28 | 90 | 44 | 16 |
| 17 | ∞ | 0 | 46 | 14 | 20 | 40 | 34 | 20 | 30 | 34 | 0 | 10 | 22 | 26 | 2 | 40 | 0 | 19 | 6 | 54 | 17 |
| 18 | 58 | 38 | 72 | 54 | 28 | 62 | 60 | 42 | 44 | 50 | 20 | 10 | 54 | ∞ | 80 | 72 | 18 | 0 | 30 | 74 | 18 |
| 19 | 40 | 40 | 40 | 28 | 30 | 48 | ∞ | 0 | 18 | 26 | 84 | 18 | 36 | 46 | 6 | 36 | 84 | 16 | 0 | 54 | 19 |
| 20 | 48 | 42 | 8 | 22 | 30 | 12 | 64 | 50 | 54 | 68 | 24 | 12 | 42 | 60 | 0 | ∞ | 32 | 10 | 38 | 0 | 20 |
| | 1 | 2 | 3 | 4 | 5 | 6 | 7 | 8 | 9 | 10 | 11 | 12 | 13 | 14 | 15 | 16 | 17 | 18 | 19 | 20 | |

We now attempt to construct a new chain. $s_{17} = (4^{-16}\ 18^{10}\ 12^6)$: 0.

Our new arcs are (4  12), (18  13), (12  14).

$H_{18}$ = ($1^3$ $17^{37}$ $2^{13}$ $4^{11}$ $12^{17}$ $14^1$ $18^{19}$ $13^9$ $16^3$ $20^{19}$ $10^{11}$ $7^1$ $19^3$ $3^{11}$ $9^3$ $6^3$ $8^3$ $5^9$ $15^1$ $11^{13}$): 190.

$$H_{18}^{-1}M$$

| I | 17 | 4 | 9 | 12 | 15 | 8 | 19 | 5 | 6 | 7 | 1 | 14 | 16 | 18 | 11 | 20 | 2 | 13 | 3 | 10 | |
|---|----|----|----|----|----|----|----|----|----|----|----|----|----|----|----|----|----|----|----|----|---|
|   | 1 | 2 | 3 | 4 | 5 | 6 | 7 | 8 | 9 | 10 | 11 | 12 | 13 | 14 | 15 | 16 | 17 | 18 | 19 | 20 | |
| 1 | 0 | 4 | 20 | 64 | 88 | 16 | 10 | 2 | 10 | 14 | ∞ | 84 | 96 | 4 | 60 | 14 | 0 | 76 | 2 | 26 | 1 |
| 2 | 56 | 0 | 26 | 27 | 54 | 20 | 66 | 6 | 8 | 18 | 84 | 48 | 58 | 64 | 50 | 76 | ∞ | 34 | 18 | 36 | 2 |
| 3 | 52 | 70 | 0 | 22 | 40 | 88 | 60 | 80 | 82 | 86 | 62 | 30 | 39 | 58 | 18 | 66 | 68 | 28 | ∞ | 12 | 3 |
| 4 | 42 | ∞ | 50 | 0 | 18 | 48 | 58 | 32 | 36 | 42 | 8 | 16 | 28 | 52 | 8 | 70 | 44 | 58 | 30 | 62 | 4 |
| 5 | 46 | 28 | 70 | 78 | 0 | 68 | 32 | ∞ | 40 | 50 | 14 | 88 | 2 | 30 | 72 | 38 | 20 | 84 | 22 | 38 | 5 |
| 6 | 26 | 84 | 0 | 4 | 16 | 0 | 44 | 92 | ∞ | 4 | 68 | 12 | 26 | 34 | 92 | 58 | 72 | 8 | 76 | 4 | 6 |
| 7 | 88 | 98 | 38 | 58 | 78 | 28 | 0 | 10 | 12 | ∞ | 82 | 72 | 82 | 96 | 52 | 6 | 72 | 76 | 88 | 48 | 7 |
| 8 | 83 | 3 | 3 | 43 | 73 | ∞ | 2 | 0 | 3 | 15 | 85 | 55 | 79 | 85 | 37 | 7 | 87 | 49 | 93 | 23 | 8 |
| 9 | 28 | 90 | ∞ | 86 | 8 | 10 | 46 | 94 | 0 | 6 | 66 | 96 | 20 | 40 | 84 | 50 | 84 | 90 | 88 | 16 | 9 |
| 10 | 48 | 82 | 8 | 26 | 36 | 2 | 56 | 88 | 4 | 0 | 72 | 32 | 42 | 50 | 20 | 36 | 76 | 30 | 78 | ∞ | 10 |
| 11 | 70 | 24 | 50 | 38 | 64 | 44 | 80 | 26 | 36 | 38 | 0 | 58 | 68 | 76 | ∞ | 86 | 8 | 44 | 18 | 54 | 11 |
| 12 | 22 | 16 | 46 | ∞ | 10 | 44 | 46 | 22 | 32 | 34 | 2 | 17 | 12 | 36 | 14 | 52 | 4 | 6 | 14 | 52 | 12 |
| 13 | 10 | 28 | 60 | 74 | 8 | 58 | 24 | 30 | 34 | 22 | 2 | 8 | 0 | 18 | 68 | 30 | 4 | ∞ | 12 | 64 | 13 |
| 14 | 92 | 96 | 28 | 62 | 0 | 23 | 12 | 2 | 8 | 16 | 86 | ∞ | 86 | 0 | 52 | 16 | 86 | 70 | 92 | 50 | 14 |
| 15 | 48 | 78 | 30 | 6 | ∞ | 20 | 72 | 82 | 96 | 20 | 58 | 30 | 32 | 50 | 0 | 78 | 66 | 12 | 70 | 78 | 15 |
| 16 | 50 | 4 | 34 | 28 | 40 | 28 | 76 | 8 | 10 | 14 | 78 | 38 | ∞ | 58 | 26 | 0 | 80 | 34 | 90 | 44 | 16 |
| 17 | ∞ | 0 | 46 | 19 | 20 | 40 | 34 | 20 | 30 | 34 | 0 | 14 | 22 | 26 | 2 | 40 | 0 | 10 | 6 | 54 | 17 |
| 18 | 48 | 28 | 62 | 10 | 18 | 52 | 50 | 32 | 34 | 40 | 10 | 44 | 44 | ∞ | 70 | 62 | 8 | 0 | 20 | 64 | 18 |
| 19 | 40 | 40 | 40 | 16 | 30 | 48 | ∞ | 0 | 18 | 26 | 84 | 28 | 36 | 46 | 6 | 36 | 84 | 18 | 0 | 54 | 19 |
| 20 | 48 | 42 | 8 | 10 | 30 | 12 | 64 | 50 | 54 | 68 | 24 | 22 | 42 | 60 | 0 | ∞ | 32 | 12 | 38 | 0 | 20 |
|   | 1 | 2 | 3 | 4 | 5 | 6 | 7 | 8 | 9 | 10 | 11 | 12 | 13 | 14 | 15 | 16 | 17 | 18 | 19 | 20 | |

$s_{19}$ = ($12^{-14}$ $15^7$ $4^{16}$): 9. Our new arcs are (12 11), (15 12), (4 14).

**H₁₉** = (1³ 17³⁷ 2¹³ 4²⁷ 14¹ 18⁹ 13⁹ 16³ 20¹⁹ 10⁷ 7¹ 19³ 3¹¹ 9³ 6³ 8³ 5⁹ 15⁷ 12³ 11¹³): 190

$$H_{19}^{-1}M$$

| II | 17 | 4 | 9 | 14 | 15 | 8 | 19 | 5 | 6 | 7 | 1 | 11 | 16 | 18 | 12 | 20 | 2 | 13 | 3 | 10 | |
|---|---|---|---|---|---|---|---|---|---|---|---|---|---|---|---|---|---|---|---|---|---|
|   | 1 | 2 | 3 | 4 | 5 | 6 | 7 | 8 | 9 | 10 | 11 | 12 | 13 | 14 | 15 | 16 | 17 | 18 | 19 | 20 | |
| 1 | 0 | 4 | 20 | 84 | 88 | 16 | 10 | 2 | 10 | 14 | ∞ | 60 | 96 | 4 | 64 | 14 | 0 | 76 | 2 | 26 | 1 |
| 2 | 56 | 0 | 26 | 48 | 54 | 20 | 66 | 6 | 8 | 18 | 84 | 50 | 58 | 64 | 27 | 76 | ∞ | 34 | 18 | 36 | 2 |
| 3 | 52 | 70 | 0 | 30 | 40 | 88 | 60 | 80 | 82 | 86 | 62 | 18 | 39 | 58 | 22 | 66 | 68 | 28 | ∞ | 12 | 3 |
| 4 | 26 | ∞ | 34 | 0 | 2 | 32 | 42 | 16 | 20 | 26 | 8 | 24 | 12 | 40 | 16 | 54 | 28 | 42 | 14 | 46 | 4 |
| 5 | 46 | 28 | 70 | 88 | 0 | 68 | 32 | ∞ | 40 | 50 | 14 | 72 | 2 | 30 | 78 | 38 | 20 | 84 | 22 | 38 | 5 |
| 6 | 26 | 84 | 0 | 12 | 16 | 0 | 44 | 92 | ∞ | 4 | 68 | 92 | 26 | 34 | 4 | 58 | 72 | 8 | 76 | 4 | 6 |
| 7 | 88 | 98 | 38 | 72 | 78 | 28 | 0 | 10 | 12 | ∞ | 82 | 52 | 82 | 96 | 58 | 6 | 72 | 76 | 88 | 48 | 7 |
| 8 | 83 | 3 | 3 | 55 | 73 | ∞ | 2 | 0 | 3 | 15 | 85 | 37 | 79 | 85 | 43 | 7 | 87 | 49 | 93 | 23 | 8 |
| 9 | 28 | 90 | ∞ | 96 | 8 | 10 | 46 | 94 | 0 | 6 | 66 | 84 | 20 | 40 | 86 | 50 | 84 | 90 | 88 | 16 | 9 |
| 10 | 48 | 82 | 8 | 32 | 36 | 2 | 56 | 88 | 4 | 0 | 72 | 20 | 42 | 50 | 26 | 36 | 76 | 30 | 78 | ∞ | 10 |
| 11 | 70 | 26 | 50 | 58 | 64 | 44 | 80 | 26 | 36 | 12 | 0 | ∞ | 68 | 76 | 38 | 86 | 8 | 44 | 18 | 54 | 11 |
| 12 | 30 | 24 | 54 | 31 | 24 | 52 | 54 | 32 | 12 | 42 | 8 | 0 | 20 | 44 | ∞ | 60 | 12 | 8 | 22 | 68 | 12 |
| 13 | 10 | 28 | 60 | 8 | 8 | 58 | 24 | 30 | 34 | 22 | 2 | 68 | 0 | 18 | 74 | 30 | 4 | ∞ | 12 | 64 | 13 |
| 14 | 92 | 96 | 28 | ∞ | 0 | 23 | 12 | 2 | 8 | 16 | 86 | 52 | 86 | 0 | 62 | 16 | 86 | 70 | 92 | 50 | 14 |
| 15 | 41 | 69 | 23 | 23 | ∞ | 13 | 65 | 75 | 89 | 13 | 51 | 6 | 25 | 43 | 0 | 71 | 59 | 5 | 62 | 71 | 15 |
| 16 | 50 | 4 | 34 | 38 | 40 | 28 | 76 | 8 | 10 | 14 | 78 | 26 | ∞ | 58 | 28 | 0 | 80 | 34 | 90 | 44 | 16 |
| 17 | ∞ | 0 | 46 | 14 | 20 | 40 | 34 | 20 | 30 | 34 | 0 | 2 | 22 | 26 | 19 | 40 | 0 | 10 | 6 | 54 | 17 |
| 18 | 48 | 28 | 62 | 44 | 18 | 52 | 50 | 32 | 34 | 40 | 10 | 70 | 44 | ∞ | 10 | 62 | 8 | 0 | 20 | 64 | 18 |
| 19 | 40 | 40 | 40 | 28 | 30 | 48 | ∞ | 0 | 18 | 26 | 84 | 6 | 36 | 46 | 16 | 36 | 84 | 18 | 0 | 54 | 19 |
| 20 | 48 | 42 | 8 | 22 | 30 | 12 | 64 | 50 | 54 | 68 | 24 | 0 | 42 | 60 | 10 | ∞ | 32 | 12 | 38 | 0 | 20 |
|   | 1 | 2 | 3 | 4 | 5 | 6 | 7 | 8 | 9 | 10 | 11 | 12 | 13 | 14 | 15 | 16 | 17 | 18 | 19 | 20 | |

We are able to obtain (18⁻¹⁰ 15⁻⁶ 12⁸): -8. This is the last permissible matrix in the chain. So far the chain yielded   0 + (9) + (-8) = 1. Thus, we have not been able to reduce the value of our 20-cycle. Therefore, $H_{17}$ is our approximation to an optimal n-cycle.

**COMMENT.** [ln(n)] + 1 was arbitrarily chosen as the length of a chain. Whether the length of a chain must increase strictly monotonically as n increases is an open question.

**Example 5.** In this example, a matrix, M, is defined in the following way: we first obtain the smallest set of 342 odd primes (each modulo 100). Then giving each remainder an ordinal number corresponding to the value of the prime from which it was derived, Ord(1) through Ord(19) (1st row), Ord(20) through Ord(38) (2nd row), …, Ord(24) … Ord(42) (19th row). Using M, our purpose is to obtain an approximation to an optimal 19-cycle.

M

| ------- | | 1 | 2 | 3 | 4 | 5 | 6 | 7 | 8 | 9 | 10 | 11 | 12 | 13 | 14 | 15 | 16 | 17 | 18 | 19 | |
|---|---|---|---|---|---|---|---|---|---|---|---|---|---|---|---|---|---|---|---|---|---|
| | 1 | ∞ | 3 | 5 | 7 | 11 | 13 | 17 | 19 | 23 | 29 | 31 | 37 | 41 | 43 | 47 | 53 | 59 | 61 | 67 | 1 |
| | 2 | 71 | ∞ | 73 | 79 | 83 | 89 | 97 | 1 | 3 | 7 | 9 | 13 | 27 | 31 | 37 | 39 | 49 | 51 | 57 | 2 |
| | 3 | 63 | 67 | ∞ | 73 | 79 | 81 | 91 | 93 | 97 | 99 | 11 | 23 | 27 | 29 | 33 | 39 | 41 | 51 | 57 | 3 |
| | 4 | 63 | 69 | 71 | ∞ | 77 | 81 | 83 | 93 | 7 | 11 | 13 | 17 | 31 | 37 | 47 | 49 | 19 | 21 | 31 | 4 |
| | 5 | 33 | 39 | 43 | 49 | ∞ | 57 | 61 | 63 | 67 | 79 | 87 | 91 | 99 | 3 | 9 | 21 | 23 | 41 | 47 | 5 |
| | 6 | 57 | 63 | 69 | 71 | 77 | ∞ | 87 | 93 | 99 | 1 | 7 | 13 | 17 | 19 | 31 | 41 | 43 | 47 | 53 | 6 |
| | 7 | 59 | 61 | 73 | 77 | 83 | 91 | ∞ | 1 | 9 | 19 | 27 | 33 | 39 | 43 | 51 | 57 | 61 | 69 | 73 | 7 |
| | 8 | 87 | 97 | 9 | 11 | 21 | 23 | 27 | ∞ | 29 | 39 | 53 | 57 | 59 | 63 | 77 | 81 | 83 | 87 | 7 | 8 |
| | 9 | 11 | 19 | 29 | 37 | 41 | 47 | 53 | 67 | ∞ | 71 | 77 | 83 | 91 | 97 | 9 | 13 | 19 | 21 | 31 | 9 |
| | 10 | 33 | 39 | 49 | 51 | 61 | 63 | 69 | 87 | 91 | ∞ | 93 | 97 | 3 | 9 | 17 | 23 | 29 | 51 | 53 | 10 |
| | 11 | 63 | 71 | 81 | 87 | 93 | 1 | 13 | 17 | 23 | 29 | ∞ | 31 | 37 | 49 | 59 | 77 | 79 | 83 | 89 | 11 |
| | 12 | 91 | 97 | 1 | 3 | 7 | 19 | 21 | 27 | 61 | 67 | 73 | ∞ | 81 | 99 | 9 | 23 | 27 | 29 | 33 | 12 |
| | 13 | 39 | 47 | 51 | 53 | 59 | 71 | 81 | 83 | 87 | 89 | 93 | 99 | ∞ | 11 | 23 | 31 | 43 | 49 | 53 | 13 |
| | 14 | 59 | 67 | 71 | 79 | 83 | 97 | 1 | 7 | 9 | 13 | 19 | 21 | 27 | ∞ | 37 | 57 | 63 | 67 | 69 | 14 |
| | 15 | 93 | 97 | 99 | 9 | 21 | 23 | 33 | 41 | 47 | 53 | 59 | 77 | 83 | 87 | ∞ | 89 | 1 | 11 | 23 | 15 |
| | 16 | 31 | 47 | 61 | 67 | 71 | 73 | 77 | 79 | 89 | 1 | 7 | 13 | 31 | 33 | 49 | ∞ | 51 | 73 | 79 | 16 |
| | 17 | 87 | 93 | 97 | 99 | 3 | 11 | 17 | 27 | 29 | 39 | 53 | 63 | 69 | 81 | 83 | 87 | ∞ | 89 | 99 | 17 |
| | 18 | 11 | 13 | 29 | 31 | 37 | 41 | 43 | 53 | 61 | 79 | 3 | 7 | 13 | 21 | 37 | 39 | 43 | ∞ | 51 | 18 |
| | 19 | 67 | 69 | 73 | 81 | 87 | 93 | 97 | 9 | 11 | 33 | 39 | 41 | 47 | 51 | 57 | 71 | 77 | 81 | ∞ | 19 |
| | | 1 | 2 | 3 | 4 | 5 | 6 | 7 | 8 | 9 | 10 | 11 | 12 | 13 | 14 | 15 | 16 | 17 | 18 | 19 | |

$H_0 = (1^3 \ 2^{73} \ 3^{73} \ 4^{77} \ 5^{57} \ 6^{87} \ 7^1 \ 8^{29} \ 9^{71} \ 10^{93} \ 11^{31} \ 12^{81} \ 13^{11} \ 14^{37} \ 15^{89} \ 16^{51} \ 17^{89} \ 18^{51} \ 19^{67})$ : **1071**.

$H_0^{-1}M$

|  |  | 2 | 3 | 4 | 5 | 6 | 7 | 8 | 9 | 10 | 11 | 12 | 13 | 14 | 15 | 16 | 17 | 18 | 19 | 1 |  |
|---|---|---|---|---|---|---|---|---|---|---|---|---|---|---|---|---|---|---|---|---|---|
| ------- |  | 1 | 2 | 3 | 4 | 5 | 6 | 7 | 8 | 9 | 10 | 11 | 12 | 13 | 14 | 15 | 16 | 17 | 18 | 19 |  |
|  | 1 | 0 | 2 | 4 | 8 | 10 | 14 | 16 | 20 | 26 | 28 | 34 | 38 | 40 | 44 | 50 | 56 | 58 | 64 | ∞ | 1 |
| 72,70,66 | 2 | ∞ | 0 | 6 | 10 | 16 | 24 | **72** | **70** | **66** | 64 | 60 | 46 | 42 | 36 | 34 | 24 | 22 | 16 | 2 | 2 |
| 62,50,46 | 3 | 6 | ∞ | 0 | 6 | 8 | 18 | 20 | 24 | 26 | **62** | **50** | **46** | 44 | 40 | 34 | 32 | 22 | 16 | 10 | 3 |
| 70,66,64 | 4 | 8 | 6 | ∞ | 0 | 4 | 6 | 16 | **70** | **66** | **64** | 60 | 46 | 40 | 30 | 28 | 58 | 56 | 46 | 14 | 4 |
| 54,48,36 | 5 | 18 | 14 | 8 | ∞ | 0 | 4 | 6 | 10 | 22 | 30 | 34 | 42 | **54** | **48** | **36** | 34 | 16 | 10 | 24 | 5 |
| 86,80,74 | 6 | 24 | 18 | 16 | 10 | ∞ | 0 | 6 | 12 | **86** | **80** | **74** | 60 | 68 | 56 | 46 | 44 | 40 | 34 | 30 | 6 |
|  | 7 | 60 | 72 | 76 | 82 | 90 | ∞ | 0 | 8 | 18 | 26 | 32 | 38 | 42 | 50 | 56 | 60 | 68 | 72 | 58 | 7 |
| 22,20,18 | 8 | 68 | **20** | **18** | 8 | 6 | **2** | ∞ | 0 | 10 | 24 | 28 | 30 | 34 | 48 | 52 | 54 | 58 | **22** | 58 | 8 |
| 62,60,58 | 9 | 52 | 42 | 34 | 30 | 24 | 18 | 4 | ∞ | 0 | 6 | 12 | 20 | 26 | **62** | **58** | **52** | 50 | 40 | 60 | 9 |
| 90,84,76 | 10 | 54 | 54 | 42 | 32 | 30 | 24 | 6 | 2 | ∞ | 0 | 4 | **90** | **84** | **76** | 70 | 64 | 42 | 40 | 60 | 10 |
| 30,18,14 | 11 | 40 | 50 | 56 | 62 | **30** | **18** | **14** | 8 | 2 | ∞ | 0 | 6 | 18 | 28 | 46 | 48 | 52 | 58 | 32 | 11 |
| 80,78,75 | 12 | 16 | **80** | **78** | 74 | 62 | 60 | 54 | 20 | 14 | 8 | ∞ | 0 | 18 | **72** | 58 | 54 | 52 | 48 | 10 | 12 |
|  | 13 | 36 | 40 | 42 | 48 | 60 | 70 | 72 | 76 | 78 | 82 | 88 | ∞ | 0 | 12 | 20 | 32 | 38 | 42 | 28 | 13 |
| 36,30,28 | 14 | 30 | 34 | 42 | 46 | 60 | **36** | **30** | **28** | 24 | 18 | 16 | 10 | ∞ | 0 | 20 | 26 | 30 | 32 | 22 | 14 |
| 88,80,78 | 15 | 8 | 10 | **80** | 68 | 66 | 56 | 48 | 42 | 36 | 30 | 12 | 6 | 2 | ∞ | 0 | **88** | **78** | 66 | 4 | 15 |
| 50,44,38 | 16 | 4 | 10 | 16 | 20 | 22 | 26 | 28 | 38 | **50** | **44** | **38** | 20 | 18 | 2 | ∞ | 0 | 22 | 28 | **20** | 16 |
| 86,78,72 | 17 | 4 | 8 | 10 | **86** | **78** | **72** | 62 | 60 | 50 | 36 | 26 | 20 | 8 | 5 | 2 | ∞ | 0 | 10 | **2** | 17 |
| 48,44,40 | 18 | 38 | 22 | 20 | 14 | 10 | 8 | 2 | 10 | 28 | **48** | **44** | **38** | 30 | 14 | 12 | 8 | ∞ | 0 | 40 | 18 |
| 58,56,34 | 19 | 2 | 6 | 14 | 20 | 26 | 30 | **58** | **56** | **34** | 28 | 26 | 20 | 16 | 10 | 4 | 10 | 14 | ∞ | 0 | 19 |
|  |  | 1 | 2 | 3 | 4 | 5 | 6 | 7 | 8 | 9 | 10 | 11 | 12 | 13 | 14 | 15 | 16 | 17 | 18 | 19 |  |

$s_1 = (10^{-90}\ 12^{-80}\ 2^{-64})$: -234. Oour new arcs are (10 13), (12 3), (2 11).

$H_1$ = ($1^3$ $2^9$ $11^{31}$ $12^1$ $3^{73}$ $4^{77}$ $5^{57}$ $6^{87}$ $7^1$ $8^{29}$ $9^{71}$ $10^3$ $13^{11}$ $14^{37}$ $15^{89}$ $16^{51}$ $17^{89}$ $18^{51}$ $19^{67}$): 837.

$$H_1^{-1}M$$

|  |  | 2 | 11 | 4 | 5 | 6 | 7 | 8 | 9 | 10 | 13 | 12 | 3 | 14 | 15 | 16 | 17 | 18 | 19 | 1 |  |
|---|---|---|---|---|---|---|---|---|---|---|---|---|---|---|---|---|---|---|---|---|---|
| ------- |  | 1 | 2 | 3 | 4 | 5 | 6 | 7 | 8 | 9 | 10 | 11 | 12 | 13 | 14 | 15 | 16 | 17 | 18 | 19 |  |
|  | 1 | 0 | 28 | 4 | 8 | 10 | 14 | 16 | 20 | 26 | 38 | 34 | 2 | 40 | 44 | 50 | 56 | 58 | 64 | ∞ | 1 |
| 8,6,2 | 2 | ∞ | 0 | 70 | 74 | 80 | 88 | 8 | 6 | 2 | 18 | 4 | 64 | 22 | 28 | 30 | 40 | 42 | 48 | 62 | 2 |
| 62,50,46 | 3 | 6 | 62 | 0 | 6 | 8 | 18 | 20 | 24 | 26 | **46** | **50** | ∞ | **44** | **40** | **34** | **32** | **22** | **16** | **10** | 3 |
| 70,66,64 | 4 | 8 | 64 | ∞ | 0 | 4 | 6 | 16 | **70** | **66** | 46 | 60 | 6 | 40 | 30 | 28 | 58 | 56 | 46 | 14 | 4 |
| 54,48,36 | 5 | 18 | 30 | 8 | ∞ | 0 | 4 | 6 | 10 | 22 | 42 | 34 | **14** | **54** | **48** | **36** | **34** | 16 | 10 | 24 | 5 |
| 86,80,74 | 6 | 24 | 80 | 16 | 10 | ∞ | 0 | 6 | 12 | **86** | 60 | **74** | 18 | 68 | 56 | 46 | 44 | 40 | 34 | 30 | 6 |
|  | 7 | 60 | 26 | 76 | 82 | 90 | ∞ | 0 | 8 | 18 | 38 | 32 | 72 | 42 | 50 | 56 | 60 | 68 | 72 | 58 | 7 |
| 22,20,18 | 8 | 68 | 24 | **18** | 8 | 6 | **2** | ∞ | 0 | 10 | 30 | 28 | **20** | 34 | 48 | 52 | 54 | 58 | **22** | 58 | 8 |
| 62,60,58 | 9 | 52 | 6 | **34** | 30 | 24 | 18 | 4 | ∞ | 0 | 20 | 12 | **42** | 26 | **62** | **58** | **52** | 50 | 40 | **60** | 9 |
|  | 10 | 36 | 90 | 48 | 58 | 60 | 66 | 84 | 88 | ∞ | 0 | 94 | 36 | 6 | 14 | 20 | 26 | 48 | 50 | 30 | 10 |
| 30,18,14 | 11 | 40 | ∞ | 56 | 62 | **30** | **18** | **14** | 8 | **2** | 6 | 0 | 50 | 18 | 28 | 46 | 48 | 52 | 58 | 32 | 11 |
|  | 12 | 96 | 72 | 2 | 6 | 18 | 20 | 26 | 60 | 66 | 80 | ∞ | 0 | 98 | 8 | 22 | 26 | 28 | 32 | 90 | 12 |
|  | 13 | 36 | 82 | 42 | 48 | 60 | 70 | 72 | 76 | 78 | ∞ | 88 | 40 | 0 | 12 | 20 | 32 | 38 | 42 | 28 | 13 |
| 36,30,28 | 14 | 30 | **18** | 42 | 46 | 60 | **36** | **30** | **28** | 24 | 10 | **16** | 34 | ∞ | 0 | 20 | 26 | 30 | 32 | 22 | 14 |
| 88,80,78 | 15 | 8 | 30 | **80** | 68 | 66 | 56 | 48 | 42 | 36 | 6 | 12 | 10 | **2** | ∞ | 0 | **88** | **78** | 66 | 4 | 15 |
| 50,44,38 | 16 | 4 | **44** | 16 | 20 | 22 | 26 | 28 | **38** | **50** | 20 | **38** | 10 | 18 | **2** | ∞ | 0 | 22 | 28 | **20** | 16 |
| 86,78,72 | 17 | 4 | 36 | 10 | **86** | **78** | **72** | 62 | 60 | 50 | 20 | 26 | 8 | 8 | 5 | **2** | ∞ | 0 | 10 | **2** | 17 |
| 48,44,40 | 18 | 38 | **48** | 20 | 14 | 10 | 8 | 2 | 10 | 28 | 38 | **44** | 22 | 30 | 14 | 12 | 8 | ∞ | 0 | **40** | 18 |
| 58,56,34 | 19 | 2 | 28 | 14 | 20 | 26 | 30 | **58** | **56** | **34** | 20 | 26 | 6 | 16 | 10 | 4 | 10 | 14 | ∞ | 0 | 19 |
|  |  | 1 | 2 | 3 | 4 | 5 | 6 | 7 | 8 | 9 | 10 | 11 | 12 | 13 | 14 | 15 | 16 | 17 | 18 | 19 |  |

$s_2$ = ($15^{-88}$ $16^{-50}$ $9^{-58}$): -196.  Our new arcs are (15 17), (16 10), (9 16).

$H_2 = (1^3\ 2^9\ 11^{31}\ 12^1\ 3^{73}\ 4^{77}\ 5^{57}\ 6^{87}\ 7^1\ 8^{29}\ 9^{13}\ 16^1\ 10^3\ 13^{11}\ 14^{37}\ 15^1\ 17^{89}\ 18^{51}\ 19^{67})$: 641

$$H_2^{-1}M$$

|  |  | 2 | 11 | 4 | 5 | 6 | 7 | 8 | 9 | 16 | 13 | 12 | 3 | 14 | 15 | 17 | 10 | 18 | 19 | 1 |  |
|---|---|---|---|---|---|---|---|---|---|---|---|---|---|---|---|---|---|---|---|---|---|
|  |  | 1 | 2 | 3 | 4 | 5 | 6 | 7 | 8 | 9 | 10 | 11 | 12 | 13 | 14 | 15 | 16 | 17 | 18 | 19 |  |
| ------- | 1 | 0 | 28 | 4 | 8 | 10 | 14 | 16 | 20 | 50 | 38 | 34 | 2 | 40 | 44 | 56 | 26 | 58 | 64 | ∞ | 1 |
| 8,6,2 | 2 | ∞ | 0 | 70 | 74 | 80 | 88 | **8** | **6** | 30 | 18 | 4 | 64 | 22 | 28 | 40 | **2** | 42 | 48 | 62 | 2 |
| 62,50,46 | 3 | 6 | 62 | 0 | 6 | 8 | 18 | 20 | 24 | **34** | **46** | **50** | ∞ | 44 | 40 | 32 | 26 | **22** | **16** | **10** | 3 |
| 70,66,64 | 4 | 8 | **64** | ∞ | 0 | 4 | 6 | 16 | **70** | 28 | 46 | 60 | 6 | 40 | 30 | 58 | **66** | 56 | 46 | 14 | 4 |
| 54,48,36 | 5 | 18 | 30 | 8 | ∞ | 0 | 4 | 6 | 10 | **36** | 42 | 34 | 14 | **54** | **48** | 34 | 22 | 16 | 10 | 24 | 5 |
| 86,80,74 | 6 | 24 | **80** | **16** | 10 | ∞ | 0 | 6 | 12 | 46 | 60 | **74** | 18 | 68 | 56 | 44 | **86** | 40 | 34 | 30 | 6 |
|  | 7 | 60 | 26 | 76 | 82 | 90 | ∞ | 0 | 8 | 56 | 38 | 32 | 72 | 42 | 50 | 60 | 18 | 68 | 72 | 58 | 7 |
| 22,20,18 | 8 | 68 | 24 | **18** | 8 | 6 | **2** | ∞ | 0 | 52 | 30 | 28 | **20** | 34 | 48 | 54 | 10 | 58 | **22** | 58 | 8 |
| 4,2 | 9 | 6 | 64 | 24 | 28 | 34 | 40 | 54 | ∞ | 0 | 78 | 70 | 16 | 84 | **4** | 6 | 0 | 8 | 18 | **2** | 9 |
|  | 10 | 36 | 90 | 48 | 58 | 60 | 66 | 84 | 88 | 20 | 0 | 94 | 36 | 6 | 14 | 26 | ∞ | 48 | 50 | 30 | 10 |
| 30,18,14 | 11 | 40 | ∞ | 56 | 62 | **30** | **18** | **14** | 8 | 46 | 6 | 0 | 50 | 18 | 28 | 48 | **2** | 52 | 58 | 32 | 11 |
|  | 12 | 96 | 72 | 2 | 6 | 18 | 20 | 26 | 60 | 22 | 80 | ∞ | 0 | 98 | 8 | 26 | 66 | 28 | 32 | 90 | 12 |
|  | 13 | 36 | 82 | 42 | 48 | 60 | 70 | 72 | 76 | 20 | ∞ | 88 | 40 | 0 | 12 | 32 | 78 | 38 | 42 | 28 | 13 |
| 36,30,28 | 14 | 30 | **18** | 42 | 46 | 60 | **36** | **30** | **28** | 20 | **10** | **16** | 34 | ∞ | 0 | 26 | **24** | 30 | 32 | 22 | 14 |
| 36 | 15 | 96 | 58 | 8 | 20 | 22 | 32 | 40 | 46 | 0 | 82 | 76 | 98 | 86 | ∞ | 0 | **36** | 10 | 22 | 92 | 15 |
|  | 16 | 46 | 6 | 66 | 70 | 72 | 76 | 78 | 88 | ∞ | 30 | 12 | 60 | 32 | 48 | 0 | 0 | 72 | 78 | 30 | 16 |
| 86,78,72 | 17 | 5 | **36** | 10 | **86** | **78** | **72** | 62 | 60 | **2** | 20 | 26 | 8 | 8 | 5 | ∞ | 50 | 0 | 10 | **2** | 17 |
| 48,44,40 | 18 | 38 | **48** | 20 | 14 | 10 | 8 | 2 | 10 | **12** | 38 | **44** | 22 | 30 | 14 | 8 | 28 | ∞ | 0 | **40** | 18 |
| 58,56,34 | 19 | 2 | **28** | 14 | 20 | 26 | 30 | **58** | **56** | 4 | **20** | **26** | 6 | **16** | 10 | 10 | **34** | 14 | ∞ | 0 | 19 |
|  |  | 1 | 2 | 3 | 4 | 5 | 6 | 7 | 8 | 9 | 10 | 11 | 12 | 13 | 14 | 15 | 16 | 17 | 18 | 19 |  |

**Note** . One approach that the author uses is to start with the *smallest* or *near smallest* negative value as a first arc. Then we have two choices. (1) The second arc is negative or small and postivie. Or (2) the third arc whose end point is the inital point of the 3-cycle is negative or small. If possible, we like the third arc to have small value. This is generally possible toward the beginning of the algorithm.

$s_3 = (17^{-72}\ 4^{-70}\ 8^{58})$: -98 . Our new arcs are (17 5), (4 9), (8 18).

$H_3$ = (1³ 2⁹ 11³¹ 12¹ 3⁷³ 4⁷ 9¹³ 16¹ 10³ 13¹¹ 14³⁷ 15¹ 17³ 5⁵⁷ 6⁸⁷ 7¹ 8⁸⁷ 18⁵¹ 19⁶⁷): 466

$$H_3^{-1}M$$

|          |    | 2 | 11 | 4 | 9 | 6 | 7 | 8 | 18 | 16 | 13 | 12 | 3 | 14 | 15 | 17 | 10 | 5 | 19 | 1 |    |
|----------|----|---|----|---|---|---|---|---|----|----|----|----|---|----|----|----|----|---|----|---|----|
| -------  |    | 1 | 2  | 3 | 4 | 5 | 6 | 7 | 8  | 9  | 10 | 11 | 12| 13 | 14 | 15 | 16 |17 | 18 |19 |    |
|          | 1  | 0 | 28 | 4 | 20| 10| 14| 16| 58 | 50 | 38 | 34 | 2 | 40 | 44 | 56 | 26 | 8 | 64 | ∞ | 1  |
| 8,6,2    | 2  | ∞ | 0  | 70| 6 | 80| 88| 8 | 42 | 30 | 18 | 4  | 64| 22 | 28 | 40 | 2  | 74| 48 | 62| 2  |
| 62,50,46 | 3  | 6 | 62 | 0 | 24| 8 | 18| 20| 22 | 34 | 46 | 50 | ∞ | 44 | 40 | 32 | 26 | 6 | 16 | 10| 3  |
|          | 4  | 62| 6  | ∞ | 0 | 74| 76| 86| 14 | 42 | 24 | 10 | 64| 30 | 40 | 12 | 4  | 70| 24 | 56| 4  |
| 54,48,36 | 5  | 18| 30 | 8 | 10| 0 | 4 | 6 | 16 | 36 | 42 | 34 | 14| 54 | 48 | 34 | 22 | ∞ | 10 | 24| 5  |
| 86,80,74 | 6  | 24| 80 | 16| 12| ∞ | 0 | 6 | 40 | 46 | 60 | 74 | 18| 68 | 56 | 44 | 86 | 10| 34 | 30| 6  |
|          | 7  | 60| 26 | 76| 8 | 90| ∞ | 0 | 68 | 56 | 38 | 32 | 72| 42 | 50 | 60 | 18 | 82| 72 | 58| 7  |
| 80,78,76 | 8  | 10| 34 | 76| 58| 64| 60| ∞ | 0  | 6  | 28 | 30 | 78| 6  | 10 | 54 | 4  | 66| 80 | 0 | 8  |
| 4,2      | 9  | 6 | 64 | 24| ∞ | 34| 40| 54| 8  | 0  | 78 | 70 | 16| 84 | 4  | 6  | 0  | 28| 18 | 2 | 9  |
|          | 10 | 36| 90 | 48| 88| 60| 66| 84| 48 | 20 | 0  | 94 | 36| 6  | 14 | 26 | ∞  | 58| 50 | 30| 10 |
| 30,18,14 | 11 | 40| ∞  | 56| 8 | 30| 18| 14| 52 | 46 | 6  | 0  | 50| 18 | 28 | 48 | 2  | 62| 58 | 32| 11 |
|          | 12 | 96| 72 | 2 | 60| 18| 20| 26| 28 | 22 | 80 | ∞  | 0 | 98 | 8  | 26 | 66 | 6 | 32 | 90| 12 |
|          | 13 | 36| 82 | 42| 76| 60| 70| 72| 38 | 20 | ∞  | 88 | 40| 0  | 12 | 32 | 78 | 48| 42 | 28| 13 |
| 36,30,28 | 14 | 30| 18 | 42| 28| 60| 36| 30| 30 | 20 | 10 | 16 | 34| ∞  | 0  | 26 | 24 | 46| 32 | 22| 14 |
| 36       | 15 | 96| 58 | 8 | 46| 22| 32| 40| 10 | 0  | 82 | 76 | 98| 86 | ∞  | 0  | 36 | 20| 22 | 92| 15 |
|          | 16 | 46| 6  | 66| 88| 72| 76| 78| 72 | ∞  | 30 | 12 | 60| 32 | 48 | 0  | 0  | 70| 78 | 30| 16 |
|          | 17 | 91| 50 | 96| 26| 8 | 14| 24| 86 | 84 | 66 | 60 | 94| 78 | 81 | ∞  | 36 | 0 | 96 | 84| 17 |
| 48,44,40 | 18 | 38| 48 | 20| 10| 10| 8 | 2 | ∞  | 12 | 38 | 44 | 22| 30 | 14 | 8  | 28 | 14| 0  | 40| 18 |
| 58,56,34 | 19 | 2 | 28 | 14| 56| 26| 30| 58| 14 | 4  | 20 | 26 | 6 | 16 | 10 | 10 | 34 | 20| ∞  | 0 | 19 |
|          |    | 1 | 2  | 3 | 4 | 5 | 6 | 7 | 8  | 9  | 10 | 11 | 12| 13 | 14 | 15 | 16 | 17| 18 | 19|    |

(8⁻⁸⁰ 18⁻²⁰ 3⁻²²): -122.  Our new arcs are (8  19), (18  4), (3  18).

$H_4 = (1^3\ 2^9\ \ 11^{31}\ 12^1\ 3^{51}\ 18^{31}\ 4^7\ 9^{13}\ 16^1\ 10^3\ 13^{11}\ 14^{37}\ 15^1\ 17^3\ 5^{57}\ 6^{87}\ 7^1\ 8^7\ 19^{67})$: 421.

$$H_4^{-1}M$$

|  |  | 2 | 11 | 18 | 9 | 6 | 7 | 8 | 19 | 16 | 13 | 12 | 3 | 14 | 15 | 17 | 10 | 5 | 4 | 1 |  |
|---|---|---|---|---|---|---|---|---|---|---|---|---|---|---|---|---|---|---|---|---|---|
| ------ |  | **1** | **2** | **3** | **4** | **5** | **6** | **7** | **8** | **9** | **10** | **11** | **12** | **13** | **14** | **15** | **16** | **17** | **18** | **19** |  |
|  | **1** | 0 | 28 | 58 | 20 | 10 | 14 | 16 | 64 | 50 | 38 | 34 | 2 | 40 | 44 | 56 | 26 | 8 | 4 | ∞ | **1** |
| 8,6,2 | **2** | ∞ | 0 | 42 | **6** | 80 | 88 | **8** | 48 | 30 | 18 | 4 | 64 | 22 | 28 | 40 | **2** | 74 | 70 | 62 | **2** |
| 40,28,24 | **3** | 16 | **40** | 0 | 46 | 30 | 40 | 42 | 6 | **12** | **24** | **28** | ∞ | **20** | **18** | **10** | 48 | 28 | 22 | 12 | **3** |
|  | **4** | 62 | 6 | 14 | 0 | 74 | 76 | 86 | 24 | 42 | 24 | 10 | 64 | 30 | 40 | 12 | 4 | 70 | ∞ | 56 | **4** |
| 54,48,36 | **5** | **18** | 30 | **16** | 10 | 0 | 4 | 6 | **10** | **36** | 42 | 34 | **14** | **54** | **48** | **34** | 22 | ∞ | **8** | 24 | **5** |
| 86,80,74 | **6** | **24** | **80** | **40** | 12 | ∞ | 0 | 6 | **34** | **46** | **60** | **74** | 18 | **68** | **56** | **44** | **86** | 10 | **16** | 30 | **6** |
|  | **7** | 60 | 26 | 68 | 8 | 90 | ∞ | 0 | 72 | 56 | 38 | 32 | 72 | 42 | 50 | 60 | 18 | 82 | 76 | 58 | **7** |
|  | **8** | 90 | 46 | 80 | 22 | 16 | 20 | ∞ | 0 | 74 | 52 | 50 | 2 | 86 | 70 | 76 | 76 | 14 | 4 | 80 | **8** |
| 4,2 | **9** | 6 | 64 | 8 | ∞ | 34 | 40 | 54 | 18 | 0 | 78 | 70 | 16 | 84 | **4** | 6 | 0 | 28 | 24 | **2** | **9** |
|  | **10** | 36 | 90 | 48 | 88 | 60 | 66 | 84 | 50 | 20 | 0 | 94 | 36 | 6 | 14 | 26 | ∞ | 58 | 48 | 30 | **10** |
| 30,18,14 | **11** | 40 | ∞ | 52 | **8** | **30** | **18** | **14** | 58 | 46 | 6 | 0 | 50 | 18 | 28 | 48 | **2** | 62 | 56 | 32 | **11** |
|  | **12** | 96 | 72 | 28 | 60 | 18 | 20 | 26 | 32 | 22 | 80 | ∞ | 0 | 98 | 8 | 26 | 66 | 6 | 2 | 90 | **12** |
|  | **13** | 36 | 82 | 38 | 76 | 60 | 70 | 72 | 42 | 20 | ∞ | 88 | 40 | 0 | 12 | 32 | 78 | 48 | 42 | 28 | **13** |
| 36,30,28 | **14** | 30 | **18** | 30 | **28** | 60 | **36** | **30** | 32 | 20 | **10** | **16** | 34 | ∞ | 0 | 26 | **24** | 46 | 42 | 22 | **14** |
| 36 | **15** | 96 | 58 | 10 | 46 | 22 | 32 | 40 | 22 | 0 | 82 | 76 | 98 | 86 | ∞ | 0 | **36** | 20 | 8 | 92 | **15** |
|  | **16** | 46 | 6 | 72 | 88 | 72 | 76 | 78 | 78 | ∞ | 30 | 12 | 60 | 32 | 48 | 0 | 0 | 70 | 66 | 30 | **16** |
|  | **17** | 91 | 50 | 86 | 26 | 8 | 14 | 24 | 96 | 84 | 66 | 60 | 94 | 78 | 81 | ∞ | 36 | 0 | 96 | 84 | **17** |
| 28,24,20 | **18** | **18** | **28** | ∞ | 30 | 10 | 12 | 22 | 20 | 8 | **18** | **24** | **2** | **10** | **4** | 12 | 48 | 6 | 0 | **20** | **18** |
| 58,56,34 | **19** | 2 | **28** | 14 | **56** | 26 | 30 | **58** | ∞ | 4 | **20** | **26** | 6 | **16** | **10** | 10 | **34** | 20 | 14 | 0 | **19** |
|  |  | **1** | **2** | **3** | **4** | **5** | **6** | **7** | **8** | **9** | **10** | **11** | **12** | **13** | **14** | **15** | **16** | **17** | **18** | **19** |  |

$(6^{-74}\ 11^{-30}\ 5^4)$: -100.  Our new arcs are (6 12), (11 6), (5 7). 421

$H_5$ = ($1^3$ $2^9$ $11^1$ $6^{13}$ $12^1$ $3^{51}$ $18^{31}$ $4^7$ $9^{13}$ $16^1$ $10^3$ $13^{11}$ $14^{37}$ $15^1$ $17^3$ $5^{61}$ $7^1$ $8^7$ $19^{67}$): 320

$$H_5^{-1}M$$

|  |  | 2 | 11 | 18 | 9 | 7 | 12 | 8 | 19 | 16 | 13 | 6 | 3 | 14 | 15 | 17 | 10 | 5 | 4 | 1 |  |
|---|---|---|---|---|---|---|---|---|---|---|---|---|---|---|---|---|---|---|---|---|---|
| ------ |  | 1 | 2 | 3 | 4 | 5 | 6 | 7 | 8 | 9 | 10 | 11 | 12 | 13 | 14 | 15 | 16 | 17 | 18 | 19 |  |
|  | 1 | 0 | 28 | 58 | 20 | 14 | 34 | 16 | 64 | 50 | 38 | 10 | 2 | 40 | 44 | 56 | 26 | 8 | 4 | ∞ | 1 |
| 8,6,2 | 2 | ∞ | 0 | 42 | 6 | 88 | 4 | 8 | 48 | 30 | 18 | 80 | 64 | 22 | 28 | 40 | 2 | 74 | 70 | 62 | 2 |
| 40,28,24 | 3 | 16 | 40 | 0 | 46 | 40 | 28 | 42 | 6 | 12 | 24 | 30 | ∞ | 20 | 18 | 10 | 48 | 28 | 22 | 12 | 3 |
|  | 4 | 62 | 6 | 14 | 0 | 76 | 10 | 86 | 24 | 42 | 24 | 74 | 64 | 30 | 40 | 12 | 4 | 70 | ∞ | 56 | 4 |
| 58,52,40 | 5 | 22 | 26 | 20 | 6 | 0 | 30 | 2 | 14 | 40 | 38 | 4 | 18 | 58 | 52 | 38 | 18 | ∞ | 12 | 28 | 5 |
| 12,6 | 6 | 50 | 6 | 34 | 86 | 74 | 0 | 80 | 40 | 28 | 14 | ∞ | 56 | 6 | 18 | 30 | 12 | 64 | 58 | 44 | 6 |
|  | 7 | 60 | 26 | 68 | 8 | ∞ | 32 | 0 | 72 | 56 | 38 | 90 | 72 | 42 | 50 | 60 | 18 | 82 | 76 | 58 | 7 |
|  | 8 | 90 | 46 | 80 | 22 | 20 | 50 | ∞ | 0 | 74 | 52 | 16 | 2 | 86 | 70 | 76 | 76 | 14 | 4 | 80 | 8 |
| 4,2 | 9 | 6 | 64 | 8 | ∞ | 40 | 70 | 54 | 18 | 0 | 78 | 34 | 16 | 84 | 4 | 6 | 0 | 28 | 24 | 2 | 9 |
|  | 10 | 36 | 90 | 48 | 88 | 66 | 94 | 84 | 50 | 20 | 0 | 60 | 36 | 6 | 14 | 26 | ∞ | 58 | 48 | 30 | 10 |
| 30,18,14 | 11 | 70 | ∞ | 82 | 22 | 12 | 30 | 16 | 88 | 76 | 36 | 0 | 80 | 48 | 58 | 78 | 28 | 92 | 86 | 62 | 11 |
|  | 12 | 96 | 72 | 28 | 60 | 20 | ∞ | 26 | 32 | 22 | 80 | 18 | 0 | 98 | 8 | 26 | 66 | 6 | 2 | 90 | 12 |
|  | 13 | 36 | 82 | 38 | 76 | 70 | 88 | 72 | 42 | 20 | ∞ | 60 | 40 | 0 | 12 | 32 | 78 | 48 | 42 | 28 | 13 |
| 36,30,28 | 14 | 30 | 18 | 30 | 28 | 36 | 16 | 30 | 32 | 20 | 10 | 60 | 34 | ∞ | 0 | 26 | 24 | 46 | 42 | 22 | 14 |
| 36 | 15 | 96 | 58 | 10 | 46 | 32 | 76 | 40 | 22 | 0 | 82 | 22 | 98 | 86 | ∞ | 0 | 36 | 20 | 8 | 92 | 15 |
|  | 16 | 46 | 6 | 72 | 88 | 76 | 12 | 78 | 78 | ∞ | 30 | 72 | 60 | 32 | 48 | 0 | 0 | 70 | 66 | 30 | 16 |
|  | 17 | 91 | 50 | 86 | 26 | 14 | 60 | 24 | 96 | 84 | 66 | 8 | 94 | 78 | 81 | ∞ | 36 | 0 | 96 | 84 | 17 |
| 28,24,20 | 18 | 18 | 28 | ∞ | 30 | 12 | 24 | 22 | 20 | 8 | 18 | 10 | 2 | 10 | 4 | 12 | 48 | 6 | 0 | 20 | 18 |
| 58,56,34 | 19 | 2 | 28 | 14 | 56 | 30 | 26 | 58 | ∞ | 4 | 20 | 26 | 6 | 16 | 10 | 10 | 34 | 20 | 14 | 0 | 19 |
|  |  | 1 | 2 | 3 | 4 | 5 | 6 | 7 | 8 | 9 | 10 | 11 | 12 | 13 | 14 | 15 | 16 | 17 | 18 | 19 |  |

$s_{61}$ = ($5^{-58}$ $13^{12}$ $14^{-36}$): -82. Our new arcs are (5 14), (13 15), (14 7).

$H_6 = (1^3 \ 2^9 \ 11^1 \ 6^{13} \ 12^1 \ 3^{51} \ 18^{31} \ 4^7 \ 9^{13} \ 16^1 \ 10^3 \ 13^{23} \ 15^1 \ 17^3 \ 5^3 \ 14^1 \ 7^1 \ 8^{21} \ 19^{67})$: 253

$$H_6^{-1}M$$

|          |    | 2  | 11 | 18 | 9  | 14 | 12 | 8  | 19 | 16 | 13 | 6  | 3  | 15 | 7  | 17 | 10 | 5  | 4  | 1  |    |
|----------|----|----|----|----|----|----|----|----|----|----|----|----|----|----|----|----|----|----|----|----|----|
| ------   |    | 1  | 2  | 3  | 4  | 5  | 6  | 7  | 8  | 9  | 10 | 11 | 12 | 13 | 14 | 15 | 16 | 17 | 18 | 19 |    |
|          | 1  | 0  | 28 | 58 | 20 | 40 | 34 | 16 | 64 | 50 | 38 | 10 | 2  | 44 | 14 | 56 | 26 | 8  | 4  | ∞  | 1  |
| 8,6,2    | 2  | ∞  | 0  | 42 | 6  | 22 | 4  | 8  | 48 | 30 | 18 | 80 | 64 | 28 | 88 | 40 | 2  | 74 | 70 | 62 | 2  |
| 40,28,24 | 3  | 16 | 40 | 0  | 46 | 20 | 28 | 42 | 6  | 12 | 24 | 30 | ∞  | 18 | 40 | 10 | 48 | 28 | 22 | 12 | 3  |
|          | 4  | 62 | 6  | 14 | 0  | 30 | 10 | 86 | 24 | 42 | 24 | 74 | 64 | 40 | 76 | 12 | 4  | 70 | ∞  | 56 | 4  |
|          | 5  | 36 | 84 | 38 | 64 | 0  | 88 | 60 | 44 | 18 | 96 | 54 | 40 | 6  | 58 | 20 | 76 | ∞  | 46 | 30 | 5  |
| 12,6     | 6  | 50 | 6  | 34 | 86 | 6  | 0  | 80 | 40 | 28 | 14 | ∞  | 56 | 18 | 74 | 30 | 12 | 64 | 58 | 44 | 6  |
|          | 7  | 60 | 26 | 68 | 8  | 42 | 32 | 0  | 72 | 56 | 38 | 90 | 72 | 50 | ∞  | 60 | 18 | 82 | 76 | 58 | 7  |
|          | 8  | 90 | 46 | 80 | 22 | 86 | 50 | ∞  | 0  | 74 | 52 | 16 | 2  | 70 | 20 | 76 | 76 | 14 | 4  | 80 | 8  |
| 4,2      | 9  | 6  | 64 | 8  | ∞  | 84 | 70 | 54 | 18 | 0  | 78 | 34 | 16 | 4  | 40 | 6  | 0  | 28 | 24 | 2  | 9  |
|          | 10 | 36 | 90 | 48 | 88 | 6  | 94 | 84 | 50 | 20 | 0  | 60 | 36 | 14 | 66 | 26 | ∞  | 58 | 48 | 30 | 10 |
|          | 11 | 70 | ∞  | 82 | 22 | 48 | 30 | 16 | 88 | 76 | 36 | 0  | 80 | 58 | 12 | 78 | 28 | 92 | 86 | 62 | 11 |
|          | 12 | 96 | 72 | 28 | 60 | 98 | ∞  | 26 | 32 | 22 | 80 | 18 | 0  | 8  | 20 | 26 | 66 | 6  | 2  | 90 | 12 |
|          | 13 | 24 | 70 | 26 | 64 | 12 | 76 | 60 | 30 | 8  | ∞  | 48 | 28 | 0  | 58 | 20 | 66 | 32 | 30 | 16 | 13 |
| 6        | 14 | 66 | 18 | 6  | 8  | ∞  | 20 | 6  | 68 | 56 | 26 | 96 | 70 | 36 | 0  | 62 | 12 | 82 | 78 | 58 | 14 |
|          | 15 | 96 | 58 | 10 | 46 | 86 | 76 | 40 | 22 | 0  | 82 | 22 | 98 | ∞  | 32 | 0  | 52 | 20 | 8  | 92 | 15 |
|          | 16 | 46 | 6  | 72 | 88 | 32 | 12 | 78 | 78 | ∞  | 30 | 72 | 60 | 48 | 76 | 0  | 0  | 70 | 66 | 30 | 16 |
|          | 17 | 91 | 50 | 86 | 26 | 78 | 60 | 24 | 96 | 84 | 66 | 8  | 94 | 81 | 14 | ∞  | 36 | 0  | 96 | 84 | 17 |
| 28,24,20 | 18 | 18 | 28 | ∞  | 30 | 10 | 24 | 22 | 20 | 8  | 18 | 10 | 2  | 4  | 12 | 12 | 48 | 6  | 0  | 20 | 18 |
| 58,56,34 | 19 | 2  | 28 | 14 | 56 | 16 | 26 | 58 | ∞  | 4  | 20 | 26 | 6  | 10 | 30 | 10 | 34 | 20 | 14 | 0  | 19 |
|          |    | 1  | 2  | 3  | 4  | 5  | 6  | 7  | 8  | 9  | 10 | 11 | 12 | 13 | 14 | 15 | 16 | 17 | 18 | 19 |    |

$s_7 = (19^{-56} \ 4^4 \ 16^{30})$: -22 .  Our new arcs are  (19  9),  (4  10),  (16  1).

$H_7 = (1^3\ 2^9\ 11^1\ 6^{13}\ 12^1\ 3^{51}\ 18^{31}\ 4^{11}\ 10^3\ 13^{23}\ 15^1\ 17^3\ 5^3\ 14^1\ 7^1\ 8^7\ 19^{11}\ 9^{13}\ 16^{31})$: 231

$$H_7^{-1}M$$

| I | | 2 | 11 | 18 | 10 | 14 | 12 | 8 | 19 | 16 | 13 | 6 | 3 | 15 | 7 | 17 | 1 | 5 | 4 | 9 | |
|---|---|---|---|---|---|---|---|---|---|---|---|---|---|---|---|---|---|---|---|---|---|
| ------- | | 1 | 2 | 3 | 4 | 5 | 6 | 7 | 8 | 9 | 10 | 11 | 12 | 13 | 14 | 15 | 16 | 17 | 18 | 19 | |
| | 1 | 0 | 28 | 58 | 26 | 40 | 34 | 16 | 64 | 50 | 38 | 10 | 2 | 44 | 14 | 56 | ∞ | 8 | 4 | 20 | 1 |
| 8,6,2 | 2 | ∞ | 0 | 42 | 2 | 22 | 4 | **8** | 48 | 30 | 18 | 80 | 64 | 28 | 88 | 40 | 62 | 74 | 70 | **6** | 2 |
| 40,28,24 | 3 | 16 | **40** | 0 | 48 | **20** | **28** | 42 | 6 | **12** | **24** | 30 | ∞ | **18** | 40 | **10** | 12 | 28 | 22 | 46 | 3 |
| 4 | 4 | 54 | **2** | 6 | 0 | 26 | 6 | 82 | 20 | 38 | 20 | 70 | 60 | 36 | 72 | 8 | 52 | 66 | ∞ | **4** | 4 |
| | 5 | 36 | 84 | 48 | 76 | 0 | 88 | 60 | 44 | 18 | 96 | 54 | 40 | 6 | 58 | 20 | 30 | ∞ | 46 | 64 | 5 |
| 12,6 | 6 | 50 | **6** | 34 | **12** | 6 | 0 | 80 | 40 | 28 | 14 | ∞ | 56 | 18 | 74 | 30 | 44 | 64 | 58 | 86 | 6 |
| | 7 | 60 | 26 | 68 | 18 | 42 | 32 | 0 | 72 | 56 | 38 | 90 | 72 | 50 | ∞ | 60 | 58 | 82 | 76 | 8 | 7 |
| | 8 | 90 | 46 | 80 | 76 | 86 | 50 | ∞ | 0 | 74 | 52 | 16 | 2 | 70 | 20 | 76 | 80 | 14 | 4 | 22 | 8 |
| 4,2 | 9 | 6 | 64 | 8 | 0 | 84 | 70 | 54 | 18 | 0 | 78 | 34 | 16 | **4** | 40 | 6 | **2** | 28 | 24 | ∞ | 9 |
| | 10 | 36 | 90 | 48 | ∞ | 6 | 94 | 84 | 50 | 20 | 0 | 60 | 36 | 14 | 66 | 26 | 30 | 58 | 48 | 88 | 10 |
| | 11 | 70 | ∞ | 82 | 28 | 48 | 30 | 16 | 88 | 76 | 36 | 0 | 80 | 58 | 12 | 78 | 62 | 92 | 86 | 22 | 11 |
| | 12 | 96 | 72 | 28 | 66 | 98 | ∞ | 26 | 32 | 22 | 80 | 18 | 0 | 8 | 20 | 26 | 90 | 6 | 2 | 60 | 12 |
| 12 | 13 | 24 | 70 | 26 | 66 | **12** | 76 | 60 | 30 | 8 | ∞ | 48 | 28 | 0 | 58 | 20 | 16 | 32 | 30 | 64 | 13 |
| 6 | 14 | 66 | 18 | **6** | 12 | ∞ | 20 | 6 | 68 | 56 | 26 | 96 | 70 | 36 | 0 | 62 | 58 | 82 | 78 | 8 | 14 |
| | 15 | 96 | 58 | 10 | 52 | 86 | 76 | 40 | 22 | 0 | 82 | 22 | 98 | ∞ | 32 | 0 | 92 | 20 | 8 | 46 | 15 |
| 30,24,18 | 16 | 16 | **24** | 42 | 0 | 2 | **18** | 48 | 48 | ∞ | 0 | 42 | 30 | 18 | 46 | 20 | 0 | 40 | 36 | 58 | 16 |
| | 17 | 91 | 50 | 86 | 36 | 78 | 60 | 24 | 96 | 84 | 66 | 8 | 94 | 81 | 14 | ∞ | 84 | 0 | 96 | 26 | 17 |
| 28,24,20 | 18 | **18** | **28** | ∞ | 48 | **10** | **24** | 22 | 20 | 8 | **18** | 10 | **2** | **4** | 12 | 12 | **20** | 6 | 0 | 30 | 18 |
| 2 | 19 | 58 | 28 | 70 | 40 | 40 | 30 | **2** | ∞ | 60 | 36 | 82 | 62 | 46 | 86 | 66 | 56 | 76 | 70 | 0 | 19 |
| | | 1 | 2 | 3 | 4 | 5 | 6 | 7 | 8 | 9 | 10 | 11 | 12 | 13 | 14 | 15 | 16 | 17 | 18 | 19 | |

When there are comparatively few negative arcs, it is worthwhile to spend time to check all possible $H_i$-admissible 3-cycles, not only first and second arcs, but first and third arcs for a possibly good 3-cycle. We now start a chain.

$s_8 = (16^{-30}\ 4^{-4}\ 19^{56})$: 22. Our new arcs are (16 10), (4 9), (19 1).

$H_8 = (1^3\ 2^9\ 11^1\ 6^{13}\ 12^1\ 3^{51}\ 18^{31}\ 4^7\ 9^{13}\ 16^1\ 10^3\ 13^{23}\ 15^1\ 17^3\ 5^3\ 14^1\ 7^1\ 8^7\ 19^{67})$: 229.

$$H_8^{-1}M$$

| II | | 2 | 11 | 18 | 9 | 14 | 12 | 8 | 19 | 16 | 13 | 6 | 3 | 15 | 7 | 17 | 10 | 5 | 4 | 1 | |
|---|---|---|---|---|---|---|---|---|---|---|---|---|---|---|---|---|---|---|---|---|---|
| ------- | | 1 | 2 | 3 | 4 | 5 | 6 | 7 | 8 | 9 | 10 | 11 | 12 | 13 | 14 | 15 | 16 | 17 | 18 | 19 | |
| | 1 | 0 | 28 | 58 | 20 | 40 | 34 | 16 | 64 | 50 | 38 | 10 | 2 | 44 | 14 | 56 | 26 | 8 | 4 | ∞ | 1 |
| 8,6,2 | 2 | ∞ | 0 | 42 | **6** | 22 | 4 | **8** | 48 | 30 | 18 | 80 | 64 | 28 | 88 | 40 | **2** | 74 | 70 | 62 | 2 |
| 40,28,24 | 3 | 16 | **40** | 0 | 46 | **22** | **28** | 42 | 6 | **12** | **24** | 30 | ∞ | **18** | 40 | **10** | 48 | 28 | 22 | 12 | 3 |
| | 4 | 58 | 2 | 10 | 0 | 30 | 10 | 86 | 24 | 42 | 24 | 74 | 64 | 42 | 76 | 12 | 4 | 70 | ∞ | 56 | 4 |
| | 5 | 36 | 84 | 48 | 64 | 0 | 88 | 60 | 44 | 18 | 96 | 54 | 40 | 6 | 58 | 20 | 76 | ∞ | 46 | 30 | 5 |
| 12,6 | 6 | 50 | **6** | 34 | 86 | 6 | 0 | 80 | 40 | 28 | 14 | ∞ | 56 | 18 | 74 | 30 | **12** | 64 | 58 | 44 | 6 |
| | 7 | 60 | 26 | 68 | 8 | 42 | 32 | 0 | 72 | 56 | 38 | 90 | 72 | 50 | ∞ | 60 | 18 | 82 | 76 | 58 | 7 |
| | 8 | 90 | 46 | 80 | 22 | 86 | 50 | ∞ | 0 | 74 | 52 | 16 | 2 | 70 | 20 | 76 | 76 | 14 | 4 | 80 | 8 |
| 4,2 | 9 | 6 | 64 | 8 | ∞ | 84 | 70 | 54 | 18 | 0 | 78 | 34 | 16 | **4** | 40 | 6 | 0 | 28 | 24 | **2** | 9 |
| | 10 | 36 | 90 | 48 | 88 | 6 | 94 | 84 | 50 | 20 | 0 | 60 | 36 | 14 | 66 | 26 | ∞ | 58 | 48 | 30 | 10 |
| | 11 | 70 | ∞ | 82 | 22 | 48 | 30 | 16 | 88 | 76 | 36 | 0 | 80 | 58 | 12 | 78 | 28 | 92 | 86 | 62 | 11 |
| | 12 | 96 | 72 | 28 | 60 | 98 | ∞ | 26 | 32 | 22 | 80 | 18 | 0 | 8 | 20 | 26 | 66 | 6 | 2 | 90 | 12 |
| 12 | 13 | 24 | 70 | 26 | 64 | **12** | 76 | 60 | 30 | 8 | ∞ | 48 | 28 | 0 | 58 | 20 | 66 | 32 | 30 | 16 | 13 |
| | 14 | 66 | 18 | 66 | 8 | ∞ | 20 | 6 | 68 | 56 | 26 | 96 | 70 | 36 | 0 | 62 | 12 | 82 | 78 | 58 | 14 |
| | 15 | 96 | 58 | 10 | 46 | 86 | 76 | 40 | 22 | 0 | 82 | 22 | 98 | ∞ | 32 | 0 | 52 | 20 | 8 | 92 | 15 |
| | 16 | 46 | 6 | 72 | 88 | 32 | 12 | 78 | 78 | ∞ | 31 | 72 | 60 | 48 | 76 | 50 | 0 | 70 | 66 | 30 | 16 |
| | 17 | 91 | 50 | 86 | 26 | 78 | 60 | 24 | 96 | 84 | 66 | 8 | 94 | 81 | 14 | ∞ | 36 | 0 | 96 | 84 | 17 |
| 28,24,20 | 18 | 18 | 28 | ∞ | 30 | **10** | **24** | 22 | 20 | 8 | **18** | 10 | **2** | 6 | 12 | 12 | 48 | 6 | 0 | **20** | 18 |
| 58,56,28 | 19 | 2 | **28** | 14 | **56** | 16 | 26 | **58** | ∞ | 4 | **20** | 26 | 6 | **10** | 30 | 10 | **16** | 20 | 14 | 0 | 19 |
| | | 1 | 2 | 3 | 4 | 5 | 6 | 7 | 8 | 9 | 10 | 11 | 12 | 13 | 14 | 15 | 16 | 17 | 18 | 19 | |

$s_9 = (19^{-26}\ 6^{-12}\ 16^{31}\ )$: -7 . Our new arcs are (19 12 ), (6 10), (16 1).

$H_9$ ($1^3$ $2^9$ $11^1$ $6^1$ $10^3$ $13^{23}$ $15^2$ $17^3$ $5^3$ $14^1$ $7^1$ $8^7$ $19^{41}$ $12^1$ $3^{51}$ $18^{31}$ $4^7$ $9^{13}$ $16^{31}$): 232.

$$H_9^{-1}M$$

| III |  | 2 | 11 | 18 | 9 | 14 | 10 | 8 | 19 | 16 | 13 | 6 | 3 | 15 | 7 | 17 | 1 | 5 | 4 | 12 |  |
|---|---|---|---|---|---|---|---|---|---|---|---|---|---|---|---|---|---|---|---|---|---|
| ------- |  | 1 | 2 | 3 | 4 | 5 | 6 | 7 | 8 | 9 | 10 | 11 | 12 | 13 | 14 | 15 | 16 | 17 | 18 | 19 |  |
|  | 1 | 0 | 28 | 58 | 20 | 40 | 26 | 16 | 64 | 50 | 38 | 10 | 2 | 44 | 14 | 56 | ∞ | 8 | 4 | 34 | 1 |
| 8,6,2 | 2 | ∞ | 0 | 42 | 6 | 22 | 2 | 8 | 48 | 30 | 18 | 80 | 64 | 28 | 88 | 40 | 62 | 74 | 70 | 4 | 2 |
| 40,28,24 | 3 | 16 | 40 | 0 | 46 | 22 | 48 | 42 | 6 | 12 | 24 | 30 | ∞ | 18 | 40 | 10 | 12 | 28 | 22 | 28 | 3 |
|  | 4 | 58 | 2 | 10 | 0 | 30 | 4 | 86 | 24 | 42 | 24 | 74 | 64 | 42 | 76 | 12 | 56 | 70 | ∞ | 10 | 4 |
|  | 5 | 36 | 84 | 48 | 64 | 0 | 76 | 60 | 44 | 18 | 96 | 54 | 40 | 6 | 58 | 20 | 30 | ∞ | 46 | 88 | 5 |
|  | 6 | 62 | 6 | 46 | 98 | 18 | 0 | 92 | 52 | 40 | 26 | ∞ | 68 | 30 | 86 | 42 | 56 | 76 | 70 | 12 | 6 |
|  | 7 | 60 | 26 | 68 | 8 | 42 | 18 | 0 | 72 | 56 | 38 | 90 | 72 | 50 | ∞ | 60 | 58 | 82 | 76 | 32 | 7 |
|  | 8 | 90 | 46 | 80 | 22 | 86 | 76 | ∞ | 0 | 74 | 52 | 16 | 2 | 70 | 20 | 76 | 80 | 14 | 4 | 50 | 8 |
| 4,2 | 9 | 6 | 64 | 8 | ∞ | 84 | 0 | 54 | 18 | 0 | 78 | 34 | 16 | 4 | 40 | 6 | 2 | 28 | 24 | 70 | 9 |
|  | 10 | 36 | 90 | 48 | 88 | 6 | ∞ | 84 | 50 | 20 | 0 | 60 | 36 | 14 | 66 | 26 | 30 | 58 | 48 | 94 | 10 |
|  | 11 | 70 | ∞ | 82 | 22 | 48 | 28 | 16 | 88 | 76 | 36 | 0 | 80 | 58 | 12 | 78 | 62 | 92 | 86 | 30 | 11 |
|  | 12 | 96 | 72 | 28 | 60 | 98 | 66 | 26 | 32 | 22 | 80 | 18 | 0 | 8 | 20 | 26 | 90 | 6 | 2 | ∞ | 12 |
| 12 | 13 | 24 | 70 | 26 | 64 | 12 | 66 | 60 | 30 | 8 | ∞ | 48 | 28 | 0 | 58 | 20 | 16 | 32 | 30 | 76 | 13 |
|  | 14 | 66 | 18 | 66 | 8 | ∞ | 12 | 6 | 68 | 56 | 26 | 96 | 70 | 36 | 0 | 62 | 58 | 82 | 78 | 20 | 14 |
|  | 15 | 96 | 58 | 10 | 46 | 86 | 52 | 40 | 22 | 0 | 82 | 22 | 98 | ∞ | 32 | 0 | 92 | 20 | 8 | 76 | 15 |
| 30,24,18 | 16 | 16 | 24 | 42 | 58 | 2 | 30 | 48 | 48 | ∞ | 1 | 42 | 30 | 18 | 46 | 20 | 0 | 40 | 36 | 18 | 16 |
|  | 17 | 91 | 50 | 86 | 26 | 78 | 36 | 24 | 96 | 84 | 66 | 8 | 94 | 81 | 14 | ∞ | 84 | 0 | 96 | 60 | 17 |
| 28,24,20 | 18 | 18 | 28 | ∞ | 30 | 10 | 48 | 22 | 20 | 8 | 18 | 10 | 2 | 6 | 12 | 12 | 20 | 6 | 0 | 24 | 18 |
| 32,30,2 | 19 | 28 | 2 | 40 | 30 | 10 | 10 | 32 | ∞ | 30 | 6 | 52 | 32 | 16 | 56 | 36 | 26 | 46 | 40 | 0 | 19 |
|  |  | 1 | 2 | 3 | 4 | 5 | 6 | 7 | 8 | 9 | 10 | 11 | 12 | 13 | 14 | 15 | 16 | 17 | 18 | 19 |  |

$s_{10}$ = ($19^{-30}$ $4^2$ $2^4$): -24. The sum of the values obtained in the chain going through $s_8$, $s_9$ and $s_{10}$ is

22 + (-7) + (-24) = -9. Thus, it was successful. Our new arcs are (19 9), (4 11), (2 12).

$H_{10}$ = ($1^3\ 2^{13}\ 12^1\ 3^{51}\ 18^{31}\ 4^{13}\ 11^1\ 6^1\ 10^3\ 13^{23}\ 15^1\ 17^3\ 5^3\ 14^1\ 7^1\ 8^7\ 19^{11}\ 9^{23}\ 16^{31}$): 221.

$$H_{10}^{-1}M$$

|  |  | 2 | 12 | 18 | 11 | 14 | 10 | 8 | 19 | 16 | 13 | 6 | 3 | 15 | 7 | 17 | 1 | 5 | 4 | 9 |  |
|---|---|---|---|---|---|---|---|---|---|---|---|---|---|---|---|---|---|---|---|---|---|
| ------- |  | 1 | 2 | 3 | 4 | 5 | 6 | 7 | 8 | 9 | 10 | 11 | 12 | 13 | 14 | 15 | 16 | 17 | 18 | 19 |  |
|  | 1 | 0 | 34 | 58 | 28 | 40 | 26 | 16 | 64 | 50 | 38 | 10 | 2 | 44 | 14 | 56 | ∞ | 8 | 4 | 20 | 1 |
| 12,10,6 | 2 | ∞ | 0 | 38 | 4 | 18 | 6 | 12 | 44 | 26 | 14 | 76 | 60 | 24 | 84 | 36 | 58 | 70 | 66 | 10 | 2 |
| 40,28,24 | 3 | 16 | 28 | 0 | 40 | 22 | 48 | 42 | 6 | 12 | 24 | 30 | ∞ | 18 | 40 | 10 | 12 | 28 | 22 | 46 | 3 |
|  | 4 | 56 | 8 | 8 | 0 | 28 | 2 | 84 | 22 | 40 | 22 | 72 | 62 | 40 | 74 | 10 | 54 | 68 | ∞ | 2 | 4 |
|  | 5 | 36 | 88 | 48 | 84 | 0 | 76 | 60 | 44 | 18 | 96 | 54 | 40 | 6 | 58 | 20 | 30 | ∞ | 46 | 64 | 5 |
|  | 6 | 62 | 12 | 46 | 6 | 18 | 0 | 92 | 52 | 40 | 26 | ∞ | 68 | 30 | 86 | 42 | 56 | 76 | 70 | 98 | 6 |
|  | 7 | 60 | 32 | 68 | 26 | 42 | 18 | 0 | 72 | 56 | 38 | 90 | 72 | 50 | ∞ | 60 | 58 | 82 | 76 | 8 | 7 |
|  | 8 | 90 | 50 | 80 | 46 | 86 | 76 | ∞ | 0 | 74 | 52 | 16 | 2 | 70 | 20 | 76 | 80 | 14 | 4 | 22 | 8 |
| 4,2 | 9 | 6 | 70 | 8 | 64 | 84 | 0 | 54 | 18 | 0 | 78 | 34 | 16 | 4 | 40 | 6 | 2 | 28 | 24 | ∞ | 9 |
|  | 10 | 36 | 94 | 48 | 90 | 6 | ∞ | 84 | 50 | 20 | 0 | 60 | 36 | 14 | 66 | 26 | 30 | 58 | 48 | 88 | 10 |
|  | 11 | 70 | 30 | 82 | ∞ | 48 | 28 | 16 | 88 | 76 | 36 | 0 | 80 | 58 | 12 | 78 | 62 | 92 | 86 | 22 | 11 |
|  | 12 | 96 | ∞ | 28 | 72 | 98 | 66 | 26 | 32 | 22 | 80 | 18 | 0 | 8 | 20 | 26 | 90 | 6 | 2 | 60 | 12 |
| 12 | 13 | 24 | 76 | 26 | 70 | 12 | 66 | 60 | 30 | 8 | ∞ | 48 | 28 | 0 | 58 | 20 | 16 | 32 | 30 | 64 | 13 |
|  | 14 | 66 | 20 | 66 | 18 | ∞ | 12 | 6 | 68 | 56 | 26 | 96 | 70 | 36 | 0 | 62 | 58 | 82 | 78 | 8 | 14 |
|  | 15 | 96 | 76 | 10 | 58 | 86 | 52 | 40 | 22 | 0 | 82 | 22 | 98 | ∞ | 32 | 0 | 92 | 20 | 8 | 46 | 15 |
| 30,24,18 | 16 | 16 | 18 | 42 | 24 | 2 | 30 | 48 | 48 | ∞ | 1 | 42 | 30 | 18 | 46 | 20 | 0 | 40 | 36 | 58 | 16 |
|  | 17 | 91 | 60 | 86 | 50 | 78 | 36 | 24 | 96 | 84 | 66 | 8 | 94 | 81 | 14 | ∞ | 84 | 0 | 96 | 26 | 17 |
| 28,24,20 | 18 | 18 | 24 | ∞ | 28 | 10 | 48 | 22 | 20 | 8 | 18 | 10 | 2 | 6 | 12 | 12 | 20 | 6 | 0 | 30 | 18 |
| 2 | 19 | 58 | 30 | 70 | 28 | 40 | 40 | 2 | ∞ | 60 | 36 | 82 | 62 | 46 | 86 | 66 | 56 | 76 | 70 | 0 | 19 |
|  |  | 1 | 2 | 3 | 4 | 5 | 6 | 7 | 8 | 9 | 10 | 11 | 12 | 13 | 14 | 15 | 16 | 17 | 18 | 19 |  |

$s_{11}$ = ($3^{-40}\ 4^{10}\ 15^{10}$): -20. Our new arcs are (3 11), (4 17), (15 18).

$H_{11}$ = ($1^3$ $2^{13}$ $12^1$ $3^{11}$ $11^1$ $6^1$ $10^3$ $13^{23}$ $15^{11}$ $18^{31}$ $4^{29}$ $17^3$ $5^3$ $14^1$ $7^1$ $8^7$ $19^{11}$ $9^{13}$ $16^{31}$): 197

$$H_{11}^{-1}M$$

|  |  | 2 | 12 | 11 | 17 | 14 | 10 | 8 | 19 | 16 | 13 | 6 | 3 | 15 | 7 | 18 | 1 | 5 | 4 | 9 |  |
|---|---|---|---|---|---|---|---|---|---|---|---|---|---|---|---|---|---|---|---|---|---|
| ------- |  | **1** | **2** | **3** | **4** | **5** | **6** | **7** | **8** | **9** | **10** | **11** | **12** | **13** | **14** | **15** | **16** | **17** | **18** | **19** |  |
|  | **1** | 0 | 34 | 28 | 56 | 40 | 26 | 16 | 64 | 50 | 38 | 10 | 2 | 44 | 14 | 58 | ∞ | 8 | 4 | 20 | **1** |
| **12,10,6** | **2** | ∞ | 0 | **4** | 36 | 18 | **6** | **12** | 44 | 26 | 14 | 76 | 60 | 24 | 84 | 38 | 58 | 70 | 66 | **10** | **2** |
|  | **3** | 56 | 12 | 0 | 30 | 18 | 88 | 82 | 46 | 28 | 16 | 70 | ∞ | 22 | 80 | 0 | 52 | 68 | 62 | 86 | **3** |
| **12,8,6** | **4** | 46 | **2** | **6** | 0 | 18 | **8** | 74 | 12 | 30 | 12 | 62 | 52 | 30 | 64 | 8 | 44 | 58 | ∞ | **12** | **4** |
|  | **5** | 36 | 88 | 84 | 20 | 0 | 76 | 60 | 44 | 18 | 96 | 54 | 40 | 6 | 58 | 48 | 30 | ∞ | 46 | 64 | **5** |
|  | **6** | 62 | 12 | 6 | 42 | 18 | 0 | 92 | 52 | 40 | 26 | ∞ | 68 | 30 | 86 | 46 | 56 | 76 | 70 | 98 | **6** |
|  | **7** | 60 | 32 | 26 | 60 | 42 | 18 | 0 | 72 | 56 | 38 | 90 | 72 | 50 | ∞ | 68 | 58 | 82 | 76 | 8 | **7** |
|  | **8** | 90 | 50 | 46 | 76 | 86 | 76 | ∞ | 0 | 74 | 52 | 16 | 2 | 70 | 20 | 80 | 80 | 14 | 4 | 22 | **8** |
| **4,2** | **9** | 6 | 70 | 64 | 6 | 84 | 0 | 54 | 18 | 0 | 78 | 34 | 16 | **4** | 40 | 8 | **2** | 28 | 24 | ∞ | **9** |
|  | **10** | 36 | 94 | 90 | 26 | 6 | ∞ | 84 | 50 | 20 | 0 | 60 | 36 | 14 | 66 | 48 | 30 | 58 | 48 | 88 | **10** |
|  | **11** | 70 | 30 | ∞ | 78 | 48 | 28 | 16 | 88 | 76 | 36 | 0 | 80 | 58 | 12 | 82 | 62 | 92 | 86 | 22 | **11** |
|  | **12** | 96 | ∞ | 72 | 26 | 98 | 66 | 26 | 32 | 22 | 80 | 18 | 0 | 8 | 20 | 28 | 90 | 6 | 2 | 60 | **12** |
| **12** | **13** | 24 | 76 | 70 | 20 | **12** | 66 | 60 | 30 | 8 | ∞ | 48 | 28 | 0 | 58 | 26 | 16 | 32 | 30 | 64 | **13** |
|  | **14** | 66 | 20 | 18 | 62 | ∞ | 12 | 6 | 68 | 56 | 26 | 96 | 70 | 36 | 0 | 66 | 58 | 82 | 78 | 8 | **14** |
|  | **15** | 86 | 66 | 48 | **10** | 76 | 42 | 30 | 12 | **10** | 72 | 12 | 88 | ∞ | 22 | 0 | 82 | 10 | **2** | 36 | **15** |
| **30,24,18** | **16** | 16 | **18** | **24** | 20 | 2 | **30** | 48 | 48 | ∞ | 1 | 42 | 30 | 18 | 46 | 42 | 0 | 40 | 36 | 58 | **16** |
|  | **17** | 91 | 60 | 50 | ∞ | 78 | 36 | 24 | 96 | 84 | 66 | 8 | 94 | 81 | 14 | 86 | 84 | 0 | 96 | 26 | **17** |
| **24,20,18** | **18** | **18** | **24** | **28** | 12 | **10** | 48 | 22 | 20 | 8 | **18** | 10 | **2** | 6 | 12 | ∞ | **20** | 6 | 0 | 30 | **18** |
| **2** | **19** | 58 | 30 | 28 | 66 | 40 | 40 | **2** | ∞ | 60 | 36 | 82 | 62 | 46 | 86 | 70 | 56 | 76 | 70 | 0 | **19** |
|  |  | **1** | **2** | **3** | **4** | **5** | **6** | **7** | **8** | **9** | **10** | **11** | **12** | **13** | **14** | **15** | **16** | **17** | **18** | **19** |  |

$s_{12}$ = ($18^{-18}$ $1^2$ $12^2$): -14. Our new arcs are (18 2), (1 3), (12 4).

$H_{12}$ = (1⁵ 3¹¹ 11¹ 6¹ 10³ 13²³ 15¹¹ 18¹³ 2¹³ 12³ 4¹⁹ 17³ 5³ 14¹ 7¹ 8⁷ 19¹¹ 9¹³ 16³¹): 173

$$H_{12}^{-1}M$$

|  |  | 3 | 12 | 11 | 17 | 14 | 10 | 8 | 19 | 16 | 13 | 6 | 4 | 15 | 7 | 18 | 1 | 5 | 2 | 9 |  |
|---|---|---|---|---|---|---|---|---|---|---|---|---|---|---|---|---|---|---|---|---|---|
|  | ------- | **1** | **2** | **3** | **4** | **5** | **6** | **7** | **8** | **9** | **10** | **11** | **12** | **13** | **14** | **15** | **16** | **17** | **18** | **19** |  |
|  | **1** | 0 | 32 | 26 | 54 | 38 | 24 | 14 | 62 | 48 | 36 | 8 | 2 | 42 | 12 | 58 | ∞ | 6 | **2** | 18 | **1** |
| 12,10,6 | **2** | 60 | 0 | **4** | 36 | 18 | **6** | **12** | 44 | 26 | 14 | 76 | 66 | 24 | 84 | 38 | 58 | 70 | ∞ | **10** | **2** |
|  | **3** | ∞ | 12 | 0 | 30 | 18 | 88 | 82 | 46 | 28 | 16 | 70 | 62 | 22 | 80 | 0 | 52 | 68 | 56 | 86 | **3** |
| 12,8,6 | **4** | 52 | **2** | **6** | 0 | 18 | **8** | 74 | 12 | 30 | 12 | 62 | ∞ | 30 | 64 | 8 | 44 | 58 | 46 | **12** | **4** |
|  | **5** | 40 | 88 | 84 | 20 | 0 | 76 | 60 | 44 | 18 | 96 | 54 | 46 | 6 | 58 | 48 | 30 | ∞ | 36 | 64 | **5** |
|  | **6** | 68 | 12 | 6 | 42 | 18 | 0 | 92 | 52 | 40 | 26 | ∞ | 70 | 30 | 86 | 46 | 56 | 76 | 62 | 98 | **6** |
|  | **7** | 72 | 32 | 26 | 60 | 42 | 18 | 0 | 72 | 56 | 38 | 90 | 76 | 50 | ∞ | 68 | 58 | 82 | 60 | 8 | **7** |
|  | **8** | 2 | 50 | 46 | 76 | 86 | 76 | ∞ | 0 | 74 | 52 | 16 | 4 | 70 | 20 | 80 | 80 | 14 | 90 | 22 | **8** |
| 4,2 | **9** | 16 | 70 | 64 | 6 | 84 | 0 | 54 | 18 | 0 | 78 | 34 | 24 | **4** | 40 | 8 | **2** | 28 | 6 | ∞ | **9** |
|  | **10** | 36 | 94 | 90 | 26 | 6 | ∞ | 84 | 50 | 20 | 0 | 60 | 48 | 14 | 66 | 48 | 30 | 58 | 36 | 88 | **10** |
|  | **11** | 80 | 30 | ∞ | 78 | 48 | 28 | 16 | 88 | 76 | 36 | 0 | 86 | 58 | 12 | 82 | 62 | 92 | 70 | 22 | **11** |
|  | **12** | **2** | ∞ | 70 | 24 | 96 | 64 | 24 | 30 | 20 | 78 | 16 | 0 | 6 | 18 | 28 | 88 | 4 | 94 | 58 | **12** |
| 12 | **13** | 28 | 76 | 70 | 20 | **12** | 66 | 60 | 30 | 8 | ∞ | 48 | 30 | 0 | 58 | 26 | 16 | 32 | 24 | 64 | **13** |
| (1^ | **14** | 70 | 20 | 18 | 62 | ∞ | 12 | 6 | 68 | 56 | 26 | 96 | 78 | 36 | 0 | 66 | 58 | 82 | 66 | 8 | **14** |
|  | **15** | 98 | 76 | 58 | 0 | 86 | 52 | 40 | 22 | 0 | 82 | 22 | 8 | ∞ | 32 | 0 | 92 | 20 | 96 | 46 | **15** |
| 30,24,18 | **16** | 30 | **18** | **24** | 20 | 2 | **30** | 48 | 48 | ∞ | 1 | 42 | 36 | 18 | 46 | 42 | 0 | 40 | 16 | 58 | **16** |
|  | **17** | 94 | 60 | 50 | ∞ | 78 | 36 | 24 | 96 | 84 | 66 | 8 | 96 | 81 | 14 | 86 | 84 | 0 | 91 | 26 | **17** |
| 10,8,2 | **18** | 16 | **8** | **10** | 30 | 8 | 66 | 40 | 38 | 26 | 0 | 28 | 18 | 24 | 30 | ∞ | **2** | 24 | 0 | 48 | **18** |
| 2 | **19** | 62 | 30 | 28 | 66 | 40 | 40 | **2** | ∞ | 60 | 36 | 82 | 70 | 46 | 86 | 70 | 56 | 76 | 58 | 0 | **19** |
|  |  | **1** | **2** | **3** | **4** | **5** | **6** | **7** | **8** | **9** | **10** | **11** | **12** | **13** | **14** | **15** | **16** | **17** | **18** | **19** |  |

We now start a new chain with $s_{13}$ = (13⁻¹² 5¹⁸ 9⁻⁴): 2. Our new arcs are (13 14), (5 16), (9 15).

$H_{13}$ = (1⁵ 3¹¹ 11¹ 6¹ 10³ 13¹¹ 14¹ 7¹ 8⁷ 19¹¹ 9⁹ 15¹¹ 18¹³ 2¹³ 12³ 4¹⁹ 17³ 5²¹ 16³¹): 175

$$H_{13}^{-1}M$$

| I |  | 3 | 12 | 11 | 17 | 16 | 10 | 8 | 19 | 15 | 13 | 6 | 4 | 14 | 7 | 18 | 1 | 5 | 2 | 9 |  |
|---|---|---|---|---|---|---|---|---|---|---|---|---|---|---|---|---|---|---|---|---|---|
| ------- |  | 1 | 2 | 3 | 4 | 5 | 6 | 7 | 8 | 9 | 10 | 11 | 12 | 13 | 14 | 15 | 16 | 17 | 18 | 19 |  |
|  | 1 | 0 | 32 | 26 | 54 | 48 | 24 | 14 | 62 | 42 | 36 | 8 | 2 | 38 | 12 | 58 | ∞ | 6 | 2 | 18 | 1 |
| 12,10,6 | 2 | 60 | 0 | 4 | 36 | 26 | 6 | 12 | 44 | 24 | 14 | 76 | 66 | 18 | 84 | 38 | 58 | 70 | ∞ | 10 | 2 |
|  | 3 | ∞ | 12 | 0 | 30 | 28 | 88 | 82 | 46 | 22 | 16 | 70 | 62 | 18 | 80 | 0 | 52 | 68 | 56 | 86 | 3 |
| 12,8,6 | 4 | 52 | 2 | 6 | 0 | 30 | 8 | 74 | 12 | 30 | 12 | 62 | ∞ | 18 | 64 | 8 | 44 | 58 | 46 | 12 | 4 |
| 12 | 5 | 22 | 70 | 66 | 2 | 0 | 58 | 42 | 26 | 12 | 78 | 36 | 28 | 0 | 40 | 48 | 12 | ∞ | 18 | 46 | 5 |
|  | 6 | 68 | 12 | 6 | 42 | 40 | 0 | 92 | 52 | 30 | 26 | ∞ | 70 | 18 | 86 | 46 | 56 | 76 | 62 | 98 | 6 |
|  | 7 | 72 | 32 | 26 | 60 | 56 | 18 | 0 | 72 | 50 | 38 | 90 | 76 | 42 | ∞ | 68 | 58 | 82 | 60 | 8 | 7 |
|  | 8 | 2 | 50 | 46 | 76 | 74 | 76 | ∞ | 0 | 70 | 52 | 16 | 4 | 86 | 20 | 80 | 80 | 14 | 90 | 22 | 8 |
|  | 9 | 20 | 74 | 68 | 10 | 4 | 4 | 58 | 22 | 0 | 82 | 38 | 28 | 84 | 44 | 8 | 2 | 32 | 10 | ∞ | 9 |
|  | 10 | 36 | 94 | 90 | 26 | 20 | ∞ | 84 | 50 | 14 | 0 | 60 | 48 | 6 | 66 | 48 | 30 | 58 | 36 | 88 | 10 |
|  | 11 | 80 | 30 | ∞ | 78 | 76 | 28 | 16 | 88 | 58 | 36 | 0 | 86 | 48 | 12 | 82 | 62 | 92 | 70 | 22 | 11 |
|  | 12 | 2 | ∞ | 70 | 24 | 20 | 64 | 24 | 30 | 6 | 78 | 16 | 0 | 96 | 18 | 28 | 88 | 4 | 94 | 58 | 12 |
|  | 13 | 40 | 88 | 82 | 32 | 20 | 78 | 72 | 42 | 12 | ∞ | 60 | 42 | 0 | 70 | 26 | 28 | 48 | 36 | 76 | 13 |
|  | 14 | 70 | 20 | 18 | 62 | 56 | 12 | 6 | 68 | 36 | 26 | 96 | 78 | ∞ | 0 | 66 | 58 | 82 | 66 | 8 | 14 |
|  | 15 | 98 | 76 | 58 | 0 | 0 | 52 | 40 | 22 | ∞ | 82 | 22 | 8 | 86 | 32 | 0 | 92 | 20 | 96 | 46 | 15 |
| 30,24,18 | 16 | 30 | 18 | 24 | 20 | ∞ | 30 | 48 | 48 | 18 | 1 | 42 | 36 | 2 | 46 | 42 | 0 | 40 | 16 | 58 | 16 |
|  | 17 | 94 | 60 | 50 | ∞ | 84 | 36 | 24 | 96 | 81 | 66 | 8 | 96 | 78 | 14 | 86 | 84 | 0 | 91 | 26 | 17 |
| 10,8,2 | 18 | 16 | 8 | 10 | 30 | 26 | 66 | 40 | 38 | 24 | 0 | 28 | 18 | 8 | 30 | ∞ | 2 | 24 | 0 | 48 | 18 |
| 2 | 19 | 62 | 30 | 28 | 66 | 60 | 40 | 2 | ∞ | 46 | 36 | 82 | 70 | 40 | 86 | 70 | 56 | 76 | 58 | 0 | 19 |
|  |  | 1 | 2 | 3 | 4 | 5 | 6 | 7 | 8 | 9 | 10 | 11 | 12 | 13 | 14 | 15 | 16 | 17 | 18 | 19 |  |

$s_{13}$ = (16⁻³⁰ 6³⁰ 9²): 2 . Our new arcs are (16 10), (6 15), (9 1).

$H_{13}$ = ($1^5$ $3^{11}$ $11^1$ $6^{31}$ $15^{11}$ $18^{13}$ $2^{13}$ $12^3$ $4^{19}$ $17^3$ $5^{21}$ $16^1$ $10^3$ $13^{11}$ $14^1$ $7^1$ $8^7$ $19^{11}$ $9^{11}$): 177

$$H_{13}^{-1}M$$

| II | | 3 | 12 | 11 | 17 | 16 | 15 | 8 | 19 | 1 | 13 | 6 | 4 | 14 | 7 | 18 | 10 | 5 | 2 | 9 | |
|---|---|---|---|---|---|---|---|---|---|---|---|---|---|---|---|---|---|---|---|---|---|
| ------- | | 1 | 2 | 3 | 4 | 5 | 6 | 7 | 8 | 9 | 10 | 11 | 12 | 13 | 14 | 15 | 16 | 17 | 18 | 19 | |
| | 1 | 0 | 32 | 26 | 54 | 48 | 42 | 14 | 62 | ∞ | 36 | 8 | 2 | 38 | 12 | 58 | 24 | 6 | 2 | 18 | 1 |
| 12,10,6 | 2 | 60 | 0 | 4 | 36 | 26 | 24 | 12 | 44 | 58 | 14 | 76 | 66 | 18 | 84 | 38 | 6 | 70 | ∞ | 10 | 2 |
| | 3 | ∞ | 12 | 0 | 30 | 28 | 22 | 82 | 46 | 52 | 16 | 70 | 62 | 18 | 80 | 0 | 88 | 68 | 56 | 86 | 3 |
| 12,8,6 | 4 | 52 | 2 | 6 | 0 | 30 | 30 | 74 | 12 | 44 | 12 | 62 | ∞ | 18 | 64 | 8 | 8 | 58 | 46 | 12 | 4 |
| 12 | 5 | 22 | 70 | 66 | 2 | 0 | 12 | 42 | 26 | 12 | 78 | 36 | 28 | 0 | 40 | 48 | 58 | ∞ | 18 | 46 | 5 |
| 30,24,18 | 6 | 38 | 18 | 24 | 12 | 10 | 0 | 62 | 22 | 26 | 4 | ∞ | 40 | 12 | 56 | 16 | 30 | 46 | 32 | 68 | 6 |
| | 7 | 72 | 32 | 26 | 60 | 56 | 50 | 0 | 72 | 58 | 38 | 90 | 76 | 42 | ∞ | 68 | 18 | 82 | 60 | 8 | 7 |
| | 8 | 2 | 50 | 46 | 76 | 74 | 70 | ∞ | 0 | 80 | 52 | 16 | 4 | 86 | 20 | 80 | 76 | 14 | 90 | 22 | 8 |
| 2 | 9 | 18 | 72 | 66 | 8 | 2 | 2 | 56 | 20 | 0 | 80 | 36 | 26 | 82 | 42 | 6 | 2 | 30 | 8 | ∞ | 9 |
| | 10 | 36 | 94 | 90 | 26 | 20 | 14 | 84 | 50 | 30 | 0 | 60 | 48 | 6 | 66 | 48 | ∞ | 58 | 36 | 88 | 10 |
| | 11 | 80 | 30 | ∞ | 78 | 76 | 58 | 16 | 88 | 62 | 36 | 0 | 86 | 48 | 12 | 82 | 28 | 92 | 70 | 22 | 11 |
| | 12 | 2 | ∞ | 70 | 24 | 20 | 6 | 24 | 30 | 88 | 78 | 16 | 0 | 96 | 18 | 28 | 64 | 4 | 94 | 58 | 12 |
| | 13 | 40 | 88 | 82 | 32 | 20 | 12 | 72 | 42 | 28 | ∞ | 60 | 42 | 0 | 70 | 26 | 78 | 48 | 36 | 76 | 13 |
| | 14 | 70 | 20 | 18 | 62 | 56 | 36 | 6 | 68 | 58 | 26 | 96 | 78 | ∞ | 0 | 66 | 12 | 82 | 66 | 8 | 14 |
| | 15 | 98 | 76 | 58 | 0 | 0 | ∞ | 40 | 22 | 92 | 82 | 22 | 8 | 86 | 32 | 0 | 52 | 20 | 96 | 46 | 15 |
| | 16 | 60 | 12 | 6 | 50 | ∞ | 48 | 78 | 78 | 30 | 31 | 72 | 66 | 32 | 76 | 72 | 0 | 70 | 46 | 88 | 16 |
| | 17 | 94 | 60 | 50 | ∞ | 84 | 81 | 24 | 96 | 84 | 66 | 8 | 96 | 78 | 14 | 86 | 36 | 0 | 91 | 26 | 17 |
| 10,8,2 | 18 | 16 | 8 | 10 | 30 | 26 | 24 | 40 | 38 | 2 | 0 | 28 | 18 | 8 | 30 | ∞ | 66 | 24 | 0 | 48 | 18 |
| 2 | 19 | 62 | 30 | 28 | 66 | 60 | 46 | 2 | ∞ | 56 | 36 | 82 | 70 | 40 | 86 | 70 | 40 | 76 | 58 | 0 | 19 |
| | | 1 | 2 | 3 | 4 | 5 | 6 | 7 | 8 | 9 | 10 | 11 | 12 | 13 | 14 | 15 | 16 | 17 | 18 | 19 | |

We obtain $s_{14}$ = ($6^{-30}$ $16^6$ $3^{22}$): -2. Thus, the sum of values for this chain is 2 +2 + (-2) = 4. Therefore, we can go no further in our algorithm. Our best approximation to an optimal 19-cycle is $H_{12}$ with a value of 173.

**COMMENT.** There are two possible weaknesses in this algorithm. (1) Unlike mpst algorithms, a new iteration may have a larger value than the previous one. (2) As we approach an optimal tour or n-cyle, the number of small-valued arcs remaining in the matrix becomes smaller.

# Chapter 6

# P VERSUS NP

In this chapter, we discuss the relationship of the FWK algorithm to the question of whether $P \neq NP$.

Definition 1. An algorithm, $A$, can be applied in polynomial time to a problem, $P(n)$, such that for all positive values of $n$ there exists a polynomial $f(n)$ such that the running time of $P(n)$ is less than $f(n)$.

Definition 2. Let $S = \{ all\ n\ X\ n\ random\ cost\ matrices\ M_{in} | i = 1,2,3,...; n = 2,3,4,... \}$.

Definition 3. An algorithm, $EX_i$, for obtaining a solution to the traveling salesman problem is said to be exact if it always obtains an optimal tour ($n$-cycle) for any $n\ X\ n$ cost matrix to which it is applied.

Defintion 4. $EX = \{ all\ exact\ algorithms\ EX_i | i = 1,2,3,4,... \}$.

Definition 5. The running time of an algorithm $A$ is denoted by $r.t.(A)$.

Defintoin 6. $HK$ is the Held-Karp lower bound for an optimal tour (n-cycle).

Defintion 7. $\sigma_{FWKOPT}$ denotes a relative optimal tour ($n$-cycle) obtained by applying a heuristic algorithm to an $n\ X\ n$ cost matrix.

Conjecture C. For i = 1,2,3, ..., there exists a fixed polynomial $p(n)$ such that $r.t.(FWK) \leq r.t.(EX_i)p(n)$.

Theorem 5.1 Assume that conjecture C is correct. Given an arbitrary element of $S$, when $FWK$ is applied to an arbitrary element of $S$, $M_{in}$, $(1)\ |\sigma_{FWKOPT}(M_{in})| = HK$, and/or
$(2)\ r.t.FWK(M_{in}) \leq p(n)\ for\ i = 1,2,3,...; n = 2,3,4,..$, or $(3)\ P \neq NP$.

*Comment.* By constructing a large number of diverse examples, we may be able to get an idea as to whether polynomial running times keep increasing in degree or whether the degrees have an upper bound that isn't exceeded no matter how large $n$ becomes. If the latter doesn't occur, that would seem to imply that $P \neq NP$. To be more precise, let $n^{i(n)}$ be the maximum number of subpaths we obtain when applying the FWK algorithm to each element of a large subset of $n\ X\ n$ cost matrices in $S$. Then if for $n = 2,3,4,...$, $i(n)$ yields a strictly monotonically increasing sequence which doesn't appear to have a fixed upper bound, this would infer that $P \neq NP$.